\documentclass[journal]{IEEEtran}
\usepackage{amssymb}
\usepackage{booktabs}
\usepackage{here}

\usepackage{cite}

\ifCLASSINFOpdf
  \usepackage[pdftex]{graphicx}
\else
  \usepackage[dvips]{graphicx}
\fi
\usepackage{amsmath}
\usepackage{rotating}
\usepackage{multirow}
\usepackage{url}
\usepackage{color}
\usepackage{diagbox}

\usepackage{algorithm, algpseudocode}

\global\long\def\Ours{RHUIDR}
\global\long\def\Ourss{RHUIDR }

\newcommand{\argmin}{\mathop{\rm argmin}\limits}

\global\long\def\ElemMatGeneral{X}
\global\long\def\MatGeneral{\mathbf{\ElemMatGeneral}}

\global\long\def\SymbTrue#1{\bar{#1}}
\global\long\def\SymbEst#1{\hat{#1}}
\global\long\def\Valbest#1{\textbf{#1}}
\global\long\def\ValSecnd#1{\underline{#1}}

\global\long\def\ValSinguMax#1{\sigma_{1}(#1)}

\global\long\def\DiffHSIReg{\mathbf{K}}

\global\long\def\MatEndmember{\mathbf{E}}

\global\long\def\ElemMatAbun{A}

\global\long\def\MatAbun{\mathbf{\ElemMatAbun}}
\global\long\def\MatAbunTrue{\SymbTrue{\MatAbun}}
\global\long\def\MatAbunEst{\SymbEst{\MatAbun}}

\global\long\def\ElemVecAbun{a}
\global\long\def\VecAbun{\mathbf{\ElemVecAbun}}
\global\long\def\VecAbunEst{\SymbEst{\VecAbun}}

\global\long\def\ElemMatHSIObsev{V}
\global\long\def\MatHSIObsev{\mathbf{\ElemMatHSIObsev}}

\global\long\def\MatSpar{\mathbf{S}}
\global\long\def\MatSparTrue{\SymbTrue{\MatSpar}}

\global\long\def\MatStripe{\mathbf{L}}
\global\long\def\MatStripeTrue{\SymbTrue{\MatStripe}}

\global\long\def\MatZero{\mathbf{O}}

\global\long\def\OpeDiffSymb{\mathbf{D}}
\global\long\def\OpeDiff#1{\OpeDiffSymb(#1)}
\global\long\def\OpeDiffAdj#1{\OpeDiffSymb^{*}(#1)}
\global\long\def\OpeDiffvSymb{\OpeDiffSymb_{v}}
\global\long\def\OpeDiffv#1{\OpeDiffvSymb(#1)}
\global\long\def\OpeDiffvAdj#1{\OpeDiffvSymb^{*}(#1)}
\global\long\def\OpeDiffhSymb{\OpeDiffSymb_{h}}
\global\long\def\OpeDiffh#1{\OpeDiffhSymb(#1)}

\global\long\def\OpeDiffbSymb{\OpeDiffSymb_{b}}
\global\long\def\OpeDiffb#1{\OpeDiffbSymb(#1)}

\global\long\def\CubeDiffb{\mathcal{D}_{b}}
\global\long\def\CubeDiffv{\mathcal{D}_{v}}
\global\long\def\CubeDiffh{\mathcal{D}_{h}}

\global\long\def\ParamHSSTV{\omega}
\global\long\def\OpeDiffHSSTVSymb{\mathbf{C}}
\global\long\def\OpeDiffHSSTV#1{\OpeDiffHSSTVSymb_{\ParamHSSTV}(#1)}

\global\long\def\ElemMatHSIGeneral{H}
\global\long\def\ElemMatHSIGeneralTrue{\SymbTrue{\ElemMatHSIGeneral}}
\global\long\def\ElemMatHSIGeneralEst{\SymbEst{\ElemMatHSIGeneral}}
\global\long\def\MatHSIGeneral{\mathbf{\ElemMatHSIGeneral}}
\global\long\def\MatHSIGeneralTrue{\SymbTrue{\MatHSIGeneral}}
\global\long\def\MatHSIGeneralEst{\SymbEst{\MatHSIGeneral}}
\global\long\def\CubeHSIGeneral{\mathcal{\ElemMatHSIGeneral}}

\global\long\def\NumPixel{n}
\global\long\def\NumEndmember{m}
\global\long\def\NumBand{l}
\global\long\def\NumPixelVer{n_{1}}
\global\long\def\NumPixelHor{n_{2}}

\global\long\def\SetRealNum{\mathbb{R}}
\global\long\def\SetNNRealNum{\SetRealNum_{+}}

\global\long\def\SetNNOrthantAbun{\SetNNRealNum^{\NumEndmember \times \NumPixel}}

\global\long\def\SetElemZero{\{\MatZero\}}

\global\long\def\ParamFidel{\varepsilon}
\global\long\def\ParamRateFidel{\alpha_{\sigma}}
\global\long\def\SetConsFidel{\mathcal{B}_{F, \ParamFidel}^{\MatHSIObsev}}

\global\long\def\ParamConsSpar{\eta}
\global\long\def\ParamRateConsSpar{\alpha_{\eta}}
\global\long\def\SetConsSpar{\mathcal{B}_{1, \ParamConsSpar}}

\global\long\def\FuncRegRec{\mathcal{R}}

\global\long\def\FuncIndSymb#1{\iota_{#1}}
\global\long\def\FuncInd#1#2{\FuncIndSymb{#1}(#2)}

\global\long\def\NumIter{t}

\makeatletter
\def\bstctlcite{\@ifnextchar[{\@bstctlcite}{\@bstctlcite[@auxout]}}
\def\@bstctlcite[#1]#2{\@bsphack
	\@for\@citeb:=#2\do{%
		\edef\@citeb{\expandafter\@firstofone\@citeb}%
		\if@filesw\immediate\write\csname #1\endcsname{\string\citation{\@citeb}}\fi}%
	\@esphack}
\makeatother

\begin{document}
\bstctlcite{IEEEexample:BSTcontrol}
%
\title{Towards Robust Hyperspectral Unmixing: Mixed Noise Modeling and Image-Domain Regularization}
%
%
%

\author{Kazuki~Naganuma,~\IEEEmembership{Student~Member,~IEEE,}
        Shunsuke~Ono,~\IEEEmembership{Member,~IEEE,}
\thanks{Manuscript received XXX, XXX; revised XXX XXX, XXX.}%
\thanks{K. Naganuma is with the Department of Computer Science, Tokyo Institute of Technology, Yokohama, 226-8503, Japan (e-mail: naganuma.k.aa@m.titech.ac.jp).}
\thanks{S. Ono is with the Department of Computer Science, Tokyo Institute of Technology, Yokohama, 226-8503, Japan (e-mail: ono@c.titech.ac.jp).}
\thanks{This work was supported in part by JST ACT-X Grant Number JPMJAX23CJ, JST PRESTO under Grant JPMJPR21C4, and JST AdCORP under Grant JPMJKB2307, in part by JSPS KAKENHI under Grant 22H03610, 22H00512, and 23H01415, and in part by Grant-in-Aid for JSPS Fellows under Grant 23KJ0912.}}

%
%

\markboth{Journal of \LaTeX\ Class Files,~Vol.~XX, No.~X, August~20XX}%
{Shell \MakeLowercase{\textit{et al.}}: Bare Demo of IEEEtran.cls for IEEE Journals}
%



\maketitle

\begin{abstract}
Hyperspectral (HS) unmixing is the process of decomposing an HS image into material-specific spectra (endmembers) and their spatial distributions (abundance maps). Existing unmixing methods have two limitations with respect to noise robustness. First, if the input HS image is highly noisy, even if the balance between sparse and piecewise-smooth regularizations for abundance maps is carefully adjusted, noise may remain in the estimated abundance maps or undesirable artifacts may appear. Second, existing methods do not explicitly account for the effects of stripe noise, which is common in HS measurements, in their formulations, resulting in significant degradation of unmixing performance when such noise is present in the input HS image. To overcome these limitations, we propose a new robust hyperspectral unmixing method based on constrained convex optimization. Our method employs, in addition to the two regularizations for the abundance maps, regularizations for the HS image reconstructed by mixing the estimated abundance maps and endmembers. This strategy makes the unmixing process much more robust in highly-noisy scenarios, under the assumption that the abundance maps used to reconstruct the HS image with desirable spatio-spectral structure are also expected to have desirable properties. Furthermore, our method is designed to accommodate a wider variety of noise including stripe noise. To solve the formulated optimization problem, we develop an efficient algorithm based on a preconditioned primal-dual splitting method, which can automatically determine appropriate stepsizes based on the problem structure. Experiments on synthetic and real HS images demonstrate the advantages of our method over existing methods.
\end{abstract}


%
\IEEEpeerreviewmaketitle

\section{Introduction}
\IEEEPARstart{H}{yperspectral (HS)} images are three-dimensional cube data consisting of two-dimensional spatial and one-dimensional spectral information.
Compared to grayscale or RGB images, HS images provide more than several hundred bands, each of which contains specific unique wavelength characteristics of materials such as minerals, soils, and liquids. 
Therefore, HS images have various applications, such as ecology, mineralogy, biotechnology, and agriculture \cite{ghamisi2017advances,chang2003hyperspectral,thenkabail2016hyperspectral,lu2020recent}.
Due to the trade-off between spatial resolution and spectral resolution, HS sensors do not have a sufficient spatial resolution, resulting in containing multiple components (called endmembers) in a pixel~\cite{keshava2002spectral}, which is referred to as a mixel. The process of decomposing the mixel into endmembers and their abundance maps is called unmixing.
Unmixing has been actively studied in the remote sensing field because it is essential for HS image analysis\cite{bioucas2012hyperspectral,ma2013signal} and other applications, such as denoising \cite{rasti2018noise,zhao2014hyperspectral} and data fusion \cite{licciardi2009decision,yokoya2017hyperspectral}.

Unmixing methods fall into two categories according to their assumptions: non-blind~\cite{iordache2011sparse,iordache2014collaborative,li2020superpixel,ren2021nonconvex,shen2022superpixel,superpixed_YLiang_2023,iordache2012total,aggarwal2016hyperspectral,wang2017hyperspectral,zhang2018spectral,wang2019row,Li_robustUnmixing_2021,Ince_doubleSpatial_2022,rs_Deng_RobustDual_2023,PnPUnm_ZhaoM_2022,SUnCNN_RastiB_2022,UnDIP_RastiB_2022,huang2018joint,BiJSpLRU_JHuang_2021,MDLRR_WuLing_2023} and blind unmixing~\cite{NMF_TV_FengXR_2018,SS_MNF_LiHC_2022,NMF_Review_FengXR_2022,ozkan2018endnet,Rasti_2022_MiSiCNet}. Non-blind unmixing methods estimate abundance maps from a given endmember library. Endmembers in the library are potentially much larger in number than endmembers included in real HS images, i.e., its corresponding abundance maps become sparse. 
On the other hand, blind unmixing methods simultaneously estimate an endmember library and abundance maps, allowing us to obtain the abundances of endmembers whose spectral libraries are unknown.

For blind unmixing, nonnegative matrix factorization-based approaches~\cite{NMF_TV_FengXR_2018,SS_MNF_LiHC_2022,NMF_Review_FengXR_2022} and learning-based approaches~\cite{ozkan2018endnet,Rasti_2022_MiSiCNet,hong2022endmember} have attracted attention.
Nonnegative matrix factorization-based methods design and solve an optimization problem that incorporates the functions of the product of an endmember matrix and an abundance map matrix. 
When solving the optimization problem, they take an approach that iterates alternate updates of the two matrices: updating the endmember library matrix by solving the subproblem with the abundance map matrix fixed, updating the abundance maps by solving the subproblem with the endmember library matrix fixed using some non-blind unmixing method.
Learning-based methods often involve the following steps: extraction of initial endmembers from an input HS image, estimation of corresponding initial abundance maps by some non-blind unmixing methods, and then learning of sophisticated unmixing and reconstruction networks based on this information. 
Therefore, non-blind unmixing is a fundamental task that must precede blind unmixing. Henceforth, non-blind unmixing will simply be referred to as unmixing.

Although very accurate unmixing can be achieved using state-of-the-art methods if a noise-free HS image is available, real-world HS images are often contaminated by various types of noise such as Gaussian noise, outliers, missing values, and stripe noise due to environmental factors and sensor failures. Such noise obviously has a negative impact on unmixing performance and needs to be dealt with appropriately. The simplest way is a two-step approach, where noise is first removed from a given HS image beforehand, followed by unmixing. However, such methods are also likely to remove even important spectral information. It is therefore essential to develop a method that can \textit{simultaneously} separate noise (without affecting spectral information) during the unmixing process, which we refer to as \textit{noise-robust} unmixing.

Many noise-robust unmixing techniques explicitly model noises and then take the approach of solving optimization problems that incorporate functions characterizing abundance maps.
Based on the fact that HS images consist of a small fraction of the endmembers in a library, the methods in~\cite{iordache2011sparse,iordache2014collaborative,li2020superpixel,ren2021nonconvex,shen2022superpixel,superpixed_YLiang_2023} employ a sparse regularization. Abundance maps are also piecewise smooth because neighboring pixels often have the same endmembers. To capture the nature, the methods in~\cite{iordache2012total,aggarwal2016hyperspectral,wang2017hyperspectral,zhang2018spectral,wang2019row} adopt a combination of sparse and piecewise-smooth regularizations. 
As a more advanced approach to promote the sparsity of abundance maps, some methods adopt a coarse abundance map-based weighted sparse 	unmixing approach~\cite{Li_robustUnmixing_2021,Ince_doubleSpatial_2022,rs_Deng_RobustDual_2023}, which includes the following three steps. 
First, this approach segments a target HS image into superpixel blocks and averages the pixels of the superpixel blocks to obtain a coarsely denoised HS image. Then, by applying a sparse unmixing method (e.g., the method in~\cite{iordache2011sparse}) to the coarse HS image, coarse abundance maps are generated. Finally, based on the coarse abundance maps, superpixel-wise and endmember-wise weights are computed to promote the weighted sparsity of abundance maps.
In addition, the methods in~\cite{PnPUnm_ZhaoM_2022,SUnCNN_RastiB_2022,UnDIP_RastiB_2022} estimate abundance maps using a regularization based on deep neural networks, and the methods in~\cite{huang2018joint,BiJSpLRU_JHuang_2021,MDLRR_WuLing_2023} adopt a combination of sparse and low-rank regularizations.

\begin{figure}[t]
	\centering
	\begin{minipage}{0.85\hsize}
		\centerline{\includegraphics[width = \hsize]{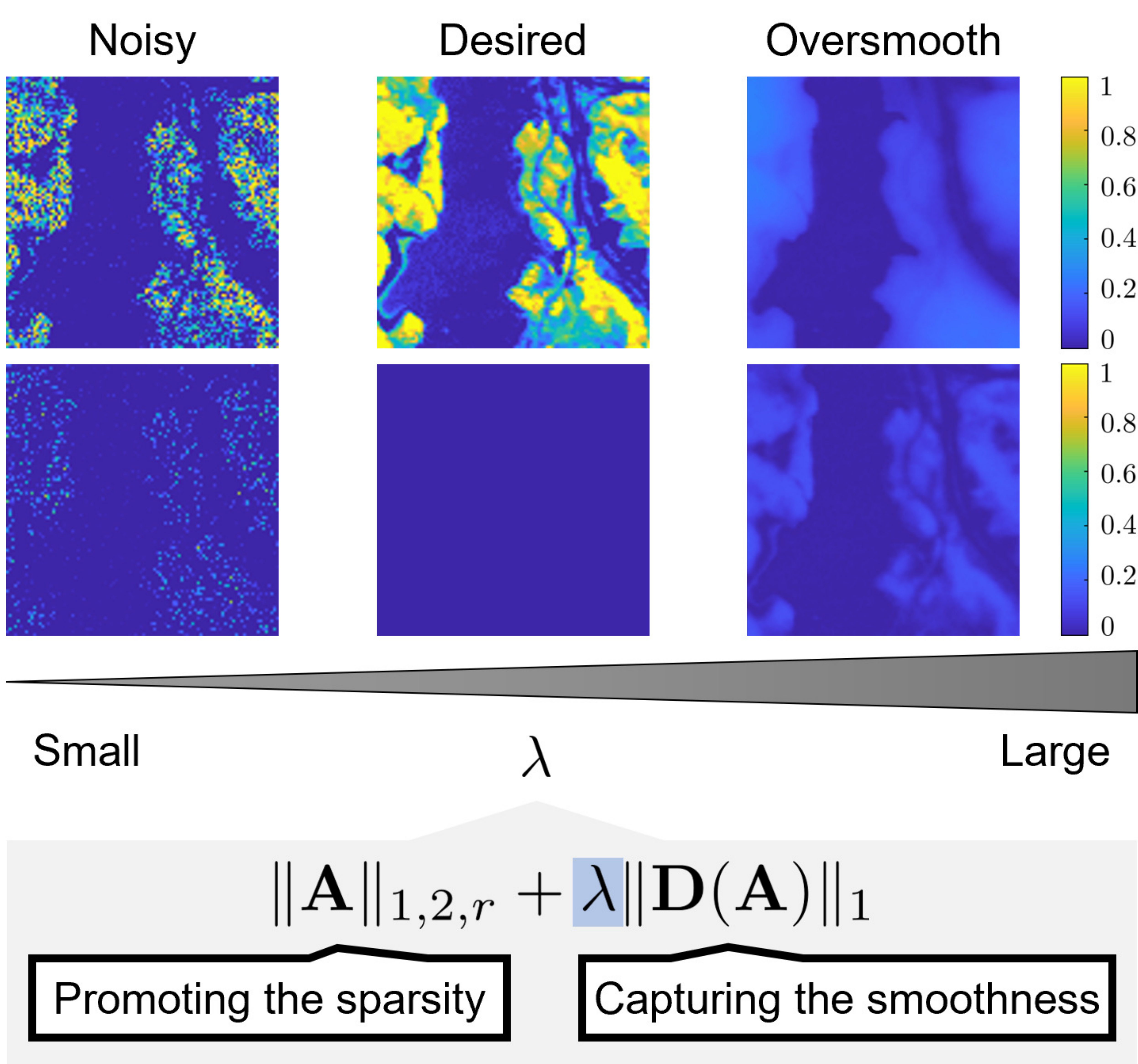}}
	\end{minipage}
	
	\caption{A Difficulty in dealing with high-level noise in unmixing.}
	\label{fig:intro}
\end{figure}

As we have discussed, various studies have been carried out to mitigate the effects of noise in unmixing, but there are still two limitations in terms of robustness to noise. The first is that the performance of unmixing is severely degraded when the input HS image is contaminated with high levels of noise. The second is that existing unmixing methods cannot adequately deal with stripe noise.

As reviewed in the previous subsection, many existing unmixing methods use a combination of sparse and piecewise-smooth regularization to characterize the abundance maps. 
However, as shown in Fig.~\ref{fig:intro}, balancing these regularizations becomes very difficult when unmixing HS images contaminated with high levels of noise. 
In fact, if the weight of the sparse regularization is increased, a large amount of noise remains in the estimated abundance maps. 
Conversely, if the weight of the piecewise-smooth regularization is increased, the estimated abundance maps will contain many inappropriate components that are not present in the original HS image. 
In existing methods, adjusting the weights to avoid both problems is a very sensitive and tedious task. 

To resolve this difficulty, we focus on the regularizations for the HS image reconstructed by mixing the estimated abundance maps and the endmembers, which we call image-domain regularizations, in addition to the regularizations for the abundance maps. 
Our assumption is the following: if the reconstructed HS image has desirable properties in its spatio-spectral structure, then the estimated abundance maps used for reconstruction should also have desirable properties. 
Therefore, we believe that incorporating spatio-spectral regularization for HS images into the unmixing formulation can improve the unmixing performance in high-noise situations where abundance maps are difficult to estimate using existing methods.
Fortunately, in the context of HS image restoration, many effective HS image regularizations have been studied~\cite{yuan2012hyperspectral,zhang2013hyperspectral,he2015total,aggarwal2016sstv,takeyama2020constrained,l0l1_Wang_2021,GSSTV_Takemoto2022}. 
By adopting them as image-domain regularizations, we can robustify the unmixing process under highly-noisy scenarios.

Regarding the second limitation in dealing with stripe noise, existing unmixing methods mainly deal with Gaussian noise and sparse noise.
However, in addition to these noises, actual HS images are often contaminated with stripe noise, mainly due to external disturbances and calibration errors~\cite{Carfantan_destriping_2010,Liu_destriping_2018,FC_naganuma_2022}. 
Since stripe noise is not Gaussian and is often not sparse~\cite{Kind_of_Stripes}, it cannot be handled by existing methods, leading to performance degradation in unmixing.

Based on the above discussion, this paper proposes robust hyperspectral unmixing using image-domain regularization (\Ours).
We formulate the unmixing problem as a constrained convex optimization problem.
In order to solve the optimization problem, we develop an efficient algorithm based on the preconditioned primal-dual splitting method (P-PDS)~\cite{pock2011diagonal} with an operator-norm-based stepsize selection method~\cite{Naganuma_PDS_TSP_2023}.
In terms of the features of \Ours, the contributions of the paper can be summarized as follows.

\begin{enumerate}
	\item \textit{(Robust to high levels of noise):} \Ourss employs not only the abundance map regularizations but also image-domain regularizations, which robustify the unmixing process under highly-noisy scenarios. 
	\item \textit{(Robust to mixed noise including stripe noise):} By explicitly modeling three types of noise (Gaussian noise, sparse noise, and stripe noise) as in~\eqref{hsi_model}, \Ourss can adequately handle mixed noise, including stripe noise, which is difficult to handle in existing methods.
	\item \textit{(Easy to adjust hyperparameters):} In the formulated optimization problem, we model data-fidelity and noise terms as hard constraints instead of adding them to the objective function. This type of constrained formulation decouples interdependent hyperparameters into independent ones, thus facilitating parameter settings, which will be detailed in Sec.~\ref{ssec:problem_formulation}.
	\item \textit{(Avoiding adjusting stepsizes):} Unlike the optimization algorithms used in existing unmixing methods, our P-PDS-based algorithm can automatically determine the appropriate stepsizes based on the problem structure.
\end{enumerate}
Experiments on synthetic and real HS images demonstrate the advantages of \Ourss over existing methods.

The paper is organized as follows.
In Section~\ref{sec:preliminaries}, we introduce mathematical tools.
In Section~\ref{sec:proposed method}, we explain the proposed method, \Ours, with its formulation and algorithm.
In Section~\ref{experience}, we conduct experiments to show the superiority of \Ourss over the existing methods.
Finally, we conclude the paper in Section~\ref{sec:conclusion}.

\section{Preliminaries}
\label{sec:preliminaries}

\subsection{Notations}
In this paper, 
we denote the sets of real numbers and non-negative real numbers as $\SetRealNum$ and $\SetNNRealNum$, respectively. Matrices are denoted by capitalized boldface letters (e.g., $\MatGeneral$), and the element at the $i$th row and $j$-th column of matrix $\MatGeneral$ is denoted by $\ElemMatGeneral_{i,j}$ or $[\MatGeneral]_{i,j}$.
An HS image with the number of bands $l$ and spatial size $n_1 \times n_2$ is treated as a matrix $\MatHSIGeneral \in \SetRealNum^{l\times n_1 n_2}$ of size $l\times n_1 n_2$ and $[\CubeHSIGeneral]_{i,j,k}$ indicates the $(i, j, k)$-th value of the cube data $\CubeHSIGeneral \in \SetRealNum^{n_1 \times n_2 \times l}$ corresponding to $\MatHSIGeneral$.
The $\ell_1$-norm $\|\cdot\|_1$, the Frobenius norm $\|\cdot\|_F$, the mixed $\ell_{1,2}$-norm grouped by row $\|\cdot\|_{1,2,r}$, and the mixed $\ell_{1,2}$-norm grouped by column $\|\cdot\|_{1,2,c}$ are defined by $\|\MatGeneral\|_1 = \sum_{i,j}|\ElemMatGeneral_{i,j}|$, $\|\MatGeneral\|_F = \sqrt{\sum_{i,j}\ElemMatGeneral_{i,j}^2}$, $\|\MatGeneral\|_{1,2,r}=\sum_i\sqrt{\sum_{j}\ElemMatGeneral_{i, j}^{2}}$, and $\|\MatGeneral\|_{1,2,c}=\sum_{j}\sqrt{\sum_{i}\ElemMatGeneral_{i, j}^{2}}$, respectively.
Let $\mathbf{G} : \SetRealNum^{M_1\times N_1} \rightarrow \SetRealNum^{M_2\times N_2}$ be a linear operator. A linear operator $\mathbf{G}^{*} : \SetRealNum^{M_2\times N_2} \rightarrow \SetRealNum^{M_1\times N_1}$ is called the adjoint operator of $\mathbf{G}$ if it is satisfied with   $\langle\mathbf{G}(\MatGeneral),\mathbf{Y}\rangle=\langle\MatGeneral,\mathbf{G}^{*}(\mathbf{Y})\rangle$ for any $\MatGeneral \in \SetRealNum^{M_1\times N_1}$ and  $\mathbf{Y}\in\SetRealNum^{M_2\times N_2}$.

\subsection{Preconditioned Primal-Dual Splitting Method (P-PDS)}
\label{sec:formula_pds}
Let $f_{1}, \ldots, f_{N}, g_{1}, \ldots, g_{M}$ be proximable\footnote{If an efficient computation of the proximity operator of $f$ is available, we call $f$ proximable.} proper lower-semicontinuous convex functions.
Consider a convex optimization problem of the following form:
\begin{align}
	\min_{\substack{\mathbf{Y}_1,\ldots,\mathbf{Y}_{N},\\ \mathbf{Z}_1,...,\mathbf{Z}_M}} \: 
	& \sum_{i=1}^{N} f_{i}(\mathbf{Y}_{i})
	+ \sum_{j=1}^{M} g_{j}(\mathbf{Z}_{j}), \nonumber \\
	\mathrm{s.t.} \: &
	\begin{cases}
		\mathbf{Z}_i=\sum_{i=1}^N \mathbf{G}_{1,i}(\mathbf{Y}_i),\\
		\vdots,\\
		\mathbf{Z}_M=\sum_{i=1}^N \mathbf{G}_{M,i}(\mathbf{Y}_i),
	\end{cases}
\label{eq:pds}
\end{align}
where $\mathbf{G}_{j,i}$ $(i=1,\ldots,N,j=1,\ldots,M)$ are linear operators.
We define the \textit{proximity operator} of $f_{i}$ (and $g_{j}$ as well) with a parameter $\gamma > 0$ by 
\begin{equation}
	\label{eq:prox}
	\mathrm{prox}_{\gamma f_{i}}(\mathbf{X}) := \argmin_{\mathbf{Y} \in \SetRealNum^{M\times N}} f_{i}(\mathbf{Y})+\frac{1}{2\gamma}\|\mathbf{X-Y}\|_{F}^2.
\end{equation}
Then, the preconditioned primal-dual splitting method (P-PDS) solves Prob. (\ref{eq:pds}) by the following iterative procedures:
\begin{equation}
	\label{eq:gene_P_PDS}
	\left\lfloor
	\begin{array}{ll}
		\widetilde{\mathbf{Y}}_{1} & \leftarrow \mathbf{Y}^{(\NumIter)}_{1}-\gamma_{1,1}\sum_{j=1}^{M} \mathbf{G}_{j,1}^*(\mathbf{Z}^{(\NumIter)}_j),\\
		\mathbf{Y}^{(\NumIter+1)}_{1} & \leftarrow \mathrm{prox}_{\gamma_{1,1}f_{1}}(\widetilde{\mathbf{Y}}_{1}),\\
		& \vdots \\
		\widetilde{\mathbf{Y}}_{N} & \leftarrow \mathbf{Y}^{(\NumIter)}_{N}-\gamma_{1,N}\sum_{j=1}^{M} \mathbf{G}_{j,N}^*(\mathbf{Z}^{(\NumIter)}_{j}),\\
		\mathbf{Y}^{(\NumIter+1)}_{N} & \leftarrow \mathrm{prox}_{\gamma_{1,N}f_{N}}(\widetilde{\mathbf{Y}}_{N}),\\
		\widetilde{\mathbf{Z}}_{1} & \leftarrow  \mathbf{Z}^{(\NumIter)}_{1}+\gamma_{2,1}\sum_{i=1}^{N}\mathbf{G}_{1,i}(2\mathbf{Y}^{(\NumIter+1)}_{i}-\mathbf{Y}^{(\NumIter)}_{i}),\\
		\mathbf{Z}^{(\NumIter+1)}_{1} & \leftarrow \widetilde{\mathbf{Z}}_{1} - \gamma_{2,1}\mathrm{prox}_{\frac{1}{\gamma_{2,1}}g_{1}}(\tfrac{1}{\gamma_{2,1}}\widetilde{\mathbf{Z}}_{1}),\\
		& \vdots \\
		\widetilde{\mathbf{Z}}_{M} & \leftarrow  \mathbf{Z}^{(\NumIter)}_{M}+\gamma_{2,M}\sum_{i=1}^{N}\mathbf{G}_{M,i}(2\mathbf{Y}^{(\NumIter+1)}_{i}-\mathbf{Y}^{(\NumIter)}_{i}), \\
		\mathbf{Z}^{(\NumIter+1)}_{M} & \leftarrow \mathbf{Z}^{(\NumIter)}_{M} - \gamma_{2,M}\mathrm{prox}_{\frac{1}{\gamma_{2,M}}g_{M}}(\tfrac{1}{\gamma_{2,M}}\widetilde{\mathbf{Z}}_{M}),
	\end{array}
	\right.
\end{equation}
where $\gamma_{1,i}$ $(i=1,\cdots,N)$ and $\gamma_{2,j}$ $(j=1,\cdots,M)$ are the stepsize parameters.
The stepsize parameters can be determined automatically as follows~\cite{Naganuma_PDS_TSP_2023}:
\begin{equation}
	\label{eq:stepsize_exact}
	\gamma_{1,i}=\frac{1}{\sum_{j=1}^{M}\|\mathbf{G}_{j,i}\|_{\mathrm{op}}^{2}},~\gamma_{2,j}=\frac{1}{N},
\end{equation}
where $\|\mathbf{G}_{j,i}\|_{\mathrm{op}}^{2}$ is the operator norm\footnote{
	Let $\mathbf{G}:\SetRealNum^{n_{1}, m_{1}} \rightarrow \SetRealNum^{n_{2}, m_{2}}$ be a linear operator.
	Then, the operator norm of $\mathbf{G}$ is defined by $\|\mathbf{G}\|_{\mathrm{op}}:=\sup_{\mathbf{X}\neq\MatZero}\|\mathbf{G(X)}\|_F / \|\mathbf{X}\|_F.$
} of $\mathbf{G}_{j,i}$. 
However, the operator norms of some linear operators cannot be easily calculated (e.g., difference operators and composite operators of linear operators).
Therefore, we can use their upper bounds $\mu_{j,i} \in [\|\mathbf{G}_{j,i}\|_{\mathrm{op}}, \infty)$ to determine the stepsize parameters as
\begin{equation}
\label{eq:stepsize_relaxed}
\gamma_{1,i}=\frac{1}{\sum_{j=1}^{M}\mu_{j,i}^2},~\gamma_{2,j}=\frac{1}{N}.
\end{equation}
From~\cite[Theorem 1]{pock2011diagonal} and~\cite[Theorem III.2]{Naganuma_PDS_TSP_2023}, the sequences generated by P-PDS~\eqref{eq:gene_P_PDS} with the stepsizes in Eq.~\eqref{eq:stepsize_relaxed} are guaranteed to converge to a solution of Prob.~\eqref{eq:pds}.

\begin{figure*}[t]
	\centering
	\begin{minipage}[t]{0.90\hsize}
		\centerline{
			\includegraphics[width=\hsize]{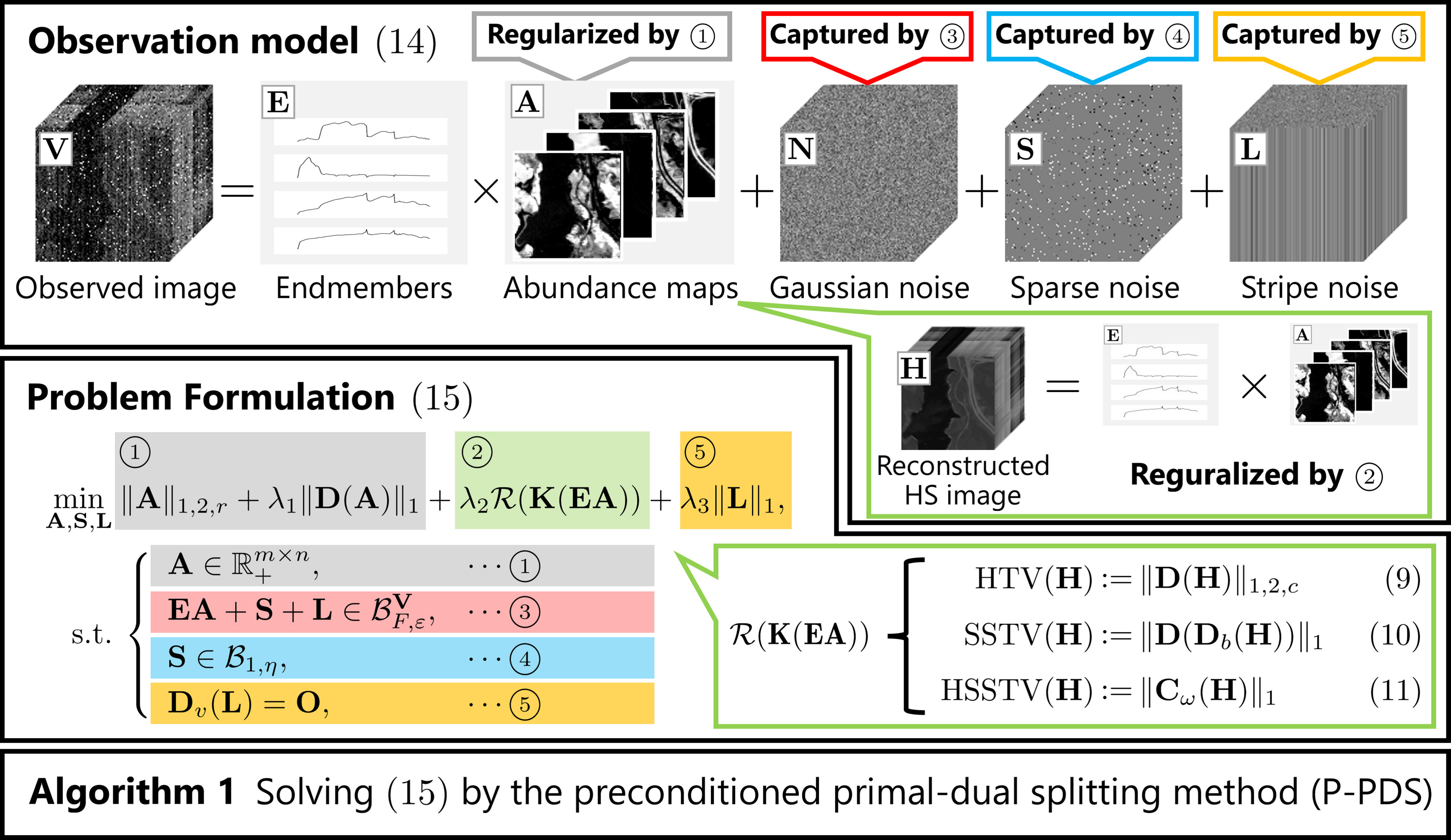}
		}
	\end{minipage}
	
	\vspace{-1mm}
	
	\caption{Illustration of the proposed method, i.e., \Ours.}
	\label{fig:Overall}
\end{figure*}

\subsection{Regularizations for an HS Image}
This section introduces the regularizations for an HS image $\MatHSIGeneral \in \SetRealNum^{\NumBand \times \NumPixelVer \NumPixelHor}$.
Let $\OpeDiffvSymb :\SetRealNum^{\NumBand \times \NumPixelVer \NumPixelHor} \rightarrow \SetRealNum^{\NumBand \times \NumPixelVer \NumPixelHor}$, $\OpeDiffhSymb :\SetRealNum^{\NumBand \times \NumPixelVer \NumPixelHor} \rightarrow \SetRealNum^{\NumBand \times \NumPixelVer \NumPixelHor}$, and $\OpeDiffbSymb: \SetRealNum^{\NumBand \times \NumPixelVer \NumPixelHor} \rightarrow \SetRealNum^{\NumBand \times \NumPixelVer \NumPixelHor}$ be respectively vertical, horizontal, and spectral difference operators, which are given by 
\begin{equation}
	[\CubeDiffv]_{i, j, k} := 
	\begin{cases}
		[\CubeHSIGeneral]_{i + 1, j, k} - [\CubeHSIGeneral]_{i, j, k}, & 
		\mathrm{if} \: i < \NumPixelVer; \\
		0, &
		\mathrm{otherwise},\\
	\end{cases}
\end{equation}
\begin{equation}
	[\CubeDiffh]_{i, j, k} := 
	\begin{cases}
		[\CubeHSIGeneral]_{i, j + 1, k} - [\CubeHSIGeneral]_{i, j, k}, & 
		\mathrm{if} \: j < \NumPixelHor; \\
		0, &
		\mathrm{otherwise},\\
	\end{cases}
\end{equation}
\begin{equation}
	[\CubeDiffb]_{i, j, k} := 
	\begin{cases}
		[\CubeHSIGeneral]_{i , j, k + 1} - [\CubeHSIGeneral]_{i, j, k}, & 
		\mathrm{if} \: k < \NumBand; \\
		0, &
		\mathrm{otherwise},\\
	\end{cases}
\end{equation}
where $\CubeHSIGeneral \in \SetRealNum^{\NumPixelVer \times \NumPixelHor \times \NumBand}$, $\CubeDiffv \in \SetRealNum^{\NumPixelVer \times \NumPixelHor \times \NumBand}$, $\CubeDiffh \in \SetRealNum^{\NumPixelVer \times \NumPixelHor \times \NumBand}$, and $\CubeDiffb \in \SetRealNum^{\NumPixelVer \times \NumPixelHor \times \NumBand}$ are the three-dimensional data corresponding to $\MatHSIGeneral$,  $\OpeDiffv{\MatHSIGeneral}$,  $\OpeDiffh{\MatHSIGeneral}$, and $\OpeDiffb{\MatHSIGeneral}$, respectively.
Then, HTV~\cite{yuan2012hyperspectral}, 
SSTV~\cite{aggarwal2016sstv}, and HSSTV~\cite{takeyama2020constrained} are defined by
\begin{align}
	\text{HTV}(\MatHSIGeneral) & := \|\OpeDiff{\MatHSIGeneral}\|_{1,2,c}, \label{eq:HTV} \\
	\text{SSTV}(\MatHSIGeneral) & := \|\OpeDiff{\OpeDiffb{\MatHSIGeneral}}\|_{1}, \label{eq:SSTV} \\
	\text{HSSTV}(\MatHSIGeneral) & :=
	\|\OpeDiffHSSTV{\MatHSIGeneral}\|_{1}, \label{eq:HSSTV} 
\end{align}
where $\OpeDiffSymb$ is the spatial difference operator:
\begin{equation}
	\OpeDiff{\MatHSIGeneral} := 
	\begin{bmatrix} 
		\OpeDiffv{\MatHSIGeneral} \\ 
		\OpeDiffh{\MatHSIGeneral}
	\end{bmatrix},
\end{equation}
and $\OpeDiffHSSTVSymb_{\omega}$ is a combination of spatial and spatio-spectral difference operators with a balancing parameter $\ParamHSSTV > 0$:
\begin{equation}
	\OpeDiffHSSTV{\MatHSIGeneral} := 
	\begin{bmatrix}
		\OpeDiff{\OpeDiffb{\MatHSIGeneral}} \\ 
		\omega \OpeDiff{\MatHSIGeneral}
	\end{bmatrix}.
\end{equation}

\begin{table}[t]
	\caption{Specific Function $\mathcal{R}$ and Linear Operator $\DiffHSIReg$ in Each Rconstructed-Image Regularization}
	\label{tab:R_K}
	\vspace{-1mm}
	\centering
	\begin{tabular}{ccc}
		\toprule
		Regularizations & $\mathcal{R}$ & $\DiffHSIReg$ \\
		\cmidrule(lr){1-1} \cmidrule(lr){2-3}
		HTV (Eq.~\eqref{eq:HTV}) & $\|\cdot\|_{1,2,c}$ & $\OpeDiffSymb$  \\
		SSTV (Eq.~\eqref{eq:SSTV}) & $\|\cdot\|_{1}$ & $\OpeDiffSymb \circ \OpeDiffbSymb$  \\
		HSSTV (Eq.~\eqref{eq:HSSTV}) & $\|\cdot\|_{1}$ & $\OpeDiffHSSTVSymb_{\ParamHSSTV}$  \\
		\bottomrule
	\end{tabular}
\end{table}

\begin{table}[t]
	\caption{Stepsizes $\gamma_1,\gamma_2, \gamma_3$, and $\gamma_4$ for Each Algorithm That Solves An Optimization Problem Incorprating Each Reconstructed-Image Regularization.}
	\vspace{-1mm}
	\label{tab:gamma}
	\centering
	\begin{tabular}{ccccc}
		\toprule
		Regularizations &  $\gamma_1$ & $\gamma_2$ & $\gamma_3$ & $\gamma_4$ \\
		\cmidrule(lr){1-1} \cmidrule(lr){2-5}
		HTV (Eq.~\eqref{eq:HTV}) & $\tfrac{1}{9 + 9\ValSinguMax{\mathbf{A}}^2}$ & $1$ & $\tfrac{1}{5}$  & $\tfrac{1}{3}$ \\
		\addlinespace[2pt]
		SSTV (Eq.~\eqref{eq:SSTV}) & $\tfrac{1}{9 + 33\ValSinguMax{\mathbf{A}}^2}$ & $1$ & $\tfrac{1}{5}$ & $\tfrac{1}{3}$ \\
		\addlinespace[2pt]
		HSSTV (Eq.~\eqref{eq:HSSTV}) & $\tfrac{1}{9 + (33 + 8\omega^2)\ValSinguMax{\mathbf{A}}^2}$ & $1$ & $\tfrac{1}{5}$  & $\tfrac{1}{3}$ \\
		\bottomrule
	\end{tabular}
\end{table}

HTV captures spectral correlations by promoting the sparsity of spatial differences grouped by the spectral direction. 
SSTV captures piecewise smoothness in the spatial and spectral directions by using the composite operator of the spatial and spectral differences (spatio-spectral difference). However, it does not sufficiently evaluate direct spatial piecewise smoothness, resulting in residual noise and artifacts.
HSSTV promotes both spatio-spectral and direct spatial smoothness and thus is a more powerful regularization in general.

\section{Proposed Method}
\label{sec:proposed method}
A general diagram of the proposed method, \Ours, is shown in Fig.~\ref{fig:Overall}. 
In the following, we first introduce an observation model with three types of noise. 
Based on the model, we then formulate the unmixing problem as a constrained convex optimization problem. 
Finally, we describe a P-PDS-based algorithm to efficiently solve the optimization problem with its computational complexity.

\begin{algorithm}[t]
	\caption{A P-PDS-based algorithm for solving Prob.~\eqref{eq_problem}}
	\label{RCHU_3DTV}
	\begin{algorithmic}[1]
		\Require $\MatHSIObsev$, $\MatEndmember$, $\lambda_1$, $\lambda_2$, $\lambda_3$, $\ParamFidel$, and $\ParamConsSpar$
		\Ensure $\MatAbun^{(\NumIter)}$, $\MatSpar^{(\NumIter)}$, $\MatStripe^{(\NumIter)}$
		
		\State Initialize $\MatAbun^{(0)}$, $\MatSpar^{(0)}$, $\MatStripe^{(0)}$, $\mathbf{Z}_1^{(0)}$, $\mathbf{Z}_2^{(0)}$, $\mathbf{Z}_3^{(0)}$, $\mathbf{Z}_4^{(0)}$, and $\mathbf{Z}_5^{(0)}$;
		\State Set $\gamma_1,\gamma_2, \gamma_3$, and $\gamma_4$ as in Tab.~\ref{tab:gamma};
		
		\While{until a stopping criterion is satisfied}
		\State $
		\widetilde{\MatAbun} \leftarrow \mathbf{Z}_1^{(\NumIter)} 
		+ \OpeDiffAdj{\mathbf{Z}_2^{(\NumIter)}} 
		+ \MatEndmember^{*}(\DiffHSIReg^*(\mathbf{Z}_3^{(\NumIter)})) 
		+ \MatEndmember^{*}\mathbf{Z}_4^{(\NumIter)}
		$;
		\vspace{1mm}
		\State $
			\widetilde{\MatAbun} \leftarrow \MatAbun^{(\NumIter)} - \gamma_1\widetilde{\MatAbun}
		$;
		\State $\MatAbun^{(\NumIter + 1)} \leftarrow \textrm{prox}_{\gamma_1 \iota_{\mathbb{R}_+^{m\times n}}}(\widetilde{\MatAbun})$  by~\eqref{prox_R};
		
		\State $\widetilde{\MatSpar} \leftarrow \MatSpar^{(\NumIter)} - \gamma_2\mathbf{Z}_4^{(\NumIter)}$;
		\State ${\MatSpar}^{(\NumIter+1)} \leftarrow \textrm{prox}_{\gamma_2 \iota_{\SetConsSpar}} (\widetilde{\MatSpar});$
		
		\State $\widetilde{\MatStripe} \leftarrow \MatStripe^{(\NumIter)} - \gamma_3(\mathbf{Z}_4^{(\NumIter)} + \OpeDiffvAdj{\mathbf{Z}_5^{(\NumIter)}});$
		\State $\MatStripe^{(\NumIter+1)} \leftarrow \textrm{prox}_{\gamma_3\lambda_3\|\cdot\|_1}(\widetilde{\MatStripe})$ by~\eqref{prox_1};
		
		\State $\widetilde{\mathbf{Z}}_1 \leftarrow \mathbf{Z}_1^{(\NumIter)}+\gamma_4(2\MatAbun^{(\NumIter+1)} - \MatAbun^{(\NumIter)})$;
		\State $\mathbf{Z}_1^{(\NumIter+1)} \leftarrow \widetilde{\mathbf{Z}}_1 - \gamma_4\textrm{prox}_{\tfrac{1}{\gamma_4}\|\cdot\|_{1,2,r}}(\tfrac{\widetilde{\mathbf{Z}}_1}{\gamma_4})$ by~\eqref{prox_12r};
		
		\State $\widetilde{\mathbf{Z}}_2 \leftarrow \mathbf{Z}_2^{(\NumIter)} + \gamma_4 \OpeDiff{2\MatAbun^{(\NumIter + 1)} - \MatAbun^{(\NumIter)}}$;
		\State $\mathbf{Z}_2^{(\NumIter+1)} \leftarrow \widetilde{\mathbf{Z}}_2-\gamma_4\textrm{prox}_{\tfrac{\lambda_1}{\gamma_4}\|\cdot\|_1}(\tfrac{\widetilde{\mathbf{Z}}_2}{\gamma_4})$ by~\eqref{prox_1};
		
		\State $\widetilde{\mathbf{Z}}_3 \leftarrow \mathbf{Z}_3^{(\NumIter)} 
		+ \gamma_4\DiffHSIReg(\MatEndmember(2\MatAbun^{(\NumIter+1)}
		- \MatAbun^{(\NumIter)}))$;
		\State $\mathbf{Z}_3^{(\NumIter+1)} \leftarrow \widetilde{\mathbf{Z}}_3-\gamma_4\textrm{prox}_{\tfrac{\lambda_2}{\gamma_4}\FuncRegRec}(\tfrac{\widetilde{\mathbf{Z}}_3}{\gamma_4})$ by~\eqref{prox_1} or~\eqref{prox_12c};
		
		\State $\widetilde{\mathbf{Z}}_{4}^{\prime} \leftarrow 2(\MatEndmember\MatAbun^{(\NumIter+1)} 
		+ \MatSpar^{(\NumIter+1)} 
		+ \MatStripe^{(\NumIter+1)})$;
		\State $\widetilde{\mathbf{Z}}_{4} \leftarrow \MatEndmember\MatAbun^{(\NumIter)} 
		+ \MatSpar^{(\NumIter)} + \MatStripe^{(\NumIter)}$;
		\State $\widetilde{\mathbf{Z}}_{4} \leftarrow \mathbf{Z}_4^{(\NumIter)}+\gamma_4(\widetilde{\mathbf{Z}}_{4}^{\prime} - \widetilde{\mathbf{Z}}_{4})$;
		\State $\mathbf{Z}_4^{(\NumIter+1)} \leftarrow \widetilde{\mathbf{Z}}_{4}-\gamma_4\textrm{prox}_{\tfrac{1}{\gamma_{4}} \FuncIndSymb{\SetConsFidel}}(\tfrac{\widetilde{\mathbf{Z}}_{4}}{\gamma_{4}})$ by~\eqref{prox_B};
		
		\State $\widetilde{\mathbf{Z}}_{5} \leftarrow \mathbf{Z}_5^{(\NumIter)}+\gamma_4\OpeDiffv{2\MatStripe^{(\NumIter+1)} - \MatStripe^{(\NumIter)}}$;
		\State $\mathbf{Z}_5^{(\NumIter+1)} \leftarrow \widetilde{\mathbf{Z}}_{5}-\gamma_4\textrm{prox}_{\tfrac{1}{\gamma_4} \FuncIndSymb{\SetElemZero}} (\tfrac{\widetilde{\mathbf{Z}}_{5}}{\gamma_4})$ by~\eqref{prox_O};
		
		\State $\NumIter \leftarrow \NumIter + 1$;
		\EndWhile
	\end{algorithmic}
\end{algorithm}

\subsection{Problem Formulation}
\label{ssec:problem_formulation}
Let $\MatEndmember \in \SetRealNum^{\NumBand \times \NumEndmember}$, $\MatAbunTrue \in \SetRealNum^{\NumEndmember \times \NumPixel}$, $\bar{\mathbf{N}} \in \SetRealNum^{\NumBand \times \NumPixel}$, $\MatSparTrue \in \SetRealNum^{\NumBand \times \NumPixel}$, and $\MatStripeTrue \in \SetRealNum^{\NumBand \times \NumPixel}$  be a given endmember library, a true abundance matrix, Gaussian noise, sparse noise, and stripe noise (need not be sparse), respectively.
Consider the following observation model:
\begin{equation}
	\label{hsi_model}
	\MatHSIObsev=\MatEndmember\MatAbunTrue+\bar{\mathbf{N}}+\MatSparTrue+\MatStripeTrue.
\end{equation}
Note that this model explicitly deals with stripe noise as an additive component $\MatStripeTrue$.
Based on Eq.~\eqref{hsi_model}, we formulate an unmixing problem as the following constrained convex optimization problem:
\begin{align}
	\min_{\MatAbun, \MatSpar, \MatStripe} \: & 
	\|\MatAbun\|_{1,2,r}+\lambda_1\|\OpeDiff{\MatAbun}\|_1+\lambda_2\FuncRegRec(\DiffHSIReg(\MatEndmember \MatAbun))+\lambda_3\|\MatStripe\|_1, \nonumber \\
	\text{s.t.} \: & 
	\begin{cases}
		\MatAbun \in \SetNNOrthantAbun,\\
		\MatEndmember \MatAbun + \MatSpar + \MatStripe \in \SetConsFidel,\\
		\MatSpar \in \SetConsSpar,\\
		\OpeDiffv{\MatStripe} \in \SetElemZero,
	\end{cases}
	\label{eq_problem}
\end{align}
where $\lambda_1>0$, $\lambda_2>0$, and $\lambda_3>0$ are hyperparameters that balance each term.
The first term is the joint-sparse regularization that evaluates the row sparsity of abundance maps $\MatAbun$.
The second term promotes the piecewise smoothness of $\MatAbun$.
The first constraint guarantees the non-negativity of $\MatAbun$.
Note that we do not explicitly adopt the abundance sum-to-one constraint. 
This is because, in real-world situations, the abundance sum-to-one constraint tends to be a strong assumption for the linear mixing model based unmixing because the spectral signatures are often affected by a positive scaling factor that varies from pixel to pixel~\cite{iordache2011sparse}.

The third term is the regularization of the reconstructed HS image.
This image-domain regularization enhances the noise robustness of unmixing beyond the capability of the abundance regularizations by capturing the desirable nature of the reconstructed HS image (e.g., spatio-spectral correlation).
In this paper, we focus on three image-domain regularizations: HTV in~\eqref{eq:HTV}, SSTV in~\eqref{eq:SSTV}, and HSSTV in~\eqref{eq:HSSTV}.
In each case, $\FuncRegRec$ and $\DiffHSIReg$ are defined as shown in Tab. \ref{tab:R_K}.
By further generalizing the third term, \Ourss can incorporate other regularizations proposed, e.g., in~\cite{l0l1_Wang_2021,GSSTV_Takemoto2022}.

\begin{table}[t]
	\caption{Computational Complexities of Each Operation.}
	\label{tab:computational_complexity}
	\vspace{-1mm}
	\centering
	\begin{tabular}{cc}
		\toprule
		Operations &  $O$-notation \\
		\cmidrule(lr){1-1} \cmidrule(lr){2-2}
		$\MatEndmember \MatAbun$, 
		($\MatEndmember \in \SetRealNum^{\NumBand \times \NumEndmember}$ 
		and $\MatAbun \in \SetRealNum^{\NumEndmember \times \NumPixel}$) 
		& $O(\NumPixel \NumEndmember \NumBand)$ \\
		
		\addlinespace[5pt]
		$\OpeDiff{\MatAbun}$, 
		($\MatAbun \in \SetRealNum^{\NumEndmember \times \NumPixel}$) 
		& $O(\NumPixel \NumEndmember)$ \\
		
		\addlinespace[5pt]
		$\OpeDiff{\MatHSIGeneral}$, 
		($\MatHSIGeneral \in \SetRealNum^{\NumBand \times \NumPixel}$) 
		& $O(\NumPixel \NumBand)$ \\
		
		\addlinespace[5pt]
		$\DiffHSIReg(\MatHSIGeneral)$, 
		$\begin{cases}
			\DiffHSIReg = \OpeDiffSymb, \: (\MatHSIGeneral \in \SetRealNum^{\NumBand \times \NumPixel}) \\
			\DiffHSIReg = \OpeDiffSymb \circ \OpeDiffbSymb, \: (\MatHSIGeneral \in \SetRealNum^{\NumBand \times \NumPixel}) \\
			\DiffHSIReg = \OpeDiffHSSTVSymb_{\ParamHSSTV}, \: (\MatHSIGeneral \in \SetRealNum^{\NumBand \times \NumPixel})
		\end{cases}$
		& $O(\NumPixel \NumBand)$ \\
		
		\addlinespace[5pt]
		$\textrm{prox}_{\gamma_1 \FuncIndSymb{\SetNNOrthantAbun}}(\MatAbun)$, ($\MatAbun \in \SetRealNum^{\NumEndmember \times \NumPixel}$) 
		& $O(\NumPixel \NumEndmember)$ \\
		
		\addlinespace[5pt]
		$\textrm{prox}_{\gamma_2 \iota_{\SetConsSpar}}(\MatSpar)$, 
		($\MatSpar \in \SetRealNum^{\NumBand \times \NumPixel}$) 
		& $O(\NumPixel \NumBand \log \NumPixel \NumBand)$ \\
		
		\addlinespace[5pt]
		$\textrm{prox}_{\gamma_3\lambda_3\|\cdot\|_1}(\MatStripe)$,
		($\MatStripe \in \SetRealNum^{\NumBand \times \NumPixel}$) 
		& $O(\NumPixel \NumBand)$ \\
		
		\addlinespace[5pt]
		$\textrm{prox}_{\tfrac{1}{\gamma_4}\|\cdot\|_{1,2,r}}(\mathbf{Z}_{1})$,
		($\mathbf{Z}_{1} \in \SetRealNum^{\NumEndmember \times \NumPixel}$) 
		& $O(\NumPixel \NumEndmember)$ \\
		
		\addlinespace[5pt]
		$\textrm{prox}_{\tfrac{\lambda_1}{\gamma_4}\|\cdot\|_1}(\mathbf{Z}_{2})$,
		($ \mathbf{Z}_{2} \in \SetRealNum^{2\NumEndmember \times \NumPixel} $) 
		& $O(\NumPixel \NumEndmember)$ \\
		
		\addlinespace[5pt]
		$\textrm{prox}_{\tfrac{\lambda_2}{\gamma_4}\FuncRegRec}(\mathbf{Z}_{3})$, 
		$\begin{cases}
			\FuncRegRec = \|\cdot\|_{1,2,c}, \: (\mathbf{Z}_{3} \in \SetRealNum^{2\NumBand \times \NumPixel}) \\
			\FuncRegRec = \|\cdot\|_{1}, \: (\mathbf{Z}_{3} \in \SetRealNum^{2\NumBand \times \NumPixel}) \\
			\FuncRegRec = \|\cdot\|_{1}, \: (\mathbf{Z}_{3} \in \SetRealNum^{4\NumBand \times \NumPixel})
		\end{cases}$
		& $O(\NumPixel \NumBand)$ \\
		
		\addlinespace[5pt]
		$\textrm{prox}_{\tfrac{1}{\gamma_{4}} \FuncIndSymb{\SetConsFidel}}(\mathbf{Z}_{4})$,
		($\mathbf{Z}_{4} \in \SetRealNum^{\NumBand \times \NumPixel}$) 
		& $O(\NumPixel \NumBand)$ \\
		\bottomrule
	\end{tabular}
\end{table}

\begin{table}[t]
	\caption{Assumptions and Noise Considered in Each Method.}
	\vspace{-1mm}
	\label{tab:noise}
	\centering
	\begin{tabular}{ccccc}
		\toprule
		\multirow{2}{*}{Methods} & \multirow{2}{*}{Assumption} & \multicolumn{3}{c}{Noise} \\
		\cmidrule(lr){3-5}
		& & Gaussian & Sparse & Stripe \\
		\cmidrule(lr){1-5}
		\vspace{0.5mm}
		CLSUnSAL~\cite{iordache2014collaborative} & non-blind & \checkmark & - & - \\
		\vspace{0.5mm}
		JSTV~\cite{aggarwal2016hyperspectral} & non-blind & \checkmark & \checkmark & - \\
		\vspace{0.5mm}
		RSSUn-TV~\cite{wang2019row} & non-blind & \checkmark & - & - \\
		\vspace{0.5mm}
		LGSU~\cite{shen2022superpixel} & non-blind & \checkmark & - & - \\
		\vspace{0.5mm}
		UnDIP~\cite{UnDIP_RastiB_2022} & non-blind & \checkmark & - & - \\
		\vspace{0.5mm}
		EGU-Net~\cite{hong2022endmember} & blind & \checkmark & - & - \\
		\vspace{0.5mm}
		RDSWSU~\cite{rs_Deng_RobustDual_2023} & non-blind & \checkmark & - & - \\
		\vspace{0.5mm}
		MdLRR~\cite{MDLRR_WuLing_2023} & non-blind & \checkmark & - & - \\
		\textbf{\Ourss (Ours)} & non-blind & \checkmark &\checkmark&\checkmark\\
		\bottomrule
	\end{tabular}
\end{table}

The second constraint serves as data-fidelity to the observed HS image $\MatHSIObsev$ with the Frobenius norm ball $\SetConsFidel$ with the center $\MatHSIObsev$ and radius $\ParamFidel$, defined by
\begin{equation}
	\SetConsFidel := \{ \mathbf{X} \in \SetRealNum^{\NumBand \times \NumPixel} \: | \: \|\MatHSIObsev - \mathbf{X}\|_{F} \leq \ParamFidel \}.
\end{equation}
The third constraint evaluates the sparsity of $\MatSpar$ with the $\ell_1$-norm ball $\SetConsSpar$ with center $\MatZero$ and radius $\ParamConsSpar$, defined by
\begin{equation}
	\SetConsSpar := \{ \mathbf{X} \in \SetRealNum^{\NumBand \times \NumPixel} \: | \: \| \mathbf{X} \|_{1} \leq \ParamConsSpar \}.
\end{equation}
As described in the third contribution, using such constraints instead of data-fidelity and sparse terms makes it easy to adjust hyperparameters since the parameters can be determined based only on noise intensity.
Indeed, this kind of constrained formulation has played an important role in facilitating parameter setup of signal recovery problems~\cite{CSALSA,EPIpre,ono_2015,ono2017primal,ono_2019}.
The detailed setting of these parameters $\ParamFidel$ and $\ParamConsSpar$ is shown in Sec.~V-B.

The fourth term controls the intensity of stripe noise $\MatStripe$ and the fourth constraint captures the vertical flatness property by imposing zero to the vertical gradient of $\MatStripe$.
The term and constraint accurately characterize stripe noise\cite{FC_naganuma_2022}. 
Therefore, our method can estimate abundance maps from HS images contaminated by mixed noise including dense stripe noise.

\begin{table}[t]
	\caption{Hyperparameter Settings in Each Method.}
	\vspace{-1mm}
	\label{tab:parameter_settings}
	\centering
	\begin{tabular}{cc}
		\toprule
		Methods & Parameter search range \\
		\cmidrule(lr){1-1} \cmidrule(lr){2-2}
		\vspace{0.5mm}
		CLSUnSAL~\cite{iordache2014collaborative} & $\lambda \in \{ 10^{-4}, 10^{-3}, 10^{-2}, 10^{-1}, 10^{0}, 10^{1}, 10^{2} \}$ \\
		\vspace{0.5mm}
		JSTV~\cite{aggarwal2016hyperspectral} & \begin{tabular}{c} $\lambda_{1} \in \{ 10^{-2}, 10^{-1}, 10^{0}, 10^{1} \}$, \\ $\lambda_{2} \in \{ 10^{-2}, 10^{-1}, 10^{0}, 10^{1} \}$, \\ $\lambda_{3} \in \{ 10^{-1}, 10^{0}, 10^{1} \}$ \end{tabular} \\
		\vspace{0.5mm}
		RSSUn-TV~\cite{wang2019row} & \begin{tabular}{c} $\lambda \in \{ 10^{-2}, 10^{-1}, 10^{0}, 10^{1} \}$, \\ $\lambda_{TV} \in \{ 10^{-4}, 10^{-3}, 10^{-2}, 10^{-1} \}$ \end{tabular} \\
		\vspace{0.5mm}
		LGSU~\cite{shen2022superpixel} & \begin{tabular}{c} $\lambda_{g} \in \{ 10^{-3}, 10^{-2}, 10^{-1}, 10^{0}, 10^{1} \}$, \\ $\lambda_{l} \in \{ 10^{-4}, 10^{-3}, 10^{-2}, 10^{-1} \}$ \end{tabular} \\
		\vspace{0.5mm}
		UnDIP~\cite{UnDIP_RastiB_2022} & -- \\
		\vspace{0.5mm}
		EGU-Net~\cite{hong2022endmember} & -- \\
		\vspace{0.5mm}
		RDSWSU~\cite{rs_Deng_RobustDual_2023} & $\lambda \in \{ 10^{-3}, 10^{-2}, 10^{-1}, 10^{0}, 10^{1} \}$ \\
		\vspace{0.5mm}
		MdLRR~\cite{MDLRR_WuLing_2023} & \begin{tabular}{c} $\lambda \in \{ 10^{-3}, 10^{-2}, 10^{-1} \}$, \\ $\tau \in \{ 10^{-3}, 10^{-2}, 10^{-1}, 10^{0} \}$ \end{tabular} \\
		\textbf{\Ourss (Ours)} & \begin{tabular}{c} $\lambda_{1} \in \{ 10^{-2}, 10^{-1}, 10^{0}, 10^{1} \}$, \\ $\lambda_{2} \in \{ 10^{-2}, 10^{-1}, 10^{0}, 10^{1} \}$, \\ $\alpha \in \{ 0.95, 0.98 \}$ \end{tabular} \\
		\bottomrule
	\end{tabular}
	\color{black}
\end{table}


\subsection{Optimization Algorithm}
To solve Prob.~\eqref{eq_problem} by an algorithm based on P-PDS, we need to transform Prob.~\eqref{eq_problem} into Prob.~\eqref{eq:pds}.
First, using the indicator functions\footnote{If $C \subset \SetRealNum^{M\times N}$ satisfies $\lambda \mathbf{X} + (1 - \lambda)\mathbf{Y} \in C$ for any $\mathbf{X, Y} \in C$ and $\lambda \in [0, 1]$, $C$ is a convex set. For a non-empty closed convex set $C$, the indicator function $\FuncIndSymb{C} : \SetRealNum^{M\times N} \rightarrow (- \infty, \infty]$ is defined by $\FuncInd{C}{\mathbf{X}} :=  0$, if  $\mathbf{X}\in C$; $\FuncInd{C}{\mathbf{X}} := \infty$, otherwise.}, Prob. (\ref{eq_problem}) are rewritten as follows:
\begin{align}
	\label{alg1}
	\min_{\MatAbun, \MatSpar, \MatStripe} \: 
	& \|\MatAbun\|_{1,2,r}+\lambda_1\|\OpeDiff{\MatAbun}\|_1+\lambda_2 \FuncRegRec(\DiffHSIReg\circ\MatEndmember(\MatAbun))+\lambda_3\|\MatStripe\|_1 \nonumber \\
	& +\FuncInd{\SetNNOrthantAbun}{\MatAbun}
	+\FuncInd{\SetConsFidel}{\MatEndmember \MatAbun + \MatSpar + \MatStripe} \nonumber \\
	& + \FuncInd{\SetConsSpar}{\MatSpar}
	+ \FuncInd{\SetElemZero}{\OpeDiffv{\MatStripe}},
\end{align}
where $\DiffHSIReg\circ\MatEndmember$ is the composite operator of $\DiffHSIReg$ and $\MatEndmember$, i.e., $\DiffHSIReg\circ\MatEndmember(\MatAbun) = \DiffHSIReg(\MatEndmember \MatAbun)$.
Introducing auxiliary variables $\mathbf{Z}_{1}$, $\mathbf{Z}_{2}$, $\mathbf{Z}_{3}$, $\mathbf{Z}_{4}$, and $\mathbf{Z}_{5}$, 
we can transform Prob.~\eqref{alg1} into the following equivalent problem:
\begin{align}
	\min_{\substack{\MatAbun, \MatSpar, \MatStripe, \\ \mathbf{Z}_{1},\ldots,\mathbf{Z}_{5}}} \:
	& \FuncInd{\SetNNOrthantAbun}{\MatAbun}
	+ \FuncInd{\SetConsSpar}{\MatSpar}
	+ \lambda_3\|\MatStripe\|_1
	+ \|\mathbf{Z}_1\|_{1,2,r} \nonumber \\
	& + \lambda_1\|\mathbf{Z}_2\|_1
	+ \lambda_2\FuncRegRec(\mathbf{Z}_3)
	+ \FuncInd{\SetConsFidel}{\mathbf{Z}_4}
	+ \FuncInd{\SetElemZero}{\mathbf{Z}_5} \nonumber \\
	\mathrm{s.t.} \:
	& \begin{cases}
		\mathbf{Z}_{1} = \MatAbun, \\
		\mathbf{Z}_{2} = \OpeDiff{\MatAbun}, \\
		\mathbf{Z}_{3} = \DiffHSIReg\circ\MatEndmember(\MatAbun), \\
		\mathbf{Z}_{4} = \MatEndmember \MatAbun + \MatSpar + \MatStripe, \\
		\mathbf{Z}_{5} = \OpeDiffv{\MatStripe}.
	\end{cases}
	\label{alg2}
\end{align}
Finally, by defining $f_{1}(\MatAbun) = \FuncInd{\SetNNOrthantAbun}{\MatAbun}$, $f_{2}(\MatSpar) = \FuncInd{\SetConsSpar}{\MatSpar}$, $f_{3}(\MatStripe) = \lambda_{3}\|\MatStripe\|_{1}$, $g_{1}(\mathbf{Z}_{1}) = \|\mathbf{Z}_{1}\|_{1,2,r}$, $g_{2}(\mathbf{Z}_{2}) = \lambda_{1}\|\mathbf{Z}_{2}\|_{1}$, $g_{3}(\mathbf{Z}_{3}) = \lambda_{2}\FuncRegRec(\mathbf{Z}_{3})$, $g_{4}(\mathbf{Z}_{4}) = \FuncInd{\SetConsFidel}{\mathbf{Z}_{4}}$, and $g_{5}(\mathbf{Z}_{5}) = \FuncInd{\SetElemZero}{\mathbf{Z}_{5}}$,
Prob.~\eqref{eq:pds} is reduced to Prob.~\eqref{alg2}, i.e., Prob.~\eqref{eq_problem}.

\begin{figure}[t]
	\centering
	\begin{minipage}{0.24\hsize}
		\centerline{
			\includegraphics[width=\hsize]{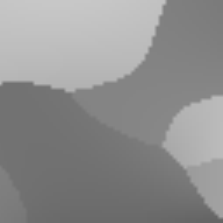}
		}
		\centerline{(a): \textit{Synth 1}}
	\end{minipage}
	\begin{minipage}{0.05\hsize}
	\centerline{~}
	\end{minipage}
	\begin{minipage}{0.24\hsize}
		\centerline{\includegraphics[width=\hsize]{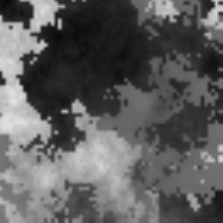}}
		\centerline{(b): \textit{Synth 2}}
	\end{minipage}
	\begin{minipage}{0.05\hsize}
	\centerline{~}
	\end{minipage}
	\begin{minipage}{0.24\hsize}
	\centerline{
		\includegraphics[width=\hsize]{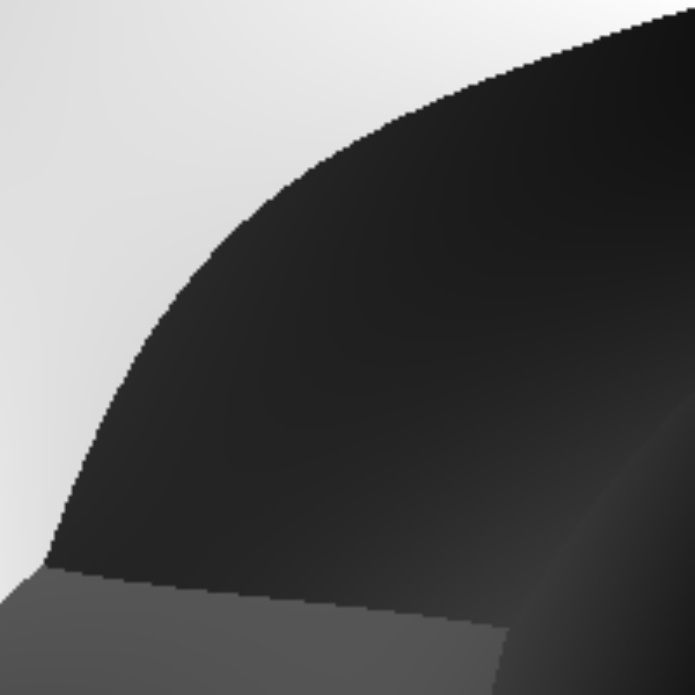}
	}
	\centerline{(c): \textit{Synth 3}}
	\end{minipage}
	
	\vspace{1mm}
	
	\begin{minipage}{0.24\hsize}
		\centerline{\includegraphics[width=\hsize]{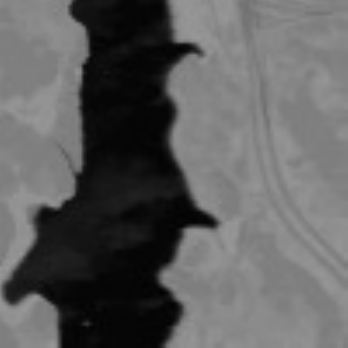}}
		\centerline{(d): \textit{Jasper Ridge}}
	\end{minipage}
	\begin{minipage}{0.05\hsize}
	\centerline{~}
	\end{minipage}
	\begin{minipage}{0.24\hsize}
		\centerline{\includegraphics[width=\hsize]{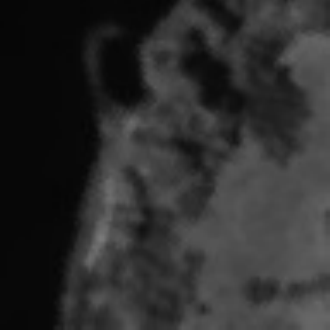}}
		\centerline{(e): \textit{Samson}}
	\end{minipage}
	\begin{minipage}{0.05\hsize}
	\centerline{~}
	\end{minipage}
	\begin{minipage}{0.24\hsize}
	\centerline{
		\includegraphics[width=\hsize]{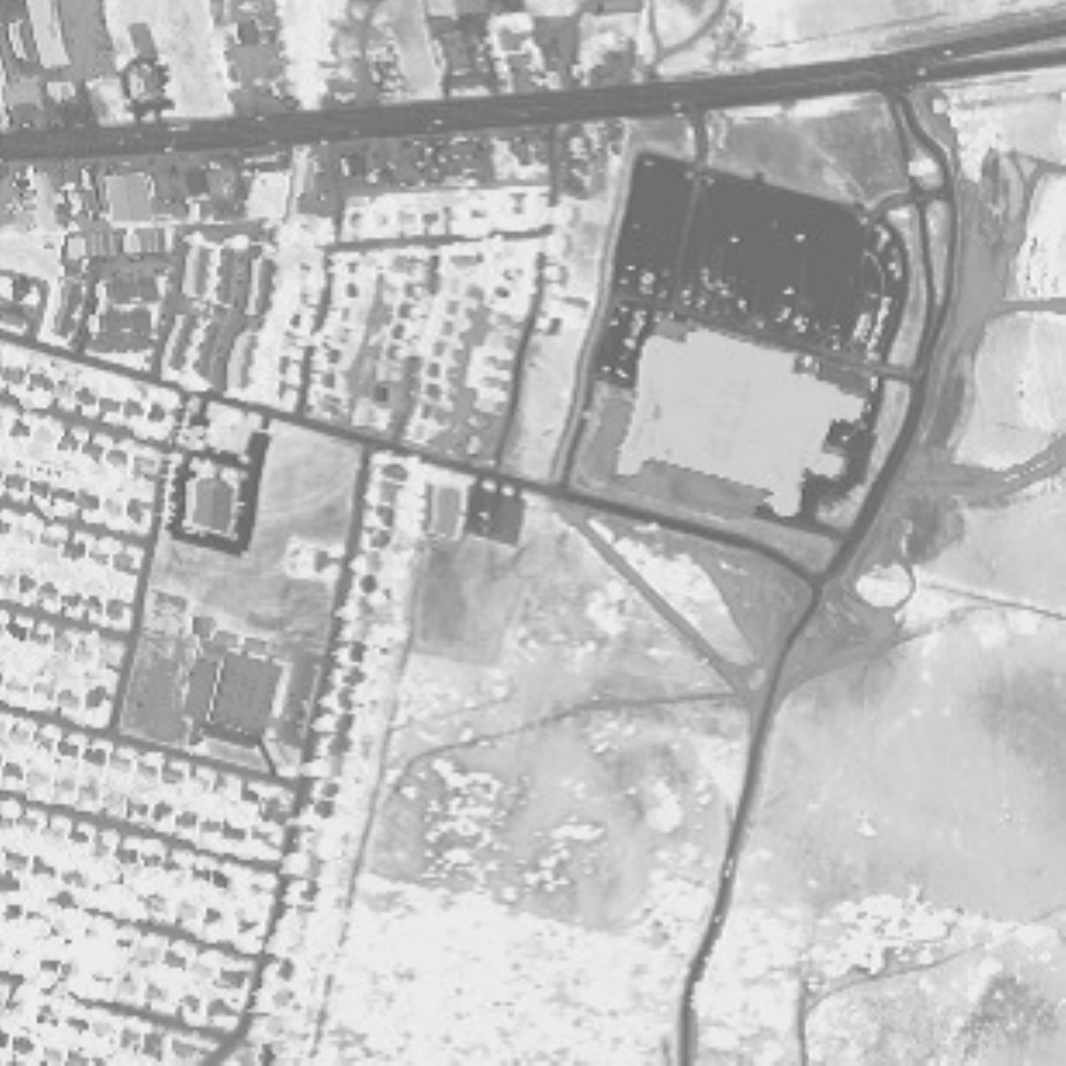}
	}
	\centerline{(f): \textit{Urban}}
	\end{minipage}
	
	\caption{Original HS images.}
	\label{fig:Org_HSimages}
\end{figure}


The algorithm for solving Prob.~\eqref{eq_problem} is summarized in Algorithm~\ref{RCHU_3DTV}.
The linear operator $\DiffHSIReg$ in steps 4 and 15, and the function $\FuncRegRec$ in step 16 depend on what regularization is adopted, as shown in Tab.~\ref{tab:R_K}.
The proximity operators are calculated as follows:
\begin{align}
	\label{prox_R}
	[\mathrm{prox}_{\gamma \FuncIndSymb{\SetNNOrthantAbun}} (\MatAbun)]_{i,j} &= \max(0, \ElemMatAbun_{i,j}),\\
	\label{prox_1}
	[\mathrm{prox}_{\gamma\|\cdot\|_1}(\MatAbun )]_{i,j}&=\mathrm{sign}(\ElemMatAbun_{i,j})\max(|\ElemMatAbun_{i,j}|-\gamma,0) ,\\
	\label{prox_12r}
	[\mathrm{prox}_{\gamma\|\cdot\|_{1,2,r}}(\MatAbun)]_{i,j}&=\max(1-\tfrac{\gamma}{\sqrt{\sum_{j}\ElemMatAbun_{i,j}^2}},0)\ElemMatAbun_{i,j}, \\
	\label{prox_12c}
	[\mathrm{prox}_{\gamma\|\cdot\|_{1,2,c}}(\MatAbun)]_{i,j}&=\max(1-\tfrac{\gamma}{\sqrt{\sum_{i}\ElemMatAbun_{i,j}^2}},0)\ElemMatAbun_{i,j}, \\
	\label{prox_B}
	\mathrm{prox}_{\gamma \FuncIndSymb{\SetConsFidel}}(\MatAbun) & = 
	\begin{cases}
		\MatAbun, & \text{if} ~ \MatAbun\in \SetConsFidel;\\
		\MatHSIObsev + \frac{\ParamFidel(\MatAbun - \MatHSIObsev)}{\|\MatAbun - \MatHSIObsev\|_F}, & \text{otherwise},
	\end{cases} \\
	\label{prox_O}
	\mathrm{prox}_{\gamma \FuncIndSymb{\SetElemZero}} (\MatAbun) & = \MatZero.
\end{align}
The proximity operator of $\iota_{\gamma \SetConsSpar}$ can be efficiently computed by the $\ell_{1}$-ball projection algorithm~\cite{fast_l1_ball_projection}.

\begin{figure}[t]
	\centering
	\begin{minipage}{0.48\hsize}
		\centerline{\includegraphics[width=\hsize]{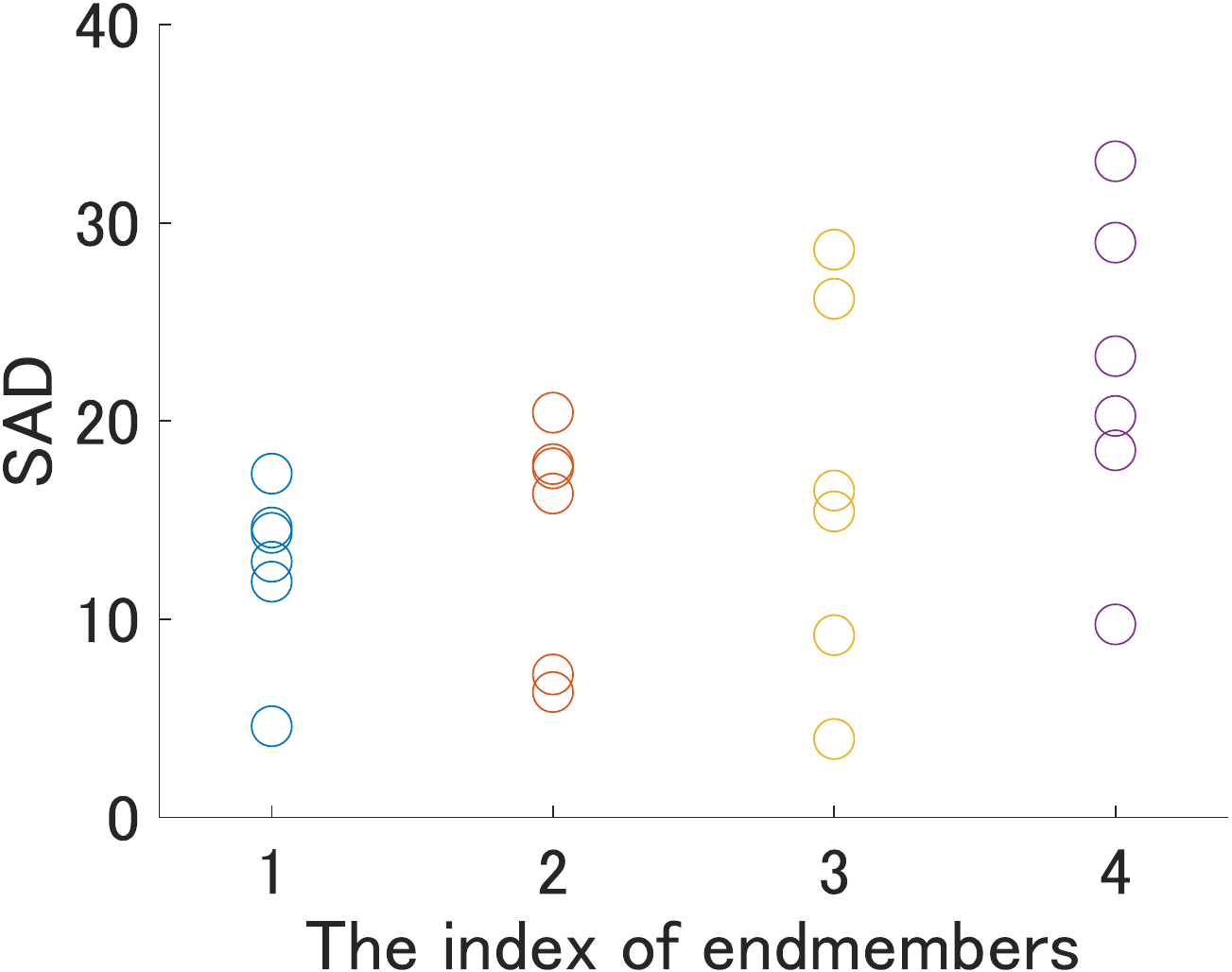}}
		\centerline{(a): \textit{Synth 1}}
	\end{minipage}
	\begin{minipage}{0.48\hsize}
		\centerline{\includegraphics[width=\hsize]{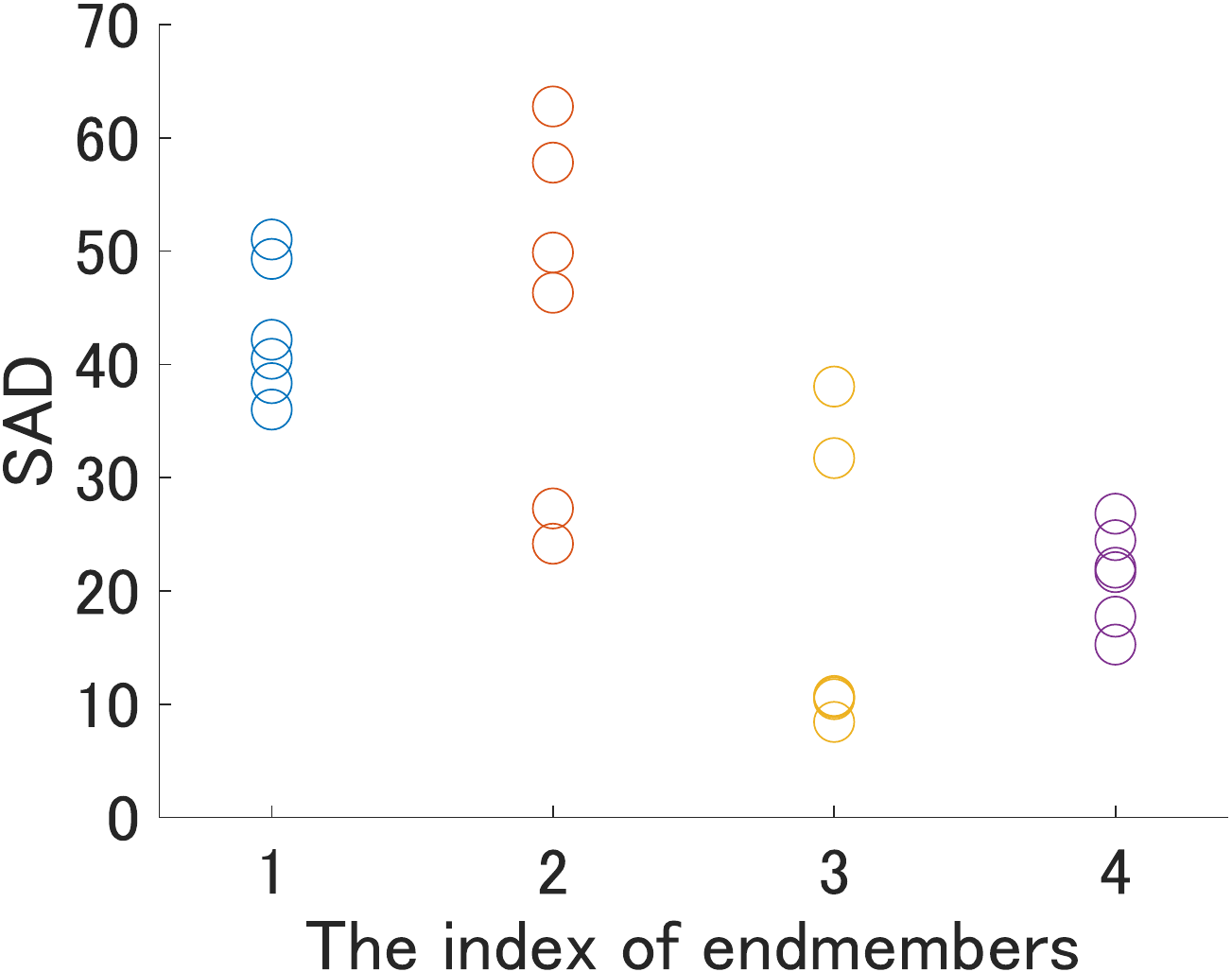}}
		\centerline{(b): \textit{Synth 2}}
	\end{minipage}
	
	\vspace{1mm}
	
	\begin{minipage}{0.48\hsize}
		\centerline{\includegraphics[width=\hsize]{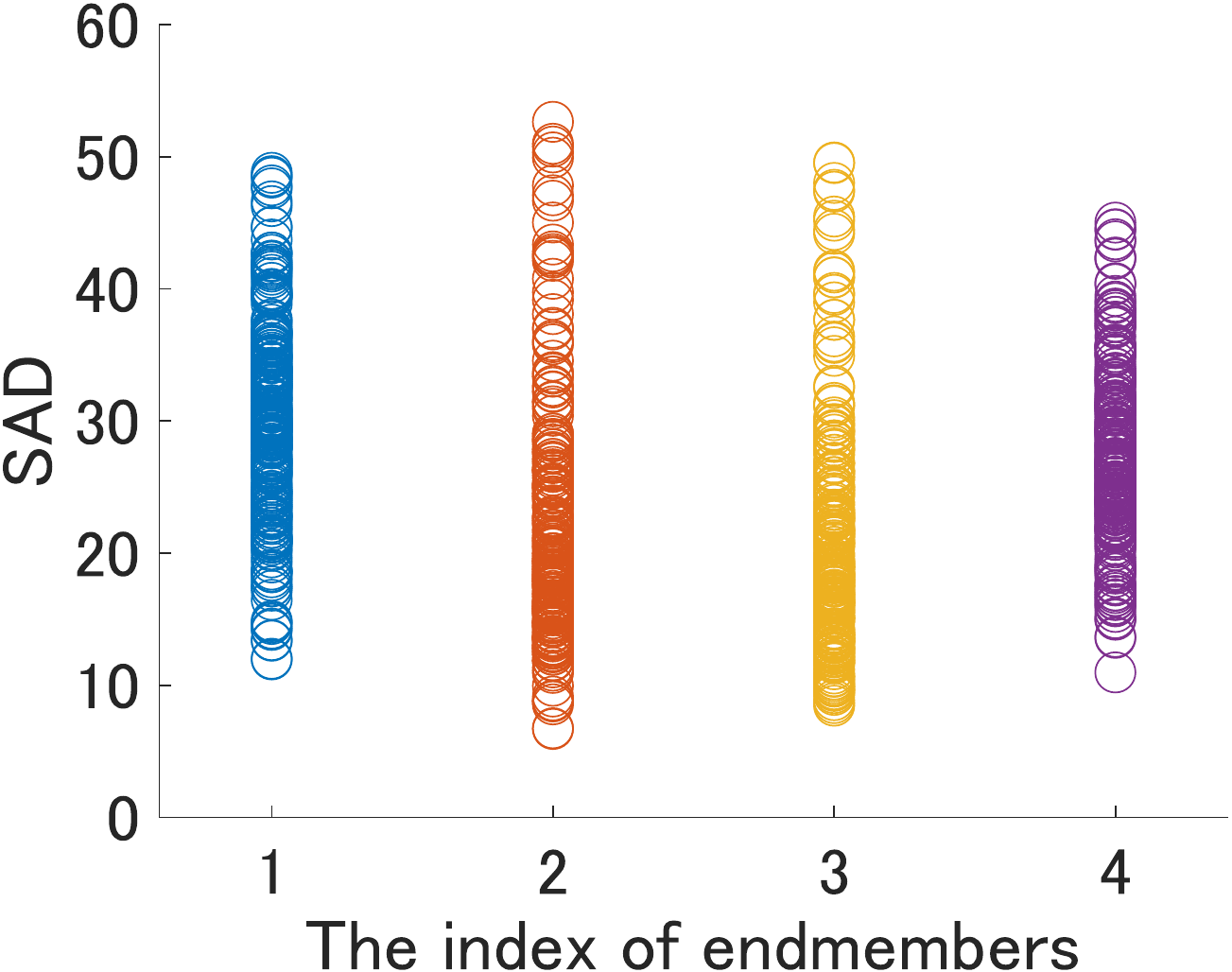}}
		\centerline{(c): \textit{Synth 3}}
	\end{minipage}
	\begin{minipage}{0.48\hsize}
		\centerline{\includegraphics[width=\hsize]{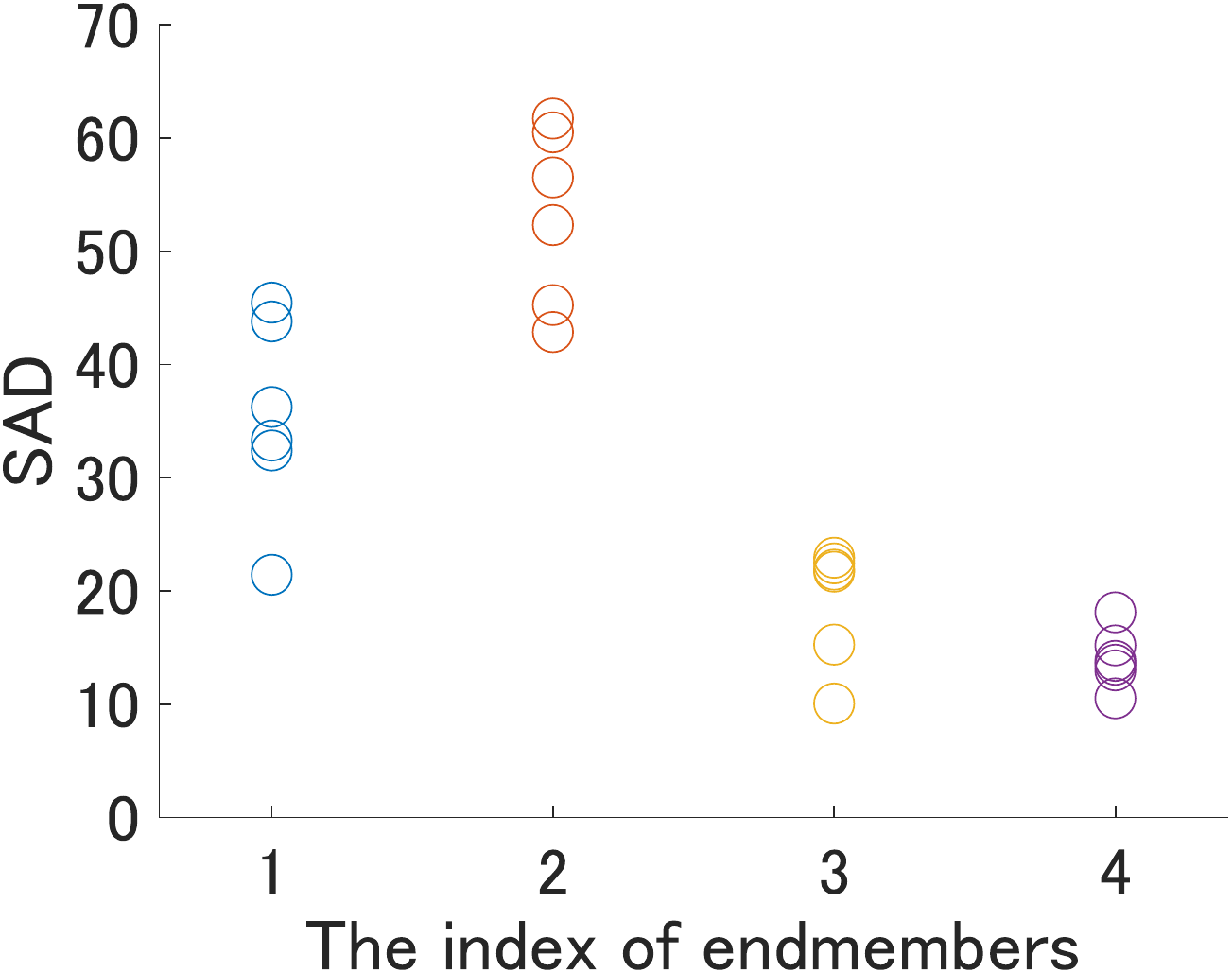}}
		\centerline{(d): \textit{Jasper Ridge}}
	\end{minipage}
	
	\vspace{1mm}
	
	\begin{minipage}{0.48\hsize}
		\centerline{\includegraphics[width=\hsize]{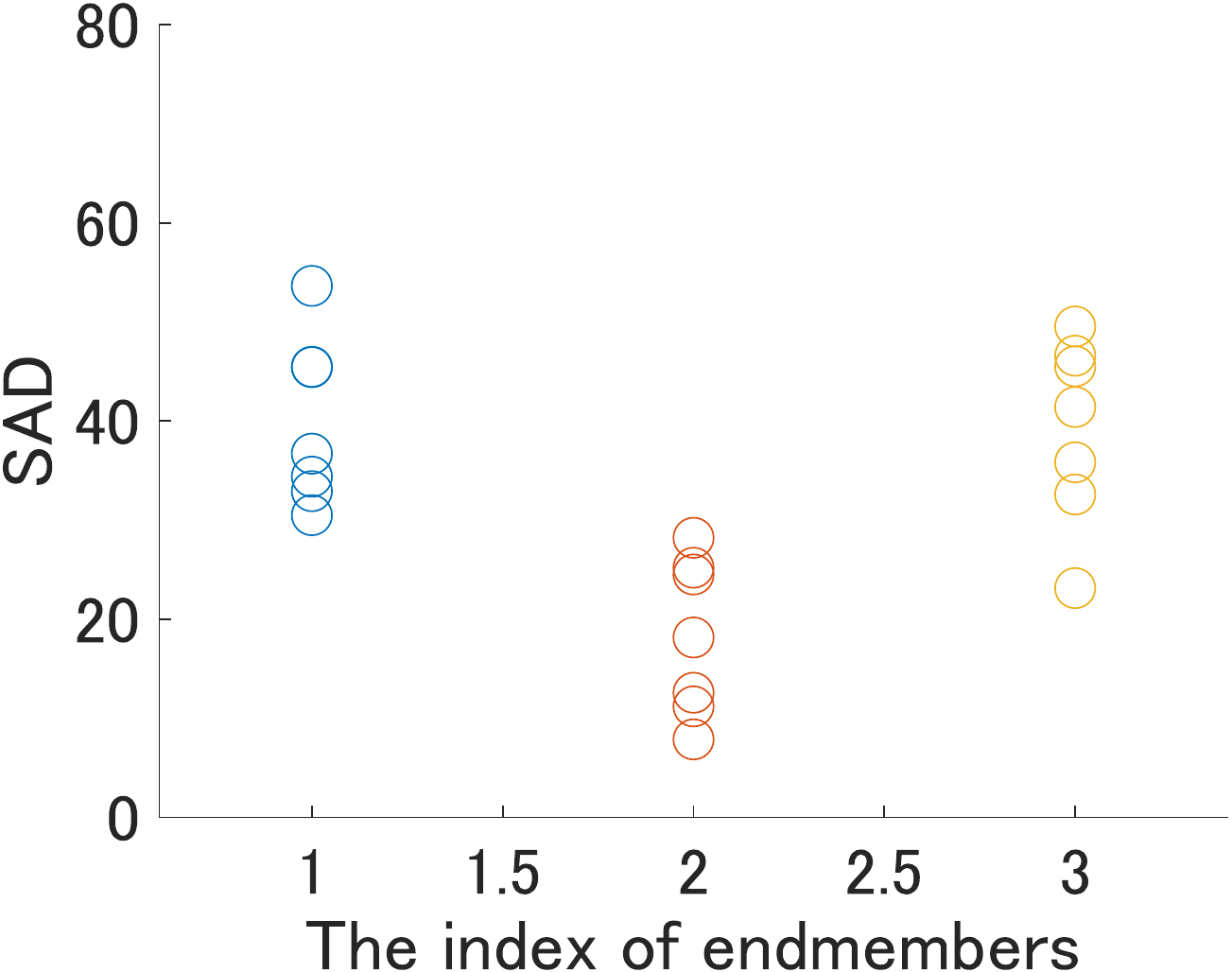}}
		\centerline{(e): \textit{Samson}}
	\end{minipage}
	\begin{minipage}{0.48\hsize}
		\centerline{\includegraphics[width=\hsize]{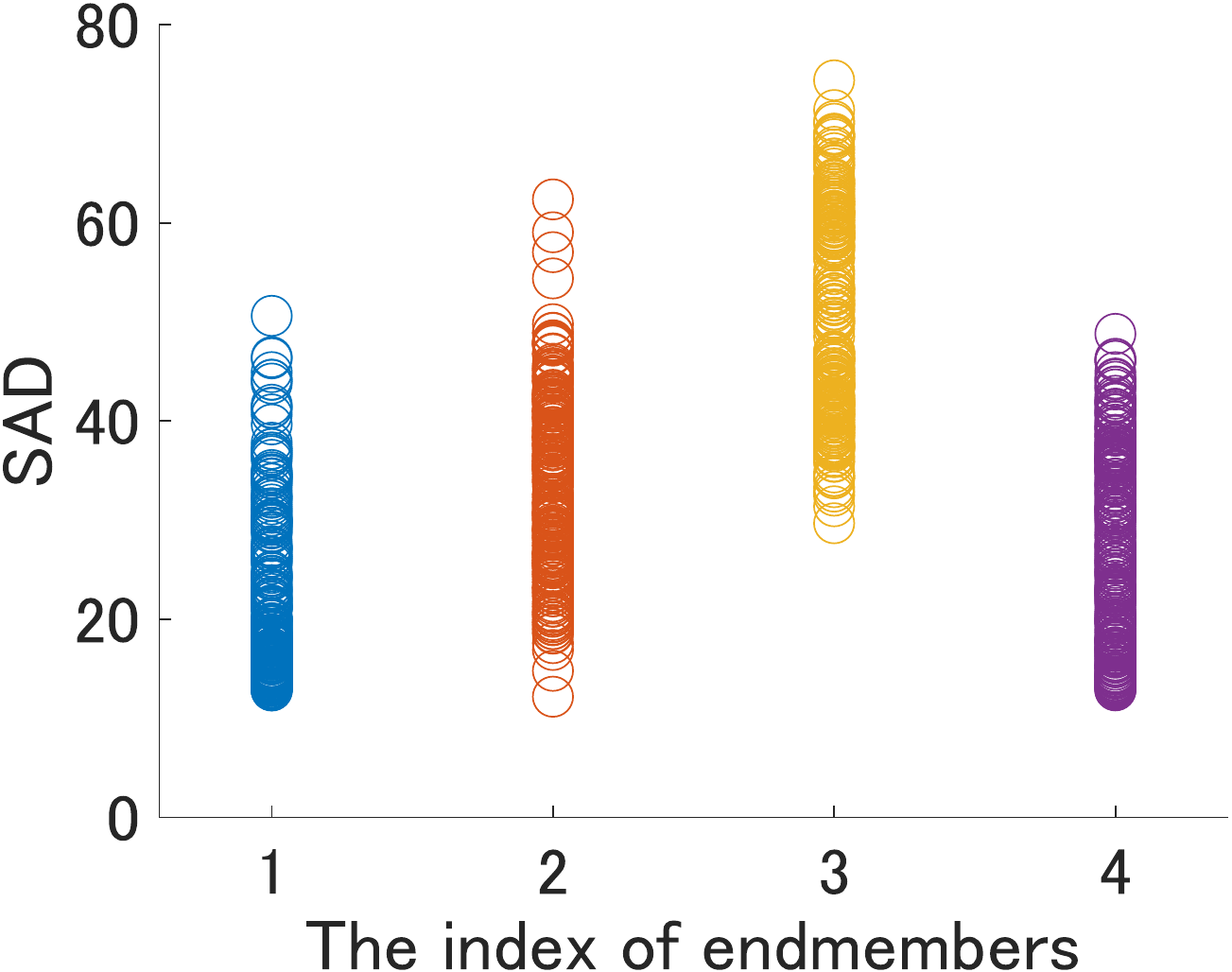}}
		\centerline{(f): \textit{Urban}}
	\end{minipage}
	
	\caption{The distributions of spectral angle distances between the spectra of endmembers present in HS images and the spectra of the other endmembers. Vertical axes indicate the spectral angle ditances. Horizontal axes indicate the indices of the endmembers present in the HS images. The SAD values tend to be the same for all data sets.}
	\label{fig:SAD}
\end{figure}


Based on Eq.~\eqref{eq:stepsize_exact}, the stepsizes of Algorithm~\ref{RCHU_3DTV} are given as
\begin{align}
	\label{eq:stepsize_ours_gen}
	& \gamma_{1} = \tfrac{1}{\|\mathbf{I}\|_{\mathrm{op}}^{2} + \|\OpeDiffSymb\|_{\mathrm{op}}^{2} + \|\DiffHSIReg \circ \MatEndmember\|_{\mathrm{op}}^{2} + \|\MatEndmember\|_{\mathrm{op}}^{2} }, 
	\gamma_{2} = \tfrac{1}{\|\mathbf{I}\|_{\mathrm{op}}^{2}}, \nonumber \\
	& \gamma_{3} = \tfrac{1}{\|\mathbf{I}\|_{\mathrm{op}}^{2} + \|\OpeDiffvSymb\|_{\mathrm{op}}^{2}}, 
	\gamma_{4} = \tfrac{1}{3}. 
\end{align}
An identity matrix of any size satisfies $\|\mathbf{I}\|_{\mathrm{op}} = 1$.
The operator norm $\| \MatEndmember \|_{\mathrm{op}}$ is equal to its maximum singular value $\ValSinguMax{\MatEndmember}$.
The operator norms of the other linear operators are not easy to obtain\footnote{Note that the difference operators are not implemented as matrices. Therefore, we cannot easily obtain the singular values of the matrices representing the difference operators.}, but they are suppressed by $\|\OpeDiffvSymb\|_{\mathrm{op}} \leq 2$, $\|\OpeDiffbSymb\|_{\mathrm{op}} \leq 2$, $\|\OpeDiffSymb\|_{\mathrm{op}} \leq 2\sqrt{2}$,
$\|\DiffHSIReg \circ \MatEndmember\|_{\mathrm{op}} \leq \|\DiffHSIReg\|_{\mathrm{op}} \|\MatEndmember\|_{\mathrm{op}}$, 
$\|\OpeDiffSymb \circ \OpeDiffbSymb\|_{\mathrm{op}} \leq \|\OpeDiffSymb\|_{\mathrm{op}} \|\OpeDiffbSymb\|_{\mathrm{op}}$, and
$\|\OpeDiffHSSTVSymb_{\ParamHSSTV}\|_{\mathrm{op}} \leq \sqrt{32 + 8\ParamHSSTV^{2}}$.
By substituting these upper bounds into Eq.~\eqref{eq:stepsize_relaxed},
the specific stepsizes are given as shown in Tab.~\ref{tab:gamma}. 
This stepsizes design method allows us to avoid the stepsize adjustment for Algorithm~\ref{RCHU_3DTV}.

\subsection{Computational Complexity}
In general, the computational complexity of our algorithm varies depending on what function and linear operator are used as an image-domain regularization.
Our method adopts three image-domain regularizations: HTV, SSTV, and HSSTV.
The computational complexities of linear operators and functions including all the image-domain regularizations are given in Tab.~\ref{tab:computational_complexity}. 
From these results, we derive the computational complexities of each step as follows:
\begin{itemize}
	\item The complexities of Steps 4, 15, 17, and 18 are $O(\NumPixel \NumEndmember \NumBand)$.
	\item The complexities of Steps 5, 6, 11, 12, 13, and 14 are $O(\NumPixel \NumEndmember)$.
	\item The complexities of Steps 7, 9, 10, 16, 19, 20, 21, and 22 are $O(\NumPixel \NumBand)$.
	\item The complexity of Step 8 is $O(\NumPixel \NumBand \log \NumPixel \NumBand)$.
\end{itemize}
Therefore, the complexity for each iteration of the algorithm is $O(\NumPixel \NumBand \max\{\NumEndmember, \log \NumPixel \NumBand\})$.

\begin{table*}[!h]
	\caption{SRE, RMSE, Ps, MPSNR, and MSSIM in the Experiments Using \textit{Synth 1}.}
	\vspace{-1mm}
	\label{tab:result_synth_Legendre}
	\centering
	\scalebox{0.75}{
		\begin{tabular}{ccccccccccccc} \toprule
			\multirow{3}{*}{ Noise } & \multirow{3}{*}{Metrics}& \multicolumn{11}{c}{Methods}\\ 
			\cmidrule(lr){3-13}
			& & CLSUnSAL & JSTV & RSSUn-TV & LGSU & UnDIP & EGU-Net & RDSWSU & MdLRR & \textbf{\Ours} & \textbf{\Ours} & \textbf{\Ours} \\ 
			& & \cite{iordache2014collaborative} & \cite{aggarwal2016hyperspectral} & \cite{wang2019row} & \cite{shen2022superpixel} & \cite{UnDIP_RastiB_2022} & \cite{hong2022endmember} & \cite{rs_Deng_RobustDual_2023} & \cite{MDLRR_WuLing_2023} & \textbf{(HTV)} & \textbf{(SSTV)} & \textbf{(HSSTV)} \\ 
			
			\midrule
\multirow{6}{*}{Case 1} 
& Setup & \begin{tabular}{c} \hspace{-4mm} $\lambda = 10^{-1}$ \hspace{-4mm}\end{tabular} & \begin{tabular}{c} \hspace{-4mm} $\lambda_{1} = 10^{1}$, \hspace{-4mm} \\\hspace{-4mm} $\lambda_{2} = 10^{1}$, \hspace{-4mm} \\\hspace{-4mm} $\lambda_{3} = 10^{1}$ \hspace{-4mm}\end{tabular} & \begin{tabular}{c} \hspace{-4mm} $\lambda = 10^{1}$, \hspace{-4mm} \\\hspace{-4mm} $\lambda_{TV} = 10^{-2}$ \hspace{-4mm}\end{tabular} & \begin{tabular}{c} \hspace{-4mm} $\lambda_{g} = 10^{-1}$, \hspace{-4mm} \\\hspace{-4mm} $\lambda_{l} = 10^{-2}$ \hspace{-4mm}\end{tabular} &  --  &  --  & \begin{tabular}{c} \hspace{-4mm} $\lambda = 10^{-2}$ \hspace{-4mm}\end{tabular} & \begin{tabular}{c} \hspace{-4mm} $\lambda = 10^{-3}$, \hspace{-4mm} \\\hspace{-4mm} $\tau = 10^{0}$ \hspace{-4mm}\end{tabular} & \begin{tabular}{c} \hspace{-4mm} $\lambda_{1} = 10^{0}$, \hspace{-4mm} \\\hspace{-4mm} $\lambda_{2} = 10^{1}$, \hspace{-4mm} \\\hspace{-4mm} $\varepsilon = 0.95$ \hspace{-4mm}\end{tabular} & \begin{tabular}{c} \hspace{-4mm} $\lambda_{1} = 10^{1}$, \hspace{-4mm} \\\hspace{-4mm} $\lambda_{2} = 10^{-2}$, \hspace{-4mm} \\\hspace{-4mm} $\varepsilon = 0.95$ \hspace{-4mm}\end{tabular} & \begin{tabular}{c} \hspace{-4mm} $\lambda_{1} = 10^{0}$, \hspace{-4mm} \\\hspace{-4mm} $\lambda_{2} = 10^{-2}$, \hspace{-4mm} \\\hspace{-4mm} $\varepsilon = 0.98$ \hspace{-4mm}\end{tabular} \\ 
& SRE  &   16.22 &   15.73 &   22.15 &   16.73 &    2.88 &   -5.52 &   16.38 &   18.96 &   \Valbest{24.31} &   21.84 &   \ValSecnd{23.52} \\ 
& RMSE &   0.0385 &   0.0388 &   0.0197 &   0.0391 &   0.1417 &   0.2489 &   0.0398 &   0.0284 &   \Valbest{0.0153} &   0.0200 &   \ValSecnd{0.0167} \\ 
& Ps   &   \Valbest{1.00} &   \Valbest{1.00} &   \Valbest{1.00} &   \Valbest{1.00} &   0.59 &   0.01 &   \Valbest{1.00} &   \Valbest{1.00} &   \Valbest{1.00} &   \Valbest{1.00} &   \Valbest{1.00} \\ 
& MPSNR &   45.23 &   38.35 &   46.53 &   45.09 &   23.92 &   15.32 &   45.17 &   45.59 &   \Valbest{54.23} &   \ValSecnd{49.34} &   47.74 \\ 
& MSSIM &   0.9775 &   0.9164 &   0.9840 &   0.9768 &   0.8096 &   0.4255 &   0.9766 &   0.9804 &   \Valbest{0.9988} &   \ValSecnd{0.9937} &   0.9886 \\ 
\midrule 

\multirow{6}{*}{Case 2} 
& Setup & \begin{tabular}{c} \hspace{-4mm} $\lambda = 10^{0}$ \hspace{-4mm}\end{tabular} & \begin{tabular}{c} \hspace{-4mm} $\lambda_{1} = 10^{1}$, \hspace{-4mm} \\\hspace{-4mm} $\lambda_{2} = 10^{1}$, \hspace{-4mm} \\\hspace{-4mm} $\lambda_{3} = 10^{1}$ \hspace{-4mm}\end{tabular} & \begin{tabular}{c} \hspace{-4mm} $\lambda = 10^{1}$, \hspace{-4mm} \\\hspace{-4mm} $\lambda_{TV} = 10^{-2}$ \hspace{-4mm}\end{tabular} & \begin{tabular}{c} \hspace{-4mm} $\lambda_{g} = 10^{0}$, \hspace{-4mm} \\\hspace{-4mm} $\lambda_{l} = 10^{-1}$ \hspace{-4mm}\end{tabular} &  --  &  --  & \begin{tabular}{c} \hspace{-4mm} $\lambda = 10^{-2}$ \hspace{-4mm}\end{tabular} & \begin{tabular}{c} \hspace{-4mm} $\lambda = 10^{-2}$, \hspace{-4mm} \\\hspace{-4mm} $\tau = 10^{0}$ \hspace{-4mm}\end{tabular} & \begin{tabular}{c} \hspace{-4mm} $\lambda_{1} = 10^{0}$, \hspace{-4mm} \\\hspace{-4mm} $\lambda_{2} = 10^{1}$, \hspace{-4mm} \\\hspace{-4mm} $\varepsilon = 0.95$ \hspace{-4mm}\end{tabular} & \begin{tabular}{c} \hspace{-4mm} $\lambda_{1} = 10^{1}$, \hspace{-4mm} \\\hspace{-4mm} $\lambda_{2} = 10^{-2}$, \hspace{-4mm} \\\hspace{-4mm} $\varepsilon = 0.95$ \hspace{-4mm}\end{tabular} & \begin{tabular}{c} \hspace{-4mm} $\lambda_{1} = 10^{0}$, \hspace{-4mm} \\\hspace{-4mm} $\lambda_{2} = 10^{-1}$, \hspace{-4mm} \\\hspace{-4mm} $\varepsilon = 0.98$ \hspace{-4mm}\end{tabular} \\ 
& SRE  &   12.57 &   11.35 &   14.22 &   14.91 &    0.42 &   -5.59 &   13.12 &   16.24 &   \Valbest{21.64} &   20.19 &   \ValSecnd{20.84} \\ 
& RMSE &   0.0569 &   0.0615 &   0.0485 &   0.0484 &   0.1892 &   0.2510 &   0.0567 &   0.0393 &   \Valbest{0.0206} &   0.0240 &   \ValSecnd{0.0226} \\ 
& Ps   &   0.98 &   \Valbest{1.00} &   \Valbest{1.00} &   0.99 &   0.50 &   0.01 &   0.97 &   \Valbest{1.00} &   \Valbest{1.00} &   \Valbest{1.00} &   \Valbest{1.00} \\ 
& MPSNR &   40.17 &   34.04 &   39.90 &   39.74 &   19.76 &   15.25 &   39.67 &   39.91 &   \Valbest{49.20} &   \ValSecnd{43.39} &   41.70 \\ 
& MSSIM &   0.9356 &   0.8270 &   0.9338 &   0.9298 &   0.6906 &   0.3899 &   0.9261 &   0.9327 &   \Valbest{0.9959} &   \ValSecnd{0.9735} &   0.9572 \\ 
\midrule 

\multirow{6}{*}{Case 3} 
& Setup & \begin{tabular}{c} \hspace{-4mm} $\lambda = 10^{0}$ \hspace{-4mm}\end{tabular} & \begin{tabular}{c} \hspace{-4mm} $\lambda_{1} = 10^{1}$, \hspace{-4mm} \\\hspace{-4mm} $\lambda_{2} = 10^{1}$, \hspace{-4mm} \\\hspace{-4mm} $\lambda_{3} = 10^{1}$ \hspace{-4mm}\end{tabular} & \begin{tabular}{c} \hspace{-4mm} $\lambda = 10^{-2}$, \hspace{-4mm} \\\hspace{-4mm} $\lambda_{TV} = 10^{-2}$ \hspace{-4mm}\end{tabular} & \begin{tabular}{c} \hspace{-4mm} $\lambda_{g} = 10^{0}$, \hspace{-4mm} \\\hspace{-4mm} $\lambda_{l} = 10^{-1}$ \hspace{-4mm}\end{tabular} &  --  &  --  & \begin{tabular}{c} \hspace{-4mm} $\lambda = 10^{0}$ \hspace{-4mm}\end{tabular} & \begin{tabular}{c} \hspace{-4mm} $\lambda = 10^{-1}$, \hspace{-4mm} \\\hspace{-4mm} $\tau = 10^{0}$ \hspace{-4mm}\end{tabular} & \begin{tabular}{c} \hspace{-4mm} $\lambda_{1} = 10^{0}$, \hspace{-4mm} \\\hspace{-4mm} $\lambda_{2} = 10^{0}$, \hspace{-4mm} \\\hspace{-4mm} $\varepsilon = 0.98$ \hspace{-4mm}\end{tabular} & \begin{tabular}{c} \hspace{-4mm} $\lambda_{1} = 10^{1}$, \hspace{-4mm} \\\hspace{-4mm} $\lambda_{2} = 10^{-2}$, \hspace{-4mm} \\\hspace{-4mm} $\varepsilon = 0.95$ \hspace{-4mm}\end{tabular} & \begin{tabular}{c} \hspace{-4mm} $\lambda_{1} = 10^{0}$, \hspace{-4mm} \\\hspace{-4mm} $\lambda_{2} = 10^{-2}$, \hspace{-4mm} \\\hspace{-4mm} $\varepsilon = 0.98$ \hspace{-4mm}\end{tabular} \\ 
& SRE  &   11.28 &   15.18 &    7.63 &   13.13 &   -1.03 &   -5.62 &   11.58 &   14.64 &   \Valbest{23.92} &   21.06 &   \ValSecnd{23.73} \\ 
& RMSE &   0.0646 &   0.0411 &   0.1051 &   0.0583 &   0.2085 &   0.2512 &   0.0790 &   0.0494 &   \Valbest{0.0159} &   0.0218 &   \ValSecnd{0.0162} \\ 
& Ps   &   0.97 &   \Valbest{1.00} &   0.85 &   0.98 &   0.32 &   0.01 &   0.96 &   \Valbest{1.00} &   \Valbest{1.00} &   \Valbest{1.00} &   \Valbest{1.00} \\ 
& MPSNR &   37.77 &   38.32 &   37.09 &   37.30 &   17.82 &   15.24 &   36.15 &   37.82 &   \Valbest{50.54} &   \ValSecnd{48.43} &   46.82 \\ 
& MSSIM &   0.9066 &   0.9204 &   0.8957 &   0.8919 &   0.6237 &   0.3827 &   0.9239 &   0.9087 &   \Valbest{0.9949} &   \ValSecnd{0.9922} &   0.9860 \\ 
\midrule 

\multirow{6}{*}{Case 4} 
& Setup & \begin{tabular}{c} \hspace{-4mm} $\lambda = 10^{0}$ \hspace{-4mm}\end{tabular} & \begin{tabular}{c} \hspace{-4mm} $\lambda_{1} = 10^{1}$, \hspace{-4mm} \\\hspace{-4mm} $\lambda_{2} = 10^{1}$, \hspace{-4mm} \\\hspace{-4mm} $\lambda_{3} = 10^{1}$ \hspace{-4mm}\end{tabular} & \begin{tabular}{c} \hspace{-4mm} $\lambda = 10^{-2}$, \hspace{-4mm} \\\hspace{-4mm} $\lambda_{TV} = 10^{-2}$ \hspace{-4mm}\end{tabular} & \begin{tabular}{c} \hspace{-4mm} $\lambda_{g} = 10^{0}$, \hspace{-4mm} \\\hspace{-4mm} $\lambda_{l} = 10^{-1}$ \hspace{-4mm}\end{tabular} &  --  &  --  & \begin{tabular}{c} \hspace{-4mm} $\lambda = 10^{0}$ \hspace{-4mm}\end{tabular} & \begin{tabular}{c} \hspace{-4mm} $\lambda = 10^{-1}$, \hspace{-4mm} \\\hspace{-4mm} $\tau = 10^{0}$ \hspace{-4mm}\end{tabular} & \begin{tabular}{c} \hspace{-4mm} $\lambda_{1} = 10^{0}$, \hspace{-4mm} \\\hspace{-4mm} $\lambda_{2} = 10^{0}$, \hspace{-4mm} \\\hspace{-4mm} $\varepsilon = 0.98$ \hspace{-4mm}\end{tabular} & \begin{tabular}{c} \hspace{-4mm} $\lambda_{1} = 10^{1}$, \hspace{-4mm} \\\hspace{-4mm} $\lambda_{2} = 10^{-2}$, \hspace{-4mm} \\\hspace{-4mm} $\varepsilon = 0.95$ \hspace{-4mm}\end{tabular} & \begin{tabular}{c} \hspace{-4mm} $\lambda_{1} = 10^{0}$, \hspace{-4mm} \\\hspace{-4mm} $\lambda_{2} = 10^{-1}$, \hspace{-4mm} \\\hspace{-4mm} $\varepsilon = 0.98$ \hspace{-4mm}\end{tabular} \\ 
& SRE  &    9.10 &   14.79 &    4.45 &   10.19 &   -1.62 &   -5.66 &   10.25 &   12.23 &   \Valbest{23.59} &   21.89 &   \ValSecnd{23.31} \\ 
& RMSE &   0.0809 &   0.0429 &   0.1558 &   0.0797 &   0.2123 &   0.2526 &   0.0918 &   0.0636 &   \Valbest{0.0166} &   0.0199 &   \ValSecnd{0.0172} \\ 
& Ps   &   0.93 &   \Valbest{1.00} &   0.48 &   0.94 &   0.31 &   0.01 &   0.91 &   0.98 &   \Valbest{1.00} &   \Valbest{1.00} &   \Valbest{1.00} \\ 
& MPSNR &   34.66 &   37.77 &   33.63 &   33.94 &   19.02 &   15.23 &   33.82 &   34.27 &   \Valbest{47.92} &   \ValSecnd{47.58 }&   45.63 \\ 
& MSSIM &   0.8556 &   0.9084 &   0.8286 &   0.8275 &   0.6136 &   0.3774 &   0.8746 &   0.8448 &   \ValSecnd{0.9893} &   \Valbest{0.9900} &   0.9810 \\ 
\midrule 

\multirow{6}{*}{Case 5} 
& Setup & \begin{tabular}{c} \hspace{-4mm} $\lambda = 10^{0}$ \hspace{-4mm}\end{tabular} & \begin{tabular}{c} \hspace{-4mm} $\lambda_{1} = 10^{0}$, \hspace{-4mm} \\\hspace{-4mm} $\lambda_{2} = 10^{1}$, \hspace{-4mm} \\\hspace{-4mm} $\lambda_{3} = 10^{0}$ \hspace{-4mm}\end{tabular} & \begin{tabular}{c} \hspace{-4mm} $\lambda = 10^{0}$, \hspace{-4mm} \\\hspace{-4mm} $\lambda_{TV} = 10^{-2}$ \hspace{-4mm}\end{tabular} & \begin{tabular}{c} \hspace{-4mm} $\lambda_{g} = 10^{0}$, \hspace{-4mm} \\\hspace{-4mm} $\lambda_{l} = 10^{-1}$ \hspace{-4mm}\end{tabular} &  --  &  --  & \begin{tabular}{c} \hspace{-4mm} $\lambda = 10^{0}$ \hspace{-4mm}\end{tabular} & \begin{tabular}{c} \hspace{-4mm} $\lambda = 10^{-1}$, \hspace{-4mm} \\\hspace{-4mm} $\tau = 10^{0}$ \hspace{-4mm}\end{tabular} & \begin{tabular}{c} \hspace{-4mm} $\lambda_{1} = 10^{0}$, \hspace{-4mm} \\\hspace{-4mm} $\lambda_{2} = 10^{1}$, \hspace{-4mm} \\\hspace{-4mm} $\varepsilon = 0.95$ \hspace{-4mm}\end{tabular} & \begin{tabular}{c} \hspace{-4mm} $\lambda_{1} = 10^{1}$, \hspace{-4mm} \\\hspace{-4mm} $\lambda_{2} = 10^{-2}$, \hspace{-4mm} \\\hspace{-4mm} $\varepsilon = 0.95$ \hspace{-4mm}\end{tabular} & \begin{tabular}{c} \hspace{-4mm} $\lambda_{1} = 10^{1}$, \hspace{-4mm} \\\hspace{-4mm} $\lambda_{2} = 10^{-2}$, \hspace{-4mm} \\\hspace{-4mm} $\varepsilon = 0.95$ \hspace{-4mm}\end{tabular} \\ 
& SRE  &   10.48 &   13.43 &    8.03 &   12.19 &   -1.11 &   -5.61 &   11.28 &   13.96 &   \Valbest{21.38} &   \ValSecnd{21.02} &   19.96 \\ 
& RMSE &   0.0703 &   0.0507 &   0.0964 &   0.0641 &   0.2058 &   0.2522 &   0.0819 &   0.0527 &   \Valbest{0.0211} &   \ValSecnd{0.0218} &   0.0244 \\ 
& Ps   &   0.96 &   \Valbest{1.00} &   0.86 &   0.97 &   0.31 &   0.01 &   0.95 &   0.99 &   \Valbest{1.00} &   \Valbest{1.00} &   \Valbest{1.00} \\ 
& MPSNR &   36.97 &   36.20 &   36.18 &   36.53 &   18.10 &   15.22 &   35.62 &   36.98 &   \Valbest{50.17} &   45.92 &   \ValSecnd{45.99} \\ 
& MSSIM &   0.8951 &   0.9374 &   0.8800 &   0.8778 &   0.6443 &   0.3822 &   0.9151 &   0.8939 &   \Valbest{0.9983} &   0.9886 &   \ValSecnd{0.9893} \\ 
\midrule 

\multirow{6}{*}{Case 6} 
& Setup & \begin{tabular}{c} \hspace{-4mm} $\lambda = 10^{0}$ \hspace{-4mm}\end{tabular} & \begin{tabular}{c} \hspace{-4mm} $\lambda_{1} = 10^{0}$, \hspace{-4mm} \\\hspace{-4mm} $\lambda_{2} = 10^{0}$, \hspace{-4mm} \\\hspace{-4mm} $\lambda_{3} = 10^{0}$ \hspace{-4mm}\end{tabular} & \begin{tabular}{c} \hspace{-4mm} $\lambda = 10^{1}$, \hspace{-4mm} \\\hspace{-4mm} $\lambda_{TV} = 10^{-2}$ \hspace{-4mm}\end{tabular} & \begin{tabular}{c} \hspace{-4mm} $\lambda_{g} = 10^{0}$, \hspace{-4mm} \\\hspace{-4mm} $\lambda_{l} = 10^{-1}$ \hspace{-4mm}\end{tabular} &  --  &  --  & \begin{tabular}{c} \hspace{-4mm} $\lambda = 10^{0}$ \hspace{-4mm}\end{tabular} & \begin{tabular}{c} \hspace{-4mm} $\lambda = 10^{-1}$, \hspace{-4mm} \\\hspace{-4mm} $\tau = 10^{0}$ \hspace{-4mm}\end{tabular} & \begin{tabular}{c} \hspace{-4mm} $\lambda_{1} = 10^{0}$, \hspace{-4mm} \\\hspace{-4mm} $\lambda_{2} = 10^{0}$, \hspace{-4mm} \\\hspace{-4mm} $\varepsilon = 0.95$ \hspace{-4mm}\end{tabular} & \begin{tabular}{c} \hspace{-4mm} $\lambda_{1} = 10^{1}$, \hspace{-4mm} \\\hspace{-4mm} $\lambda_{2} = 10^{-2}$, \hspace{-4mm} \\\hspace{-4mm} $\varepsilon = 0.95$ \hspace{-4mm}\end{tabular} & \begin{tabular}{c} \hspace{-4mm} $\lambda_{1} = 10^{0}$, \hspace{-4mm} \\\hspace{-4mm} $\lambda_{2} = 10^{0}$, \hspace{-4mm} \\\hspace{-4mm} $\varepsilon = 0.95$ \hspace{-4mm}\end{tabular} \\ 
& SRE  &    9.72 &   10.91 &    6.50 &   12.21 &    0.22 &   -5.65 &   11.21 &   13.31 &   \Valbest{17.14} &   14.86 &   \ValSecnd{16.80} \\ 
& RMSE &   0.0764 &   0.0670 &   0.1173 &   0.0656 &   0.1763 &   0.2521 &   0.0812 &   0.0563 &   \Valbest{0.0339} &   0.0425 &   \ValSecnd{0.0350} \\ 
& Ps   &   0.93 &   \Valbest{1.00} &   0.71 &   0.97 &   0.37 &   0.01 &   0.95 &   0.99 &   \Valbest{1.00} &   \Valbest{1.00} &   \Valbest{1.00} \\ 
& MPSNR &   35.82 &   31.08 &   35.01 &   35.58 &   19.28 &   15.23 &   35.23 &   35.76 &   41.22 &   \Valbest{42.52} &   \ValSecnd{42.03} \\ 
& MSSIM &   0.8661 &   0.8674 &   0.8454 &   0.8515 &   0.5572 &   0.3818 &   0.8849 &   0.8610 &   0.9560 &   \Valbest{0.9741} &   \ValSecnd{0.9660} \\ 
\midrule

\multirow{6}{*}{Case 7} 
& Setup & \begin{tabular}{c} \hspace{-4mm} $\lambda = 10^{0}$ \hspace{-4mm}\end{tabular} & \begin{tabular}{c} \hspace{-4mm} $\lambda_{1} = 10^{0}$, \hspace{-4mm} \\\hspace{-4mm} $\lambda_{2} = 10^{-1}$, \hspace{-4mm} \\\hspace{-4mm} $\lambda_{3} = 10^{0}$ \hspace{-4mm}\end{tabular} & \begin{tabular}{c} \hspace{-4mm} $\lambda = 10^{-2}$, \hspace{-4mm} \\\hspace{-4mm} $\lambda_{TV} = 10^{-2}$ \hspace{-4mm}\end{tabular} & \begin{tabular}{c} \hspace{-4mm} $\lambda_{g} = 10^{-1}$, \hspace{-4mm} \\\hspace{-4mm} $\lambda_{l} = 10^{-1}$ \hspace{-4mm}\end{tabular} &  --  &  --  & \begin{tabular}{c} \hspace{-4mm} $\lambda = 10^{0}$ \hspace{-4mm}\end{tabular} & \begin{tabular}{c} \hspace{-4mm} $\lambda = 10^{-1}$, \hspace{-4mm} \\\hspace{-4mm} $\tau = 10^{0}$ \hspace{-4mm}\end{tabular} & \begin{tabular}{c} \hspace{-4mm} $\lambda_{1} = 10^{0}$, \hspace{-4mm} \\\hspace{-4mm} $\lambda_{2} = 10^{1}$, \hspace{-4mm} \\\hspace{-4mm} $\varepsilon = 0.98$ \hspace{-4mm}\end{tabular} & \begin{tabular}{c} \hspace{-4mm} $\lambda_{1} = 10^{1}$, \hspace{-4mm} \\\hspace{-4mm} $\lambda_{2} = 10^{-2}$, \hspace{-4mm} \\\hspace{-4mm} $\varepsilon = 0.95$ \hspace{-4mm}\end{tabular} & \begin{tabular}{c} \hspace{-4mm} $\lambda_{1} = 10^{1}$, \hspace{-4mm} \\\hspace{-4mm} $\lambda_{2} = 10^{-1}$, \hspace{-4mm} \\\hspace{-4mm} $\varepsilon = 0.95$ \hspace{-4mm}\end{tabular} \\ 
& SRE  &   10.49 &   10.68 &    9.73 &   12.92 &   -0.87 &   -5.63 &   12.19 &   14.60 &   \ValSecnd{16.71} &   16.66 &   \Valbest{17.48} \\ 
& RMSE &   0.0715 &   0.0700 &   0.0803 &   0.0607 &   0.1991 &   0.2516 &   0.0734 &   0.0493 &   0.0357 &   \ValSecnd{0.0351} &   \Valbest{0.0323} \\ 
& Ps   &   0.94 &   0.99 &   0.93 &   0.98 &   0.34 &   0.01 &   0.97 &   \Valbest{1.00} &   \Valbest{1.00} &   \Valbest{1.00} &   \Valbest{1.00} \\ 
& MPSNR &   36.88 &   30.34 &   35.99 &   36.31 &   20.45 &   15.28 &   35.56 &   36.73 &   \ValSecnd{39.64} &   39.58 &   \Valbest{39.78} \\ 
& MSSIM &   0.8779 &   0.8063 &   0.8578 &   0.8612 &   0.5501 &   0.3838 &   0.9005 &   0.8740 &   0.9339 &   \ValSecnd{0.9378} &   \Valbest0.9406 \\ 
\midrule 

\multirow{6}{*}{Case 8} 
& Setup & \begin{tabular}{c} \hspace{-4mm} $\lambda = 10^{0}$ \hspace{-4mm}\end{tabular} & \begin{tabular}{c} \hspace{-4mm} $\lambda_{1} = 10^{1}$, \hspace{-4mm} \\\hspace{-4mm} $\lambda_{2} = 10^{1}$, \hspace{-4mm} \\\hspace{-4mm} $\lambda_{3} = 10^{1}$ \hspace{-4mm}\end{tabular} & \begin{tabular}{c} \hspace{-4mm} $\lambda = 10^{-2}$, \hspace{-4mm} \\\hspace{-4mm} $\lambda_{TV} = 10^{-2}$ \hspace{-4mm}\end{tabular} & \begin{tabular}{c} \hspace{-4mm} $\lambda_{g} = 10^{0}$, \hspace{-4mm} \\\hspace{-4mm} $\lambda_{l} = 10^{-1}$ \hspace{-4mm}\end{tabular} &  --  &  --  & \begin{tabular}{c} \hspace{-4mm} $\lambda = 10^{0}$ \hspace{-4mm}\end{tabular} & \begin{tabular}{c} \hspace{-4mm} $\lambda = 10^{-1}$, \hspace{-4mm} \\\hspace{-4mm} $\tau = 10^{0}$ \hspace{-4mm}\end{tabular} & \begin{tabular}{c} \hspace{-4mm} $\lambda_{1} = 10^{0}$, \hspace{-4mm} \\\hspace{-4mm} $\lambda_{2} = 10^{0}$, \hspace{-4mm} \\\hspace{-4mm} $\varepsilon = 0.95$ \hspace{-4mm}\end{tabular} & \begin{tabular}{c} \hspace{-4mm} $\lambda_{1} = 10^{0}$, \hspace{-4mm} \\\hspace{-4mm} $\lambda_{2} = 10^{0}$, \hspace{-4mm} \\\hspace{-4mm} $\varepsilon = 0.95$ \hspace{-4mm}\end{tabular} & \begin{tabular}{c} \hspace{-4mm} $\lambda_{1} = 10^{0}$, \hspace{-4mm} \\\hspace{-4mm} $\lambda_{2} = 10^{0}$, \hspace{-4mm} \\\hspace{-4mm} $\varepsilon = 0.95$ \hspace{-4mm}\end{tabular} \\ 
& SRE  &    8.78 &    8.81 &    5.53 &   11.15 &   -1.87 &   -5.64 &   11.01 &   12.05 &   \Valbest{14.96} &   11.89 &   \ValSecnd{14.08} \\ 
& RMSE &   0.0851 &   0.0798 &   0.1314 &   0.0732 &   0.2121 &   0.2527 &   0.0839 &   0.0642 &   \Valbest{0.0427} &   0.0592 &   \ValSecnd{0.0468} \\ 
& Ps   &   0.91 &   0.99 &   0.65 &   0.96 &   0.23 &   0.01 &   0.94 &   0.98 &   \Valbest{1.00} &   \Valbest{1.00} &   \Valbest{1.00} \\ 
& MPSNR &   34.38 &   31.16 &   33.42 &   34.02 &   19.49 &   15.25 &   34.03 &   34.15 &   \ValSecnd{40.36} &   37.76 &   \Valbest{41.30} \\ 
& MSSIM &   0.8292 &   0.7416 &   0.7969 &   0.8067 &   0.6387 &   0.3791 &   0.8517 &   0.8148 &   \ValSecnd{0.9518} &   0.9118 &   \Valbest0.9653 \\ 
\bottomrule
\end{tabular}
}
\end{table*}

\begin{table*}[!h]
	\caption{SRE, RMSE, Ps, MPSNR, and MSSIM in the Experiments Using \textit{Synth 2}.}
	\vspace{-1mm}
	\label{tab:result_synth}
	\centering
	\scalebox{0.75}{
	\begin{tabular}{ccccccccccccc} \toprule
		\multirow{3}{*}{ Noise } & \multirow{3}{*}{Metrics}& \multicolumn{11}{c}{Methods}\\ 
		\cmidrule(lr){3-13}
		& & CLSUnSAL & JSTV & RSSUn-TV & LGSU & UnDIP & EGU-Net & RDSWSU & MdLRR & \textbf{\Ours} & \textbf{\Ours} & \textbf{\Ours} \\ 
		& & \cite{iordache2014collaborative} & \cite{aggarwal2016hyperspectral} & \cite{wang2019row} & \cite{shen2022superpixel} & \cite{UnDIP_RastiB_2022} & \cite{hong2022endmember} & \cite{rs_Deng_RobustDual_2023} & \cite{MDLRR_WuLing_2023} & \textbf{(HTV)} & \textbf{(SSTV)} & \textbf{(HSSTV)} \\ 
		\midrule 
		
\multirow{6}{*}{Case 1} 
& Setup & \begin{tabular}{c} \hspace{-4mm} $\lambda = 10^{1}$ \hspace{-4mm}\end{tabular} & \begin{tabular}{c} \hspace{-4mm} $\lambda_{1} = 10^{0}$, \hspace{-4mm} \\\hspace{-4mm} $\lambda_{2} = 10^{1}$, \hspace{-4mm} \\\hspace{-4mm} $\lambda_{3} = 10^{0}$ \hspace{-4mm}\end{tabular} & \begin{tabular}{c} \hspace{-4mm} $\lambda = 10^{1}$, \hspace{-4mm} \\\hspace{-4mm} $\lambda_{TV} = 10^{-2}$ \hspace{-4mm}\end{tabular} & \begin{tabular}{c} \hspace{-4mm} $\lambda_{g} = 10^{0}$, \hspace{-4mm} \\\hspace{-4mm} $\lambda_{l} = 10^{-3}$ \hspace{-4mm}\end{tabular} &  --  &  --  & \begin{tabular}{c} \hspace{-4mm} $\lambda = 10^{-2}$ \hspace{-4mm}\end{tabular} & \begin{tabular}{c} \hspace{-4mm} $\lambda = 10^{-1}$, \hspace{-4mm} \\\hspace{-4mm} $\tau = 10^{0}$ \hspace{-4mm}\end{tabular} & \begin{tabular}{c} \hspace{-4mm} $\lambda_{1} = 10^{0}$, \hspace{-4mm} \\\hspace{-4mm} $\lambda_{2} = 10^{0}$, \hspace{-4mm} \\\hspace{-4mm} $\varepsilon = 0.98$ \hspace{-4mm}\end{tabular} & \begin{tabular}{c} \hspace{-4mm} $\lambda_{1} = 10^{0}$, \hspace{-4mm} \\\hspace{-4mm} $\lambda_{2} = 10^{-1}$, \hspace{-4mm} \\\hspace{-4mm} $\varepsilon = 0.98$ \hspace{-4mm}\end{tabular} & \begin{tabular}{c} \hspace{-4mm} $\lambda_{1} = 10^{0}$, \hspace{-4mm} \\\hspace{-4mm} $\lambda_{2} = 10^{-1}$, \hspace{-4mm} \\\hspace{-4mm} $\varepsilon = 0.98$ \hspace{-4mm}\end{tabular} \\ 
& SRE  &   18.06 &   16.65 &   19.03 &   19.78 &    1.88 &   -5.02 &   20.80 &   19.81 &   \Valbest{21.11} &   20.78 &   \ValSecnd{20.99} \\ 
& RMSE &   0.0270 &   0.0318 &   0.0252 &   0.0236 &   0.1518 &   0.2272 &   0.0212 &   0.0235 &   \Valbest{0.0198} &   0.0206 &   \ValSecnd{0.0201} \\ 
& Ps   &   \Valbest{1.00} &   \Valbest{1.00} &   \Valbest{1.00} &   \Valbest{1.00} &   0.55 &   0.00 &   \Valbest{1.00} &   \Valbest{1.00} &   \Valbest{1.00} &   \Valbest{1.00} &   \Valbest{1.00} \\ 
& MPSNR &   41.37 &   32.54 &   41.80 &   41.60 &   27.09 &   15.79 &   41.90 &   41.71 &   \Valbest{42.84} &   42.39 &   \ValSecnd{42.55} \\ 
& MSSIM &   0.9889 &   0.9435 &   0.9870 &   0.9864 &   0.8702 &   0.3498 &   0.9871 &   0.9876 &   \Valbest{0.9911} &   0.9891 &   \ValSecnd{0.9897} \\ 
\midrule 

\multirow{6}{*}{Case 2} 
& Setup & \begin{tabular}{c} \hspace{-4mm} $\lambda = 10^{1}$ \hspace{-4mm}\end{tabular} & \begin{tabular}{c} \hspace{-4mm} $\lambda_{1} = 10^{0}$, \hspace{-4mm} \\\hspace{-4mm} $\lambda_{2} = 10^{1}$, \hspace{-4mm} \\\hspace{-4mm} $\lambda_{3} = 10^{0}$ \hspace{-4mm}\end{tabular} & \begin{tabular}{c} \hspace{-4mm} $\lambda = 10^{-2}$, \hspace{-4mm} \\\hspace{-4mm} $\lambda_{TV} = 10^{-2}$ \hspace{-4mm}\end{tabular} & \begin{tabular}{c} \hspace{-4mm} $\lambda_{g} = 10^{0}$, \hspace{-4mm} \\\hspace{-4mm} $\lambda_{l} = 10^{-4}$ \hspace{-4mm}\end{tabular} &  --  &  --  & \begin{tabular}{c} \hspace{-4mm} $\lambda = 10^{-2}$ \hspace{-4mm}\end{tabular} & \begin{tabular}{c} \hspace{-4mm} $\lambda = 10^{-1}$, \hspace{-4mm} \\\hspace{-4mm} $\tau = 10^{0}$ \hspace{-4mm}\end{tabular} & \begin{tabular}{c} \hspace{-4mm} $\lambda_{1} = 10^{0}$, \hspace{-4mm} \\\hspace{-4mm} $\lambda_{2} = 10^{0}$, \hspace{-4mm} \\\hspace{-4mm} $\varepsilon = 0.98$ \hspace{-4mm}\end{tabular} & \begin{tabular}{c} \hspace{-4mm} $\lambda_{1} = 10^{0}$, \hspace{-4mm} \\\hspace{-4mm} $\lambda_{2} = 10^{-1}$, \hspace{-4mm} \\\hspace{-4mm} $\varepsilon = 0.98$ \hspace{-4mm}\end{tabular} & \begin{tabular}{c} \hspace{-4mm} $\lambda_{1} = 10^{0}$, \hspace{-4mm} \\\hspace{-4mm} $\lambda_{2} = 10^{-1}$, \hspace{-4mm} \\\hspace{-4mm} $\varepsilon = 0.98$ \hspace{-4mm}\end{tabular} \\ 
& SRE  &   15.49 &   13.97 &   13.99 &   16.01 &    0.60 &   -5.08 &   16.68 &   15.62 &   \Valbest{17.50} &   17.23 &   \ValSecnd{17.38} \\ 
& RMSE &   0.0358 &   0.0425 &   0.0449 &   0.0365 &   0.1760 &   0.2308 &   0.0342 &   0.0377 &   \Valbest{0.0297} &   0.0308 &   \ValSecnd{0.0302} \\ 
& Ps   &   \Valbest{1.00} &   0.99 &   \Valbest{1.00} &   \Valbest{1.00} &   0.48 &   0.00 &   \Valbest{1.00} &   \Valbest{1.00} &   \Valbest{1.00} &   \Valbest{1.00} &   \Valbest{1.00} \\ 
& MPSNR &   38.11 &   30.06 &   37.29 &   37.46 &   24.74 &   15.25 &   37.58 &   37.52 &   \Valbest{39.00} &   38.35 &   \ValSecnd{38.51} \\ 
& MSSIM &   0.9735 &   0.8983 &   0.9645 &   0.9654 &   0.8380 &   0.2937 &   0.9667 &   0.9665 &   \Valbest{0.9787} &   0.9731 &   \ValSecnd{0.9743} \\ 
\midrule 

\multirow{6}{*}{Case 3} 
& Setup & \begin{tabular}{c} \hspace{-4mm} $\lambda = 10^{1}$ \hspace{-4mm}\end{tabular} & \begin{tabular}{c} \hspace{-4mm} $\lambda_{1} = 10^{0}$, \hspace{-4mm} \\\hspace{-4mm} $\lambda_{2} = 10^{1}$, \hspace{-4mm} \\\hspace{-4mm} $\lambda_{3} = 10^{0}$ \hspace{-4mm}\end{tabular} & \begin{tabular}{c} \hspace{-4mm} $\lambda = 10^{-2}$, \hspace{-4mm} \\\hspace{-4mm} $\lambda_{TV} = 10^{-2}$ \hspace{-4mm}\end{tabular} & \begin{tabular}{c} \hspace{-4mm} $\lambda_{g} = 10^{0}$, \hspace{-4mm} \\\hspace{-4mm} $\lambda_{l} = 10^{-2}$ \hspace{-4mm}\end{tabular} &  --  &  --  & \begin{tabular}{c} \hspace{-4mm} $\lambda = 10^{-1}$ \hspace{-4mm}\end{tabular} & \begin{tabular}{c} \hspace{-4mm} $\lambda = 10^{-1}$, \hspace{-4mm} \\\hspace{-4mm} $\tau = 10^{0}$ \hspace{-4mm}\end{tabular} & \begin{tabular}{c} \hspace{-4mm} $\lambda_{1} = 10^{0}$, \hspace{-4mm} \\\hspace{-4mm} $\lambda_{2} = 10^{0}$, \hspace{-4mm} \\\hspace{-4mm} $\varepsilon = 0.98$ \hspace{-4mm}\end{tabular} & \begin{tabular}{c} \hspace{-4mm} $\lambda_{1} = 10^{0}$, \hspace{-4mm} \\\hspace{-4mm} $\lambda_{2} = 10^{-1}$, \hspace{-4mm} \\\hspace{-4mm} $\varepsilon = 0.98$ \hspace{-4mm}\end{tabular} & \begin{tabular}{c} \hspace{-4mm} $\lambda_{1} = 10^{0}$, \hspace{-4mm} \\\hspace{-4mm} $\lambda_{2} = 10^{-1}$, \hspace{-4mm} \\\hspace{-4mm} $\varepsilon = 0.98$ \hspace{-4mm}\end{tabular} \\ 
& SRE  &   11.62 &   16.37 &   10.81 &   12.36 &    1.69 &   -5.08 &   14.32 &   12.50 &   \Valbest{20.80} &   20.46 &   \ValSecnd{20.71} \\ 
& RMSE &   0.0529 &   0.0326 &   0.0620 &   0.0529 &   0.1586 &   0.2319 &   0.0436 &   0.0514 &   \Valbest{0.0204} &   0.0213 &   \ValSecnd{0.0207} \\ 
& Ps   &   \Valbest{1.00} &   \Valbest{1.00} &   0.97 &   0.98 &   0.62 &   0.00 &   \Valbest{1.00} &   0.99 &   \Valbest{1.00} &   \Valbest{1.00} &   \Valbest{1.00} \\ 
& MPSNR &   33.78 &   32.31 &   33.71 &   33.78 &   27.65 &   14.98 &   34.19 &   33.89 &   \Valbest{42.24} &   41.90 &   \ValSecnd{42.06} \\ 
& MSSIM &   0.9525 &   0.9444 &   0.9419 &   0.9421 &   0.8395 &   0.2635 &   0.9477 &   0.9445 &   \Valbest{0.9902} &   0.9883 &   \ValSecnd{0.9889} \\ 
\midrule 

\multirow{6}{*}{Case 4} 
& Setup & \begin{tabular}{c} \hspace{-4mm} $\lambda = 10^{1}$ \hspace{-4mm}\end{tabular} & \begin{tabular}{c} \hspace{-4mm} $\lambda_{1} = 10^{0}$, \hspace{-4mm} \\\hspace{-4mm} $\lambda_{2} = 10^{1}$, \hspace{-4mm} \\\hspace{-4mm} $\lambda_{3} = 10^{0}$ \hspace{-4mm}\end{tabular} & \begin{tabular}{c} \hspace{-4mm} $\lambda = 10^{-2}$, \hspace{-4mm} \\\hspace{-4mm} $\lambda_{TV} = 10^{-2}$ \hspace{-4mm}\end{tabular} & \begin{tabular}{c} \hspace{-4mm} $\lambda_{g} = 10^{0}$, \hspace{-4mm} \\\hspace{-4mm} $\lambda_{l} = 10^{-2}$ \hspace{-4mm}\end{tabular} &  --  &  --  & \begin{tabular}{c} \hspace{-4mm} $\lambda = 10^{-1}$ \hspace{-4mm}\end{tabular} & \begin{tabular}{c} \hspace{-4mm} $\lambda = 10^{-1}$, \hspace{-4mm} \\\hspace{-4mm} $\tau = 10^{0}$ \hspace{-4mm}\end{tabular} & \begin{tabular}{c} \hspace{-4mm} $\lambda_{1} = 10^{0}$, \hspace{-4mm} \\\hspace{-4mm} $\lambda_{2} = 10^{0}$, \hspace{-4mm} \\\hspace{-4mm} $\varepsilon = 0.98$ \hspace{-4mm}\end{tabular} & \begin{tabular}{c} \hspace{-4mm} $\lambda_{1} = 10^{0}$, \hspace{-4mm} \\\hspace{-4mm} $\lambda_{2} = 10^{-1}$, \hspace{-4mm} \\\hspace{-4mm} $\varepsilon = 0.98$ \hspace{-4mm}\end{tabular} & \begin{tabular}{c} \hspace{-4mm} $\lambda_{1} = 10^{0}$, \hspace{-4mm} \\\hspace{-4mm} $\lambda_{2} = 10^{-1}$, \hspace{-4mm} \\\hspace{-4mm} $\varepsilon = 0.98$ \hspace{-4mm}\end{tabular} \\ 
& SRE  &    8.49 &   15.93 &    7.29 &    9.18 &   -0.08 &   -5.08 &   11.89 &    9.17 &   \Valbest{19.02} &   18.65 &   \ValSecnd{18.92} \\ 
& RMSE &   0.0723 &   0.0344 &   0.0910 &   0.0732 &   0.1784 &   0.2332 &   0.0565 &   0.0722 &   \Valbest{0.0251} &   0.0262 &   \ValSecnd{0.0254} \\ 
& Ps   &   0.97 &   \Valbest{1.00} &   0.88 &   0.94 &   0.43 &   0.00 &   0.98 &   0.95 &   \Valbest{1.00} &   \Valbest{1.00} &   \Valbest{1.00} \\ 
& MPSNR &   30.06 &   31.42 &   29.91 &   30.03 &   24.84 &   14.71 &   30.48 &   30.07 &   \Valbest{41.30} &   40.80 &   \ValSecnd{40.97} \\ 
& MSSIM &   0.9143 &   0.9299 &   0.8993 &   0.9001 &   0.7894 &   0.2472 &   0.9096 &   0.9025 &   \Valbest{0.9871} &   0.9844 &   \ValSecnd{0.9852} \\ 
\midrule 

\multirow{6}{*}{Case 5} 
& Setup & \begin{tabular}{c} \hspace{-4mm} $\lambda = 10^{1}$ \hspace{-4mm}\end{tabular} & \begin{tabular}{c} \hspace{-4mm} $\lambda_{1} = 10^{0}$, \hspace{-4mm} \\\hspace{-4mm} $\lambda_{2} = 10^{1}$, \hspace{-4mm} \\\hspace{-4mm} $\lambda_{3} = 10^{0}$ \hspace{-4mm}\end{tabular} & \begin{tabular}{c} \hspace{-4mm} $\lambda = 10^{-2}$, \hspace{-4mm} \\\hspace{-4mm} $\lambda_{TV} = 10^{-2}$ \hspace{-4mm}\end{tabular} & \begin{tabular}{c} \hspace{-4mm} $\lambda_{g} = 10^{0}$, \hspace{-4mm} \\\hspace{-4mm} $\lambda_{l} = 10^{-2}$ \hspace{-4mm}\end{tabular} &  --  &  --  & \begin{tabular}{c} \hspace{-4mm} $\lambda = 10^{-1}$ \hspace{-4mm}\end{tabular} & \begin{tabular}{c} \hspace{-4mm} $\lambda = 10^{-1}$, \hspace{-4mm} \\\hspace{-4mm} $\tau = 10^{0}$ \hspace{-4mm}\end{tabular} & \begin{tabular}{c} \hspace{-4mm} $\lambda_{1} = 10^{0}$, \hspace{-4mm} \\\hspace{-4mm} $\lambda_{2} = 10^{0}$, \hspace{-4mm} \\\hspace{-4mm} $\varepsilon = 0.98$ \hspace{-4mm}\end{tabular} & \begin{tabular}{c} \hspace{-4mm} $\lambda_{1} = 10^{0}$, \hspace{-4mm} \\\hspace{-4mm} $\lambda_{2} = 10^{-1}$, \hspace{-4mm} \\\hspace{-4mm} $\varepsilon = 0.98$ \hspace{-4mm}\end{tabular} & \begin{tabular}{c} \hspace{-4mm} $\lambda_{1} = 10^{0}$, \hspace{-4mm} \\\hspace{-4mm} $\lambda_{2} = 10^{-1}$, \hspace{-4mm} \\\hspace{-4mm} $\varepsilon = 0.98$ \hspace{-4mm}\end{tabular} \\ 
& SRE  &   11.95 &   17.36 &   11.27 &   12.68 &    1.63 &   -5.09 &   14.25 &   12.86 &   \Valbest{21.52} &   21.27 &   \ValSecnd{21.42} \\ 
& RMSE &   0.0513 &   0.0294 &   0.0590 &   0.0512 &   0.1553 &   0.2323 &   0.0439 &   0.0495 &   \Valbest{0.0189} &   0.0195 &   \ValSecnd{0.0191} \\ 
& Ps   &   \Valbest{1.00} &   \Valbest{1.00} &   0.98 &   0.99 &   0.53 &   0.00 &   \Valbest{1.00} &   0.99 &   \Valbest{1.00} &   \Valbest{1.00} &   \Valbest{1.00} \\ 
& MPSNR &   33.88 &   32.87 &   33.84 &   33.90 &   25.62 &   14.95 &   34.24 &   34.02 &   \Valbest{42.83} &   42.51 &   \ValSecnd{42.65} \\ 
& MSSIM &   0.9530 &   0.9509 &   0.9428 &   0.9429 &   0.8397 &   0.2607 &   0.9484 &   0.9453 &   \Valbest{0.9909} &   0.9890 &   \ValSecnd{0.9895} \\ 
\midrule 

\multirow{6}{*}{Case 6} 
& Setup & \begin{tabular}{c} \hspace{-4mm} $\lambda = 10^{1}$ \hspace{-4mm}\end{tabular} & \begin{tabular}{c} \hspace{-4mm} $\lambda_{1} = 10^{0}$, \hspace{-4mm} \\\hspace{-4mm} $\lambda_{2} = 10^{1}$, \hspace{-4mm} \\\hspace{-4mm} $\lambda_{3} = 10^{0}$ \hspace{-4mm}\end{tabular} & \begin{tabular}{c} \hspace{-4mm} $\lambda = 10^{-2}$, \hspace{-4mm} \\\hspace{-4mm} $\lambda_{TV} = 10^{-2}$ \hspace{-4mm}\end{tabular} & \begin{tabular}{c} \hspace{-4mm} $\lambda_{g} = 10^{-1}$, \hspace{-4mm} \\\hspace{-4mm} $\lambda_{l} = 10^{-1}$ \hspace{-4mm}\end{tabular} &  --  &  --  & \begin{tabular}{c} \hspace{-4mm} $\lambda = 10^{-1}$ \hspace{-4mm}\end{tabular} & \begin{tabular}{c} \hspace{-4mm} $\lambda = 10^{-1}$, \hspace{-4mm} \\\hspace{-4mm} $\tau = 10^{0}$ \hspace{-4mm}\end{tabular} & \begin{tabular}{c} \hspace{-4mm} $\lambda_{1} = 10^{0}$, \hspace{-4mm} \\\hspace{-4mm} $\lambda_{2} = 10^{0}$, \hspace{-4mm} \\\hspace{-4mm} $\varepsilon = 0.98$ \hspace{-4mm}\end{tabular} & \begin{tabular}{c} \hspace{-4mm} $\lambda_{1} = 10^{0}$, \hspace{-4mm} \\\hspace{-4mm} $\lambda_{2} = 10^{-2}$, \hspace{-4mm} \\\hspace{-4mm} $\varepsilon = 0.98$ \hspace{-4mm}\end{tabular} & \begin{tabular}{c} \hspace{-4mm} $\lambda_{1} = 10^{0}$, \hspace{-4mm} \\\hspace{-4mm} $\lambda_{2} = 10^{-2}$, \hspace{-4mm} \\\hspace{-4mm} $\varepsilon = 0.98$ \hspace{-4mm}\end{tabular} \\ 
& SRE  &   11.08 &   13.75 &    9.03 &   11.44 &   -0.74 &   -5.10 &   13.62 &   11.43 &   \Valbest{17.96} &   17.55 &   \ValSecnd{17.81} \\ 
& RMSE &   0.0564 &   0.0434 &   0.0776 &   0.0599 &   0.1962 &   0.2331 &   0.0477 &   0.0584 &   \Valbest{0.0282} &   0.0296 &   \ValSecnd{0.0288} \\ 
& Ps   &   0.99 &   0.99 &   0.92 &   0.97 &   0.39 &   0.00 &   0.99 &   0.98 &   \Valbest{1.00} &   \Valbest{1.00} &   \Valbest{1.00} \\ 
& MPSNR &   33.06 &   29.63 &   32.71 &   32.84 &   26.02 &   14.78 &   33.28 &   32.93 &   \Valbest{38.55} &   38.14 &   \ValSecnd{38.24} \\ 
& MSSIM &   0.9401 &   0.8945 &   0.9254 &   0.9282 &   0.8051 &   0.2515 &   0.9327 &   0.9284 &   \Valbest{0.9798} &   0.9724 &   \ValSecnd{0.9732} \\ 
\midrule 

\multirow{6}{*}{Case 7} 
& Setup & \begin{tabular}{c} \hspace{-4mm} $\lambda = 10^{1}$ \hspace{-4mm}\end{tabular} & \begin{tabular}{c} \hspace{-4mm} $\lambda_{1} = 10^{0}$, \hspace{-4mm} \\\hspace{-4mm} $\lambda_{2} = 10^{1}$, \hspace{-4mm} \\\hspace{-4mm} $\lambda_{3} = 10^{0}$ \hspace{-4mm}\end{tabular} & \begin{tabular}{c} \hspace{-4mm} $\lambda = 10^{-2}$, \hspace{-4mm} \\\hspace{-4mm} $\lambda_{TV} = 10^{-2}$ \hspace{-4mm}\end{tabular} & \begin{tabular}{c} \hspace{-4mm} $\lambda_{g} = 10^{0}$, \hspace{-4mm} \\\hspace{-4mm} $\lambda_{l} = 10^{-2}$ \hspace{-4mm}\end{tabular} &  --  &  --  & \begin{tabular}{c} \hspace{-4mm} $\lambda = 10^{-1}$ \hspace{-4mm}\end{tabular} & \begin{tabular}{c} \hspace{-4mm} $\lambda = 10^{-1}$, \hspace{-4mm} \\\hspace{-4mm} $\tau = 10^{0}$ \hspace{-4mm}\end{tabular} & \begin{tabular}{c} \hspace{-4mm} $\lambda_{1} = 10^{0}$, \hspace{-4mm} \\\hspace{-4mm} $\lambda_{2} = 10^{-1}$, \hspace{-4mm} \\\hspace{-4mm} $\varepsilon = 0.98$ \hspace{-4mm}\end{tabular} & \begin{tabular}{c} \hspace{-4mm} $\lambda_{1} = 10^{0}$, \hspace{-4mm} \\\hspace{-4mm} $\lambda_{2} = 10^{-1}$, \hspace{-4mm} \\\hspace{-4mm} $\varepsilon = 0.98$ \hspace{-4mm}\end{tabular} & \begin{tabular}{c} \hspace{-4mm} $\lambda_{1} = 10^{0}$, \hspace{-4mm} \\\hspace{-4mm} $\lambda_{2} = 10^{-1}$, \hspace{-4mm} \\\hspace{-4mm} $\varepsilon = 0.98$ \hspace{-4mm}\end{tabular} \\ 
& SRE  &   12.85 &   11.74 &   10.75 &   13.58 &    0.68 &   -5.10 &   14.65 &   13.25 &   15.49 &   \ValSecnd{15.68} &   \Valbest{16.05} \\ 
& RMSE &   0.0477 &   0.0531 &   0.0649 &   0.0483 &   0.1714 &   0.2326 &   0.0437 &   0.0495 &   0.0370 &   \ValSecnd{0.0362} &   \Valbest{0.0348} \\ 
& Ps   &   \Valbest{1.00} &   0.99 &   0.96 &   0.99 &   0.46 &   0.00 &   0.99 &   0.99 &   \Valbest{1.00} &   \Valbest{1.00} &   \Valbest{1.00} \\ 
& MPSNR &   35.24 &   27.62 &   34.22 &   34.53 &   25.30 &   14.92 &   34.93 &   34.63 &   \Valbest{36.76} &   35.77 &   \ValSecnd{36.11} \\ 
& MSSIM &   0.9488 &   0.8577 &   0.9334 &   0.9365 &   0.8282 &   0.2464 &   0.9425 &   0.9382 &   \Valbest{0.9664} &   0.9531 &   \ValSecnd{0.9570} \\ 
\midrule 

\multirow{6}{*}{Case 8} 
& Setup & \begin{tabular}{c} \hspace{-4mm} $\lambda = 10^{1}$ \hspace{-4mm}\end{tabular} & \begin{tabular}{c} \hspace{-4mm} $\lambda_{1} = 10^{0}$, \hspace{-4mm} \\\hspace{-4mm} $\lambda_{2} = 10^{1}$, \hspace{-4mm} \\\hspace{-4mm} $\lambda_{3} = 10^{0}$ \hspace{-4mm}\end{tabular} & \begin{tabular}{c} \hspace{-4mm} $\lambda = 10^{-2}$, \hspace{-4mm} \\\hspace{-4mm} $\lambda_{TV} = 10^{-2}$ \hspace{-4mm}\end{tabular} & \begin{tabular}{c} \hspace{-4mm} $\lambda_{g} = 10^{0}$, \hspace{-4mm} \\\hspace{-4mm} $\lambda_{l} = 10^{-1}$ \hspace{-4mm}\end{tabular} &  --  &  --  & \begin{tabular}{c} \hspace{-4mm} $\lambda = 10^{-1}$ \hspace{-4mm}\end{tabular} & \begin{tabular}{c} \hspace{-4mm} $\lambda = 10^{-1}$, \hspace{-4mm} \\\hspace{-4mm} $\tau = 10^{0}$ \hspace{-4mm}\end{tabular} & \begin{tabular}{c} \hspace{-4mm} $\lambda_{1} = 10^{0}$, \hspace{-4mm} \\\hspace{-4mm} $\lambda_{2} = 10^{-1}$, \hspace{-4mm} \\\hspace{-4mm} $\varepsilon = 0.98$ \hspace{-4mm}\end{tabular} & \begin{tabular}{c} \hspace{-4mm} $\lambda_{1} = 10^{0}$, \hspace{-4mm} \\\hspace{-4mm} $\lambda_{2} = 10^{-2}$, \hspace{-4mm} \\\hspace{-4mm} $\varepsilon = 0.98$ \hspace{-4mm}\end{tabular} & \begin{tabular}{c} \hspace{-4mm} $\lambda_{1} = 10^{0}$, \hspace{-4mm} \\\hspace{-4mm} $\lambda_{2} = 10^{-2}$, \hspace{-4mm} \\\hspace{-4mm} $\varepsilon = 0.98$ \hspace{-4mm}\end{tabular} \\ 
& SRE  &    9.67 &   10.96 &    7.13 &   10.72 &    3.34 &   -5.09 &   12.50 &   10.12 &   12.99 &   \ValSecnd{14.26} &   \Valbest{14.60} \\ 
& RMSE &   0.0655 &   0.0576 &   0.0976 &   0.0650 &   0.1275 &   0.2333 &   0.0547 &   0.0682 &   0.0471 &   \ValSecnd{0.0415} &   \Valbest{0.0405} \\ 
& Ps   &   0.98 &   0.98 &   0.85 &   0.96 &   0.66 &   0.00 &   0.98 &   0.95 &   0.98 &   0.99 &   0.99 \\ 
& MPSNR &   31.72 &   26.70 &   31.17 &   31.50 &   27.93 &   14.73 &   32.01 &   31.48 &   34.17 &   \ValSecnd{35.12} &   \Valbest{35.21} \\ 
& MSSIM &   0.9172 &   0.8258 &   0.8977 &   0.9054 &   0.8409 &   0.2389 &   0.9102 &   0.9029 &   \Valbest{0.9575} &   0.9517 &   \ValSecnd{0.9525} \\ 
		\bottomrule
	\end{tabular}
}
\end{table*}

\begin{table*}[!h]
	\caption{SRE, RMSE, Ps, MPSNR, and MSSIM in the Experiments Using \textit{Synth 3}.}
	\vspace{-1mm}
	\label{tab:result_synth_3}
	\centering
	\scalebox{0.75}{
		\begin{tabular}{ccccccccccccc} \toprule
			\multirow{3}{*}{ Noise } & \multirow{3}{*}{Metrics}& \multicolumn{11}{c}{Methods}\\ 
			\cmidrule(lr){3-13}
			& & CLSUnSAL & JSTV & RSSUn-TV & LGSU & UnDIP & EGU-Net & RDSWSU & MdLRR & \textbf{\Ours} & \textbf{\Ours} & \textbf{\Ours} \\ 
			& & \cite{iordache2014collaborative} & \cite{aggarwal2016hyperspectral} & \cite{wang2019row} & \cite{shen2022superpixel} & \cite{UnDIP_RastiB_2022} & \cite{hong2022endmember} & \cite{rs_Deng_RobustDual_2023} & \cite{MDLRR_WuLing_2023} & \textbf{(HTV)} & \textbf{(SSTV)} & \textbf{(HSSTV)} \\ 
			\midrule 
			
			\multirow{6}{*}{Case 1} 
			& Setup & \begin{tabular}{c} \hspace{-4mm} $\lambda = 10^{0}$ \hspace{-4mm}\end{tabular} & \begin{tabular}{c} \hspace{-4mm} $\lambda_{1} = 10^{1}$, \hspace{-4mm} \\\hspace{-4mm} $\lambda_{2} = 10^{-2}$, \hspace{-4mm} \\\hspace{-4mm} $\lambda_{3} = 10^{1}$ \hspace{-4mm}\end{tabular} & \begin{tabular}{c} \hspace{-4mm} $\lambda = 10^{0}$, \hspace{-4mm} \\\hspace{-4mm} $\lambda_{TV} = 10^{-2}$ \hspace{-4mm}\end{tabular} & \begin{tabular}{c} \hspace{-4mm} $\lambda_{g} = 10^{0}$, \hspace{-4mm} \\\hspace{-4mm} $\lambda_{l} = 10^{-2}$ \hspace{-4mm}\end{tabular} &  --  &  --  & \begin{tabular}{c} \hspace{-4mm} $\lambda = 10^{-1}$ \hspace{-4mm}\end{tabular} & \begin{tabular}{c} \hspace{-4mm} $\lambda = 10^{-1}$, \hspace{-4mm} \\\hspace{-4mm} $\tau = 10^{0}$ \hspace{-4mm}\end{tabular} & \begin{tabular}{c} \hspace{-4mm} $\lambda_{1} = 10^{-1}$, \hspace{-4mm} \\\hspace{-4mm} $\lambda_{2} = 10^{0}$, \hspace{-4mm} \\\hspace{-4mm} $\varepsilon = 0.95$ \hspace{-4mm}\end{tabular} & \begin{tabular}{c} \hspace{-4mm} $\lambda_{1} = 10^{0}$, \hspace{-4mm} \\\hspace{-4mm} $\lambda_{2} = 10^{-1}$, \hspace{-4mm} \\\hspace{-4mm} $\varepsilon = 0.95$ \hspace{-4mm}\end{tabular} & \begin{tabular}{c} \hspace{-4mm} $\lambda_{1} = 10^{-1}$, \hspace{-4mm} \\\hspace{-4mm} $\lambda_{2} = 10^{-1}$, \hspace{-4mm} \\\hspace{-4mm} $\varepsilon = 0.95$ \hspace{-4mm}\end{tabular} \\ 
			& SRE  &   21.68 &   -0.00 &   23.03 &   22.90 &   -0.49 &   -3.64 &   25.54 &   24.21 & \Valbest{  30.26} &   26.38 & \ValSecnd{  27.58} \\ 
			& RMSE &   0.0045 &   1.8879 &   0.0039 &   0.0040 &   0.0380 &   0.0520 &   0.0030 &   0.0034 & \Valbest{  0.0017} &   0.0026 & \ValSecnd{  0.0023} \\ 
			& Ps   & \Valbest{  1.00} &   0.00 & \Valbest{  1.00} & \Valbest{  1.00} & {  0.52} &   0.02 & \Valbest{  1.00} & \Valbest{  1.00} & \Valbest{  1.00} & \Valbest{  1.00} & \Valbest{  1.00} \\ 
			& MPSNR &   42.76 &  -44.42 &   42.72 &   43.14 &   18.91 &   12.43 &   45.54 &   43.10 & \Valbest{  56.00} &   46.05 & \ValSecnd{  46.49} \\ 
			& MSSIM &   0.9614 &   0.7841 &   0.9556 &   0.9587 &   0.6812 &   0.6928 &   0.9721 &   0.9647 & \Valbest{  0.9997} &   0.9822 & \ValSecnd{  0.9834} \\ 
			\midrule 
			
			\multirow{6}{*}{Case 2} 
			& Setup & \begin{tabular}{c} \hspace{-4mm} $\lambda = 10^{0}$ \hspace{-4mm}\end{tabular} & \begin{tabular}{c} \hspace{-4mm} $\lambda_{1} = 10^{-2}$, \hspace{-4mm} \\\hspace{-4mm} $\lambda_{2} = 10^{-1}$, \hspace{-4mm} \\\hspace{-4mm} $\lambda_{3} = 10^{1}$ \hspace{-4mm}\end{tabular} & \begin{tabular}{c} \hspace{-4mm} $\lambda = 10^{1}$, \hspace{-4mm} \\\hspace{-4mm} $\lambda_{TV} = 10^{-2}$ \hspace{-4mm}\end{tabular} & \begin{tabular}{c} \hspace{-4mm} $\lambda_{g} = 10^{0}$, \hspace{-4mm} \\\hspace{-4mm} $\lambda_{l} = 10^{-2}$ \hspace{-4mm}\end{tabular} &  --  &  --  & \begin{tabular}{c} \hspace{-4mm} $\lambda = 10^{-1}$ \hspace{-4mm}\end{tabular} & \begin{tabular}{c} \hspace{-4mm} $\lambda = 10^{-1}$, \hspace{-4mm} \\\hspace{-4mm} $\tau = 10^{0}$ \hspace{-4mm}\end{tabular} & \begin{tabular}{c} \hspace{-4mm} $\lambda_{1} = 10^{-1}$, \hspace{-4mm} \\\hspace{-4mm} $\lambda_{2} = 10^{0}$, \hspace{-4mm} \\\hspace{-4mm} $\varepsilon = 0.95$ \hspace{-4mm}\end{tabular} & \begin{tabular}{c} \hspace{-4mm} $\lambda_{1} = 10^{0}$, \hspace{-4mm} \\\hspace{-4mm} $\lambda_{2} = 10^{-1}$, \hspace{-4mm} \\\hspace{-4mm} $\varepsilon = 0.95$ \hspace{-4mm}\end{tabular} & \begin{tabular}{c} \hspace{-4mm} $\lambda_{1} = 10^{-1}$, \hspace{-4mm} \\\hspace{-4mm} $\lambda_{2} = 10^{-1}$, \hspace{-4mm} \\\hspace{-4mm} $\varepsilon = 0.98$ \hspace{-4mm}\end{tabular} \\ 
			& SRE  &   16.96 &   -0.00 &   15.10 &   16.82 &   -3.02 &   -0.35 &   20.97 &   18.83 & \Valbest{  28.67} &   22.34 & \ValSecnd{  22.72} \\ 
			& RMSE &   0.0076 &   1.6608 &   0.0095 &   0.0080 &   0.0410 &   0.0346 &   0.0050 &   0.0063 & \Valbest{  0.0020} &   0.0041 & \ValSecnd{  0.0040} \\ 
			& Ps   & \Valbest{  1.00} &   0.00 & \Valbest{  1.00} & \Valbest{  1.00} &   0.02 & {  0.53} & \Valbest{  1.00} & \Valbest{  1.00} & \Valbest{  1.00} & \Valbest{  1.00} & \Valbest{  1.00} \\ 
			& MPSNR &   37.22 &  -43.74 &   36.07 &   36.67 &   18.26 &   15.57 &   40.13 &   35.15 & \Valbest{  52.34} &   40.45 & \ValSecnd{  42.56} \\ 
			& MSSIM &   0.8721 &   0.4809 &   0.8288 &   0.8415 &   0.7765 &   0.8055 &   0.9052 &   0.8377 & \Valbest{  0.9991} &   0.9361 & \ValSecnd{  0.9624} \\ 
			\midrule 
			
			\multirow{6}{*}{Case 3} 
			& Setup & \begin{tabular}{c} \hspace{-4mm} $\lambda = 10^{0}$ \hspace{-4mm}\end{tabular} & \begin{tabular}{c} \hspace{-4mm} $\lambda_{1} = 10^{-2}$, \hspace{-4mm} \\\hspace{-4mm} $\lambda_{2} = 10^{-2}$, \hspace{-4mm} \\\hspace{-4mm} $\lambda_{3} = 10^{1}$ \hspace{-4mm}\end{tabular} & \begin{tabular}{c} \hspace{-4mm} $\lambda = 10^{1}$, \hspace{-4mm} \\\hspace{-4mm} $\lambda_{TV} = 10^{-2}$ \hspace{-4mm}\end{tabular} & \begin{tabular}{c} \hspace{-4mm} $\lambda_{g} = 10^{1}$, \hspace{-4mm} \\\hspace{-4mm} $\lambda_{l} = 10^{-1}$ \hspace{-4mm}\end{tabular} &  --  &  --  & \begin{tabular}{c} \hspace{-4mm} $\lambda = 10^{-1}$ \hspace{-4mm}\end{tabular} & \begin{tabular}{c} \hspace{-4mm} $\lambda = 10^{-1}$, \hspace{-4mm} \\\hspace{-4mm} $\tau = 10^{0}$ \hspace{-4mm}\end{tabular} & \begin{tabular}{c} \hspace{-4mm} $\lambda_{1} = 10^{-1}$, \hspace{-4mm} \\\hspace{-4mm} $\lambda_{2} = 10^{0}$, \hspace{-4mm} \\\hspace{-4mm} $\varepsilon = 0.95$ \hspace{-4mm}\end{tabular} & \begin{tabular}{c} \hspace{-4mm} $\lambda_{1} = 10^{0}$, \hspace{-4mm} \\\hspace{-4mm} $\lambda_{2} = 10^{-1}$, \hspace{-4mm} \\\hspace{-4mm} $\varepsilon = 0.95$ \hspace{-4mm}\end{tabular} & \begin{tabular}{c} \hspace{-4mm} $\lambda_{1} = 10^{-1}$, \hspace{-4mm} \\\hspace{-4mm} $\lambda_{2} = 10^{-1}$, \hspace{-4mm} \\\hspace{-4mm} $\varepsilon = 0.98$ \hspace{-4mm}\end{tabular} \\ 
			& SRE  &   13.12 &   -0.00 &   10.65 &   13.66 &   -3.83 &   -4.33 &   15.63 &   13.97 & \Valbest{  29.66} &   25.91 & \ValSecnd{  26.93} \\ 
			& RMSE &   0.0111 &   2.1084 &   0.0152 &   0.0112 &   0.0429 &   0.0479 &   0.0090 &   0.0105 & \Valbest{  0.0018} &   0.0028 & \ValSecnd{  0.0025} \\ 
			& Ps   & \ValSecnd{  0.99} &   0.00 &   0.98 &   0.97 &   0.02 &   0.02 & \Valbest{  1.00} & {  0.99} & \Valbest{  1.00} & \Valbest{  1.00} & \Valbest{  1.00} \\ 
			& MPSNR &   32.00 &  -45.39 &   31.33 &   32.49 &   20.44 &   12.87 &   34.41 &   29.86 & \Valbest{  55.27} &   45.23 & \ValSecnd{  47.14} \\ 
			& MSSIM &   0.7730 &   0.5410 &   0.7136 &   0.7836 &   0.5486 &   0.6956 &   0.8264 &   0.6874 & \Valbest{  0.9996} &   0.9783 & \ValSecnd{  0.9870} \\ 
			\midrule 
			
			\multirow{6}{*}{Case 4} 
			& Setup & \begin{tabular}{c} \hspace{-4mm} $\lambda = 10^{0}$ \hspace{-4mm}\end{tabular} & \begin{tabular}{c} \hspace{-4mm} $\lambda_{1} = 10^{-2}$, \hspace{-4mm} \\\hspace{-4mm} $\lambda_{2} = 10^{-2}$, \hspace{-4mm} \\\hspace{-4mm} $\lambda_{3} = 10^{1}$ \hspace{-4mm}\end{tabular} & \begin{tabular}{c} \hspace{-4mm} $\lambda = 10^{1}$, \hspace{-4mm} \\\hspace{-4mm} $\lambda_{TV} = 10^{-2}$ \hspace{-4mm}\end{tabular} & \begin{tabular}{c} \hspace{-4mm} $\lambda_{g} = 10^{1}$, \hspace{-4mm} \\\hspace{-4mm} $\lambda_{l} = 10^{-1}$ \hspace{-4mm}\end{tabular} &  --  &  --  & \begin{tabular}{c} \hspace{-4mm} $\lambda = 10^{1}$ \hspace{-4mm}\end{tabular} & \begin{tabular}{c} \hspace{-4mm} $\lambda = 10^{-1}$, \hspace{-4mm} \\\hspace{-4mm} $\tau = 10^{0}$ \hspace{-4mm}\end{tabular} & \begin{tabular}{c} \hspace{-4mm} $\lambda_{1} = 10^{-1}$, \hspace{-4mm} \\\hspace{-4mm} $\lambda_{2} = 10^{0}$, \hspace{-4mm} \\\hspace{-4mm} $\varepsilon = 0.95$ \hspace{-4mm}\end{tabular} & \begin{tabular}{c} \hspace{-4mm} $\lambda_{1} = 10^{0}$, \hspace{-4mm} \\\hspace{-4mm} $\lambda_{2} = 10^{-1}$, \hspace{-4mm} \\\hspace{-4mm} $\varepsilon = 0.95$ \hspace{-4mm}\end{tabular} & \begin{tabular}{c} \hspace{-4mm} $\lambda_{1} = 10^{-1}$, \hspace{-4mm} \\\hspace{-4mm} $\lambda_{2} = 10^{-1}$, \hspace{-4mm} \\\hspace{-4mm} $\varepsilon = 0.98$ \hspace{-4mm}\end{tabular} \\ 
			& SRE  &    9.59 &   -0.00 &    6.88 &    9.73 &   -6.48 &   -3.74 &   10.14 &    9.44 & \Valbest{  30.92} &   25.52 & \ValSecnd{  26.68} \\ 
			& RMSE &   0.0156 &   2.0454 &   0.0225 &   0.0161 &   0.0489 &   0.0432 &   0.0167 &   0.0169 & \Valbest{  0.0016} &   0.0029 & \ValSecnd{  0.0025} \\ 
			& Ps   & {  0.96} &   0.00 &   0.87 &   0.94 &   0.00 &   0.04 &   0.92 &   0.95 & \Valbest{  1.00} & \Valbest{  1.00} & \Valbest{  1.00} \\ 
			& MPSNR &   27.91 &  -45.13 &   27.34 &   28.16 &   16.13 &   15.71 &   30.11 &   25.25 & \Valbest{  55.21} &   44.34 & \ValSecnd{  45.62} \\ 
			& MSSIM &   0.6530 &   0.4900 &   0.5783 &   0.6521 &   0.4271 &   0.7979 &   0.8082 &   0.5151 & \Valbest{  0.9996} &   0.9730 & \ValSecnd{  0.9803} \\ 
			\midrule 
			
			\multirow{6}{*}{Case 5} 
			& Setup & \begin{tabular}{c} \hspace{-4mm} $\lambda = 10^{0}$ \hspace{-4mm}\end{tabular} & \begin{tabular}{c} \hspace{-4mm} $\lambda_{1} = 10^{-1}$, \hspace{-4mm} \\\hspace{-4mm} $\lambda_{2} = 10^{0}$, \hspace{-4mm} \\\hspace{-4mm} $\lambda_{3} = 10^{1}$ \hspace{-4mm}\end{tabular} & \begin{tabular}{c} \hspace{-4mm} $\lambda = 10^{1}$, \hspace{-4mm} \\\hspace{-4mm} $\lambda_{TV} = 10^{-2}$ \hspace{-4mm}\end{tabular} & \begin{tabular}{c} \hspace{-4mm} $\lambda_{g} = 10^{1}$, \hspace{-4mm} \\\hspace{-4mm} $\lambda_{l} = 10^{-1}$ \hspace{-4mm}\end{tabular} &  --  &  --  & \begin{tabular}{c} \hspace{-4mm} $\lambda = 10^{-1}$ \hspace{-4mm}\end{tabular} & \begin{tabular}{c} \hspace{-4mm} $\lambda = 10^{-1}$, \hspace{-4mm} \\\hspace{-4mm} $\tau = 10^{0}$ \hspace{-4mm}\end{tabular} & \begin{tabular}{c} \hspace{-4mm} $\lambda_{1} = 10^{0}$, \hspace{-4mm} \\\hspace{-4mm} $\lambda_{2} = 10^{1}$, \hspace{-4mm} \\\hspace{-4mm} $\varepsilon = 0.95$ \hspace{-4mm}\end{tabular} & \begin{tabular}{c} \hspace{-4mm} $\lambda_{1} = 10^{0}$, \hspace{-4mm} \\\hspace{-4mm} $\lambda_{2} = 10^{-1}$, \hspace{-4mm} \\\hspace{-4mm} $\varepsilon = 0.98$ \hspace{-4mm}\end{tabular} & \begin{tabular}{c} \hspace{-4mm} $\lambda_{1} = 10^{0}$, \hspace{-4mm} \\\hspace{-4mm} $\lambda_{2} = 10^{-1}$, \hspace{-4mm} \\\hspace{-4mm} $\varepsilon = 0.98$ \hspace{-4mm}\end{tabular} \\ 
			& SRE  &   12.80 &   -0.00 &   10.15 &   13.66 &   -5.29 &   -4.01 &   15.24 &   13.47 & \Valbest{  24.96} &   21.46 & \ValSecnd{  21.54} \\ 
			& RMSE &   0.0115 &   1.7139 &   0.0161 &   0.0111 &   0.0461 &   0.0470 &   0.0094 &   0.0112 & \Valbest{  0.0031} &   0.0046 & \ValSecnd{  0.0045} \\ 
			& Ps   & \ValSecnd{  0.99} &   0.00 &   0.97 &   0.97 &   0.00 &   0.04 & \Valbest{  1.00} & {  0.99} & \Valbest{  1.00} & \Valbest{  1.00} & \Valbest{  1.00} \\ 
			& MPSNR &   31.76 &  -44.03 &   31.01 &   32.21 &   15.25 &   13.15 &   34.15 &   29.38 & \Valbest{  47.52} &   41.18 & \ValSecnd{  41.51} \\ 
			& MSSIM &   0.7597 &   0.4738 &   0.6970 &   0.7688 &   0.4620 &   0.7016 &   0.8153 &   0.6696 & \Valbest{  0.9980} &   0.9507 & \ValSecnd{  0.9550} \\ 
			\midrule 
			
			\multirow{6}{*}{Case 6} 
			& Setup & \begin{tabular}{c} \hspace{-4mm} $\lambda = 10^{0}$ \hspace{-4mm}\end{tabular} & \begin{tabular}{c} \hspace{-4mm} $\lambda_{1} = 10^{-2}$, \hspace{-4mm} \\\hspace{-4mm} $\lambda_{2} = 10^{-1}$, \hspace{-4mm} \\\hspace{-4mm} $\lambda_{3} = 10^{1}$ \hspace{-4mm}\end{tabular} & \begin{tabular}{c} \hspace{-4mm} $\lambda = 10^{1}$, \hspace{-4mm} \\\hspace{-4mm} $\lambda_{TV} = 10^{-2}$ \hspace{-4mm}\end{tabular} & \begin{tabular}{c} \hspace{-4mm} $\lambda_{g} = 10^{1}$, \hspace{-4mm} \\\hspace{-4mm} $\lambda_{l} = 10^{-1}$ \hspace{-4mm}\end{tabular} &  --  &  --  & \begin{tabular}{c} \hspace{-4mm} $\lambda = 10^{-1}$ \hspace{-4mm}\end{tabular} & \begin{tabular}{c} \hspace{-4mm} $\lambda = 10^{-1}$, \hspace{-4mm} \\\hspace{-4mm} $\tau = 10^{0}$ \hspace{-4mm}\end{tabular} & \begin{tabular}{c} \hspace{-4mm} $\lambda_{1} = 10^{0}$, \hspace{-4mm} \\\hspace{-4mm} $\lambda_{2} = 10^{0}$, \hspace{-4mm} \\\hspace{-4mm} $\varepsilon = 0.98$ \hspace{-4mm}\end{tabular} & \begin{tabular}{c} \hspace{-4mm} $\lambda_{1} = 10^{0}$, \hspace{-4mm} \\\hspace{-4mm} $\lambda_{2} = 10^{-1}$, \hspace{-4mm} \\\hspace{-4mm} $\varepsilon = 0.98$ \hspace{-4mm}\end{tabular} & \begin{tabular}{c} \hspace{-4mm} $\lambda_{1} = 10^{0}$, \hspace{-4mm} \\\hspace{-4mm} $\lambda_{2} = 10^{-1}$, \hspace{-4mm} \\\hspace{-4mm} $\varepsilon = 0.98$ \hspace{-4mm}\end{tabular} \\ 
			& SRE  &   11.84 &   -0.00 &    8.75 &   12.21 &   -6.97 &   -4.64 &   14.26 &   11.85 & \Valbest{  24.39} & \ValSecnd{  20.38} &   20.36 \\ 
			& RMSE &   0.0127 &   1.5402 &   0.0188 &   0.0126 &   0.0530 &   0.0520 &   0.0105 &   0.0134 & \Valbest{  0.0033} & \ValSecnd{  0.0051} & \ValSecnd{  0.0051} \\ 
			& Ps   & {  0.98} &   0.00 &   0.93 &   0.96 &   0.00 &   0.01 & \Valbest{  1.00} & {  0.98} & \Valbest{  1.00} & \Valbest{  1.00} & \Valbest{  1.00} \\ 
			& MPSNR &   30.96 &  -43.10 &   30.01 &   31.31 &   17.37 &   12.17 &   33.31 &   27.65 & \Valbest{  47.02} &   38.23 & \ValSecnd{  38.86} \\ 
			& MSSIM &   0.7132 &   0.2564 &   0.6349 &   0.7118 &   0.4976 &   0.6436 &   0.7763 &   0.5911 & \Valbest{  0.9956} &   0.9029 & \ValSecnd{  0.9176} \\ 
			\midrule

			\multirow{6}{*}{Case 7} 
			& Setup & \begin{tabular}{c} \hspace{-4mm} $\lambda = 10^{0}$ \hspace{-4mm}\end{tabular} & \begin{tabular}{c} \hspace{-4mm} $\lambda_{1} = 10^{-1}$, \hspace{-4mm} \\\hspace{-4mm} $\lambda_{2} = 10^{0}$, \hspace{-4mm} \\\hspace{-4mm} $\lambda_{3} = 10^{1}$ \hspace{-4mm}\end{tabular} & \begin{tabular}{c} \hspace{-4mm} $\lambda = 10^{1}$, \hspace{-4mm} \\\hspace{-4mm} $\lambda_{TV} = 10^{-2}$ \hspace{-4mm}\end{tabular} & \begin{tabular}{c} \hspace{-4mm} $\lambda_{g} = 10^{1}$, \hspace{-4mm} \\\hspace{-4mm} $\lambda_{l} = 10^{-2}$ \hspace{-4mm}\end{tabular} &  --  &  --  & \begin{tabular}{c} \hspace{-4mm} $\lambda = 10^{-1}$ \hspace{-4mm}\end{tabular} & \begin{tabular}{c} \hspace{-4mm} $\lambda = 10^{-1}$, \hspace{-4mm} \\\hspace{-4mm} $\tau = 10^{0}$ \hspace{-4mm}\end{tabular} & \begin{tabular}{c} \hspace{-4mm} $\lambda_{1} = 10^{-1}$, \hspace{-4mm} \\\hspace{-4mm} $\lambda_{2} = 10^{0}$, \hspace{-4mm} \\\hspace{-4mm} $\varepsilon = 0.95$ \hspace{-4mm}\end{tabular} & \begin{tabular}{c} \hspace{-4mm} $\lambda_{1} = 10^{0}$, \hspace{-4mm} \\\hspace{-4mm} $\lambda_{2} = 10^{-1}$, \hspace{-4mm} \\\hspace{-4mm} $\varepsilon = 0.95$ \hspace{-4mm}\end{tabular} & \begin{tabular}{c} \hspace{-4mm} $\lambda_{1} = 10^{0}$, \hspace{-4mm} \\\hspace{-4mm} $\lambda_{2} = 10^{-1}$, \hspace{-4mm} \\\hspace{-4mm} $\varepsilon = 0.95$ \hspace{-4mm}\end{tabular} \\ 
			& SRE  &   14.14 &   -0.00 &   10.63 &   13.86 &   -4.27 &   -4.21 &   17.18 &   14.05 & \Valbest{  25.57} & \ValSecnd{  20.01} &   20.00 \\ 
			& RMSE &   0.0103 &   1.4475 &   0.0158 &   0.0113 &   0.0464 &   0.0527 &   0.0077 &   0.0109 & \Valbest{  0.0029} & \ValSecnd{  0.0054} & \ValSecnd{  0.0054} \\ 
			& Ps   & \ValSecnd{  0.99} &   0.00 &   0.97 &   0.97 &   0.01 &   0.01 & \Valbest{  1.00} & \ValSecnd{  0.99} & \Valbest{  1.00} & \Valbest{  1.00} & \Valbest{  1.00} \\ 
			& MPSNR &   33.96 &  -42.56 &   32.28 &   33.57 &   19.63 &   12.39 &   36.47 &   29.24 & \Valbest{  48.64} &   37.09 & \ValSecnd{  38.06} \\ 
			& MSSIM &   0.7684 &   0.2893 &   0.6830 &   0.7356 &   0.4921 &   0.6608 &   0.8178 &   0.6416 & \Valbest{  0.9969} &   0.8716 & \ValSecnd{  0.8980} \\ 
			\midrule 
			
			\multirow{6}{*}{Case 8} 
			& Setup & \begin{tabular}{c} \hspace{-4mm} $\lambda = 10^{0}$ \hspace{-4mm}\end{tabular} & \begin{tabular}{c} \hspace{-4mm} $\lambda_{1} = 10^{1}$, \hspace{-4mm} \\\hspace{-4mm} $\lambda_{2} = 10^{-2}$, \hspace{-4mm} \\\hspace{-4mm} $\lambda_{3} = 10^{1}$ \hspace{-4mm}\end{tabular} & \begin{tabular}{c} \hspace{-4mm} $\lambda = 10^{1}$, \hspace{-4mm} \\\hspace{-4mm} $\lambda_{TV} = 10^{-2}$ \hspace{-4mm}\end{tabular} & \begin{tabular}{c} \hspace{-4mm} $\lambda_{g} = 10^{1}$, \hspace{-4mm} \\\hspace{-4mm} $\lambda_{l} = 10^{-3}$ \hspace{-4mm}\end{tabular} &  --  &  --  & \begin{tabular}{c} \hspace{-4mm} $\lambda = 10^{-1}$ \hspace{-4mm}\end{tabular} & \begin{tabular}{c} \hspace{-4mm} $\lambda = 10^{-1}$, \hspace{-4mm} \\\hspace{-4mm} $\tau = 10^{0}$ \hspace{-4mm}\end{tabular} & \begin{tabular}{c} \hspace{-4mm} $\lambda_{1} = 10^{0}$, \hspace{-4mm} \\\hspace{-4mm} $\lambda_{2} = 10^{0}$, \hspace{-4mm} \\\hspace{-4mm} $\varepsilon = 0.98$ \hspace{-4mm}\end{tabular} & \begin{tabular}{c} \hspace{-4mm} $\lambda_{1} = 10^{0}$, \hspace{-4mm} \\\hspace{-4mm} $\lambda_{2} = 10^{-1}$, \hspace{-4mm} \\\hspace{-4mm} $\varepsilon = 0.98$ \hspace{-4mm}\end{tabular} & \begin{tabular}{c} \hspace{-4mm} $\lambda_{1} = 10^{0}$, \hspace{-4mm} \\\hspace{-4mm} $\lambda_{2} = 10^{-1}$, \hspace{-4mm} \\\hspace{-4mm} $\varepsilon = 0.98$ \hspace{-4mm}\end{tabular} \\ 
			& SRE  &   10.67 &   -0.00 &    6.95 &    9.74 &   -6.72 &   -3.78 &   12.74 &    9.49 & \Valbest{  21.12} &   18.67 & \ValSecnd{  18.69} \\ 
			& RMSE &   0.0144 &   1.0417 &   0.0234 &   0.0167 &   0.0517 &   0.0474 &   0.0126 &   0.0176 & \Valbest{  0.0047} & \ValSecnd{  0.0062} & \ValSecnd{  0.0062} \\ 
			& Ps   &   0.96 &   0.01 &   0.84 &   0.94 &   0.00 &   0.04 & \ValSecnd{  0.99} &   0.94 & \Valbest{  1.00} & \Valbest{  1.00} & \Valbest{  1.00} \\ 
			& MPSNR &   29.84 &  -39.41 &   28.58 &   29.37 &   14.55 &   13.37 &   32.13 &   24.96 & \Valbest{  43.66} &   35.56 & \ValSecnd{  36.44} \\ 
			& MSSIM &   0.6476 &   0.4875 &   0.5494 &   0.5894 &   0.4098 &   0.7328 &   0.7146 &   0.4711 & \Valbest{  0.9925} &   0.8375 & \ValSecnd{  0.8696} \\ 
			\bottomrule
		\end{tabular}
	}
\end{table*}

\begin{figure*}[t]
	\centering
	\begin{minipage}[t]{0.075\hsize}
		\centerline{
			\includegraphics[height = 40pt]{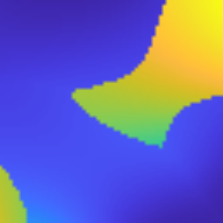}
		}
	\end{minipage}
	\begin{minipage}[t]{0.075\hsize}
		\centerline{
			\includegraphics[height = 40pt]{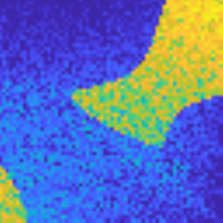}
		}
	\end{minipage}
	\begin{minipage}[t]{0.075\hsize}
		\centerline{
			\includegraphics[height = 40pt]{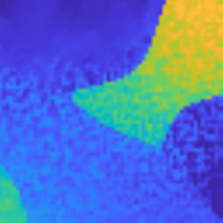}
		}
	\end{minipage}
	\begin{minipage}[t]{0.075\hsize}
		\centerline{
			\includegraphics[height = 40pt]{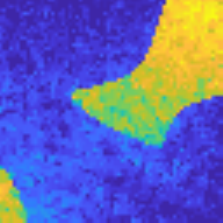}
		}
	\end{minipage}
	\begin{minipage}[t]{0.075\hsize}
		\centerline{
			\includegraphics[height = 40pt]{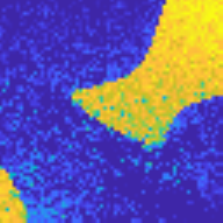}
		}
	\end{minipage}
	\begin{minipage}[t]{0.075\hsize}
		\centerline{
			\includegraphics[height = 40pt]{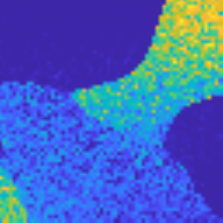}
		}
	\end{minipage}
	\begin{minipage}[t]{0.075\hsize}
	\centerline{
		\includegraphics[height = 40pt]{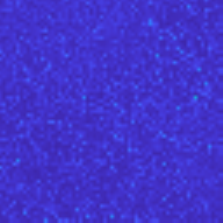}
	}
	\end{minipage}
	\begin{minipage}[t]{0.075\hsize}
	\centerline{
		\includegraphics[height = 40pt]{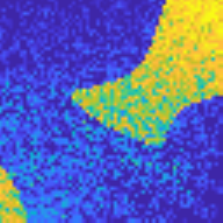}
	}
	\end{minipage}
	\begin{minipage}[t]{0.075\hsize}
		\centerline{
			\includegraphics[height = 40pt]{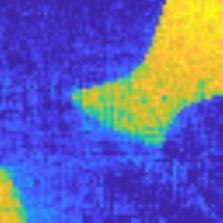}
		}
	\end{minipage}
	\begin{minipage}[t]{0.075\hsize}
		\centerline{
			\includegraphics[height = 40pt]{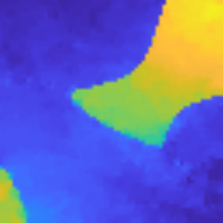}
		}
	\end{minipage}
	\begin{minipage}[t]{0.075\hsize}
		\centerline{
			\includegraphics[height = 40pt]{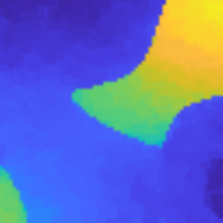}
		}
	\end{minipage}
	\begin{minipage}[t]{0.075\hsize}
		\centerline{
			\includegraphics[height = 40pt]{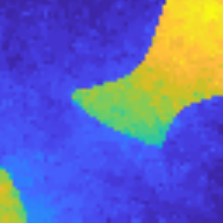}
		}
	\end{minipage}
	\begin{minipage}[t]{0.02\hsize}
		\centerline{
			\includegraphics[height = 40pt]{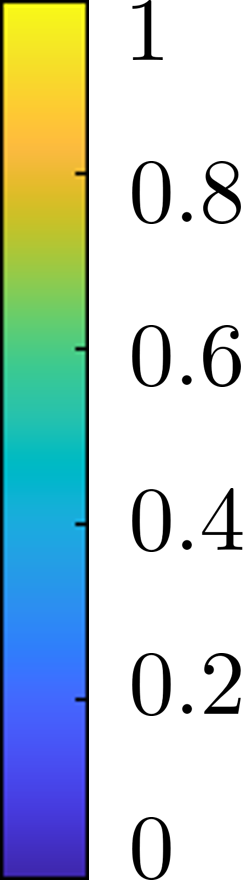}
		}
	\end{minipage}
	
	\vspace{1mm}
	
	\begin{minipage}[t]{0.075\hsize}
		\centerline{
			\includegraphics[height = 40pt]{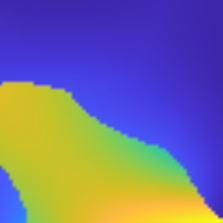}
		}
	\end{minipage}
	\begin{minipage}[t]{0.075\hsize}
		\centerline{
			\includegraphics[height = 40pt]{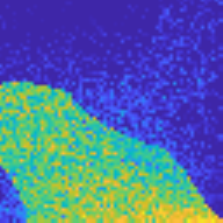}
		}
	\end{minipage}
	\begin{minipage}[t]{0.075\hsize}
		\centerline{
			\includegraphics[height = 40pt]{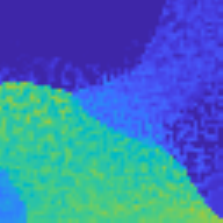}
		}
	\end{minipage}
	\begin{minipage}[t]{0.075\hsize}
		\centerline{
			\includegraphics[height = 40pt]{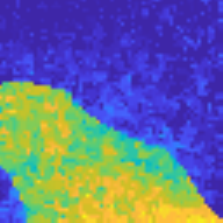}
		}
	\end{minipage}        
	\begin{minipage}[t]{0.075\hsize}
		\centerline{
			\includegraphics[height = 40pt]{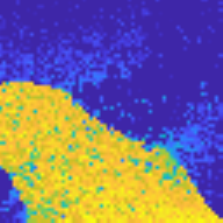}
		}
	\end{minipage}
	\begin{minipage}[t]{0.075\hsize}
		\centerline{
			\includegraphics[height = 40pt]{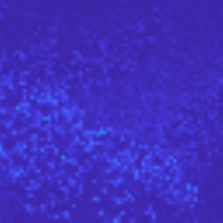}
		}
	\end{minipage}
	\begin{minipage}[t]{0.075\hsize}
	\centerline{
		\includegraphics[height = 40pt]{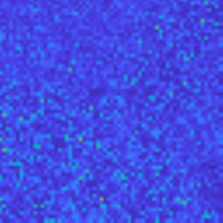}
	}
	\end{minipage}
	\begin{minipage}[t]{0.075\hsize}
	\centerline{
		\includegraphics[height = 40pt]{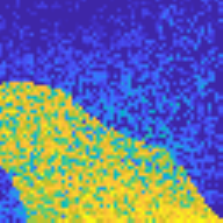}
	}
	\end{minipage}
	\begin{minipage}[t]{0.075\hsize}
		\centerline{
			\includegraphics[height = 40pt]{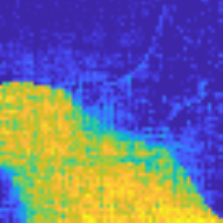}
		}
	\end{minipage}
	\begin{minipage}[t]{0.075\hsize}
		\centerline{
			\includegraphics[height = 40pt]{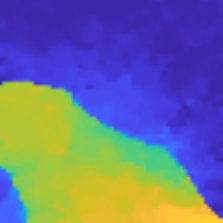}
		}
	\end{minipage}
	\begin{minipage}[t]{0.075\hsize}
		\centerline{
			\includegraphics[height = 40pt]{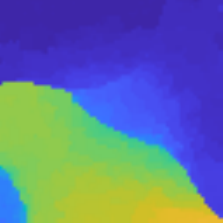}
		}
	\end{minipage}
	\begin{minipage}[t]{0.075\hsize}
		\centerline{
			\includegraphics[height = 40pt]{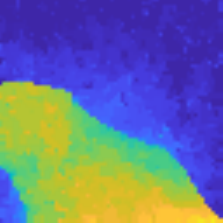}
		}
	\end{minipage}
	\begin{minipage}[t]{0.02\hsize}
		\centerline{
			\includegraphics[height = 40pt]{./fig/colorbar_20.png}
		}
	\end{minipage}
	
	\vspace{1mm}
	
	\begin{minipage}[t]{0.075\hsize}
		\centerline{
			\includegraphics[height = 40pt]{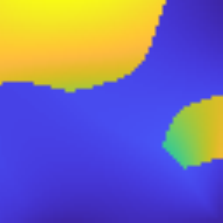}
		}
	\end{minipage}
	\begin{minipage}[t]{0.075\hsize}
		\centerline{
			\includegraphics[height = 40pt]{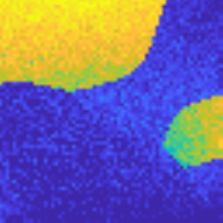}
		}
	\end{minipage}
	\begin{minipage}[t]{0.075\hsize}
		\centerline{
			\includegraphics[height = 40pt]{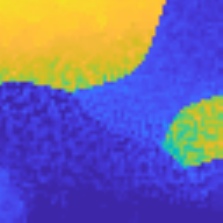}
		}
	\end{minipage}
	\begin{minipage}[t]{0.075\hsize}
		\centerline{
			\includegraphics[height = 40pt]{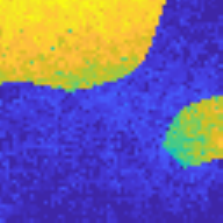}
		}
	\end{minipage}
	\begin{minipage}[t]{0.075\hsize}
		\centerline{
			\includegraphics[height = 40pt]{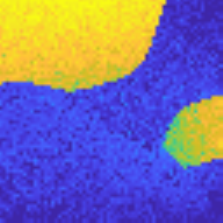}
		}
	\end{minipage}
	\begin{minipage}[t]{0.075\hsize}
		\centerline{
			\includegraphics[height = 40pt]{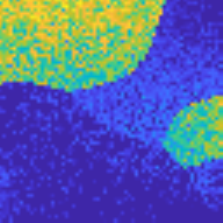}
		}
	\end{minipage}
	\begin{minipage}[t]{0.075\hsize}
	\centerline{
		\includegraphics[height = 40pt]{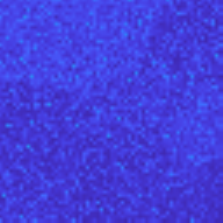}
	}
	\end{minipage}
	\begin{minipage}[t]{0.075\hsize}
	\centerline{
		\includegraphics[height = 40pt]{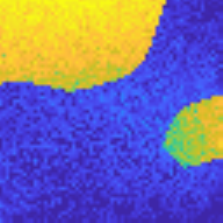}
	}
	\end{minipage}
	\begin{minipage}[t]{0.075\hsize}
		\centerline{
			\includegraphics[height = 40pt]{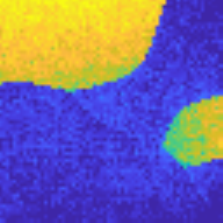}
		}
	\end{minipage}
	\begin{minipage}[t]{0.075\hsize}
		\centerline{
			\includegraphics[height = 40pt]{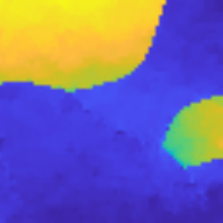}
		}
	\end{minipage}
	\begin{minipage}[t]{0.075\hsize}
		\centerline{
			\includegraphics[height = 40pt]{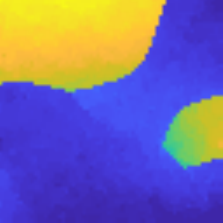}
		}
	\end{minipage}
	\begin{minipage}[t]{0.075\hsize}
		\centerline{
			\includegraphics[height = 40pt]{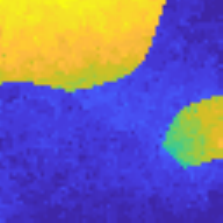}
		}
	\end{minipage}
	\begin{minipage}[t]{0.02\hsize}
		\centerline{
			\includegraphics[height = 40pt]{./fig/colorbar_20.png}
		}
	\end{minipage}
	
	\vspace{1mm}
	
	\begin{minipage}[t]{0.075\hsize}
		\centerline{
			\includegraphics[height = 40pt]{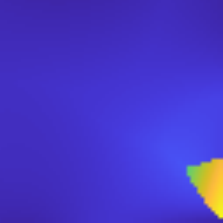}
		}
	\end{minipage}
	\begin{minipage}[t]{0.075\hsize}
		\centerline{
			\includegraphics[height = 40pt]{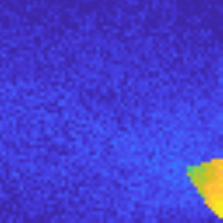}
		}
	\end{minipage}
	\begin{minipage}[t]{0.075\hsize}
		\centerline{
			\includegraphics[height = 40pt]{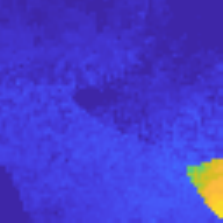}
		}
	\end{minipage}
	\begin{minipage}[t]{0.075\hsize}
		\centerline{
			\includegraphics[height = 40pt]{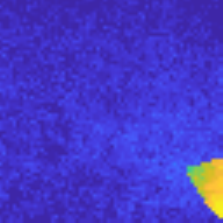}
		}
	\end{minipage}
	\begin{minipage}[t]{0.075\hsize}
		\centerline{
			\includegraphics[height = 40pt]{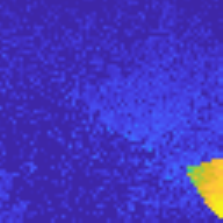}
		}
	\end{minipage}
	\begin{minipage}[t]{0.075\hsize}
		\centerline{
			\includegraphics[height = 40pt]{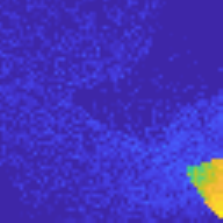}
		}
	\end{minipage}
	\begin{minipage}[t]{0.075\hsize}
	\centerline{
		\includegraphics[height = 40pt]{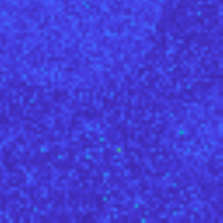}
	}
	\end{minipage}
	\begin{minipage}[t]{0.075\hsize}
	\centerline{
		\includegraphics[height = 40pt]{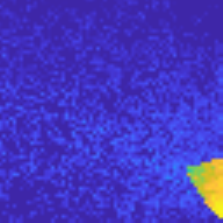}
	}
	\end{minipage}
	\begin{minipage}[t]{0.075\hsize}
		\centerline{
			\includegraphics[height = 40pt]{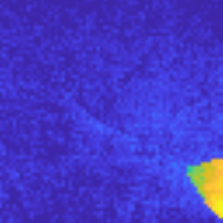}
		}
	\end{minipage}
	\begin{minipage}[t]{0.075\hsize}
		\centerline{
			\includegraphics[height = 40pt]{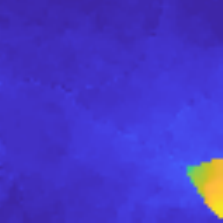}
		}
	\end{minipage}
	\begin{minipage}[t]{0.075\hsize}
		\centerline{
			\includegraphics[height = 40pt]{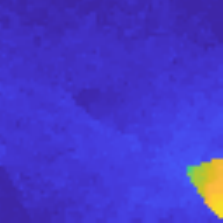}
		}
	\end{minipage}
	\begin{minipage}[t]{0.075\hsize}
		\centerline{
			\includegraphics[height = 40pt]{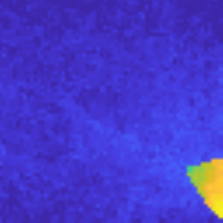}
		}
	\end{minipage}
	\begin{minipage}[t]{0.02\hsize}
		\centerline{
			\includegraphics[height = 40pt]{./fig/colorbar_20.png}
		}
	\end{minipage}
	
	\begin{minipage}[t]{0.075\hsize}
		\centerline{
			(a)
		}
	\end{minipage}
	\begin{minipage}[t]{0.075\hsize}
		\centerline{
			(b)
		}
	\end{minipage}
	\begin{minipage}[t]{0.075\hsize}
		\centerline{
			(c)
		}
	\end{minipage}
	\begin{minipage}[t]{0.075\hsize}
		\centerline{
			(d)
		}
	\end{minipage}
	\begin{minipage}[t]{0.075\hsize}
		\centerline{
			(e)
		}
	\end{minipage}
	\begin{minipage}[t]{0.075\hsize}
		\centerline{
			(f)
		}
	\end{minipage}
	\begin{minipage}[t]{0.075\hsize}
		\centerline{
			(g)
		}
	\end{minipage}
	\begin{minipage}[t]{0.075\hsize}
		\centerline{
			{(h)}
		}
	\end{minipage}
	\begin{minipage}[t]{0.075\hsize}
		\centerline{
			{(i)}
		}
	\end{minipage}
	\begin{minipage}[t]{0.075\hsize}
		\centerline{
			\textbf{(j)}
		}
	\end{minipage}
	\begin{minipage}[t]{0.075\hsize}
	\centerline{
		\textbf{(k)}
	}
	\end{minipage}
	\begin{minipage}[t]{0.075\hsize}
	\centerline{
		\textbf{(l)}
	}
	\end{minipage}
	\begin{minipage}[t]{0.02\hsize}
		\centerline{
			~
		}
	\end{minipage}
	
	\vspace{-1mm}
	
	\caption{Unmixing results of abundance maps for the \textit{Synth 1} experiments in Case 2. (a): Original abundance maps. (b): CLSUnSAL~\cite{iordache2014collaborative}. (c): JSTV~\cite{aggarwal2016hyperspectral}. (d): RSSUn-TV~\cite{wang2019row}. (e): LGSU~\cite{shen2022superpixel}. (f): UnDIP~\cite{UnDIP_RastiB_2022}. (g): EGU-Net~\cite{hong2022endmember}. (h): RDSWSU~\cite{rs_Deng_RobustDual_2023}. (i): MdLRR~\cite{MDLRR_WuLing_2023}. (j): \textbf{\Ourss (HTV)}. (k): \textbf{\Ourss (SSTV)}. (l): \textbf{\Ourss (HSSTV)}.}
	\label{fig:synth_legendre_0.1_0_0}
\end{figure*}

\begin{figure*}[t]
\centering
        \begin{minipage}[t]{0.075\hsize}
            \centerline{
            \includegraphics[height = 40pt]{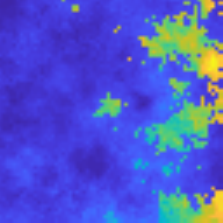}
            }
        \end{minipage}
        \begin{minipage}[t]{0.075\hsize}
            \centerline{
            \includegraphics[height = 40pt]{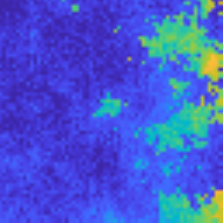}
            }
        \end{minipage}
        \begin{minipage}[t]{0.075\hsize}
            \centerline{
            \includegraphics[height = 40pt]{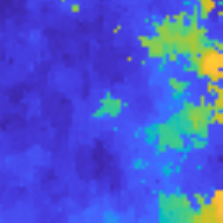}
            }
        \end{minipage}
        \begin{minipage}[t]{0.075\hsize}
            \centerline{
            \includegraphics[height = 40pt]{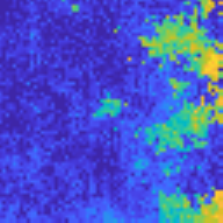}
            }
        \end{minipage}
        \begin{minipage}[t]{0.075\hsize}
            \centerline{
            \includegraphics[height = 40pt]{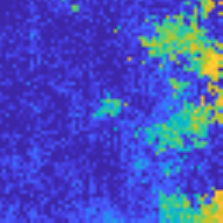}
            }
        \end{minipage}
	    \begin{minipage}[t]{0.075\hsize}
	    	\centerline{
	    		\includegraphics[height = 40pt]{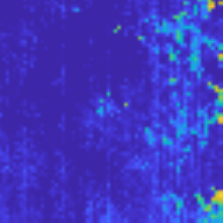}
	    	}
	    \end{minipage}
	    \begin{minipage}[t]{0.075\hsize}
	    \centerline{
	    	\includegraphics[height = 40pt]{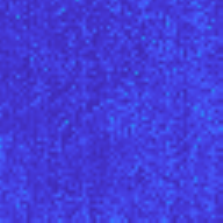}
	    }
	    \end{minipage}
	    \begin{minipage}[t]{0.075\hsize}
	    \centerline{
	    	\includegraphics[height = 40pt]{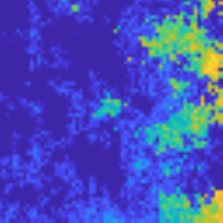}
	    }
	    \end{minipage}
	    \begin{minipage}[t]{0.075\hsize}
		    \centerline{
		    	\includegraphics[height = 40pt]{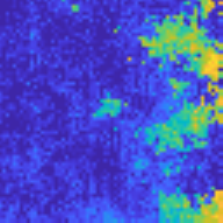}
		    }
	    \end{minipage}
        \begin{minipage}[t]{0.075\hsize}
            \centerline{
            \includegraphics[height = 40pt]{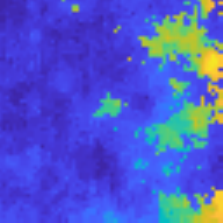}
            }
        \end{minipage}
        \begin{minipage}[t]{0.075\hsize}
            \centerline{
            \includegraphics[height = 40pt]{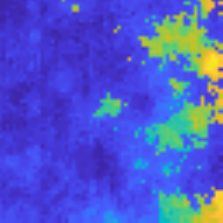}
            }
        \end{minipage}
        \begin{minipage}[t]{0.075\hsize}
            \centerline{
            \includegraphics[height = 40pt]{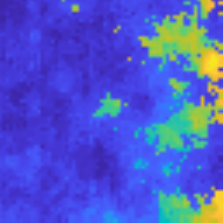}
            }
        \end{minipage}
	    \begin{minipage}[t]{0.02\hsize}
	    	\centerline{
	    		\includegraphics[height = 40pt]{./fig/colorbar_20.png}
	    	}
	    \end{minipage}
    
\vspace{1mm}
        
        \begin{minipage}[t]{0.075\hsize}
            \centerline{
            \includegraphics[height = 40pt]{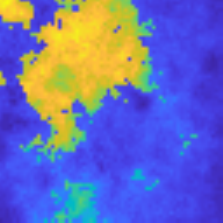}
            }
        \end{minipage}
        \begin{minipage}[t]{0.075\hsize}
            \centerline{
            \includegraphics[height = 40pt]{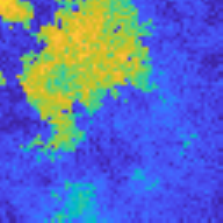}
            }
        \end{minipage}
        \begin{minipage}[t]{0.075\hsize}
            \centerline{
            \includegraphics[height = 40pt]{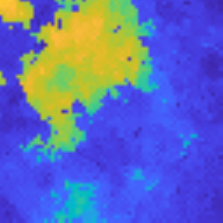}
            }
        \end{minipage}
        \begin{minipage}[t]{0.075\hsize}
            \centerline{
            \includegraphics[height = 40pt]{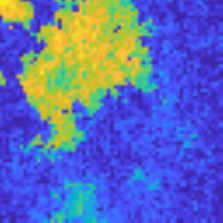}
            }
        \end{minipage}
        \begin{minipage}[t]{0.075\hsize}
            \centerline{
            \includegraphics[height = 40pt]{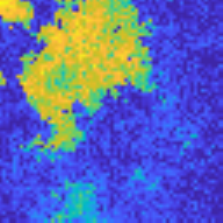}
            }
        \end{minipage}
	    \begin{minipage}[t]{0.075\hsize}
	    	\centerline{
	    		\includegraphics[height = 40pt]{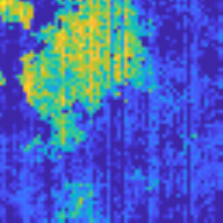}
	    	}
	    \end{minipage}
	    \begin{minipage}[t]{0.075\hsize}
	    \centerline{
	    	\includegraphics[height = 40pt]{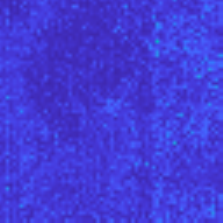}
	    }
	    \end{minipage}
	    \begin{minipage}[t]{0.075\hsize}
	    \centerline{
	    	\includegraphics[height = 40pt]{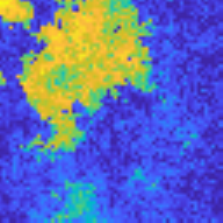}
	    }
	    \end{minipage}
	    \begin{minipage}[t]{0.075\hsize}
		    \centerline{
		    	\includegraphics[height = 40pt]{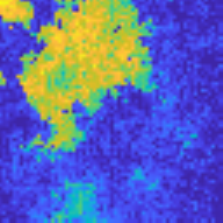}
		    }
	    \end{minipage}
        \begin{minipage}[t]{0.075\hsize}
            \centerline{
            \includegraphics[height = 40pt]{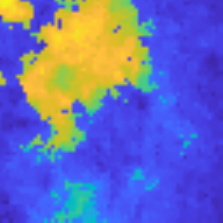}
            }
        \end{minipage}
        \begin{minipage}[t]{0.075\hsize}
            \centerline{
            \includegraphics[height = 40pt]{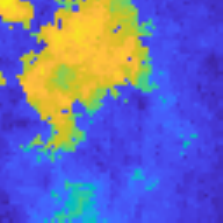}
            }
        \end{minipage}
        \begin{minipage}[t]{0.075\hsize}
            \centerline{
            \includegraphics[height = 40pt]{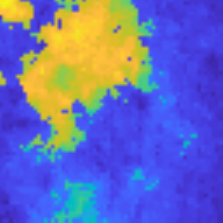}
            }
        \end{minipage}
	    \begin{minipage}[t]{0.02\hsize}
	    	\centerline{
	    		\includegraphics[height = 40pt]{./fig/colorbar_20.png}
	    	}
	    \end{minipage}
    
\vspace{1mm}
        
        \begin{minipage}[t]{0.075\hsize}
            \centerline{
            \includegraphics[height = 40pt]{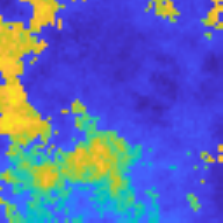}
            }
        \end{minipage}
        \begin{minipage}[t]{0.075\hsize}
            \centerline{
            \includegraphics[height = 40pt]{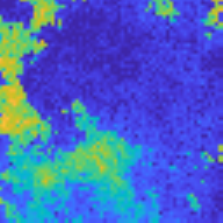}
            }
        \end{minipage}
        \begin{minipage}[t]{0.075\hsize}
            \centerline{
            \includegraphics[height = 40pt]{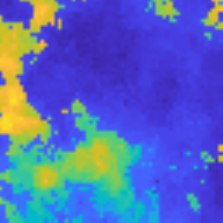}
            }
        \end{minipage}
        \begin{minipage}[t]{0.075\hsize}
            \centerline{
            \includegraphics[height = 40pt]{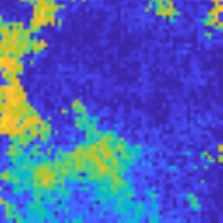}
            }
        \end{minipage}
        \begin{minipage}[t]{0.075\hsize}
            \centerline{
            \includegraphics[height = 40pt]{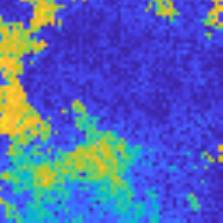}
            }
        \end{minipage}
	    \begin{minipage}[t]{0.075\hsize}
	    	\centerline{
	    		\includegraphics[height = 40pt]{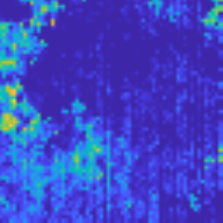}
	    	}
	    \end{minipage}
	    \begin{minipage}[t]{0.075\hsize}
	    \centerline{
	    	\includegraphics[height = 40pt]{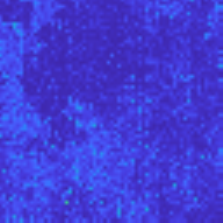}
	    }
	    \end{minipage}
	    \begin{minipage}[t]{0.075\hsize}
	    \centerline{
	    	\includegraphics[height = 40pt]{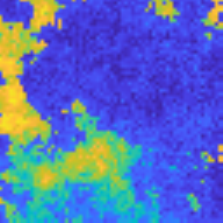}
	    }
	    \end{minipage}
	    \begin{minipage}[t]{0.075\hsize}
		    \centerline{
		    	\includegraphics[height = 40pt]{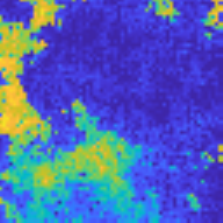}
		    }
	    \end{minipage}
        \begin{minipage}[t]{0.075\hsize}
            \centerline{
            \includegraphics[height = 40pt]{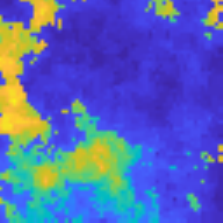}
            }
        \end{minipage}
        \begin{minipage}[t]{0.075\hsize}
            \centerline{
            \includegraphics[height = 40pt]{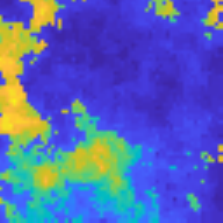}
            }
        \end{minipage}
        \begin{minipage}[t]{0.075\hsize}
            \centerline{
            \includegraphics[height = 40pt]{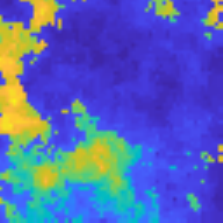}
            }
        \end{minipage}
	    \begin{minipage}[t]{0.02\hsize}
	    	\centerline{
	    		\includegraphics[height = 40pt]{./fig/colorbar_20.png}
	    	}
	    \end{minipage}
    
\vspace{1mm}
        
        \begin{minipage}[t]{0.075\hsize}
            \centerline{
            \includegraphics[height = 40pt]{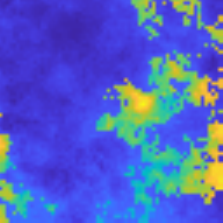}
            }
        \end{minipage}
        \begin{minipage}[t]{0.075\hsize}
            \centerline{
            \includegraphics[height = 40pt]{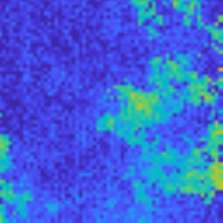}
            }
        \end{minipage}
        \begin{minipage}[t]{0.075\hsize}
            \centerline{
            \includegraphics[height = 40pt]{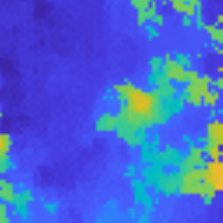}
            }
        \end{minipage}
        \begin{minipage}[t]{0.075\hsize}
            \centerline{
            \includegraphics[height = 40pt]{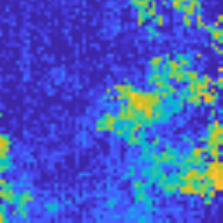}
            }
        \end{minipage}
        \begin{minipage}[t]{0.075\hsize}
            \centerline{
            \includegraphics[height = 40pt]{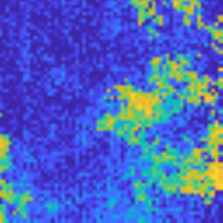}
            }
        \end{minipage}
	    \begin{minipage}[t]{0.075\hsize}
	    	\centerline{
	    		\includegraphics[height = 40pt]{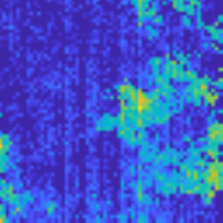}
	    	}
	    \end{minipage}
	    \begin{minipage}[t]{0.075\hsize}
	    \centerline{
	    	\includegraphics[height = 40pt]{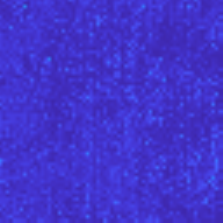}
	    }
	    \end{minipage}
	    \begin{minipage}[t]{0.075\hsize}
	    \centerline{
	    	\includegraphics[height = 40pt]{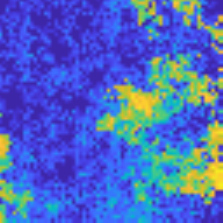}
	    }
	    \end{minipage}
	    \begin{minipage}[t]{0.075\hsize}
		    \centerline{
		    	\includegraphics[height = 40pt]{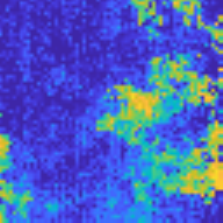}
		    }
	    \end{minipage}
        \begin{minipage}[t]{0.075\hsize}
            \centerline{
            \includegraphics[height = 40pt]{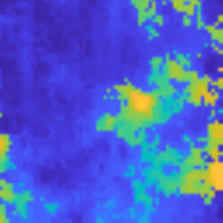}
            }
        \end{minipage}
        \begin{minipage}[t]{0.075\hsize}
            \centerline{
            \includegraphics[height = 40pt]{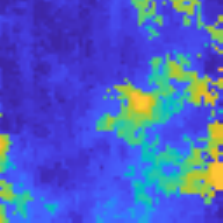}
            }
        \end{minipage}
        \begin{minipage}[t]{0.075\hsize}
            \centerline{
            \includegraphics[height = 40pt]{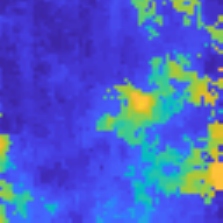}
            }
        \end{minipage}
	    \begin{minipage}[t]{0.02\hsize}
	    	\centerline{
	    		\includegraphics[height = 40pt]{./fig/colorbar_20.png}
	    	}
	    \end{minipage}
        
        \begin{minipage}[t]{0.075\hsize}
            \centerline{
            (a)
            }
        \end{minipage}
        \begin{minipage}[t]{0.075\hsize}
            \centerline{
            (b)
            }
        \end{minipage}
        \begin{minipage}[t]{0.075\hsize}
            \centerline{
            (c)
            }
        \end{minipage}
        \begin{minipage}[t]{0.075\hsize}
            \centerline{
            (d)
            }
        \end{minipage}
        \begin{minipage}[t]{0.075\hsize}
            \centerline{
            (e)
            }
        \end{minipage}
        \begin{minipage}[t]{0.075\hsize}
            \centerline{
            (f)
            }
        \end{minipage}
        \begin{minipage}[t]{0.075\hsize}
            \centerline{
            (g)
            }
        \end{minipage}
        \begin{minipage}[t]{0.075\hsize}
            \centerline{
            {(h)}
            }
        \end{minipage}
	    \begin{minipage}[t]{0.075\hsize}
	    	\centerline{
	    		{(i)}
	    	}
	    \end{minipage}
	    \begin{minipage}[t]{0.075\hsize}
	    \centerline{
	    	\textbf{(j)}
	    }
	    \end{minipage}
	    \begin{minipage}[t]{0.075\hsize}
	    \centerline{
	    	\textbf{(k)}
	    }
	    \end{minipage}
	    \begin{minipage}[t]{0.075\hsize}
	    \centerline{
	    	\textbf{(l)}
	    }
	    \end{minipage}
	    \begin{minipage}[t]{0.02\hsize}
	    	\centerline{
	    		~
	    	}
	    \end{minipage}
	    
	    \vspace{-1mm}

\caption{Unmixing results of abundance maps for the \textit{Synth 2} experiments in Case 5. (a): Original abundance maps. (b): CLSUnSAL~\cite{iordache2014collaborative}. (c): JSTV~\cite{aggarwal2016hyperspectral}. (d): RSSUn-TV~\cite{wang2019row}. (e): LGSU~\cite{shen2022superpixel}. (f): UnDIP~\cite{UnDIP_RastiB_2022}. (g): EGU-Net~\cite{hong2022endmember}. (h): RDSWSU~\cite{rs_Deng_RobustDual_2023}. (i): MdLRR~\cite{MDLRR_WuLing_2023}. (j): \textbf{\Ourss (HTV)}. (k): \textbf{\Ourss (SSTV)}. (l): \textbf{\Ourss (HSSTV)}.}
\label{synth_0.05_0.05_0.05}
\end{figure*}

\begin{figure*}[t]
	\centering
	\begin{minipage}[t]{0.075\hsize}
		\centerline{
			\includegraphics[height = 40pt]{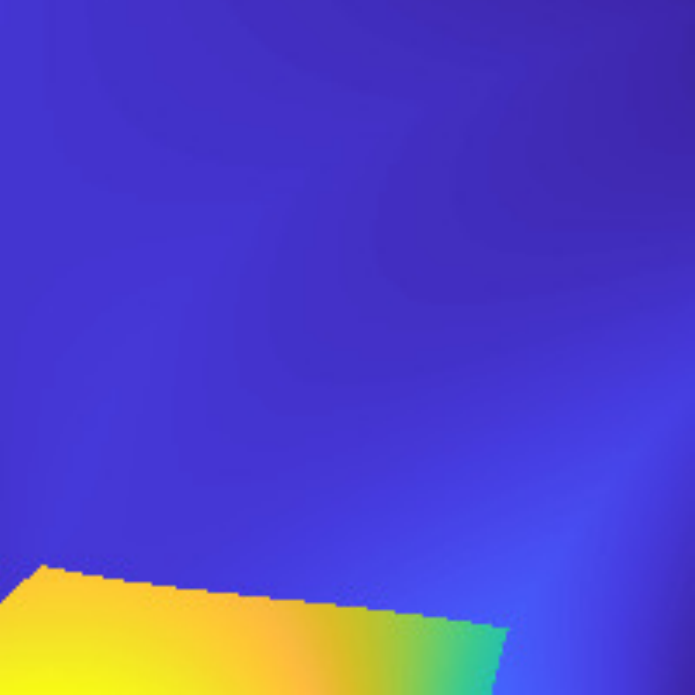}
		}
	\end{minipage}
	\begin{minipage}[t]{0.075\hsize}
		\centerline{
			\includegraphics[height = 40pt]{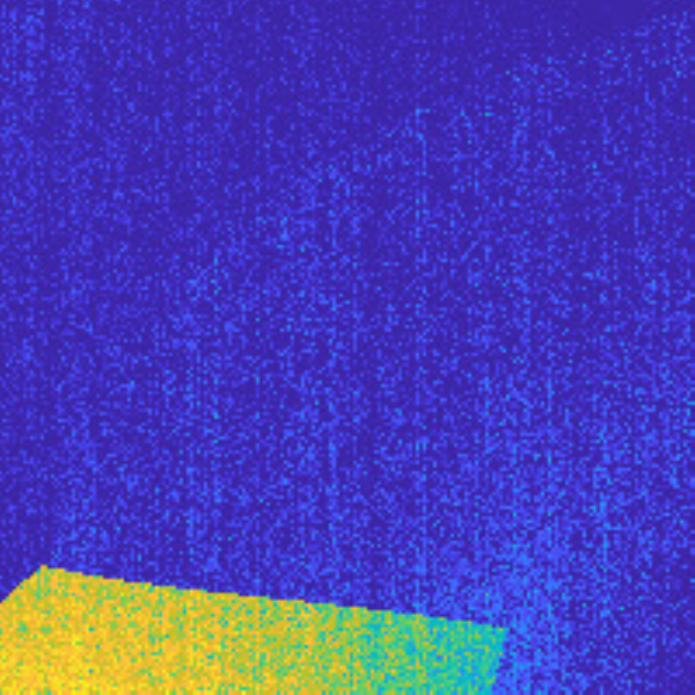}
		}
	\end{minipage}
	\begin{minipage}[t]{0.075\hsize}
		\centerline{
			\includegraphics[height = 40pt]{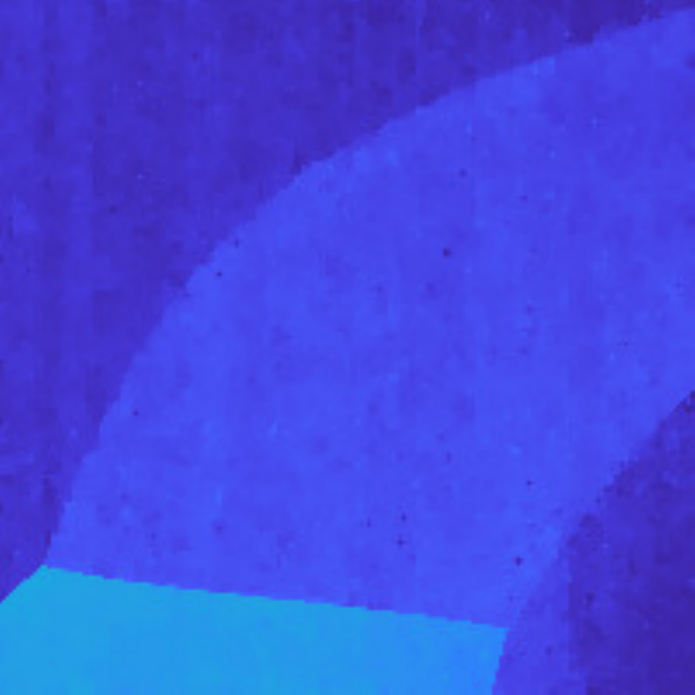}
		}
	\end{minipage}
	\begin{minipage}[t]{0.075\hsize}
		\centerline{
			\includegraphics[height = 40pt]{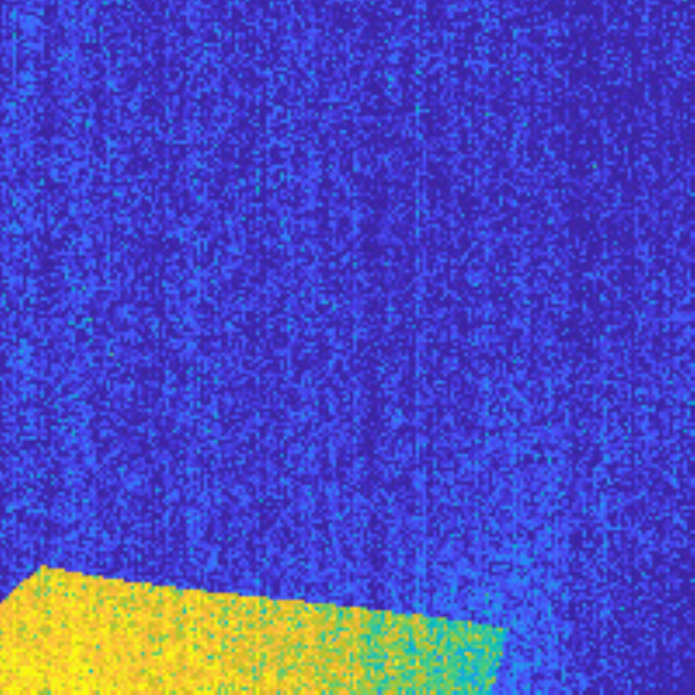}
		}
	\end{minipage}
	\begin{minipage}[t]{0.075\hsize}
		\centerline{
			\includegraphics[height = 40pt]{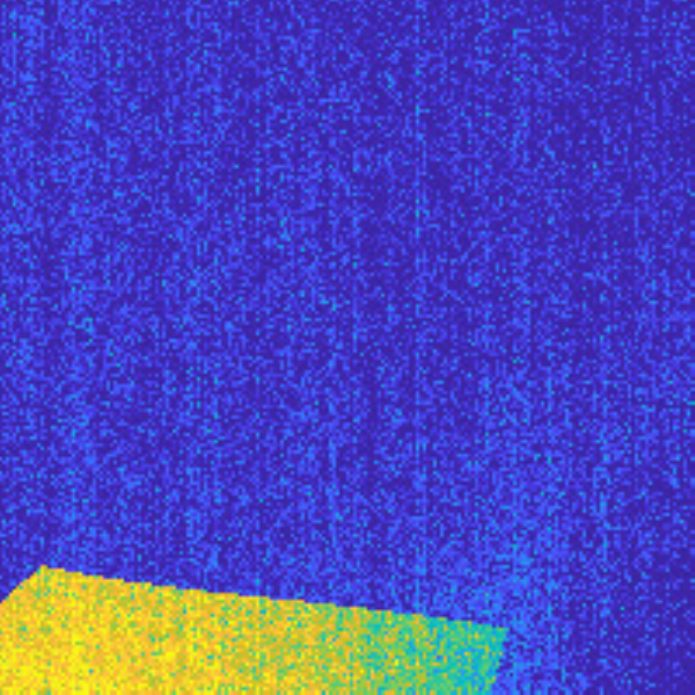}
		}
	\end{minipage}
	\begin{minipage}[t]{0.075\hsize}
		\centerline{
			\includegraphics[height = 40pt]{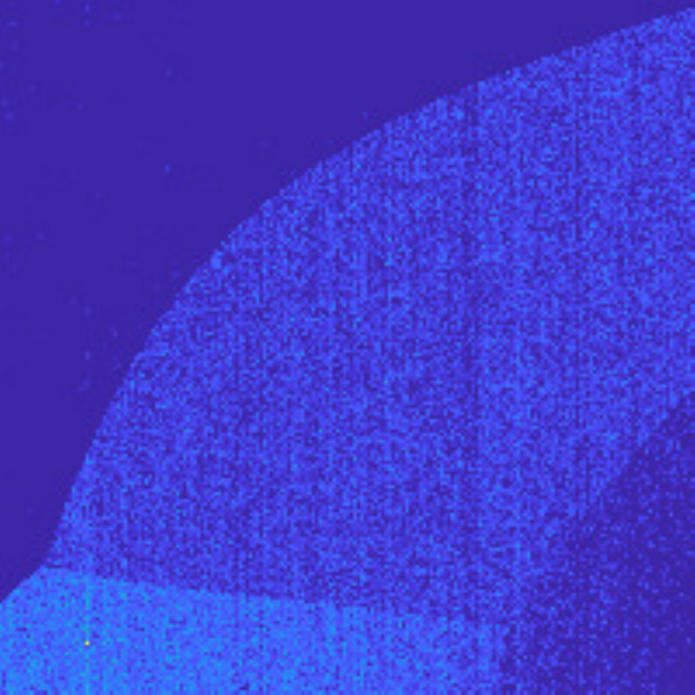}
		}
	\end{minipage}
	\begin{minipage}[t]{0.075\hsize}
		\centerline{
			\includegraphics[height = 40pt]{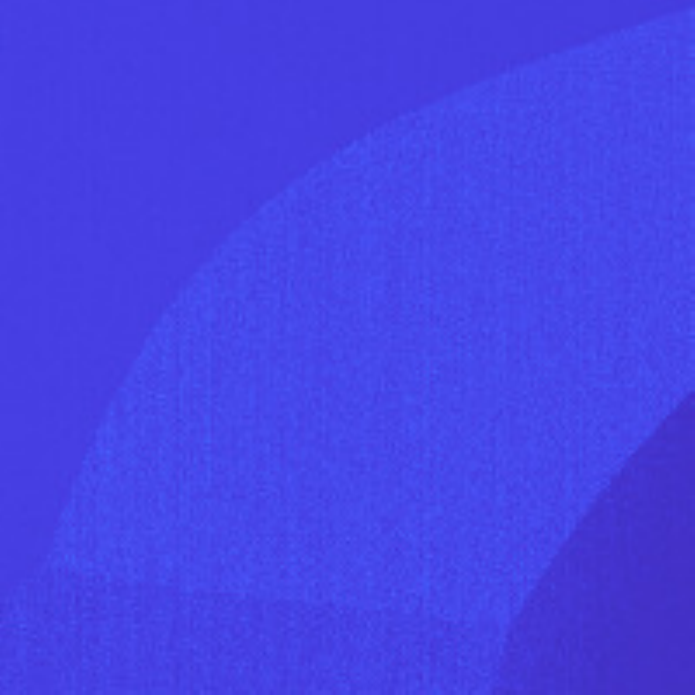}
		}
	\end{minipage}
	\begin{minipage}[t]{0.075\hsize}
		\centerline{
			\includegraphics[height = 40pt]{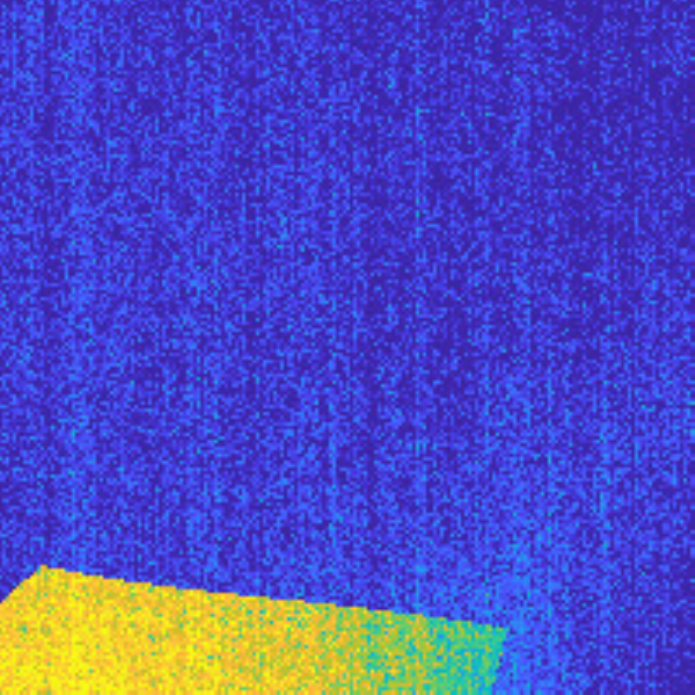}
		}
	\end{minipage}
	\begin{minipage}[t]{0.075\hsize}
		\centerline{
			\includegraphics[height = 40pt]{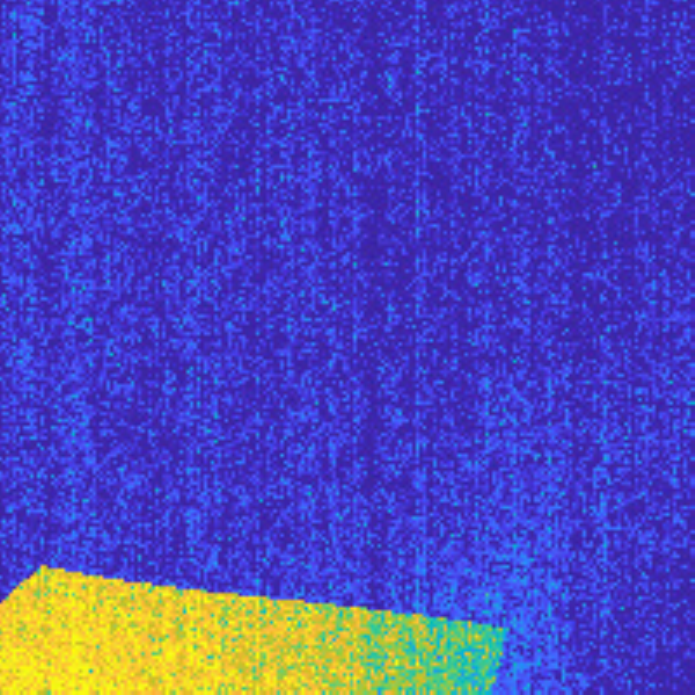}
		}
	\end{minipage}
	\begin{minipage}[t]{0.075\hsize}
		\centerline{
			\includegraphics[height = 40pt]{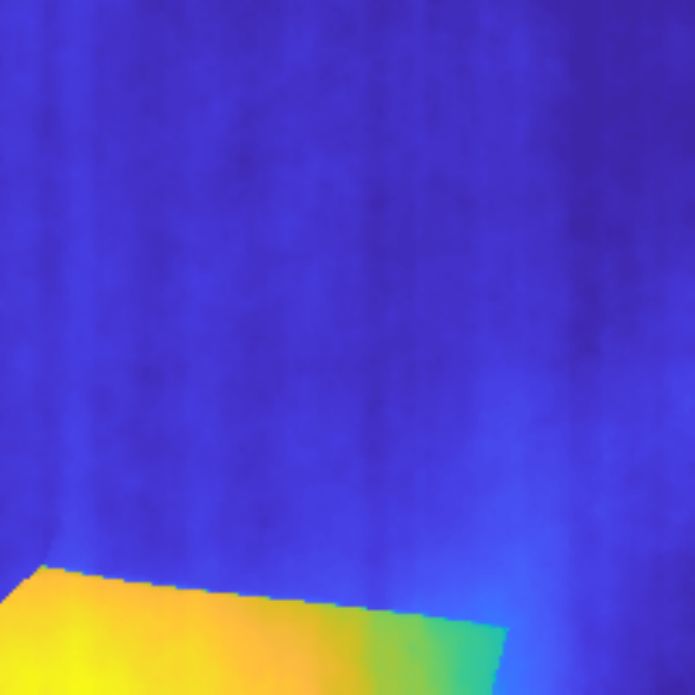}
		}
	\end{minipage}
	\begin{minipage}[t]{0.075\hsize}
		\centerline{
			\includegraphics[height = 40pt]{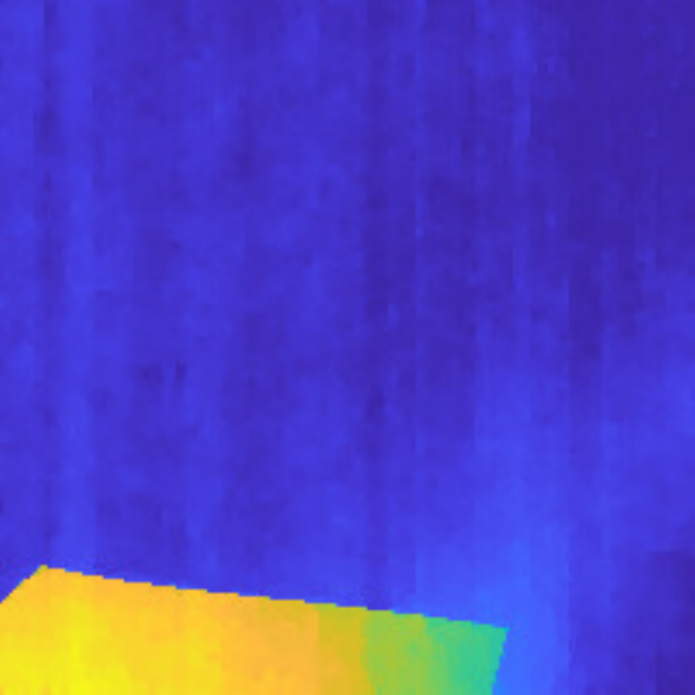}
		}
	\end{minipage}
	\begin{minipage}[t]{0.075\hsize}
		\centerline{
			\includegraphics[height = 40pt]{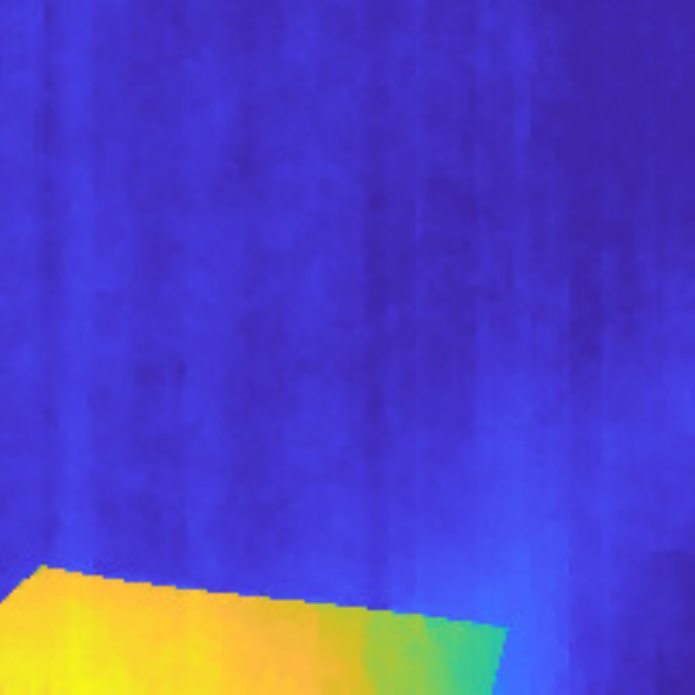}
		}
	\end{minipage}
	\begin{minipage}[t]{0.02\hsize}
		\centerline{
			\includegraphics[height = 40pt]{./fig/colorbar_20.png}
		}
	\end{minipage}
	
	\vspace{1mm}
	
	\begin{minipage}[t]{0.075\hsize}
		\centerline{
			\includegraphics[height = 40pt]{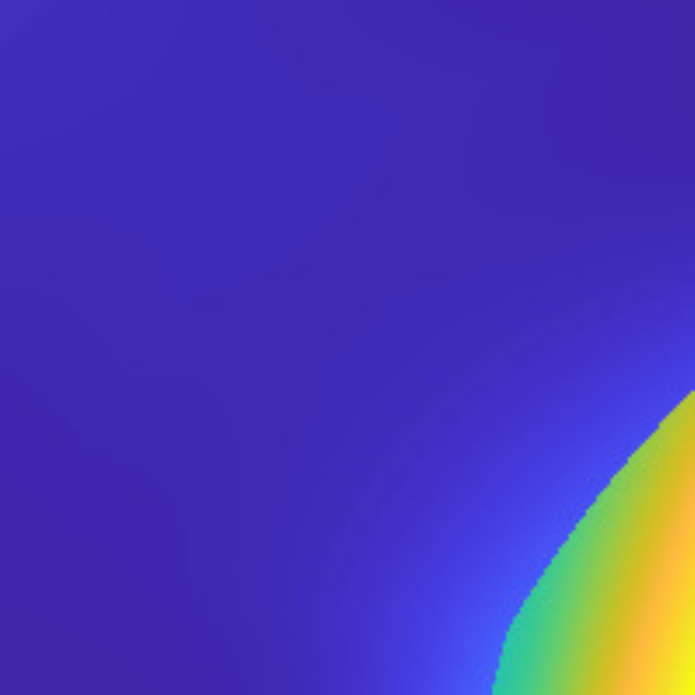}
		}
	\end{minipage}
	\begin{minipage}[t]{0.075\hsize}
		\centerline{
			\includegraphics[height = 40pt]{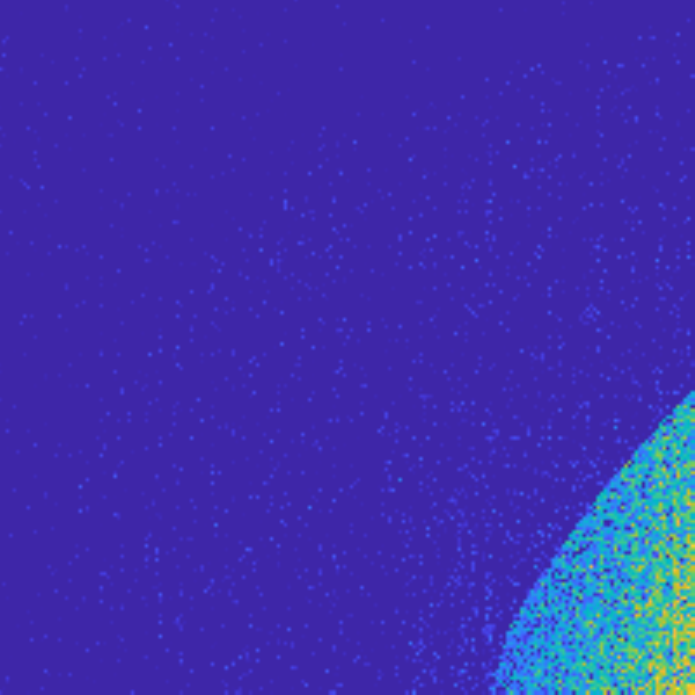}
		}
	\end{minipage}
	\begin{minipage}[t]{0.075\hsize}
		\centerline{
			\includegraphics[height = 40pt]{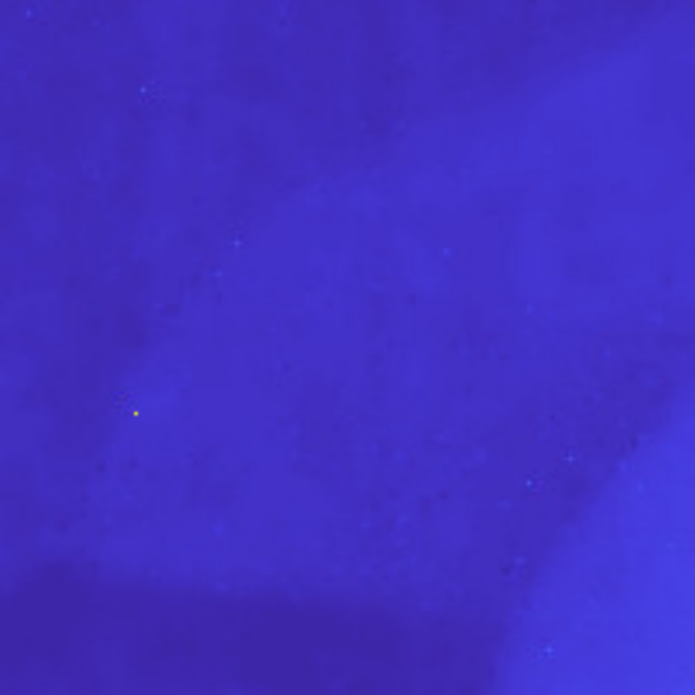}
		}
	\end{minipage}
	\begin{minipage}[t]{0.075\hsize}
		\centerline{
			\includegraphics[height = 40pt]{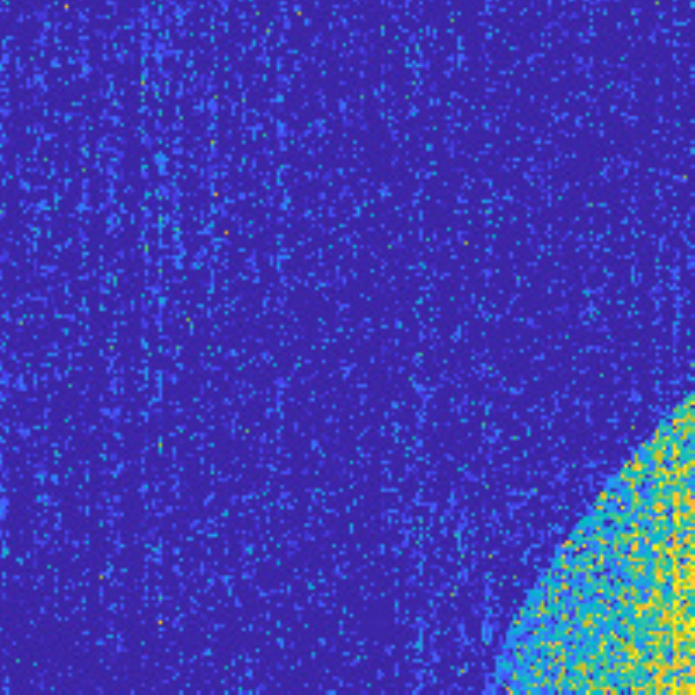}
		}
	\end{minipage}        
	\begin{minipage}[t]{0.075\hsize}
		\centerline{
			\includegraphics[height = 40pt]{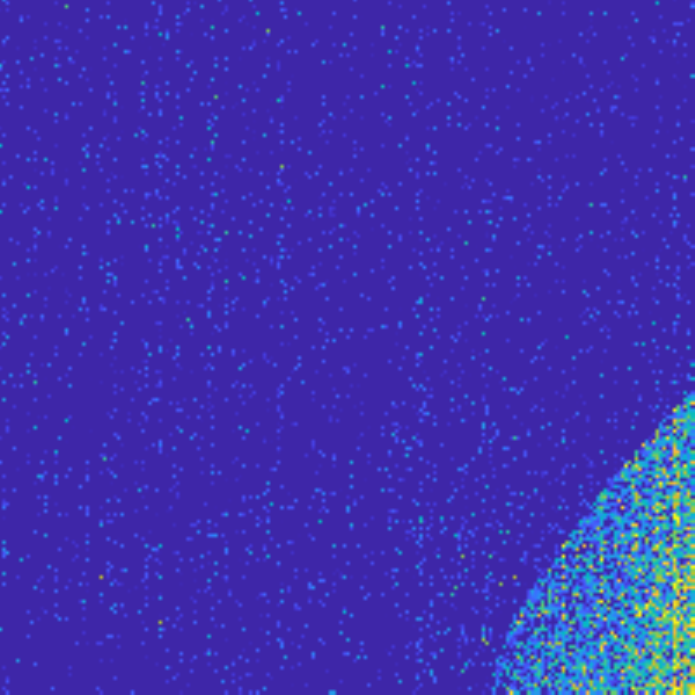}
		}
	\end{minipage}
	\begin{minipage}[t]{0.075\hsize}
		\centerline{
			\includegraphics[height = 40pt]{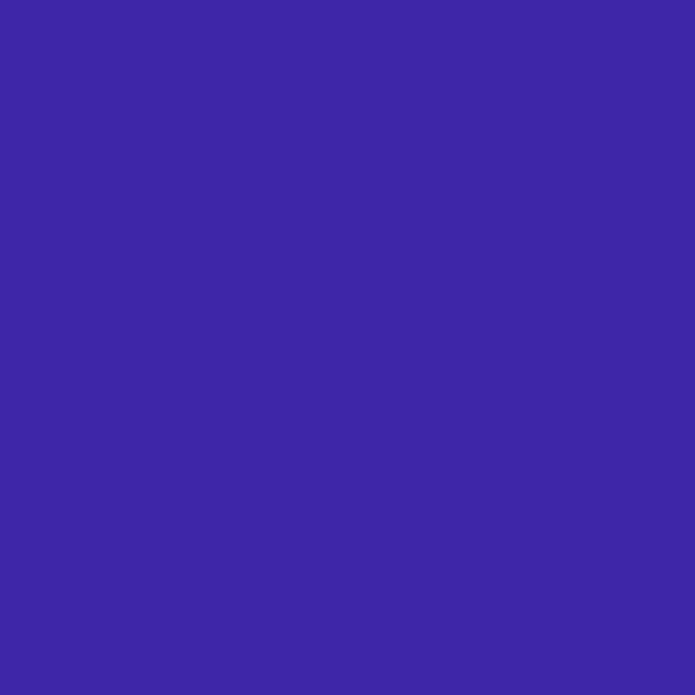}
		}
	\end{minipage}
	\begin{minipage}[t]{0.075\hsize}
		\centerline{
			\includegraphics[height = 40pt]{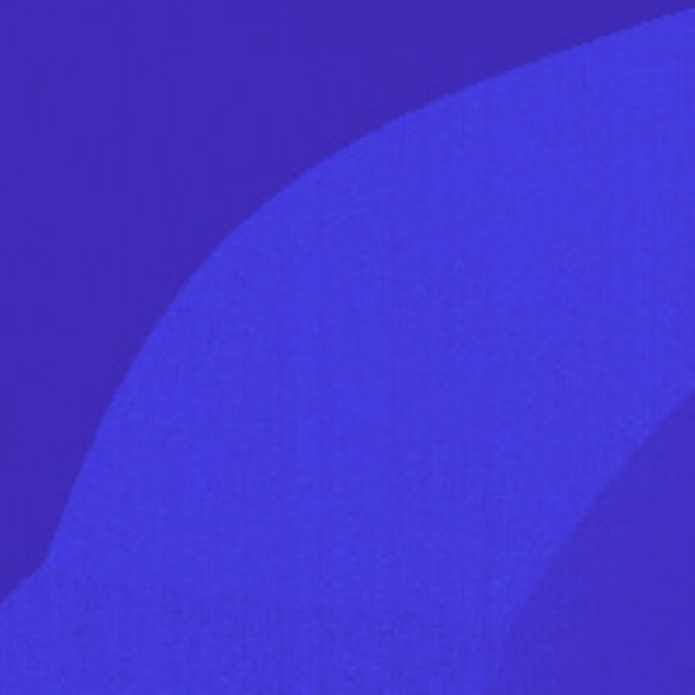}
		}
	\end{minipage}
	\begin{minipage}[t]{0.075\hsize}
		\centerline{
			\includegraphics[height = 40pt]{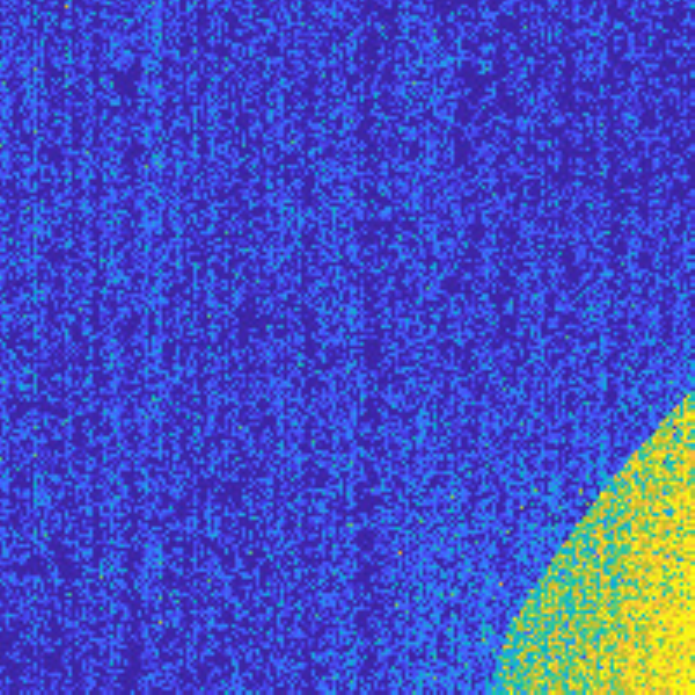}
		}
	\end{minipage}
	\begin{minipage}[t]{0.075\hsize}
		\centerline{
			\includegraphics[height = 40pt]{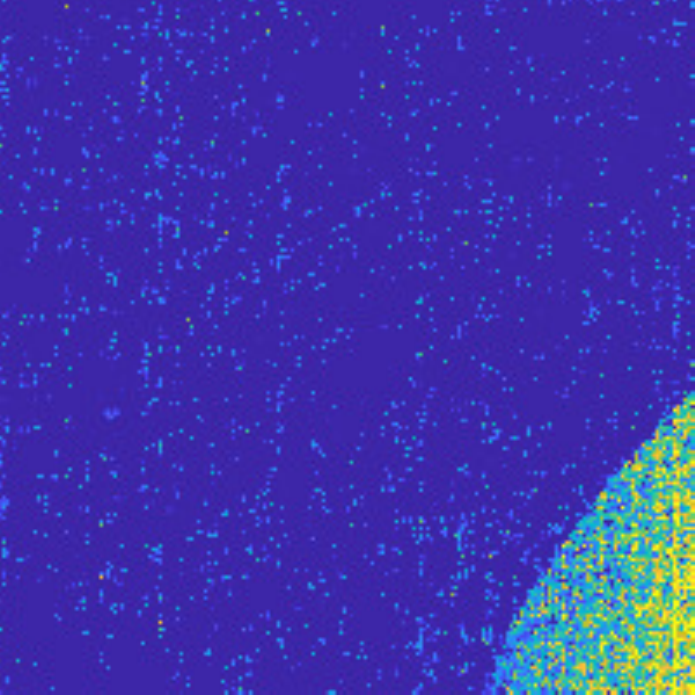}
		}
	\end{minipage}
	\begin{minipage}[t]{0.075\hsize}
		\centerline{
			\includegraphics[height = 40pt]{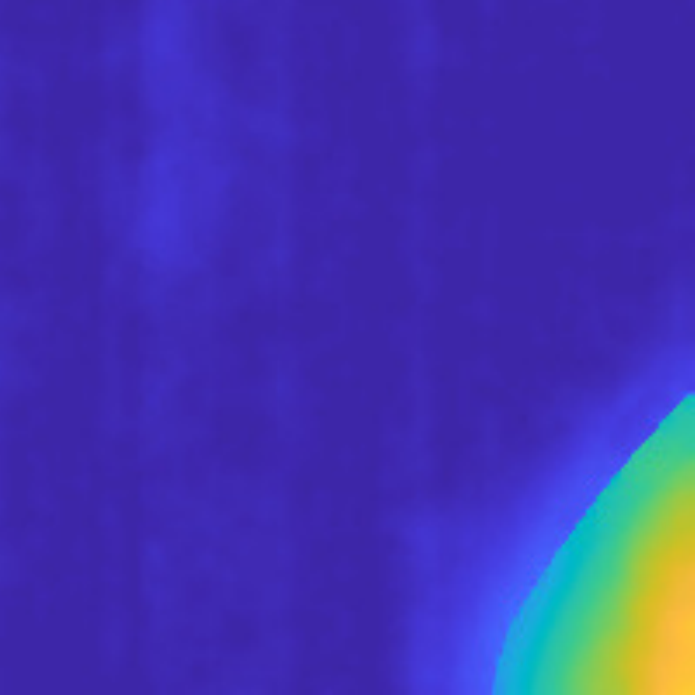}
		}
	\end{minipage}
	\begin{minipage}[t]{0.075\hsize}
		\centerline{
			\includegraphics[height = 40pt]{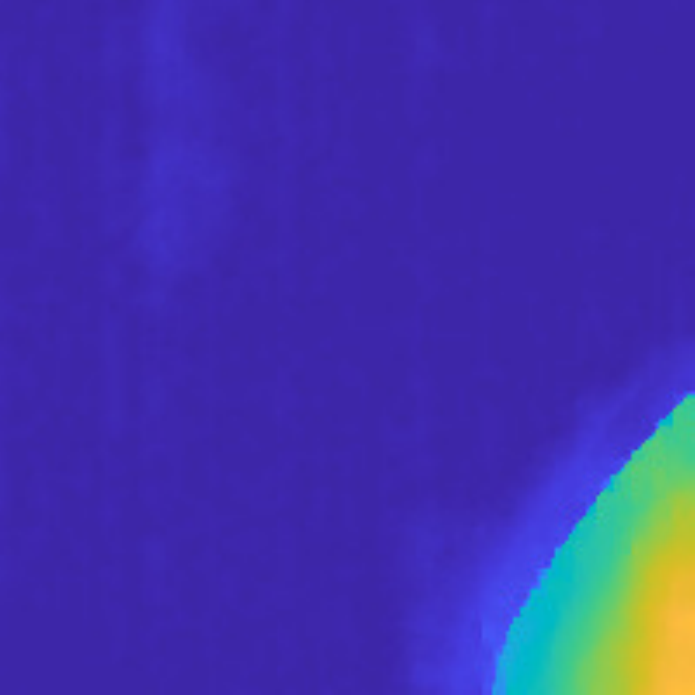}
		}
	\end{minipage}
	\begin{minipage}[t]{0.075\hsize}
		\centerline{
			\includegraphics[height = 40pt]{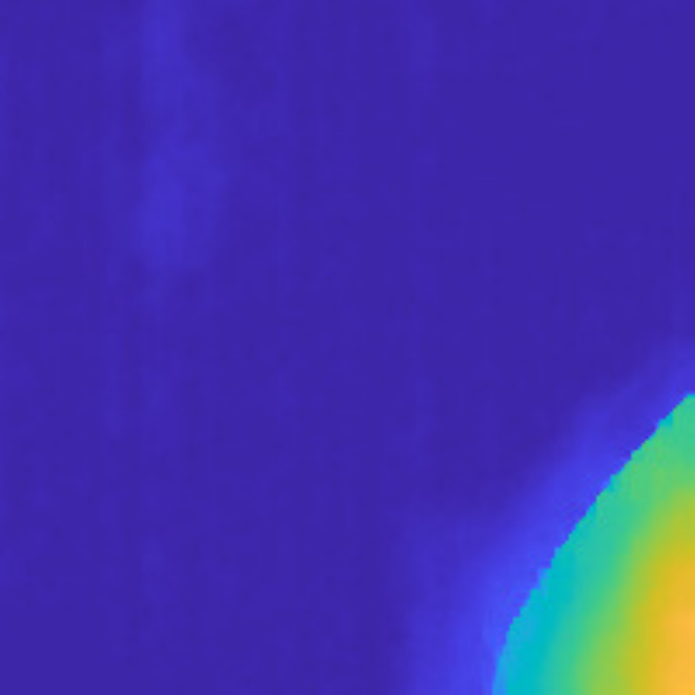}
		}
	\end{minipage}
	\begin{minipage}[t]{0.02\hsize}
		\centerline{
			\includegraphics[height = 40pt]{./fig/colorbar_20.png}
		}
	\end{minipage}
	
	\vspace{1mm}
	
	\begin{minipage}[t]{0.075\hsize}
		\centerline{
			\includegraphics[height = 40pt]{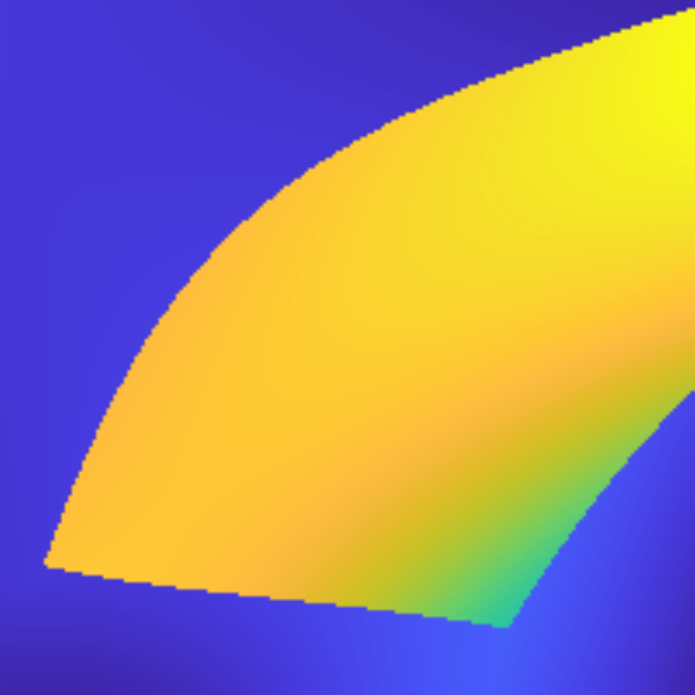}
		}
	\end{minipage}
	\begin{minipage}[t]{0.075\hsize}
		\centerline{
			\includegraphics[height = 40pt]{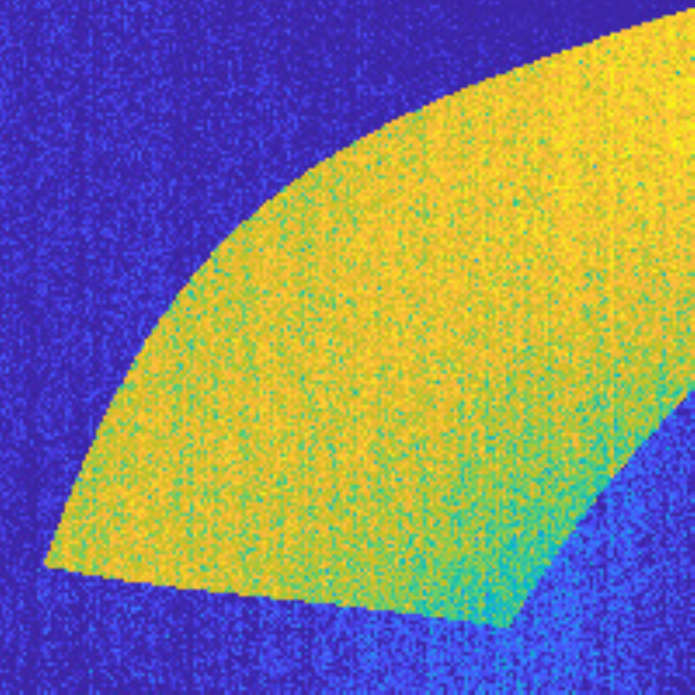}
		}
	\end{minipage}
	\begin{minipage}[t]{0.075\hsize}
		\centerline{
			\includegraphics[height = 40pt]{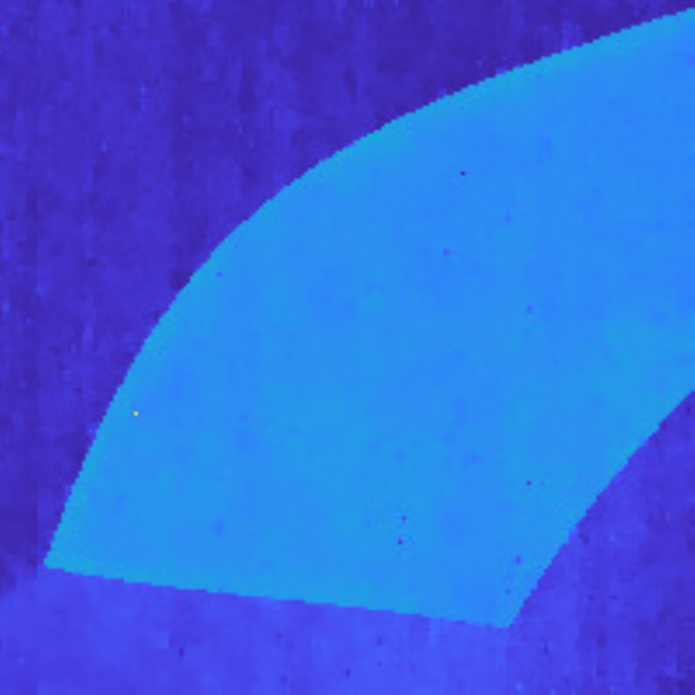}
		}
	\end{minipage}
	\begin{minipage}[t]{0.075\hsize}
		\centerline{
			\includegraphics[height = 40pt]{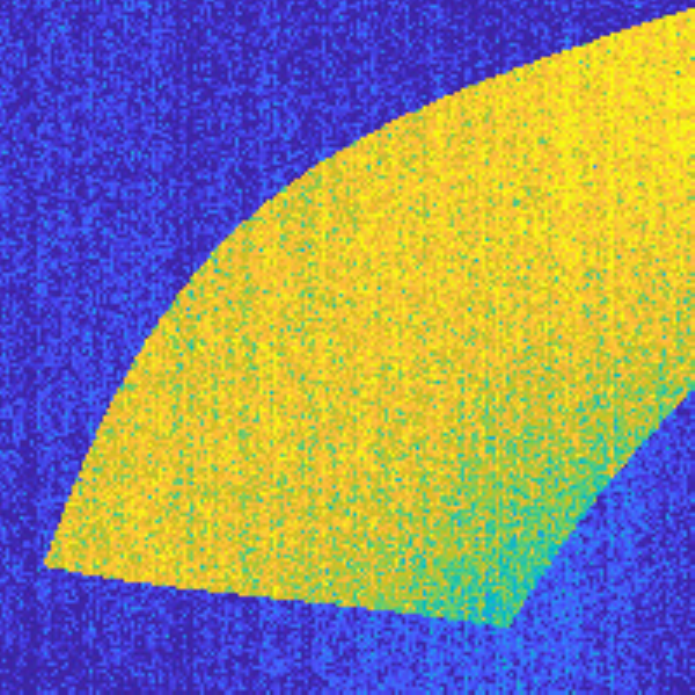}
		}
	\end{minipage}
	\begin{minipage}[t]{0.075\hsize}
		\centerline{
			\includegraphics[height = 40pt]{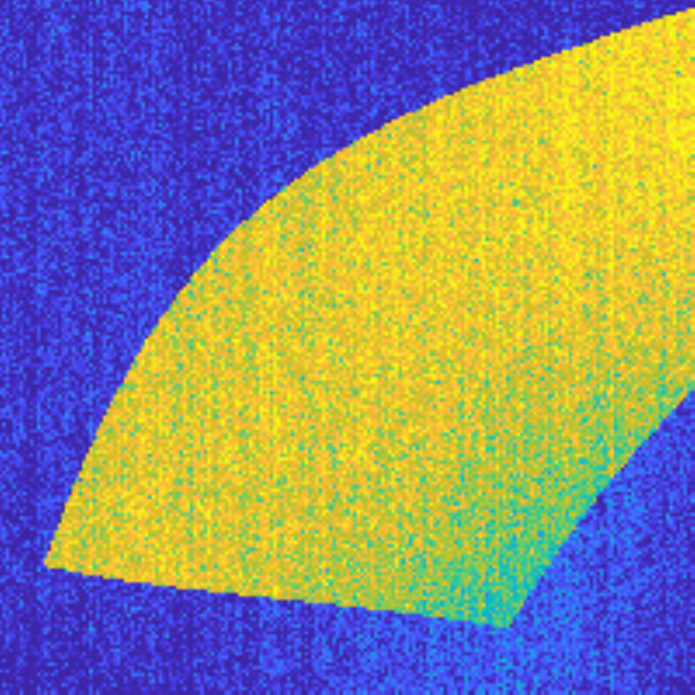}
		}
	\end{minipage}
	\begin{minipage}[t]{0.075\hsize}
		\centerline{
			\includegraphics[height = 40pt]{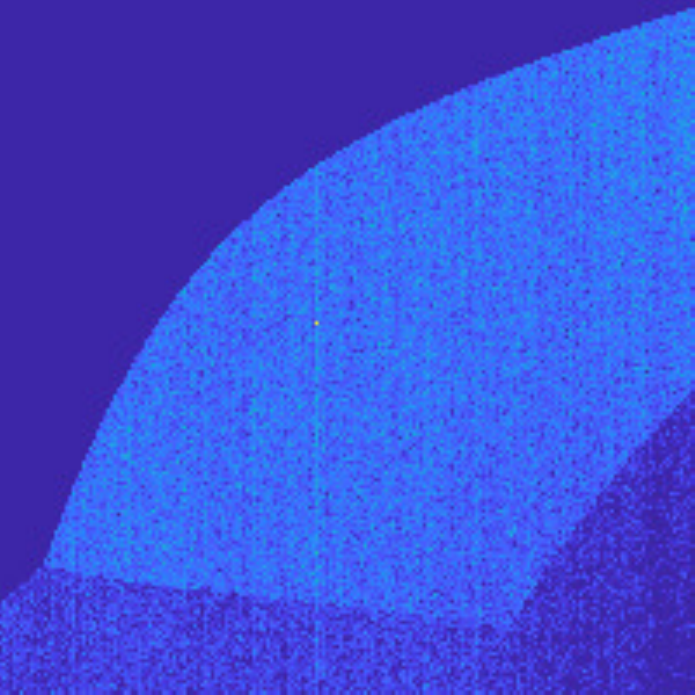}
		}
	\end{minipage}
	\begin{minipage}[t]{0.075\hsize}
		\centerline{
			\includegraphics[height = 40pt]{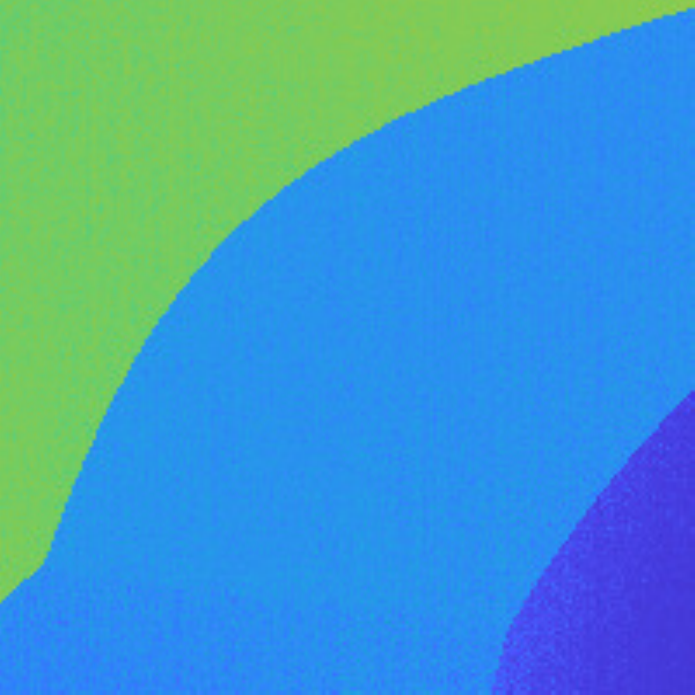}
		}
	\end{minipage}
	\begin{minipage}[t]{0.075\hsize}
		\centerline{
			\includegraphics[height = 40pt]{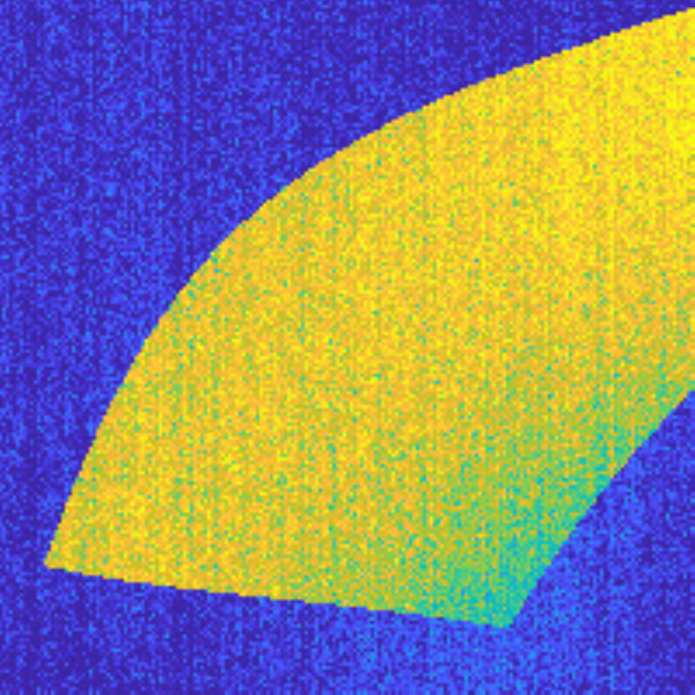}
		}
	\end{minipage}
	\begin{minipage}[t]{0.075\hsize}
		\centerline{
			\includegraphics[height = 40pt]{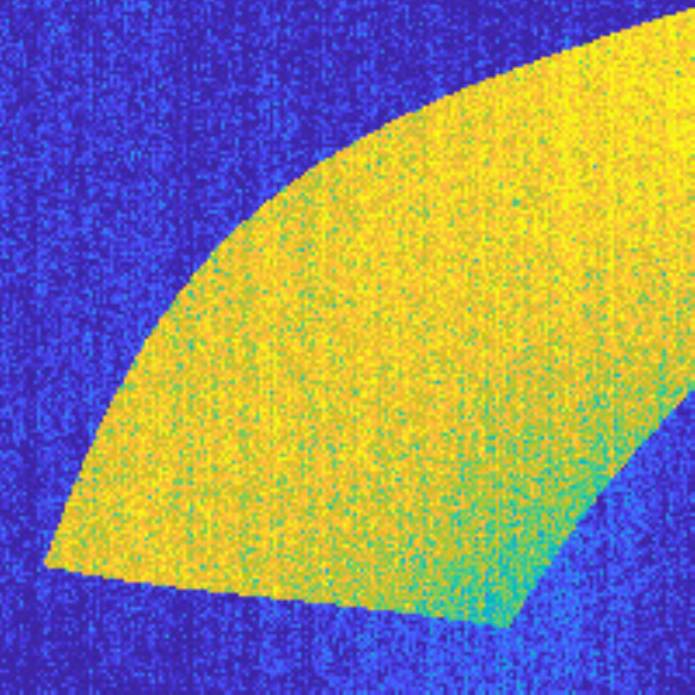}
		}
	\end{minipage}
	\begin{minipage}[t]{0.075\hsize}
		\centerline{
			\includegraphics[height = 40pt]{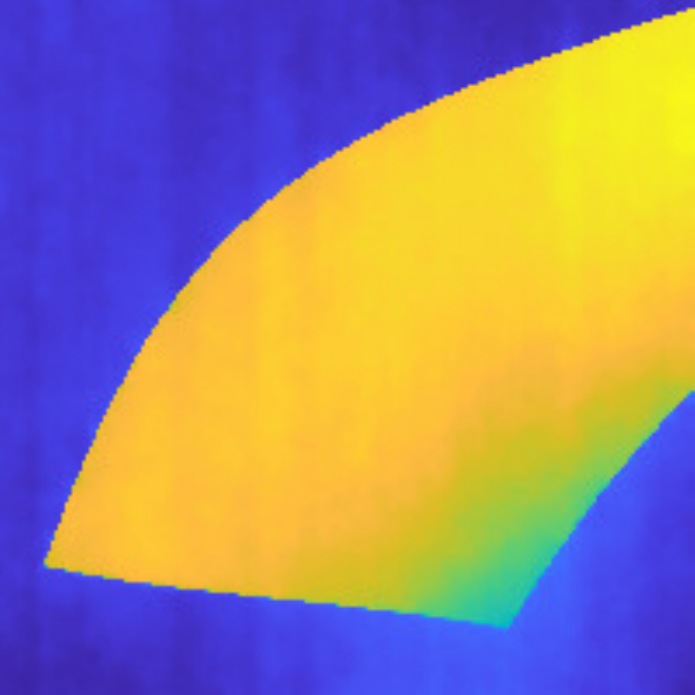}
		}
	\end{minipage}
	\begin{minipage}[t]{0.075\hsize}
		\centerline{
			\includegraphics[height = 40pt]{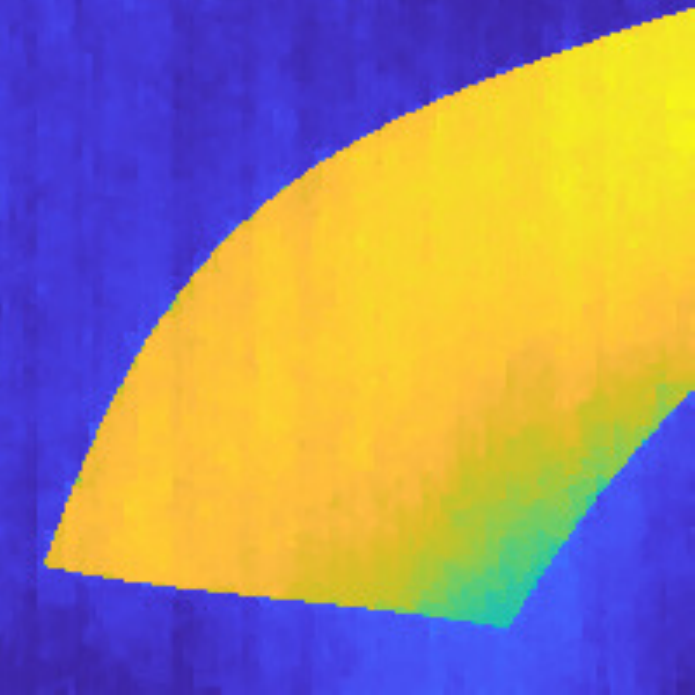}
		}
	\end{minipage}
	\begin{minipage}[t]{0.075\hsize}
		\centerline{
			\includegraphics[height = 40pt]{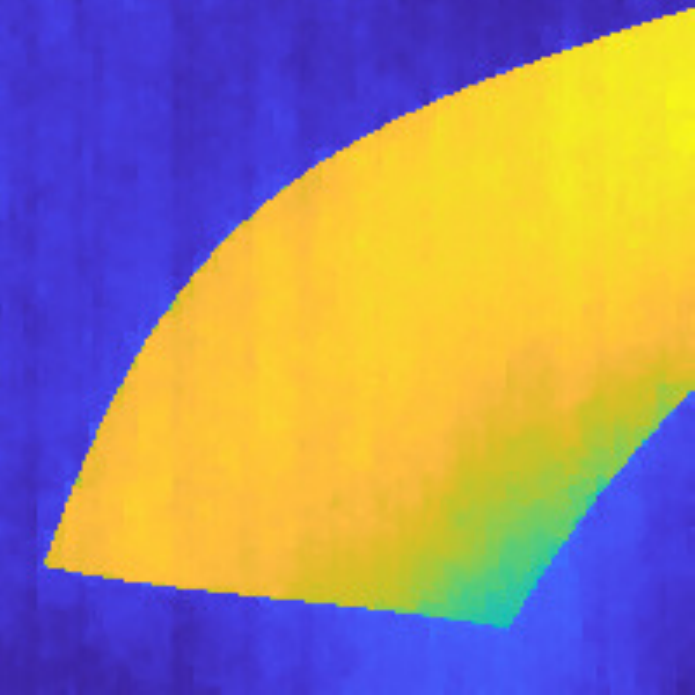}
		}
	\end{minipage}
	\begin{minipage}[t]{0.02\hsize}
		\centerline{
			\includegraphics[height = 40pt]{./fig/colorbar_20.png}
		}
	\end{minipage}
	
	\vspace{1mm}
	
	\begin{minipage}[t]{0.075\hsize}
		\centerline{
			\includegraphics[height = 40pt]{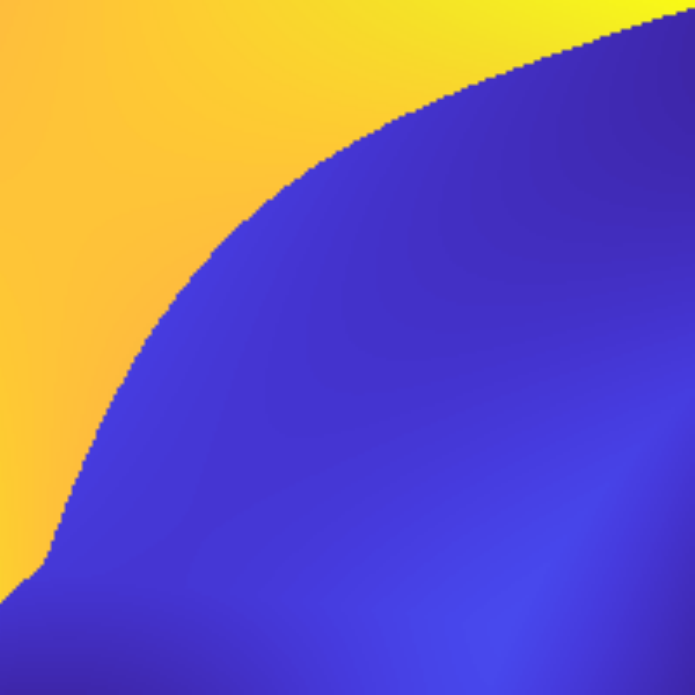}
		}
	\end{minipage}
	\begin{minipage}[t]{0.075\hsize}
		\centerline{
			\includegraphics[height = 40pt]{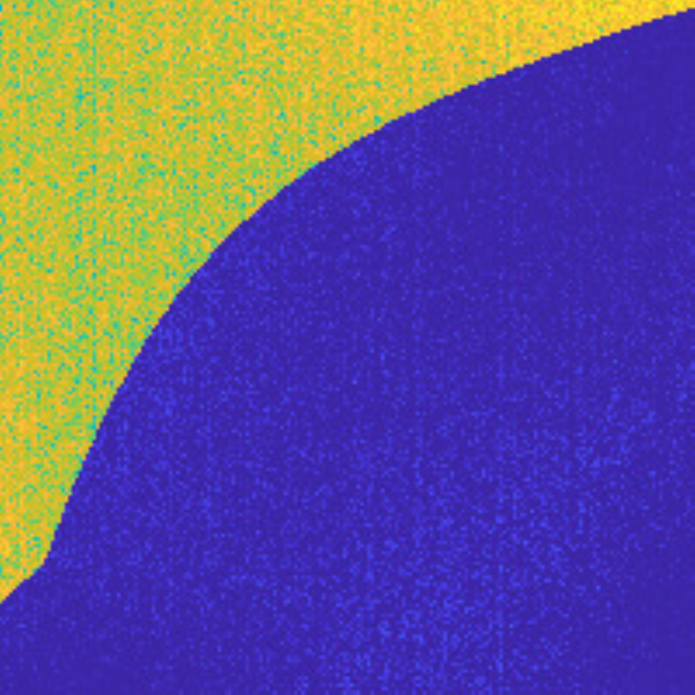}
		}
	\end{minipage}
	\begin{minipage}[t]{0.075\hsize}
		\centerline{
			\includegraphics[height = 40pt]{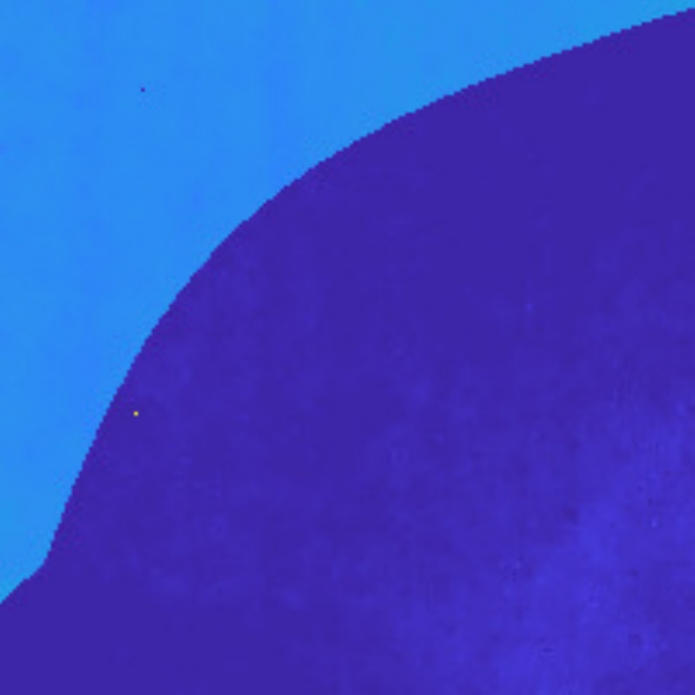}
		}
	\end{minipage}
	\begin{minipage}[t]{0.075\hsize}
		\centerline{
			\includegraphics[height = 40pt]{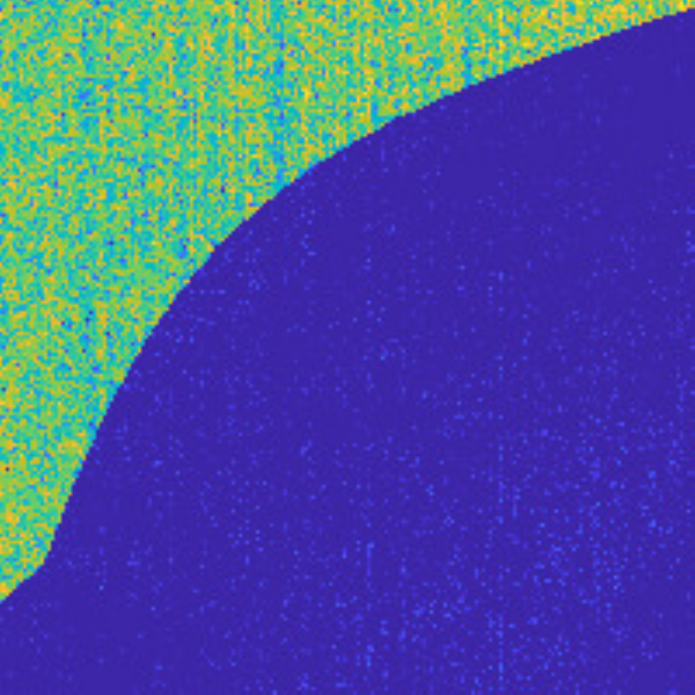}
		}
	\end{minipage}
	\begin{minipage}[t]{0.075\hsize}
		\centerline{
			\includegraphics[height = 40pt]{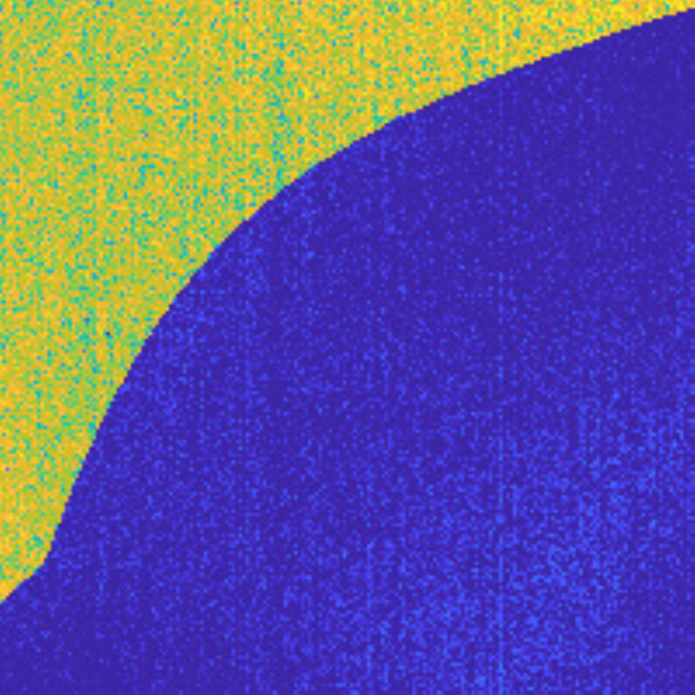}
		}
	\end{minipage}
	\begin{minipage}[t]{0.075\hsize}
		\centerline{
			\includegraphics[height = 40pt]{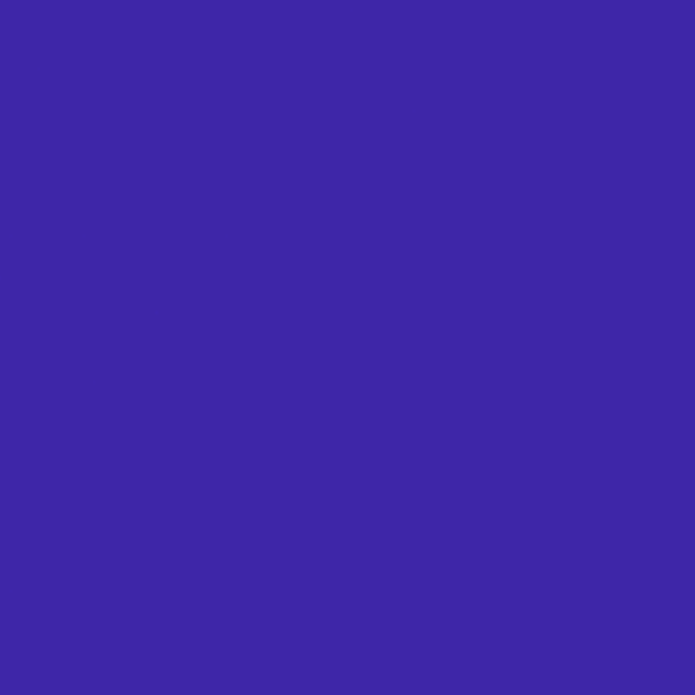}
		}
	\end{minipage}
	\begin{minipage}[t]{0.075\hsize}
		\centerline{
			\includegraphics[height = 40pt]{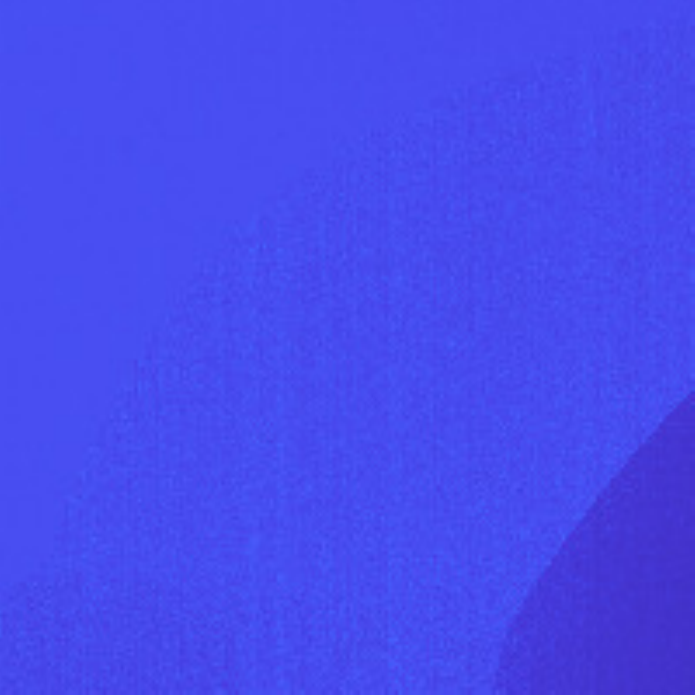}
		}
	\end{minipage}
	\begin{minipage}[t]{0.075\hsize}
		\centerline{
			\includegraphics[height = 40pt]{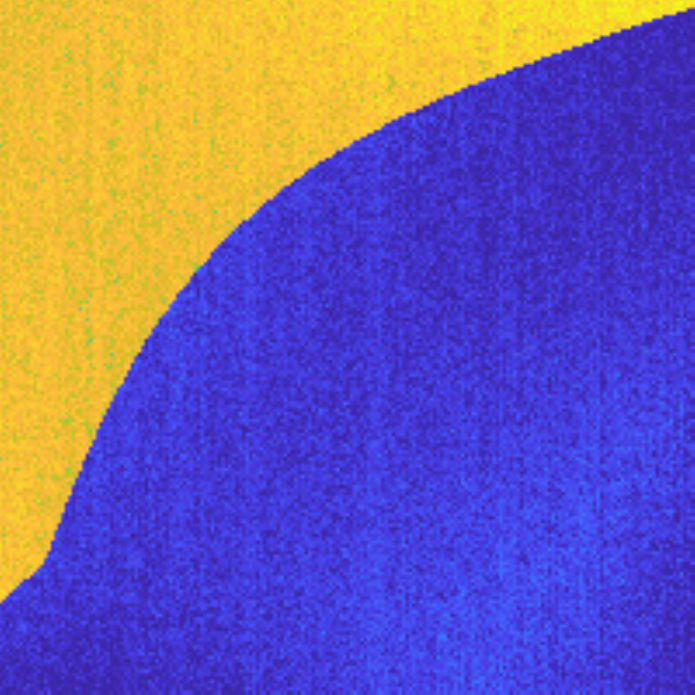}
		}
	\end{minipage}
	\begin{minipage}[t]{0.075\hsize}
		\centerline{
			\includegraphics[height = 40pt]{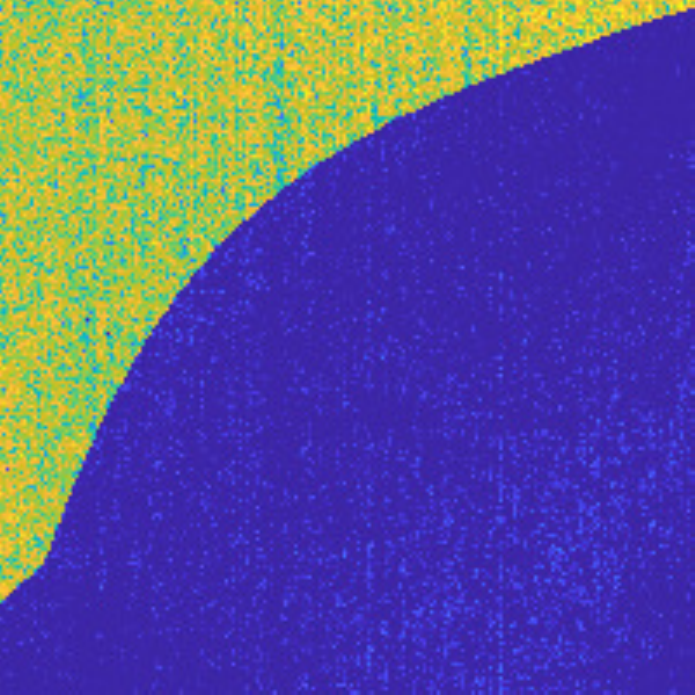}
		}
	\end{minipage}
	\begin{minipage}[t]{0.075\hsize}
		\centerline{
			\includegraphics[height = 40pt]{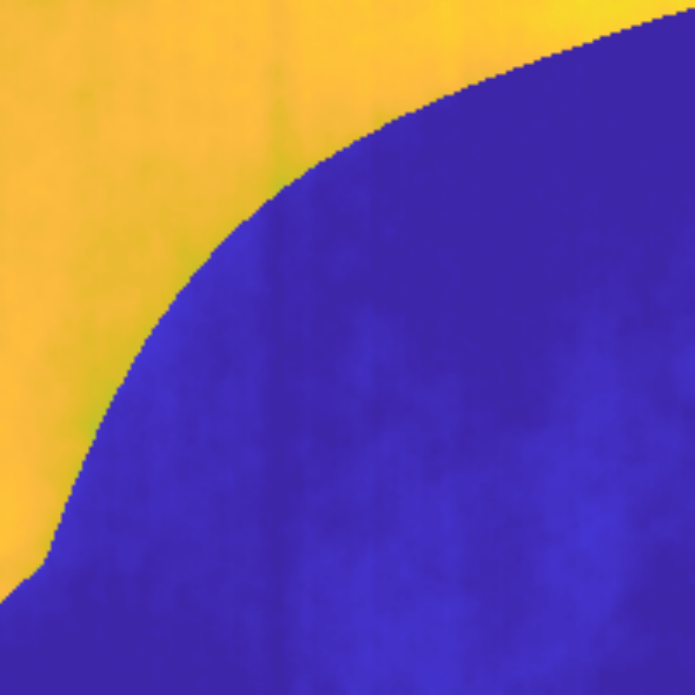}
		}
	\end{minipage}
	\begin{minipage}[t]{0.075\hsize}
		\centerline{
			\includegraphics[height = 40pt]{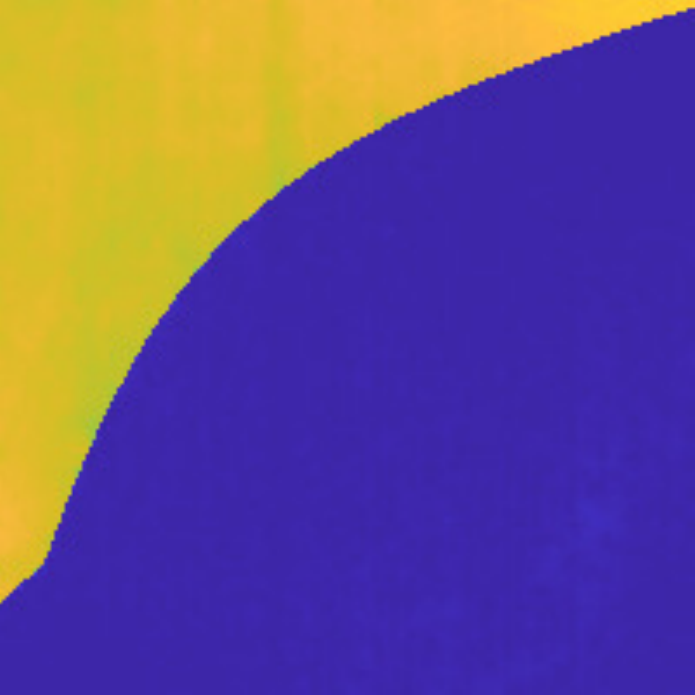}
		}
	\end{minipage}
	\begin{minipage}[t]{0.075\hsize}
		\centerline{
			\includegraphics[height = 40pt]{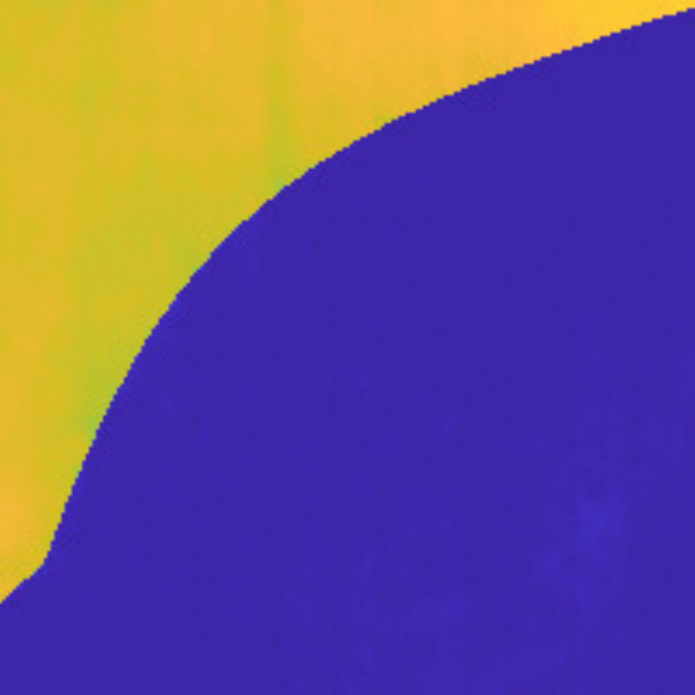}
		}
	\end{minipage}
	\begin{minipage}[t]{0.02\hsize}
		\centerline{
			\includegraphics[height = 40pt]{./fig/colorbar_20.png}
		}
	\end{minipage}
	
	\begin{minipage}[t]{0.075\hsize}
		\centerline{
			(a)
		}
	\end{minipage}
	\begin{minipage}[t]{0.075\hsize}
		\centerline{
			(b)
		}
	\end{minipage}
	\begin{minipage}[t]{0.075\hsize}
		\centerline{
			(c)
		}
	\end{minipage}
	\begin{minipage}[t]{0.075\hsize}
		\centerline{
			(d)
		}
	\end{minipage}
	\begin{minipage}[t]{0.075\hsize}
		\centerline{
			(e)
		}
	\end{minipage}
	\begin{minipage}[t]{0.075\hsize}
		\centerline{
			(f)
		}
	\end{minipage}
	\begin{minipage}[t]{0.075\hsize}
		\centerline{
			(g)
		}
	\end{minipage}
	\begin{minipage}[t]{0.075\hsize}
		\centerline{
			{(h)}
		}
	\end{minipage}
	\begin{minipage}[t]{0.075\hsize}
		\centerline{
			{(i)}
		}
	\end{minipage}
	\begin{minipage}[t]{0.075\hsize}
		\centerline{
			\textbf{(j)}
		}
	\end{minipage}
	\begin{minipage}[t]{0.075\hsize}
		\centerline{
			\textbf{(k)}
		}
	\end{minipage}
	\begin{minipage}[t]{0.075\hsize}
		\centerline{
			\textbf{(l)}
		}
	\end{minipage}
	\begin{minipage}[t]{0.02\hsize}
		\centerline{
			~
		}
	\end{minipage}
	
	\vspace{-1mm}
	
	\caption{Unmixing results of abundance maps for the \textit{Synth 3} experiments in Case 8. (a): Original abundance maps. (b): CLSUnSAL~\cite{iordache2014collaborative}. (c): JSTV~\cite{aggarwal2016hyperspectral}. (d): RSSUn-TV~\cite{wang2019row}. (e): LGSU~\cite{shen2022superpixel}. (f): UnDIP~\cite{UnDIP_RastiB_2022}. (g): EGU-Net~\cite{hong2022endmember}. (h): RDSWSU~\cite{rs_Deng_RobustDual_2023}. (i): MdLRR~\cite{MDLRR_WuLing_2023}. (j): \textbf{\Ourss (HTV)}. (k): \textbf{\Ourss (SSTV)}. (l): \textbf{\Ourss (HSSTV)}.}
	\label{fig:synth_legendre_200_noniid_Noniid_0.05_0.05}
\end{figure*}

\begin{figure*}[!h]
\centering
    \begin{minipage}[t]{0.07\hsize}
        \centerline{
        \includegraphics[width=\hsize]{./fig/synth_legendre/original/0.1_0_0/Orig_HSI-eps-converted-to.pdf}
        }
    \end{minipage}
    \begin{minipage}[t]{0.07\hsize}
        \centerline{
        \includegraphics[width=\hsize]{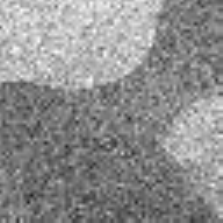}
        }
    \end{minipage}
    \begin{minipage}[t]{0.07\hsize}
        \centerline{
        \includegraphics[width=\hsize]{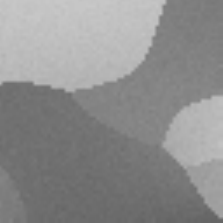}
        }
    \end{minipage}
 \begin{minipage}[t]{0.07\hsize}
  \centerline{
	    	\includegraphics[width=\hsize]{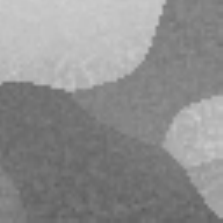}
  }
	\end{minipage}       
	\begin{minipage}[t]{0.07\hsize}
	\centerline{
		\includegraphics[width=\hsize]{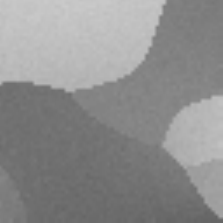}
	}
	\end{minipage}
	\begin{minipage}[t]{0.07\hsize}
	\centerline{
		\includegraphics[width=\hsize]{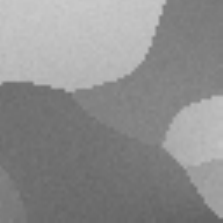}
	}
	\end{minipage}
	\begin{minipage}[t]{0.07\hsize}
		\centerline{
			\includegraphics[width=\hsize]{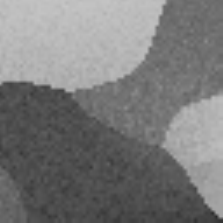}
		}
	\end{minipage}
	\begin{minipage}[t]{0.07\hsize}
	\centerline{
		\includegraphics[width=\hsize]{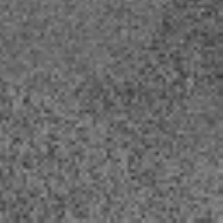}
	}
	\end{minipage}
	\begin{minipage}[t]{0.07\hsize}
	\centerline{
		\includegraphics[width=\hsize]{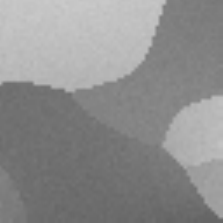}
	}
	\end{minipage}
	\begin{minipage}[t]{0.07\hsize}
		\centerline{
			\includegraphics[width=\hsize]{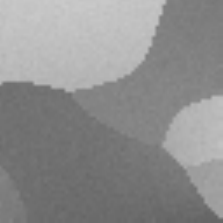}
		}
	\end{minipage}
	\begin{minipage}[t]{0.07\hsize}
		\centerline{
			\includegraphics[width=\hsize]{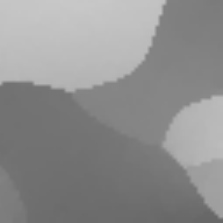}
		}
	\end{minipage}
	\begin{minipage}[t]{0.07\hsize}
		\centerline{
			\includegraphics[width=\hsize]{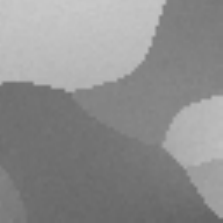}
		}
	\end{minipage}
	\begin{minipage}[t]{0.07\hsize}
		\centerline{
			\includegraphics[width=\hsize]{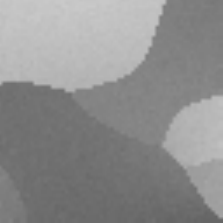}
		}
	\end{minipage}

    \begin{minipage}[t]{0.07\hsize}
        \centerline{
        (a)
        }
    \end{minipage}
    \begin{minipage}[t]{0.07\hsize}
        \centerline{
        (b)
        }
    \end{minipage}
    \begin{minipage}[t]{0.07\hsize}
        \centerline{
        (c)
        }
    \end{minipage}
    \begin{minipage}[t]{0.07\hsize}
        \centerline{
        (d)
        }
    \end{minipage}        
    \begin{minipage}[t]{0.07\hsize}
        \centerline{
        (e)
        }
    \end{minipage}
    \begin{minipage}[t]{0.07\hsize}
        \centerline{
        (f)
        }
    \end{minipage}
    \begin{minipage}[t]{0.07\hsize}
        \centerline{
        (g)
        }
    \end{minipage}
    \begin{minipage}[t]{0.07\hsize}
        \centerline{
        (h)
        }
    \end{minipage}
    \begin{minipage}[t]{0.07\hsize}
        \centerline{
        {(i)}
        }
    \end{minipage}
	\begin{minipage}[t]{0.07\hsize}
		\centerline{
			{(j)}
		}
	\end{minipage}
	\begin{minipage}[t]{0.07\hsize}
		\centerline{
			\textbf{(k)}
		}
	\end{minipage}
	\begin{minipage}[t]{0.07\hsize}
	\centerline{
		\textbf{(l)}
	}
	\end{minipage}
	\begin{minipage}[t]{0.07\hsize}
	\centerline{
		\textbf{(m)}
	}
	\end{minipage}
	
	\vspace{-1mm}

\caption{Reconstructed HS image results for the \textit{Synth 1} experiments in Case 2.  (a): Original HS image. (b): Noisy image. (c): CLSUnSAL \cite{iordache2014collaborative}.  (d): JSTV~\cite{aggarwal2016hyperspectral}. (e): RSSUn-TV~\cite{wang2019row}. (f): LGSU~\cite{shen2022superpixel}. (g): UnDIP~\cite{UnDIP_RastiB_2022}. (h): EGU-Net~\cite{hong2022endmember}. (i): RDSWSU~\cite{rs_Deng_RobustDual_2023}. (j): MdLRR~\cite{MDLRR_WuLing_2023}. (k): \textbf{\Ourss (HTV)}. (l): \textbf{\Ourss (SSTV)}. (m): \textbf{\Ourss (HSSTV)}.}
\label{fig:synth_legendre_HSI_0.1_0_0}
\end{figure*}

\begin{figure*}[!h]
\centering
    \begin{minipage}[t]{0.07\hsize}
        \centerline{
        \includegraphics[width=\hsize]{./fig/synth/original/0.05_0.05_0.05/Orig_HSI-eps-converted-to.pdf}
        }
    \end{minipage}
    \begin{minipage}[t]{0.07\hsize}
        \centerline{
        \includegraphics[width=\hsize]{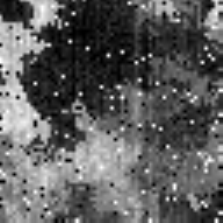}
        }
    \end{minipage}
    \begin{minipage}[t]{0.07\hsize}
        \centerline{
        \includegraphics[width=\hsize]{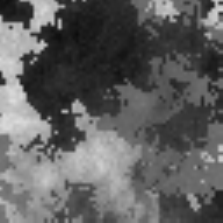}
        }
    \end{minipage}    
    \begin{minipage}[t]{0.07\hsize}
    	\centerline{
    		\includegraphics[width=\hsize]{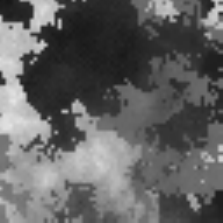}
    	}
    \end{minipage}        
    \begin{minipage}[t]{0.07\hsize}
    	\centerline{
    		\includegraphics[width=\hsize]{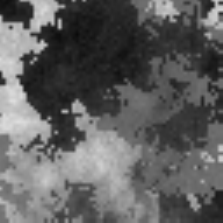}
    	}
    \end{minipage}        
    \begin{minipage}[t]{0.07\hsize}
    	\centerline{
    		\includegraphics[width=\hsize]{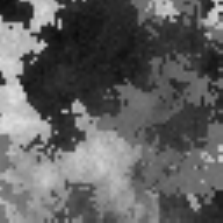}
    	}
    \end{minipage}
	\begin{minipage}[t]{0.07\hsize}
		\centerline{
			\includegraphics[width=\hsize]{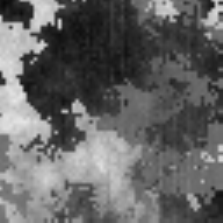}
		}
	\end{minipage}
	\begin{minipage}[t]{0.07\hsize}
	\centerline{
		\includegraphics[width=\hsize]{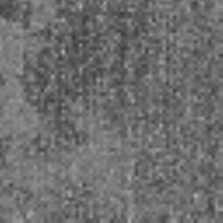}
	}
	\end{minipage}
	\begin{minipage}[t]{0.07\hsize}
	\centerline{
		\includegraphics[width=\hsize]{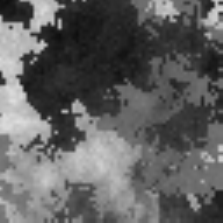}
	}
	\end{minipage}
	\begin{minipage}[t]{0.07\hsize}
		\centerline{
			\includegraphics[width=\hsize]{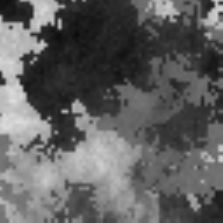}
		}
	\end{minipage}
	\begin{minipage}[t]{0.07\hsize}
		\centerline{
			\includegraphics[width=\hsize]{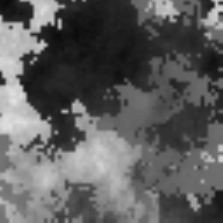}
		}
	\end{minipage}
	\begin{minipage}[t]{0.07\hsize}
		\centerline{
			\includegraphics[width=\hsize]{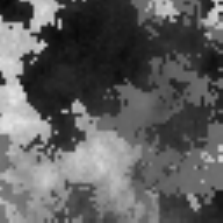}
		}
	\end{minipage}
	\begin{minipage}[t]{0.07\hsize}
		\centerline{
			\includegraphics[width=\hsize]{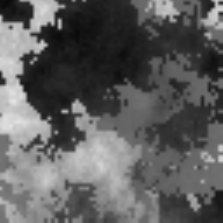}
		}
	\end{minipage}
        
    \begin{minipage}[t]{0.07\hsize}
        \centerline{
        (a)
        }
    \end{minipage}
    \begin{minipage}[t]{0.07\hsize}
        \centerline{
        (b)
        }
    \end{minipage}
    \begin{minipage}[t]{0.07\hsize}
        \centerline{
        (c)
        }
    \end{minipage}
    \begin{minipage}[t]{0.07\hsize}
        \centerline{
        (d)
        }
    \end{minipage}        
    \begin{minipage}[t]{0.07\hsize}
        \centerline{
        (e)
        }
    \end{minipage}
    \begin{minipage}[t]{0.07\hsize}
        \centerline{
        (f)
        }
    \end{minipage}
    \begin{minipage}[t]{0.07\hsize}
        \centerline{
        (g)
        }
    \end{minipage}
    \begin{minipage}[t]{0.07\hsize}
        \centerline{
        (h)
        }
    \end{minipage}
    \begin{minipage}[t]{0.07\hsize}
        \centerline{
        {(i)}
        }
    \end{minipage}
	\begin{minipage}[t]{0.07\hsize}
		\centerline{
			{(j)}
		}
	\end{minipage}
	\begin{minipage}[t]{0.07\hsize}
		\centerline{
			\textbf{(k)}
		}
	\end{minipage}
	\begin{minipage}[t]{0.07\hsize}
	\centerline{
		\textbf{(l)}
	}
	\end{minipage}
	\begin{minipage}[t]{0.07\hsize}
	\centerline{
		\textbf{(m)}
	}
	\end{minipage}

        \vspace{-1mm}

\caption{Reconstructed HS image results for
the \textit{Synth 2} experiments in Case 5. (a): Original HS image. (b): Noisy image. (c): CLSUnSAL \cite{iordache2014collaborative}.  (d): JSTV~\cite{aggarwal2016hyperspectral}. (e): RSSUn-TV~\cite{wang2019row}. (f): LGSU~\cite{shen2022superpixel}. (g): UnDIP~\cite{UnDIP_RastiB_2022}. (h): EGU-Net~\cite{hong2022endmember}. (i): RDSWSU~\cite{rs_Deng_RobustDual_2023}. (j): MdLRR~\cite{MDLRR_WuLing_2023}. (k): \textbf{\Ourss (HTV)}. (l): \textbf{\Ourss (SSTV)}. (m): \textbf{\Ourss (HSSTV)}.}
\label{synth_HSI_0.05_0.05_0.05}
\end{figure*}

\begin{figure*}[!h]
	\centering
	\begin{minipage}[t]{0.07\hsize}
		\centerline{
			\includegraphics[width=\hsize]{./fig/synth_legendre_200_noniid/original/Noniid_0.05_0.05/Orig_HSI-eps-converted-to.pdf}
		}
	\end{minipage}
	\begin{minipage}[t]{0.07\hsize}
		\centerline{
			\includegraphics[width=\hsize]{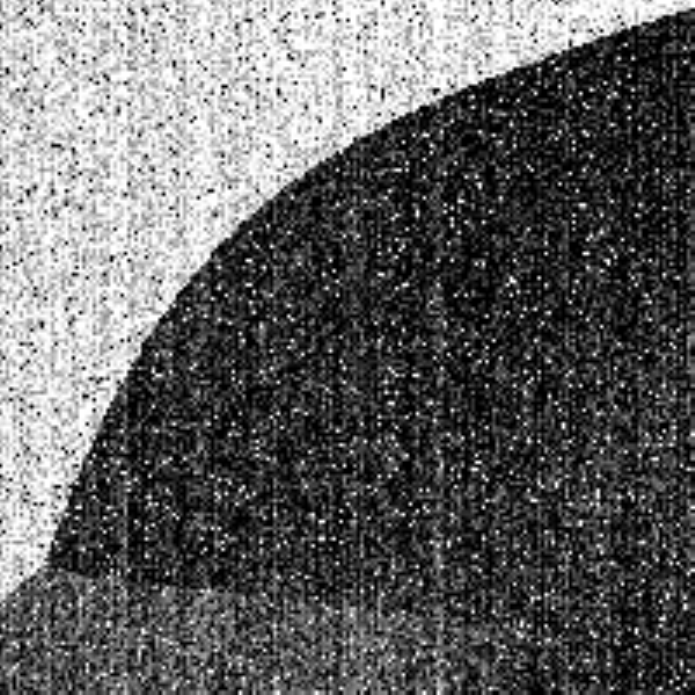}
		}
	\end{minipage}
	\begin{minipage}[t]{0.07\hsize}
		\centerline{
			\includegraphics[width=\hsize]{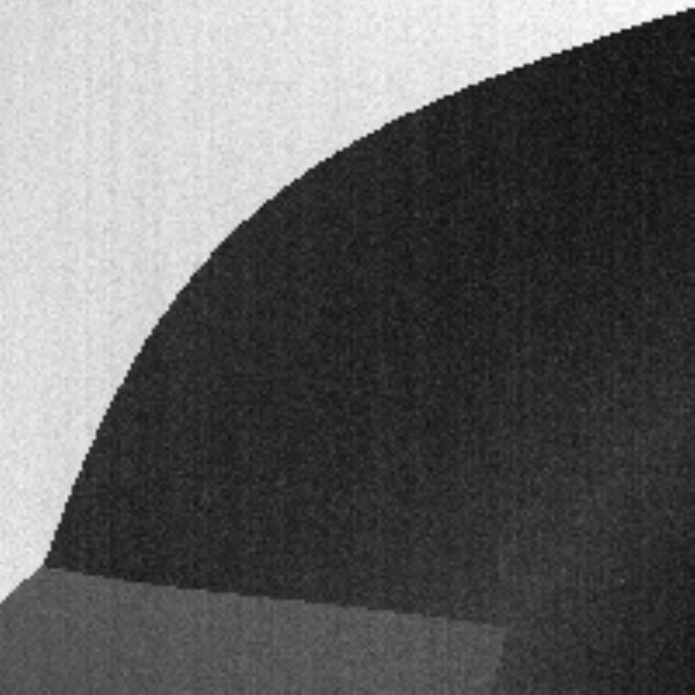}
		}
	\end{minipage}
	\begin{minipage}[t]{0.07\hsize}
		\centerline{
			\includegraphics[width=\hsize]{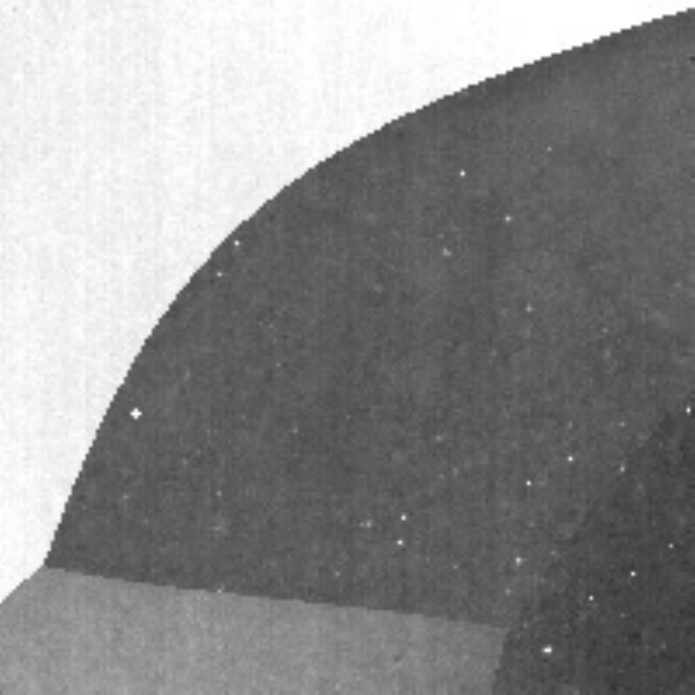}
		}
	\end{minipage}       
	\begin{minipage}[t]{0.07\hsize}
		\centerline{
			\includegraphics[width=\hsize]{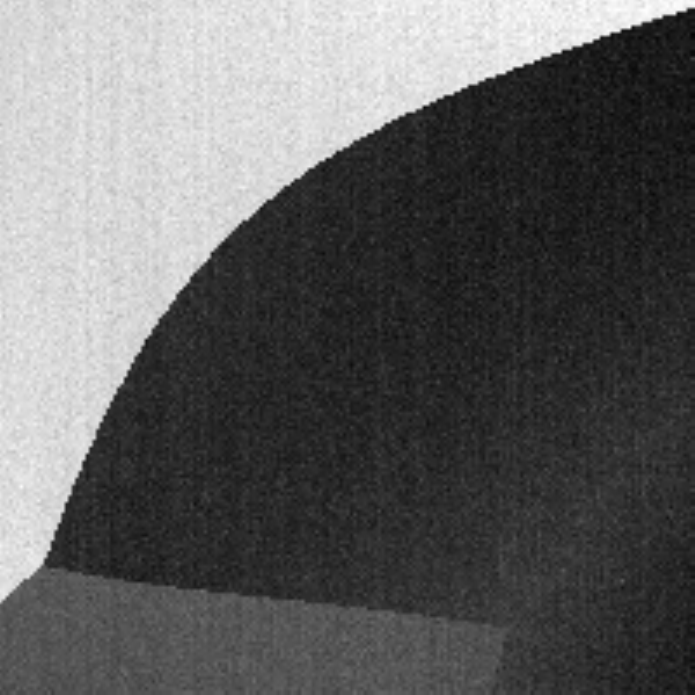}
		}
	\end{minipage}
	\begin{minipage}[t]{0.07\hsize}
		\centerline{
			\includegraphics[width=\hsize]{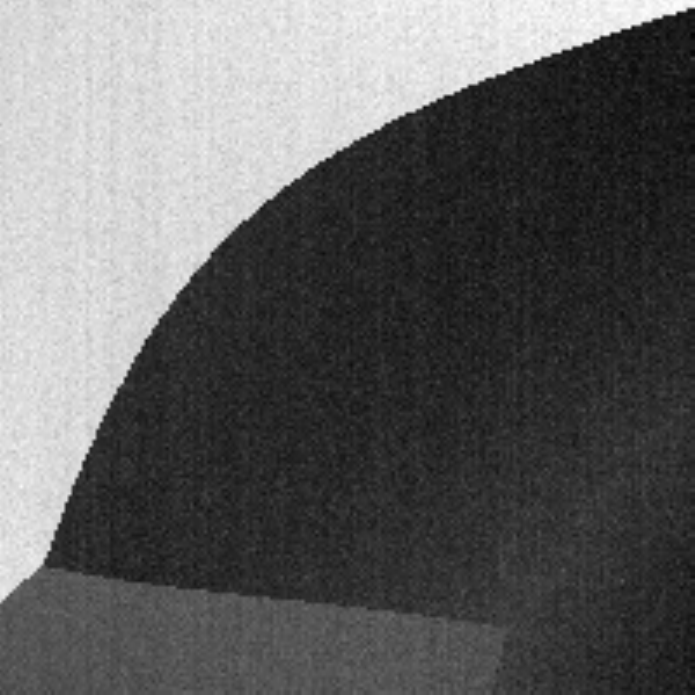}
		}
	\end{minipage}
	\begin{minipage}[t]{0.07\hsize}
		\centerline{
			\includegraphics[width=\hsize]{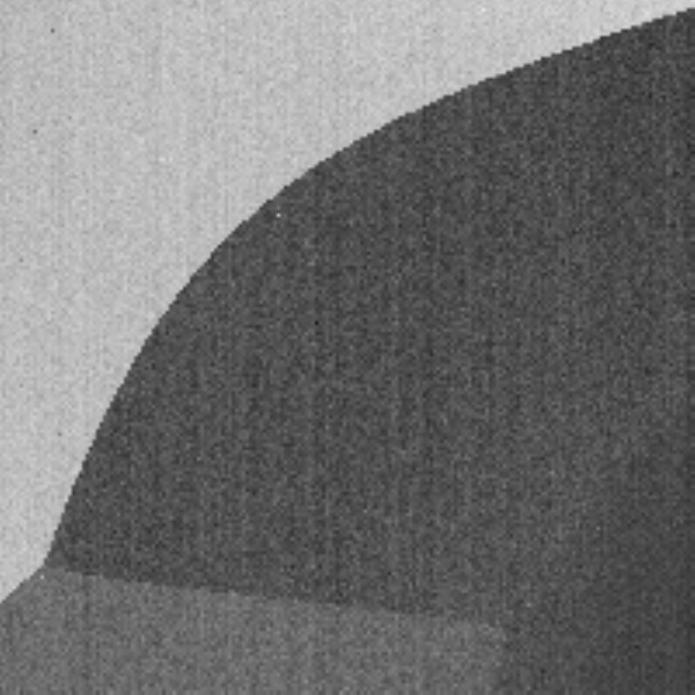}
		}
	\end{minipage}
	\begin{minipage}[t]{0.07\hsize}
		\centerline{
			\includegraphics[width=\hsize]{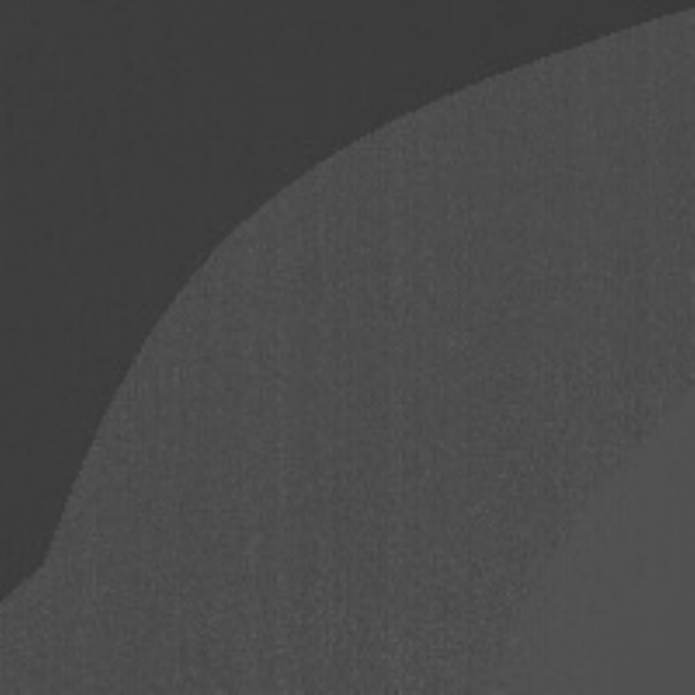}
		}
	\end{minipage}
	\begin{minipage}[t]{0.07\hsize}
		\centerline{
			\includegraphics[width=\hsize]{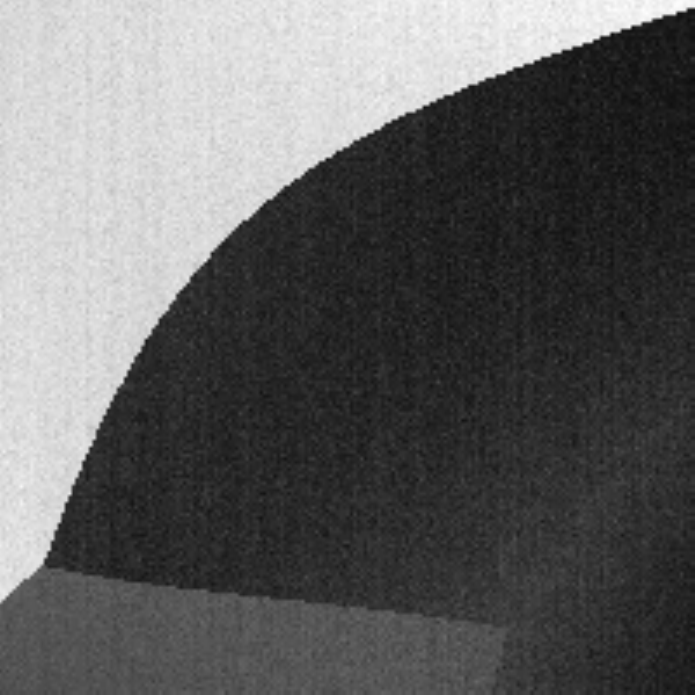}
		}
	\end{minipage}
	\begin{minipage}[t]{0.07\hsize}
		\centerline{
			\includegraphics[width=\hsize]{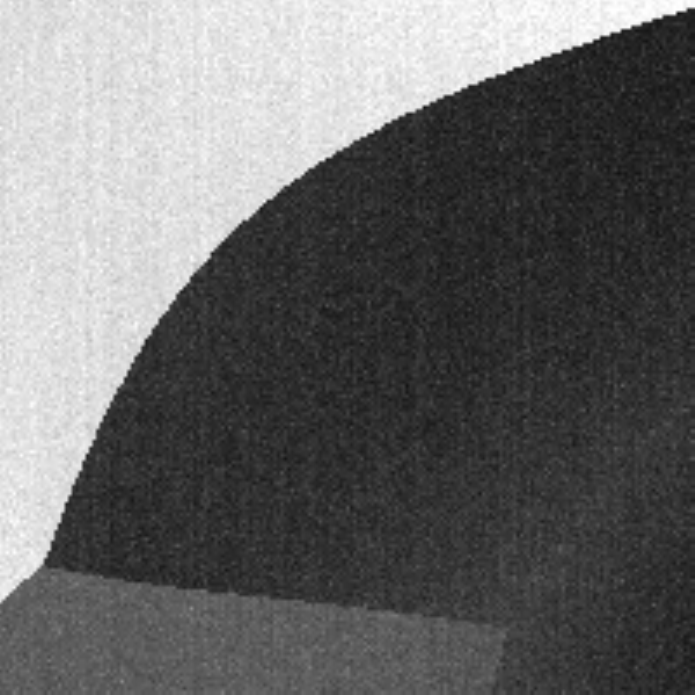}
		}
	\end{minipage}
	\begin{minipage}[t]{0.07\hsize}
		\centerline{
			\includegraphics[width=\hsize]{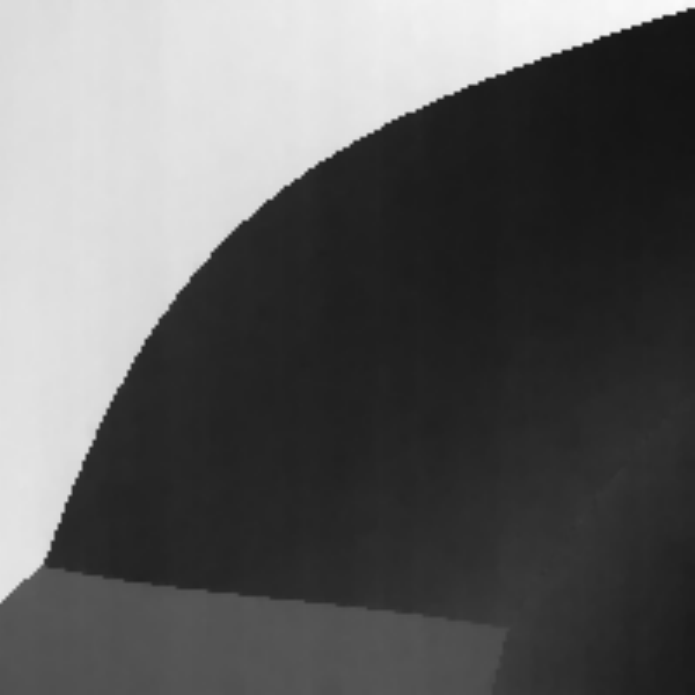}
		}
	\end{minipage}
	\begin{minipage}[t]{0.07\hsize}
		\centerline{
			\includegraphics[width=\hsize]{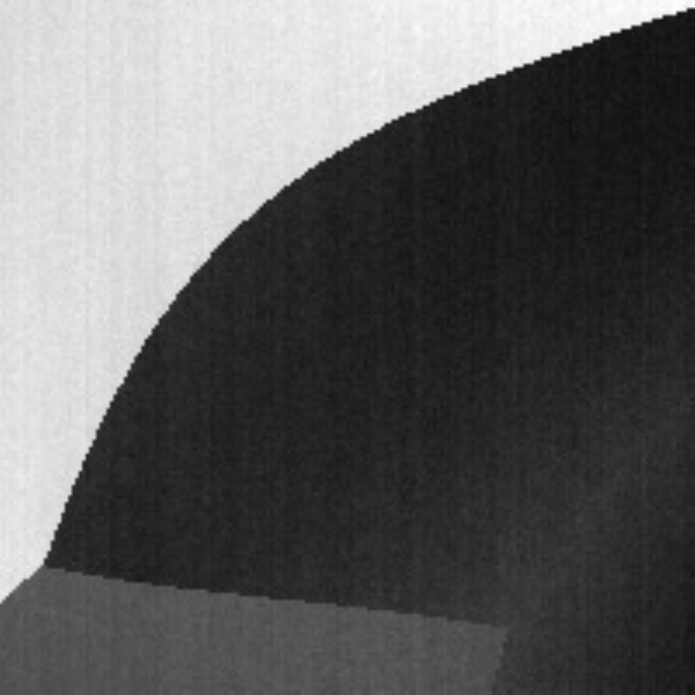}
		}
	\end{minipage}
	\begin{minipage}[t]{0.07\hsize}
		\centerline{
			\includegraphics[width=\hsize]{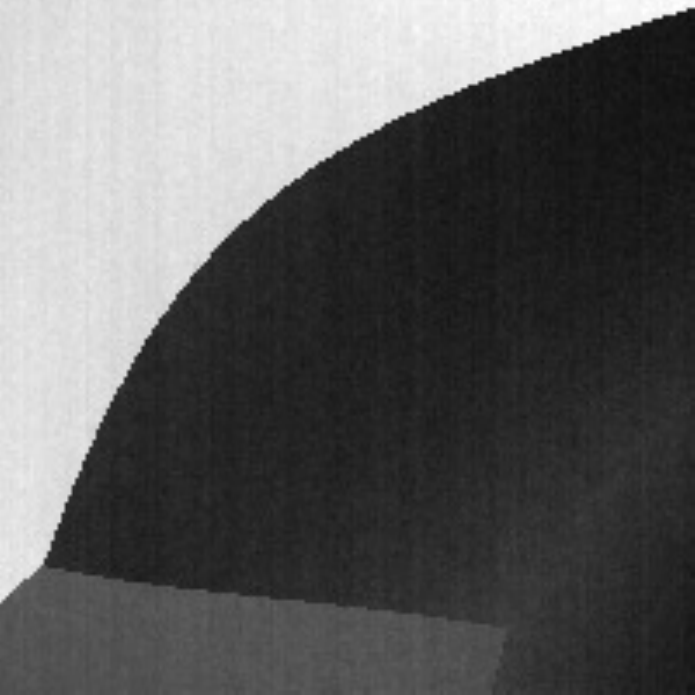}
		}
	\end{minipage}

	\begin{minipage}[t]{0.07\hsize}
		\centerline{
			(a)
		}
	\end{minipage}
	\begin{minipage}[t]{0.07\hsize}
		\centerline{
			(b)
		}
	\end{minipage}
	\begin{minipage}[t]{0.07\hsize}
		\centerline{
			(c)
		}
	\end{minipage}
	\begin{minipage}[t]{0.07\hsize}
		\centerline{
			(d)
		}
	\end{minipage}        
	\begin{minipage}[t]{0.07\hsize}
		\centerline{
			(e)
		}
	\end{minipage}
	\begin{minipage}[t]{0.07\hsize}
		\centerline{
			(f)
		}
	\end{minipage}
	\begin{minipage}[t]{0.07\hsize}
		\centerline{
			(g)
		}
	\end{minipage}
	\begin{minipage}[t]{0.07\hsize}
		\centerline{
			(h)
		}
	\end{minipage}
	\begin{minipage}[t]{0.07\hsize}
		\centerline{
			{(i)}
		}
	\end{minipage}
	\begin{minipage}[t]{0.07\hsize}
		\centerline{
			{(j)}
		}
	\end{minipage}
	\begin{minipage}[t]{0.07\hsize}
		\centerline{
			\textbf{(k)}
		}
	\end{minipage}
	\begin{minipage}[t]{0.07\hsize}
		\centerline{
			\textbf{(l)}
		}
	\end{minipage}
	\begin{minipage}[t]{0.07\hsize}
		\centerline{
			\textbf{(m)}
		}
	\end{minipage}
	
	\vspace{-1mm}

	\caption{Reconstructed HS image results for the \textit{Synth 3} experiments in Case 8.  (a): Original HS image. (b): Noisy image. (c): CLSUnSAL \cite{iordache2014collaborative}.  (d): JSTV~\cite{aggarwal2016hyperspectral}. (e): RSSUn-TV~\cite{wang2019row}. (f): LGSU~\cite{shen2022superpixel}. (g): UnDIP~\cite{UnDIP_RastiB_2022}. (h): EGU-Net~\cite{hong2022endmember}. (i): RDSWSU~\cite{rs_Deng_RobustDual_2023}. (j): MdLRR~\cite{MDLRR_WuLing_2023}. (k): \textbf{\Ourss (HTV)}. (l): \textbf{\Ourss (SSTV)}. (m): \textbf{\Ourss (HSSTV)}.}
	\label{fig:synth_legendre_200_noniid200_noniid_HSI_0.1_0.1}
\end{figure*}

\begin{table*}[!h]
	\caption{SRE, RMSE, Ps, MPSNR, and MSSIM in the Experiments Using \textit{Jasper Ridge}.}
	\vspace{-1mm}
	\label{tab:result_jasper}
	\centering
	\scalebox{0.75}{
		\begin{tabular}{ccccccccccccc} \toprule
			\multirow{3}{*}{ Noise } & \multirow{3}{*}{Metrics}& \multicolumn{11}{c}{Methods}\\ 
			\cmidrule(lr){3-13}
			& & CLSUnSAL & JSTV & RSSUn-TV & LGSU & UnDIP & EGU-Net & RDSWSU & MdLRR & \textbf{\Ours} & \textbf{\Ours} & \textbf{\Ours} \\ 
			& & \cite{iordache2014collaborative} & \cite{aggarwal2016hyperspectral} & \cite{wang2019row} & \cite{shen2022superpixel} & \cite{UnDIP_RastiB_2022} & \cite{hong2022endmember} & \cite{rs_Deng_RobustDual_2023} & \cite{MDLRR_WuLing_2023} & \textbf{(HTV)} & \textbf{(SSTV)} & \textbf{(HSSTV)} \\ 
			\midrule
			
\multirow{6}{*}{Case 1} 
& Setup & \begin{tabular}{c} \hspace{-4mm} $\lambda = 10^{0}$ \hspace{-4mm}\end{tabular} & \begin{tabular}{c} \hspace{-4mm} $\lambda_{1} = 10^{0}$, \hspace{-4mm} \\\hspace{-4mm} $\lambda_{2} = 10^{-2}$, \hspace{-4mm} \\\hspace{-4mm} $\lambda_{3} = 10^{1}$ \hspace{-4mm}\end{tabular} & \begin{tabular}{c} \hspace{-4mm} $\lambda = 10^{1}$, \hspace{-4mm} \\\hspace{-4mm} $\lambda_{TV} = 10^{-2}$ \hspace{-4mm}\end{tabular} & \begin{tabular}{c} \hspace{-4mm} $\lambda_{g} = 10^{-1}$, \hspace{-4mm} \\\hspace{-4mm} $\lambda_{l} = 10^{-2}$ \hspace{-4mm}\end{tabular} &  --  &  --  & \begin{tabular}{c} \hspace{-4mm} $\lambda = 10^{-1}$ \hspace{-4mm}\end{tabular} & \begin{tabular}{c} \hspace{-4mm} $\lambda = 10^{-1}$, \hspace{-4mm} \\\hspace{-4mm} $\tau = 10^{0}$ \hspace{-4mm}\end{tabular} & \begin{tabular}{c} \hspace{-4mm} $\lambda_{1} = 10^{0}$, \hspace{-4mm} \\\hspace{-4mm} $\lambda_{2} = 10^{-2}$, \hspace{-4mm} \\\hspace{-4mm} $\varepsilon = 0.95$ \hspace{-4mm}\end{tabular} & \begin{tabular}{c} \hspace{-4mm} $\lambda_{1} = 10^{0}$, \hspace{-4mm} \\\hspace{-4mm} $\lambda_{2} = 10^{-2}$, \hspace{-4mm} \\\hspace{-4mm} $\varepsilon = 0.95$ \hspace{-4mm}\end{tabular} & \begin{tabular}{c} \hspace{-4mm} $\lambda_{1} = 10^{0}$, \hspace{-4mm} \\\hspace{-4mm} $\lambda_{2} = 10^{-2}$, \hspace{-4mm} \\\hspace{-4mm} $\varepsilon = 0.95$ \hspace{-4mm}\end{tabular} \\ 
& SRE  &   \ValSecnd{20.24} &   15.99 &   20.12 &   19.45 &    3.59 &   -1.42 &   20.13 &   \Valbest{20.48} &   19.07 &   19.33 &   19.06 \\ 
& RMSE &   \ValSecnd{0.0256} &   0.0412 &   0.0263 &   0.0286 &   0.1518 &   0.2002 &   0.0270 &   \Valbest{0.0252} &   0.0294 &   0.0286 &   0.0295 \\ 
& Ps   &   \Valbest{1.00} &   \Valbest{1.00} &   \Valbest{1.00} &   \Valbest{1.00} &   0.64 &   0.07 &   \Valbest{1.00} &   \Valbest{1.00} &   \Valbest{1.00} &   \Valbest{1.00} &   \Valbest{1.00} \\ 
& MPSNR &   45.22 &   35.09 &   45.50 &   45.60 &   32.81 &   18.56 &   \Valbest{45.78} &   \ValSecnd{45.65} &   45.45 &   45.47 &   45.45 \\ 
& MSSIM &   0.9877 &   0.9234 &   0.9878 &   0.9871 &   0.9330 &   0.4523 &   0.9877 &   0.9876 &   \Valbest{0.9882} &   0.9881 &   \Valbest{0.9882} \\ 
\midrule 

\multirow{6}{*}{Case 2} 
& Setup & \begin{tabular}{c} \hspace{-4mm} $\lambda = 10^{0}$ \hspace{-4mm}\end{tabular} & \begin{tabular}{c} \hspace{-4mm} $\lambda_{1} = 10^{0}$, \hspace{-4mm} \\\hspace{-4mm} $\lambda_{2} = 10^{-2}$, \hspace{-4mm} \\\hspace{-4mm} $\lambda_{3} = 10^{0}$ \hspace{-4mm}\end{tabular} & \begin{tabular}{c} \hspace{-4mm} $\lambda = 10^{1}$, \hspace{-4mm} \\\hspace{-4mm} $\lambda_{TV} = 10^{-2}$ \hspace{-4mm}\end{tabular} & \begin{tabular}{c} \hspace{-4mm} $\lambda_{g} = 10^{0}$, \hspace{-4mm} \\\hspace{-4mm} $\lambda_{l} = 10^{-2}$ \hspace{-4mm}\end{tabular} &  --  &  --  & \begin{tabular}{c} \hspace{-4mm} $\lambda = 10^{-1}$ \hspace{-4mm}\end{tabular} & \begin{tabular}{c} \hspace{-4mm} $\lambda = 10^{-1}$, \hspace{-4mm} \\\hspace{-4mm} $\tau = 10^{0}$ \hspace{-4mm}\end{tabular} & \begin{tabular}{c} \hspace{-4mm} $\lambda_{1} = 10^{0}$, \hspace{-4mm} \\\hspace{-4mm} $\lambda_{2} = 10^{0}$, \hspace{-4mm} \\\hspace{-4mm} $\varepsilon = 0.98$ \hspace{-4mm}\end{tabular} & \begin{tabular}{c} \hspace{-4mm} $\lambda_{1} = 10^{0}$, \hspace{-4mm} \\\hspace{-4mm} $\lambda_{2} = 10^{-2}$, \hspace{-4mm} \\\hspace{-4mm} $\varepsilon = 0.98$ \hspace{-4mm}\end{tabular} & \begin{tabular}{c} \hspace{-4mm} $\lambda_{1} = 10^{0}$, \hspace{-4mm} \\\hspace{-4mm} $\lambda_{2} = 10^{-2}$, \hspace{-4mm} \\\hspace{-4mm} $\varepsilon = 0.98$ \hspace{-4mm}\end{tabular} \\ 
& SRE  &   14.81 &   10.14 &   13.91 &   13.93 &   -0.13 &   -3.23 &   15.35 &   14.32 &   \ValSecnd{15.49} &   \Valbest{15.53} &   15.37 \\ 
& RMSE &   0.0468 &   0.0743 &   0.0535 &   0.0527 &   0.2247 &   0.2664 &   0.0470 &   0.0510 &   \ValSecnd{0.0436} &   \Valbest{0.0432} &   0.0441 \\ 
& Ps   &   \Valbest{0.99} &   0.96 &   0.98 &   0.98 &   0.58 &   0.00 &   \Valbest{0.99} &   \Valbest{0.99} &   \Valbest{0.99} &   \Valbest{0.99} &   \Valbest{0.99} \\ 
& MPSNR &   39.53 &   32.25 &   39.45 &   39.77 &   27.85 &   17.65 &   40.15 &   39.46 &   \Valbest{40.97} &   40.31 &   \ValSecnd{40.36} \\ 
& MSSIM &   0.9582 &   0.8809 &   0.9533 &   0.9541 &   0.8343 &   0.4552 &   0.9569 &   0.9512 &   \Valbest{0.9732} &   0.9634 &   \ValSecnd{0.9641} \\ 
\midrule 

\multirow{6}{*}{Case 3} 
& Setup & \begin{tabular}{c} \hspace{-4mm} $\lambda = 10^{0}$ \hspace{-4mm}\end{tabular} & \begin{tabular}{c} \hspace{-4mm} $\lambda_{1} = 10^{0}$, \hspace{-4mm} \\\hspace{-4mm} $\lambda_{2} = 10^{1}$, \hspace{-4mm} \\\hspace{-4mm} $\lambda_{3} = 10^{1}$ \hspace{-4mm}\end{tabular} & \begin{tabular}{c} \hspace{-4mm} $\lambda = 10^{1}$, \hspace{-4mm} \\\hspace{-4mm} $\lambda_{TV} = 10^{-2}$ \hspace{-4mm}\end{tabular} & \begin{tabular}{c} \hspace{-4mm} $\lambda_{g} = 10^{0}$, \hspace{-4mm} \\\hspace{-4mm} $\lambda_{l} = 10^{-2}$ \hspace{-4mm}\end{tabular} &  --  &  --  & \begin{tabular}{c} \hspace{-4mm} $\lambda = 10^{-1}$ \hspace{-4mm}\end{tabular} & \begin{tabular}{c} \hspace{-4mm} $\lambda = 10^{-1}$, \hspace{-4mm} \\\hspace{-4mm} $\tau = 10^{0}$ \hspace{-4mm}\end{tabular} & \begin{tabular}{c} \hspace{-4mm} $\lambda_{1} = 10^{0}$, \hspace{-4mm} \\\hspace{-4mm} $\lambda_{2} = 10^{-2}$, \hspace{-4mm} \\\hspace{-4mm} $\varepsilon = 0.95$ \hspace{-4mm}\end{tabular} & \begin{tabular}{c} \hspace{-4mm} $\lambda_{1} = 10^{0}$, \hspace{-4mm} \\\hspace{-4mm} $\lambda_{2} = 10^{-2}$, \hspace{-4mm} \\\hspace{-4mm} $\varepsilon = 0.95$ \hspace{-4mm}\end{tabular} & \begin{tabular}{c} \hspace{-4mm} $\lambda_{1} = 10^{0}$, \hspace{-4mm} \\\hspace{-4mm} $\lambda_{2} = 10^{-2}$, \hspace{-4mm} \\\hspace{-4mm} $\varepsilon = 0.95$ \hspace{-4mm}\end{tabular} \\ 
& SRE  &   10.14 &   15.82 &    9.06 &    8.78 &   -1.88 &   -1.89 &   10.25 &    8.87 &   \ValSecnd{18.76} &   \Valbest{18.98} &   18.75 \\ 
& RMSE &   0.0751 &   0.0419 &   0.0961 &   0.0948 &   0.2327 &   0.1952 &   0.0872 &   0.0969 &   \ValSecnd{0.0304} &   \Valbest{0.0296} &   \ValSecnd{0.0304} \\ 
& Ps   &   0.94 &   \Valbest{1.00} &   0.88 &   0.89 &   0.35 &   0.33 &   0.92 &   0.88 &   \Valbest{1.00} &   \Valbest{1.00} &   \Valbest{1.00} \\ 
& MPSNR &   32.74 &   34.89 &   32.66 &   32.74 &   24.90 &   19.56 &   33.50 &   32.54 &   \Valbest{44.47} &   \Valbest{44.47} &   \Valbest{44.47} \\ 
& MSSIM &   0.8733 &   0.9195 &   0.8628 &   0.8584 &   0.7162 &   0.5420 &   0.8761 &   0.8561 &   \ValSecnd{0.9849} &   0.9848 &   \Valbest{0.9850} \\ 
\midrule 

\multirow{6}{*}{Case 4} 
& Setup & \begin{tabular}{c} \hspace{-4mm} $\lambda = 10^{1}$ \hspace{-4mm}\end{tabular} & \begin{tabular}{c} \hspace{-4mm} $\lambda_{1} = 10^{0}$, \hspace{-4mm} \\\hspace{-4mm} $\lambda_{2} = 10^{-1}$, \hspace{-4mm} \\\hspace{-4mm} $\lambda_{3} = 10^{0}$ \hspace{-4mm}\end{tabular} & \begin{tabular}{c} \hspace{-4mm} $\lambda = 10^{1}$, \hspace{-4mm} \\\hspace{-4mm} $\lambda_{TV} = 10^{-2}$ \hspace{-4mm}\end{tabular} & \begin{tabular}{c} \hspace{-4mm} $\lambda_{g} = 10^{0}$, \hspace{-4mm} \\\hspace{-4mm} $\lambda_{l} = 10^{-2}$ \hspace{-4mm}\end{tabular} &  --  &  --  & \begin{tabular}{c} \hspace{-4mm} $\lambda = 10^{-1}$ \hspace{-4mm}\end{tabular} & \begin{tabular}{c} \hspace{-4mm} $\lambda = 10^{-1}$, \hspace{-4mm} \\\hspace{-4mm} $\tau = 10^{0}$ \hspace{-4mm}\end{tabular} & \begin{tabular}{c} \hspace{-4mm} $\lambda_{1} = 10^{0}$, \hspace{-4mm} \\\hspace{-4mm} $\lambda_{2} = 10^{0}$, \hspace{-4mm} \\\hspace{-4mm} $\varepsilon = 0.95$ \hspace{-4mm}\end{tabular} & \begin{tabular}{c} \hspace{-4mm} $\lambda_{1} = 10^{0}$, \hspace{-4mm} \\\hspace{-4mm} $\lambda_{2} = 10^{-2}$, \hspace{-4mm} \\\hspace{-4mm} $\varepsilon = 0.95$ \hspace{-4mm}\end{tabular} & \begin{tabular}{c} \hspace{-4mm} $\lambda_{1} = 10^{0}$, \hspace{-4mm} \\\hspace{-4mm} $\lambda_{2} = 10^{-2}$, \hspace{-4mm} \\\hspace{-4mm} $\varepsilon = 0.95$ \hspace{-4mm}\end{tabular} \\ 
& SRE  &    6.70 &   13.03 &    5.70 &    5.75 &   -2.60 &   -2.05 &    6.39 &    5.61 &   \Valbest{18.31} &   \ValSecnd{18.29} &   18.26 \\ 
& RMSE &   0.1005 &   0.0559 &   0.1490 &   0.1377 &   0.2527 &   0.1926 &   0.1420 &   0.1471 &   \Valbest{0.0319} &   \ValSecnd{0.0321} &   \ValSecnd{0.0321} \\ 
& Ps   &   0.85 &   0.98 &   0.66 &   0.72 &   0.21 &   0.33 &   0.74 &   0.70 &   \Valbest{1.00} &   \Valbest{1.00} &   \Valbest{1.00} \\ 
& MPSNR &   28.01 &   35.05 &   27.98 &   28.03 &   23.94 &   19.71 &   28.39 &   27.93 &   \Valbest{43.89} &   \ValSecnd{43.39} &   \ValSecnd{43.39} \\ 
& MSSIM &   0.7871 &   0.9343 &   0.7589 &   0.7560 &   0.7056 &   0.5584 &   0.7704 &   0.7532 &   \Valbest{0.9836} &   \ValSecnd{0.9803} &   0.9802 \\ 
\midrule 

\multirow{6}{*}{Case 5} 
& Setup & \begin{tabular}{c} \hspace{-4mm} $\lambda = 10^{0}$ \hspace{-4mm}\end{tabular} & \begin{tabular}{c} \hspace{-4mm} $\lambda_{1} = 10^{0}$, \hspace{-4mm} \\\hspace{-4mm} $\lambda_{2} = 10^{-2}$, \hspace{-4mm} \\\hspace{-4mm} $\lambda_{3} = 10^{0}$ \hspace{-4mm}\end{tabular} & \begin{tabular}{c} \hspace{-4mm} $\lambda = 10^{1}$, \hspace{-4mm} \\\hspace{-4mm} $\lambda_{TV} = 10^{-2}$ \hspace{-4mm}\end{tabular} & \begin{tabular}{c} \hspace{-4mm} $\lambda_{g} = 10^{0}$, \hspace{-4mm} \\\hspace{-4mm} $\lambda_{l} = 10^{-2}$ \hspace{-4mm}\end{tabular} &  --  &  --  & \begin{tabular}{c} \hspace{-4mm} $\lambda = 10^{-1}$ \hspace{-4mm}\end{tabular} & \begin{tabular}{c} \hspace{-4mm} $\lambda = 10^{-1}$, \hspace{-4mm} \\\hspace{-4mm} $\tau = 10^{0}$ \hspace{-4mm}\end{tabular} & \begin{tabular}{c} \hspace{-4mm} $\lambda_{1} = 10^{0}$, \hspace{-4mm} \\\hspace{-4mm} $\lambda_{2} = 10^{0}$, \hspace{-4mm} \\\hspace{-4mm} $\varepsilon = 0.95$ \hspace{-4mm}\end{tabular} & \begin{tabular}{c} \hspace{-4mm} $\lambda_{1} = 10^{0}$, \hspace{-4mm} \\\hspace{-4mm} $\lambda_{2} = 10^{-2}$, \hspace{-4mm} \\\hspace{-4mm} $\varepsilon = 0.98$ \hspace{-4mm}\end{tabular} & \begin{tabular}{c} \hspace{-4mm} $\lambda_{1} = 10^{0}$, \hspace{-4mm} \\\hspace{-4mm} $\lambda_{2} = 10^{-2}$, \hspace{-4mm} \\\hspace{-4mm} $\varepsilon = 0.95$ \hspace{-4mm}\end{tabular} \\ 
& SRE  &    9.56 &   10.65 &    8.27 &    8.12 &    0.13 &   -0.50 &    9.41 &    8.21 &   \Valbest{16.36} &   16.16 &   \ValSecnd{16.19} \\ 
& RMSE &   0.0799 &   0.0716 &   0.1054 &   0.1020 &   0.1976 &   0.1761 &   0.0968 &   0.1043 &   \Valbest{0.0394} &   0.0401 &   \ValSecnd{0.0400} \\ 
& Ps   &   0.92 &   0.96 &   0.84 &   0.86 &   0.50 &   0.37 &   0.89 &   0.85 &   \ValSecnd{0.99} &   \ValSecnd{0.99} &   \Valbest{1.00} \\ 
& MPSNR &   32.39 &   32.40 &   32.26 &   32.34 &   27.32 &   20.03 &   33.10 &   32.15 &   \Valbest{42.26} &   \ValSecnd{41.63} &   41.49 \\ 
& MSSIM &   0.8645 &   0.9018 &   0.8529 &   0.8486 &   0.7606 &   0.6114 &   0.8666 &   0.8464 &   \Valbest{0.9791 }&   \ValSecnd{0.9744} &   0.9735 \\ 
\midrule 

\multirow{6}{*}{Case 6} 
& Setup & \begin{tabular}{c} \hspace{-4mm} $\lambda = 10^{1}$ \hspace{-4mm}\end{tabular} & \begin{tabular}{c} \hspace{-4mm} $\lambda_{1} = 10^{0}$, \hspace{-4mm} \\\hspace{-4mm} $\lambda_{2} = 10^{-2}$, \hspace{-4mm} \\\hspace{-4mm} $\lambda_{3} = 10^{0}$ \hspace{-4mm}\end{tabular} & \begin{tabular}{c} \hspace{-4mm} $\lambda = 10^{1}$, \hspace{-4mm} \\\hspace{-4mm} $\lambda_{TV} = 10^{-2}$ \hspace{-4mm}\end{tabular} & \begin{tabular}{c} \hspace{-4mm} $\lambda_{g} = 10^{0}$, \hspace{-4mm} \\\hspace{-4mm} $\lambda_{l} = 10^{-2}$ \hspace{-4mm}\end{tabular} &  --  &  --  & \begin{tabular}{c} \hspace{-4mm} $\lambda = 10^{-1}$ \hspace{-4mm}\end{tabular} & \begin{tabular}{c} \hspace{-4mm} $\lambda = 10^{-1}$, \hspace{-4mm} \\\hspace{-4mm} $\tau = 10^{0}$ \hspace{-4mm}\end{tabular} & \begin{tabular}{c} \hspace{-4mm} $\lambda_{1} = 10^{0}$, \hspace{-4mm} \\\hspace{-4mm} $\lambda_{2} = 10^{0}$, \hspace{-4mm} \\\hspace{-4mm} $\varepsilon = 0.95$ \hspace{-4mm}\end{tabular} & \begin{tabular}{c} \hspace{-4mm} $\lambda_{1} = 10^{0}$, \hspace{-4mm} \\\hspace{-4mm} $\lambda_{2} = 10^{-2}$, \hspace{-4mm} \\\hspace{-4mm} $\varepsilon = 0.95$ \hspace{-4mm}\end{tabular} & \begin{tabular}{c} \hspace{-4mm} $\lambda_{1} = 10^{0}$, \hspace{-4mm} \\\hspace{-4mm} $\lambda_{2} = 10^{-1}$, \hspace{-4mm} \\\hspace{-4mm} $\varepsilon = 0.95$ \hspace{-4mm}\end{tabular} \\ 
& SRE  &    9.16 &    9.13 &    7.26 &    7.87 &    1.51 &   -1.03 &    8.60 &    7.28 &   \Valbest{14.39} &   14.03 &   \ValSecnd{14.19} \\ 
& RMSE &   0.0807 &   0.0818 &   0.1204 &   0.1070 &   0.1629 &   0.1836 &   0.1075 &   0.1180 &   \Valbest{0.0489} &   0.0507 &   \ValSecnd{0.0500} \\ 
& Ps   &   0.93 &   0.95 &   0.78 &   0.84 &   0.61 &   0.34 &   0.85 &   0.80 &   \Valbest{0.99} &   \Valbest{0.99} &   \Valbest{0.99} \\ 
& MPSNR &   31.64 &   30.93 &   31.58 &   31.83 &   26.89 &   20.02 &   32.42 &   31.47 &   \Valbest{38.82} &   37.78 &   \ValSecnd{38.04} \\ 
& MSSIM &   0.8543 &   0.8573 &   0.8291 &   0.8287 &   0.7650 &   0.5836 &   0.8441 &   0.8224 &   \Valbest{0.9563} &   0.9396 &   \ValSecnd{0.9435} \\ 
\midrule 

\multirow{6}{*}{Case 7} 
& Setup & \begin{tabular}{c} \hspace{-4mm} $\lambda = 10^{1}$ \hspace{-4mm}\end{tabular} & \begin{tabular}{c} \hspace{-4mm} $\lambda_{1} = 10^{0}$, \hspace{-4mm} \\\hspace{-4mm} $\lambda_{2} = 10^{-1}$, \hspace{-4mm} \\\hspace{-4mm} $\lambda_{3} = 10^{1}$ \hspace{-4mm}\end{tabular} & \begin{tabular}{c} \hspace{-4mm} $\lambda = 10^{1}$, \hspace{-4mm} \\\hspace{-4mm} $\lambda_{TV} = 10^{-2}$ \hspace{-4mm}\end{tabular} & \begin{tabular}{c} \hspace{-4mm} $\lambda_{g} = 10^{0}$, \hspace{-4mm} \\\hspace{-4mm} $\lambda_{l} = 10^{-2}$ \hspace{-4mm}\end{tabular} &  --  &  --  & \begin{tabular}{c} \hspace{-4mm} $\lambda = 10^{-1}$ \hspace{-4mm}\end{tabular} & \begin{tabular}{c} \hspace{-4mm} $\lambda = 10^{-1}$, \hspace{-4mm} \\\hspace{-4mm} $\tau = 10^{0}$ \hspace{-4mm}\end{tabular} & \begin{tabular}{c} \hspace{-4mm} $\lambda_{1} = 10^{0}$, \hspace{-4mm} \\\hspace{-4mm} $\lambda_{2} = 10^{0}$, \hspace{-4mm} \\\hspace{-4mm} $\varepsilon = 0.98$ \hspace{-4mm}\end{tabular} & \begin{tabular}{c} \hspace{-4mm} $\lambda_{1} = 10^{0}$, \hspace{-4mm} \\\hspace{-4mm} $\lambda_{2} = 10^{-2}$, \hspace{-4mm} \\\hspace{-4mm} $\varepsilon = 0.98$ \hspace{-4mm}\end{tabular} & \begin{tabular}{c} \hspace{-4mm} $\lambda_{1} = 10^{0}$, \hspace{-4mm} \\\hspace{-4mm} $\lambda_{2} = 10^{-1}$, \hspace{-4mm} \\\hspace{-4mm} $\varepsilon = 0.98$ \hspace{-4mm}\end{tabular} \\ 
& SRE  &   11.51 &    8.85 &    9.75 &   10.24 &   -0.93 &   -1.06 &   11.75 &   10.19 &   \Valbest{13.67} &   13.27 &   \ValSecnd{13.35} \\ 
& RMSE &   0.0653 &   0.0911 &   0.0870 &   0.0805 &   0.2345 &   0.2025 &   0.0718 &   0.0822 &   \Valbest{0.0531} &   0.0551 &   \ValSecnd{0.0547} \\ 
& Ps   &   \Valbest{0.98} &   0.91 &   0.90 &   0.92 &   0.47 &   0.35 &   0.94 &   0.91 &   \Valbest{0.98} &   \Valbest{0.98} &   \Valbest{0.98} \\ 
& MPSNR &   35.46 &   26.18 &   35.69 &   35.83 &   25.69 &   18.70 &   36.49 &   35.68 &   \Valbest{38.22} &   36.94 &   \ValSecnd{37.25} \\ 
& MSSIM &   0.9131 &   0.6888 &   0.9011 &   0.9000 &   0.7413 &   0.3965 &   0.9110 &   0.8988 &   \Valbest{0.9502} &   0.9250 &   \ValSecnd{0.9310} \\ 
\midrule 

\multirow{6}{*}{Case 8} 
& Setup & \begin{tabular}{c} \hspace{-4mm} $\lambda = 10^{1}$ \hspace{-4mm}\end{tabular} & \begin{tabular}{c} \hspace{-4mm} $\lambda_{1} = 10^{0}$, \hspace{-4mm} \\\hspace{-4mm} $\lambda_{2} = 10^{-1}$, \hspace{-4mm} \\\hspace{-4mm} $\lambda_{3} = 10^{1}$ \hspace{-4mm}\end{tabular} & \begin{tabular}{c} \hspace{-4mm} $\lambda = 10^{-2}$, \hspace{-4mm} \\\hspace{-4mm} $\lambda_{TV} = 10^{-2}$ \hspace{-4mm}\end{tabular} & \begin{tabular}{c} \hspace{-4mm} $\lambda_{g} = 10^{0}$, \hspace{-4mm} \\\hspace{-4mm} $\lambda_{l} = 10^{-2}$ \hspace{-4mm}\end{tabular} &  --  &  --  & \begin{tabular}{c} \hspace{-4mm} $\lambda = 10^{-1}$ \hspace{-4mm}\end{tabular} & \begin{tabular}{c} \hspace{-4mm} $\lambda = 10^{-1}$, \hspace{-4mm} \\\hspace{-4mm} $\tau = 10^{0}$ \hspace{-4mm}\end{tabular} & \begin{tabular}{c} \hspace{-4mm} $\lambda_{1} = 10^{0}$, \hspace{-4mm} \\\hspace{-4mm} $\lambda_{2} = 10^{0}$, \hspace{-4mm} \\\hspace{-4mm} $\varepsilon = 0.95$ \hspace{-4mm}\end{tabular} & \begin{tabular}{c} \hspace{-4mm} $\lambda_{1} = 10^{0}$, \hspace{-4mm} \\\hspace{-4mm} $\lambda_{2} = 10^{-2}$, \hspace{-4mm} \\\hspace{-4mm} $\varepsilon = 0.95$ \hspace{-4mm}\end{tabular} & \begin{tabular}{c} \hspace{-4mm} $\lambda_{1} = 10^{0}$, \hspace{-4mm} \\\hspace{-4mm} $\lambda_{2} = 10^{-1}$, \hspace{-4mm} \\\hspace{-4mm} $\varepsilon = 0.95$ \hspace{-4mm}\end{tabular} \\ 
& SRE  &    8.39 &    7.73 &    6.20 &    6.75 &   -0.65 &   -3.94 &    7.34 &    6.14 &   \Valbest{12.46} &   12.12 &   \ValSecnd{12.21} \\ 
& RMSE &   0.0880 &   0.1014 &   0.1398 &   0.1257 &   0.2172 &   0.2594 &   0.1302 &   0.1387 &   \Valbest{0.0604} &   0.0623 &   \ValSecnd{0.0619} \\ 
& Ps   &   0.89 &   0.87 &   0.71 &   0.77 &   0.40 &   0.00 &   0.77 &   0.73 &   \Valbest{0.97} &   {0.97} &   {0.97} \\ 
& MPSNR &   31.05 &   25.28 &   30.86 &   31.08 &   25.74 &   15.92 &   31.74 &   30.74 &   \Valbest{36.93} &   35.67 &   \ValSecnd{35.97} \\ 
& MSSIM &   0.8282 &   0.6531 &   0.7953 &   0.7949 &   0.7198 &   0.2987 &   0.8119 &   0.7891 &   \Valbest{0.9371} &   0.9055 &   \ValSecnd{0.9124} \\ 
\bottomrule
\end{tabular}
}
\end{table*}

\begin{table*}[!h]
	\caption{SRE, RMSE, Ps, MPSNR, and MSSIM in the Experiments Using \textit{Samson}.}
	\vspace{-1mm}
	\label{tab:result_samson}
	\centering
	\scalebox{0.75}{
		\begin{tabular}{ccccccccccccc} \toprule
			\multirow{3}{*}{ Noise } & \multirow{3}{*}{Metrics}& \multicolumn{11}{c}{Methods}\\ 
			\cmidrule(lr){3-13}
			& & CLSUnSAL & JSTV & RSSUn-TV & LGSU & UnDIP & EGU-Net & RDSWSU & MdLRR & \textbf{\Ours} & \textbf{\Ours} & \textbf{\Ours} \\ 
			& & \cite{iordache2014collaborative} & \cite{aggarwal2016hyperspectral} & \cite{wang2019row} & \cite{shen2022superpixel} & \cite{UnDIP_RastiB_2022} & \cite{hong2022endmember} & \cite{rs_Deng_RobustDual_2023} & \cite{MDLRR_WuLing_2023} & \textbf{(HTV)} & \textbf{(SSTV)} & \textbf{(HSSTV)} \\ 
			\midrule
			
\multirow{6}{*}{Case 1} 
& Setup & \begin{tabular}{c} \hspace{-4mm} $\lambda = 10^{1}$ \hspace{-4mm}\end{tabular} & \begin{tabular}{c} \hspace{-4mm} $\lambda_{1} = 10^{1}$, \hspace{-4mm} \\\hspace{-4mm} $\lambda_{2} = 10^{0}$, \hspace{-4mm} \\\hspace{-4mm} $\lambda_{3} = 10^{1}$ \hspace{-4mm}\end{tabular} & \begin{tabular}{c} \hspace{-4mm} $\lambda = 10^{1}$, \hspace{-4mm} \\\hspace{-4mm} $\lambda_{TV} = 10^{-2}$ \hspace{-4mm}\end{tabular} & \begin{tabular}{c} \hspace{-4mm} $\lambda_{g} = 10^{-2}$, \hspace{-4mm} \\\hspace{-4mm} $\lambda_{l} = 10^{-4}$ \hspace{-4mm}\end{tabular} &  --  &  --  & \begin{tabular}{c} \hspace{-4mm} $\lambda = 10^{-2}$ \hspace{-4mm}\end{tabular} & \begin{tabular}{c} \hspace{-4mm} $\lambda = 10^{-3}$, \hspace{-4mm} \\\hspace{-4mm} $\tau = 10^{0}$ \hspace{-4mm}\end{tabular} & \begin{tabular}{c} \hspace{-4mm} $\lambda_{1} = 10^{-1}$, \hspace{-4mm} \\\hspace{-4mm} $\lambda_{2} = 10^{1}$, \hspace{-4mm} \\\hspace{-4mm} $\varepsilon = 0.98$ \hspace{-4mm}\end{tabular} & \begin{tabular}{c} \hspace{-4mm} $\lambda_{1} = 10^{0}$, \hspace{-4mm} \\\hspace{-4mm} $\lambda_{2} = 10^{0}$, \hspace{-4mm} \\\hspace{-4mm} $\varepsilon = 0.95$ \hspace{-4mm}\end{tabular} & \begin{tabular}{c} \hspace{-4mm} $\lambda_{1} = 10^{-1}$, \hspace{-4mm} \\\hspace{-4mm} $\lambda_{2} = 10^{0}$, \hspace{-4mm} \\\hspace{-4mm} $\varepsilon = 0.98$ \hspace{-4mm}\end{tabular} \\ 
& SRE  &   16.82 &    0.01 &   17.52 &   10.89 &    2.52 &    7.27 &   13.37 &   13.96 &   \Valbest{19.51} &   16.48 &   \ValSecnd{18.03} \\ 
& RMSE &   0.0338 &   4.0921 &   0.0309 &   0.0678 &   0.1491 &   0.0940 &   0.0539 &   0.0474 &   \Valbest{0.0254} &   0.0345 &   \ValSecnd{0.0296} \\ 
& Ps   &   \Valbest{1.00} &   0.97 &   \Valbest{1.00} &   0.94 &   0.56 &   0.80 &   0.97 &   0.99 &   \Valbest{1.00} &   \Valbest{1.00} &   \Valbest{1.00} \\ 
& MPSNR &   34.42 &  -11.99 &   45.31 &   43.54 &   28.92 &   24.85 &   44.90 &   44.01 &   43.75 &   45.52 &   46.74 \\ 
& MSSIM &   0.9218 &   0.9600 &   0.9839 &   0.9731 &   0.8409 &   0.7107 &   0.9772 &   0.9774 &   0.9836 &   0.9855 &   0.9907 \\ 
\midrule 

\multirow{6}{*}{Case 2} 
& Setup & \begin{tabular}{c} \hspace{-4mm} $\lambda = 10^{1}$ \hspace{-4mm}\end{tabular} & \begin{tabular}{c} \hspace{-4mm} $\lambda_{1} = 10^{1}$, \hspace{-4mm} \\\hspace{-4mm} $\lambda_{2} = 10^{0}$, \hspace{-4mm} \\\hspace{-4mm} $\lambda_{3} = 10^{1}$ \hspace{-4mm}\end{tabular} & \begin{tabular}{c} \hspace{-4mm} $\lambda = 10^{1}$, \hspace{-4mm} \\\hspace{-4mm} $\lambda_{TV} = 10^{-2}$ \hspace{-4mm}\end{tabular} & \begin{tabular}{c} \hspace{-4mm} $\lambda_{g} = 10^{-1}$, \hspace{-4mm} \\\hspace{-4mm} $\lambda_{l} = 10^{-4}$ \hspace{-4mm}\end{tabular} &  --  &  --  & \begin{tabular}{c} \hspace{-4mm} $\lambda = 10^{-2}$ \hspace{-4mm}\end{tabular} & \begin{tabular}{c} \hspace{-4mm} $\lambda = 10^{-2}$, \hspace{-4mm} \\\hspace{-4mm} $\tau = 10^{0}$ \hspace{-4mm}\end{tabular} & \begin{tabular}{c} \hspace{-4mm} $\lambda_{1} = 10^{-1}$, \hspace{-4mm} \\\hspace{-4mm} $\lambda_{2} = 10^{1}$, \hspace{-4mm} \\\hspace{-4mm} $\varepsilon = 0.98$ \hspace{-4mm}\end{tabular} & \begin{tabular}{c} \hspace{-4mm} $\lambda_{1} = 10^{0}$, \hspace{-4mm} \\\hspace{-4mm} $\lambda_{2} = 10^{0}$, \hspace{-4mm} \\\hspace{-4mm} $\varepsilon = 0.98$ \hspace{-4mm}\end{tabular} & \begin{tabular}{c} \hspace{-4mm} $\lambda_{1} = 10^{0}$, \hspace{-4mm} \\\hspace{-4mm} $\lambda_{2} = 10^{1}$, \hspace{-4mm} \\\hspace{-4mm} $\varepsilon = 0.98$ \hspace{-4mm}\end{tabular} \\ 
& SRE  &   \Valbest{18.36} &    6.79 &   11.64 &    6.46 &   -1.91 &    6.42 &    8.60 &    7.12 &   \ValSecnd{14.51} &   11.09 &   12.86 \\ 
& RMSE &   \Valbest{0.0291} &   0.0902 &   0.0576 &   0.1120 &   0.2408 &   0.0976 &   0.0960 &   0.1032 &   \ValSecnd{0.0435} &   0.0590 &   0.0504 \\ 
& Ps   &   \Valbest{1.00} &   0.97 &   \ValSecnd{0.99} &   0.74 &   0.31 &   0.77 &   0.83 &   0.78 &   \ValSecnd{0.99} &   \ValSecnd{0.99} &   \ValSecnd{0.99} \\ 
& MPSNR &   38.12 &   29.09 &   39.23 &   38.01 &   23.33 &   25.39 &   39.20 &   37.96 &   \Valbest{40.84} &   \ValSecnd{40.45} &   40.44 \\ 
& MSSIM &   0.9402 &   0.6970 &   0.9387 &   0.9147 &   0.6756 &   0.6802 &   0.9255 &   0.9166 &   \Valbest{0.9719} &   0.9561 &   \ValSecnd{0.9725} \\ 
\midrule 

\multirow{6}{*}{Case 3} 
& Setup & \begin{tabular}{c} \hspace{-4mm} $\lambda = 10^{1}$ \hspace{-4mm}\end{tabular} & \begin{tabular}{c} \hspace{-4mm} $\lambda_{1} = 10^{1}$, \hspace{-4mm} \\\hspace{-4mm} $\lambda_{2} = 10^{-1}$, \hspace{-4mm} \\\hspace{-4mm} $\lambda_{3} = 10^{1}$ \hspace{-4mm}\end{tabular} & \begin{tabular}{c} \hspace{-4mm} $\lambda = 10^{1}$, \hspace{-4mm} \\\hspace{-4mm} $\lambda_{TV} = 10^{-2}$ \hspace{-4mm}\end{tabular} & \begin{tabular}{c} \hspace{-4mm} $\lambda_{g} = 10^{-1}$, \hspace{-4mm} \\\hspace{-4mm} $\lambda_{l} = 10^{-3}$ \hspace{-4mm}\end{tabular} &  --  &  --  & \begin{tabular}{c} \hspace{-4mm} $\lambda = 10^{-1}$ \hspace{-4mm}\end{tabular} & \begin{tabular}{c} \hspace{-4mm} $\lambda = 10^{-2}$, \hspace{-4mm} \\\hspace{-4mm} $\tau = 10^{0}$ \hspace{-4mm}\end{tabular} & \begin{tabular}{c} \hspace{-4mm} $\lambda_{1} = 10^{-1}$, \hspace{-4mm} \\\hspace{-4mm} $\lambda_{2} = 10^{1}$, \hspace{-4mm} \\\hspace{-4mm} $\varepsilon = 0.98$ \hspace{-4mm}\end{tabular} & \begin{tabular}{c} \hspace{-4mm} $\lambda_{1} = 10^{0}$, \hspace{-4mm} \\\hspace{-4mm} $\lambda_{2} = 10^{0}$, \hspace{-4mm} \\\hspace{-4mm} $\varepsilon = 0.95$ \hspace{-4mm}\end{tabular} & \begin{tabular}{c} \hspace{-4mm} $\lambda_{1} = 10^{-1}$, \hspace{-4mm} \\\hspace{-4mm} $\lambda_{2} = 10^{0}$, \hspace{-4mm} \\\hspace{-4mm} $\varepsilon = 0.98$ \hspace{-4mm}\end{tabular} \\ 
& SRE  &   14.35 &   10.96 &    8.78 &    4.48 &   -2.66 &    4.13 &    6.49 &    4.31 &   \Valbest{18.16} &   15.61 &   \ValSecnd{16.74} \\ 
& RMSE &   0.0446 &   0.0624 &   0.0913 &   0.1695 &   0.2501 &   0.1159 &   0.1660 &   0.1715 &   \Valbest{0.0294} &   0.0378 &   \ValSecnd{0.0341} \\ 
& Ps   &   \Valbest{1.00} &   0.99 &   0.88 &   0.53 &   0.14 &   0.58 &   0.62 &   0.51 &   \Valbest{1.00} &   \Valbest{1.00} &   \Valbest{1.00} \\ 
& MPSNR &   31.57 &   32.84 &   31.68 &   31.21 &   23.22 &   21.63 &   32.82 &   30.89 &   43.09 &   \ValSecnd{44.34} &   \Valbest{45.45} \\ 
& MSSIM &   0.8178 &   0.8224 &   0.8019 &   0.7713 &   0.6394 &   0.5828 &   0.8190 &   0.7644 &   \ValSecnd{0.9812} &   0.9807 &   \Valbest{0.9868} \\ 
\midrule 

\multirow{6}{*}{Case 4} 
& Setup & \begin{tabular}{c} \hspace{-4mm} $\lambda = 10^{1}$ \hspace{-4mm}\end{tabular} & \begin{tabular}{c} \hspace{-4mm} $\lambda_{1} = 10^{1}$, \hspace{-4mm} \\\hspace{-4mm} $\lambda_{2} = 10^{1}$, \hspace{-4mm} \\\hspace{-4mm} $\lambda_{3} = 10^{1}$ \hspace{-4mm}\end{tabular} & \begin{tabular}{c} \hspace{-4mm} $\lambda = 10^{1}$, \hspace{-4mm} \\\hspace{-4mm} $\lambda_{TV} = 10^{-2}$ \hspace{-4mm}\end{tabular} & \begin{tabular}{c} \hspace{-4mm} $\lambda_{g} = 10^{-1}$, \hspace{-4mm} \\\hspace{-4mm} $\lambda_{l} = 10^{-3}$ \hspace{-4mm}\end{tabular} &  --  &  --  & \begin{tabular}{c} \hspace{-4mm} $\lambda = 10^{-1}$ \hspace{-4mm}\end{tabular} & \begin{tabular}{c} \hspace{-4mm} $\lambda = 10^{-2}$, \hspace{-4mm} \\\hspace{-4mm} $\tau = 10^{0}$ \hspace{-4mm}\end{tabular} & \begin{tabular}{c} \hspace{-4mm} $\lambda_{1} = 10^{-1}$, \hspace{-4mm} \\\hspace{-4mm} $\lambda_{2} = 10^{1}$, \hspace{-4mm} \\\hspace{-4mm} $\varepsilon = 0.98$ \hspace{-4mm}\end{tabular} & \begin{tabular}{c} \hspace{-4mm} $\lambda_{1} = 10^{0}$, \hspace{-4mm} \\\hspace{-4mm} $\lambda_{2} = 10^{0}$, \hspace{-4mm} \\\hspace{-4mm} $\varepsilon = 0.95$ \hspace{-4mm}\end{tabular} & \begin{tabular}{c} \hspace{-4mm} $\lambda_{1} = 10^{0}$, \hspace{-4mm} \\\hspace{-4mm} $\lambda_{2} = 10^{0}$, \hspace{-4mm} \\\hspace{-4mm} $\varepsilon = 0.95$ \hspace{-4mm}\end{tabular} \\ 
& SRE  &   10.51 &    0.02 &    5.16 &    3.10 &   -1.86 &   -1.41 &    3.97 &    2.90 &   \Valbest{16.09} &   14.57 &   \ValSecnd{15.24} \\ 
& RMSE &   0.0653 &   3.0269 &   0.1542 &   0.2343 &   0.2358 &   0.1998 &   0.2263 &   0.2362 &   \Valbest{0.0364} &   0.0417 &   \ValSecnd{0.0390} \\ 
& Ps   &   0.96 &   0.95 &   0.60 &   0.33 &   0.19 &   0.18 &   0.35 &   0.32 &   \Valbest{1.00} &   \Valbest{1.00} &   \Valbest{1.00} \\ 
& MPSNR &   27.06 &   -8.81 &   27.00 &   26.74 &   24.75 &   16.26 &   27.47 &   26.59 &   42.80 &   \ValSecnd{43.41} &   \Valbest{44.25} \\ 
& MSSIM &   0.6876 &   0.9504 &   0.6646 &   0.6377 &   0.6187 &   0.2036 &   0.6772 &   0.6351 &   \ValSecnd{0.9805} &   0.9761 &   \Valbest{0.9815} \\ 
\midrule 

\multirow{6}{*}{Case 5} 
& Setup & \begin{tabular}{c} \hspace{-4mm} $\lambda = 10^{1}$ \hspace{-4mm}\end{tabular} & \begin{tabular}{c} \hspace{-4mm} $\lambda_{1} = 10^{1}$, \hspace{-4mm} \\\hspace{-4mm} $\lambda_{2} = 10^{-2}$, \hspace{-4mm} \\\hspace{-4mm} $\lambda_{3} = 10^{1}$ \hspace{-4mm}\end{tabular} & \begin{tabular}{c} \hspace{-4mm} $\lambda = 10^{1}$, \hspace{-4mm} \\\hspace{-4mm} $\lambda_{TV} = 10^{-2}$ \hspace{-4mm}\end{tabular} & \begin{tabular}{c} \hspace{-4mm} $\lambda_{g} = 10^{-1}$, \hspace{-4mm} \\\hspace{-4mm} $\lambda_{l} = 10^{-3}$ \hspace{-4mm}\end{tabular} &  --  &  --  & \begin{tabular}{c} \hspace{-4mm} $\lambda = 10^{-1}$ \hspace{-4mm}\end{tabular} & \begin{tabular}{c} \hspace{-4mm} $\lambda = 10^{-2}$, \hspace{-4mm} \\\hspace{-4mm} $\tau = 10^{0}$ \hspace{-4mm}\end{tabular} & \begin{tabular}{c} \hspace{-4mm} $\lambda_{1} = 10^{0}$, \hspace{-4mm} \\\hspace{-4mm} $\lambda_{2} = 10^{1}$, \hspace{-4mm} \\\hspace{-4mm} $\varepsilon = 0.98$ \hspace{-4mm}\end{tabular} & \begin{tabular}{c} \hspace{-4mm} $\lambda_{1} = 10^{0}$, \hspace{-4mm} \\\hspace{-4mm} $\lambda_{2} = 10^{-1}$, \hspace{-4mm} \\\hspace{-4mm} $\varepsilon = 0.98$ \hspace{-4mm}\end{tabular} & \begin{tabular}{c} \hspace{-4mm} $\lambda_{1} = 10^{0}$, \hspace{-4mm} \\\hspace{-4mm} $\lambda_{2} = 10^{0}$, \hspace{-4mm} \\\hspace{-4mm} $\varepsilon = 0.98$ \hspace{-4mm}\end{tabular} \\ 
& SRE  &   \Valbest{13.60} &    7.85 &    7.63 &    4.19 &   -0.17 &    3.94 &    6.34 &    4.01 &   \ValSecnd{12.91} &   10.04 &   12.12 \\ 
& RMSE &   \Valbest{0.0478} &   0.0816 &   0.1031 &   0.1754 &   0.1974 &   0.1176 &   0.1667 &   0.1770 &   \ValSecnd{0.0489} &   0.0648 &   0.0532 \\ 
& Ps   &   0.99 &   0.98 &   0.82 &   0.50 &   0.36 &   0.59 &   0.61 &   0.49 &   \Valbest{1.00} &   0.99 &   \Valbest{1.00} \\ 
& MPSNR &   31.31 &   30.73 &   31.35 &   30.90 &   29.99 &   22.42 &   32.49 &   30.60 &   \Valbest{42.74} &   40.95 &   \ValSecnd{42.48} \\ 
& MSSIM &   0.8107 &   0.7485 &   0.7920 &   0.7622 &   0.7643 &   0.6111 &   0.8128 &   0.7561 &   \Valbest{0.9811} &   0.9625 &   \ValSecnd{0.9759} \\ 
\midrule 

\multirow{6}{*}{Case 6} 
& Setup & \begin{tabular}{c} \hspace{-4mm} $\lambda = 10^{1}$ \hspace{-4mm}\end{tabular} & \begin{tabular}{c} \hspace{-4mm} $\lambda_{1} = 10^{1}$, \hspace{-4mm} \\\hspace{-4mm} $\lambda_{2} = 10^{1}$, \hspace{-4mm} \\\hspace{-4mm} $\lambda_{3} = 10^{1}$ \hspace{-4mm}\end{tabular} & \begin{tabular}{c} \hspace{-4mm} $\lambda = 10^{1}$, \hspace{-4mm} \\\hspace{-4mm} $\lambda_{TV} = 10^{-2}$ \hspace{-4mm}\end{tabular} & \begin{tabular}{c} \hspace{-4mm} $\lambda_{g} = 10^{-1}$, \hspace{-4mm} \\\hspace{-4mm} $\lambda_{l} = 10^{-3}$ \hspace{-4mm}\end{tabular} &  --  &  --  & \begin{tabular}{c} \hspace{-4mm} $\lambda = 10^{-1}$ \hspace{-4mm}\end{tabular} & \begin{tabular}{c} \hspace{-4mm} $\lambda = 10^{-2}$, \hspace{-4mm} \\\hspace{-4mm} $\tau = 10^{0}$ \hspace{-4mm}\end{tabular} & \begin{tabular}{c} \hspace{-4mm} $\lambda_{1} = 10^{-1}$, \hspace{-4mm} \\\hspace{-4mm} $\lambda_{2} = 10^{1}$, \hspace{-4mm} \\\hspace{-4mm} $\varepsilon = 0.98$ \hspace{-4mm}\end{tabular} & \begin{tabular}{c} \hspace{-4mm} $\lambda_{1} = 10^{0}$, \hspace{-4mm} \\\hspace{-4mm} $\lambda_{2} = 10^{-1}$, \hspace{-4mm} \\\hspace{-4mm} $\varepsilon = 0.98$ \hspace{-4mm}\end{tabular} & \begin{tabular}{c} \hspace{-4mm} $\lambda_{1} = 10^{0}$, \hspace{-4mm} \\\hspace{-4mm} $\lambda_{2} = 10^{0}$, \hspace{-4mm} \\\hspace{-4mm} $\varepsilon = 0.98$ \hspace{-4mm}\end{tabular} \\ 
& SRE  &   \Valbest{13.59} &    5.51 &    6.56 &    3.78 &   -2.24 &    4.30 &    5.63 &    3.67 &   \ValSecnd{12.49} &    9.33 &   11.97 \\ 
& RMSE &   \Valbest{0.0484} &   0.0993 &   0.1219 &   0.1929 &   0.2439 &   0.1156 &   0.1898 &   0.1932 &   \ValSecnd{0.0525} &   0.0689 &   0.0543 \\ 
& Ps   &   \ValSecnd{0.99} &   0.98 &   0.74 &   0.46 &   0.26 &   0.59 &   0.56 &   0.44 &   \ValSecnd{0.99} &   0.98 &   \Valbest{1.00} \\ 
& MPSNR &   30.84 &   30.86 &   30.77 &   30.33 &   25.92 &   22.21 &   31.85 &   30.07 &   \ValSecnd{38.41} &   37.76 &   \Valbest{39.70} \\ 
& MSSIM &   0.7877 &   0.7506 &   0.7611 &   0.7326 &   0.6786 &   0.5919 &   0.7804 &   0.7277 &   0.9559 &   0.9222 &   0.9573 \\ 
\midrule 

\multirow{6}{*}{Case 7} 
& Setup & \begin{tabular}{c} \hspace{-4mm} $\lambda = 10^{1}$ \hspace{-4mm}\end{tabular} & \begin{tabular}{c} \hspace{-4mm} $\lambda_{1} = 10^{1}$, \hspace{-4mm} \\\hspace{-4mm} $\lambda_{2} = 10^{-2}$, \hspace{-4mm} \\\hspace{-4mm} $\lambda_{3} = 10^{1}$ \hspace{-4mm}\end{tabular} & \begin{tabular}{c} \hspace{-4mm} $\lambda = 10^{1}$, \hspace{-4mm} \\\hspace{-4mm} $\lambda_{TV} = 10^{-2}$ \hspace{-4mm}\end{tabular} & \begin{tabular}{c} \hspace{-4mm} $\lambda_{g} = 10^{-1}$, \hspace{-4mm} \\\hspace{-4mm} $\lambda_{l} = 10^{-3}$ \hspace{-4mm}\end{tabular} &  --  &  --  & \begin{tabular}{c} \hspace{-4mm} $\lambda = 10^{-1}$ \hspace{-4mm}\end{tabular} & \begin{tabular}{c} \hspace{-4mm} $\lambda = 10^{-2}$, \hspace{-4mm} \\\hspace{-4mm} $\tau = 10^{0}$ \hspace{-4mm}\end{tabular} & \begin{tabular}{c} \hspace{-4mm} $\lambda_{1} = 10^{0}$, \hspace{-4mm} \\\hspace{-4mm} $\lambda_{2} = 10^{1}$, \hspace{-4mm} \\\hspace{-4mm} $\varepsilon = 0.98$ \hspace{-4mm}\end{tabular} & \begin{tabular}{c} \hspace{-4mm} $\lambda_{1} = 10^{0}$, \hspace{-4mm} \\\hspace{-4mm} $\lambda_{2} = 10^{1}$, \hspace{-4mm} \\\hspace{-4mm} $\varepsilon = 0.98$ \hspace{-4mm}\end{tabular} & \begin{tabular}{c} \hspace{-4mm} $\lambda_{1} = 10^{0}$, \hspace{-4mm} \\\hspace{-4mm} $\lambda_{2} = 10^{1}$, \hspace{-4mm} \\\hspace{-4mm} $\varepsilon = 0.98$ \hspace{-4mm}\end{tabular} \\ 
& SRE  & \Valbest{  16.71} &    4.71 &    7.68 &    4.59 &   -2.67 &   -0.10 &    6.43 &    4.83 & \ValSecnd{  13.61} &    8.38 &   11.28 \\ 
& RMSE & \Valbest{  0.0350} &   0.1076 &   0.0892 &   0.1420 &   0.2622 &   0.1946 &   0.1197 &   0.1382 & \ValSecnd{  0.0466} &   0.0759 &   0.0588 \\ 
& Ps   & \Valbest{  1.00} &   0.93 &   0.87 &   0.61 &   0.14 &   0.23 &   0.67 &   0.62 & \Valbest{  1.00} &   0.98 & \ValSecnd{  0.99} \\ 
& MPSNR &   35.59 &   27.13 &   35.75 &   34.90 &   26.96 &   17.60 &   36.39 &   34.79 & \Valbest{  39.74} &   37.62 & \ValSecnd{  38.83} \\ 
& MSSIM &   0.8968 &   0.6114 &   0.8757 &   0.8482 &   0.7173 &   0.2423 &   0.8831 &   0.8475 & \Valbest{  0.9651} &   0.9317 & \ValSecnd{  0.9612} \\ 
\midrule 

\multirow{6}{*}{Case 8} 
& Setup & \begin{tabular}{c} \hspace{-4mm} $\lambda = 10^{1}$ \hspace{-4mm}\end{tabular} & \begin{tabular}{c} \hspace{-4mm} $\lambda_{1} = 10^{1}$, \hspace{-4mm} \\\hspace{-4mm} $\lambda_{2} = 10^{1}$, \hspace{-4mm} \\\hspace{-4mm} $\lambda_{3} = 10^{1}$ \hspace{-4mm}\end{tabular} & \begin{tabular}{c} \hspace{-4mm} $\lambda = 10^{1}$, \hspace{-4mm} \\\hspace{-4mm} $\lambda_{TV} = 10^{-2}$ \hspace{-4mm}\end{tabular} & \begin{tabular}{c} \hspace{-4mm} $\lambda_{g} = 10^{-1}$, \hspace{-4mm} \\\hspace{-4mm} $\lambda_{l} = 10^{-3}$ \hspace{-4mm}\end{tabular} &  --  &  --  & \begin{tabular}{c} \hspace{-4mm} $\lambda = 10^{-1}$ \hspace{-4mm}\end{tabular} & \begin{tabular}{c} \hspace{-4mm} $\lambda = 10^{-2}$, \hspace{-4mm} \\\hspace{-4mm} $\tau = 10^{0}$ \hspace{-4mm}\end{tabular} & \begin{tabular}{c} \hspace{-4mm} $\lambda_{1} = 10^{0}$, \hspace{-4mm} \\\hspace{-4mm} $\lambda_{2} = 10^{1}$, \hspace{-4mm} \\\hspace{-4mm} $\varepsilon = 0.95$ \hspace{-4mm}\end{tabular} & \begin{tabular}{c} \hspace{-4mm} $\lambda_{1} = 10^{0}$, \hspace{-4mm} \\\hspace{-4mm} $\lambda_{2} = 10^{-1}$, \hspace{-4mm} \\\hspace{-4mm} $\varepsilon = 0.95$ \hspace{-4mm}\end{tabular} & \begin{tabular}{c} \hspace{-4mm} $\lambda_{1} = 10^{-1}$, \hspace{-4mm} \\\hspace{-4mm} $\lambda_{2} = 10^{0}$, \hspace{-4mm} \\\hspace{-4mm} $\varepsilon = 0.98$ \hspace{-4mm}\end{tabular} \\ 
& SRE  & \Valbest{  11.52} &    2.72 &    5.16 &    3.30 &   -2.20 &   -1.28 &    5.23 &    3.23 & \ValSecnd{  10.39} &    6.26 &    9.40 \\ 
& RMSE & \Valbest{  0.0589} &   0.1208 &   0.1400 &   0.2047 &   0.2283 &   0.1958 &   0.1979 &   0.2037 & \ValSecnd{  0.0630} &   0.0922 &   0.0714 \\ 
& Ps   & \ValSecnd{  0.97} &   0.76 &   0.63 &   0.40 &   0.12 &   0.23 &   0.53 &   0.39 & \Valbest{  0.99} &   0.84 &   0.96 \\ 
& MPSNR &   30.31 &   29.70 &   30.13 &   29.70 &   23.63 &   16.86 &   31.25 &   29.52 & \ValSecnd{  37.22} &   35.18 & \Valbest{  37.76} \\ 
& MSSIM &   0.7595 &   0.7024 &   0.7230 &   0.6970 &   0.5958 &   0.2474 &   0.7493 &   0.6943 & \Valbest{  0.9447} &   0.8707 & \ValSecnd{  0.9405} \\ 
\bottomrule
\end{tabular}
}
\end{table*}

\begin{table*}[!h]
	\caption{SRE, RMSE, Ps, MPSNR, and MSSIM in the Experiments Using \textit{Urban}.}
	\vspace{-1mm}
	\label{tab:result_real_Urban}
	\centering
	\scalebox{0.75}{
		\begin{tabular}{ccccccccccccc} \toprule
			\multirow{3}{*}{ Noise } & \multirow{3}{*}{Metrics}& \multicolumn{11}{c}{Methods}\\ 
			\cmidrule(lr){3-13}
			& & CLSUnSAL & JSTV & RSSUn-TV & LGSU & UnDIP & EGU-Net & RDSWSU & MdLRR & \textbf{\Ours} & \textbf{\Ours} & \textbf{\Ours} \\ 
			& & \cite{iordache2014collaborative} & \cite{aggarwal2016hyperspectral} & \cite{wang2019row} & \cite{shen2022superpixel} & \cite{UnDIP_RastiB_2022} & \cite{hong2022endmember} & \cite{rs_Deng_RobustDual_2023} & \cite{MDLRR_WuLing_2023} & \textbf{(HTV)} & \textbf{(SSTV)} & \textbf{(HSSTV)} \\ 
			
			\midrule
			\multirow{6}{*}{Case 1} 
			& Setup & \begin{tabular}{c} \hspace{-4mm} $\lambda = 10^{0}$ \hspace{-4mm}\end{tabular} & \begin{tabular}{c} \hspace{-4mm} $\lambda_{1} = 10^{-2}$, \hspace{-4mm} \\\hspace{-4mm} $\lambda_{2} = 10^{-2}$, \hspace{-4mm} \\\hspace{-4mm} $\lambda_{3} = 10^{1}$ \hspace{-4mm}\end{tabular} & \begin{tabular}{c} \hspace{-4mm} $\lambda = 10^{0}$, \hspace{-4mm} \\\hspace{-4mm} $\lambda_{TV} = 10^{-2}$ \hspace{-4mm}\end{tabular} & \begin{tabular}{c} \hspace{-4mm} $\lambda_{g} = 10^{1}$, \hspace{-4mm} \\\hspace{-4mm} $\lambda_{l} = 10^{-3}$ \hspace{-4mm}\end{tabular} &  --  &  --  & \begin{tabular}{c} \hspace{-4mm} $\lambda = 10^{-1}$ \hspace{-4mm}\end{tabular} & \begin{tabular}{c} \hspace{-4mm} $\lambda = 10^{-2}$, \hspace{-4mm} \\\hspace{-4mm} $\tau = 10^{0}$ \hspace{-4mm}\end{tabular} & \begin{tabular}{c} \hspace{-4mm} $\lambda_{1} = 10^{-1}$, \hspace{-4mm} \\\hspace{-4mm} $\lambda_{2} = 10^{0}$, \hspace{-4mm} \\\hspace{-4mm} $\varepsilon = 0.95$ \hspace{-4mm}\end{tabular} & \begin{tabular}{c} \hspace{-4mm} $\lambda_{1} = 10^{-1}$, \hspace{-4mm} \\\hspace{-4mm} $\lambda_{2} = 10^{-2}$, \hspace{-4mm} \\\hspace{-4mm} $\varepsilon = 0.95$ \hspace{-4mm}\end{tabular} & \begin{tabular}{c} \hspace{-4mm} $\lambda_{1} = 10^{-1}$, \hspace{-4mm} \\\hspace{-4mm} $\lambda_{2} = 10^{-2}$, \hspace{-4mm} \\\hspace{-4mm} $\varepsilon = 0.95$ \hspace{-4mm}\end{tabular} \\ 
			& SRE  &   13.07 &   -0.00 &   15.62 & \ValSecnd{  20.91} &    0.42 &   -2.81 & \Valbest{  24.94} &   18.28 &   15.92 &   14.49 &   14.74 \\ 
			& RMSE &   0.0105 &   2.1963 &   0.0081 & \ValSecnd{  0.0047} &   0.0336 &   0.0409 & \Valbest{  0.0029} &   0.0062 &   0.0078 &   0.0090 &   0.0088 \\ 
			& Ps   & \ValSecnd{  0.99} &   0.11 & \Valbest{  1.00} & \Valbest{  1.00} &   0.53 &   0.07 & \Valbest{  1.00} & \Valbest{  1.00} & \ValSecnd{  0.99} & \ValSecnd{  0.99} & \ValSecnd{  0.99} \\ 
			& MPSNR &   40.67 &  -45.31 &   40.88 & \ValSecnd{  42.44} &   24.71 &   15.70 & \Valbest{  43.78} &   40.05 &   37.93 &   41.74 &   41.91 \\ 
			& MSSIM &   0.9813 &   0.6266 &   0.9806 &   0.9859 &   0.8480 &   0.2748 & \Valbest{  0.9894} &   0.9791 &   0.9741 &   0.9854 & \ValSecnd{  0.9863} \\ 
			\midrule 
			
			\multirow{6}{*}{Case 2} 
			& Setup & \begin{tabular}{c} \hspace{-4mm} $\lambda = 10^{0}$ \hspace{-4mm}\end{tabular} & \begin{tabular}{c} \hspace{-4mm} $\lambda_{1} = 10^{-2}$, \hspace{-4mm} \\\hspace{-4mm} $\lambda_{2} = 10^{-2}$, \hspace{-4mm} \\\hspace{-4mm} $\lambda_{3} = 10^{1}$ \hspace{-4mm}\end{tabular} & \begin{tabular}{c} \hspace{-4mm} $\lambda = 10^{1}$, \hspace{-4mm} \\\hspace{-4mm} $\lambda_{TV} = 10^{-2}$ \hspace{-4mm}\end{tabular} & \begin{tabular}{c} \hspace{-4mm} $\lambda_{g} = 10^{1}$, \hspace{-4mm} \\\hspace{-4mm} $\lambda_{l} = 10^{-4}$ \hspace{-4mm}\end{tabular} &  --  &  --  & \begin{tabular}{c} \hspace{-4mm} $\lambda = 10^{-1}$ \hspace{-4mm}\end{tabular} & \begin{tabular}{c} \hspace{-4mm} $\lambda = 10^{-1}$, \hspace{-4mm} \\\hspace{-4mm} $\tau = 10^{0}$ \hspace{-4mm}\end{tabular} & \begin{tabular}{c} \hspace{-4mm} $\lambda_{1} = 10^{-1}$, \hspace{-4mm} \\\hspace{-4mm} $\lambda_{2} = 10^{0}$, \hspace{-4mm} \\\hspace{-4mm} $\varepsilon = 0.95$ \hspace{-4mm}\end{tabular} & \begin{tabular}{c} \hspace{-4mm} $\lambda_{1} = 10^{-1}$, \hspace{-4mm} \\\hspace{-4mm} $\lambda_{2} = 10^{-2}$, \hspace{-4mm} \\\hspace{-4mm} $\varepsilon = 0.95$ \hspace{-4mm}\end{tabular} & \begin{tabular}{c} \hspace{-4mm} $\lambda_{1} = 10^{-1}$, \hspace{-4mm} \\\hspace{-4mm} $\lambda_{2} = 10^{-2}$, \hspace{-4mm} \\\hspace{-4mm} $\varepsilon = 0.95$ \hspace{-4mm}\end{tabular} \\ 
			& SRE  &    7.33 &   -0.00 &    8.91 &   12.88 &   -2.49 &   -7.34 & \Valbest{  19.15} & \ValSecnd{  12.98} &   12.62 &   10.90 &   11.37 \\ 
			& RMSE &   0.0191 &   2.1447 &   0.0165 &   0.0111 &   0.0389 &   0.0569 & \Valbest{  0.0057} &   0.0113 & \ValSecnd{  0.0110} &   0.0131 &   0.0126 \\ 
			& Ps   &   0.81 &   0.12 &   0.92 & \ValSecnd{  0.98} &   0.29 &   0.00 & \Valbest{  1.00} &   0.96 & \ValSecnd{  0.98} & \ValSecnd{  0.98} & \ValSecnd{  0.98} \\ 
			& MPSNR &   34.87 &  -45.10 &   34.39 &   35.69 &   21.98 &   13.21 & \Valbest{  37.83} &   33.83 &   35.25 &   35.77 & \ValSecnd{  36.05} \\ 
			& MSSIM &   0.9360 &   0.5726 &   0.9232 &   0.9382 &   0.7388 &   0.3409 & \Valbest{  0.9607} &   0.9306 & \ValSecnd{  0.9571} &   0.9448 &   0.9483 \\ 
			\midrule 
			
			\multirow{6}{*}{Case 3} 
			& Setup & \begin{tabular}{c} \hspace{-4mm} $\lambda = 10^{0}$ \hspace{-4mm}\end{tabular} & \begin{tabular}{c} \hspace{-4mm} $\lambda_{1} = 10^{1}$, \hspace{-4mm} \\\hspace{-4mm} $\lambda_{2} = 10^{-2}$, \hspace{-4mm} \\\hspace{-4mm} $\lambda_{3} = 10^{1}$ \hspace{-4mm}\end{tabular} & \begin{tabular}{c} \hspace{-4mm} $\lambda = 10^{1}$, \hspace{-4mm} \\\hspace{-4mm} $\lambda_{TV} = 10^{-2}$ \hspace{-4mm}\end{tabular} & \begin{tabular}{c} \hspace{-4mm} $\lambda_{g} = 10^{1}$, \hspace{-4mm} \\\hspace{-4mm} $\lambda_{l} = 10^{-3}$ \hspace{-4mm}\end{tabular} &  --  &  --  & \begin{tabular}{c} \hspace{-4mm} $\lambda = 10^{0}$ \hspace{-4mm}\end{tabular} & \begin{tabular}{c} \hspace{-4mm} $\lambda = 10^{-1}$, \hspace{-4mm} \\\hspace{-4mm} $\tau = 10^{0}$ \hspace{-4mm}\end{tabular} & \begin{tabular}{c} \hspace{-4mm} $\lambda_{1} = 10^{-1}$, \hspace{-4mm} \\\hspace{-4mm} $\lambda_{2} = 10^{0}$, \hspace{-4mm} \\\hspace{-4mm} $\varepsilon = 0.95$ \hspace{-4mm}\end{tabular} & \begin{tabular}{c} \hspace{-4mm} $\lambda_{1} = 10^{-1}$, \hspace{-4mm} \\\hspace{-4mm} $\lambda_{2} = 10^{-2}$, \hspace{-4mm} \\\hspace{-4mm} $\varepsilon = 0.95$ \hspace{-4mm}\end{tabular} & \begin{tabular}{c} \hspace{-4mm} $\lambda_{1} = 10^{-1}$, \hspace{-4mm} \\\hspace{-4mm} $\lambda_{2} = 10^{-2}$, \hspace{-4mm} \\\hspace{-4mm} $\varepsilon = 0.95$ \hspace{-4mm}\end{tabular} \\ 
			& SRE  &    5.06 &   -0.00 &    5.46 &   10.01 &   -5.38 &   -7.92 & \Valbest{  15.64} &    9.47 & \ValSecnd{  15.29} &   13.79 &   14.11 \\ 
			& RMSE &   0.0233 &   1.6567 &   0.0235 &   0.0149 &   0.0463 &   0.0561 & \ValSecnd{  0.0085} &   0.0161 & \Valbest{  0.0083} &   0.0097 &   0.0094 \\ 
			& Ps   &   0.72 &   0.23 &   0.74 & \ValSecnd{  0.93} &   0.02 &   0.00 & \Valbest{  0.99} &   0.90 & \Valbest{  0.99} & \Valbest{  0.99} & \Valbest{  0.99} \\ 
			& MPSNR &   31.03 &  -42.73 &   30.72 &   31.73 &   20.09 &   13.05 &   33.91 &   29.55 &   37.38 & \ValSecnd{  40.78} & \Valbest{  40.98} \\ 
			& MSSIM &   0.8842 &   0.8106 &   0.8656 &   0.8876 &   0.7273 &   0.3206 &   0.9320 &   0.8551 &   0.9712 & \ValSecnd{  0.9819} & \Valbest{  0.9831} \\ 
			\midrule 
			
			\multirow{6}{*}{Case 4} 
			& Setup & \begin{tabular}{c} \hspace{-4mm} $\lambda = 10^{0}$ \hspace{-4mm}\end{tabular} & \begin{tabular}{c} \hspace{-4mm} $\lambda_{1} = 10^{1}$, \hspace{-4mm} \\\hspace{-4mm} $\lambda_{2} = 10^{-2}$, \hspace{-4mm} \\\hspace{-4mm} $\lambda_{3} = 10^{-1}$ \hspace{-4mm}\end{tabular} & \begin{tabular}{c} \hspace{-4mm} $\lambda = 10^{1}$, \hspace{-4mm} \\\hspace{-4mm} $\lambda_{TV} = 10^{-2}$ \hspace{-4mm}\end{tabular} & \begin{tabular}{c} \hspace{-4mm} $\lambda_{g} = 10^{1}$, \hspace{-4mm} \\\hspace{-4mm} $\lambda_{l} = 10^{-2}$ \hspace{-4mm}\end{tabular} &  --  &  --  & \begin{tabular}{c} \hspace{-4mm} $\lambda = 10^{0}$ \hspace{-4mm}\end{tabular} & \begin{tabular}{c} \hspace{-4mm} $\lambda = 10^{-1}$, \hspace{-4mm} \\\hspace{-4mm} $\tau = 10^{0}$ \hspace{-4mm}\end{tabular} & \begin{tabular}{c} \hspace{-4mm} $\lambda_{1} = 10^{-1}$, \hspace{-4mm} \\\hspace{-4mm} $\lambda_{2} = 10^{0}$, \hspace{-4mm} \\\hspace{-4mm} $\varepsilon = 0.95$ \hspace{-4mm}\end{tabular} & \begin{tabular}{c} \hspace{-4mm} $\lambda_{1} = 10^{-1}$, \hspace{-4mm} \\\hspace{-4mm} $\lambda_{2} = 10^{-2}$, \hspace{-4mm} \\\hspace{-4mm} $\varepsilon = 0.95$ \hspace{-4mm}\end{tabular} & \begin{tabular}{c} \hspace{-4mm} $\lambda_{1} = 10^{-1}$, \hspace{-4mm} \\\hspace{-4mm} $\lambda_{2} = 10^{-2}$, \hspace{-4mm} \\\hspace{-4mm} $\varepsilon = 0.95$ \hspace{-4mm}\end{tabular} \\ 
			& SRE  &    2.56 &    0.00 &    2.08 &    6.05 &   -5.08 &   -8.40 &   12.42 &    5.33 & \Valbest{  14.95} &   13.76 & \ValSecnd{  14.05} \\ 
			& RMSE &   0.0290 &   8.8367 &   0.0339 &   0.0219 &   0.0437 &   0.0555 &   0.0119 &   0.0243 & \Valbest{  0.0086} &   0.0097 & \ValSecnd{  0.0095} \\ 
			& Ps   &   0.60 &   0.72 &   0.43 &   0.79 &   0.11 &   0.00 & \ValSecnd{  0.97} &   0.72 & \Valbest{  0.99} & \Valbest{  0.99} & \Valbest{  0.99} \\ 
			& MPSNR &   27.58 &  -57.72 &   27.20 &   28.08 &   19.23 &   13.01 &   29.85 &   25.67 &   37.34 & \ValSecnd{  39.87} & \Valbest{  40.12} \\ 
			& MSSIM &   0.8119 &   0.5146 &   0.7824 &   0.8139 &   0.6811 &   0.2982 &   0.8767 &   0.7507 &   0.9713 & \ValSecnd{  0.9776} & \Valbest{  0.9792} \\ 
			\midrule 
			
			\multirow{6}{*}{Case 5} 
			& Setup & \begin{tabular}{c} \hspace{-4mm} $\lambda = 10^{0}$ \hspace{-4mm}\end{tabular} & \begin{tabular}{c} \hspace{-4mm} $\lambda_{1} = 10^{1}$, \hspace{-4mm} \\\hspace{-4mm} $\lambda_{2} = 10^{-2}$, \hspace{-4mm} \\\hspace{-4mm} $\lambda_{3} = 10^{1}$ \hspace{-4mm}\end{tabular} & \begin{tabular}{c} \hspace{-4mm} $\lambda = 10^{1}$, \hspace{-4mm} \\\hspace{-4mm} $\lambda_{TV} = 10^{-2}$ \hspace{-4mm}\end{tabular} & \begin{tabular}{c} \hspace{-4mm} $\lambda_{g} = 10^{1}$, \hspace{-4mm} \\\hspace{-4mm} $\lambda_{l} = 10^{-3}$ \hspace{-4mm}\end{tabular} &  --  &  --  & \begin{tabular}{c} \hspace{-4mm} $\lambda = 10^{0}$ \hspace{-4mm}\end{tabular} & \begin{tabular}{c} \hspace{-4mm} $\lambda = 10^{-1}$, \hspace{-4mm} \\\hspace{-4mm} $\tau = 10^{0}$ \hspace{-4mm}\end{tabular} & \begin{tabular}{c} \hspace{-4mm} $\lambda_{1} = 10^{-1}$, \hspace{-4mm} \\\hspace{-4mm} $\lambda_{2} = 10^{1}$, \hspace{-4mm} \\\hspace{-4mm} $\varepsilon = 0.95$ \hspace{-4mm}\end{tabular} & \begin{tabular}{c} \hspace{-4mm} $\lambda_{1} = 10^{0}$, \hspace{-4mm} \\\hspace{-4mm} $\lambda_{2} = 10^{-2}$, \hspace{-4mm} \\\hspace{-4mm} $\varepsilon = 0.95$ \hspace{-4mm}\end{tabular} & \begin{tabular}{c} \hspace{-4mm} $\lambda_{1} = 10^{0}$, \hspace{-4mm} \\\hspace{-4mm} $\lambda_{2} = 10^{-2}$, \hspace{-4mm} \\\hspace{-4mm} $\varepsilon = 0.95$ \hspace{-4mm}\end{tabular} \\ 
			& SRE  &    4.47 &    0.00 &    4.44 &    8.99 &   -5.94 &   -8.13 & \Valbest{  15.21} &    8.34 & \ValSecnd{  10.38} &    9.59 &    9.59 \\ 
			& RMSE &   0.0247 &   9.6721 &   0.0262 &   0.0170 &   0.0466 &   0.0558 & \Valbest{  0.0089} &   0.0181 & \ValSecnd{  0.0138} &   0.0149 &   0.0148 \\ 
			& Ps   &   0.69 &   0.62 &   0.65 &   0.89 &   0.01 &   0.00 & \Valbest{  0.99} &   0.86 & \ValSecnd{  0.96} &   0.94 &   0.94 \\ 
			& MPSNR &   30.57 &  -58.27 &   30.16 &   31.14 &   17.69 &   13.03 &   33.48 &   28.88 &   30.36 & \ValSecnd{  37.42} & \Valbest{  37.43} \\ 
			& MSSIM &   0.8743 &   0.7935 &   0.8522 &   0.8736 &   0.6354 &   0.3098 &   0.9244 &   0.8400 &   0.8906 & \ValSecnd{  0.9677} & \Valbest{  0.9679} \\ 
			\midrule 
			
			\multirow{6}{*}{Case 6} 
			& Setup & \begin{tabular}{c} \hspace{-4mm} $\lambda = 10^{0}$ \hspace{-4mm}\end{tabular} & \begin{tabular}{c} \hspace{-4mm} $\lambda_{1} = 10^{1}$, \hspace{-4mm} \\\hspace{-4mm} $\lambda_{2} = 10^{-2}$, \hspace{-4mm} \\\hspace{-4mm} $\lambda_{3} = 10^{1}$ \hspace{-4mm}\end{tabular} & \begin{tabular}{c} \hspace{-4mm} $\lambda = 10^{1}$, \hspace{-4mm} \\\hspace{-4mm} $\lambda_{TV} = 10^{-2}$ \hspace{-4mm}\end{tabular} & \begin{tabular}{c} \hspace{-4mm} $\lambda_{g} = 10^{1}$, \hspace{-4mm} \\\hspace{-4mm} $\lambda_{l} = 10^{-3}$ \hspace{-4mm}\end{tabular} &  --  &  --  & \begin{tabular}{c} \hspace{-4mm} $\lambda = 10^{0}$ \hspace{-4mm}\end{tabular} & \begin{tabular}{c} \hspace{-4mm} $\lambda = 10^{-1}$, \hspace{-4mm} \\\hspace{-4mm} $\tau = 10^{0}$ \hspace{-4mm}\end{tabular} & \begin{tabular}{c} \hspace{-4mm} $\lambda_{1} = 10^{-1}$, \hspace{-4mm} \\\hspace{-4mm} $\lambda_{2} = 10^{0}$, \hspace{-4mm} \\\hspace{-4mm} $\varepsilon = 0.98$ \hspace{-4mm}\end{tabular} & \begin{tabular}{c} \hspace{-4mm} $\lambda_{1} = 10^{-1}$, \hspace{-4mm} \\\hspace{-4mm} $\lambda_{2} = 10^{-1}$, \hspace{-4mm} \\\hspace{-4mm} $\varepsilon = 0.98$ \hspace{-4mm}\end{tabular} & \begin{tabular}{c} \hspace{-4mm} $\lambda_{1} = 10^{-1}$, \hspace{-4mm} \\\hspace{-4mm} $\lambda_{2} = 10^{-1}$, \hspace{-4mm} \\\hspace{-4mm} $\varepsilon = 0.98$ \hspace{-4mm}\end{tabular} \\ 
			& SRE  &    3.65 &    0.00 &    3.20 &    7.48 &   -5.80 &   -8.10 & \Valbest{  14.37} &    6.85 & \ValSecnd{  10.02} &    8.59 &    8.69 \\ 
			& RMSE &   0.0269 &   9.5123 &   0.0303 &   0.0199 &   0.0484 &   0.0559 & \Valbest{  0.0098} &   0.0212 & \ValSecnd{  0.0144} &   0.0166 &   0.0165 \\ 
			& Ps   &   0.64 &   0.47 &   0.53 &   0.83 &   0.06 &   0.00 & \Valbest{  0.98} &   0.79 & \ValSecnd{  0.95} &   0.93 &   0.93 \\ 
			& MPSNR &   29.63 &  -58.13 &   29.08 &   30.01 &   18.89 &   13.26 &   32.55 &   27.29 & \Valbest{  34.34} &   33.87 & \ValSecnd{  34.18} \\ 
			& MSSIM &   0.8490 &   0.7094 &   0.8194 &   0.8423 &   0.6723 &   0.3455 &   0.9066 &   0.7968 & \Valbest{  0.9496} &   0.9227 & \ValSecnd{  0.9294} \\ 
			\midrule
			
			\multirow{6}{*}{Case 7} 
			& Setup & \begin{tabular}{c} \hspace{-4mm} $\lambda = 10^{0}$ \hspace{-4mm}\end{tabular} & \begin{tabular}{c} \hspace{-4mm} $\lambda_{1} = 10^{-2}$, \hspace{-4mm} \\\hspace{-4mm} $\lambda_{2} = 10^{0}$, \hspace{-4mm} \\\hspace{-4mm} $\lambda_{3} = 10^{1}$ \hspace{-4mm}\end{tabular} & \begin{tabular}{c} \hspace{-4mm} $\lambda = 10^{1}$, \hspace{-4mm} \\\hspace{-4mm} $\lambda_{TV} = 10^{-2}$ \hspace{-4mm}\end{tabular} & \begin{tabular}{c} \hspace{-4mm} $\lambda_{g} = 10^{1}$, \hspace{-4mm} \\\hspace{-4mm} $\lambda_{l} = 10^{-3}$ \hspace{-4mm}\end{tabular} &  --  &  --  & \begin{tabular}{c} \hspace{-4mm} $\lambda = 10^{0}$ \hspace{-4mm}\end{tabular} & \begin{tabular}{c} \hspace{-4mm} $\lambda = 10^{-1}$, \hspace{-4mm} \\\hspace{-4mm} $\tau = 10^{0}$ \hspace{-4mm}\end{tabular} & \begin{tabular}{c} \hspace{-4mm} $\lambda_{1} = 10^{-1}$, \hspace{-4mm} \\\hspace{-4mm} $\lambda_{2} = 10^{0}$, \hspace{-4mm} \\\hspace{-4mm} $\varepsilon = 0.98$ \hspace{-4mm}\end{tabular} & \begin{tabular}{c} \hspace{-4mm} $\lambda_{1} = 10^{-1}$, \hspace{-4mm} \\\hspace{-4mm} $\lambda_{2} = 10^{-1}$, \hspace{-4mm} \\\hspace{-4mm} $\varepsilon = 0.95$ \hspace{-4mm}\end{tabular} & \begin{tabular}{c} \hspace{-4mm} $\lambda_{1} = 10^{-1}$, \hspace{-4mm} \\\hspace{-4mm} $\lambda_{2} = 10^{-1}$, \hspace{-4mm} \\\hspace{-4mm} $\varepsilon = 0.95$ \hspace{-4mm}\end{tabular} \\ 
			& SRE  &    4.61 &   -0.00 &    4.46 &    9.06 &   -6.68 &   -7.88 & \Valbest{  16.19} &    8.43 & \ValSecnd{  10.55} &    7.97 &    8.32 \\ 
			& RMSE &   0.0253 &   2.0093 &   0.0270 &   0.0171 &   0.0582 &   0.0561 & \Valbest{  0.0082} &   0.0184 & \ValSecnd{  0.0136} &   0.0177 &   0.0171 \\ 
			& Ps   &   0.67 &   0.09 &   0.61 &   0.88 &   0.00 &   0.00 & \Valbest{  0.99} &   0.85 & \ValSecnd{  0.96} &   0.91 &   0.92 \\ 
			& MPSNR &   31.60 &  -44.48 &   30.64 &   31.88 &   12.53 &   12.99 & \Valbest{  34.67} &   28.67 &   32.38 &   32.70 & \ValSecnd{  33.19} \\ 
			& MSSIM &   0.8806 &   0.4875 &   0.8485 &   0.8723 &   0.1260 &   0.3170 & \Valbest{  0.9295} &   0.8358 & \ValSecnd{  0.9250} &   0.8999 &   0.9117 \\ 
			\midrule 
			
			\multirow{6}{*}{Case 8} 
			& Setup & \begin{tabular}{c} \hspace{-4mm} $\lambda = 10^{0}$ \hspace{-4mm}\end{tabular} & \begin{tabular}{c} \hspace{-4mm} $\lambda_{1} = 10^{-1}$, \hspace{-4mm} \\\hspace{-4mm} $\lambda_{2} = 10^{-1}$, \hspace{-4mm} \\\hspace{-4mm} $\lambda_{3} = 10^{1}$ \hspace{-4mm}\end{tabular} & \begin{tabular}{c} \hspace{-4mm} $\lambda = 10^{1}$, \hspace{-4mm} \\\hspace{-4mm} $\lambda_{TV} = 10^{-2}$ \hspace{-4mm}\end{tabular} & \begin{tabular}{c} \hspace{-4mm} $\lambda_{g} = 10^{1}$, \hspace{-4mm} \\\hspace{-4mm} $\lambda_{l} = 10^{-2}$ \hspace{-4mm}\end{tabular} &  --  &  --  & \begin{tabular}{c} \hspace{-4mm} $\lambda = 10^{0}$ \hspace{-4mm}\end{tabular} & \begin{tabular}{c} \hspace{-4mm} $\lambda = 10^{-1}$, \hspace{-4mm} \\\hspace{-4mm} $\tau = 10^{0}$ \hspace{-4mm}\end{tabular} & \begin{tabular}{c} \hspace{-4mm} $\lambda_{1} = 10^{-1}$, \hspace{-4mm} \\\hspace{-4mm} $\lambda_{2} = 10^{0}$, \hspace{-4mm} \\\hspace{-4mm} $\varepsilon = 0.95$ \hspace{-4mm}\end{tabular} & \begin{tabular}{c} \hspace{-4mm} $\lambda_{1} = 10^{-1}$, \hspace{-4mm} \\\hspace{-4mm} $\lambda_{2} = 10^{-2}$, \hspace{-4mm} \\\hspace{-4mm} $\varepsilon = 0.95$ \hspace{-4mm}\end{tabular} & \begin{tabular}{c} \hspace{-4mm} $\lambda_{1} = 10^{-1}$, \hspace{-4mm} \\\hspace{-4mm} $\lambda_{2} = 10^{-2}$, \hspace{-4mm} \\\hspace{-4mm} $\varepsilon = 0.95$ \hspace{-4mm}\end{tabular} \\ 
			& SRE  &    2.55 &   -0.00 &    1.55 &    5.20 &   -6.84 &   -8.13 & \Valbest{  13.14} &    4.47 & \ValSecnd{   8.92} &    6.64 &    7.32 \\ 
			& RMSE &   0.0302 &   2.1167 &   0.0380 &   0.0246 &   0.0578 &   0.0559 & \Valbest{  0.0113} &   0.0274 & \ValSecnd{  0.0160} &   0.0199 &   0.0186 \\ 
			& Ps   &   0.58 &   0.06 &   0.38 &   0.71 &   0.00 &   0.00 & \Valbest{  0.97} &   0.65 & \ValSecnd{  0.92} &   0.85 &   0.88 \\ 
			& MPSNR &   28.26 &  -44.97 &   27.45 &   28.50 &   13.55 &   13.19 &   31.20 &   25.12 & \Valbest{  31.82} &   31.34 & \ValSecnd{  31.62} \\ 
			& MSSIM &   0.8057 &   0.3782 &   0.7635 &   0.7943 &   0.1835 &   0.3380 &   0.8736 &   0.7267 & \Valbest{  0.9192} &   0.8727 & \ValSecnd{  0.8796} \\
			
			\bottomrule
		\end{tabular}
	}
\end{table*}

\section{Experiments}
\label{experience}
We demonstrate the effectiveness of the proposed non-blind unmixing method, i.e., \Ourss through comprehensive experiments using two synthetic and two real HS images.
Specifically, these experiments aim to validate that 
\begin{itemize}
	\item \Ourss achieves good unmixing performance due to image-domain regularizations,
	\item \Ourss is robust to mixed noise, including stripe noise.
\end{itemize}
As described in the introduction, existing unmixing methods are classified into blind and non-blind, depending on whether the endmember library is given or not. 
Due to the different assumptions and the fact that blind unmixing methods require a non-blind unmixing step to obtain an initial estimate, it is difficult to fairly compare non-blind unmixing methods with blind ones.
Therefore, we mainly compare \Ourss with non-blind unmixing methods. Specifically, we compare \Ourss with seven state-of-the-art non-blind unmixing methods and one state-of-the-art blind unmixing method: the collaborative sparse unmixing by variable splitting and augmented lagrangian (CLSUnSAL)~\cite{iordache2014collaborative}, the hyperspectral unmixing using joint-sparsity and total variation (JSTV)~\cite{aggarwal2016hyperspectral}\footnote{The code is available at \url{https://jp.mathworks.com/matlabcentral/fileexchange/56831-hyperspectral-unmixing-and-denoising?s_tid=FX_rc1_behav}}, the row-sparsity spectral unmixing via total variation (RSSUn-TV)~\cite{wang2019row},
the local-global-based sparse regression unmixing (LGSU)~\cite{shen2022superpixel}, the hyperspectral unmixing using deep image prior (UnDIP)~\cite{UnDIP_RastiB_2022}\footnote{The code is available at \url{https://github.com/BehnoodRasti/UnDIP}}, the endmember-guided unmixing network (EGU-Net)~\cite{hong2022endmember}\footnote{The code is available at \url{https://github.com/danfenghong/IEEE_TNNLS_EGU-Net}}, the robust dual spatial
weighted sparse unmixing (RDSWSU)~\cite{rs_Deng_RobustDual_2023}, and the multidimensional low-rank representation-based sparse hyperspectral unmixing (MdLRR)~\cite{MDLRR_WuLing_2023}\footnote{The code is available at \url{https://huangjie-uestc.github.io/}}.
To perform the experiments, we reimplemented the program codes of CLSUnSAL, RSSUn-TV, and RDSWSU. The program code of LGSU was downloaded from a web page, which is no longer accessible as of January 2024.
EGU-Net uses the number of endmembers in a target HS image. In our experiments, since the number of endmembers in a target HS image is assumed to be unknown, we set it as the number of endmembers in the endmember libraries of datasets described later.
Tab.~\ref{tab:noise} shows the assumption and the types of noise considered in each method.

\subsection{Datasets Description}
We used six datasets for experiments. 
In all datasets, their endmember libraries were composed of the spectral signatures of the endmembers in ground-truth HS images and other spectral signatures. 
This is to simulate the real-world situation where we give an endmember library by including more spectral signatures than the components of the target HS image, as assumed in many references of non-blind unmixing.

\subsubsection{Synthetic HS Image 1 (\textit{Synth 1})}
We generated the first synthetic HS image with a size of $64 \times 64 \times 224$ using the HYperspectral Data Retrieval and Analysis (HYDRA) toolbox\footnote{\url{https://www.ehu.eus/ccwintco/index.php?title=Hyperspectral_Imagery_Synthesis_tools_for_MATLAB}, accessed on Feb. 5, 2023},
which is developed by the Computational Intelligence group at the University of the Basque Country.
An endmember library consists of $10$ spectral signatures with $224$ bands from the U.S. Geological Survey (USGS) Spectral Library\footnote{\url{https://www.usgs.gov/programs/usgs-library}, accessed on Aug. 7, 2023}. 
From the endmember library, we randomly selected four endmembers and generated four original abundance maps with the spatial size of $64 \times 64$ using the Legendre method.
Fig.~\ref{fig:Org_HSimages} (a) shows one band of the generated image.

\subsubsection{Synthetic HS Image 2 (\textit{Synth 2})}
We also generated the second synthetic HS image with a size of $64 \times 64 \times 224$ using the HYDRA toolbox.
An endmember library consists of $10$ spectral signatures with $224$ bands from the USGS Spectral Library. 
From the endmember library, we randomly selected four endmembers and generated four original abundance maps with a spatial size of $64 \times 64$ using the spherical Gaussian method.
Fig.~\ref{fig:Org_HSimages} (b) shows one band of the generated image.

\subsubsection{Synthetic HS Image 3 (\textit{Synth 3})}
We generated the third synthetic HS image with a size of $64 \times 64 \times 224$ using the HYDRA toolbox.
An endmember library consists of $240$ spectral signatures with $224$ bands from the USGS Spectral Library. 
From the endmember library, we randomly selected four endmembers and generated four original abundance maps with a spatial size of $200 \times 200$ using the Legendre method.
Fig.~\ref{fig:Org_HSimages} (c) shows one band of the generated image.

\subsubsection{Real HS Image 1 (\textit{Jasper Ridge})}
\textit{Jasper Ridge} image (see Fig.~\ref{fig:Org_HSimages} (d)) is captured using an AVIRIS sensor in a rural area in California, USA.
The spatial size of the original data is $512 \times 614$ pixels, and each pixel holds spectral information in $224$ bands ranging from $380$ nm to $2500$ nm.
After removing several noisy bands and cropping the image, we obtained the image with $100 \times 100$ pixels and $198$ bands.
\textit{Jasper Ridge} contains four major endmembers: ``road,'' ``soil,'' ``water,'' and ``tree''~\cite{rodarmel2002principal}.
Adding the six endmembers from the USGS Spectral Library, we used $10$ endmembers for the experiments.

\subsubsection{Real HS Image 2 (\textit{Samson})}
\textit{Samson} (see Fig.~\ref{fig:Org_HSimages} (e)) is often used for unmixing.
The spatial size of the original data is $952 \times 952$ pixels, and each pixel holds spectral information in $156$ bands covering the wavelengths from $401$ nm to $889$ nm.
After cropping the image, we obtained the image with $95 \times 95$ pixels.
\textit{Samson} contains three major endmembers: ``soil,'' ``tree,'' and ``water.''
Adding the seven endmembers from the USGS Spectral Library, we used $10$ endmembers for the experiments.

\subsubsection{Real HS Image 3 (\textit{Urban})}
\textit{Urban} (see Fig.~\ref{fig:Org_HSimages} (f)) was collected by the Hyperspectral Digital Imagery Collection Experiment (HYDICE) over an urban area at Copperas Cove, Texas, USA. The dataset has been widely used in the field of hyperspectral unmixing. The latest data version was issued by the Geospatial Research Laboratory and Engineer Research and Development Center in 2015.3 The image consists of $307\times 307$ pixels with $210$ spectral bands in the wavelength from $400$ nm to $2500$ nm with a spectral resolution of $10$ nm at a ground sampling distance GSD of $2$ m. Due to water absorption and atmospheric effects, we reduced 210 bands to 162 bands by removing several noisy bands.
\textit{Samson} contains four major endmembers: ``Asphalt,'' ``Grass,'' ``Tree,'' and ``Roof.''
Adding the $236$ endmembers from the USGS Spectral Library, we used $240$ endmembers for the experiments.

Figure~\ref{fig:SAD} plots the distributions of the spectral angle distance (SAD) values between the spectra of endmembers present in target HS images and the spectra of the other endmembers.
The SAD values tend to be the same for all the datasets. 
In this regard, the difficulty of unmixing is not expected to change.

\subsection{Exprimental Setup}
HS images are often degraded by mixed noise in real-noise scenarios. Thus, we consider the following eight combinations of i.i.d and non-i.i.d. Gaussian noise with different standard deviations $\sigma$, salt-and-pepper noise with different rate $p_{\MatSpar}$, and stripe noise in both synthetic and real data experiments.

\textit{Case 1 (i.i.d. Gaussian noise):} The observed HS image is contaminated by white Gaussian noise with the standard deviation $\sigma = 0.05$.

\textit{Case 2 (higher-level i.i.d. Gaussian noise):} The observed HS image is contaminated by white Gaussian noise with the standard deviation $\sigma = 0.1$.

\textit{Case 3 (i.i.d. Gaussian noise + salt-and-pepper noise):} The observed HS image is contaminated by white Gaussian noise with the standard deviation $\sigma = 0.05$ and salt-and-pepper noise with the rate $p_{\MatSpar} = 0.05$.

\textit{Case 4 (i.i.d. Gaussian noise + higher-rate salt-and-pepper noise):} The observed HS image is contaminated by white Gaussian noise with the standard deviation $\sigma = 0.05$ and salt-and-pepper noise with the rate $p_{\MatSpar} = 0.1$.

\textit{Case 5 (i.i.d. Gaussian noise + salt-and-pepper noise + stripe noise):} The observed HS image is contaminated by white Gaussian noise with the standard deviation $\sigma = 0.05$ and salt-and-pepper noise with the rate $p_{\MatSpar} = 0.05$. In addition, the observed HS image is corrupted by vertical stripe noise whose intensity is random in the range $[-0.3, 0.3]$.

\textit{Case 6 (i.i.d. higher-level Gaussian noise + salt-and-pepper noise + stripe noise):} The observed HS image is contaminated by white Gaussian noise with the standard deviation $\sigma = 0.1$ and salt-and-pepper noise with the rate $p_{\MatSpar} = 0.05$. In addition, the observed HS image is corrupted by vertical stripe noise whose intensity is random in the range $[-0.3, 0.3]$.

\textit{Case 7 (non-i.i.d. Gaussian noise):} The observed HS image is contaminated by non-i.i.d. white Gaussian noise. Specifically, we corrupt each band $\mathbf{h}_{i}$ by the standard deviation $\sigma_{i}$ randomly chosen in the range $[0.1, 0.2]$.

\textit{Case 8 (non-i.i.d. Gaussian noise + salt-and-pepper noise + stripe noise):} The observed HS image is contaminated by non-i.i.d. white Gaussian noise and salt-and-pepper noise with the rate $p_{\MatSpar} = 0.05$. In addition, the observed HS image is corrupted by vertical stripe noise whose intensity is random in the range $[-0.3, 0.3]$. The non-i.i.d. white Gaussian noise is the same as in Case 7.

The hyperparameters of CLSUnSAL, JSTV, RSSUn-TV, LGSU, RDSWSU, and MdLRR were adjusted to obtain the highest SRE value for each noise case and for each dataset in the ranges shown in Tab.~\ref{tab:parameter_settings}. 
UnDIP and EGU-Net have no parameters to adjust, but we followed the experimental procedure in shown their references.
The stopping criteria of the existing methods were determined according to their references.
\Ourss has hyperparameters $\lambda_1$, $\lambda_2$, $\lambda_3$, $\ParamConsSpar$, and $\ParamFidel$. The hyperparameters  $\lambda_1$ and $\lambda_2$ were adjusted to obtain the highest SRE value in the range shown in Tab~\ref{tab:parameter_settings}.
The parameter $\lambda_3$ was set to a constant value $1$ because it is not very sensitive to performance thanks to the flatness constraint (the fourth constraint of Eq.~\eqref{eq_problem}).
The hyperparameter $\ParamConsSpar$ was set as $\ParamConsSpar = 0.5 \ParamRateConsSpar p_{\MatSpar} nl$ with $\ParamRateConsSpar = 0.9$.
The hyperparameter $\ParamFidel$ was set as $\ParamFidel = \ParamRateFidel\sigma\sqrt{(1 - p_{\MatSpar})nl}$ for i.i.d. Gaussian noise cases and set as $\ParamFidel = \ParamRateFidel\sqrt{(1 - p_{\MatSpar})nl\sum_{i}^{l}\sigma_{i}}$ for non-i.i.d. Gaussian noise cases with $\ParamRateFidel$ adjusted in the range shown in Tab.~\ref{tab:parameter_settings}.
As the parameter of HSSTV, we adopted $\ParamHSSTV = 0.05$, which is recommended in~\cite{takeyama2020constrained}. 
The maximum iteration and the stopping criterion were set to $50,000$ and $\|\MatAbun^{(\NumIter+1)}-\MatAbun^{(\NumIter)}\|_{F} / \|\MatAbun^{(\NumIter+1)}\|_{F} \le 10^{-5}$, respectively.

For the quantitative evaluation of abundance maps, we used the signal reconstruction error (SRE):
\begin{equation}
	\text{SRE[dB]} = 10\log_{10}\left(\tfrac{\|\MatAbunTrue\|_F^2}{\|\MatAbunTrue-\MatAbunEst\|_F^2}\right),
\end{equation}
the root-mean-square error (RMSE):
\begin{equation}
	\text{RMSE} = \sqrt{\tfrac{1}{\NumEndmember \NumPixel}\|\MatAbunTrue-\MatAbunEst\|_F^2},
\end{equation}
and the probability of success (Ps):
\begin{equation}
	\text{Ps} = P\left(\tfrac{\|\bar{\VecAbun}_{i}-\VecAbunEst_{i,j}\|_{2}^{2}}{\|\bar{\VecAbun}_{i}\|_{2}^2}\le\text{threshold}\right),
\end{equation}
where $\MatAbunTrue$ and $\MatAbunEst$ denote the true and estimated abundance maps, respectively.
In addition,  $\VecAbun_{i}$ is $i$-th pixel of $\MatAbun$ (i.e., $\VecAbun_{i}$ is the $i$-th column vector of $\MatAbun$).
SRE and RMSE evaluate the difference between the true and estimated abundance maps, with larger SRE or smaller RMSE indicating better-estimated performance.
Ps is the probability that the relative error is less than a certain threshold.
This threshold is a criterion for how close the true and estimated abundance should be to be considered successful. In setting the threshold, most of the literature, e.g. in~\cite{iordache2014collaborative,Li_robustUnmixing_2021,Ince_doubleSpatial_2022,rs_Deng_RobustDual_2023}, regards unmixing to be successful when $\|\VecAbun_{i}-\VecAbunEst_{i,j}\|_{2}^{2} / \|\VecAbun_{i}\|_{2}^2 < 3.16$ (i.e., $10*\log_{10}(\|\VecAbun_{i}-\VecAbunEst_{i,j}\|_{2}^{2} / \|\VecAbun_{i}\|_{2}^2) < 5$ [dB]). Therefore, in this research, the threshold was also set as $3.16$.

For the quantitative evaluation of the reconstructed HS images, we used the mean peak signal-to-noise ratio overall bands (MPSNR):
\begin{equation}
	\mathrm{MPSNR [dB]} = \frac{1}{\NumBand}\sum_{i=1}^{\NumBand}10\log_{10}\left(\tfrac{\NumPixel}{\|\ElemMatHSIGeneralTrue_{i,j} - \ElemMatHSIGeneralEst_{i,j}\|_{F}^{2}}\right),
\end{equation}
where $\MatHSIGeneralTrue$ and $\MatHSIGeneralEst$ are the ground-truth and reconstructed HS images, respectively.
In addition, we adopted the mean structural similarity overall bands (MSSIM)~\cite{MSSIM}:
\begin{equation}
\label{eq:MSSIM}
\mathrm{MSSIM} = \frac{1}{l} \sum_{i=1}^{l} \mathrm{SSIM}(\MatHSIGeneralTrue_{i}, \MatHSIGeneralEst_{i}), 
\end{equation}
where $\MatHSIGeneralTrue_{i}$ are $\MatHSIGeneralEst_{i}$ are the $i$th bands of $\MatHSIGeneralTrue$ and $\MatHSIGeneralEst$, respectively.
Higher MPSNR and MSSIM values indicate better reconstruction results.

\begin{figure*}[t]
\centering
        \begin{minipage}[t]{0.075\hsize}
            \centerline{
            \includegraphics[height = 40pt]{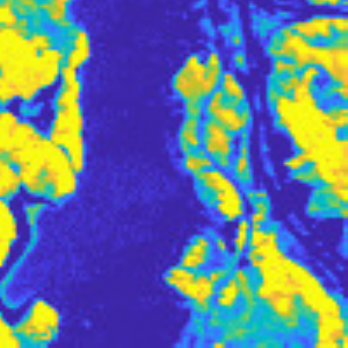}
            }
        \end{minipage}
        \begin{minipage}[t]{0.075\hsize}
            \centerline{
            \includegraphics[height = 40pt]{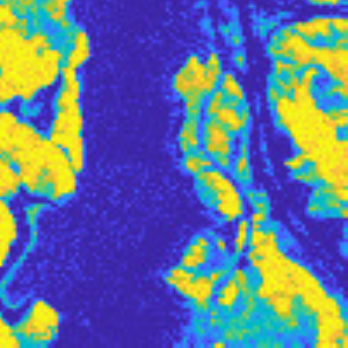}
            }
        \end{minipage}
        \begin{minipage}[t]{0.075\hsize}
            \centerline{
            \includegraphics[height = 40pt]{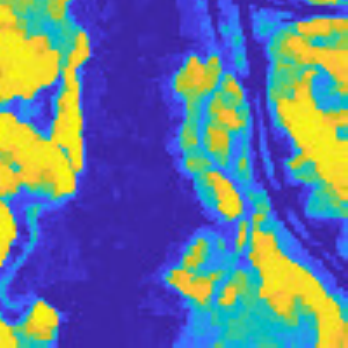}
            }
        \end{minipage}
        \begin{minipage}[t]{0.075\hsize}
            \centerline{
            \includegraphics[height = 40pt]{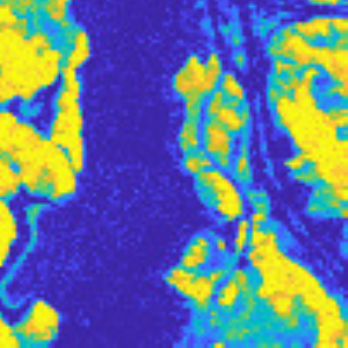}
            }
        \end{minipage}
        \begin{minipage}[t]{0.075\hsize}
            \centerline{
            \includegraphics[height = 40pt]{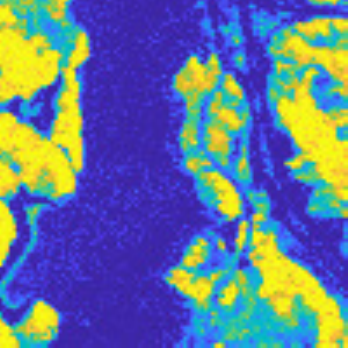}
            }
        \end{minipage}
	    \begin{minipage}[t]{0.075\hsize}
		    \centerline{
		    	\includegraphics[height = 40pt]{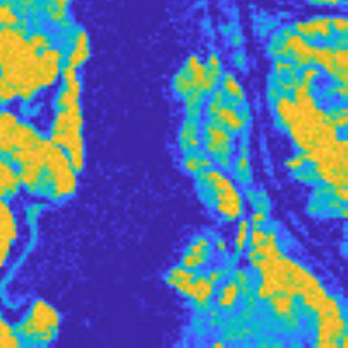}
		    }
		\end{minipage}
		\begin{minipage}[t]{0.075\hsize}
		\centerline{
			\includegraphics[height = 40pt]{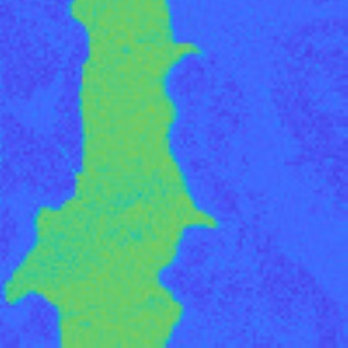}
		}
		\end{minipage}
		\begin{minipage}[t]{0.075\hsize}
		\centerline{
			\includegraphics[height = 40pt]{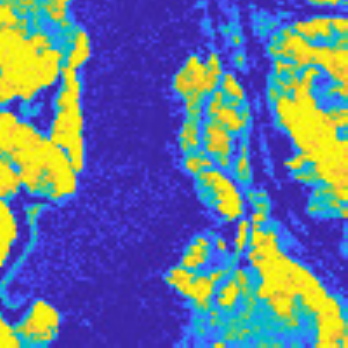}
		}
		\end{minipage}
		\begin{minipage}[t]{0.075\hsize}
			\centerline{
				\includegraphics[height = 40pt]{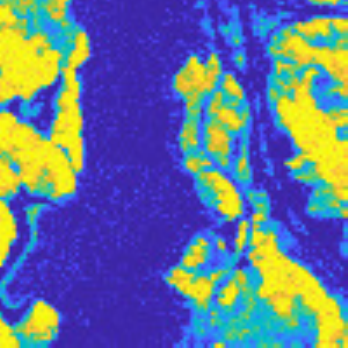}
			}
		\end{minipage}
        \begin{minipage}[t]{0.075\hsize}
            \centerline{
            \includegraphics[height = 40pt]{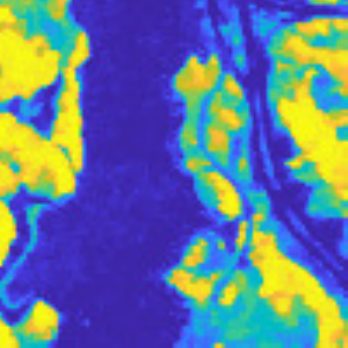}
            }
        \end{minipage}
        \begin{minipage}[t]{0.075\hsize}
            \centerline{
            \includegraphics[height = 40pt]{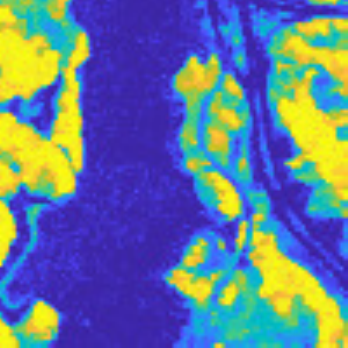}
            }
        \end{minipage}
        \begin{minipage}[t]{0.075\hsize}
            \centerline{
            \includegraphics[height = 40pt]{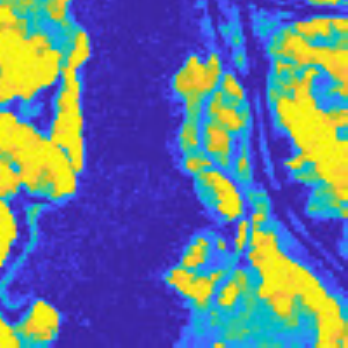}
            }
        \end{minipage}
	    \begin{minipage}[t]{0.02\hsize}
	    	\centerline{
	    		\includegraphics[height = 40pt]{./fig/colorbar_20.png}
	    	}
	    \end{minipage}
    
\vspace{1mm}
        
        \begin{minipage}[t]{0.075\hsize}
            \centerline{
            \includegraphics[height = 40pt]{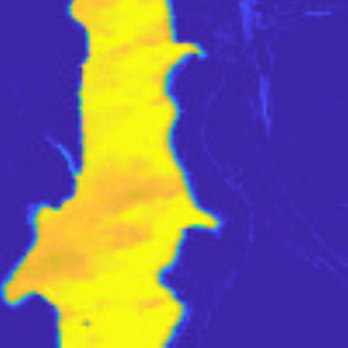}
            }
        \end{minipage}
        \begin{minipage}[t]{0.075\hsize}
            \centerline{
            \includegraphics[height = 40pt]{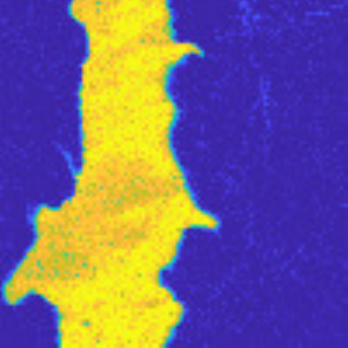}
            }
        \end{minipage}
        \begin{minipage}[t]{0.075\hsize}
            \centerline{
            \includegraphics[height = 40pt]{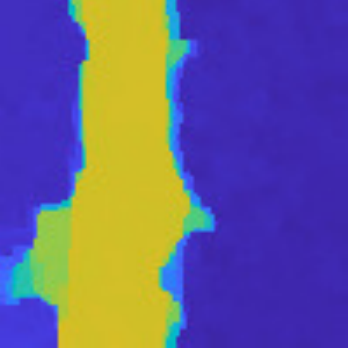}
            }
        \end{minipage}
        \begin{minipage}[t]{0.075\hsize}
            \centerline{
            \includegraphics[height = 40pt]{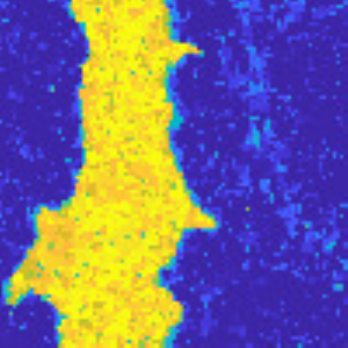}
            }
        \end{minipage}
        \begin{minipage}[t]{0.075\hsize}
            \centerline{
            \includegraphics[height = 40pt]{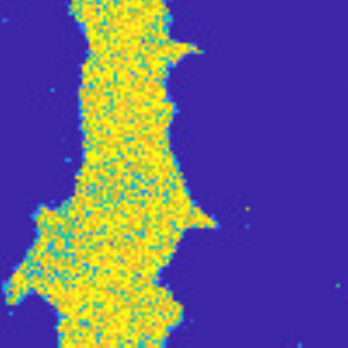}
            }
        \end{minipage}
	    \begin{minipage}[t]{0.075\hsize}
	    	\centerline{
	    		\includegraphics[height = 40pt]{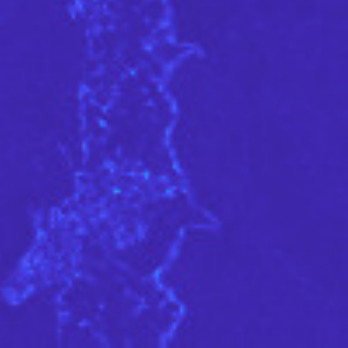}
	    	}
	    \end{minipage}
	    \begin{minipage}[t]{0.075\hsize}
	    \centerline{
	    	\includegraphics[height = 40pt]{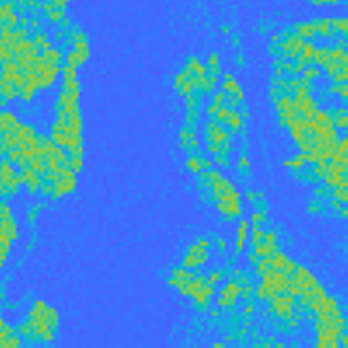}
	    }
	    \end{minipage}
	    \begin{minipage}[t]{0.075\hsize}
	    \centerline{
	    	\includegraphics[height = 40pt]{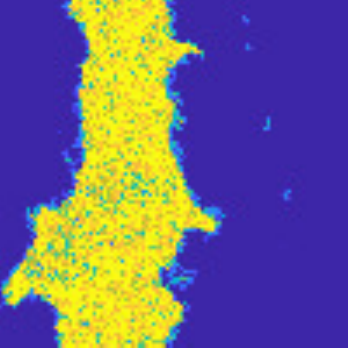}
	    }
	    \end{minipage}
	    \begin{minipage}[t]{0.075\hsize}
		    \centerline{
		    	\includegraphics[height = 40pt]{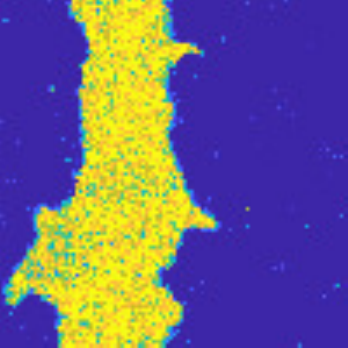}
		    }
	    \end{minipage}
        \begin{minipage}[t]{0.075\hsize}
            \centerline{
            \includegraphics[height = 40pt]{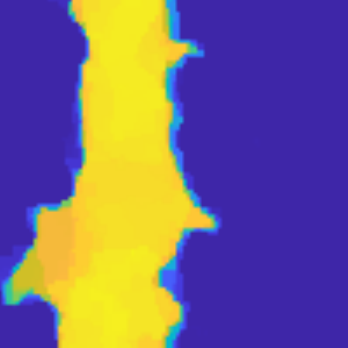}
            }
        \end{minipage}
        \begin{minipage}[t]{0.075\hsize}
            \centerline{
            \includegraphics[height = 40pt]{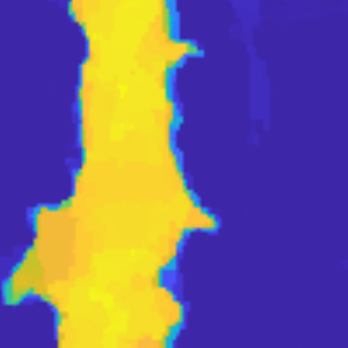}
            }
        \end{minipage}
        \begin{minipage}[t]{0.075\hsize}
            \centerline{
            \includegraphics[height = 40pt]{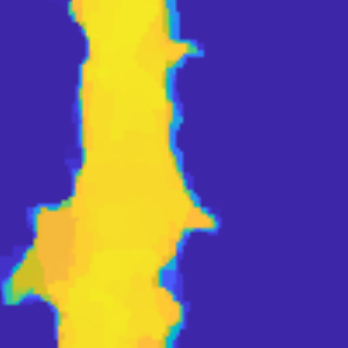}
            }
        \end{minipage}
	    \begin{minipage}[t]{0.02\hsize}
	    	\centerline{
	    		\includegraphics[height = 40pt]{./fig/colorbar_20.png}
	    	}
	    \end{minipage}
    
\vspace{1mm}
        
        \begin{minipage}[t]{0.075\hsize}
            \centerline{
            \includegraphics[height = 40pt]{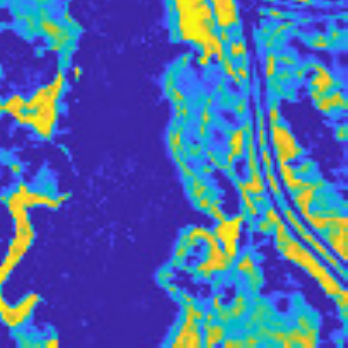}
            }
        \end{minipage}
        \begin{minipage}[t]{0.075\hsize}
            \centerline{
            \includegraphics[height = 40pt]{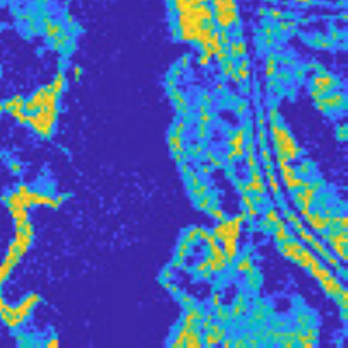}
            }
        \end{minipage}
        \begin{minipage}[t]{0.075\hsize}
            \centerline{
            \includegraphics[height = 40pt]{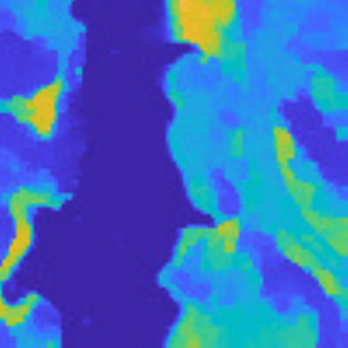}
            }
        \end{minipage}
        \begin{minipage}[t]{0.075\hsize}
            \centerline{
            \includegraphics[height = 40pt]{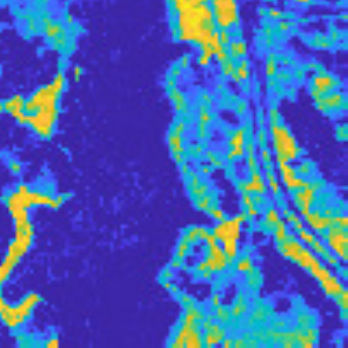}
            }
        \end{minipage}
        \begin{minipage}[t]{0.075\hsize}
            \centerline{
            \includegraphics[height = 40pt]{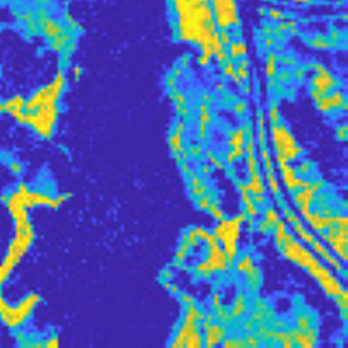}
            }
        \end{minipage}
	    \begin{minipage}[t]{0.075\hsize}
	    	\centerline{
	    		\includegraphics[height = 40pt]{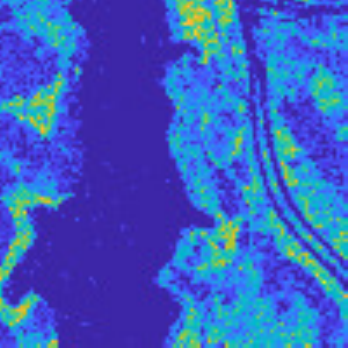}
	    	}
	    \end{minipage}
	    \begin{minipage}[t]{0.075\hsize}
	    \centerline{
	    	\includegraphics[height = 40pt]{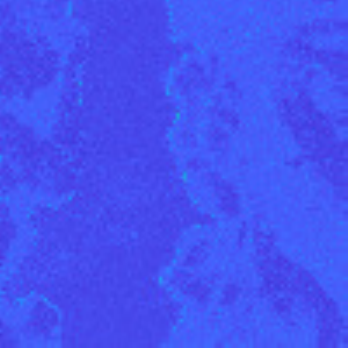}
	    }
	    \end{minipage}
	    \begin{minipage}[t]{0.075\hsize}
	    \centerline{
	    	\includegraphics[height = 40pt]{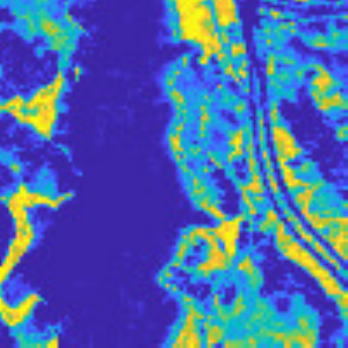}
	    }
	    \end{minipage}
	    \begin{minipage}[t]{0.075\hsize}
		    \centerline{
		    	\includegraphics[height = 40pt]{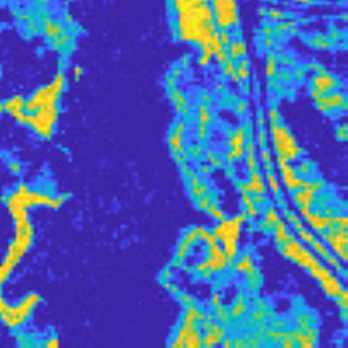}
		    }
	    \end{minipage}
        \begin{minipage}[t]{0.075\hsize}
            \centerline{
            \includegraphics[height = 40pt]{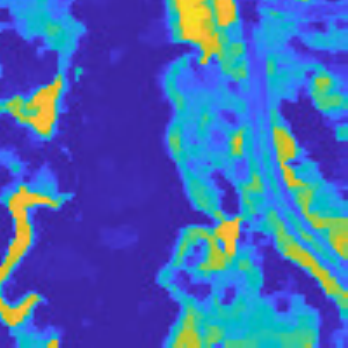}
            }
        \end{minipage}
        \begin{minipage}[t]{0.075\hsize}
            \centerline{
            \includegraphics[height = 40pt]{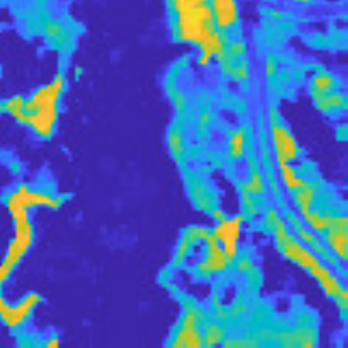}
            }
        \end{minipage}
        \begin{minipage}[t]{0.075\hsize}
            \centerline{
            \includegraphics[height = 40pt]{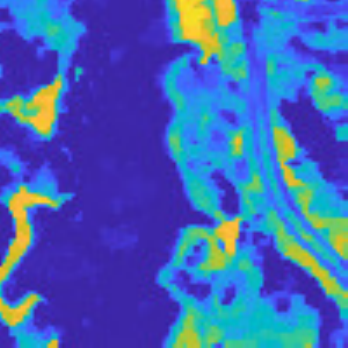}
            }
        \end{minipage}
	    \begin{minipage}[t]{0.02\hsize}
	    	\centerline{
	    		\includegraphics[height = 40pt]{./fig/colorbar_20.png}
	    	}
	    \end{minipage}
    
\vspace{1mm}
        
        \begin{minipage}[t]{0.075\hsize}
            \centerline{
            \includegraphics[height = 40pt]{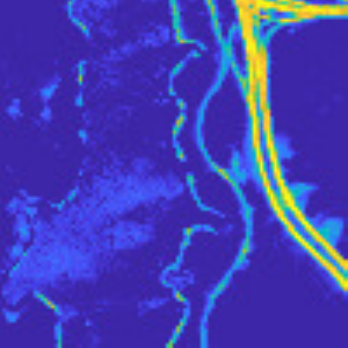}
            }
        \end{minipage}
        \begin{minipage}[t]{0.075\hsize}
            \centerline{
            \includegraphics[height = 40pt]{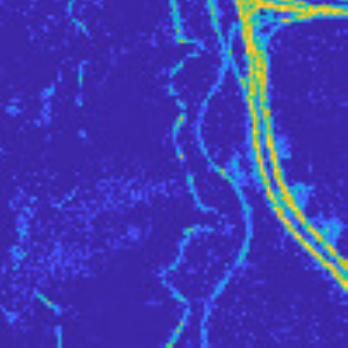}
            }
        \end{minipage}
        \begin{minipage}[t]{0.075\hsize}
            \centerline{
            \includegraphics[height = 40pt]{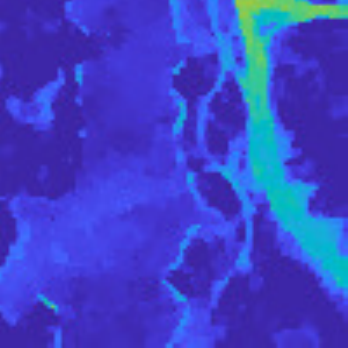}
            }
        \end{minipage}
        \begin{minipage}[t]{0.075\hsize}
            \centerline{
            \includegraphics[height = 40pt]{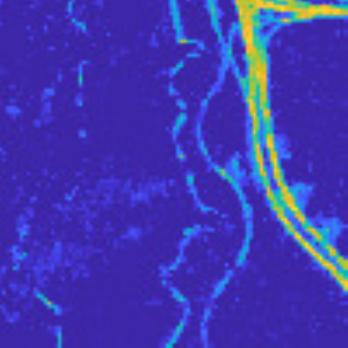}
            }
        \end{minipage}
        \begin{minipage}[t]{0.075\hsize}
            \centerline{
            \includegraphics[height = 40pt]{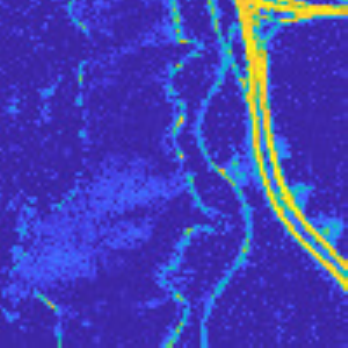}
            }
        \end{minipage}
	    \begin{minipage}[t]{0.075\hsize}
	    	\centerline{
	    		\includegraphics[height = 40pt]{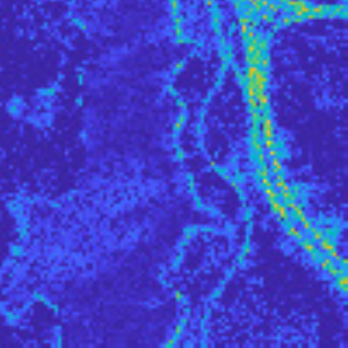}
	    	}
	    \end{minipage}
	    \begin{minipage}[t]{0.075\hsize}
	    \centerline{
	    	\includegraphics[height = 40pt]{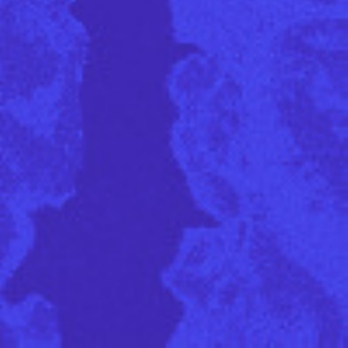}
	    }
	    \end{minipage}
	    \begin{minipage}[t]{0.075\hsize}
	    \centerline{
	    	\includegraphics[height = 40pt]{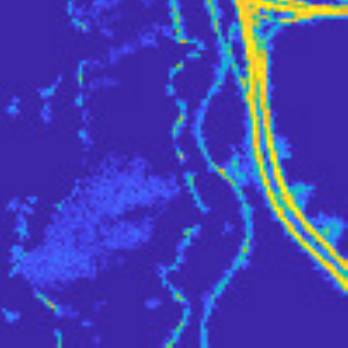}
	    }
	    \end{minipage}
	    \begin{minipage}[t]{0.075\hsize}
		    \centerline{
		    	\includegraphics[height = 40pt]{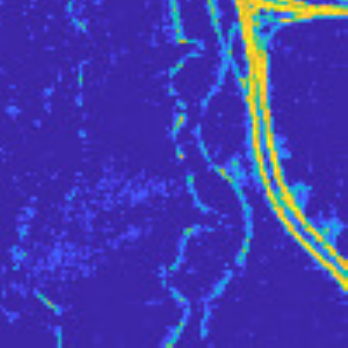}
		    }
	    \end{minipage}
        \begin{minipage}[t]{0.075\hsize}
            \centerline{
            \includegraphics[height = 40pt]{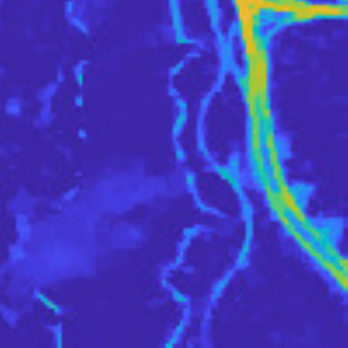}
            }
        \end{minipage}
        \begin{minipage}[t]{0.075\hsize}
            \centerline{
            \includegraphics[height = 40pt]{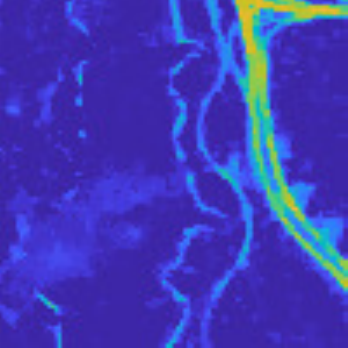}
            }
        \end{minipage}
        \begin{minipage}[t]{0.075\hsize}
            \centerline{
            \includegraphics[height = 40pt]{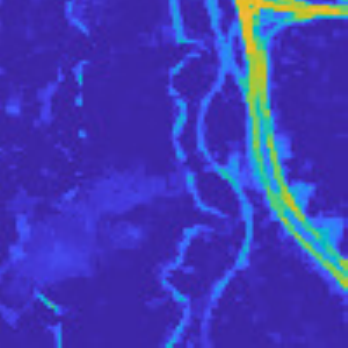}
            }
        \end{minipage}
	    \begin{minipage}[t]{0.02\hsize}
	    	\centerline{
	    		\includegraphics[height = 40pt]{./fig/colorbar_20.png}
	    	}
	    \end{minipage}
        
        \begin{minipage}[t]{0.075\hsize}
            \centerline{
            (a)
            }
        \end{minipage}
        \begin{minipage}[t]{0.075\hsize}
            \centerline{
            (b)
            }
        \end{minipage}
        \begin{minipage}[t]{0.075\hsize}
            \centerline{
            (c)
            }
        \end{minipage}
        \begin{minipage}[t]{0.075\hsize}
            \centerline{
            (d)
            }
        \end{minipage}
        \begin{minipage}[t]{0.075\hsize}
            \centerline{
            (e)
            }
        \end{minipage}
        \begin{minipage}[t]{0.075\hsize}
            \centerline{
            (f)
            }
        \end{minipage}
        \begin{minipage}[t]{0.075\hsize}
            \centerline{
            (g)
            }
        \end{minipage}
        \begin{minipage}[t]{0.075\hsize}
            \centerline{
            {(h)}
            }
        \end{minipage}
	    \begin{minipage}[t]{0.075\hsize}
	    	\centerline{
	    		{(i)}
	    	}
	    \end{minipage}
	    \begin{minipage}[t]{0.075\hsize}
	    \centerline{
	    	\textbf{(j)}
	    }
	    \end{minipage}
	    \begin{minipage}[t]{0.075\hsize}
	    \centerline{
	    	\textbf{(k)}
	    }
	    \end{minipage}
	    \begin{minipage}[t]{0.075\hsize}
	    \centerline{
	    	\textbf{(l)}
	    }
	    \end{minipage}
	    \begin{minipage}[t]{0.02\hsize}
	    	\centerline{
	    		~
	    	}
	    \end{minipage}

\caption{Unmixing results of abundance maps for the \textit{Jasper Ridge} experiments in Case 2. (a): Original abundance maps. (b): CLSUnSAL~\cite{iordache2014collaborative}. (c): JSTV~\cite{aggarwal2016hyperspectral}. (d): RSSUn-TV~\cite{wang2019row}. (e): LGSU~\cite{shen2022superpixel}. (f): UnDIP~\cite{UnDIP_RastiB_2022}. (g): EGU-Net~\cite{hong2022endmember}. (h): RDSWSU~\cite{rs_Deng_RobustDual_2023}. (i): MdLRR~\cite{MDLRR_WuLing_2023}. (j): \textbf{\Ourss (HTV)}. (k): \textbf{\Ourss (SSTV)}. (l): \textbf{\Ourss (HSSTV)}.}
\label{real_0.1_0_0}
\end{figure*}

\begin{figure*}[t]
	\centering
	\begin{minipage}[t]{0.075\hsize}
		\centerline{
			\includegraphics[height = 40pt]{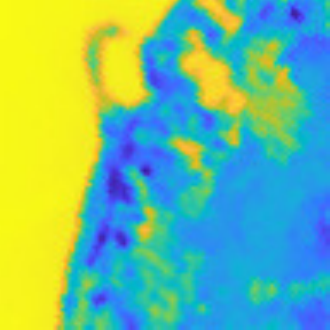}
		}
	\end{minipage}
	\begin{minipage}[t]{0.075\hsize}
		\centerline{
			\includegraphics[height = 40pt]{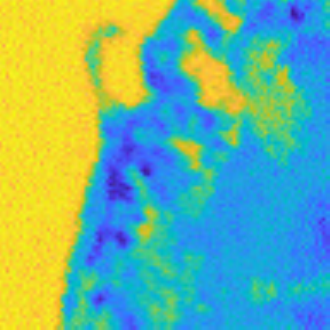}
		}
	\end{minipage}
	\begin{minipage}[t]{0.075\hsize}
		\centerline{
			\includegraphics[height = 40pt]{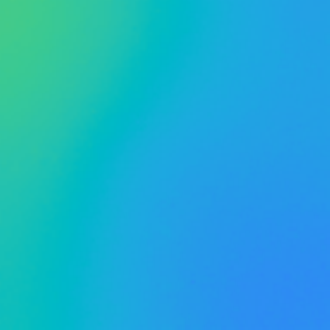}
		}
	\end{minipage}
	\begin{minipage}[t]{0.075\hsize}
		\centerline{
			\includegraphics[height = 40pt]{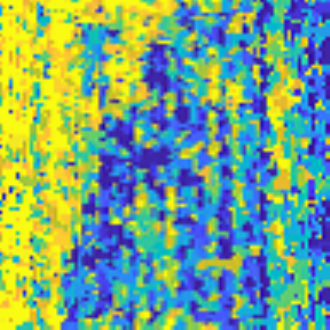}
		}
	\end{minipage}
	\begin{minipage}[t]{0.075\hsize}
		\centerline{
			\includegraphics[height = 40pt]{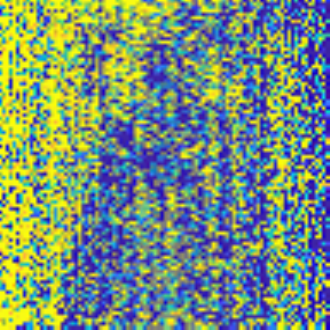}
		}
	\end{minipage}
	\begin{minipage}[t]{0.075\hsize}
		\centerline{
			\includegraphics[height = 40pt]{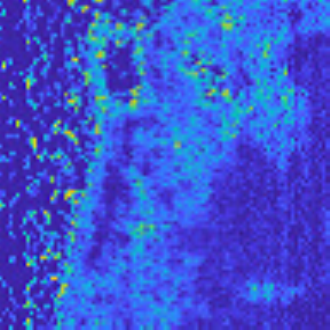}
		}
	\end{minipage}
	\begin{minipage}[t]{0.075\hsize}
	\centerline{
		\includegraphics[height = 40pt]{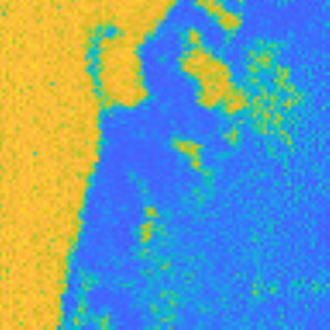}
	}
	\end{minipage}
	\begin{minipage}[t]{0.075\hsize}
	\centerline{
		\includegraphics[height = 40pt]{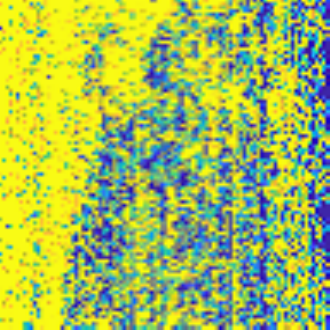}
	}
	\end{minipage}
	\begin{minipage}[t]{0.075\hsize}
		\centerline{
			\includegraphics[height = 40pt]{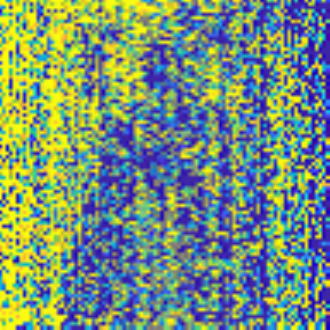}
		}
	\end{minipage}
	\begin{minipage}[t]{0.075\hsize}
		\centerline{
			\includegraphics[height = 40pt]{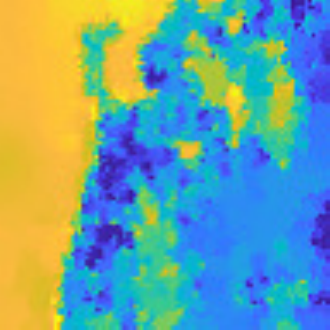}
		}
	\end{minipage}
	\begin{minipage}[t]{0.075\hsize}
		\centerline{
			\includegraphics[height = 40pt]{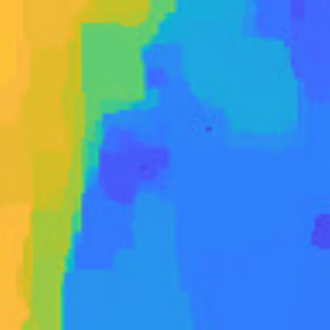}
		}
	\end{minipage}
	\begin{minipage}[t]{0.075\hsize}
		\centerline{
			\includegraphics[height = 40pt]{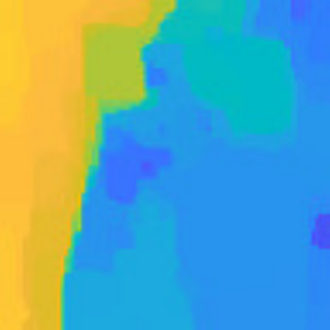}
		}
	\end{minipage}
	\begin{minipage}[t]{0.02\hsize}
		\centerline{
			\includegraphics[height = 40pt]{./fig/colorbar_20.png}
		}
	\end{minipage}
	
	\vspace{1mm}
	
	\begin{minipage}[t]{0.075\hsize}
		\centerline{
			\includegraphics[height = 40pt]{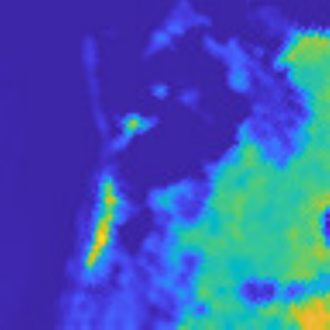}
		}
	\end{minipage}
	\begin{minipage}[t]{0.075\hsize}
		\centerline{
			\includegraphics[height = 40pt]{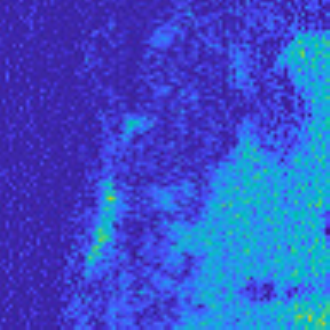}
		}
	\end{minipage}
	\begin{minipage}[t]{0.075\hsize}
		\centerline{
			\includegraphics[height = 40pt]{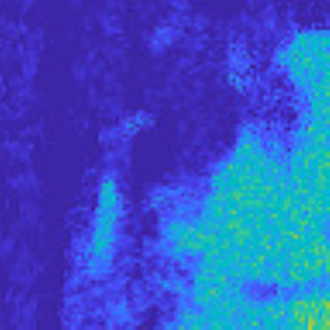}
		}
	\end{minipage}
	\begin{minipage}[t]{0.075\hsize}
		\centerline{
			\includegraphics[height = 40pt]{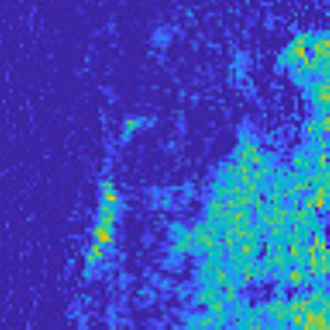}
		}
	\end{minipage}
	\begin{minipage}[t]{0.075\hsize}
		\centerline{
			\includegraphics[height = 40pt]{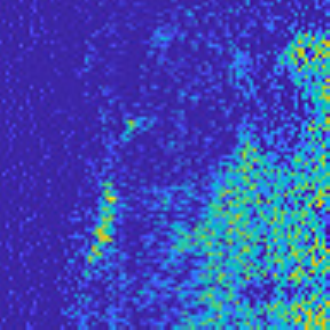}
		}
	\end{minipage}
	\begin{minipage}[t]{0.075\hsize}
		\centerline{
			\includegraphics[height = 40pt]{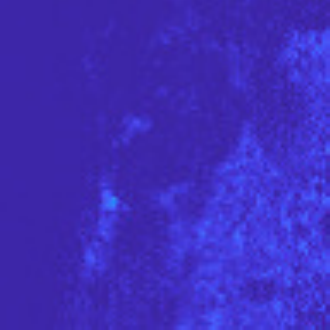}
		}
	\end{minipage}
	\begin{minipage}[t]{0.075\hsize}
	\centerline{
		\includegraphics[height = 40pt]{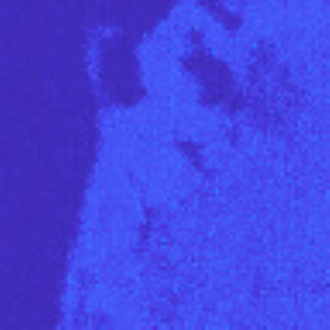}
	}
	\end{minipage}
	\begin{minipage}[t]{0.075\hsize}
	\centerline{
		\includegraphics[height = 40pt]{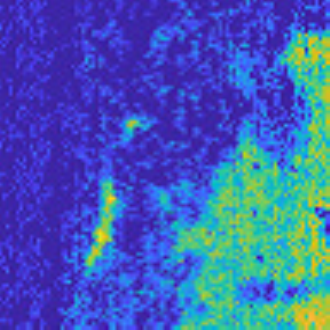}
	}
	\end{minipage}
	\begin{minipage}[t]{0.075\hsize}
		\centerline{
			\includegraphics[height = 40pt]{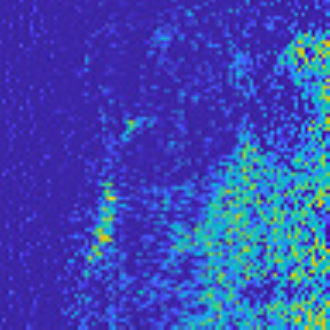}
		}
	\end{minipage}
	\begin{minipage}[t]{0.075\hsize}
		\centerline{
			\includegraphics[height = 40pt]{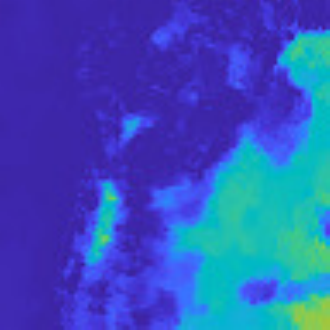}
		}
	\end{minipage}
	\begin{minipage}[t]{0.075\hsize}
		\centerline{
			\includegraphics[height = 40pt]{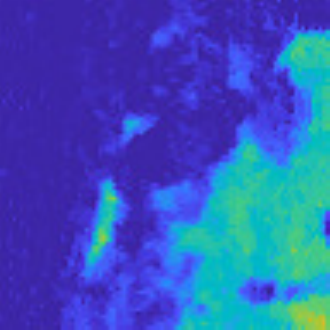}
		}
	\end{minipage}
	\begin{minipage}[t]{0.075\hsize}
		\centerline{
			\includegraphics[height = 40pt]{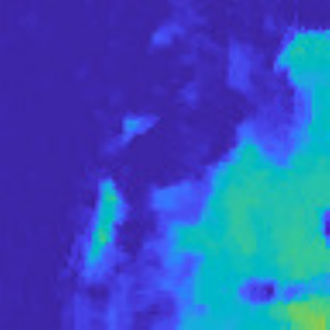}
		}
	\end{minipage}
	\begin{minipage}[t]{0.02\hsize}
		\centerline{
			\includegraphics[height = 40pt]{./fig/colorbar_20.png}
		}
	\end{minipage}
	
	\vspace{1mm}
	
	\begin{minipage}[t]{0.075\hsize}
		\centerline{
			\includegraphics[height = 40pt]{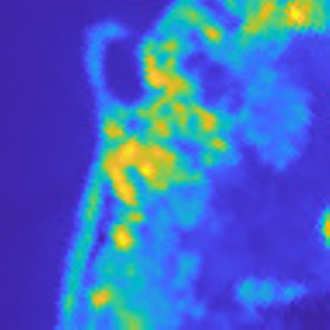}
		}
	\end{minipage}
	\begin{minipage}[t]{0.075\hsize}
		\centerline{
			\includegraphics[height = 40pt]{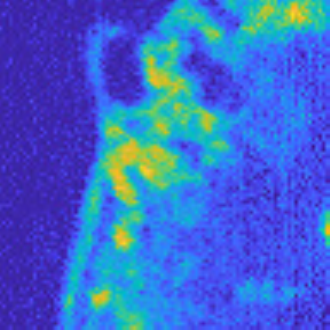}
		}
	\end{minipage}
	\begin{minipage}[t]{0.075\hsize}
		\centerline{
			\includegraphics[height = 40pt]{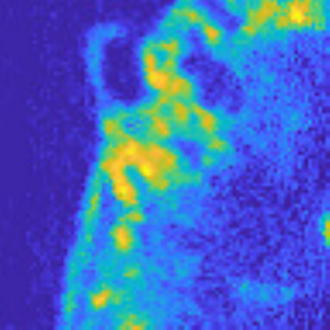}
		}
	\end{minipage}
	\begin{minipage}[t]{0.075\hsize}
		\centerline{
			\includegraphics[height = 40pt]{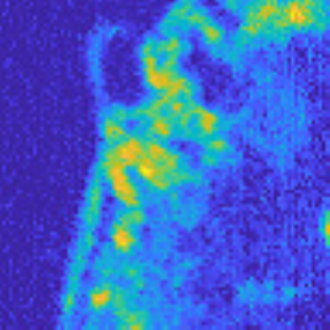}
		}
	\end{minipage}
	\begin{minipage}[t]{0.075\hsize}
		\centerline{
			\includegraphics[height = 40pt]{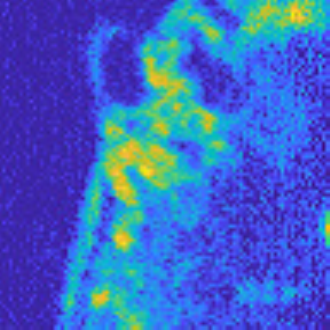}
		}
	\end{minipage}
	\begin{minipage}[t]{0.075\hsize}
		\centerline{
			\includegraphics[height = 40pt]{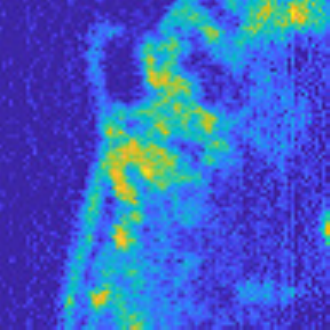}
		}
	\end{minipage}
	\begin{minipage}[t]{0.075\hsize}
	\centerline{
		\includegraphics[height = 40pt]{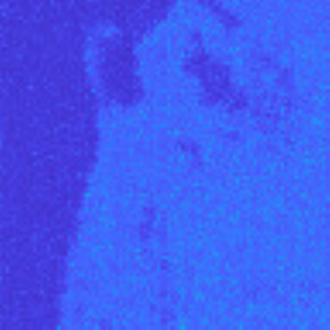}
	}
	\end{minipage}
	\begin{minipage}[t]{0.075\hsize}
	\centerline{
		\includegraphics[height = 40pt]{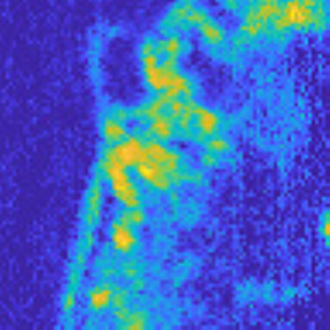}
	}
	\end{minipage}
	\begin{minipage}[t]{0.075\hsize}
		\centerline{
			\includegraphics[height = 40pt]{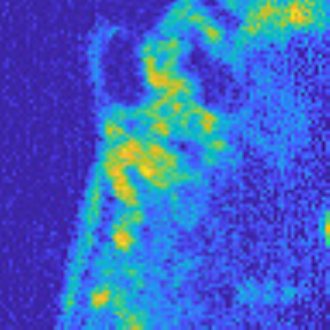}
		}
	\end{minipage}
	\begin{minipage}[t]{0.075\hsize}
		\centerline{
			\includegraphics[height = 40pt]{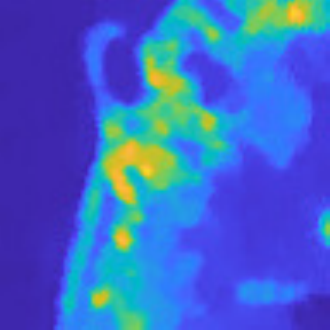}
		}
	\end{minipage}
	\begin{minipage}[t]{0.075\hsize}
		\centerline{
			\includegraphics[height = 40pt]{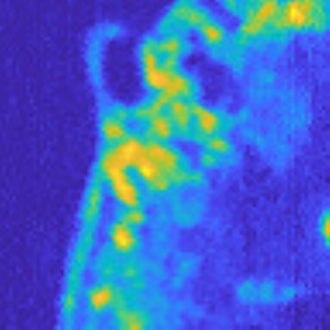}
		}
	\end{minipage}
	\begin{minipage}[t]{0.075\hsize}
		\centerline{
			\includegraphics[height = 40pt]{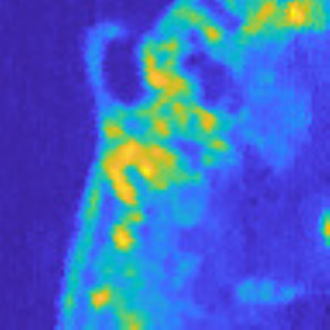}
		}
	\end{minipage}
	\begin{minipage}[t]{0.02\hsize}
		\centerline{
			\includegraphics[height = 40pt]{./fig/colorbar_20.png}
		}
	\end{minipage}
	
	\begin{minipage}[t]{0.075\hsize}
		\centerline{
			(a)
		}
	\end{minipage}
	\begin{minipage}[t]{0.075\hsize}
		\centerline{
			(b)
		}
	\end{minipage}
	\begin{minipage}[t]{0.075\hsize}
		\centerline{
			(c)
		}
	\end{minipage}
	\begin{minipage}[t]{0.075\hsize}
		\centerline{
			(d)
		}
	\end{minipage}
	\begin{minipage}[t]{0.075\hsize}
		\centerline{
			(e)
		}
	\end{minipage}
	\begin{minipage}[t]{0.075\hsize}
		\centerline{
			(f)
		}
	\end{minipage}
	\begin{minipage}[t]{0.075\hsize}
		\centerline{
			(g)
		}
	\end{minipage}
	\begin{minipage}[t]{0.075\hsize}
		\centerline{
			{(h)}
		}
	\end{minipage}
	\begin{minipage}[t]{0.075\hsize}
		\centerline{
			{(i)}
		}
	\end{minipage}
	\begin{minipage}[t]{0.075\hsize}
		\centerline{
			\textbf{(j)}
		}
	\end{minipage}
	\begin{minipage}[t]{0.075\hsize}
	\centerline{
		\textbf{(k)}
	}
	\end{minipage}
	\begin{minipage}[t]{0.075\hsize}
	\centerline{
		\textbf{(l)}
	}
	\end{minipage}
	\begin{minipage}[t]{0.02\hsize}
		\centerline{
			~
		}
	\end{minipage}
	
	\caption{Unmixing results of abundance maps for the \textit{Samson} experiments in Case 6. (a): Original abundance maps. (b): CLSUnSAL~\cite{iordache2014collaborative}. (c): JSTV~\cite{aggarwal2016hyperspectral}. (d): RSSUn-TV~\cite{wang2019row}. (e): LGSU~\cite{shen2022superpixel}. (f): UnDIP~\cite{UnDIP_RastiB_2022}. (g): EGU-Net~\cite{hong2022endmember}. (h): RDSWSU~\cite{rs_Deng_RobustDual_2023}. (i): MdLRR~\cite{MDLRR_WuLing_2023}. (j): \textbf{\Ourss (HTV)}. (k): \textbf{\Ourss (SSTV)}. (l): \textbf{\Ourss (HSSTV)}.}
	\label{real_samson_0.1_0.05_0.05}
\end{figure*}

\begin{figure*}[!h]
	\centering
	\begin{minipage}[t]{0.075\hsize}
		\centerline{
			\includegraphics[height = 40pt]{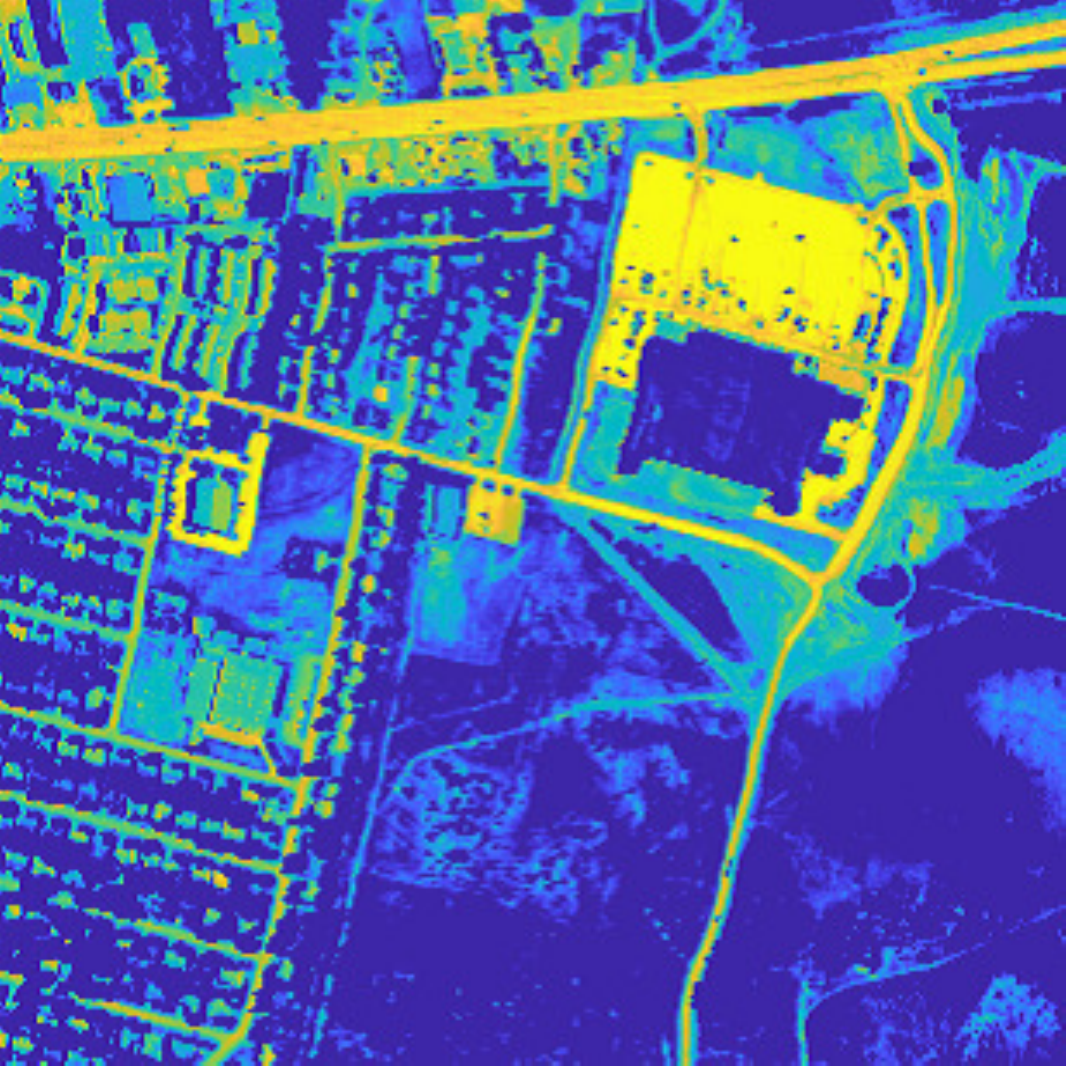}
		}
	\end{minipage}
	\begin{minipage}[t]{0.075\hsize}
		\centerline{
			\includegraphics[height = 40pt]{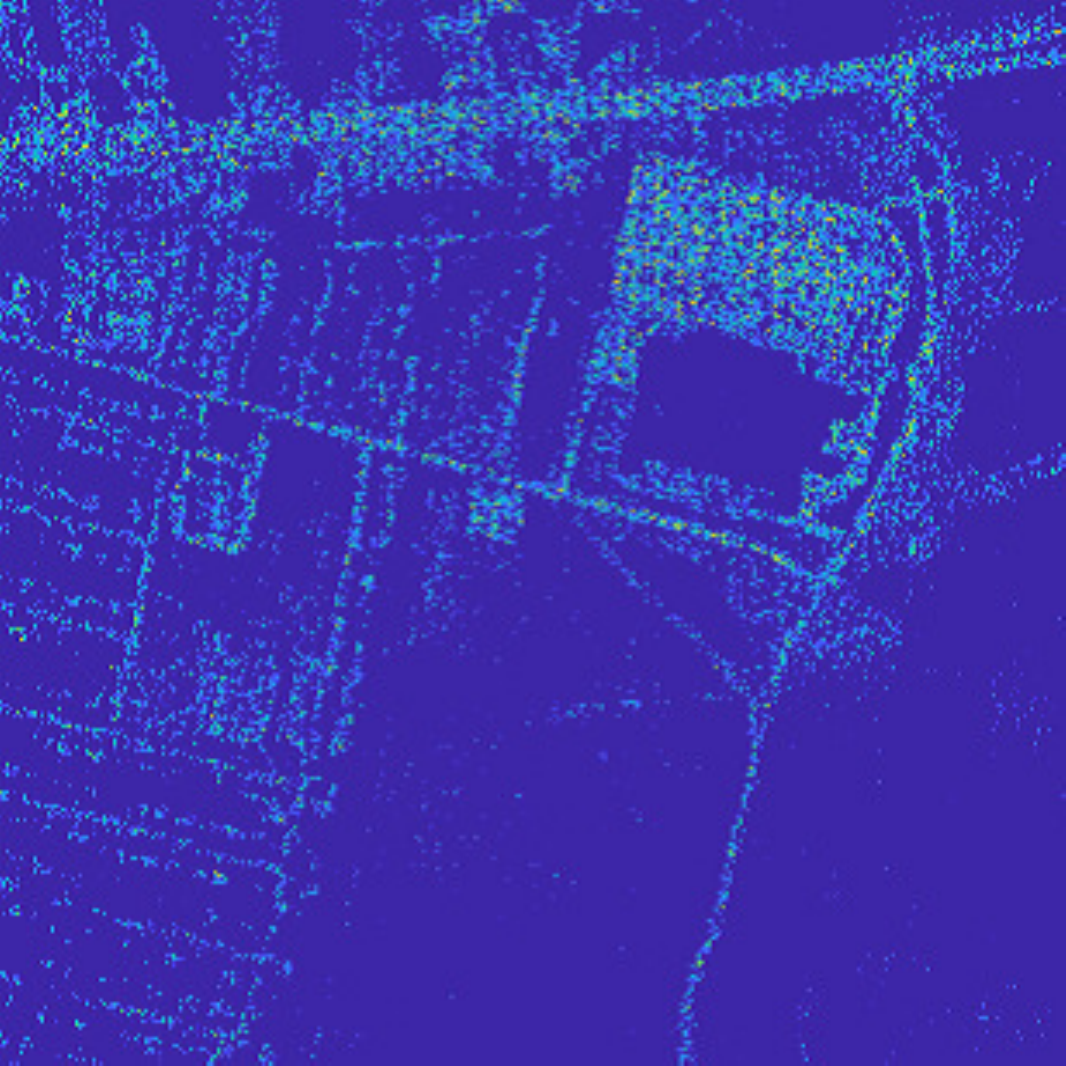}
		}
	\end{minipage}
	\begin{minipage}[t]{0.075\hsize}
		\centerline{
			\includegraphics[height = 40pt]{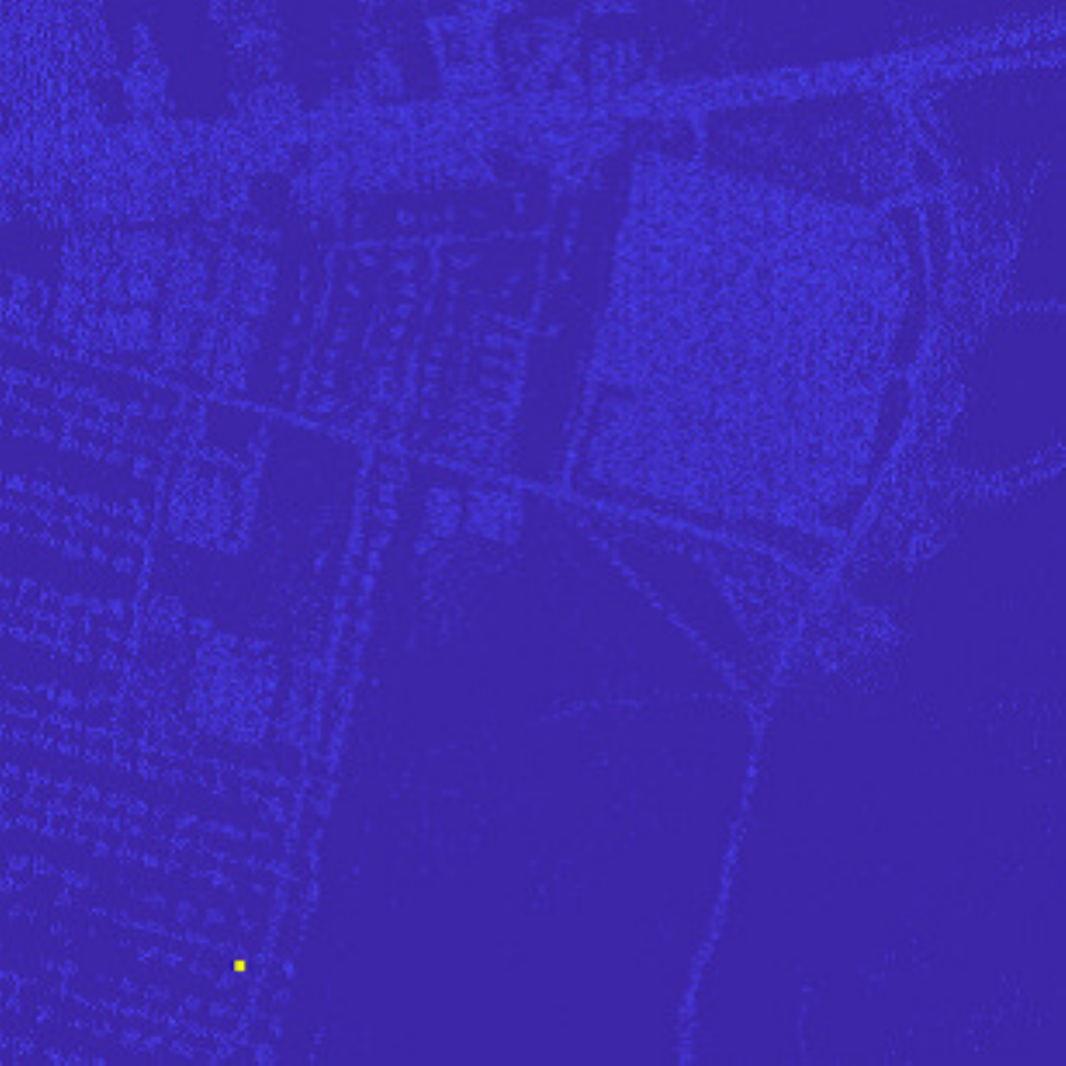}
		}
	\end{minipage}
	\begin{minipage}[t]{0.075\hsize}
		\centerline{
			\includegraphics[height = 40pt]{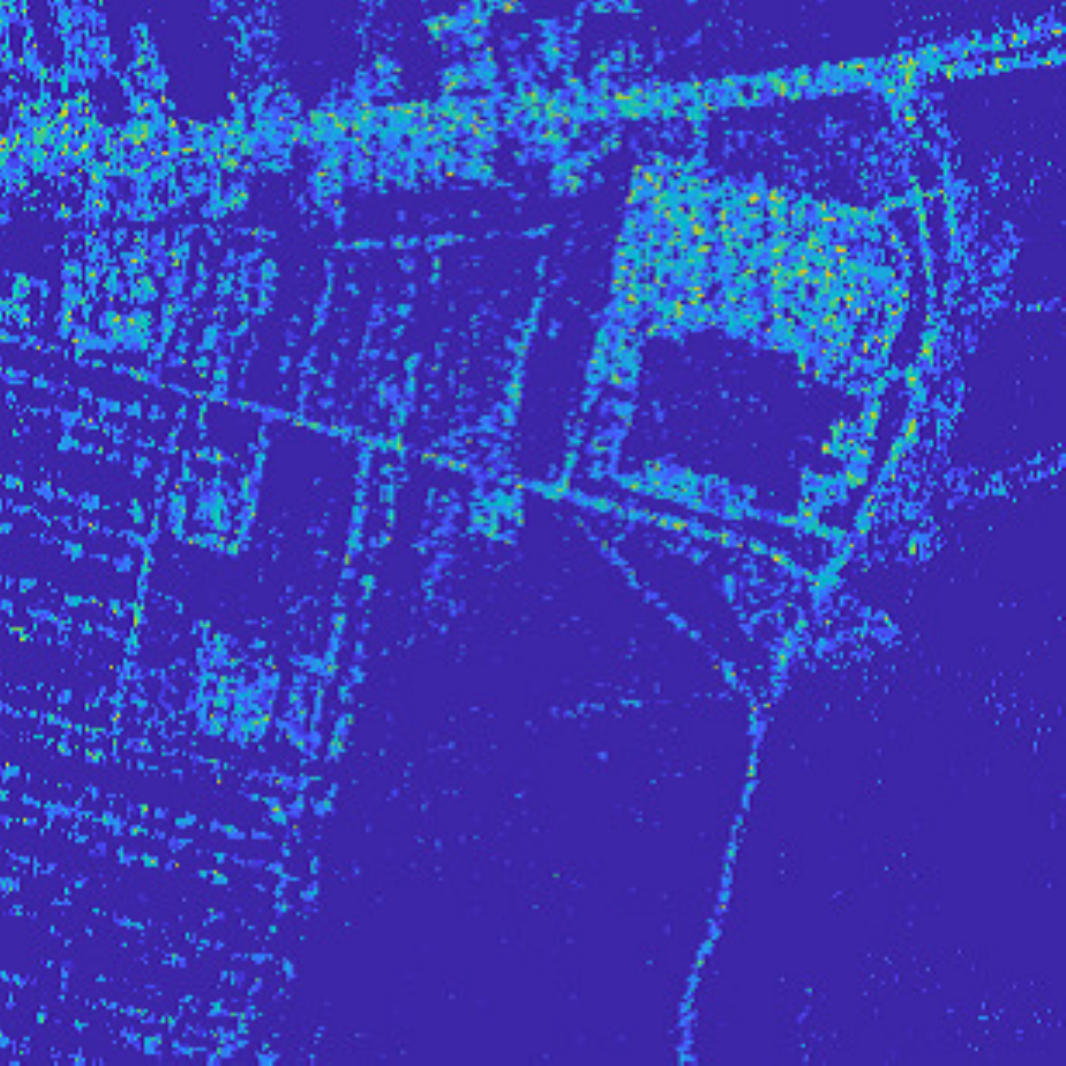}
		}
	\end{minipage}
	\begin{minipage}[t]{0.075\hsize}
		\centerline{
			\includegraphics[height = 40pt]{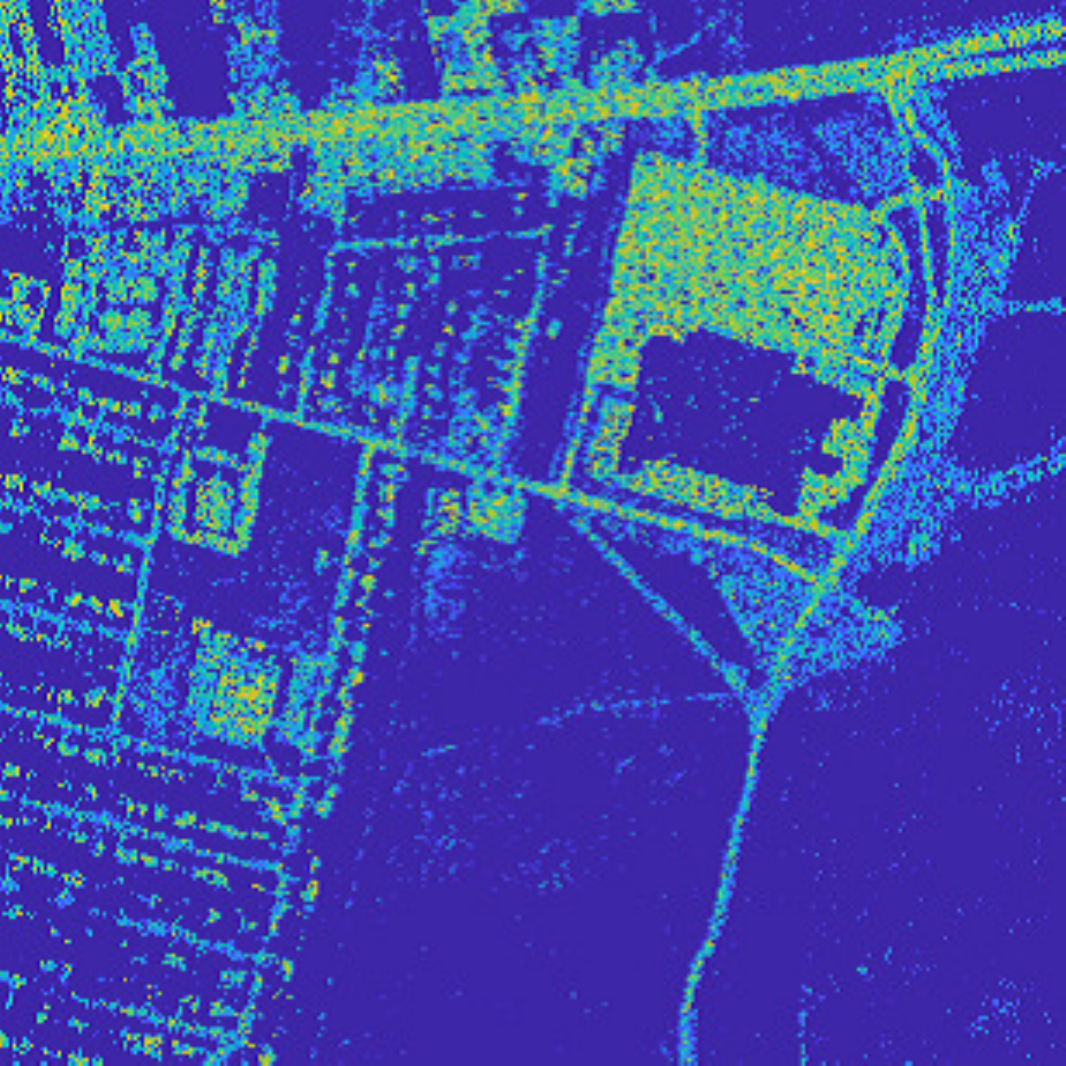}
		}
	\end{minipage}
	\begin{minipage}[t]{0.075\hsize}
		\centerline{
			\includegraphics[height = 40pt]{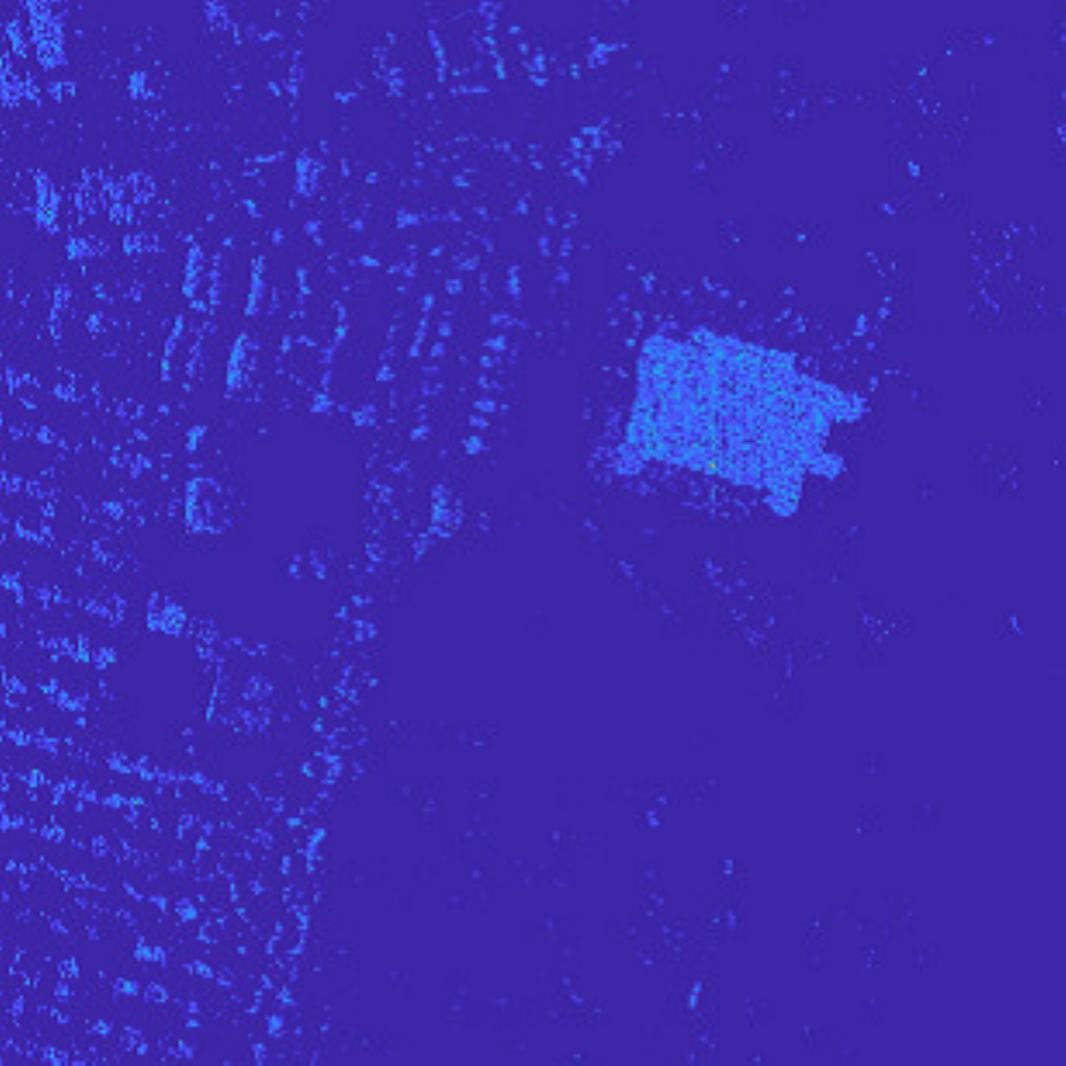}
		}
	\end{minipage}
	\begin{minipage}[t]{0.075\hsize}
		\centerline{
			\includegraphics[height = 40pt]{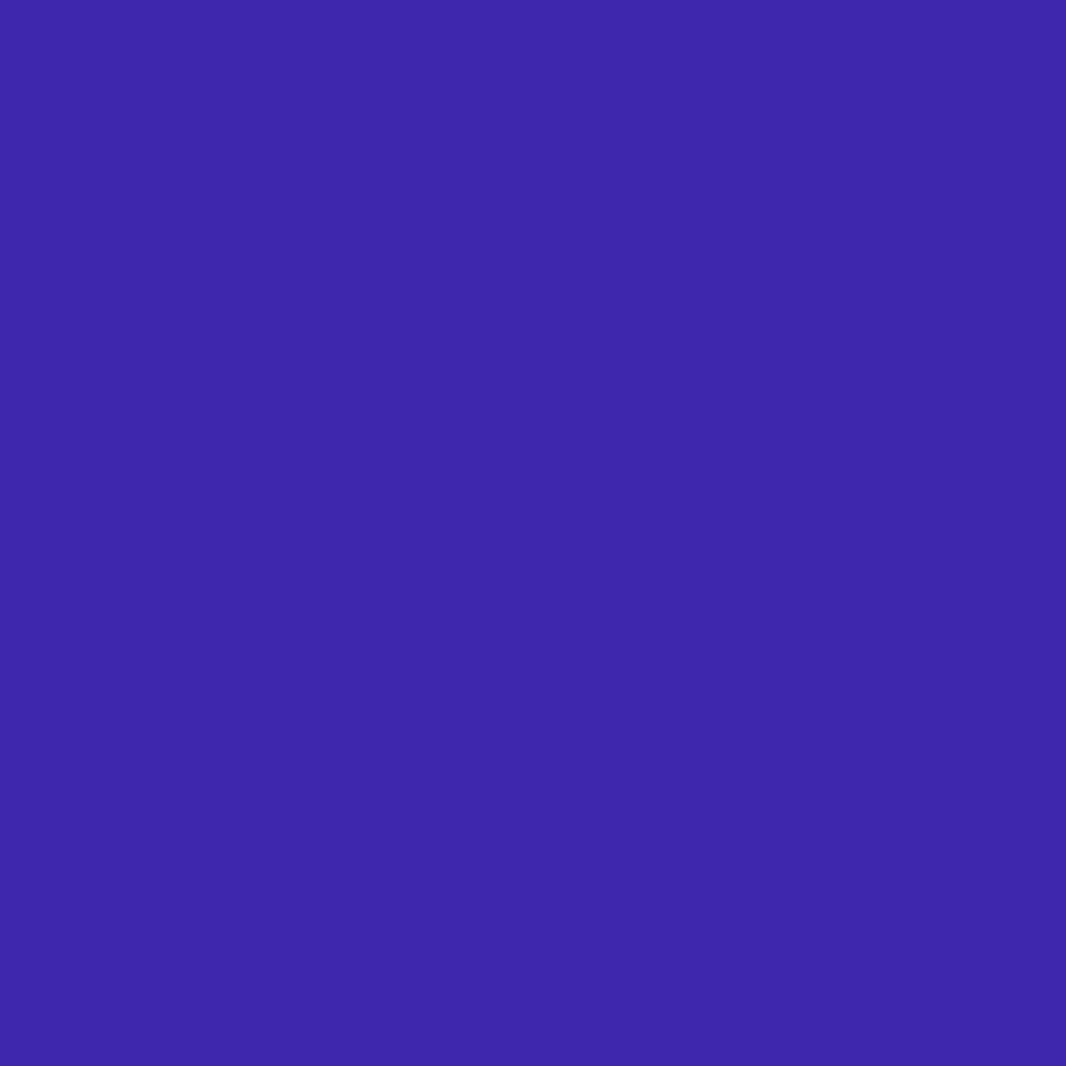}
		}
	\end{minipage}
	\begin{minipage}[t]{0.075\hsize}
		\centerline{
			\includegraphics[height = 40pt]{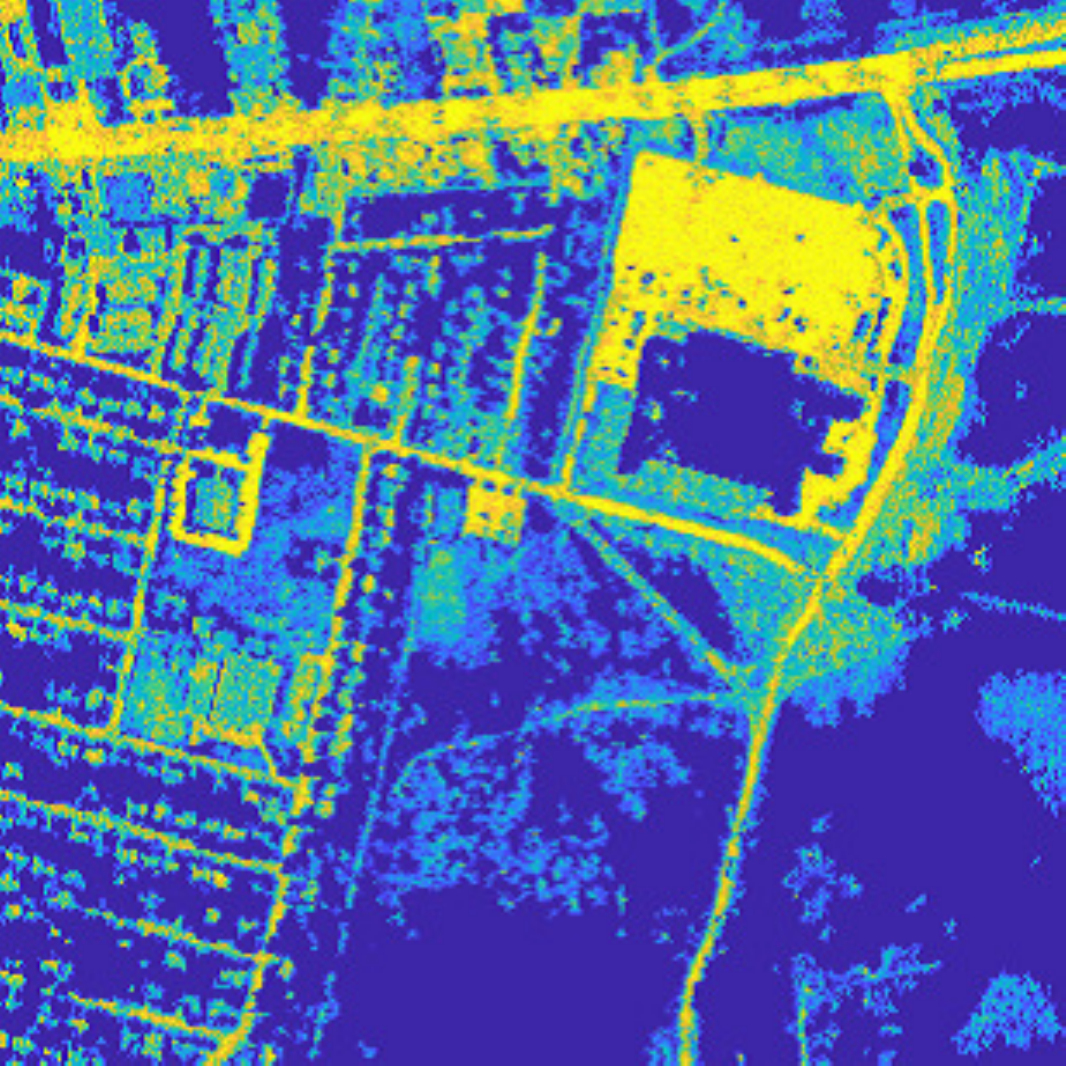}
		}
	\end{minipage}
	\begin{minipage}[t]{0.075\hsize}
		\centerline{
			\includegraphics[height = 40pt]{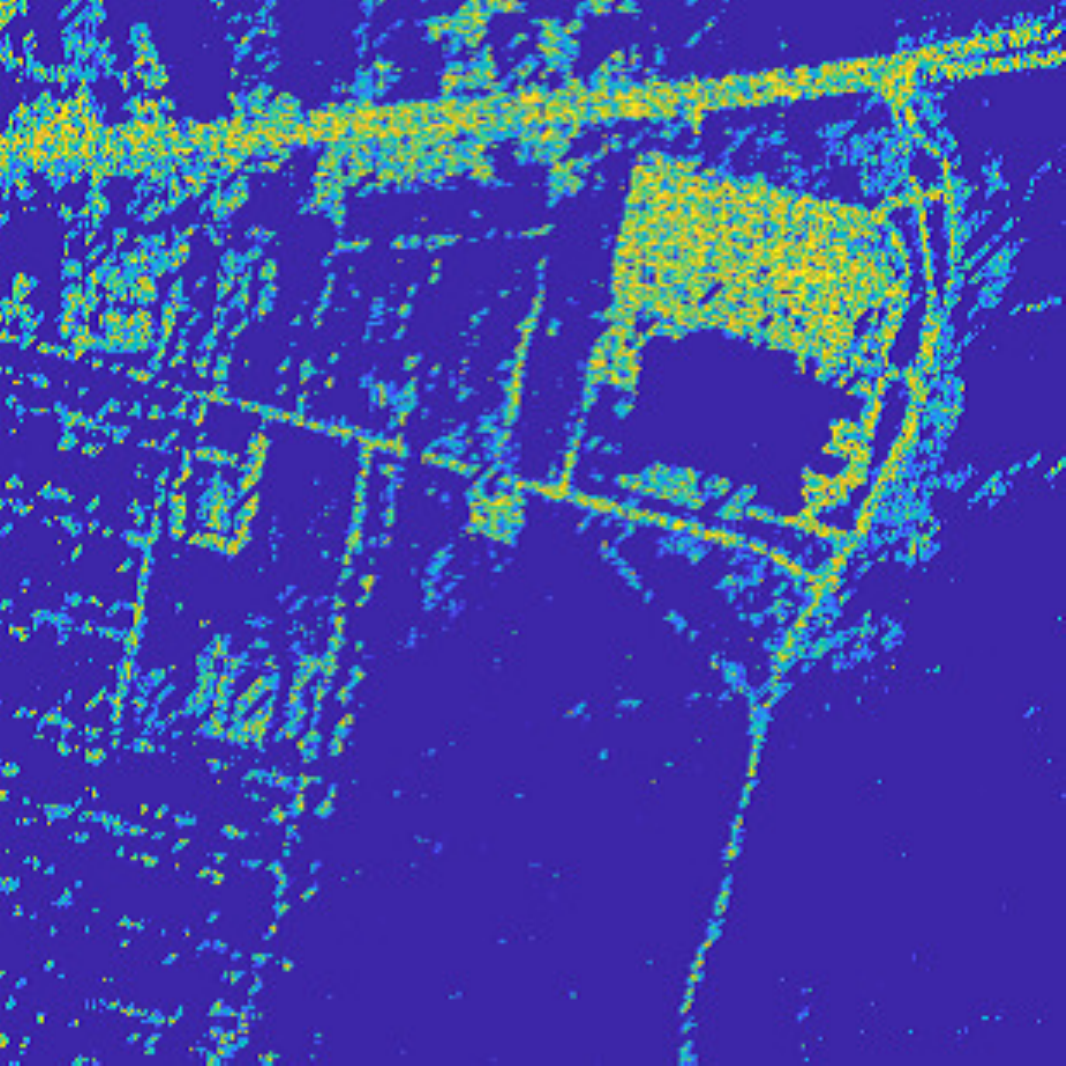}
		}
	\end{minipage}
	\begin{minipage}[t]{0.075\hsize}
		\centerline{
			\includegraphics[height = 40pt]{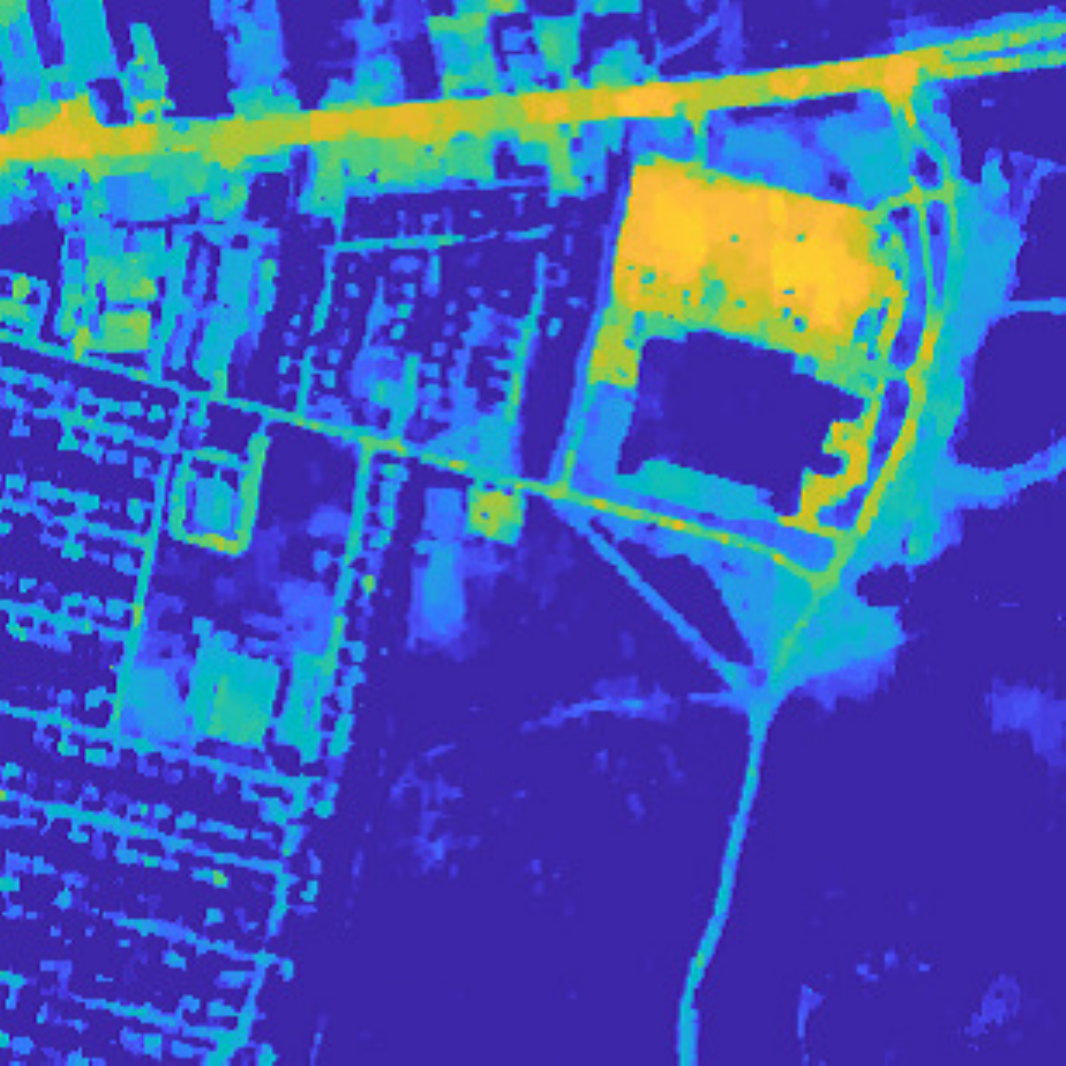}
		}
	\end{minipage}
	\begin{minipage}[t]{0.075\hsize}
		\centerline{
			\includegraphics[height = 40pt]{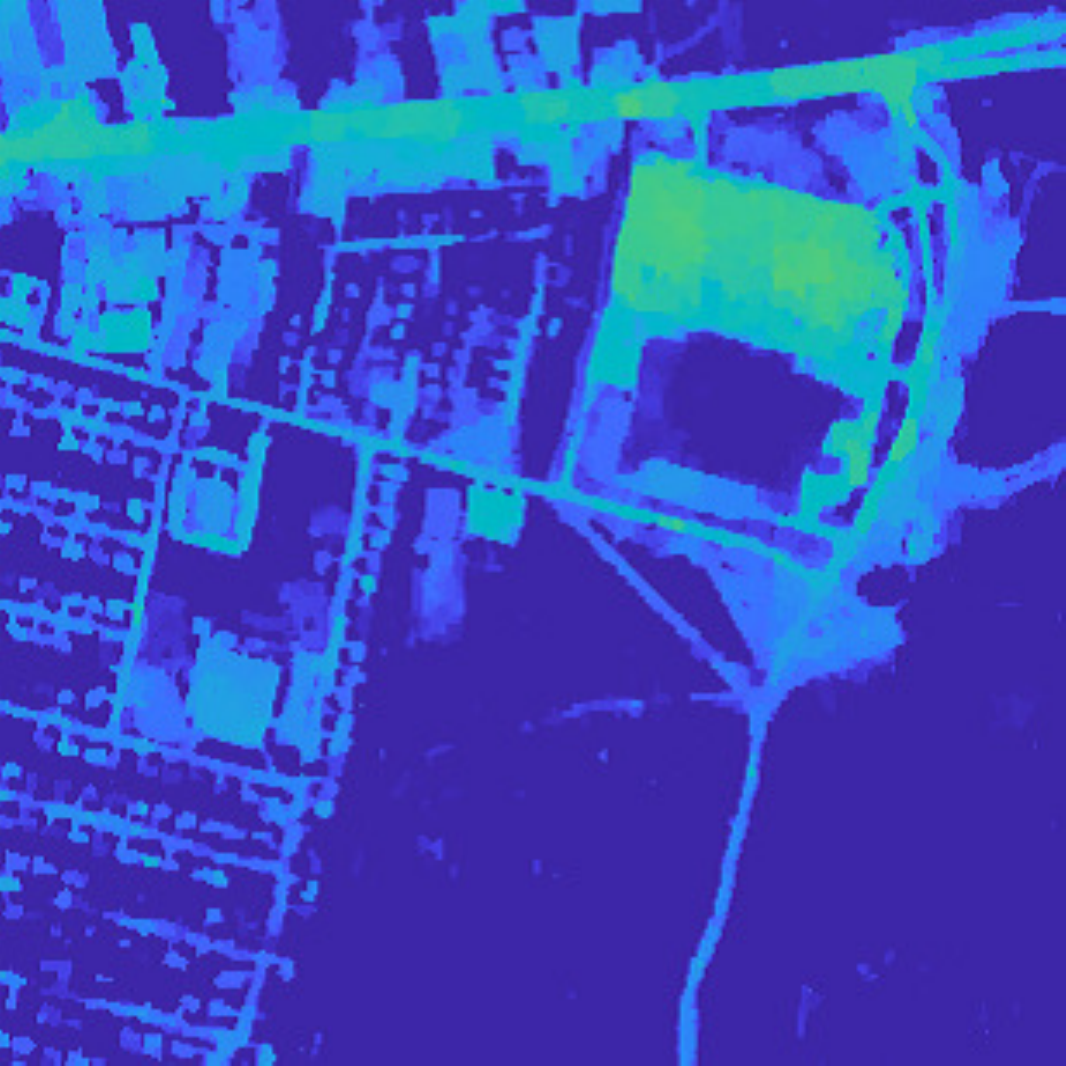}
		}
	\end{minipage}
	\begin{minipage}[t]{0.075\hsize}
		\centerline{
			\includegraphics[height = 40pt]{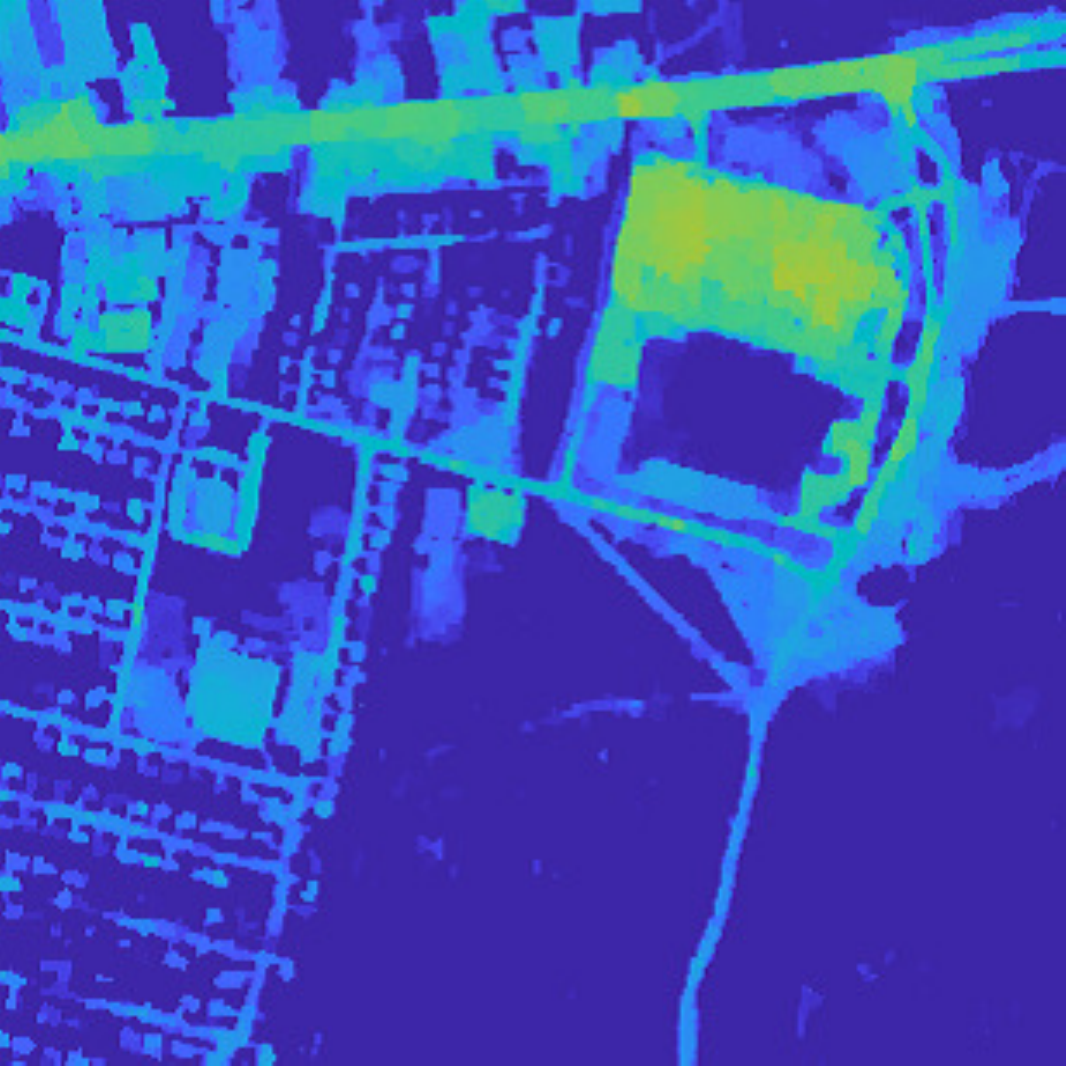}
		}
	\end{minipage}
	\begin{minipage}[t]{0.02\hsize}
		\centerline{
			\includegraphics[height = 40pt]{./fig/colorbar_20.png}
		}
	\end{minipage}
	
	\vspace{1mm}
	
	\begin{minipage}[t]{0.075\hsize}
		\centerline{
			\includegraphics[height = 40pt]{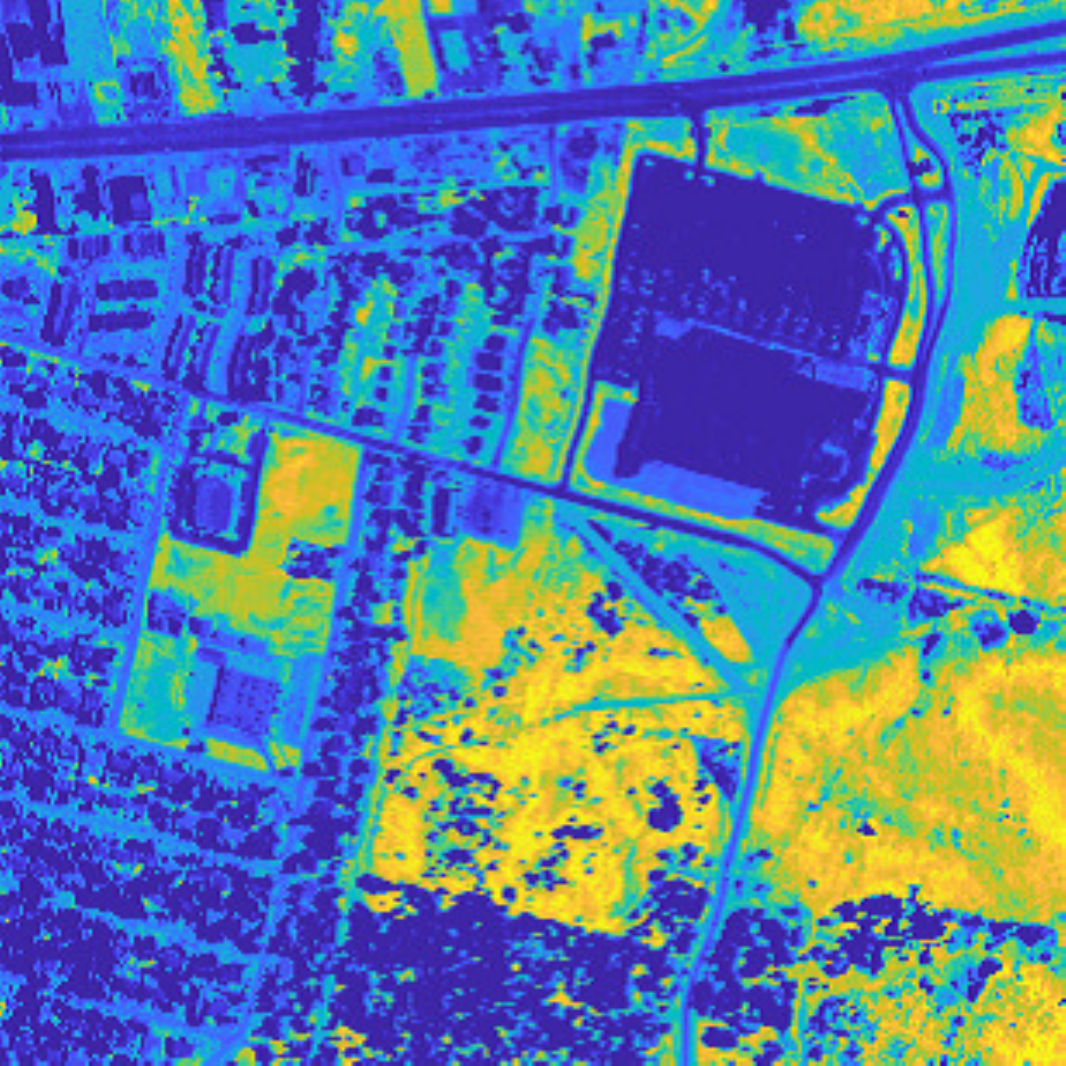}
		}
	\end{minipage}
	\begin{minipage}[t]{0.075\hsize}
		\centerline{
			\includegraphics[height = 40pt]{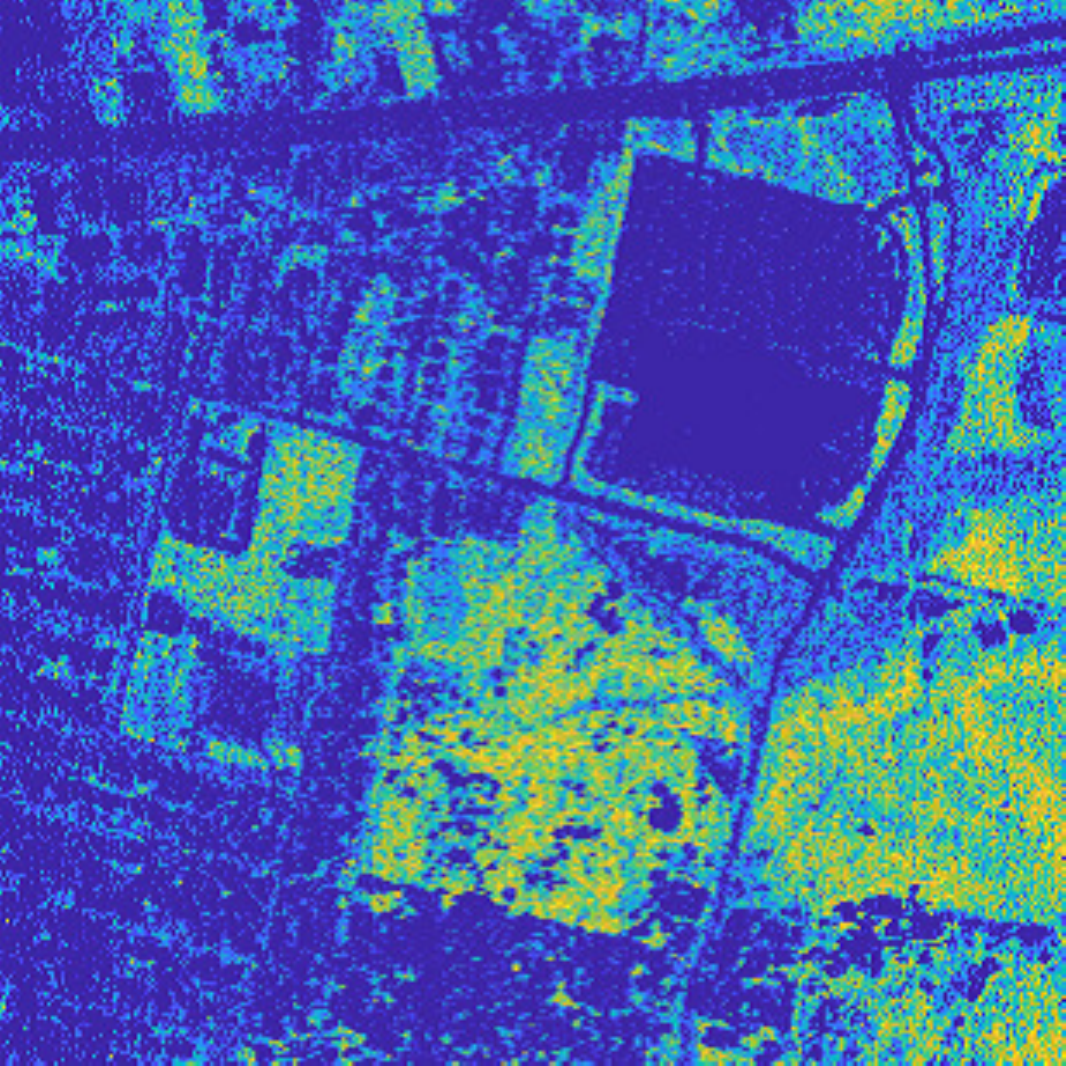}
		}
	\end{minipage}
	\begin{minipage}[t]{0.075\hsize}
		\centerline{
			\includegraphics[height = 40pt]{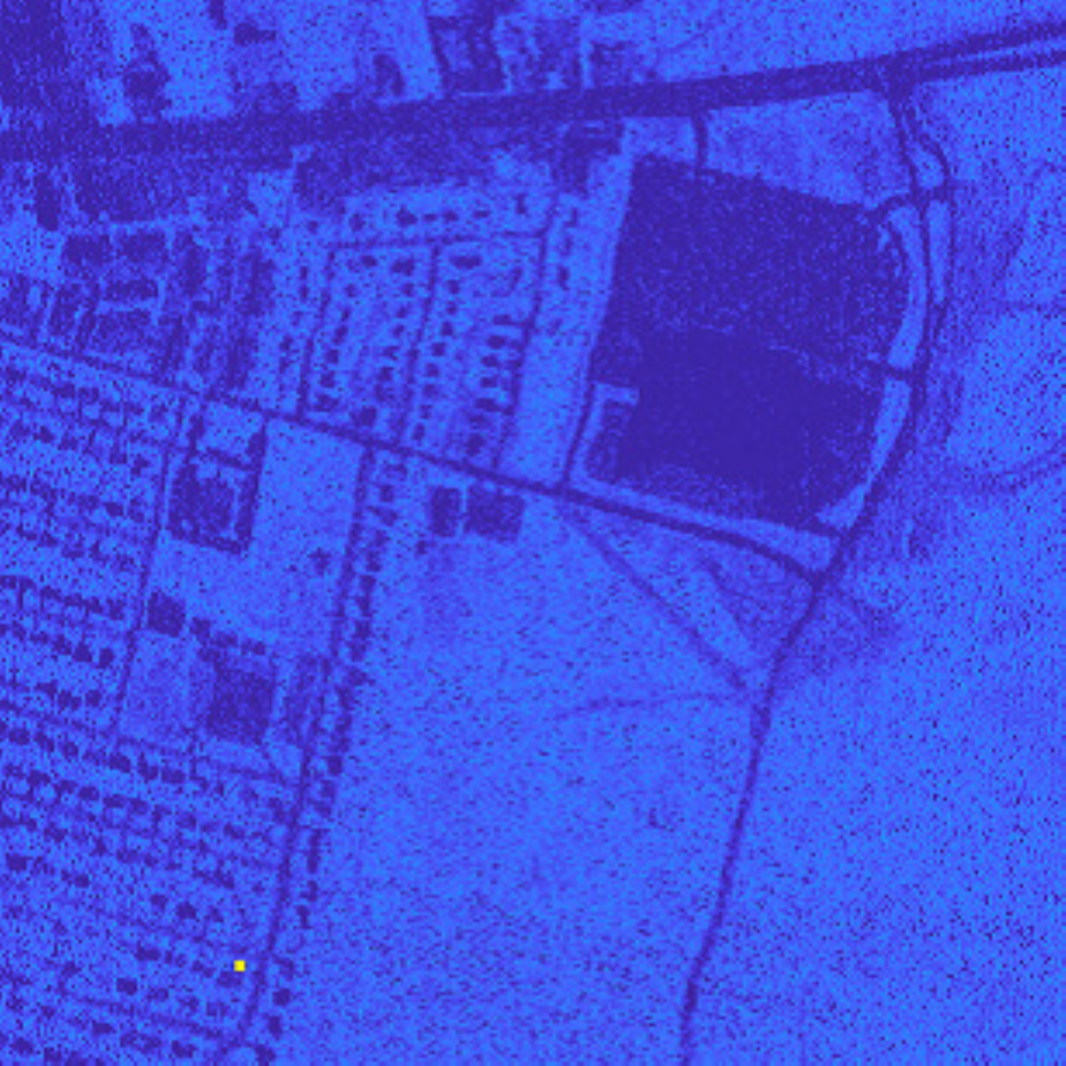}
		}
	\end{minipage}
	\begin{minipage}[t]{0.075\hsize}
		\centerline{
			\includegraphics[height = 40pt]{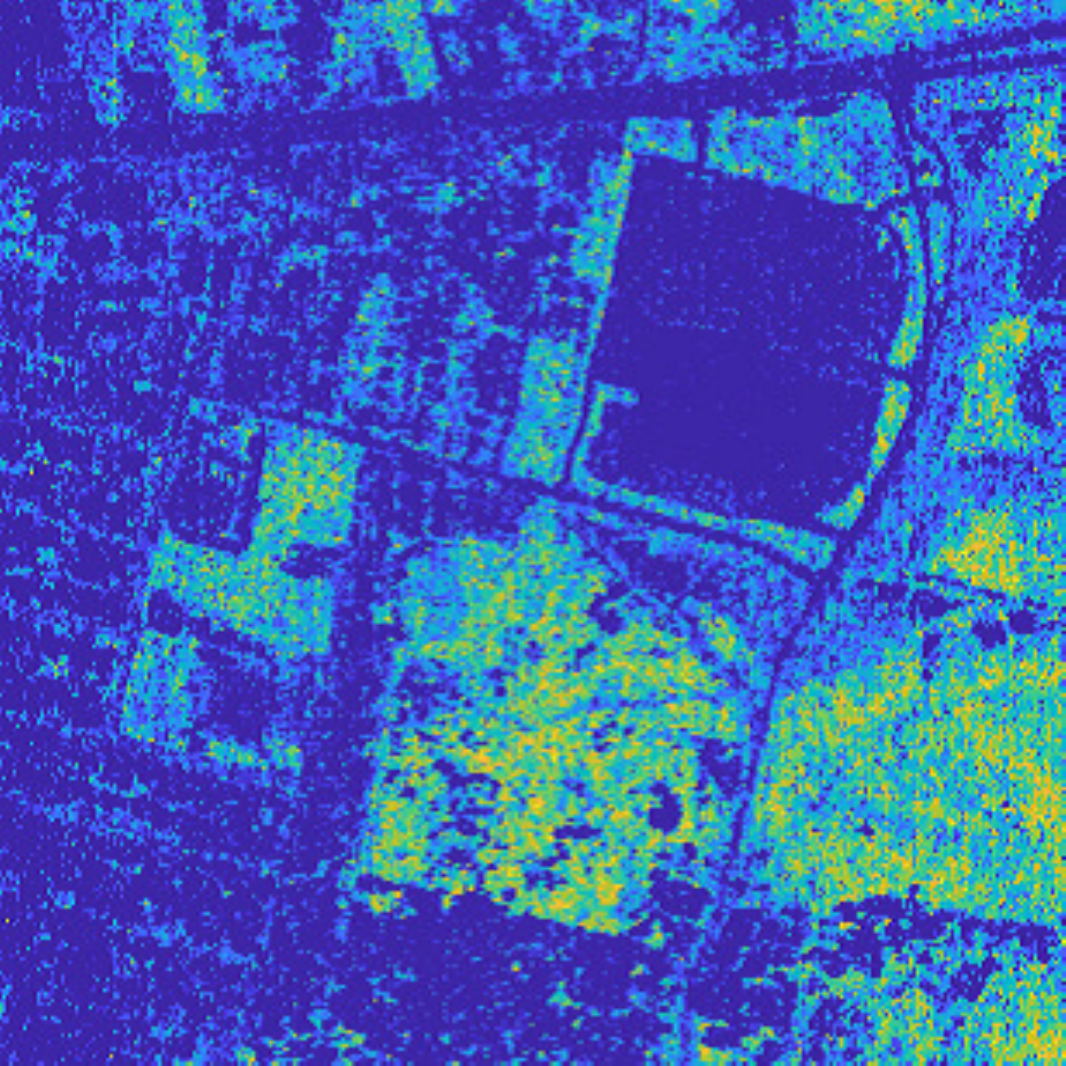}
		}
	\end{minipage}
	\begin{minipage}[t]{0.075\hsize}
		\centerline{
			\includegraphics[height = 40pt]{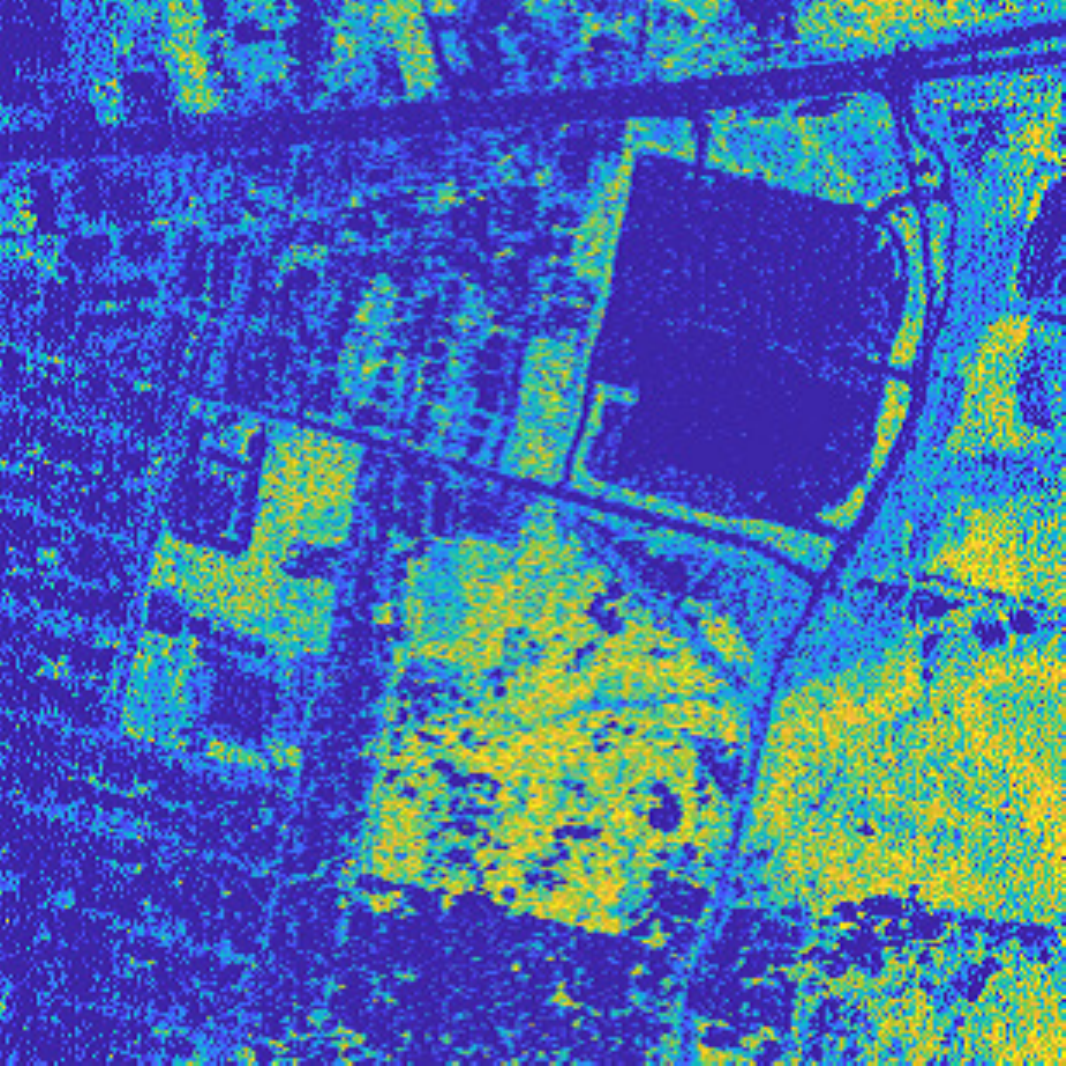}
		}
	\end{minipage}
	\begin{minipage}[t]{0.075\hsize}
		\centerline{
			\includegraphics[height = 40pt]{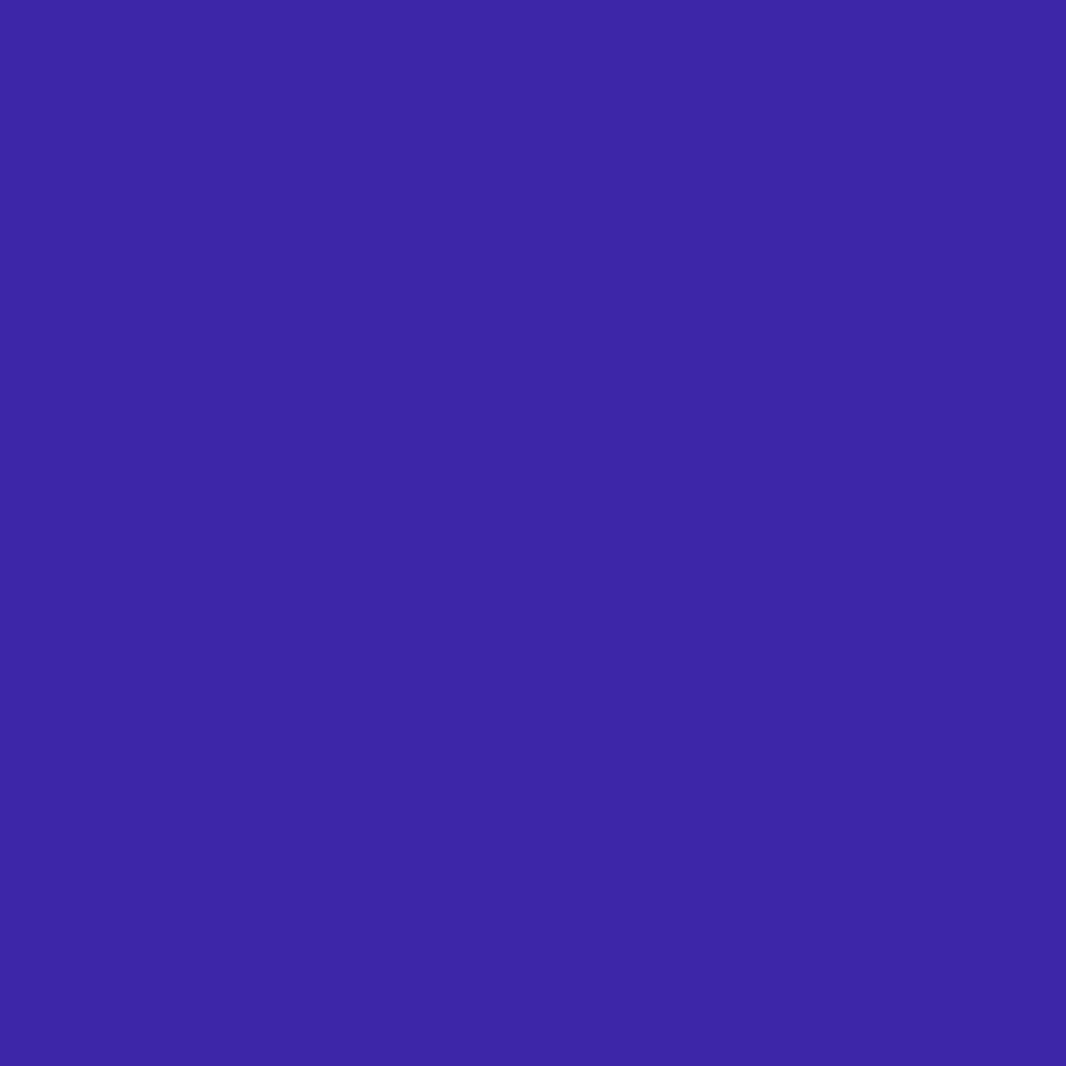}
		}
	\end{minipage}
	\begin{minipage}[t]{0.075\hsize}
		\centerline{
			\includegraphics[height = 40pt]{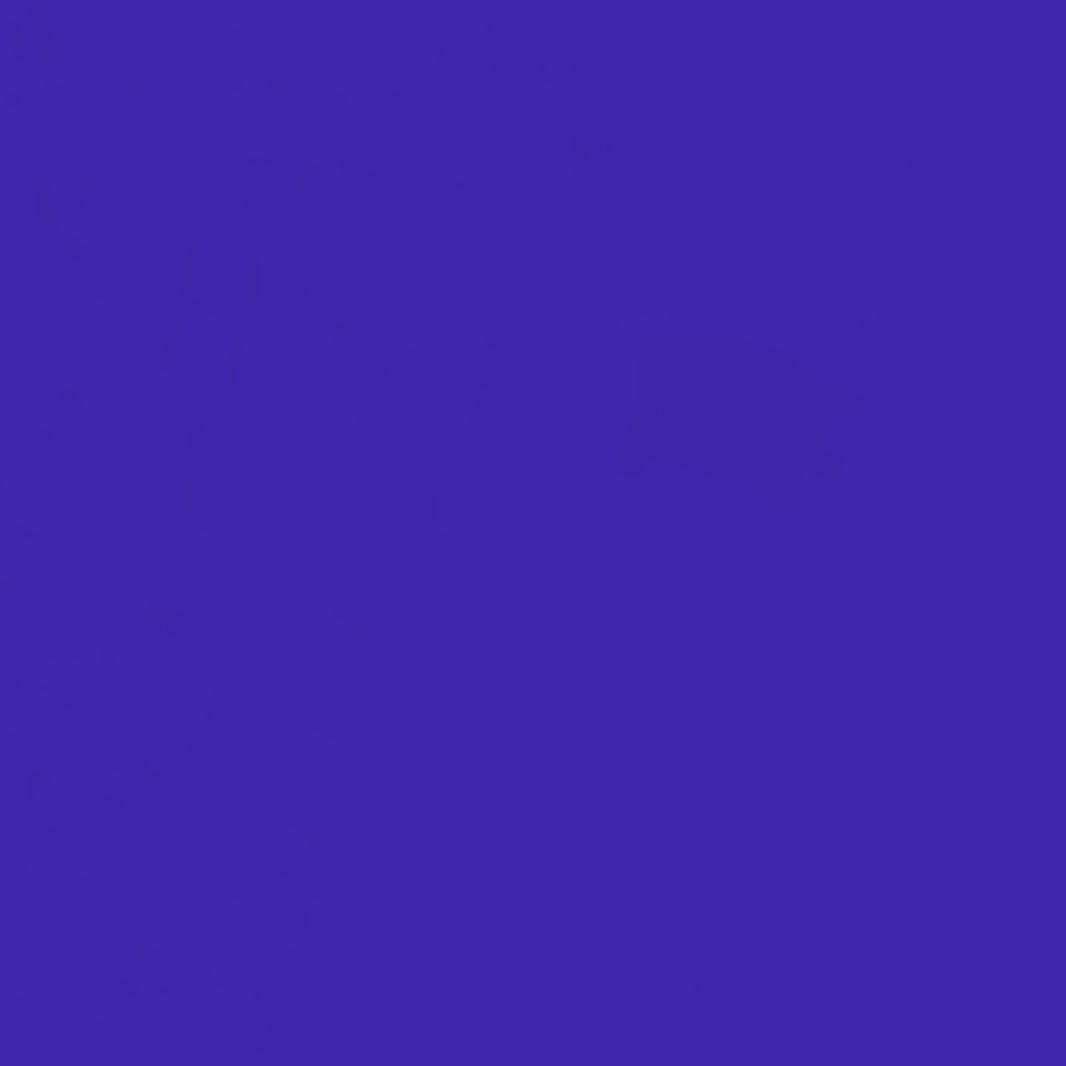}
		}
	\end{minipage}
	\begin{minipage}[t]{0.075\hsize}
		\centerline{
			\includegraphics[height = 40pt]{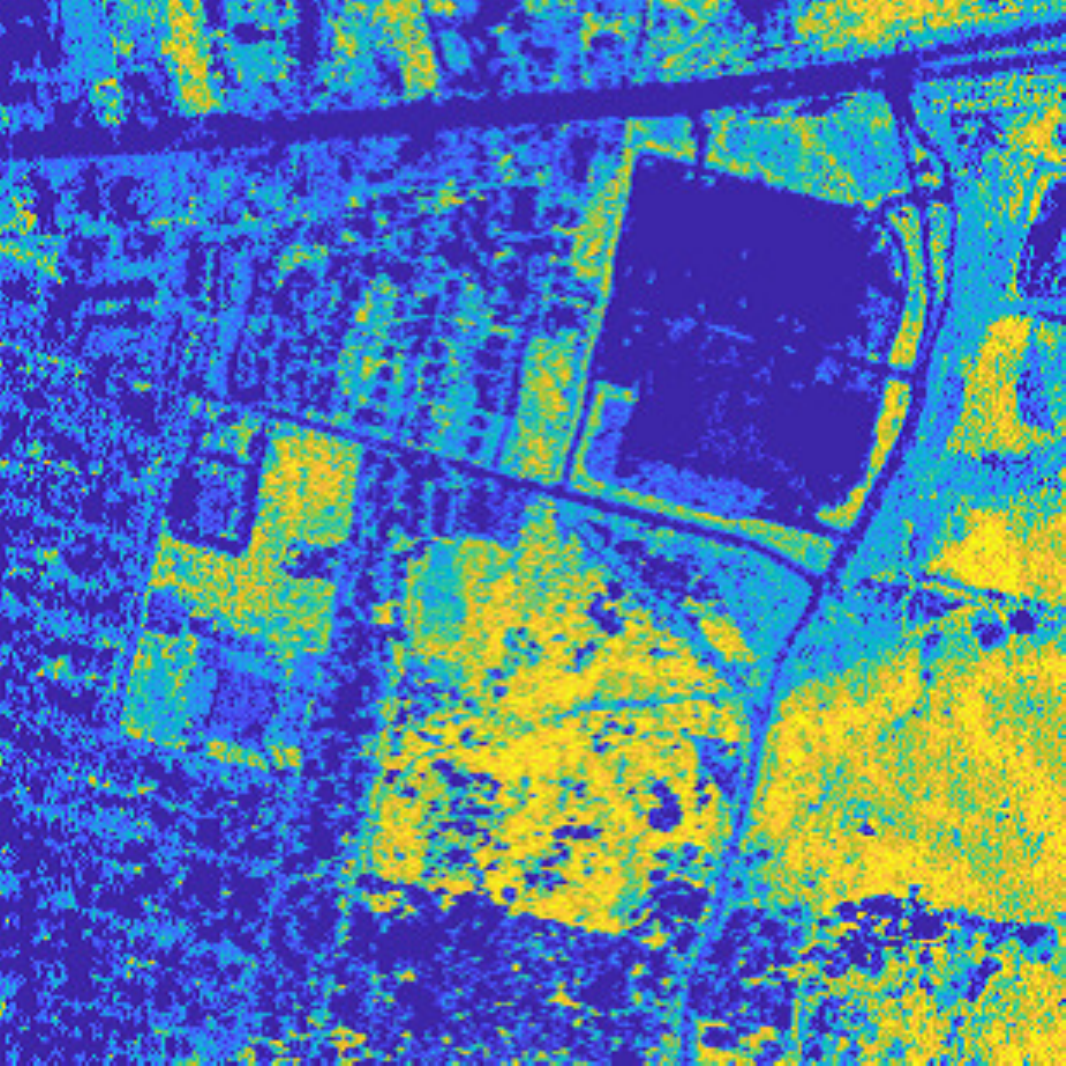}
		}
	\end{minipage}
	\begin{minipage}[t]{0.075\hsize}
		\centerline{
			\includegraphics[height = 40pt]{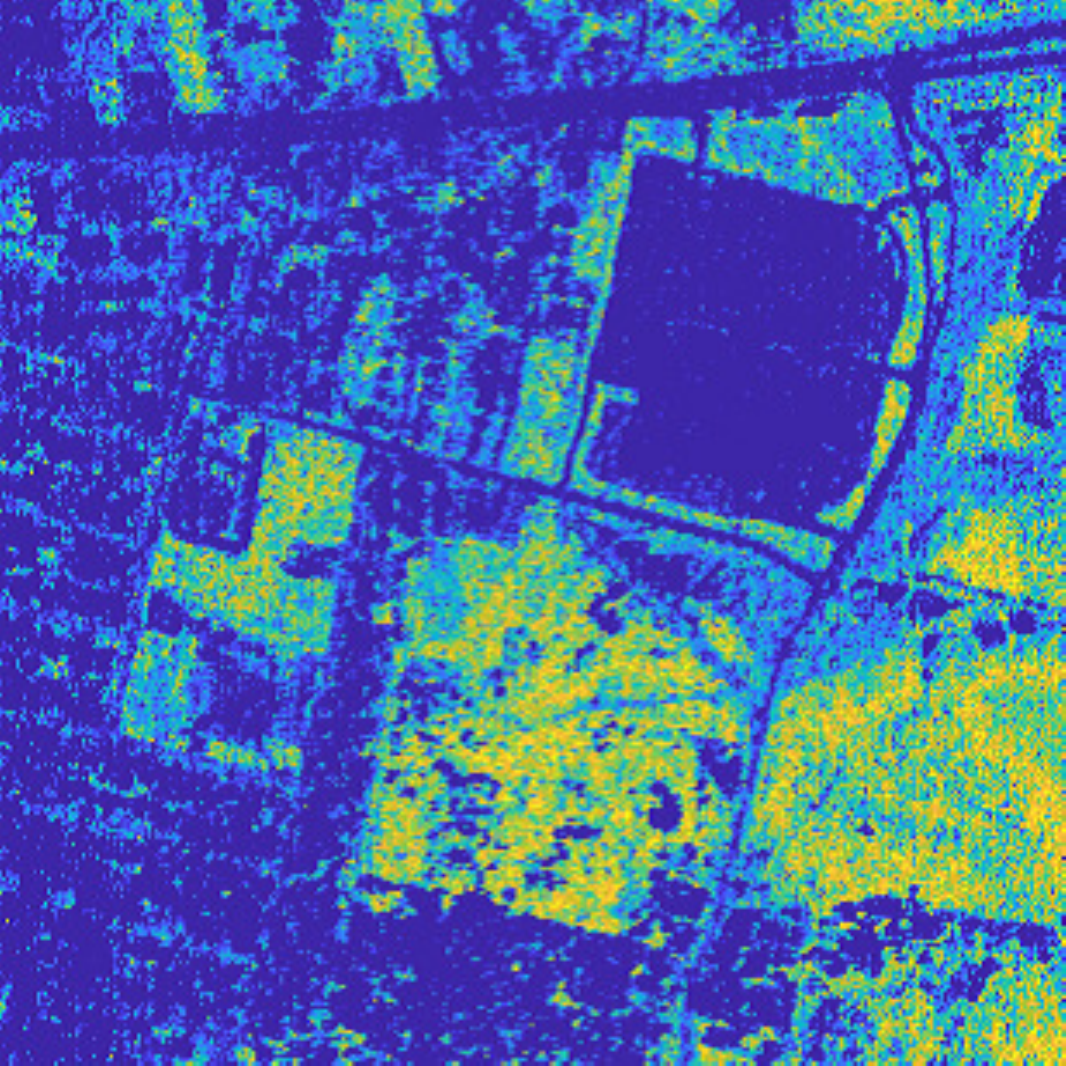}
		}
	\end{minipage}
	\begin{minipage}[t]{0.075\hsize}
		\centerline{
			\includegraphics[height = 40pt]{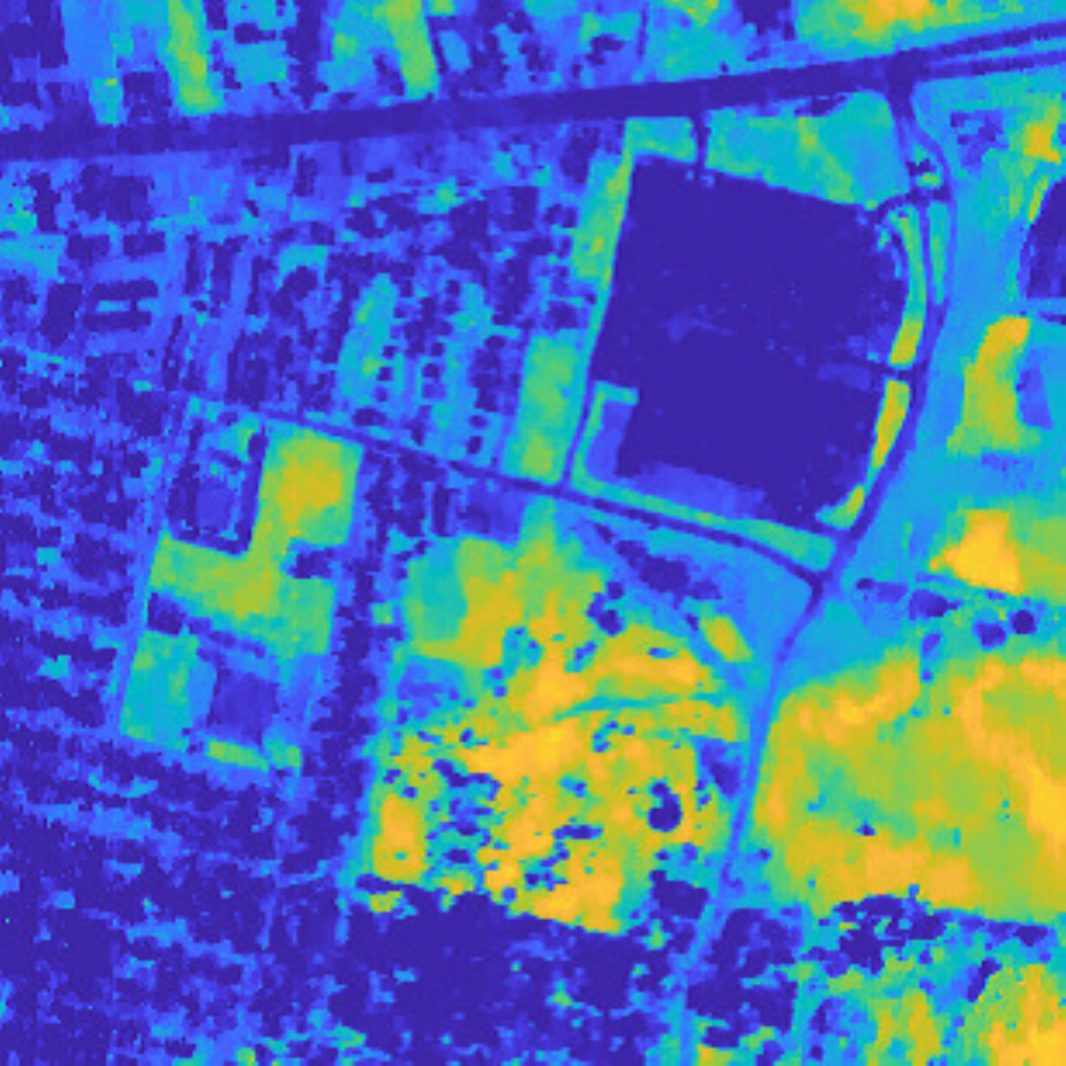}
		}
	\end{minipage}
	\begin{minipage}[t]{0.075\hsize}
		\centerline{
			\includegraphics[height = 40pt]{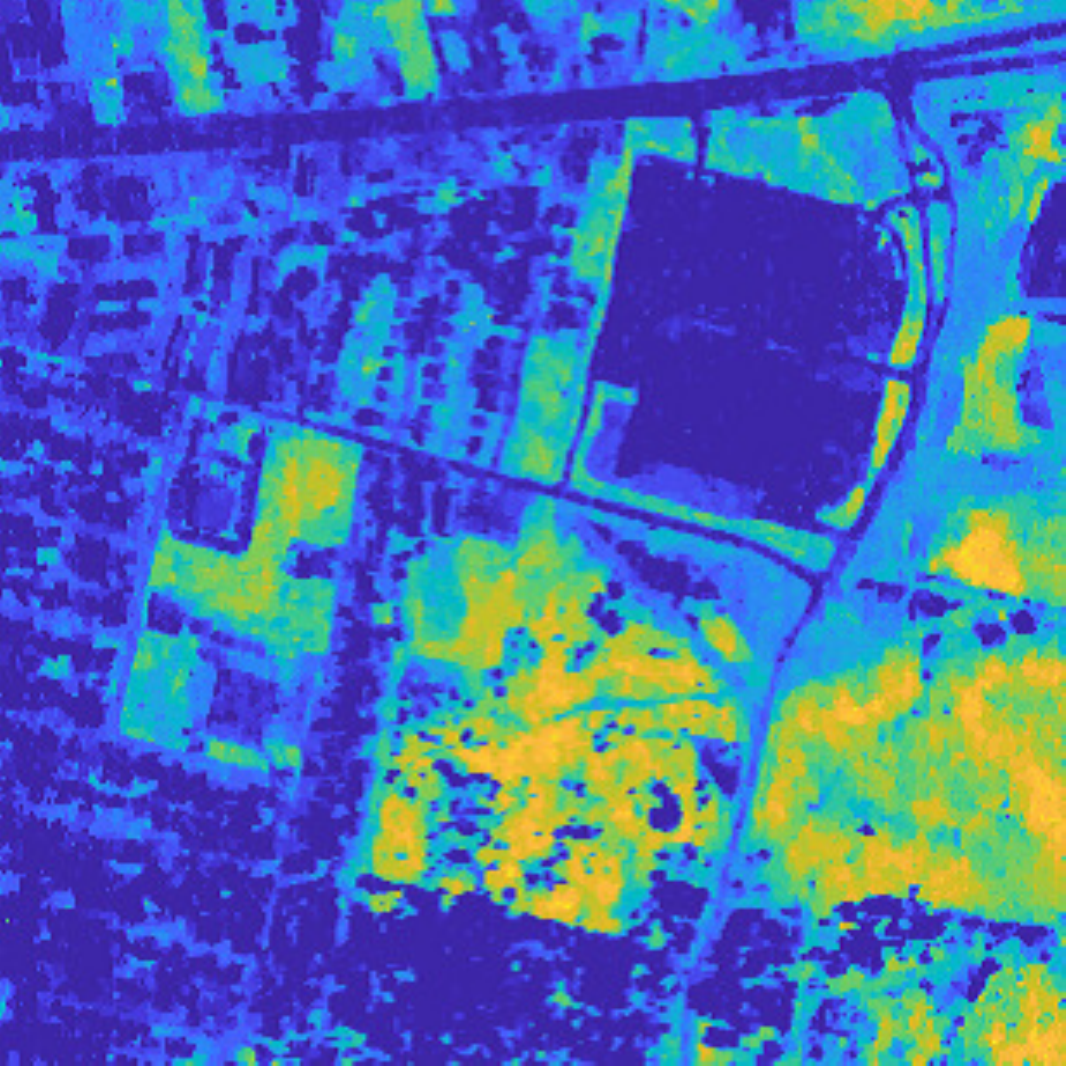}
		}
	\end{minipage}
	\begin{minipage}[t]{0.075\hsize}
		\centerline{
			\includegraphics[height = 40pt]{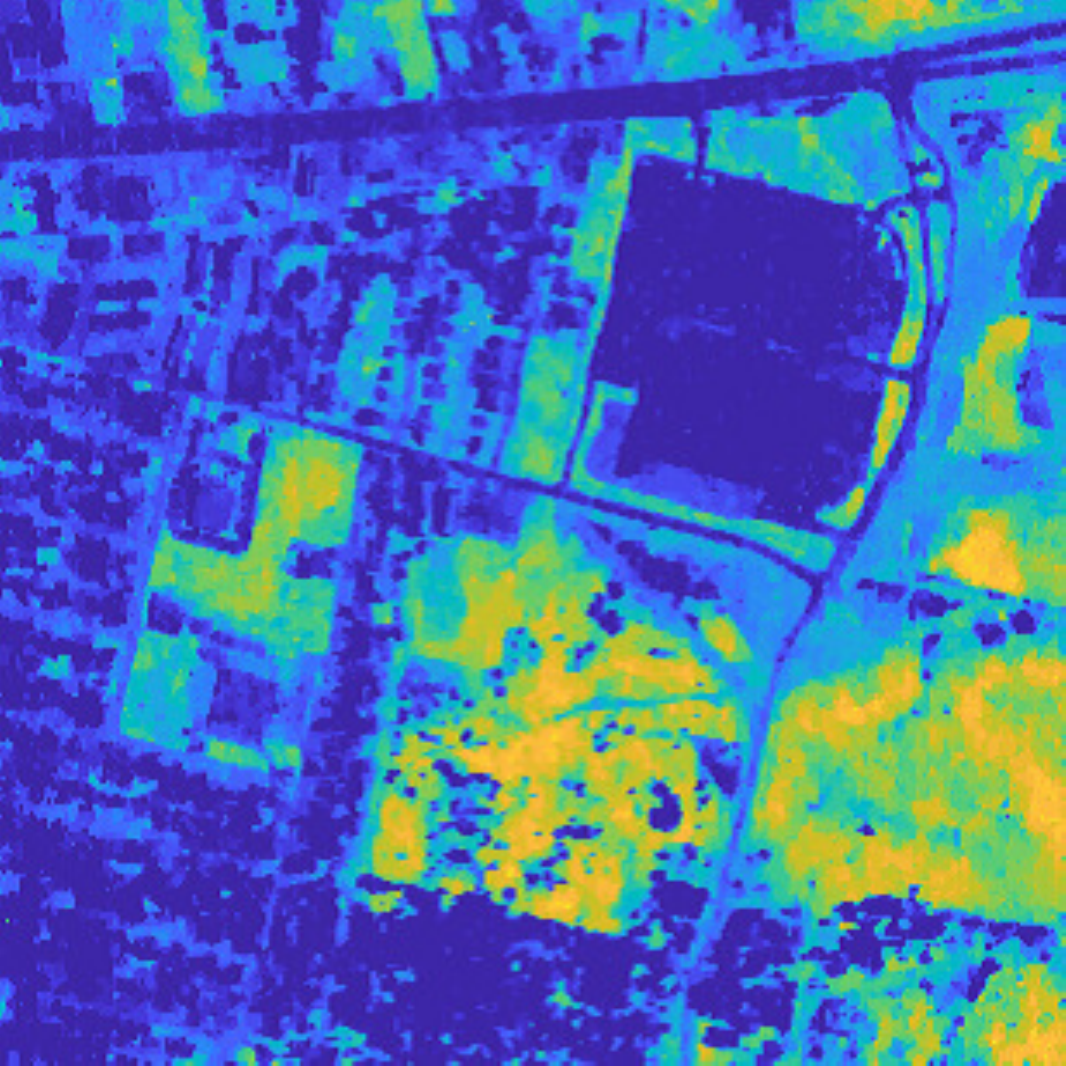}
		}
	\end{minipage}
	\begin{minipage}[t]{0.02\hsize}
		\centerline{
			\includegraphics[height = 40pt]{./fig/colorbar_20.png}
		}
	\end{minipage}
	
	\vspace{1mm}
	
	\begin{minipage}[t]{0.075\hsize}
		\centerline{
			\includegraphics[height = 40pt]{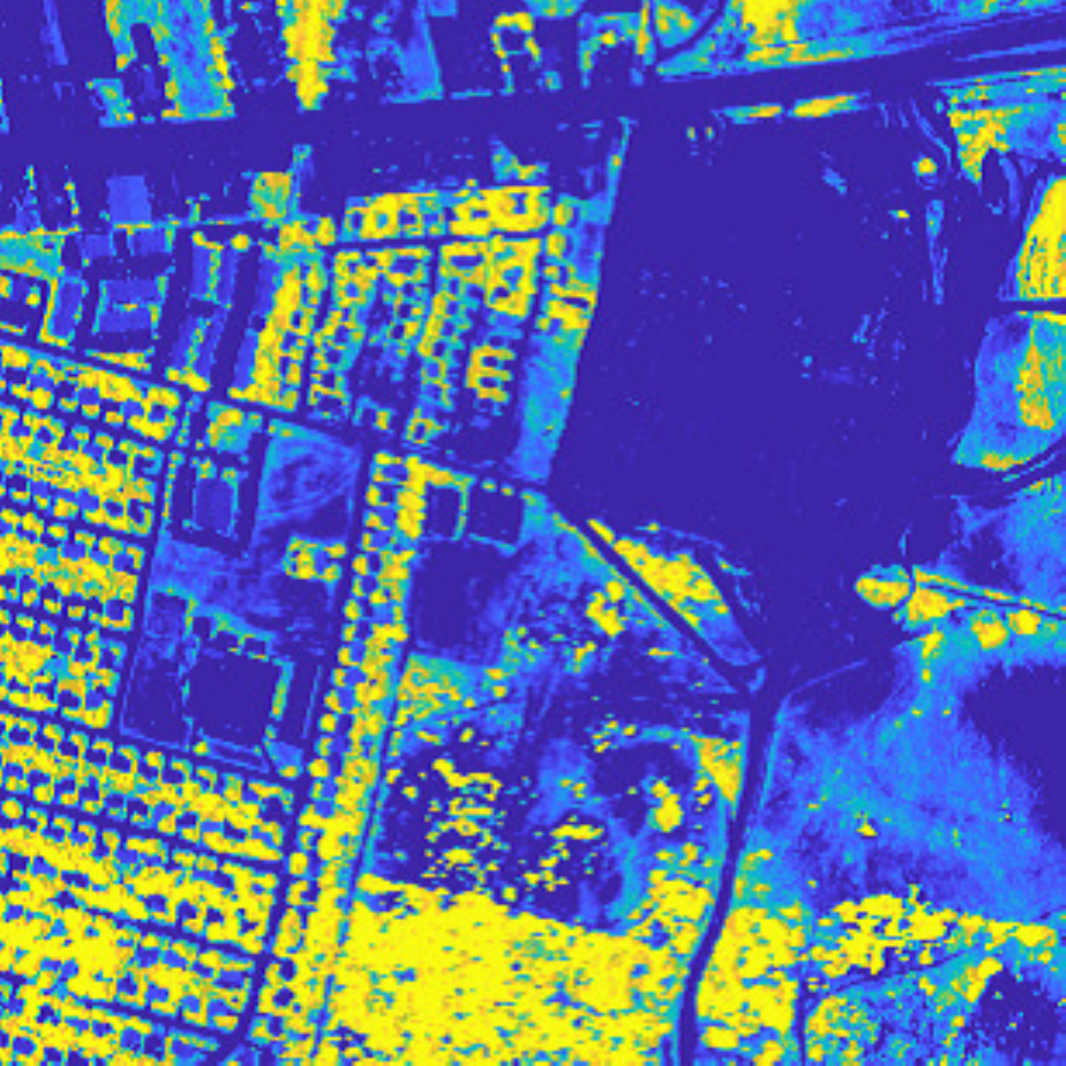}
		}
	\end{minipage}
	\begin{minipage}[t]{0.075\hsize}
		\centerline{
			\includegraphics[height = 40pt]{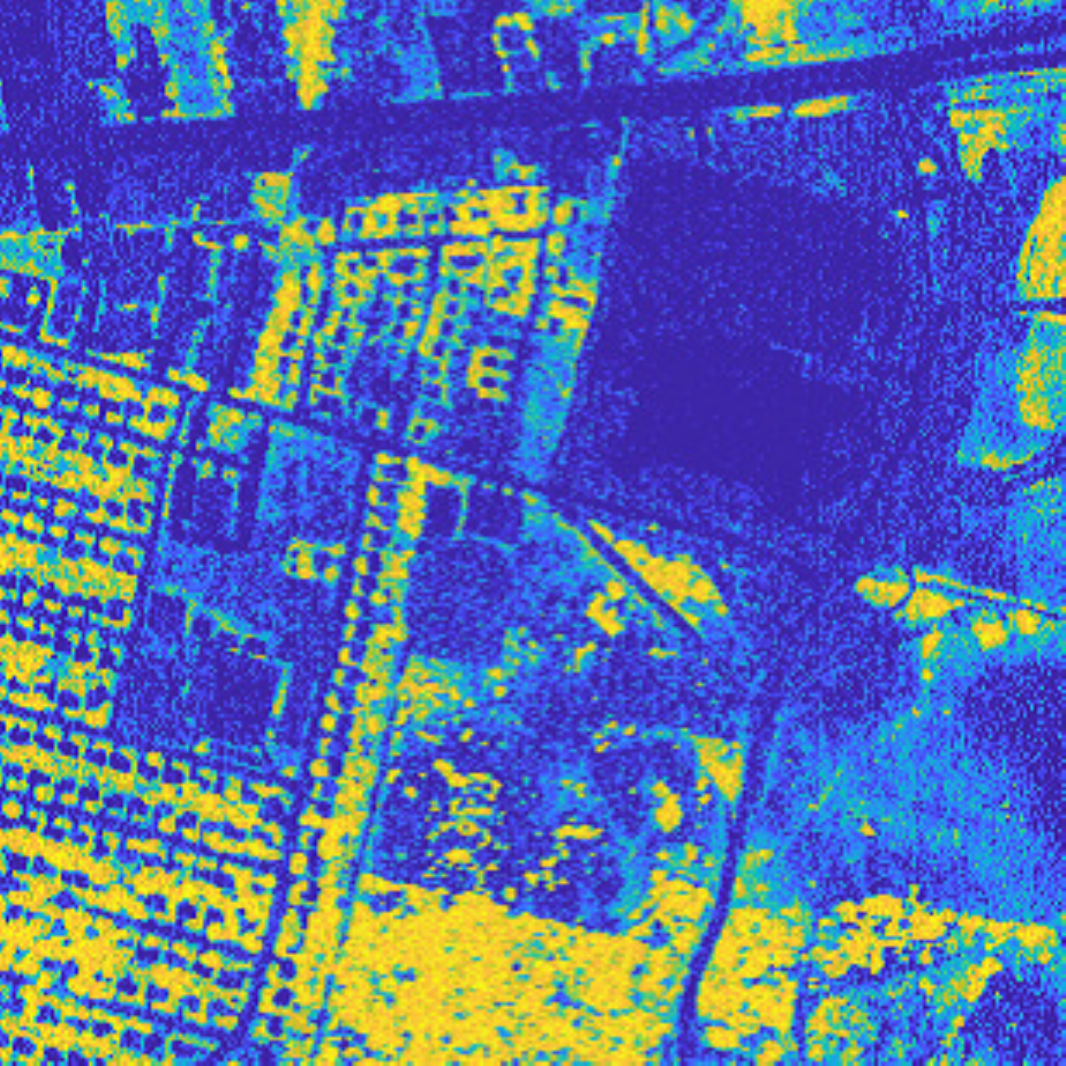}
		}
	\end{minipage}
	\begin{minipage}[t]{0.075\hsize}
		\centerline{
			\includegraphics[height = 40pt]{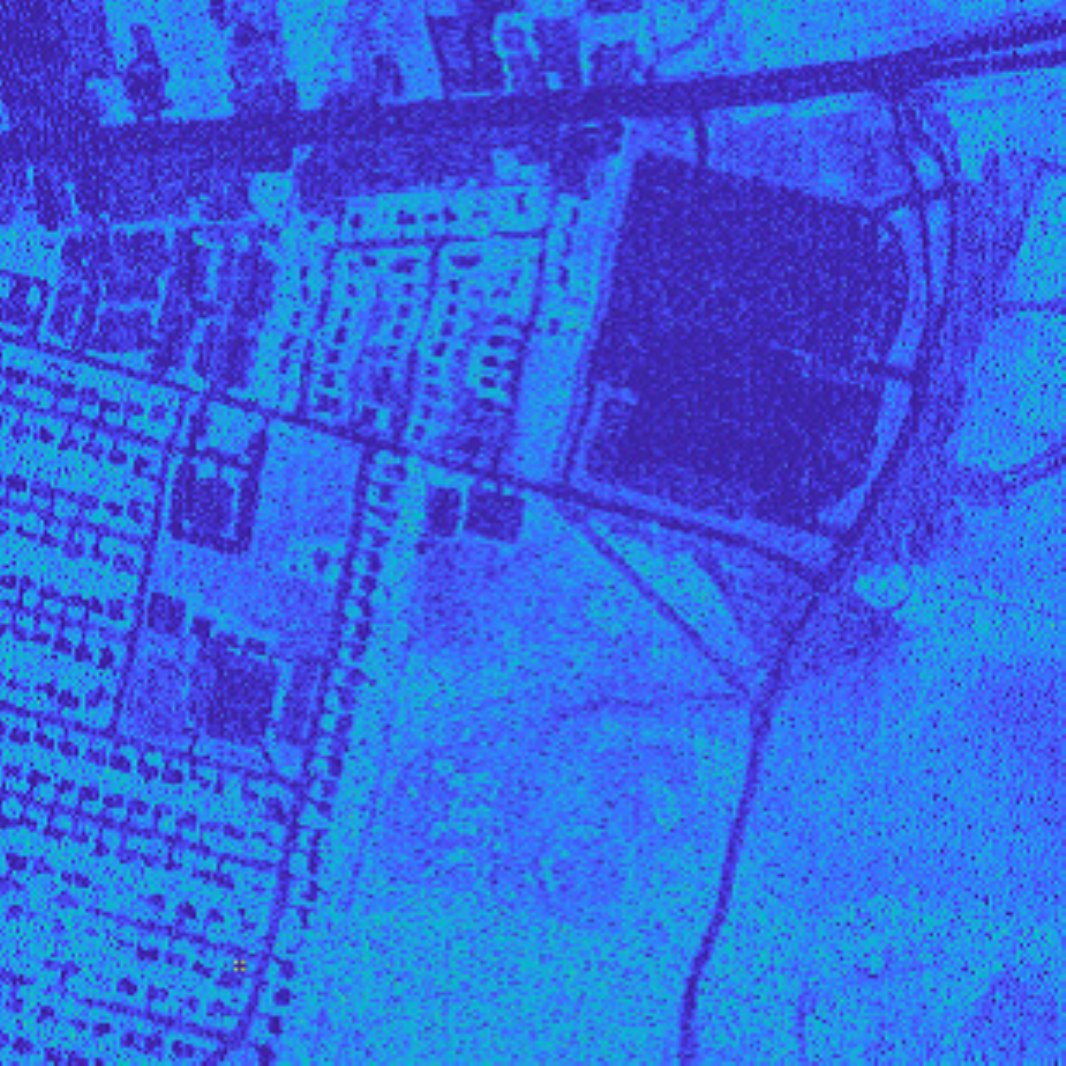}
		}
	\end{minipage}
	\begin{minipage}[t]{0.075\hsize}
		\centerline{
			\includegraphics[height = 40pt]{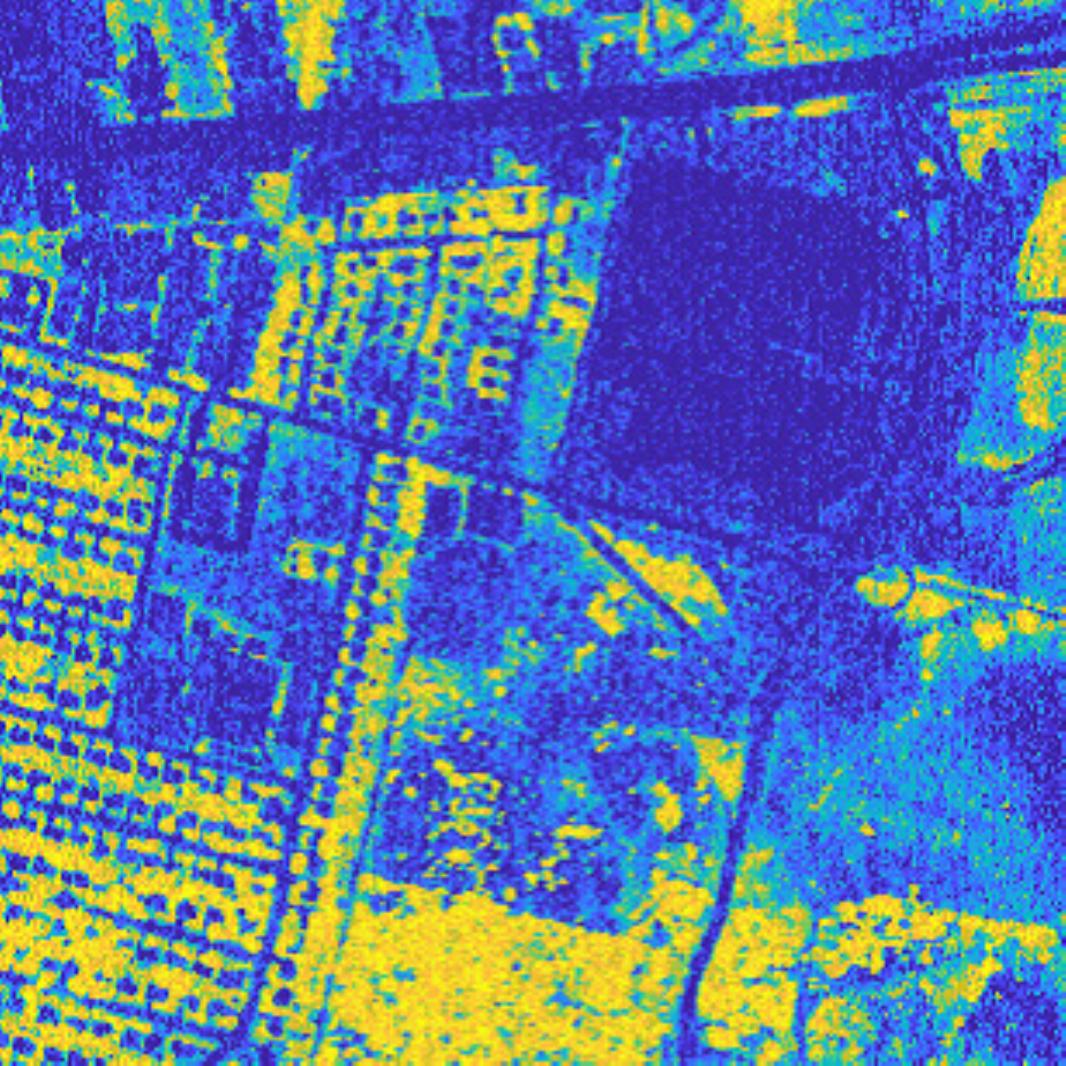}
		}
	\end{minipage}
	\begin{minipage}[t]{0.075\hsize}
		\centerline{
			\includegraphics[height = 40pt]{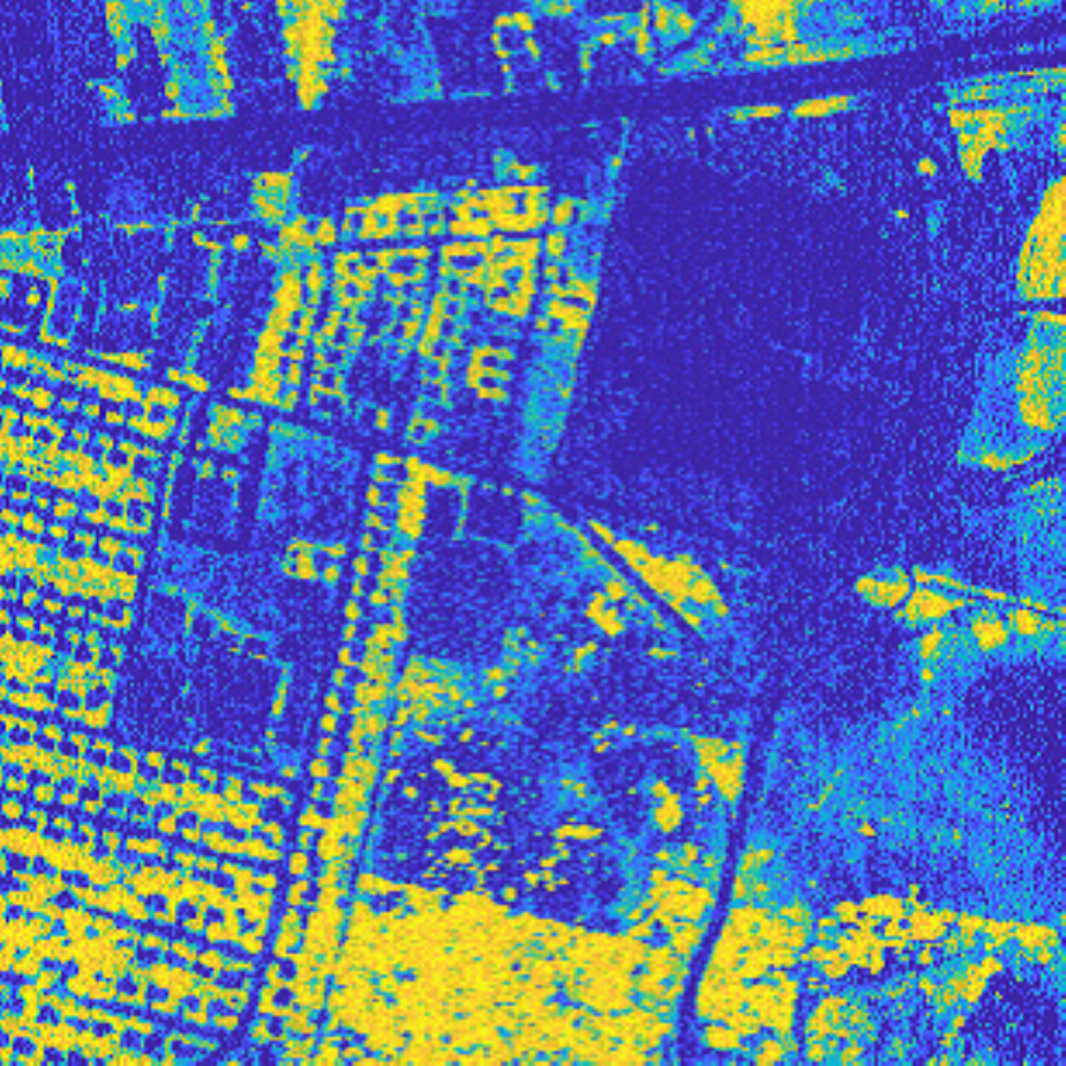}
		}
	\end{minipage}
	\begin{minipage}[t]{0.075\hsize}
		\centerline{
			\includegraphics[height = 40pt]{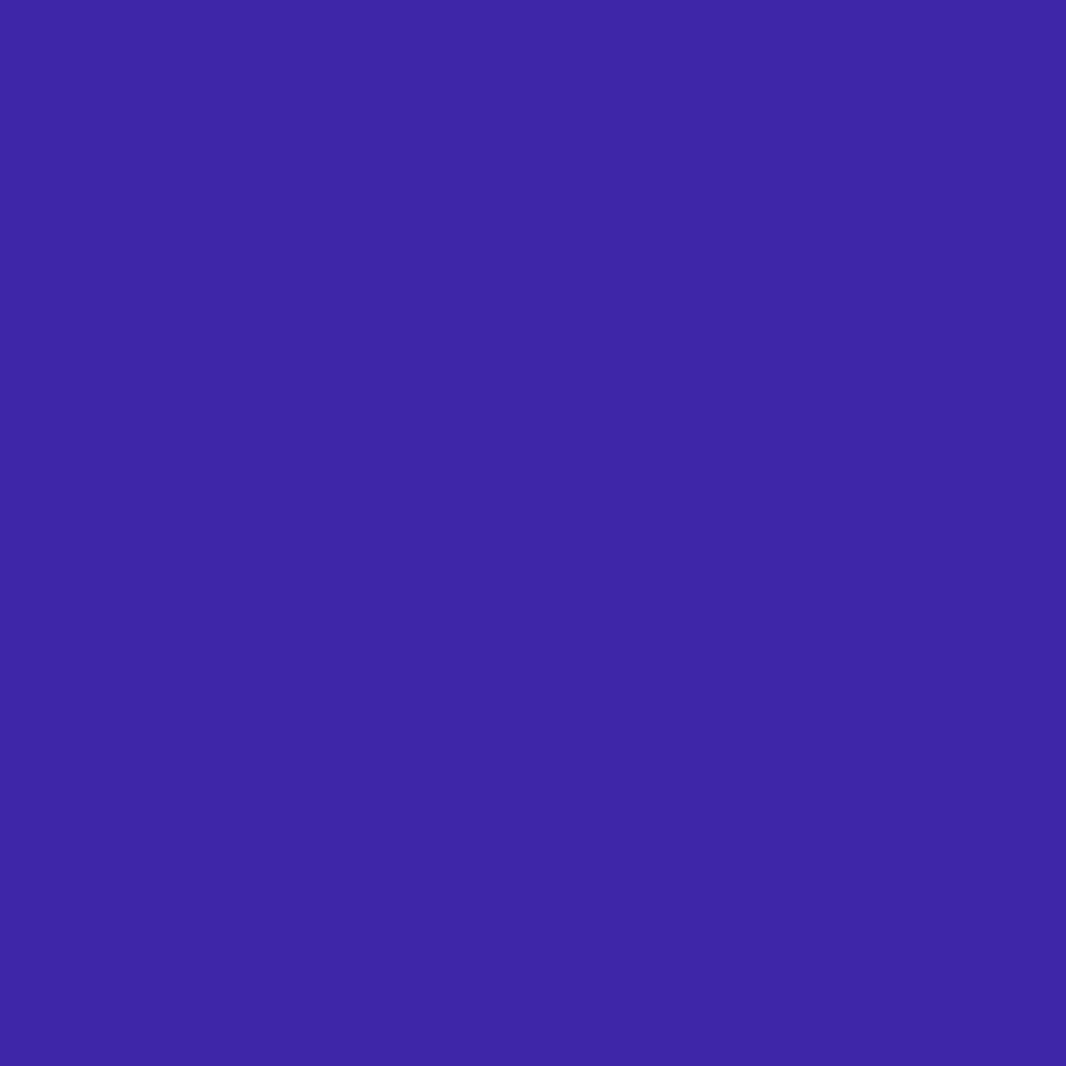}
		}
	\end{minipage}
	\begin{minipage}[t]{0.075\hsize}
		\centerline{
			\includegraphics[height = 40pt]{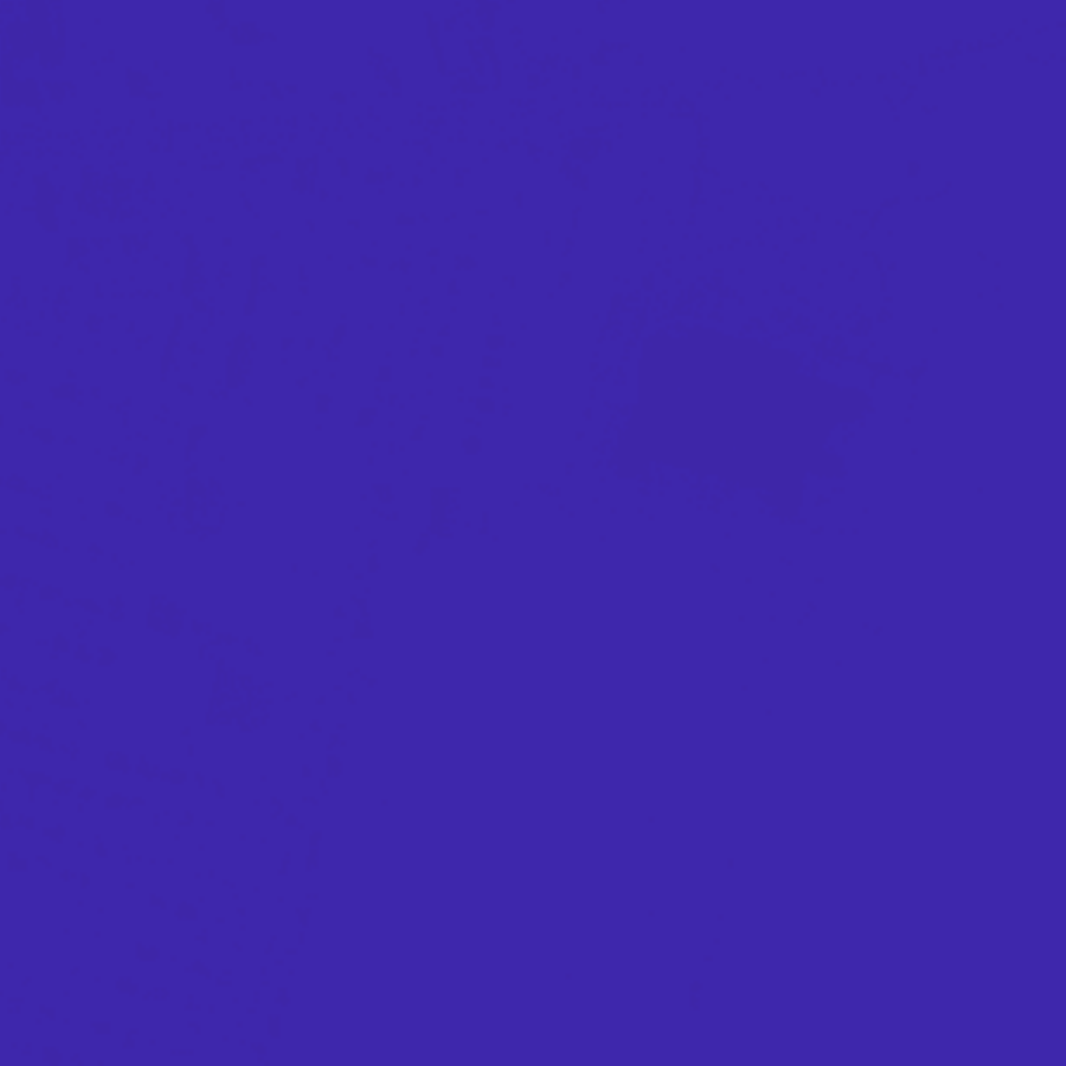}
		}
	\end{minipage}
	\begin{minipage}[t]{0.075\hsize}
		\centerline{
			\includegraphics[height = 40pt]{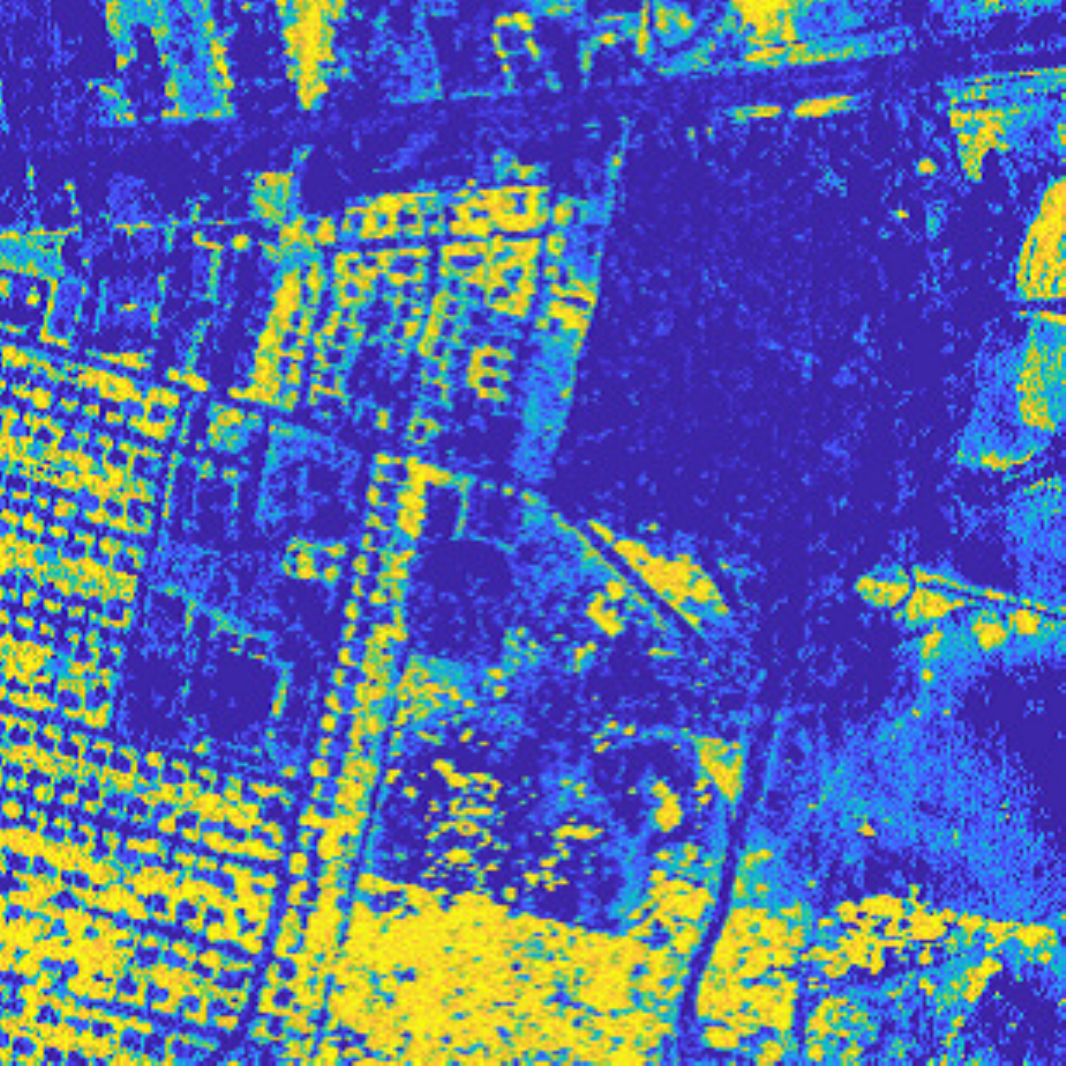}
		}
	\end{minipage}
	\begin{minipage}[t]{0.075\hsize}
		\centerline{
			\includegraphics[height = 40pt]{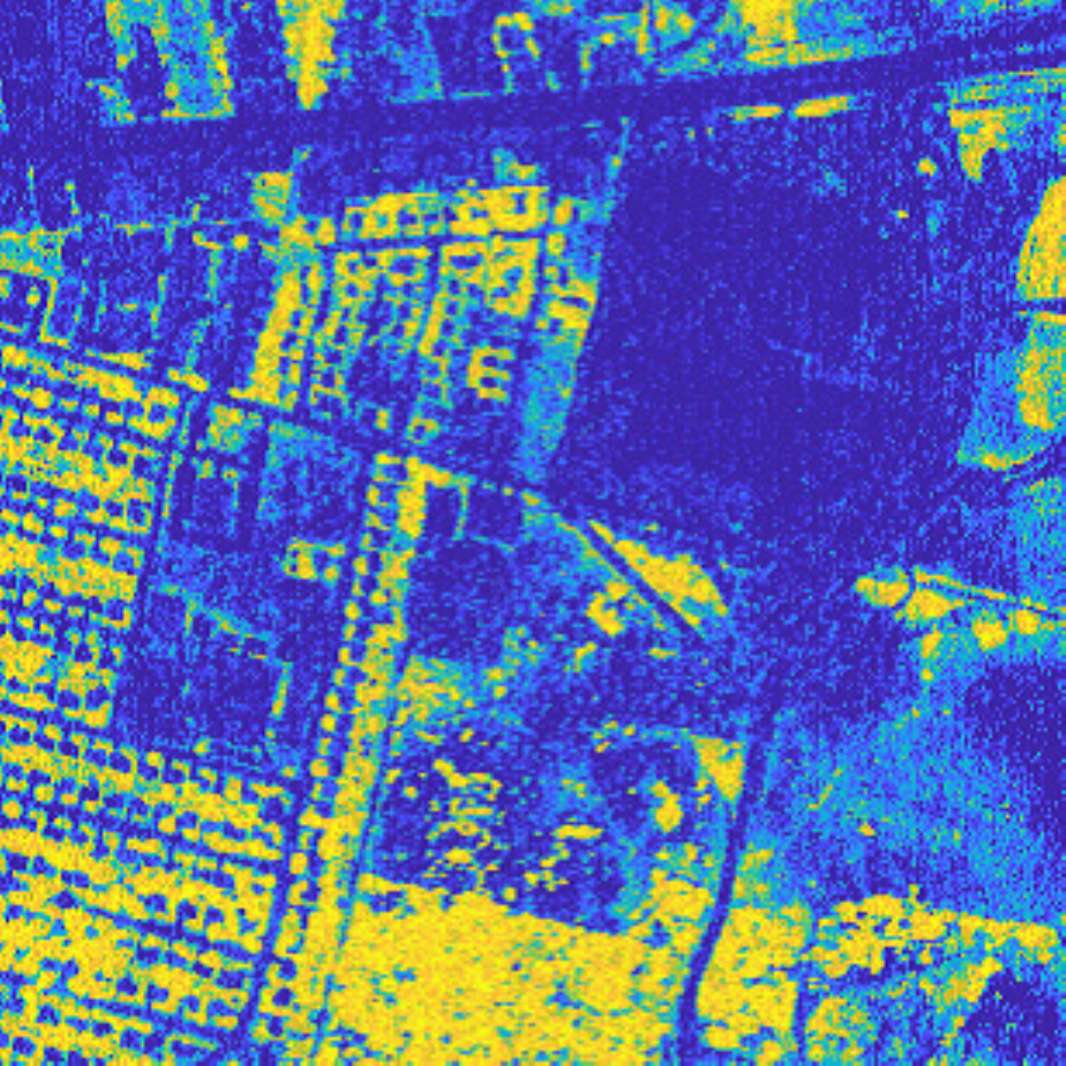}
		}
	\end{minipage}
	\begin{minipage}[t]{0.075\hsize}
		\centerline{
			\includegraphics[height = 40pt]{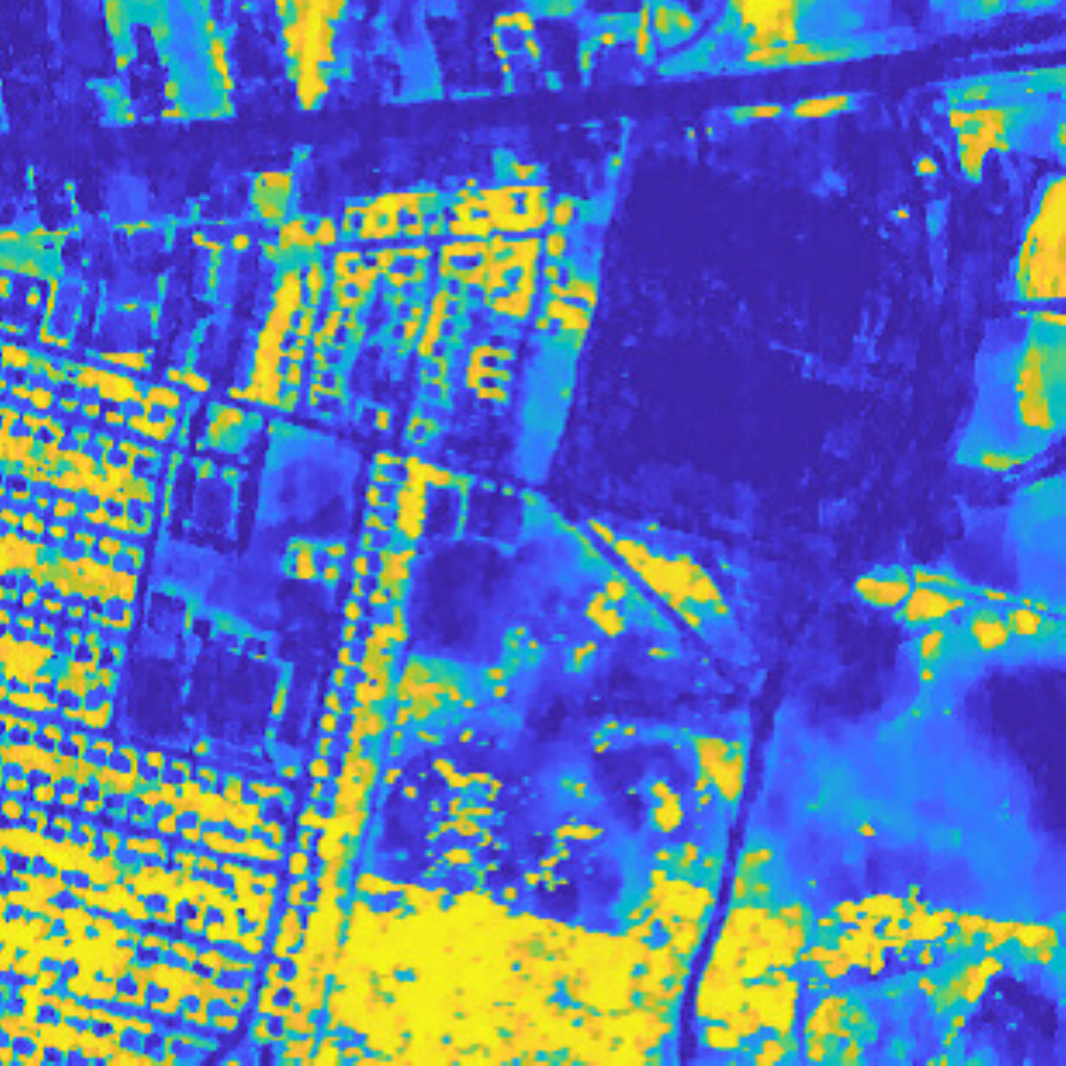}
		}
	\end{minipage}
	\begin{minipage}[t]{0.075\hsize}
		\centerline{
			\includegraphics[height = 40pt]{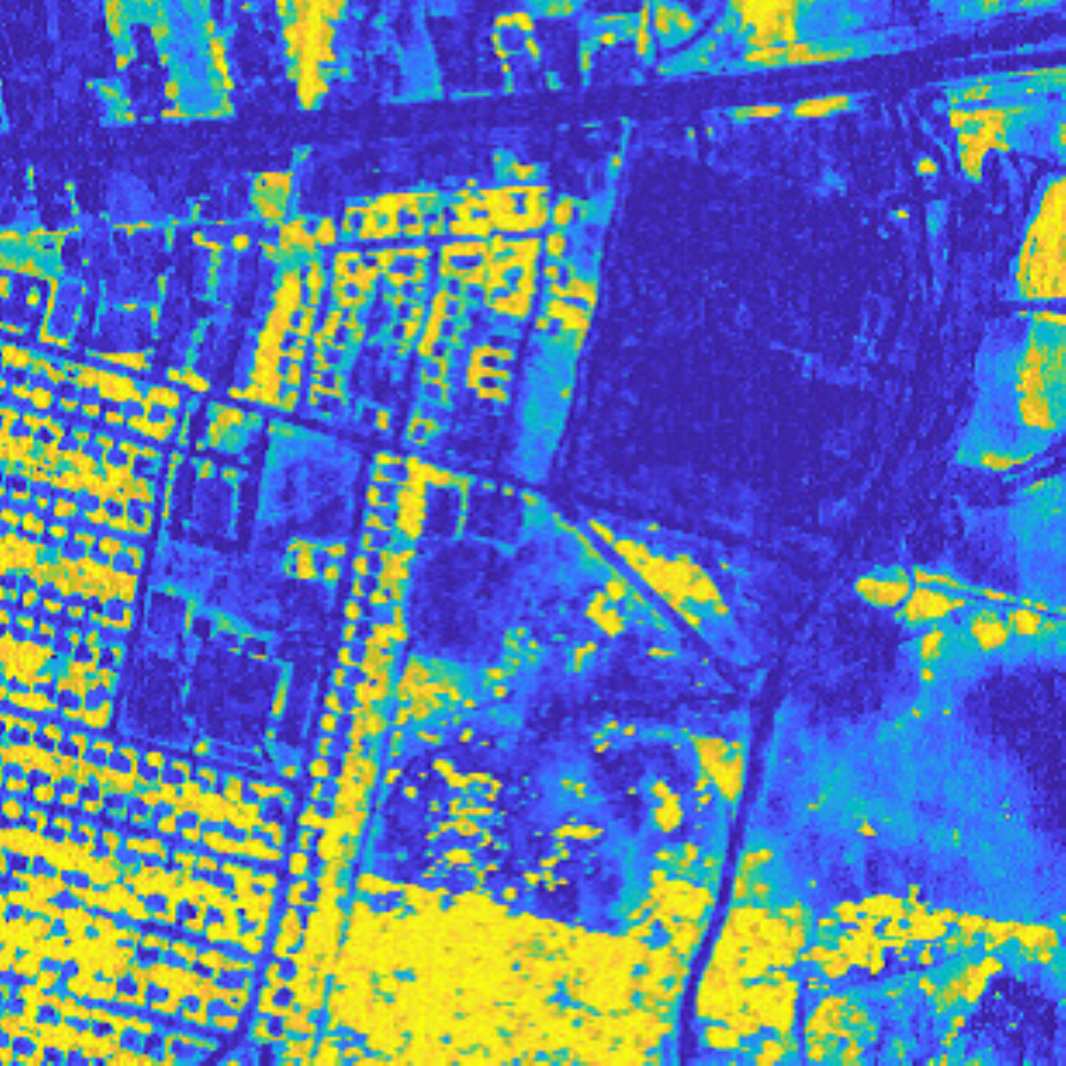}
		}
	\end{minipage}
	\begin{minipage}[t]{0.075\hsize}
		\centerline{
			\includegraphics[height = 40pt]{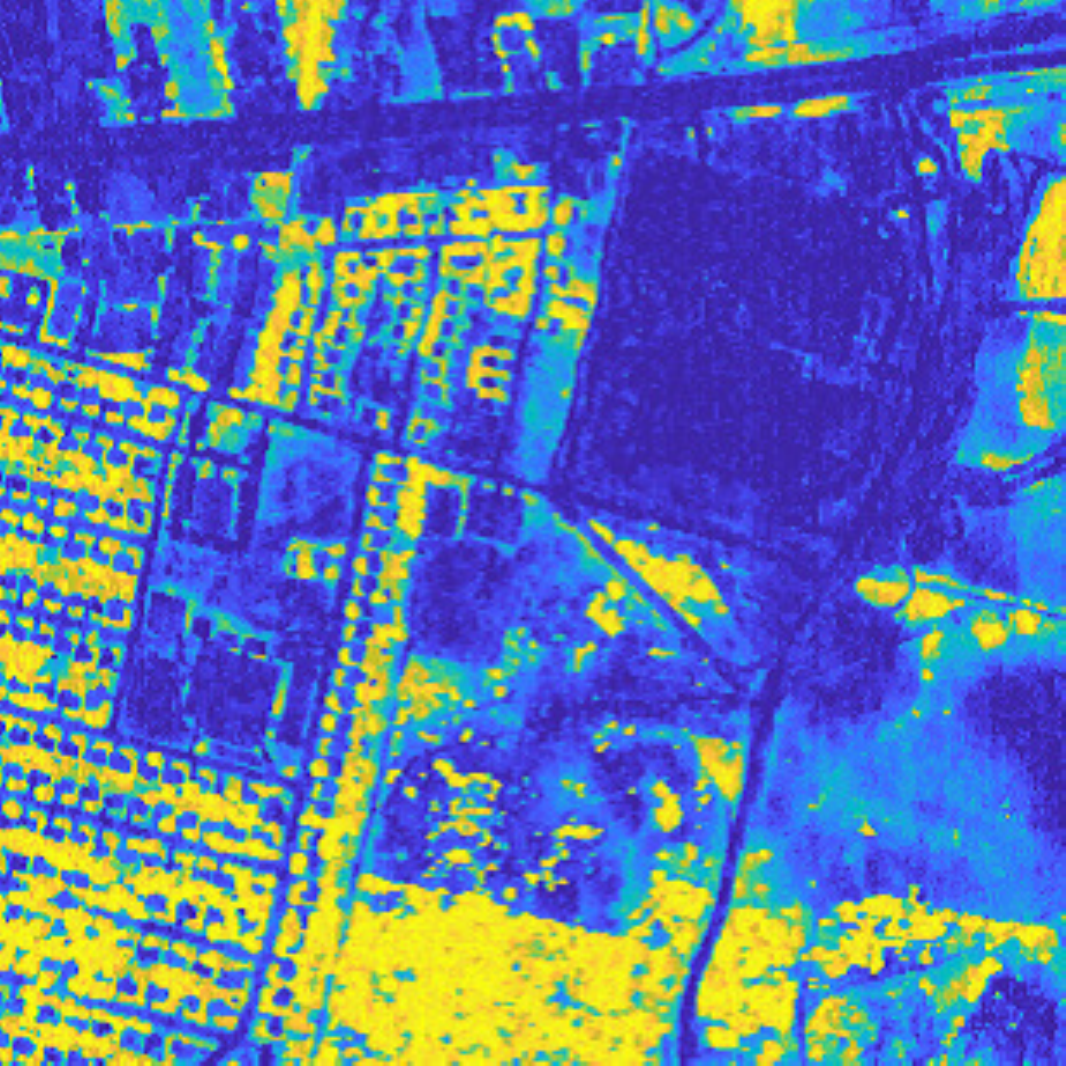}
		}
	\end{minipage}
	\begin{minipage}[t]{0.02\hsize}
		\centerline{
			\includegraphics[height = 40pt]{./fig/colorbar_20.png}
		}
	\end{minipage}
	
	\vspace{1mm}
	
	\begin{minipage}[t]{0.075\hsize}
		\centerline{
			\includegraphics[height = 40pt]{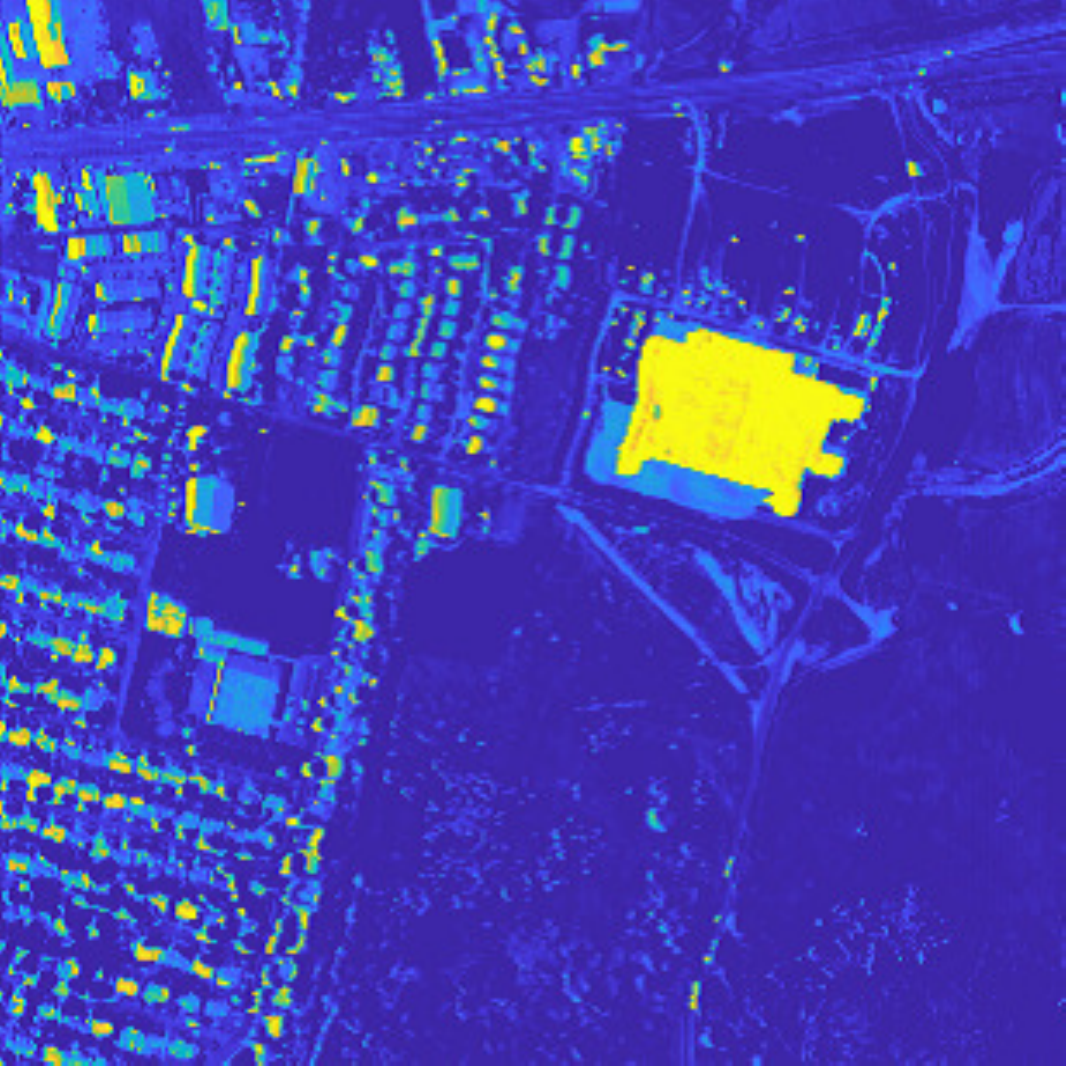}
		}
	\end{minipage}
	\begin{minipage}[t]{0.075\hsize}
		\centerline{
			\includegraphics[height = 40pt]{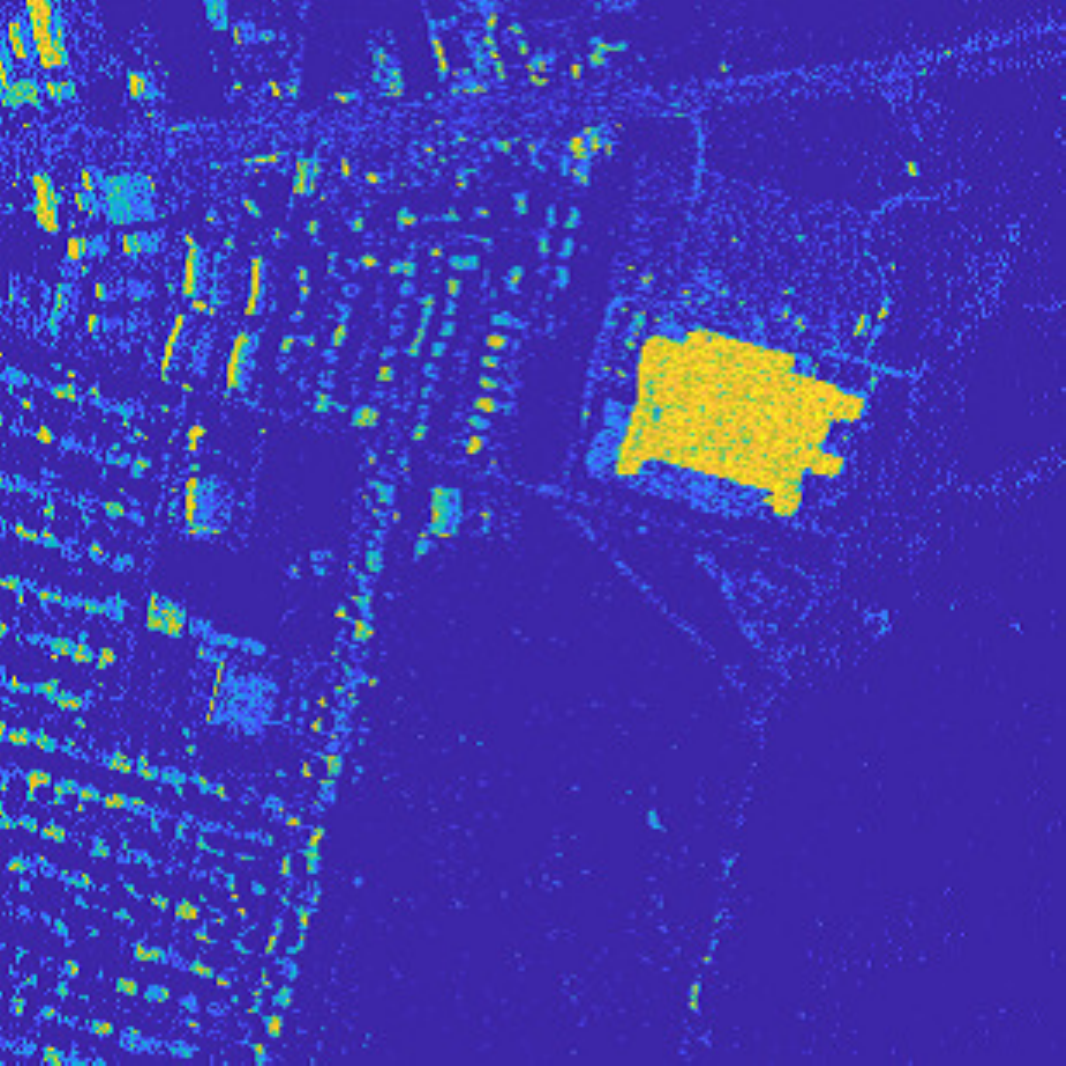}
		}
	\end{minipage}
	\begin{minipage}[t]{0.075\hsize}
		\centerline{
			\includegraphics[height = 40pt]{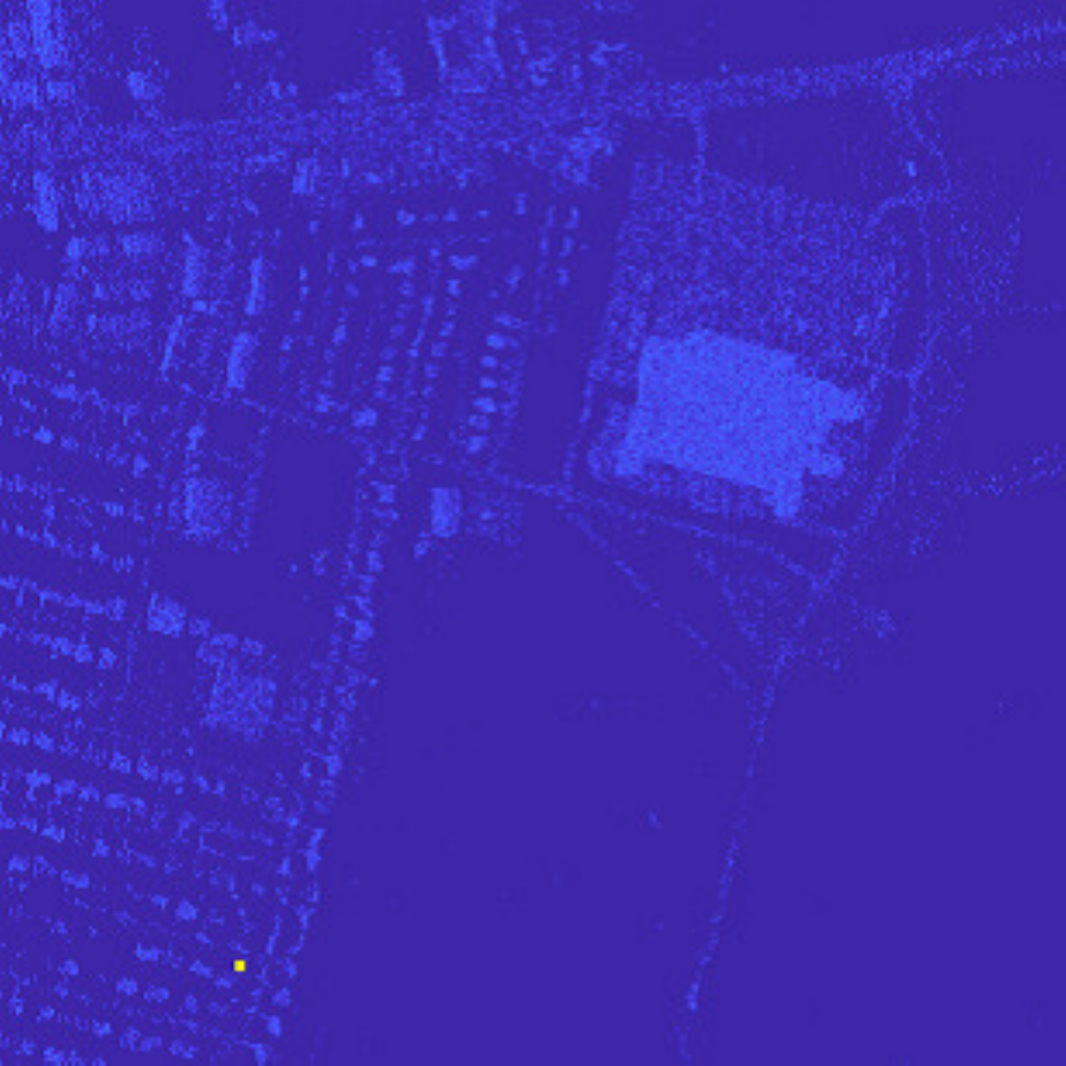}
		}
	\end{minipage}
	\begin{minipage}[t]{0.075\hsize}
		\centerline{
			\includegraphics[height = 40pt]{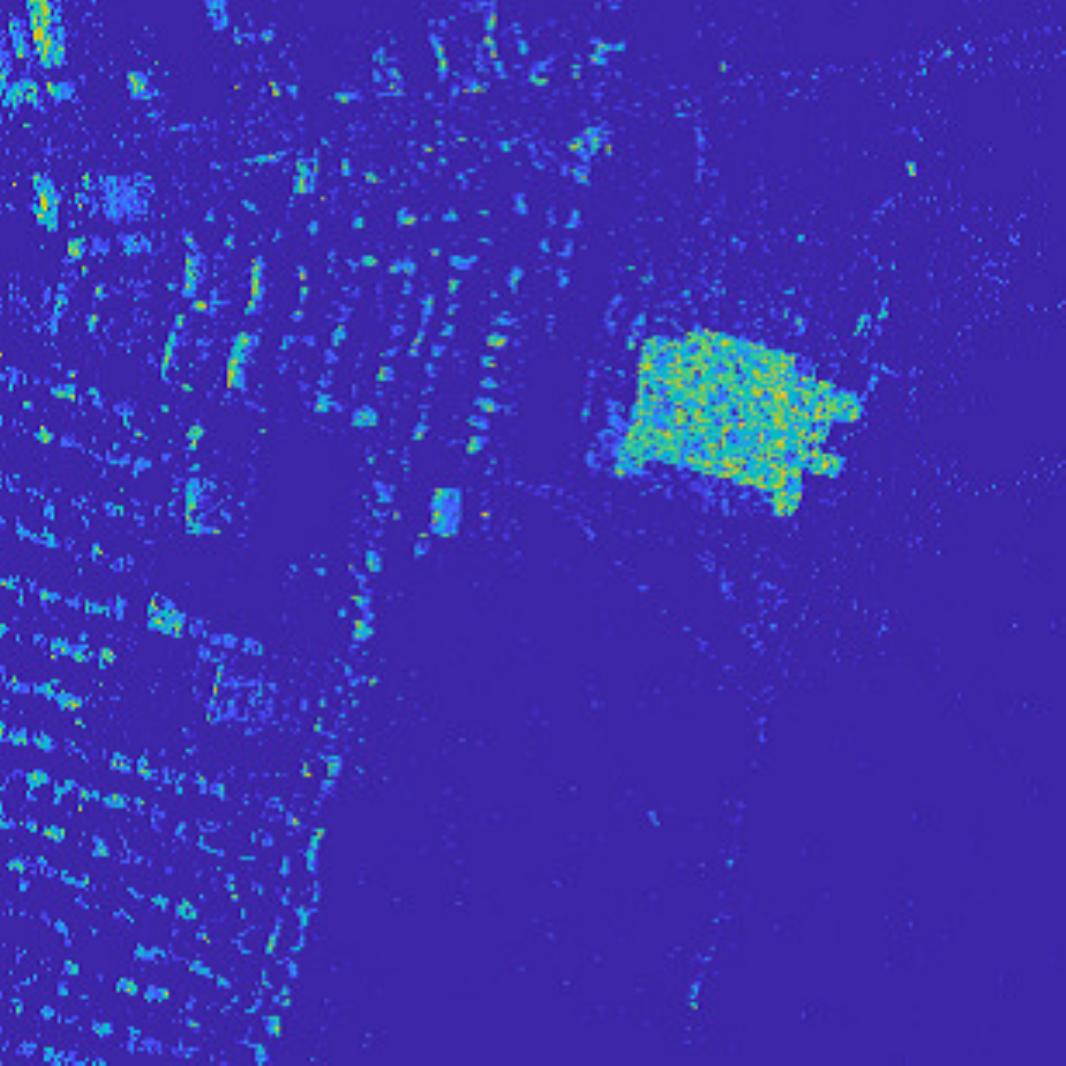}
		}
	\end{minipage}
	\begin{minipage}[t]{0.075\hsize}
		\centerline{
			\includegraphics[height = 40pt]{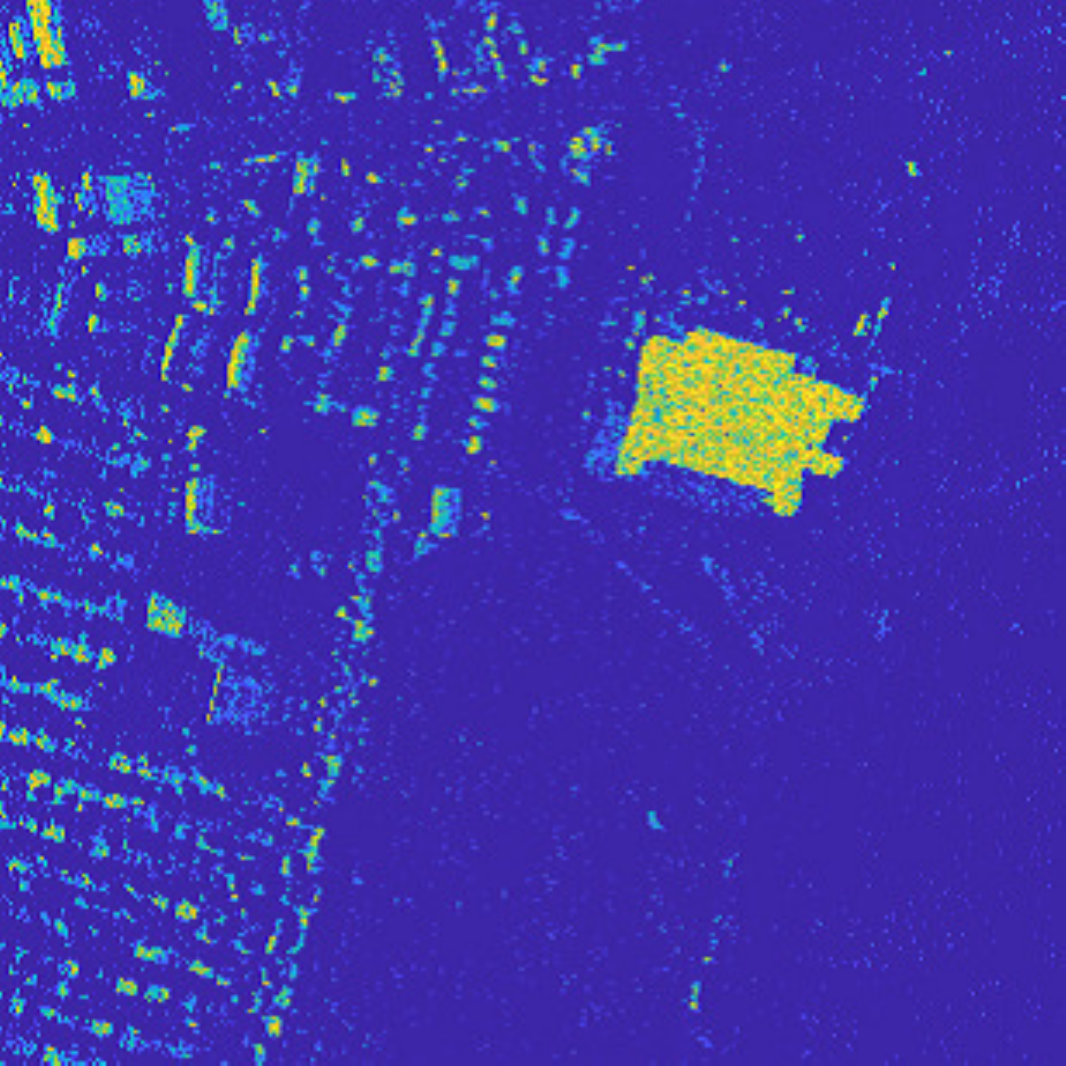}
		}
	\end{minipage}
	\begin{minipage}[t]{0.075\hsize}
		\centerline{
			\includegraphics[height = 40pt]{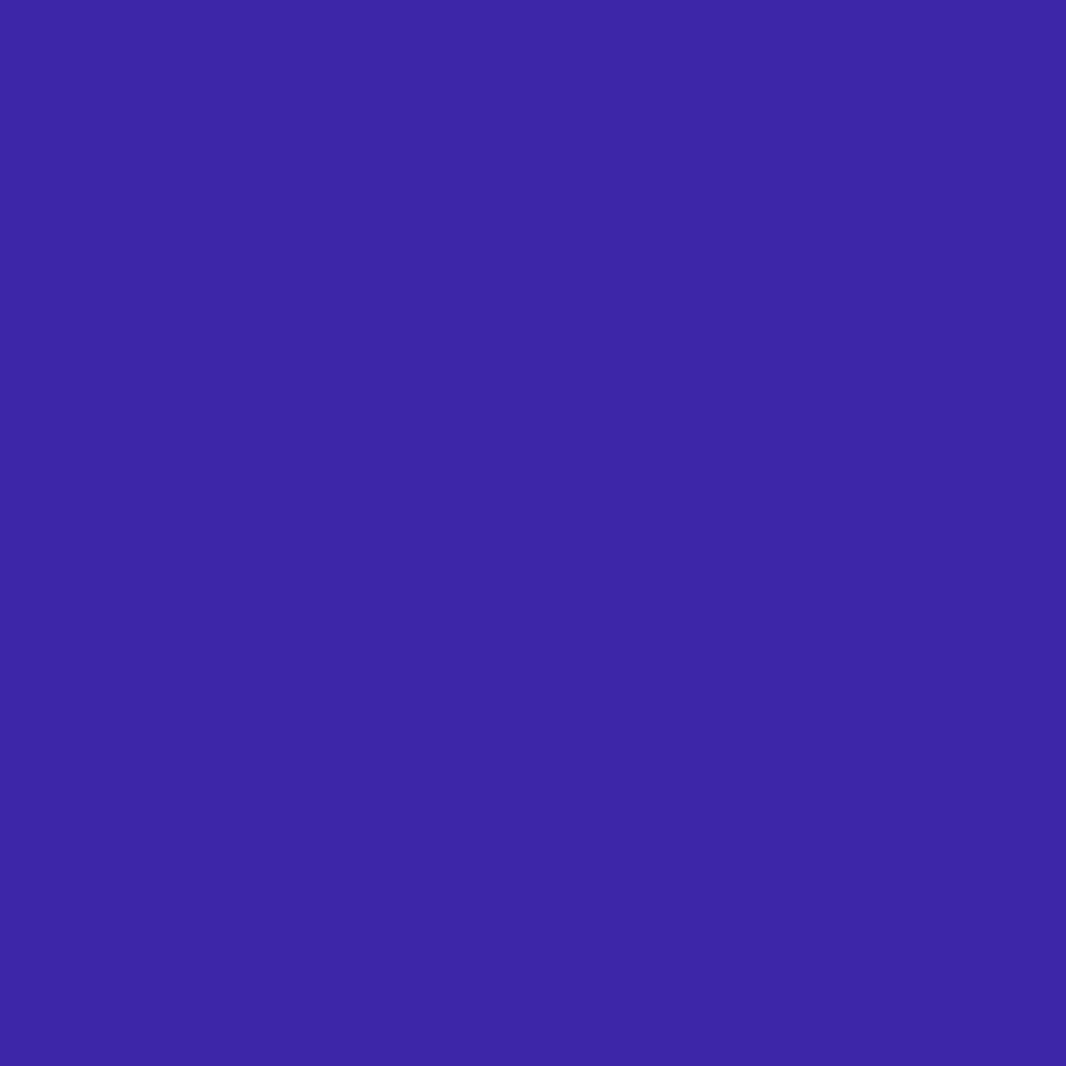}
		}
	\end{minipage}
	\begin{minipage}[t]{0.075\hsize}
		\centerline{
			\includegraphics[height = 40pt]{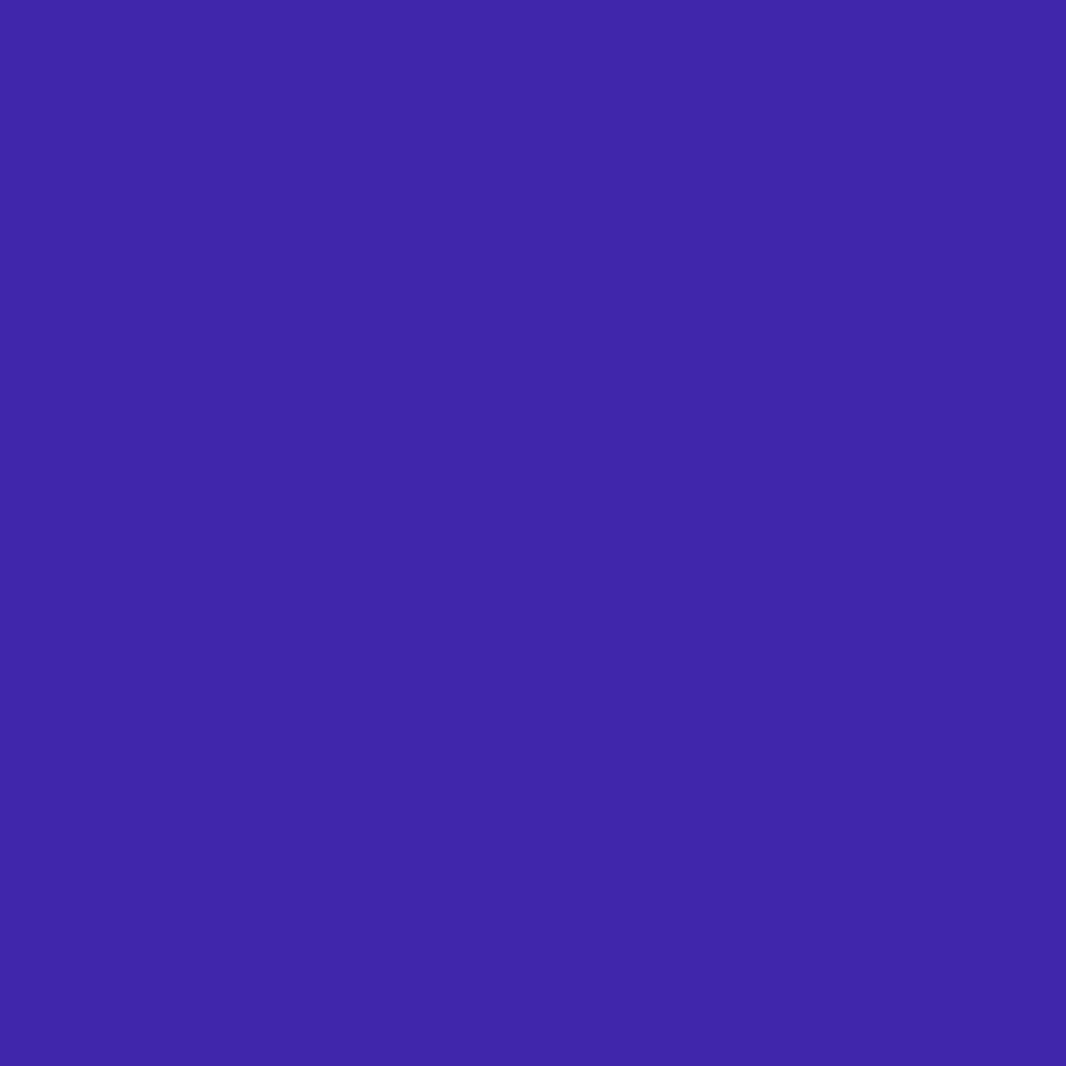}
		}
	\end{minipage}
	\begin{minipage}[t]{0.075\hsize}
		\centerline{
			\includegraphics[height = 40pt]{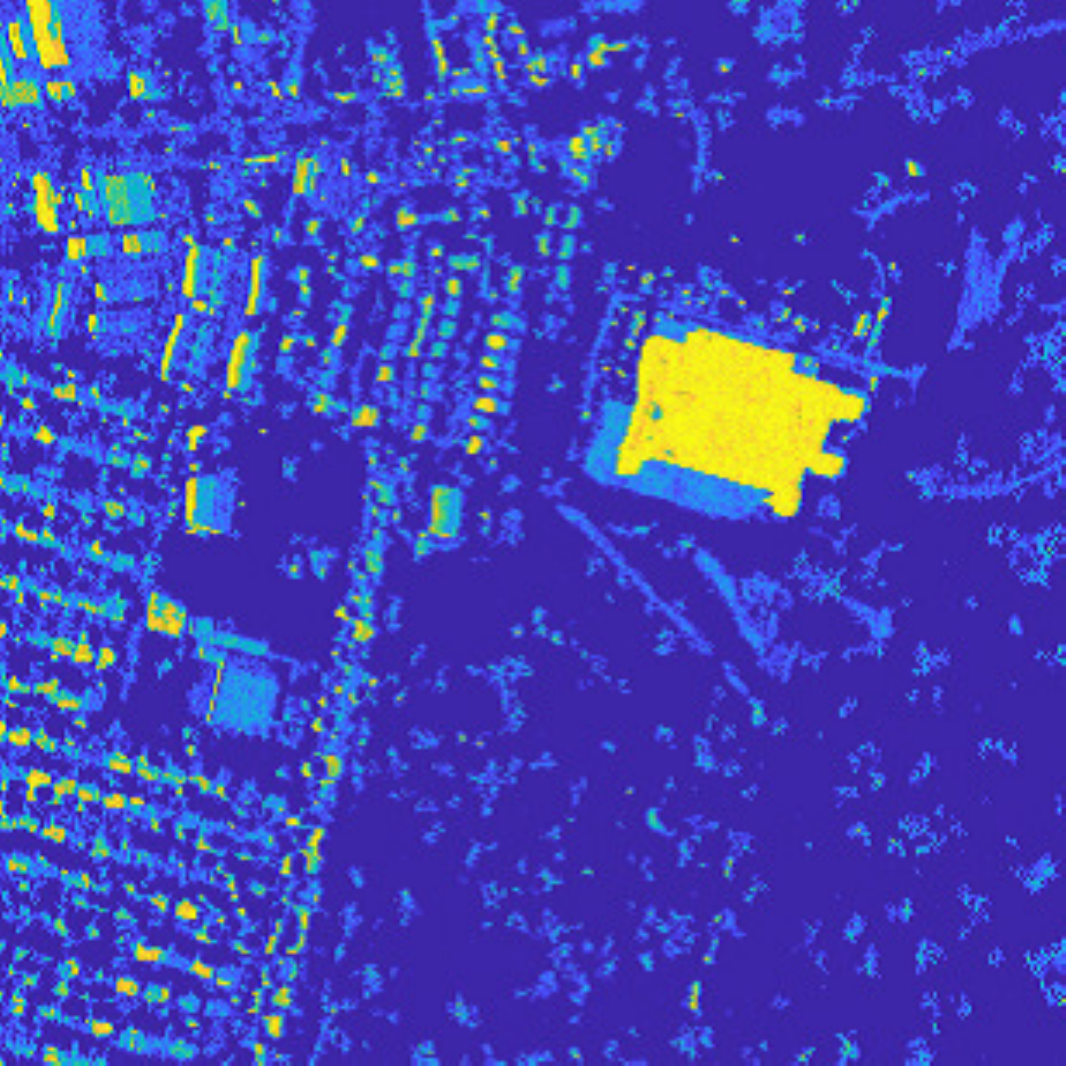}
		}
	\end{minipage}
	\begin{minipage}[t]{0.075\hsize}
		\centerline{
			\includegraphics[height = 40pt]{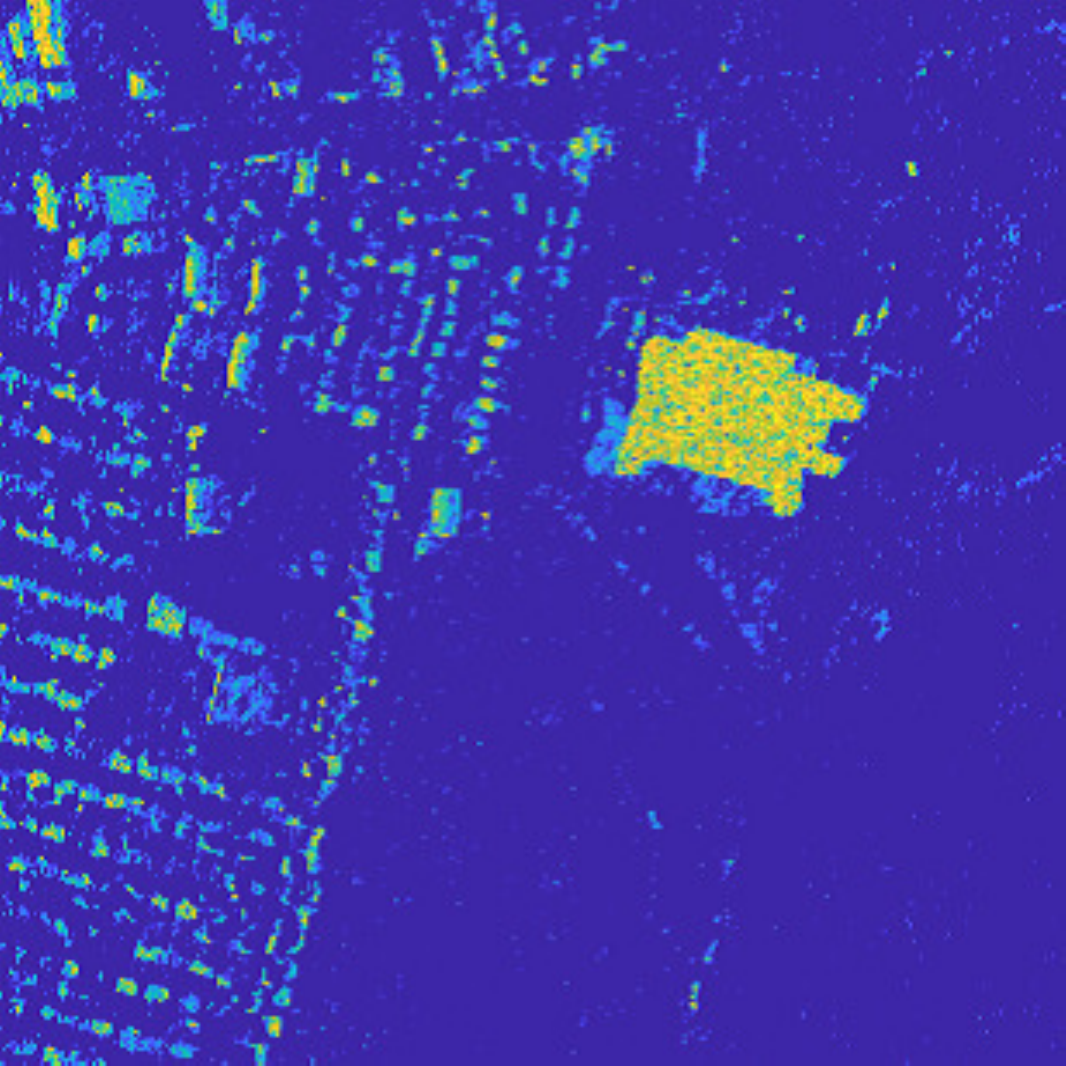}
		}
	\end{minipage}
	\begin{minipage}[t]{0.075\hsize}
		\centerline{
			\includegraphics[height = 40pt]{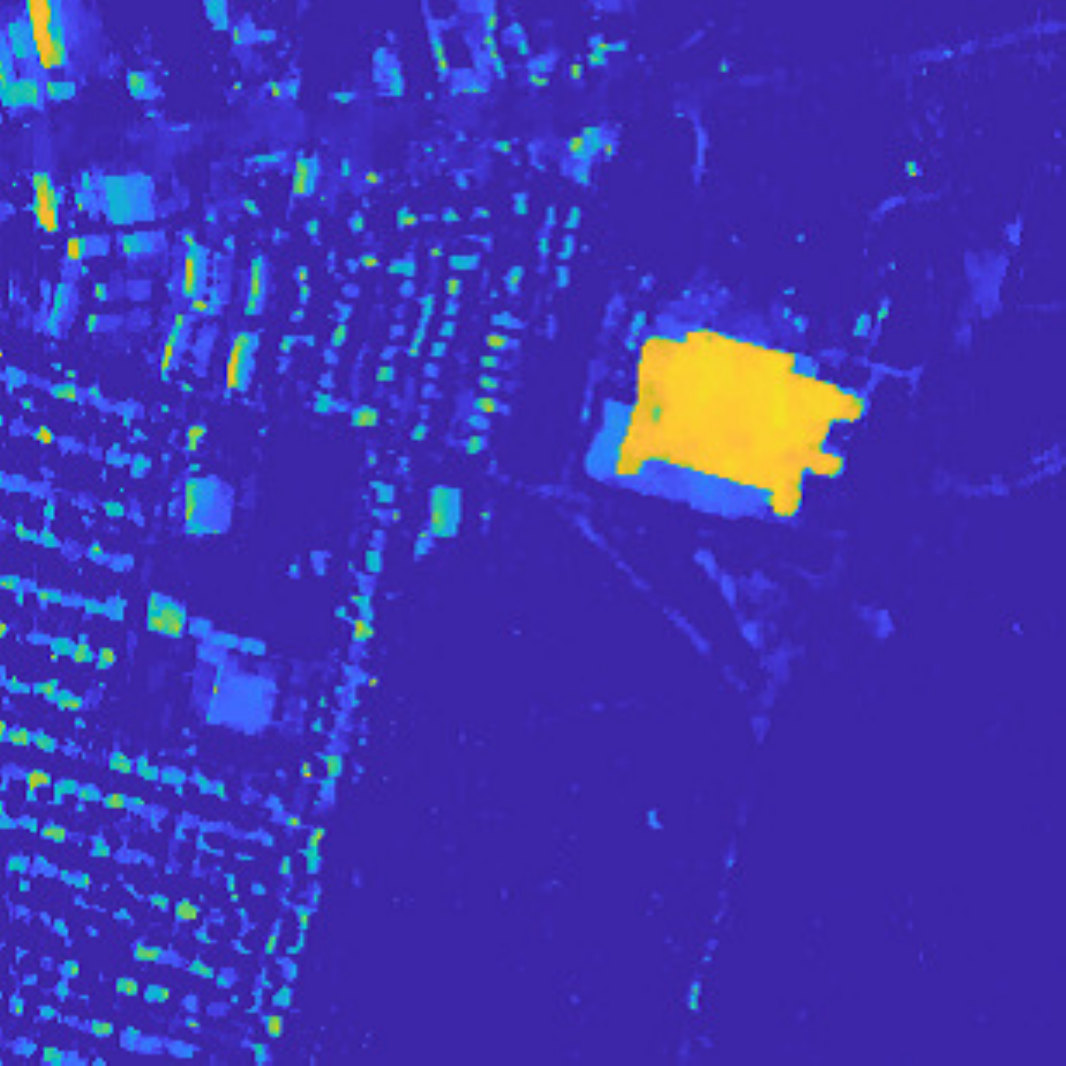}
		}
	\end{minipage}
	\begin{minipage}[t]{0.075\hsize}
		\centerline{
			\includegraphics[height = 40pt]{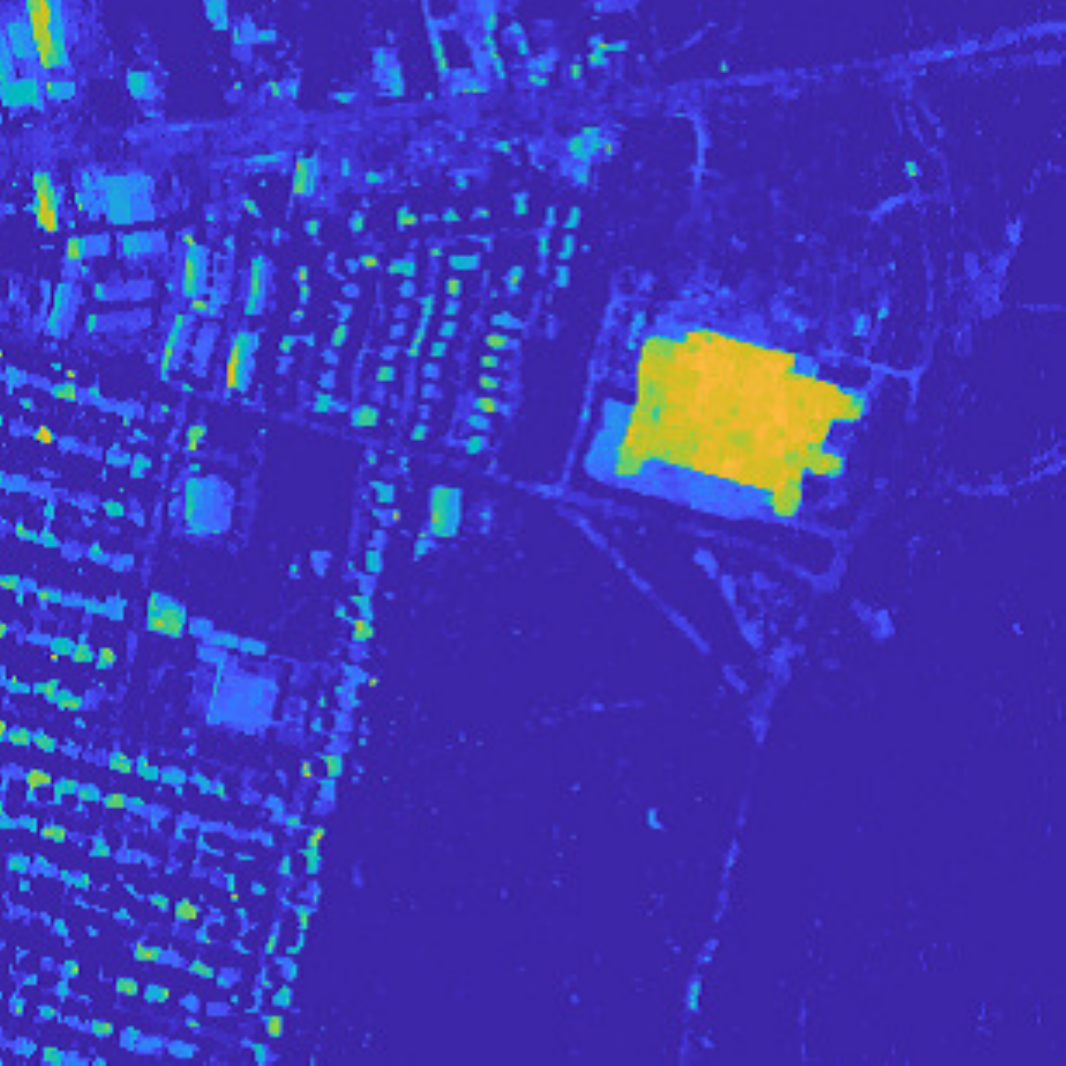}
		}
	\end{minipage}
	\begin{minipage}[t]{0.075\hsize}
		\centerline{
			\includegraphics[height = 40pt]{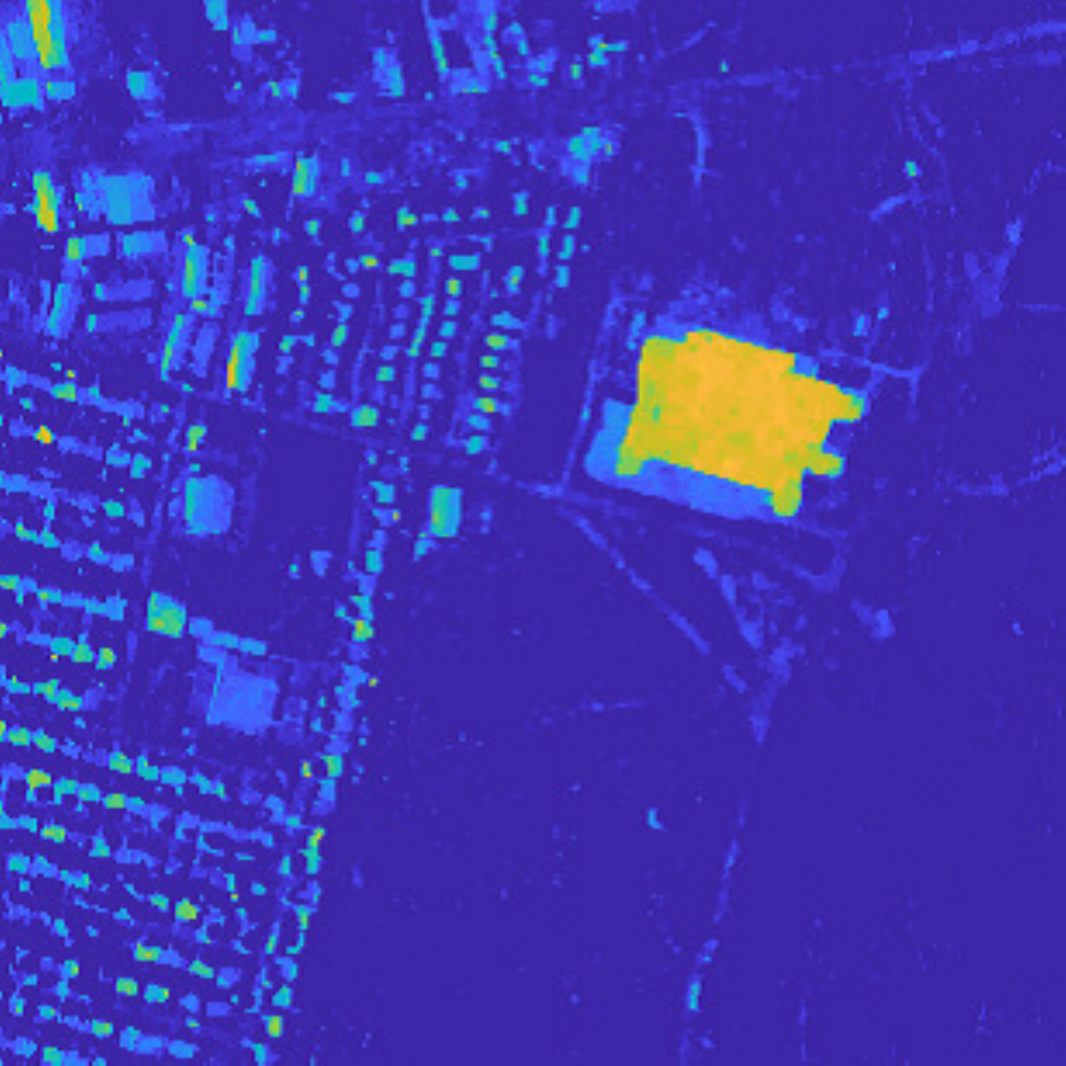}
		}
	\end{minipage}
	\begin{minipage}[t]{0.02\hsize}
		\centerline{
			\includegraphics[height = 40pt]{./fig/colorbar_20.png}
		}
	\end{minipage}
	
	\begin{minipage}[t]{0.075\hsize}
		\centerline{
			(a)
		}
	\end{minipage}
	\begin{minipage}[t]{0.075\hsize}
		\centerline{
			(b)
		}
	\end{minipage}
	\begin{minipage}[t]{0.075\hsize}
		\centerline{
			(c)
		}
	\end{minipage}
	\begin{minipage}[t]{0.075\hsize}
		\centerline{
			(d)
		}
	\end{minipage}
	\begin{minipage}[t]{0.075\hsize}
		\centerline{
			(e)
		}
	\end{minipage}
	\begin{minipage}[t]{0.075\hsize}
		\centerline{
			(f)
		}
	\end{minipage}
	\begin{minipage}[t]{0.075\hsize}
		\centerline{
			(g)
		}
	\end{minipage}
	\begin{minipage}[t]{0.075\hsize}
		\centerline{
			{(h)}
		}
	\end{minipage}
	\begin{minipage}[t]{0.075\hsize}
		\centerline{
			{(i)}
		}
	\end{minipage}
	\begin{minipage}[t]{0.075\hsize}
		\centerline{
			\textbf{(j)}
		}
	\end{minipage}
	\begin{minipage}[t]{0.075\hsize}
		\centerline{
			\textbf{(k)}
		}
	\end{minipage}
	\begin{minipage}[t]{0.075\hsize}
		\centerline{
			\textbf{(l)}
		}
	\end{minipage}
	\begin{minipage}[t]{0.02\hsize}
		\centerline{
			~
		}
	\end{minipage}
	
	\caption{Unmixing results of abundance maps for the \textit{Urban} experiments in Case 8. (a): Original abundance maps. (b): CLSUnSAL~\cite{iordache2014collaborative}. (c): JSTV~\cite{aggarwal2016hyperspectral}. (d): RSSUn-TV~\cite{wang2019row}. (e): LGSU~\cite{shen2022superpixel}. (f): UnDIP~\cite{UnDIP_RastiB_2022}. (g): EGU-Net~\cite{hong2022endmember}. (h): RDSWSU~\cite{rs_Deng_RobustDual_2023}. (i): MdLRR~\cite{MDLRR_WuLing_2023}. (j): \textbf{\Ourss (HTV)}. (k): \textbf{\Ourss (SSTV)}. (l): \textbf{\Ourss (HSSTV)}.}
	\label{real_urban_noniid_urban_Noniid_0.05_0.05}
\end{figure*}

\begin{figure*}[t]
\centering
    \begin{minipage}[t]{0.07\hsize}
        \centerline{
        \includegraphics[width=\hsize]{./fig/real/original/0.1_0_0/Orig_HSI-eps-converted-to.pdf}
        }
    \end{minipage}
    \begin{minipage}[t]{0.07\hsize}
        \centerline{
        \includegraphics[width=\hsize]{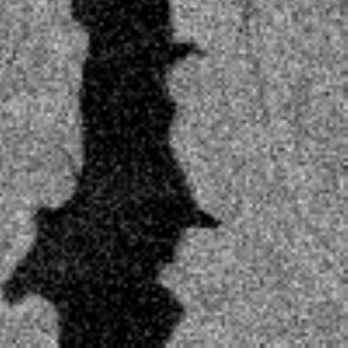}
        }
    \end{minipage}
    \begin{minipage}[t]{0.07\hsize}
        \centerline{
        \includegraphics[width=\hsize]{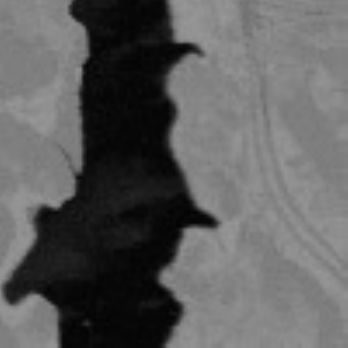}
        }
    \end{minipage}
    \begin{minipage}[t]{0.07\hsize}
    	\centerline{
    		\includegraphics[width=\hsize]{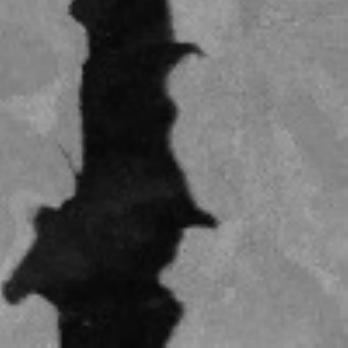}
    	}
    \end{minipage}
    \begin{minipage}[t]{0.07\hsize}
    	\centerline{
    		\includegraphics[width=\hsize]{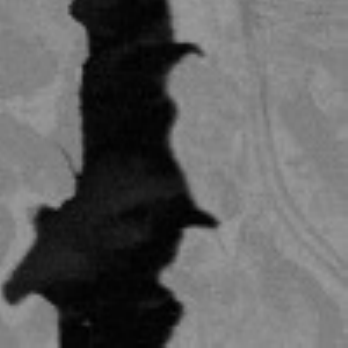}
    	}
    \end{minipage}
    \begin{minipage}[t]{0.07\hsize}
    	\centerline{
    		\includegraphics[width=\hsize]{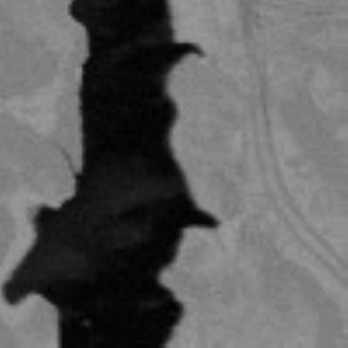}
    	}
    \end{minipage}
	\begin{minipage}[t]{0.07\hsize}
		\centerline{
			\includegraphics[width=\hsize]{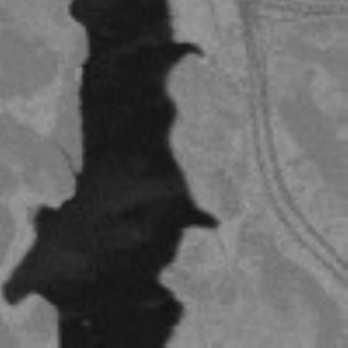}
		}
	\end{minipage}
	\begin{minipage}[t]{0.07\hsize}
	\centerline{
		\includegraphics[width=\hsize]{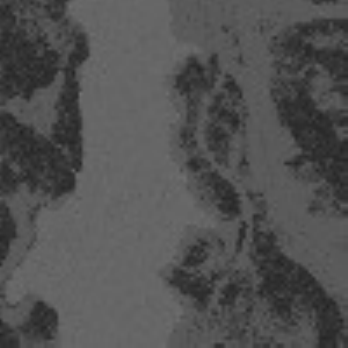}
	}
	\end{minipage}
	\begin{minipage}[t]{0.07\hsize}
	\centerline{
		\includegraphics[width=\hsize]{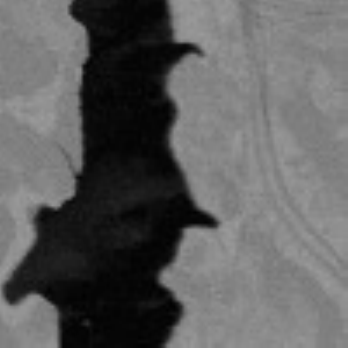}
	}
	\end{minipage}
	\begin{minipage}[t]{0.07\hsize}
		\centerline{
			\includegraphics[width=\hsize]{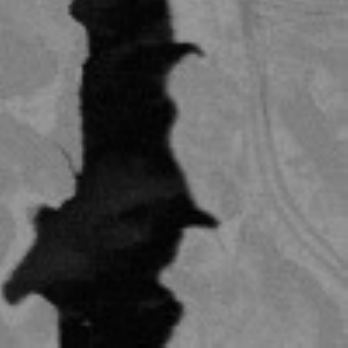}
		}
	\end{minipage}
	\begin{minipage}[t]{0.07\hsize}
		\centerline{
			\includegraphics[width=\hsize]{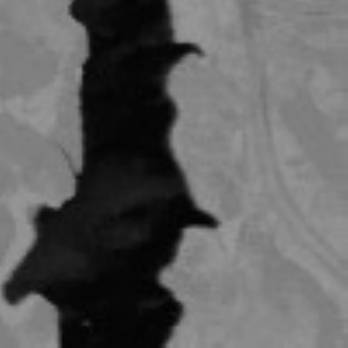}
		}
	\end{minipage}
	\begin{minipage}[t]{0.07\hsize}
		\centerline{
			\includegraphics[width=\hsize]{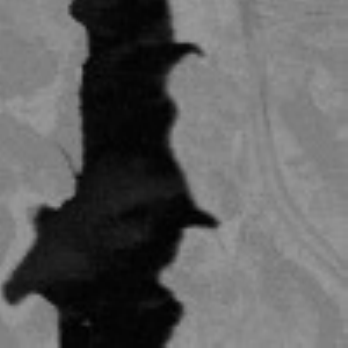}
		}
	\end{minipage}
	\begin{minipage}[t]{0.07\hsize}
		\centerline{
			\includegraphics[width=\hsize]{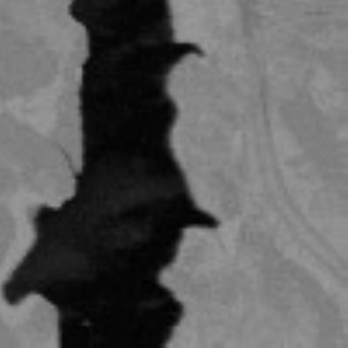}
		}
	\end{minipage}

    \begin{minipage}[t]{0.07\hsize}
        \centerline{
        (a)
        }
    \end{minipage}
    \begin{minipage}[t]{0.07\hsize}
        \centerline{
        (b)
        }
    \end{minipage}
    \begin{minipage}[t]{0.07\hsize}
        \centerline{
        (c)
        }
    \end{minipage}
    \begin{minipage}[t]{0.07\hsize}
        \centerline{
        (d)
        }
    \end{minipage}        
    \begin{minipage}[t]{0.07\hsize}
        \centerline{
        (e)
        }
    \end{minipage}
    \begin{minipage}[t]{0.07\hsize}
        \centerline{
        (f)
        }
    \end{minipage}
    \begin{minipage}[t]{0.07\hsize}
        \centerline{
        (g)
        }
    \end{minipage}
    \begin{minipage}[t]{0.07\hsize}
        \centerline{
        (h)
        }
    \end{minipage}
    \begin{minipage}[t]{0.07\hsize}
        \centerline{
        {(i)}
        }
    \end{minipage}
	\begin{minipage}[t]{0.07\hsize}
		\centerline{
			{(j)}
		}
	\end{minipage}
	\begin{minipage}[t]{0.07\hsize}
	\centerline{
		\textbf{(k)}
	}
	\end{minipage}
	\begin{minipage}[t]{0.07\hsize}
	\centerline{
		\textbf{(l)}
	}
	\end{minipage}
	\begin{minipage}[t]{0.07\hsize}
	\centerline{
		\textbf{(m)}
	}
	\end{minipage}

\caption{Reconstructed HS image results for the \textit{Jasper Ridge} experiments in Case 2. (a): Original HS image. (b): Noisy image. (c): CLSUnSAL \cite{iordache2014collaborative}.  (d): JSTV~\cite{aggarwal2016hyperspectral}. (e): RSSUn-TV~\cite{wang2019row}. (f): LGSU~\cite{shen2022superpixel}. (g): UnDIP~\cite{UnDIP_RastiB_2022}. (h): EGU-Net~\cite{hong2022endmember}. (i): RDSWSU~\cite{rs_Deng_RobustDual_2023}. (j): MdLRR~\cite{MDLRR_WuLing_2023}. (k): \textbf{\Ourss (HTV)}. (l): \textbf{\Ourss (SSTV)}. (m): \textbf{\Ourss (HSSTV)}.}
\label{real_HSI_0.1_0_0}
\end{figure*}

\begin{figure*}[!h]
	\centering
	\begin{minipage}[t]{0.07\hsize}
		\centerline{
			\includegraphics[width=\hsize]{./fig/real_samson/original/0.1_0.05_0.05/Orig_HSI-eps-converted-to.pdf}
		}
	\end{minipage}
	\begin{minipage}[t]{0.07\hsize}
		\centerline{
			\includegraphics[width=\hsize]{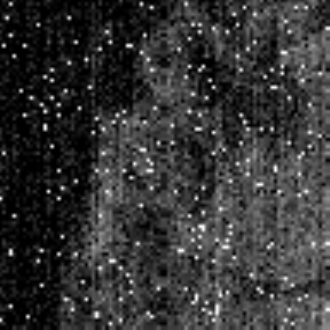}
		}
	\end{minipage}
	\begin{minipage}[t]{0.07\hsize}
		\centerline{
			\includegraphics[width=\hsize]{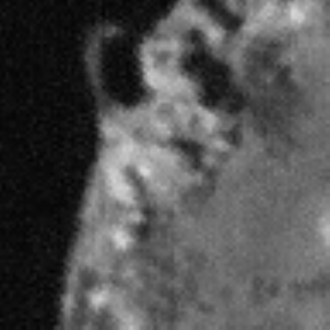}
		}
	\end{minipage}
	\begin{minipage}[t]{0.07\hsize}
		\centerline{
			\includegraphics[width=\hsize]{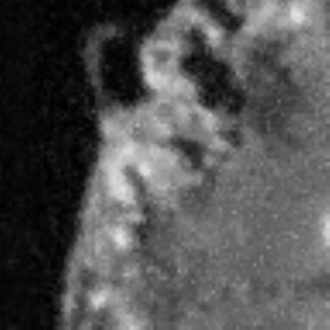}
		}
	\end{minipage}
	\begin{minipage}[t]{0.07\hsize}
		\centerline{
			\includegraphics[width=\hsize]{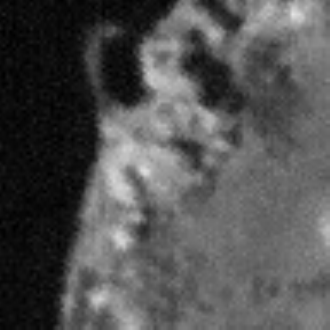}
		}
	\end{minipage}
	\begin{minipage}[t]{0.07\hsize}
		\centerline{
			\includegraphics[width=\hsize]{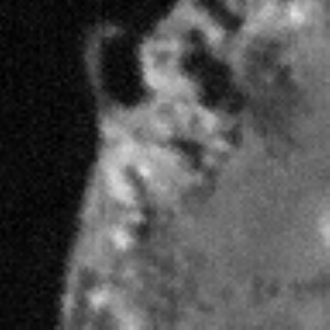}
		}
	\end{minipage}
	\begin{minipage}[t]{0.07\hsize}
		\centerline{
			\includegraphics[width=\hsize]{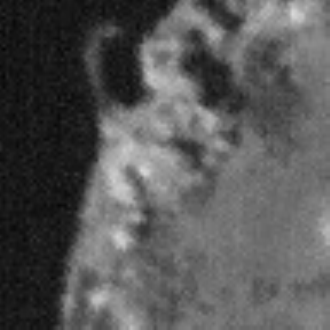}
		}
	\end{minipage}
	\begin{minipage}[t]{0.07\hsize}
	\centerline{
		\includegraphics[width=\hsize]{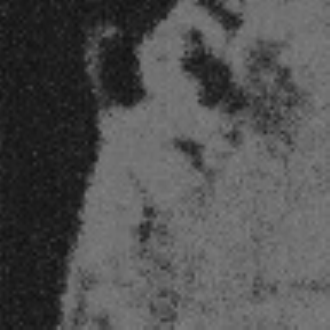}
	}
	\end{minipage}
	\begin{minipage}[t]{0.07\hsize}
	\centerline{
		\includegraphics[width=\hsize]{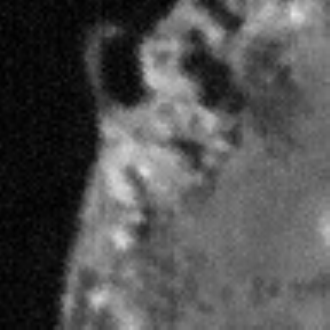}
	}
	\end{minipage}
	\begin{minipage}[t]{0.07\hsize}
		\centerline{
			\includegraphics[width=\hsize]{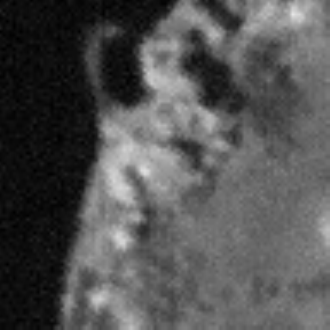}
		}
	\end{minipage}
	\begin{minipage}[t]{0.07\hsize}
		\centerline{
			\includegraphics[width=\hsize]{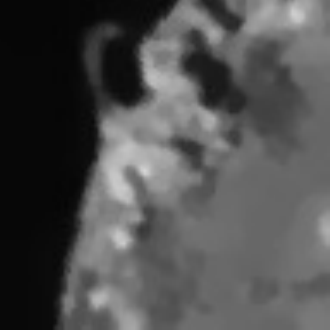}
		}
	\end{minipage}
	\begin{minipage}[t]{0.07\hsize}
		\centerline{
			\includegraphics[width=\hsize]{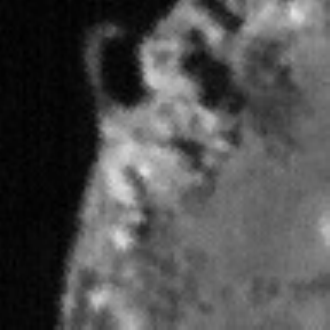}
		}
	\end{minipage}
	\begin{minipage}[t]{0.07\hsize}
		\centerline{
			\includegraphics[width=\hsize]{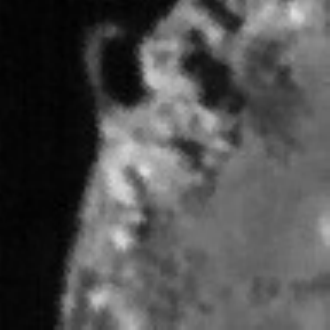}
		}
	\end{minipage}
	
	\begin{minipage}[t]{0.07\hsize}
		\centerline{
			(a)
		}
	\end{minipage}
	\begin{minipage}[t]{0.07\hsize}
		\centerline{
			(b)
		}
	\end{minipage}
	\begin{minipage}[t]{0.07\hsize}
		\centerline{
			(c)
		}
	\end{minipage}
	\begin{minipage}[t]{0.07\hsize}
		\centerline{
			(d)
		}
	\end{minipage}
	\begin{minipage}[t]{0.07\hsize}
		\centerline{
			(e)
		}
	\end{minipage}
	\begin{minipage}[t]{0.07\hsize}
		\centerline{
			(f)
		}
	\end{minipage}
	\begin{minipage}[t]{0.07\hsize}
		\centerline{
			(g)
		}
	\end{minipage}
	\begin{minipage}[t]{0.07\hsize}
		\centerline{
			(h)
		}
	\end{minipage}
	\begin{minipage}[t]{0.07\hsize}
		\centerline{
			{(i)}
		}
	\end{minipage}
	\begin{minipage}[t]{0.07\hsize}
		\centerline{
			{(j)}
		}
	\end{minipage}
	\begin{minipage}[t]{0.07\hsize}
		\centerline{
			\textbf{(k)}
		}
	\end{minipage}
	\begin{minipage}[t]{0.07\hsize}
	\centerline{
		\textbf{(l)}
	}
	\end{minipage}
	\begin{minipage}[t]{0.07\hsize}
	\centerline{
		\textbf{(m)}
	}
	\end{minipage}

	\caption{Reconstructed HS image results for the \textit{samson} experiments in Case 6. (a): Original HS image. (b): Noisy image. (c): CLSUnSAL \cite{iordache2014collaborative}.  (d): JSTV~\cite{aggarwal2016hyperspectral}. (e): RSSUn-TV~\cite{wang2019row}. (f): LGSU~\cite{shen2022superpixel}. (g): UnDIP~\cite{UnDIP_RastiB_2022}. (h): EGU-Net~\cite{hong2022endmember}. (i): RDSWSU~\cite{rs_Deng_RobustDual_2023}. (j): MdLRR~\cite{MDLRR_WuLing_2023}. (k): \textbf{\Ourss (HTV)}. (l): \textbf{\Ourss (SSTV)}. (m): \textbf{\Ourss (HSSTV)}.}
	\label{real_samson_HSI_0.1_0.05_0.05}
\end{figure*}

\begin{figure*}[t]
	\centering
	\begin{minipage}[t]{0.07\hsize}
		\centerline{
			\includegraphics[width=\hsize]{./fig/real_urban_noniid/original/Noniid_0.05_0.05/Orig_HSI-eps-converted-to.pdf}
		}
	\end{minipage}
	\begin{minipage}[t]{0.07\hsize}
		\centerline{
			\includegraphics[width=\hsize]{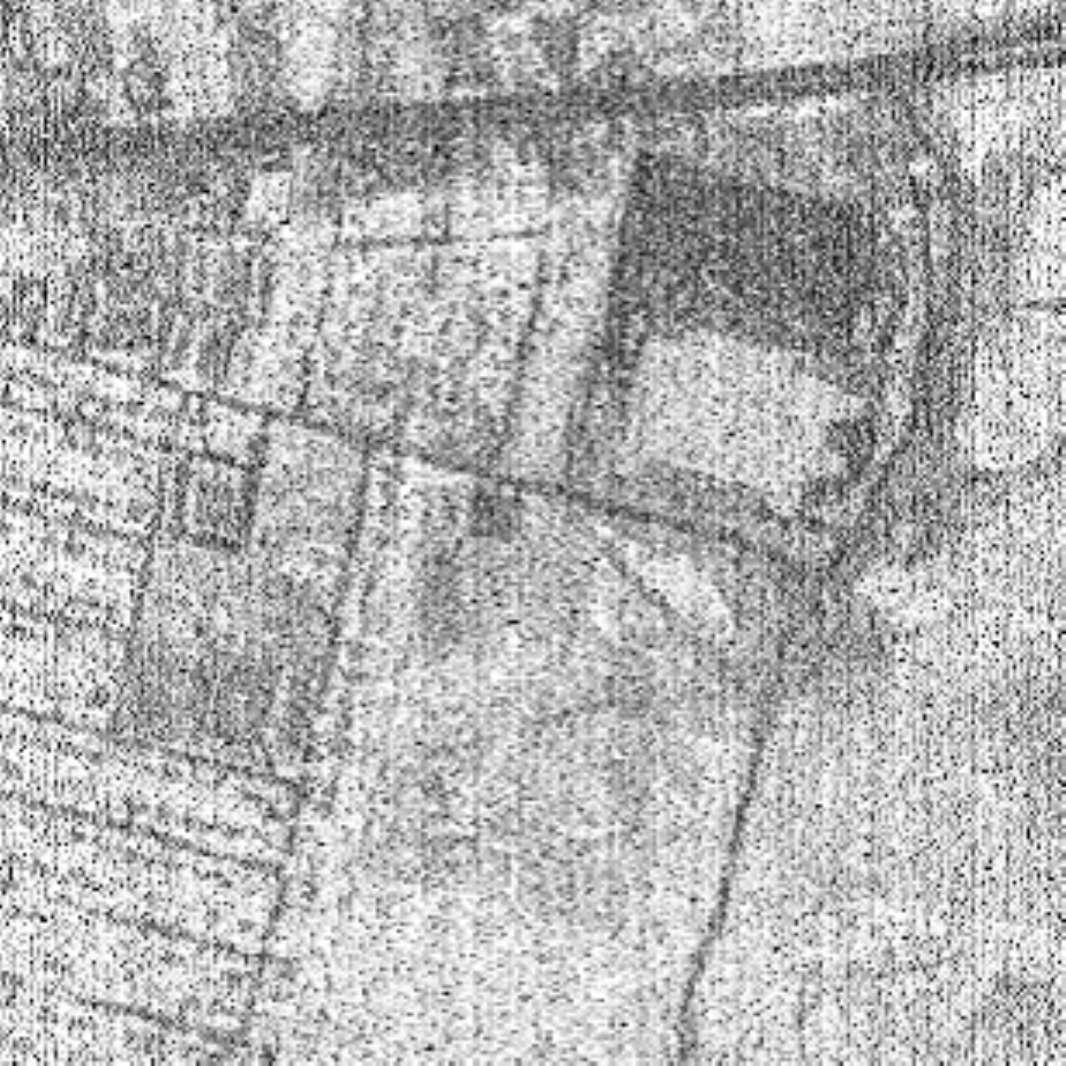}
		}
	\end{minipage}
	\begin{minipage}[t]{0.07\hsize}
		\centerline{
			\includegraphics[width=\hsize]{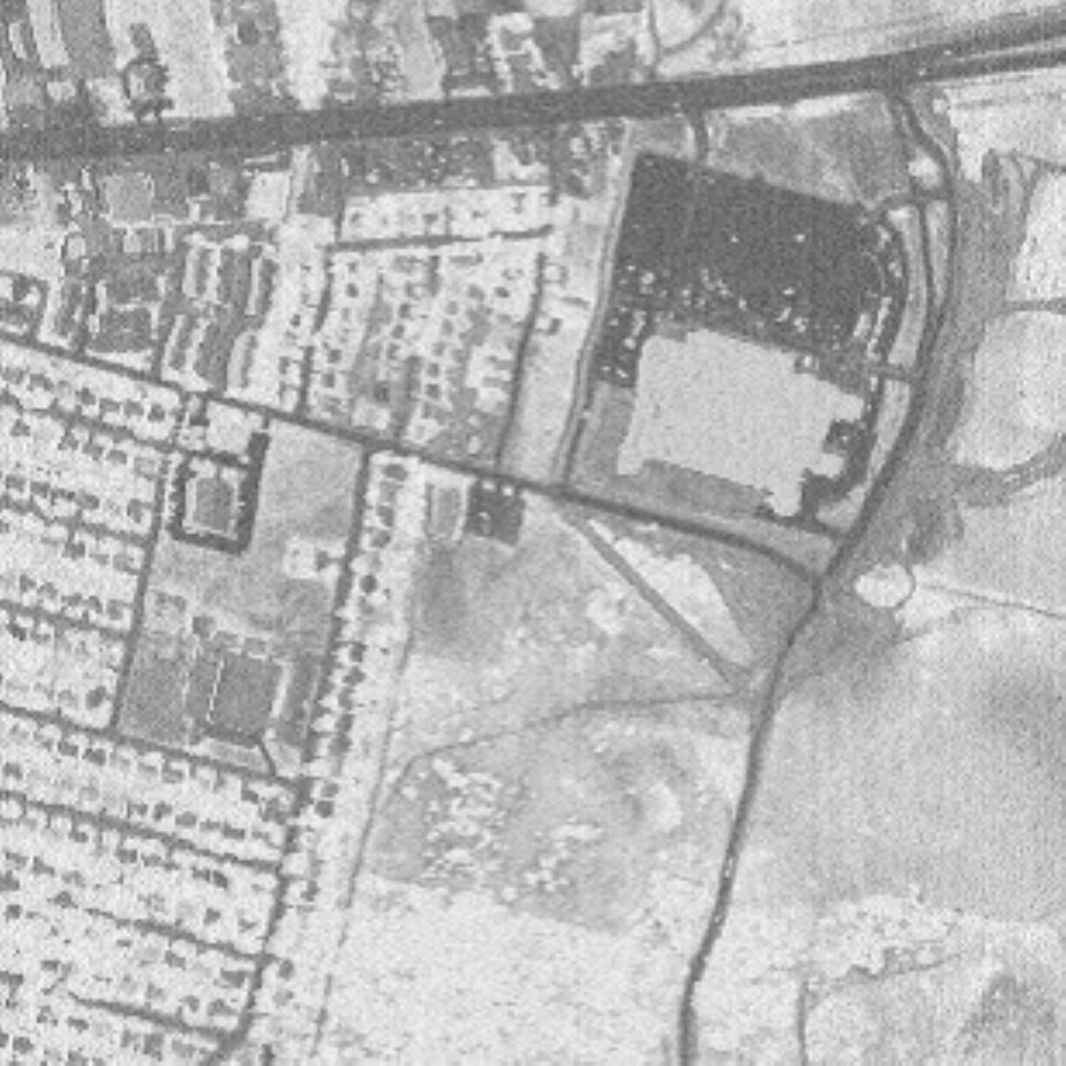}
		}
	\end{minipage}
	\begin{minipage}[t]{0.07\hsize}
		\centerline{
			\includegraphics[width=\hsize]{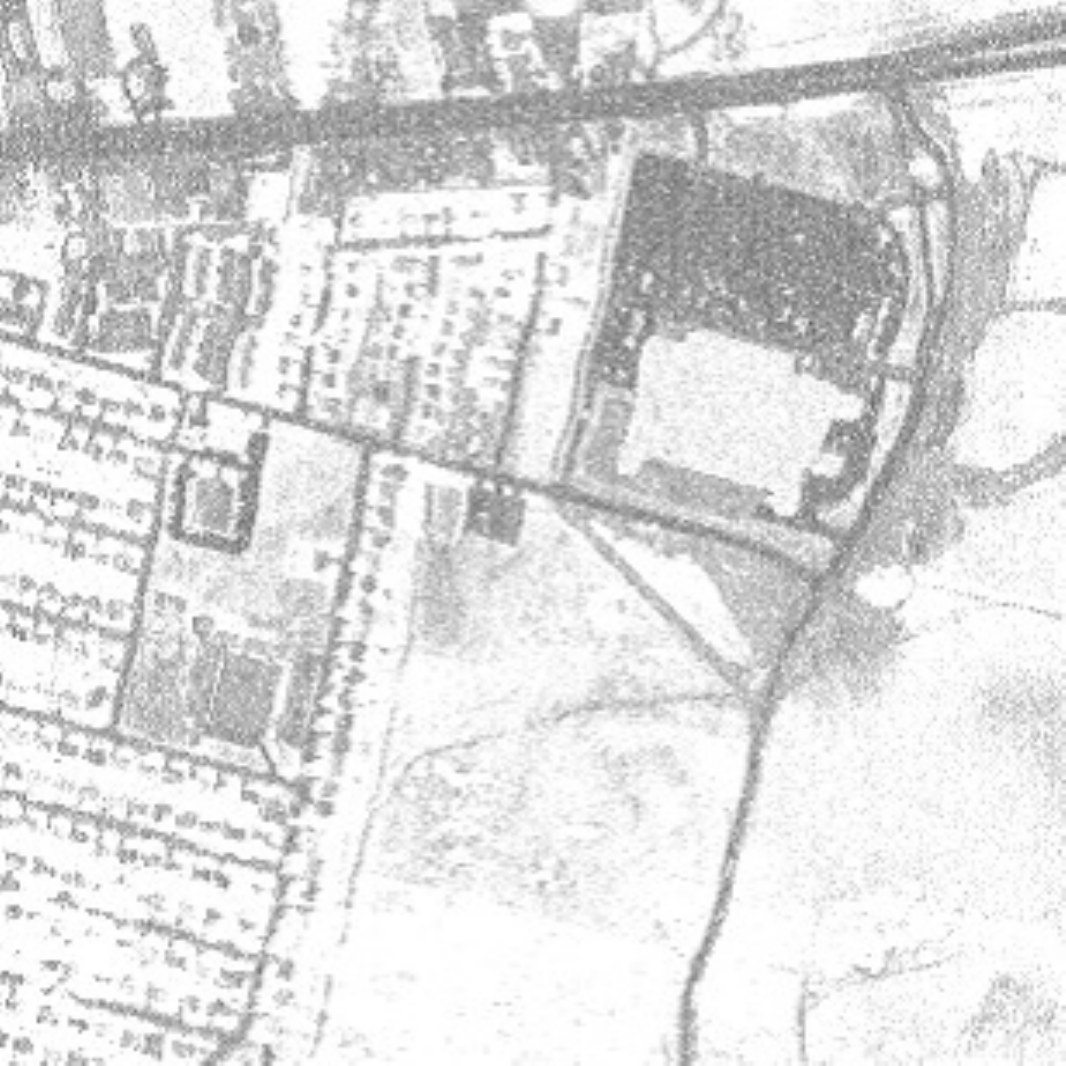}
		}
	\end{minipage}
	\begin{minipage}[t]{0.07\hsize}
		\centerline{
			\includegraphics[width=\hsize]{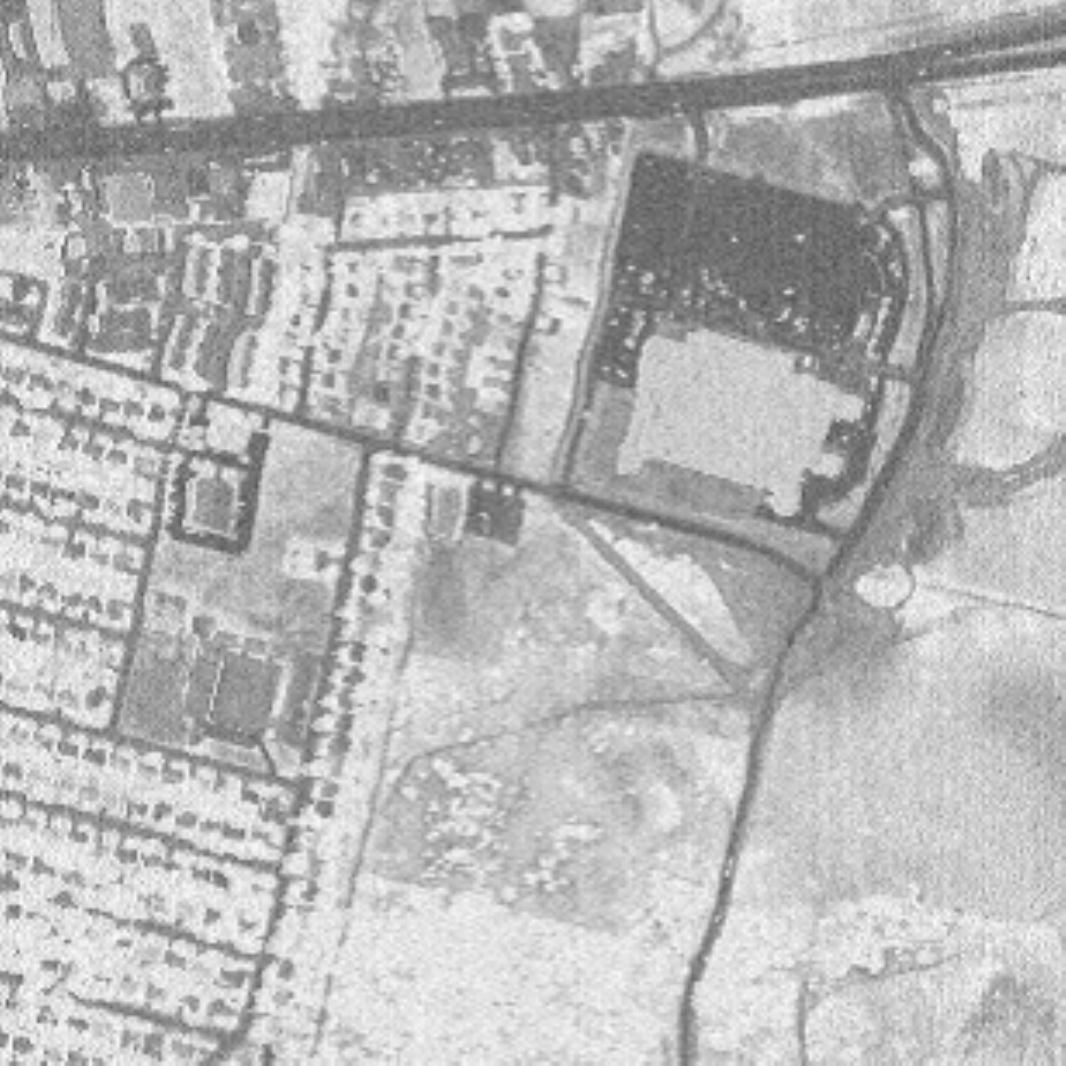}
		}
	\end{minipage}
	\begin{minipage}[t]{0.07\hsize}
		\centerline{
			\includegraphics[width=\hsize]{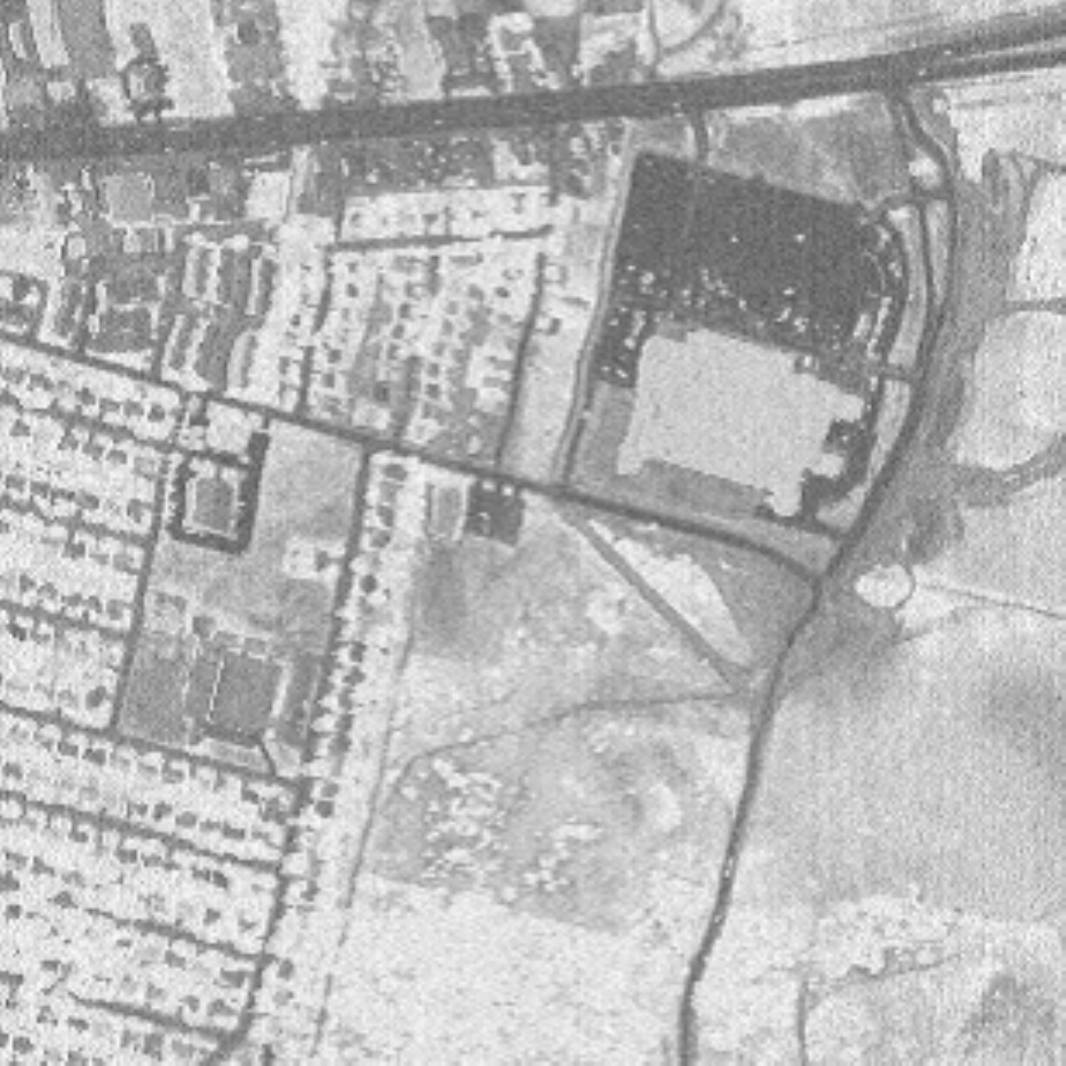}
		}
	\end{minipage}
	\begin{minipage}[t]{0.07\hsize}
		\centerline{
			\includegraphics[width=\hsize]{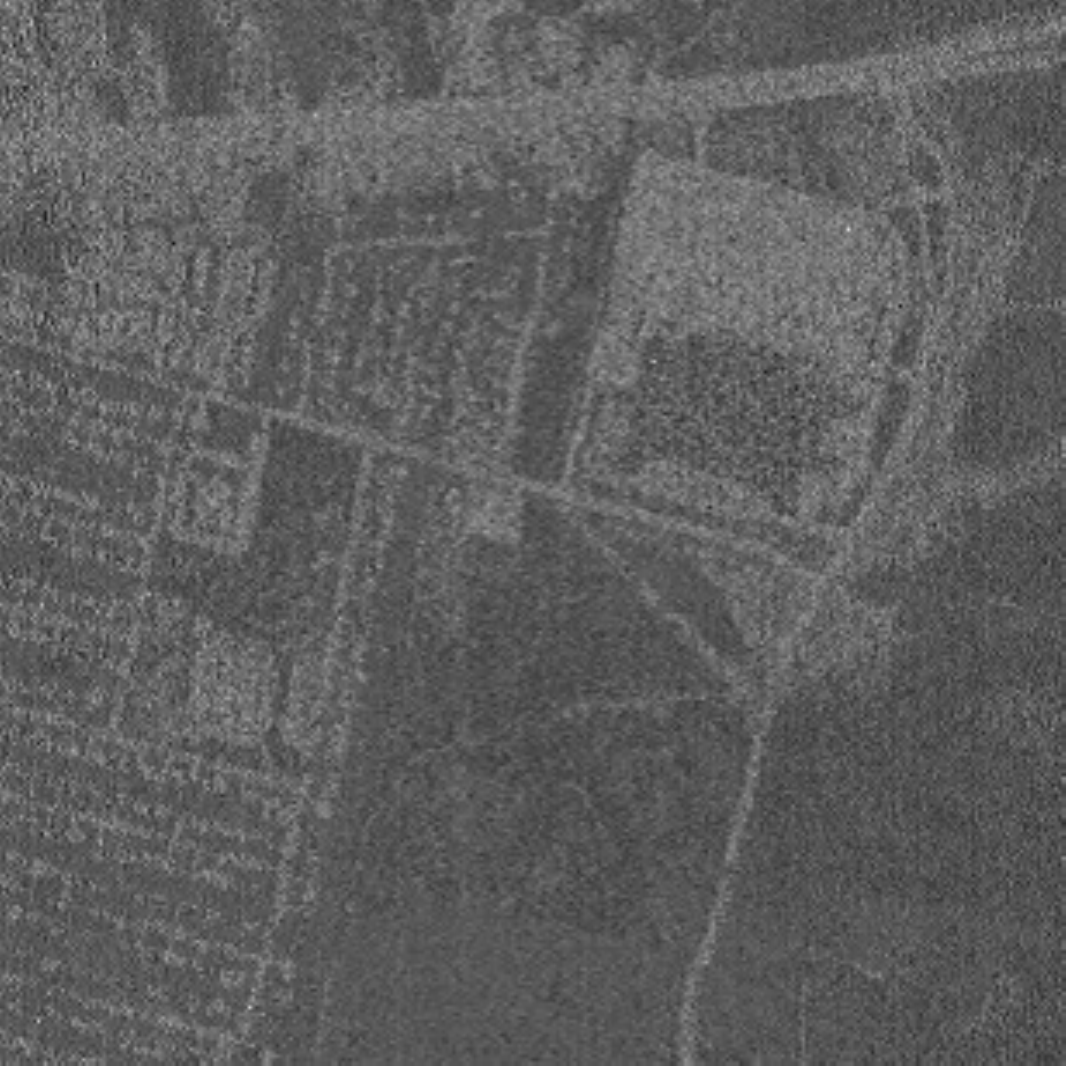}
		}
	\end{minipage}
	\begin{minipage}[t]{0.07\hsize}
		\centerline{
			\includegraphics[width=\hsize]{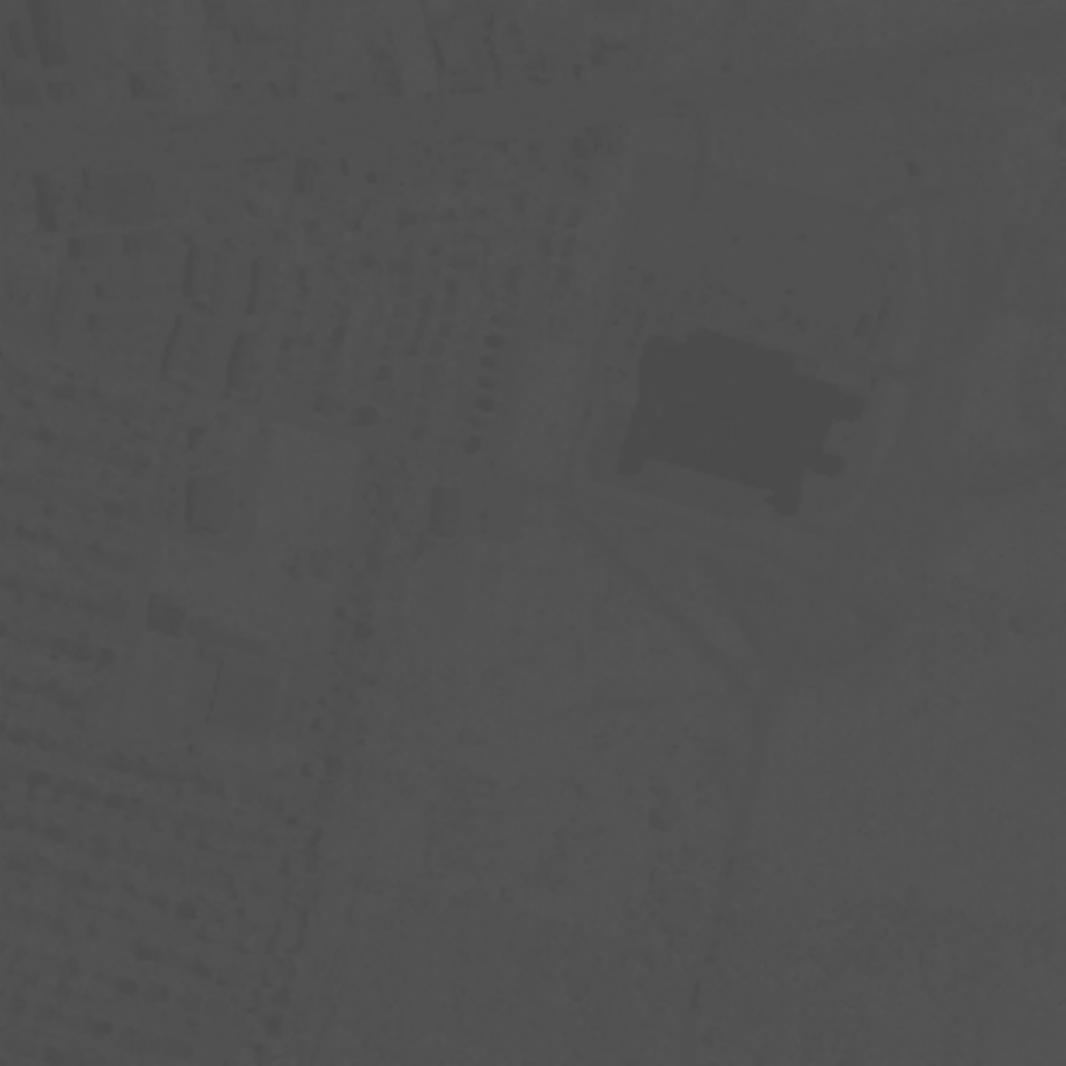}
		}
	\end{minipage}
	\begin{minipage}[t]{0.07\hsize}
		\centerline{
			\includegraphics[width=\hsize]{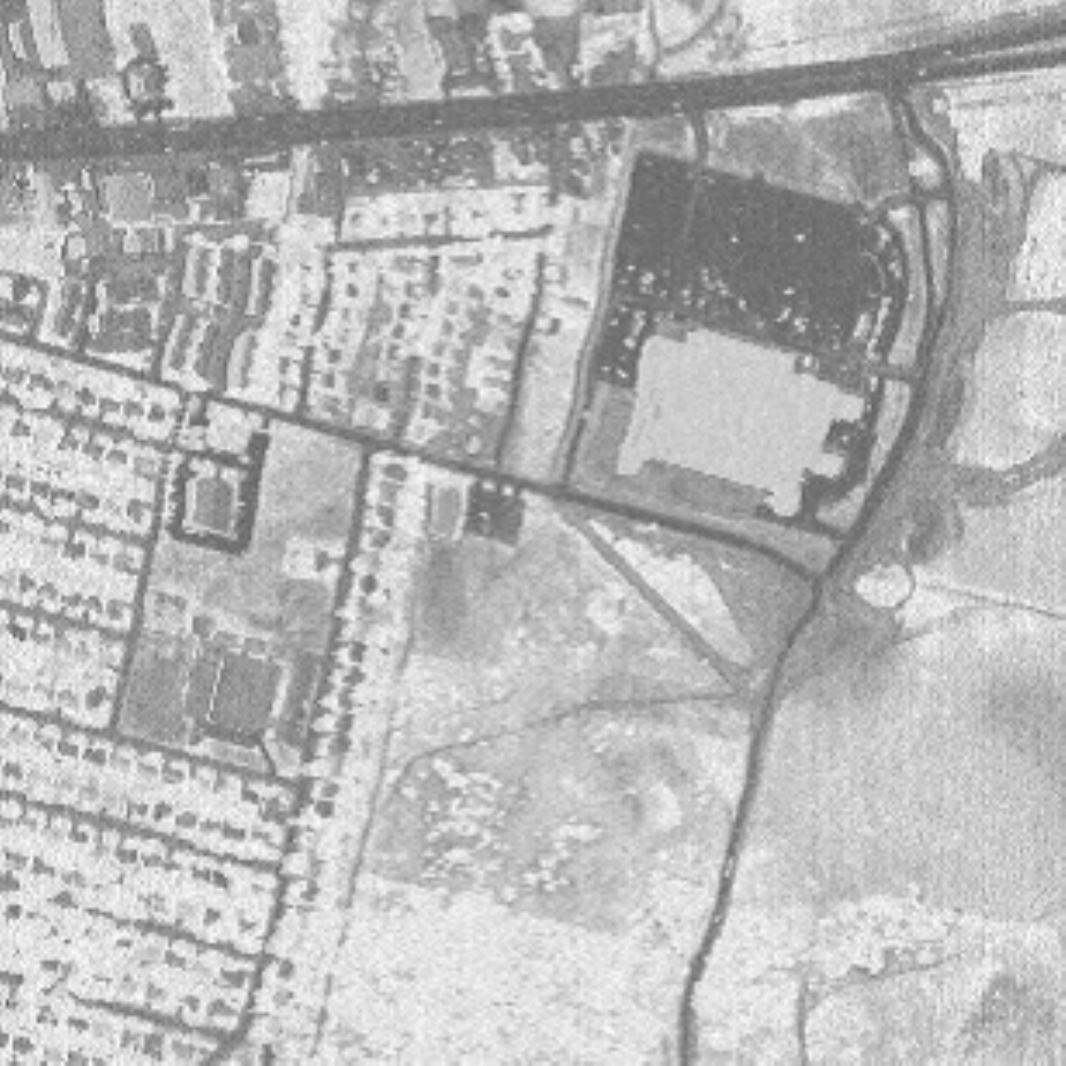}
		}
	\end{minipage}
	\begin{minipage}[t]{0.07\hsize}
		\centerline{
			\includegraphics[width=\hsize]{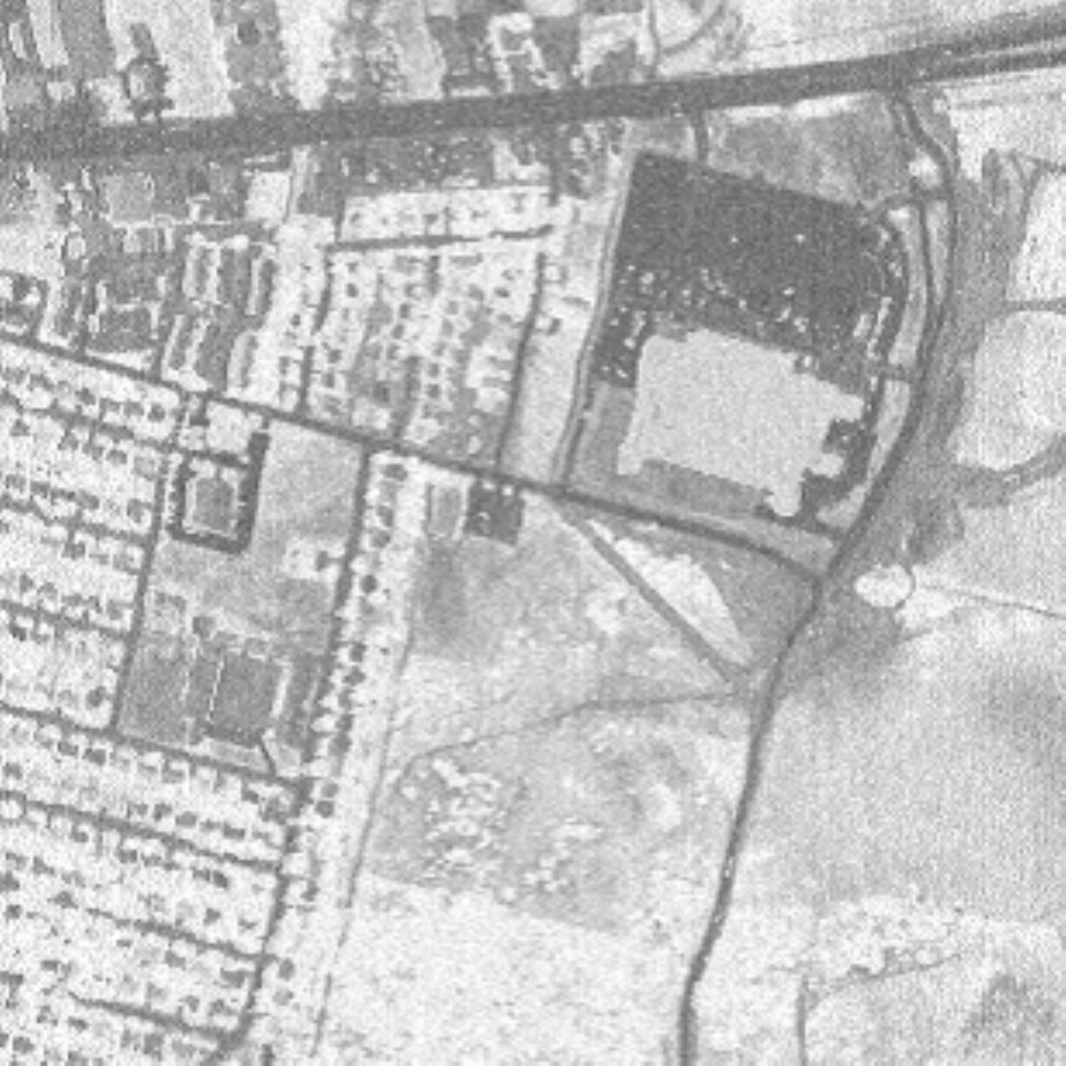}
		}
	\end{minipage}
	\begin{minipage}[t]{0.07\hsize}
		\centerline{
			\includegraphics[width=\hsize]{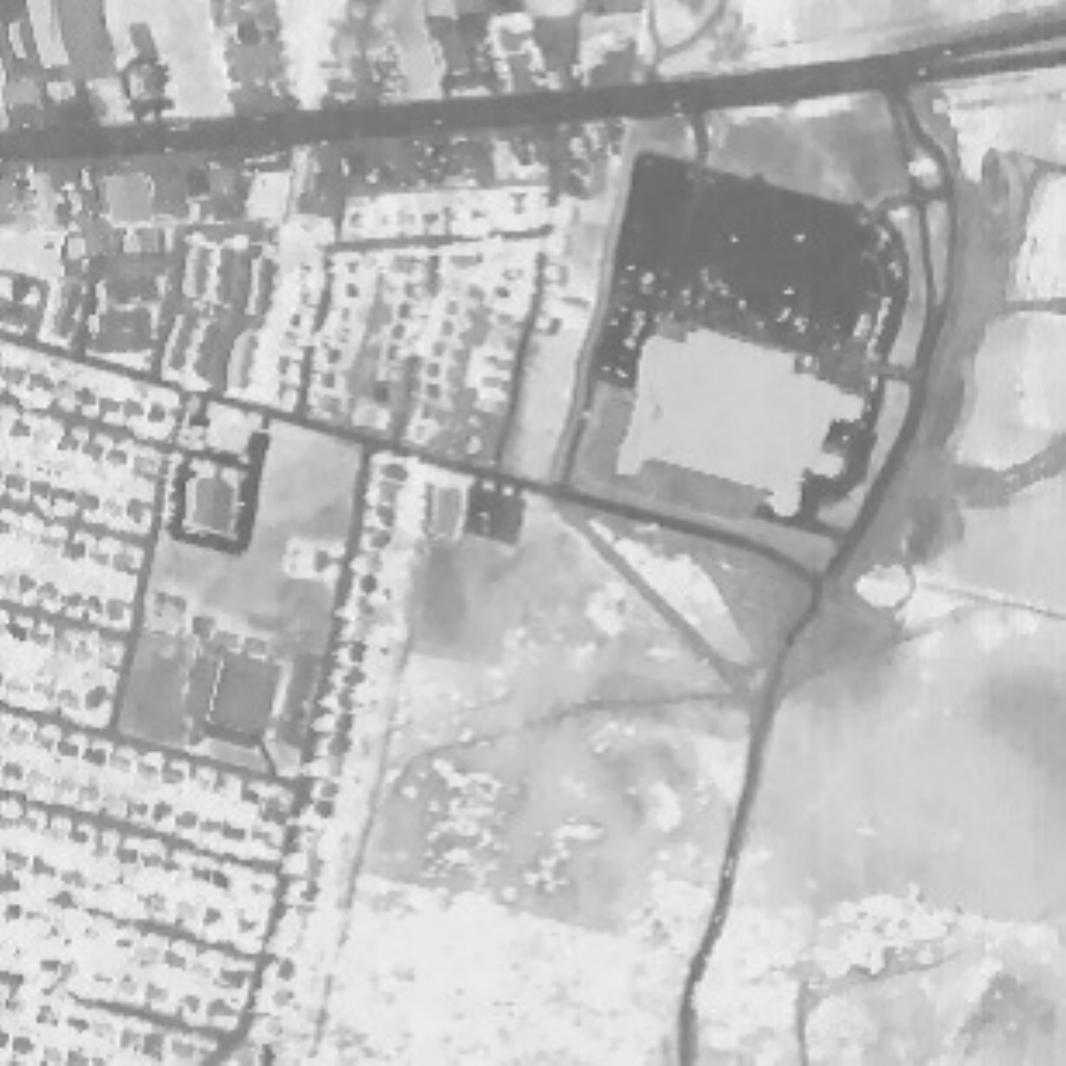}
		}
	\end{minipage}
	\begin{minipage}[t]{0.07\hsize}
		\centerline{
			\includegraphics[width=\hsize]{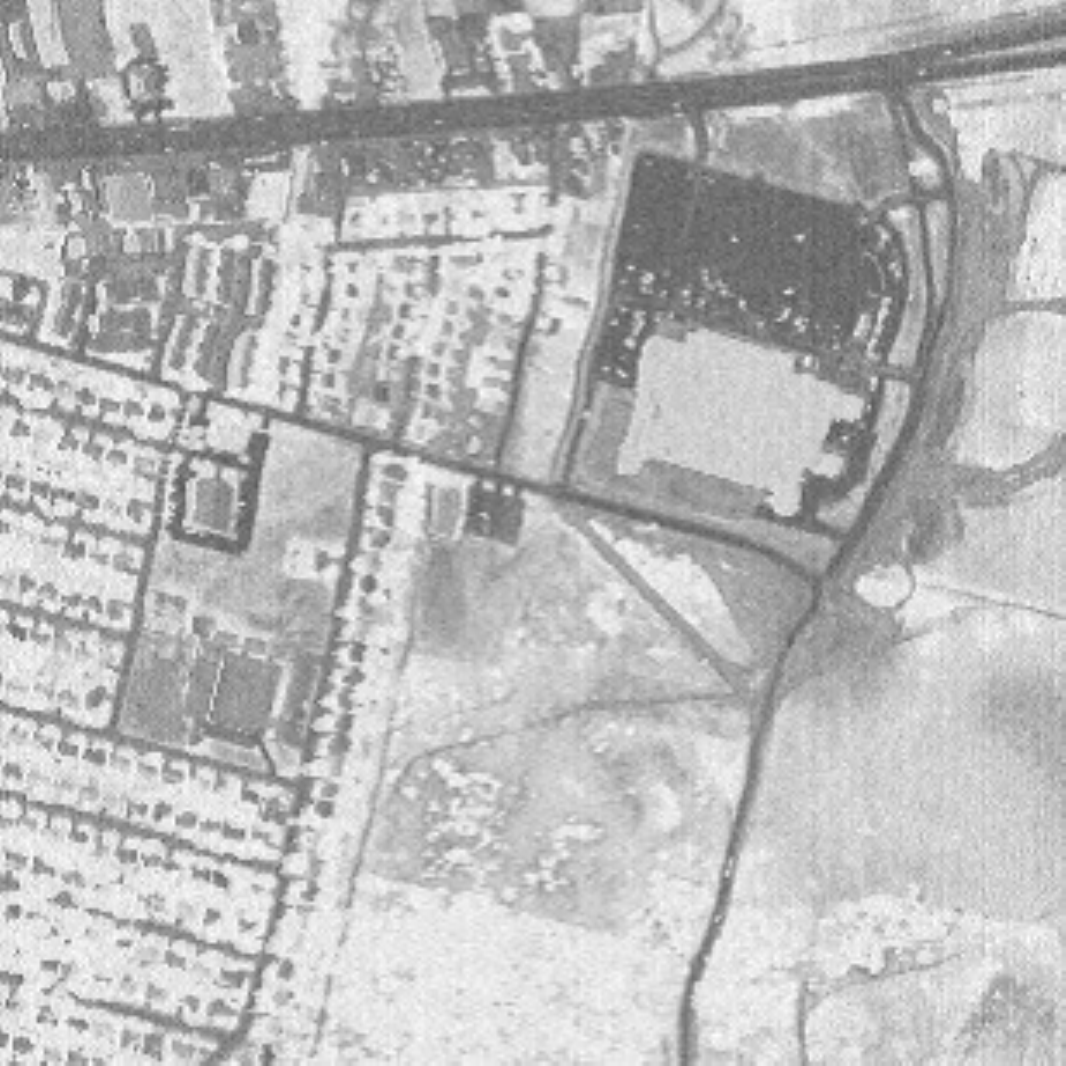}
		}
	\end{minipage}
	\begin{minipage}[t]{0.07\hsize}
		\centerline{
			\includegraphics[width=\hsize]{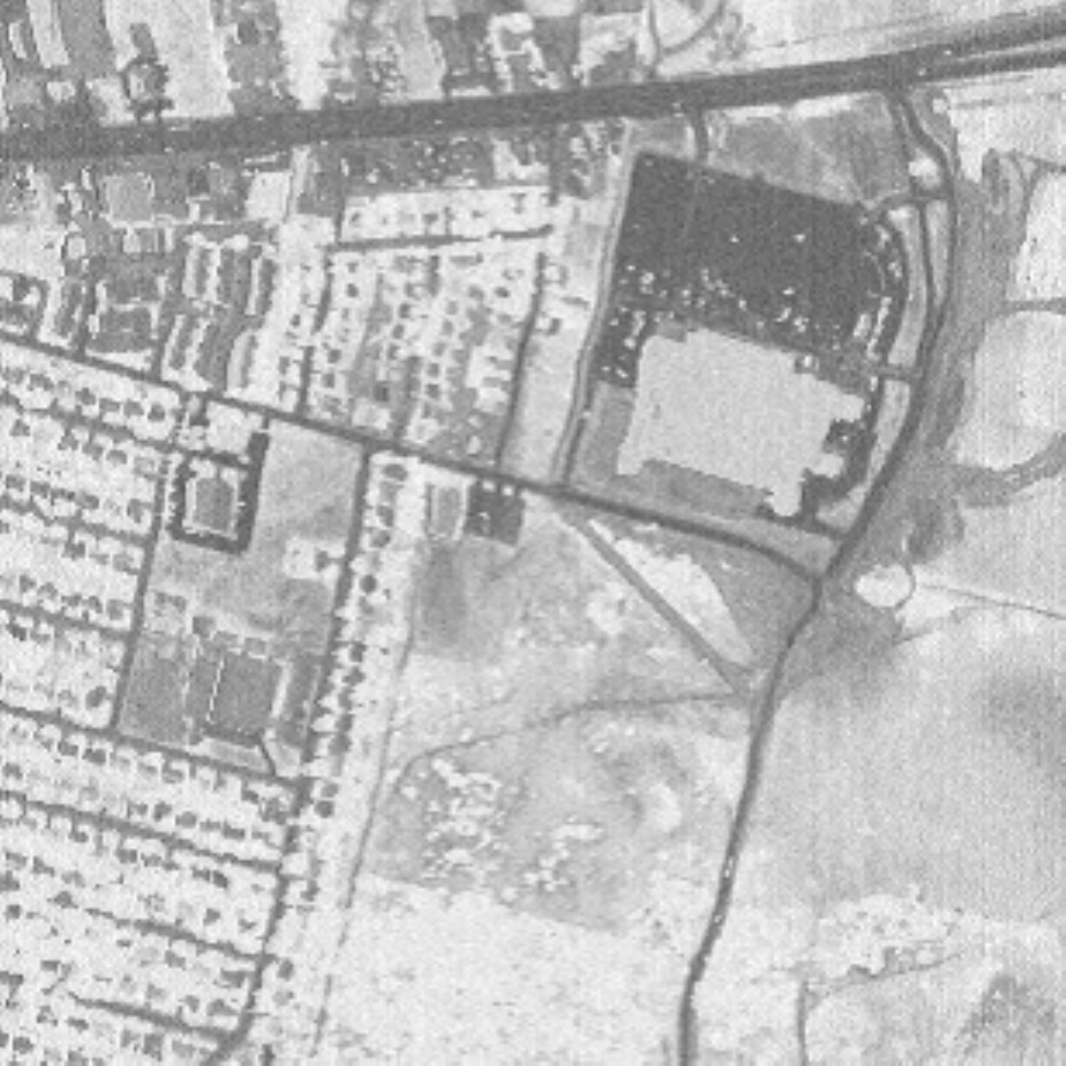}
		}
	\end{minipage}
	
	\begin{minipage}[t]{0.07\hsize}
		\centerline{
			(a)
		}
	\end{minipage}
	\begin{minipage}[t]{0.07\hsize}
		\centerline{
			(b)
		}
	\end{minipage}
	\begin{minipage}[t]{0.07\hsize}
		\centerline{
			(c)
		}
	\end{minipage}
	\begin{minipage}[t]{0.07\hsize}
		\centerline{
			(d)
		}
	\end{minipage}        
	\begin{minipage}[t]{0.07\hsize}
		\centerline{
			(e)
		}
	\end{minipage}
	\begin{minipage}[t]{0.07\hsize}
		\centerline{
			(f)
		}
	\end{minipage}
	\begin{minipage}[t]{0.07\hsize}
		\centerline{
			(g)
		}
	\end{minipage}
	\begin{minipage}[t]{0.07\hsize}
		\centerline{
			(h)
		}
	\end{minipage}
	\begin{minipage}[t]{0.07\hsize}
		\centerline{
			{(i)}
		}
	\end{minipage}
	\begin{minipage}[t]{0.07\hsize}
		\centerline{
			{(j)}
		}
	\end{minipage}
	\begin{minipage}[t]{0.07\hsize}
		\centerline{
			\textbf{(k)}
		}
	\end{minipage}
	\begin{minipage}[t]{0.07\hsize}
		\centerline{
			\textbf{(l)}
		}
	\end{minipage}
	\begin{minipage}[t]{0.07\hsize}
		\centerline{
			\textbf{(m)}
		}
	\end{minipage}

	\caption{Reconstructed HS image results for the \textit{Urban} experiments in Case 8. (a): Original HS image. (b): Noisy image. (c): CLSUnSAL \cite{iordache2014collaborative}.  (d): JSTV~\cite{aggarwal2016hyperspectral}. (e): RSSUn-TV~\cite{wang2019row}. (f): LGSU~\cite{shen2022superpixel}. (g): UnDIP~\cite{UnDIP_RastiB_2022}. (h): EGU-Net~\cite{hong2022endmember}. (i): RDSWSU~\cite{rs_Deng_RobustDual_2023}. (j): MdLRR~\cite{MDLRR_WuLing_2023}. (k): \textbf{\Ourss (HTV)}. (l): \textbf{\Ourss (SSTV)}. (m): \textbf{\Ourss (HSSTV)}.}
	\label{real_urban_noniid_HSI_Noniid_0.05_0.05}
\end{figure*}

\begin{table}[t]
	\caption{Averages of Running Times [s] in All Noise Cases for Each Dataset.}
	\vspace{-1mm}
	\label{tab:results_times}
	\centering
	\begin{tabular}{cccc}
		\toprule
		Methods &  \textit{Jasper Ridge} & \textit{Samson} & \textit{Urban} \\
		\cmidrule(lr){1-1} \cmidrule(lr){2-4}
		CLSUnSAL~\cite{iordache2014collaborative} & $1.359*10^{0}$ & $9.739*10^{-1}$ & $9.192*10^{1}$ \\ 
		\vspace{0.5mm}
		JSTV~\cite{aggarwal2016hyperspectral} & $7.776*10^{1}$ & $7.787*10^{1}$ & $5.709*10^{2}$ \\ 
		\vspace{0.5mm}
		RSSUn-TV~\cite{wang2019row} & $1.870*10^{0}$ & $1.615*10^{0}$ & $1.111*10^{2}$ \\ 
		\vspace{0.5mm}
		LGSU~\cite{shen2022superpixel} & $2.165*10^{1}$ & $1.720*10^{1}$ & $6.655*10^{2}$ \\ 
		\vspace{0.5mm}
		UnDIP~\cite{UnDIP_RastiB_2022} & $1.436*10^{2}$ & $1.030*10^{2}$ & $2.349*10^{3}$ \\ 
		\vspace{0.5mm}
		EGU-Net~\cite{hong2022endmember} & $1.669*10^{1}$ & $1.420*10^{1}$ & $2.360*10^{2}$ \\ 
		\vspace{0.5mm}
		RDSWSU~\cite{rs_Deng_RobustDual_2023} & $1.239*10^{1}$ & $9.790*10^{0}$ & $4.633*10^{2}$ \\ 
		\vspace{0.5mm}
		MdLRR~\cite{MDLRR_WuLing_2023} & $7.268*10^{0}$ & $6.592*10^{0}$ & $4.776*10^{2}$ \\ 
		\vspace{0.5mm}
		\textbf{\Ours (HTV)} & $2.252*10^{1}$ & $4.674*10^{1}$ & $5.768*10^{2}$ \\ 
		\vspace{0.5mm}
		\textbf{\Ours (SSTV)} & $2.822*10^{1}$ & $5.372*10^{1}$ & $8.269*10^{2}$ \\ 
		\vspace{0.5mm}
		\textbf{\Ours (HSSTV)} & $2.440*10^{1}$ & $5.141*10^{1}$ & $8.265*10^{2}$ \\ 
		\bottomrule
	\end{tabular}
\end{table}

\subsection{Experimental Results With Synthetic HS Images}
Tables~\ref{tab:result_synth_Legendre},~\ref{tab:result_synth}, and~\ref{tab:result_synth_3} show the SRE, RMSE, Ps, MPSNR, and MSSIM results for \textit{Synth 1}, \textit{Synth 2}, and \textit{Synth 3}, respectively. 
The best and second-best results are highlighted in bold and underlined, respectively. 
CLSUnSAL, JSTV, RSSUn-TV, and LGSU were not good in all cases.
For \textit{Synth 3}, the unmixing performance of JSTV was degraded. This may have been an issue with the algorithm because the value of the optimization problem was not fully reduced.\footnote{We used the program code implemented by the authors. Neither varying the step size nor increasing the maximum number of iterations improved the results.}
The results of RDSWSU and MdLRR were better than CLSUnSAL, JSTV, RSSUn-TV, and LGSU.
However, their performance dropped when HS images were contaminated with sparse noise and stripe noise (Cases 3, 4, 5, 6, and 8).
UnDIP and EGU-Net yielded worse results than the other existing methods.
This is because UnDIP and EGU-Net do not capture the sparsity of abundance maps.
In contrast, \Ourss yielded the best SRE, RMSE, Ps, MPSNR, and MSSIM values in the cases where the HS image is contaminated with noise that can be handled by the existing methods (Cases 1 and 2 for CLSUnSAL, RSSUn-TV, and LGSU, and Cases 1, 2, 3, and 4 for JSTV), except for Case 1 using \textit{Synth 1}. 
This indicates that the image-domain regularizations can improve the unmixing performance. 
In addition, \Ourss achieved the best performance in the other cases (Cases 5 and 6). This is due to the fact that \Ourss can handle all three types of noise. 
In the comparison of the image-domain regularizations, HTV performed better in almost all cases, and HSSTV performed better in the \textit{Synth 1} and \textit{Synth 2} experiments when HS images were contaminated with non-i.i.d. Gaussian noise (Cases 7 and 8).

\begin{figure*}[t]
	\centering
	
	\begin{minipage}{0.19\hsize}
		\centerline{\includegraphics[width=\hsize]{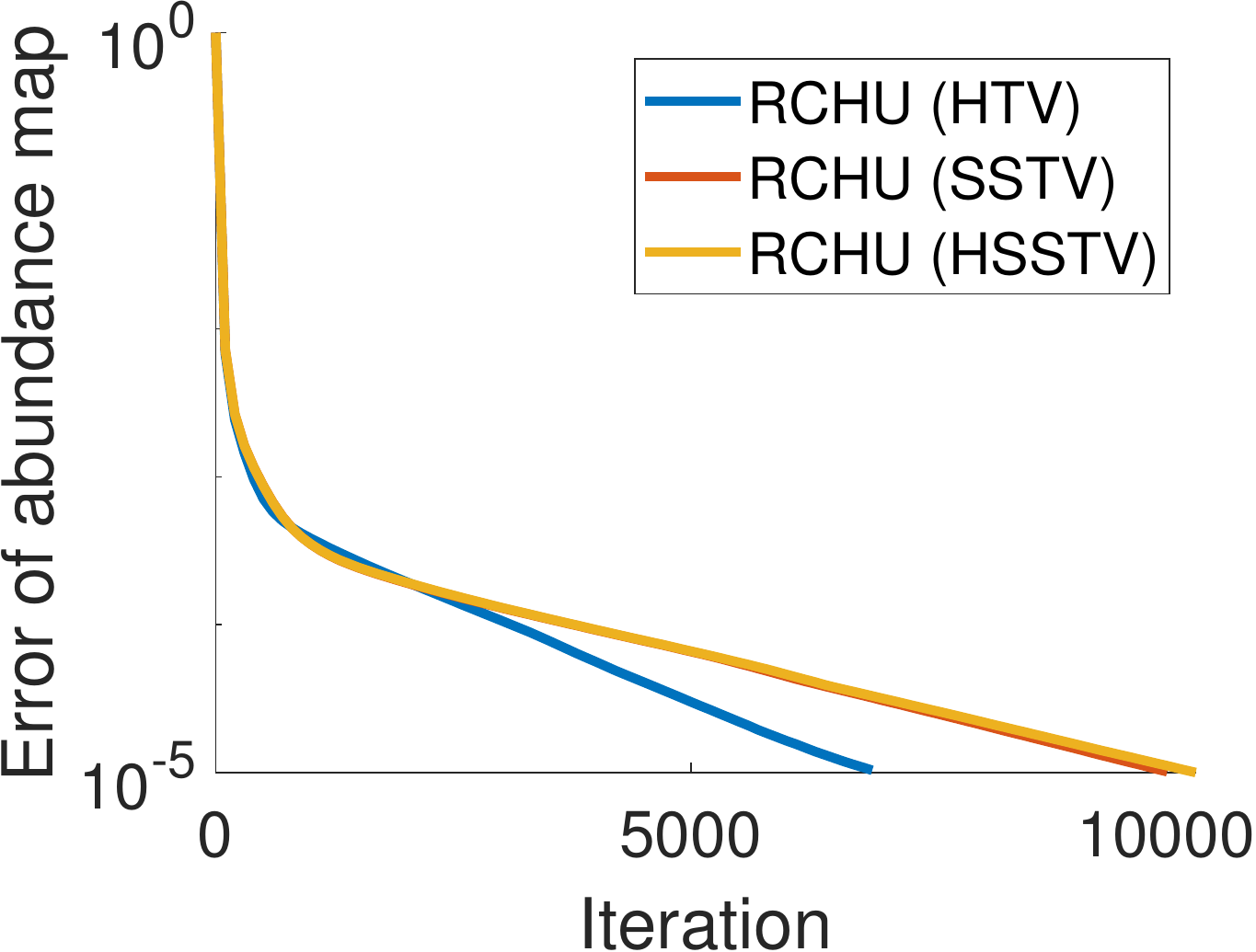}}
	\end{minipage}
	\begin{minipage}{0.19\hsize}
		\centerline{\includegraphics[width=\hsize]{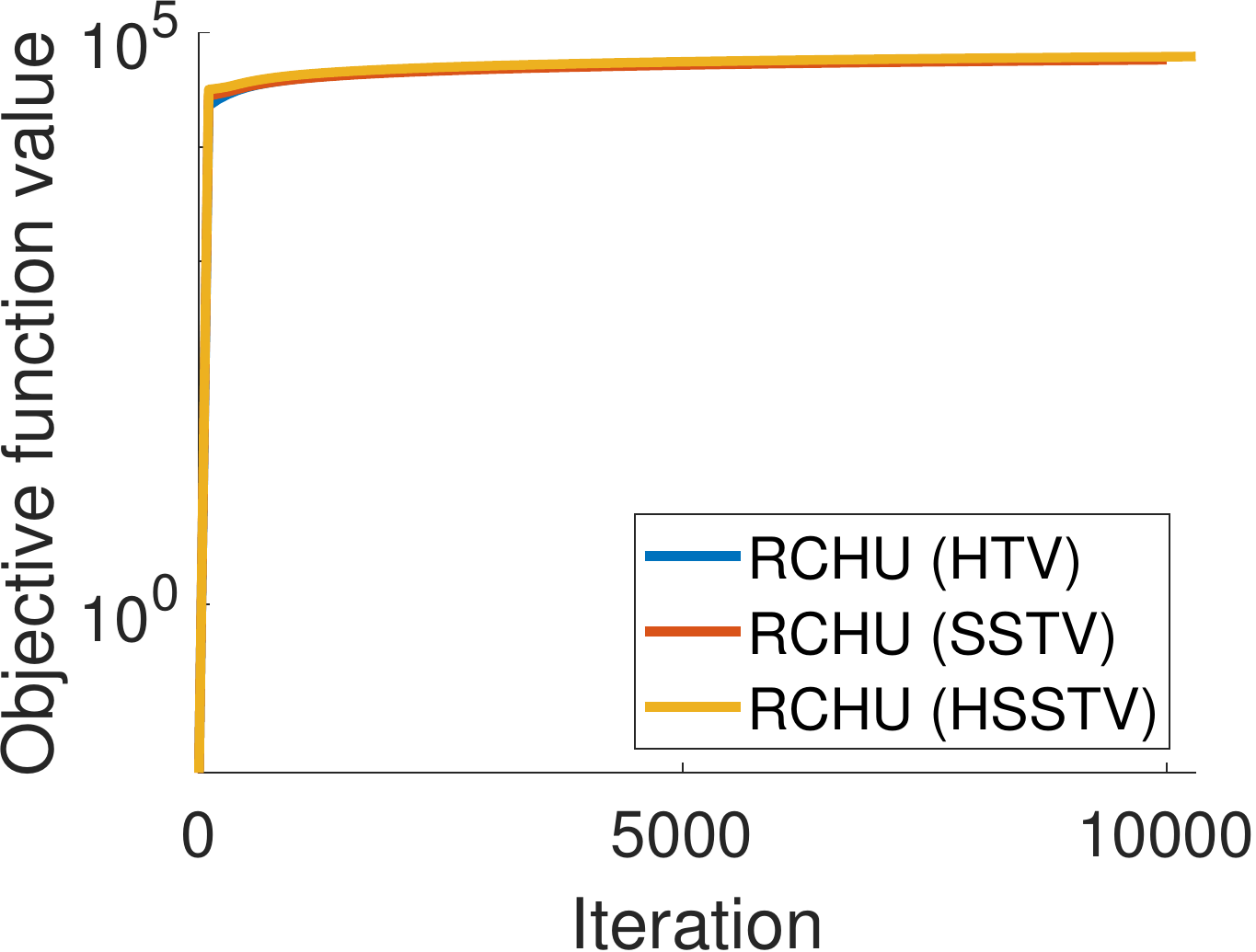}}
	\end{minipage}
	\begin{minipage}{0.19\hsize}
		\centerline{\includegraphics[width=\hsize]{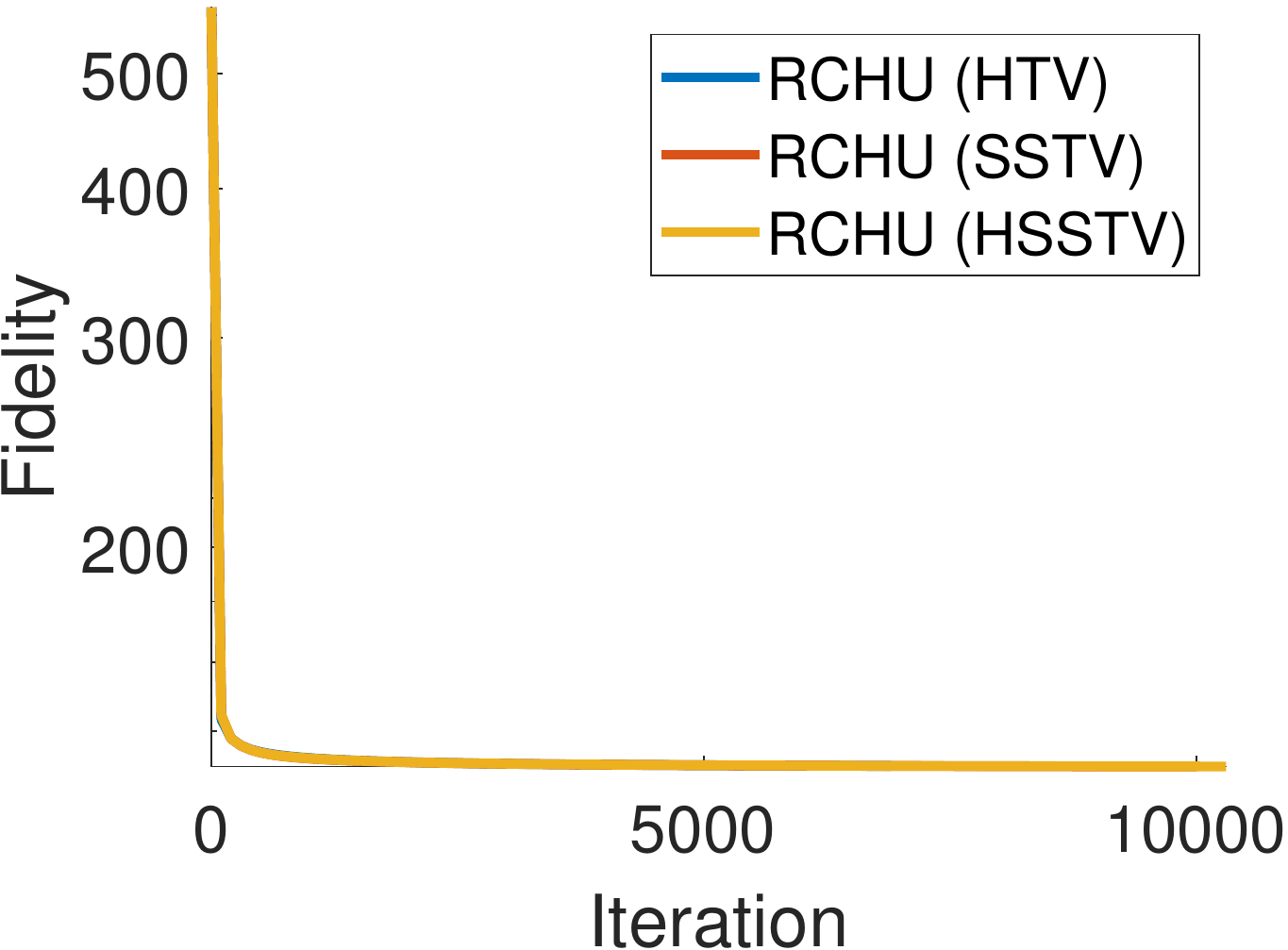}}
	\end{minipage}
	\begin{minipage}{0.19\hsize}
		\centerline{\includegraphics[width=\hsize]{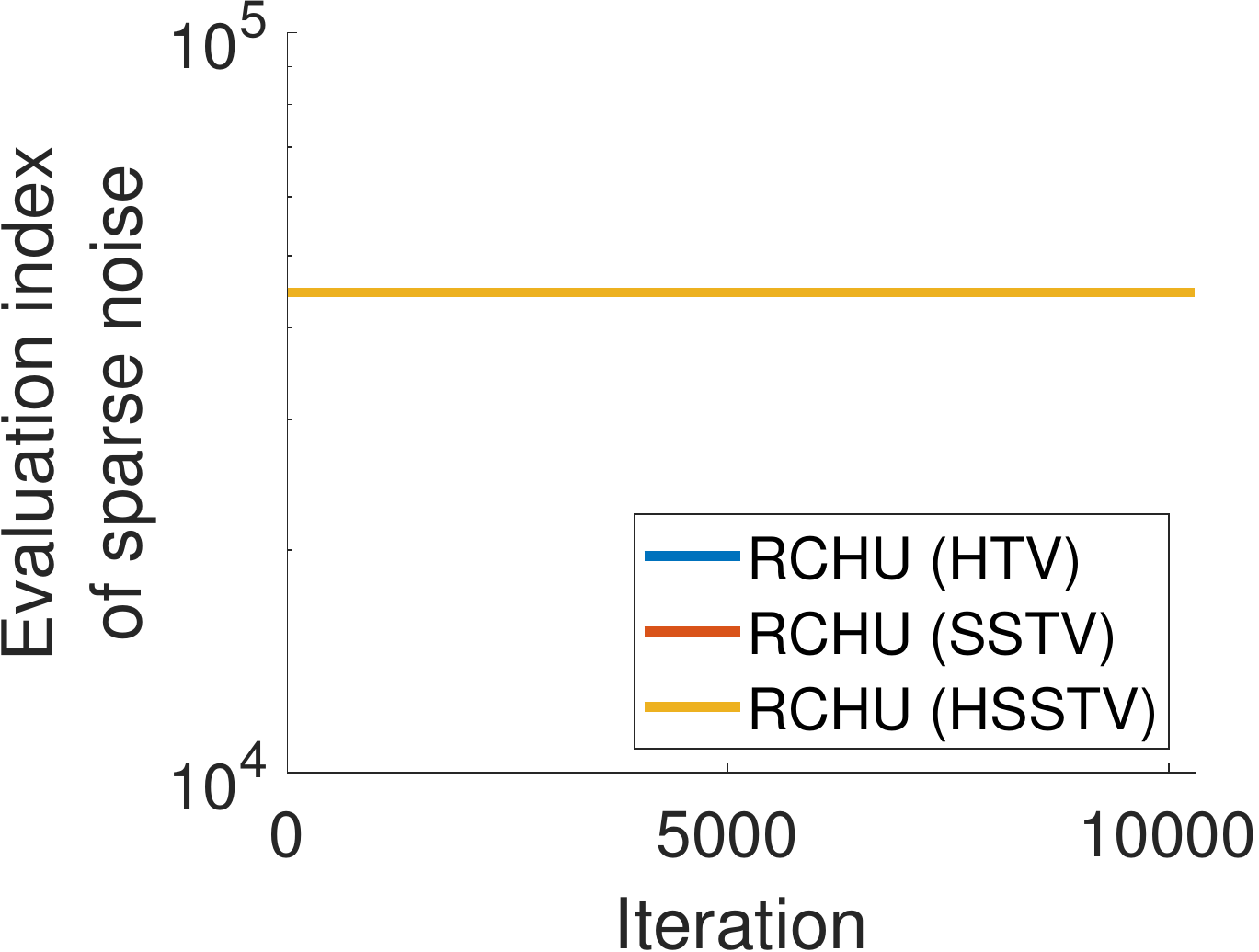}}
	\end{minipage}
	\begin{minipage}{0.19\hsize}
		\centerline{\includegraphics[width=\hsize]{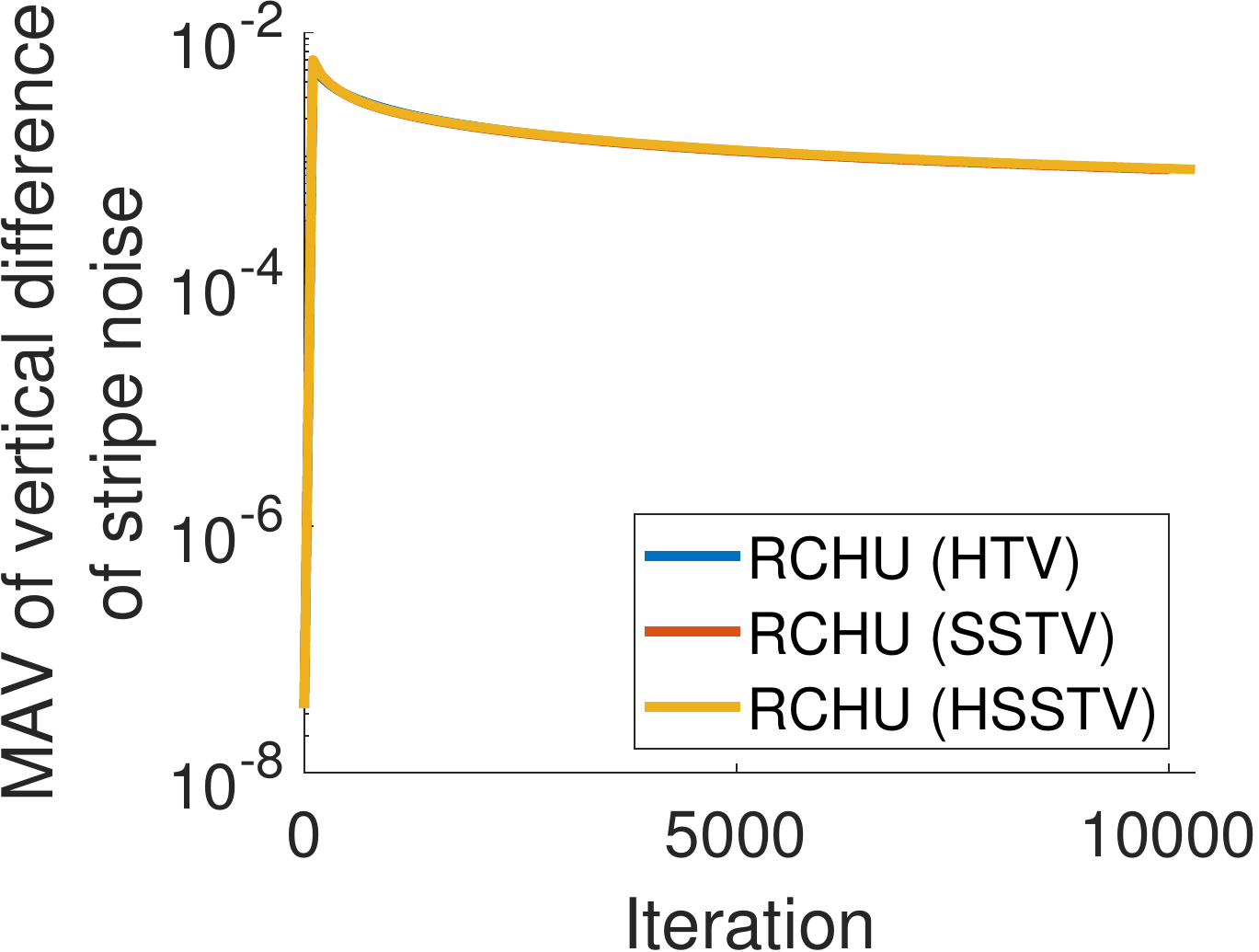}}
	\end{minipage}

\vspace{1mm}

\begin{minipage}{0.19\hsize}
	\centerline{\includegraphics[width=\hsize]{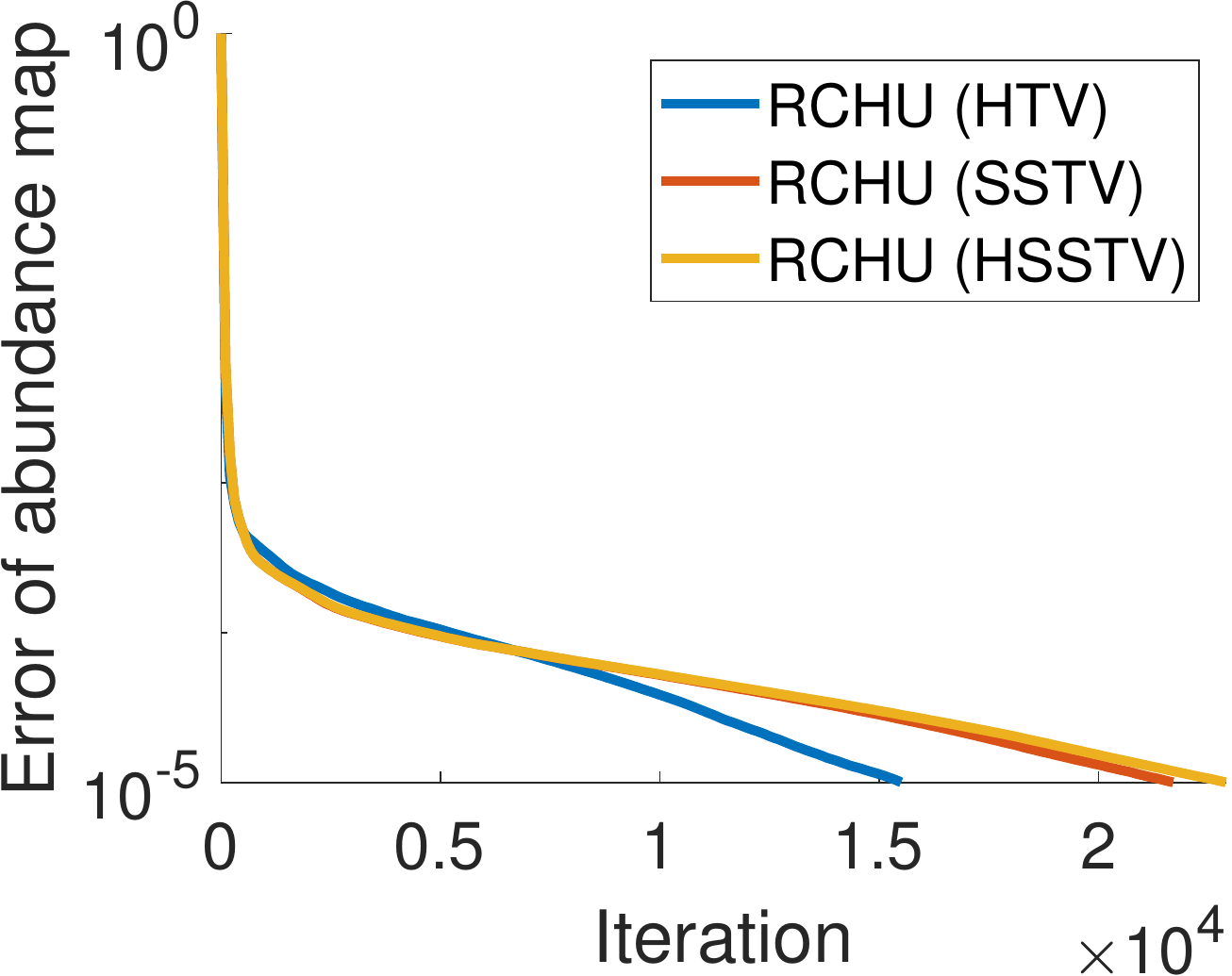}}
\end{minipage}
\begin{minipage}{0.19\hsize}
	\centerline{\includegraphics[width=\hsize]{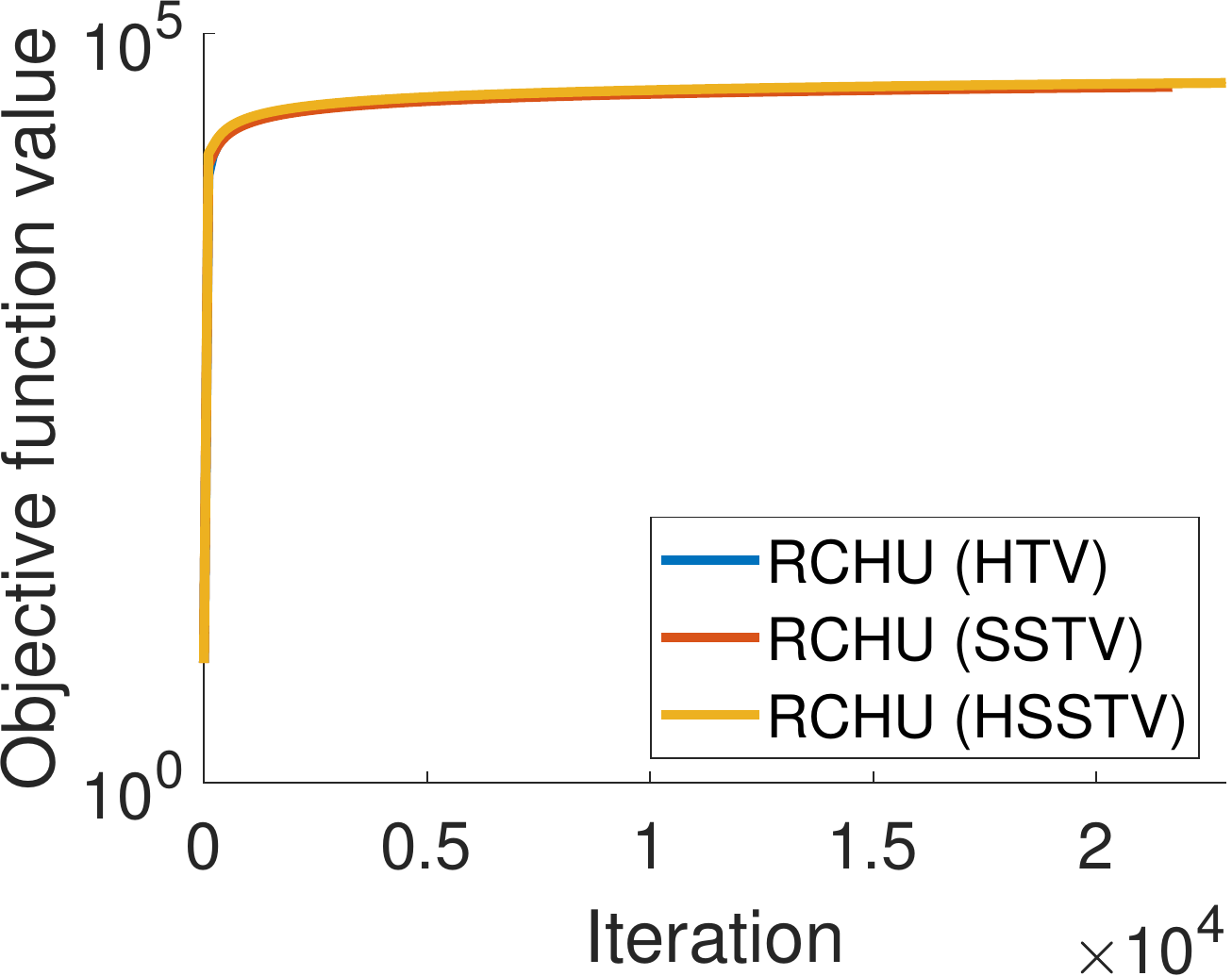}}
\end{minipage}
\begin{minipage}{0.19\hsize}
	\centerline{\includegraphics[width=\hsize]{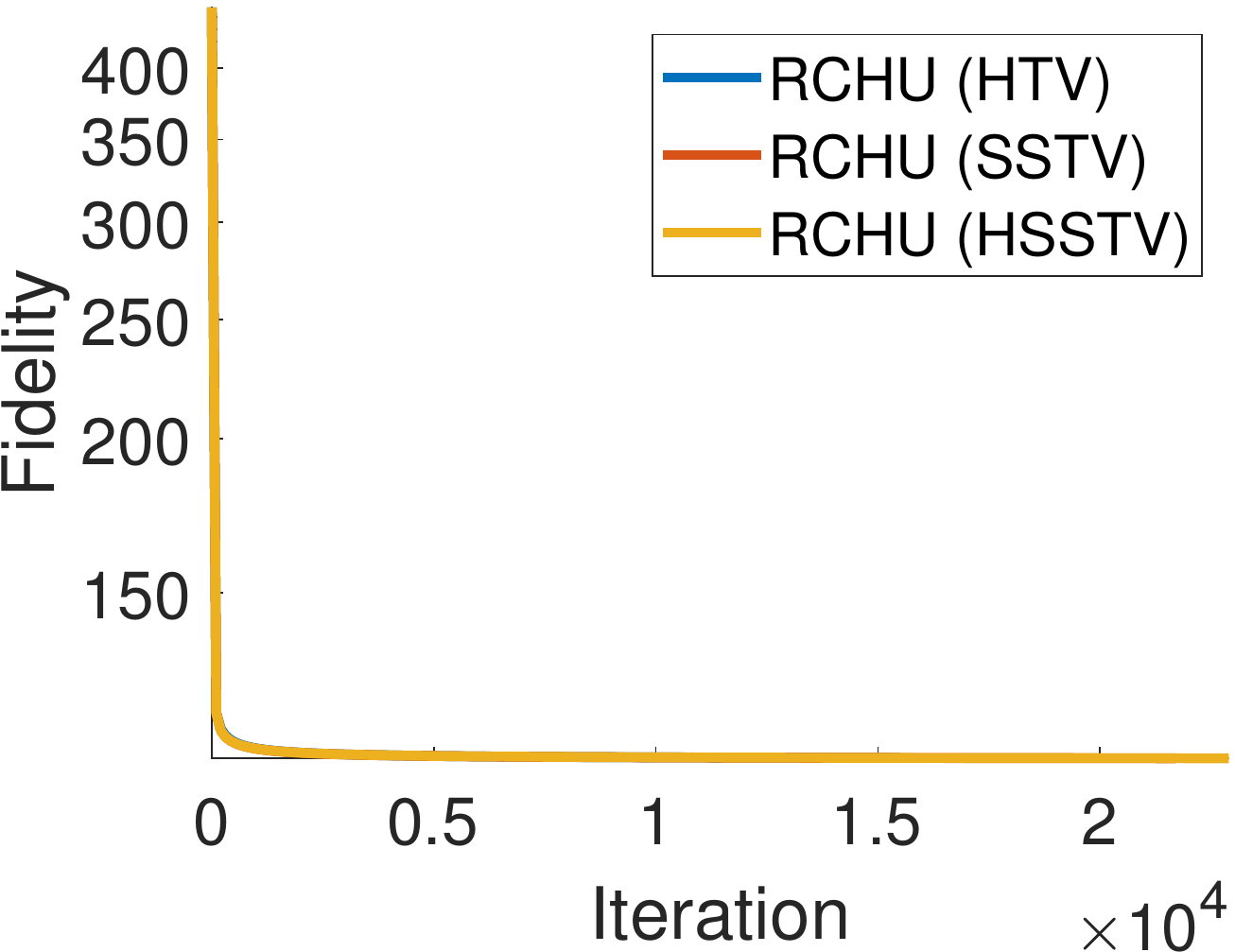}}
\end{minipage}
\begin{minipage}{0.19\hsize}
	\centerline{\includegraphics[width=\hsize]{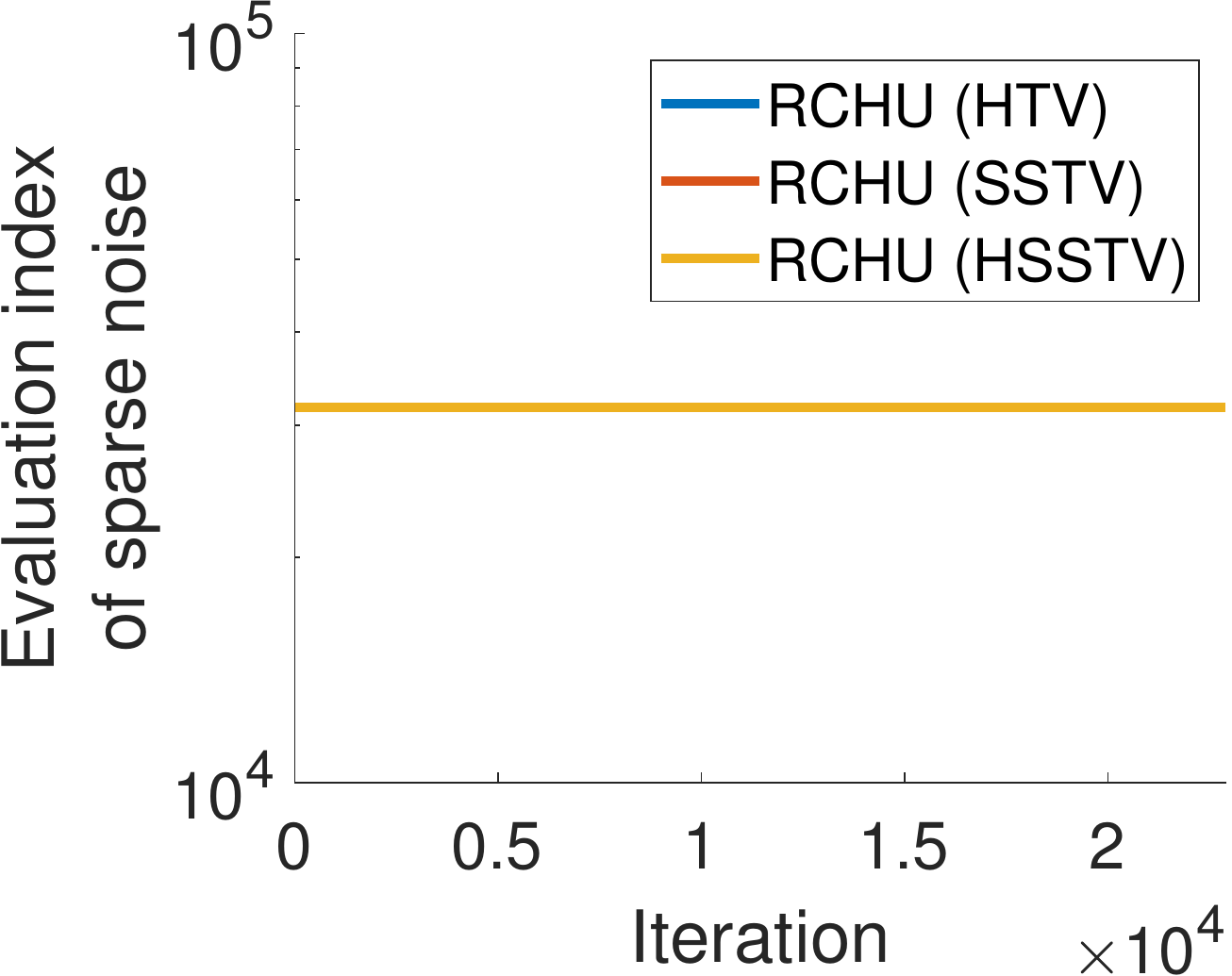}}
\end{minipage}
\begin{minipage}{0.19\hsize}
	\centerline{\includegraphics[width=\hsize]{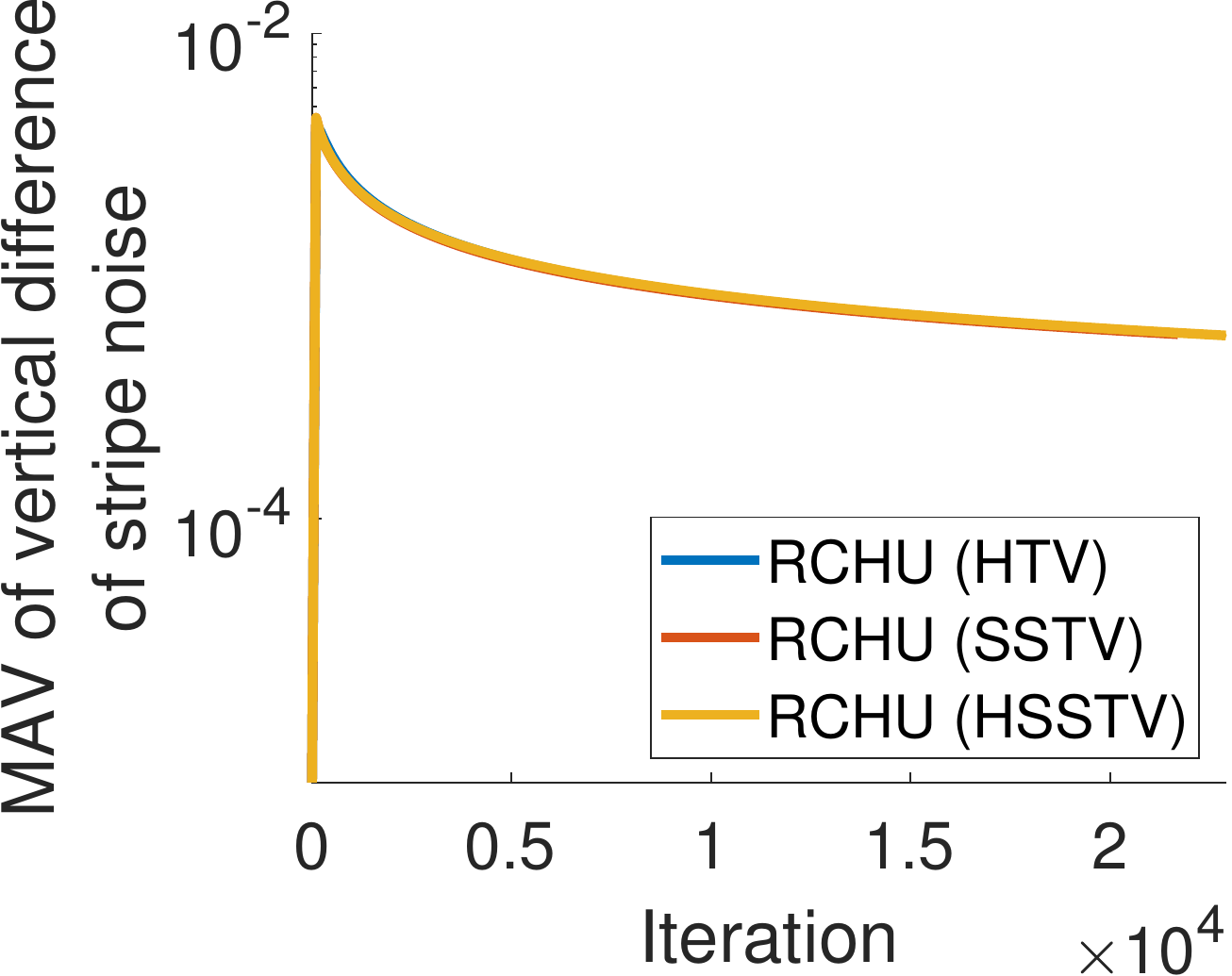}}
\end{minipage}

\vspace{1mm}

	\begin{minipage}{0.19\hsize}
		\centerline{(a)}
	\end{minipage}
	\begin{minipage}{0.19\hsize}
		\centerline{(b)}
	\end{minipage}
	\begin{minipage}{0.19\hsize}
		\centerline{(c)}
	\end{minipage}	
	\begin{minipage}{0.19\hsize}
		\centerline{(d)}
	\end{minipage}
	\begin{minipage}{0.19\hsize}
		\centerline{(e)}
	\end{minipage}
	
	\vspace{-1mm}

	\caption{Convergence analysis using the real images. The top row shows the results of experiments using \textit{Jasper Ridge}. The bottom row shows the results of experiments using \textit{Samson}. (a): The relative error of abundance maps $\|\MatAbun^{(\NumIter+1)} - \MatAbun^{(\NumIter)}\|_F/\|\MatAbun^{(\NumIter+1)}\|_{F}$ versus iteration $\NumIter$. (b): Objective function value $\|\MatAbun^{(\NumIter)}\|_{1,2,r}+\lambda_1\|\mathbf{D}(\MatAbun^{(\NumIter)})\|_1+\lambda_2\mathcal{R}(\DiffHSIReg(\MatEndmember\MatAbun^{(\NumIter)}))+\lambda_3\|\MatStripe^{(\NumIter)}\|_1$ versus iteration $\NumIter$.  (c): The $\ell_{2}$ distance between $\MatHSIObsev$ and $\MatEndmember\MatAbun^{(\NumIter)}+\MatSpar^{(\NumIter)}+\MatStripe^{(\NumIter)}$ versus iteration $\NumIter$. (d): The $\ell_{1}$ norm of $\MatSpar^{(\NumIter)}$ versus iteration $\NumIter$. (e): The mean absolute values (MAV) of $\mathbf{D}_{v}(\MatStripe^{(\NumIter)})$ versus iteration $\NumIter$.} 
	\label{fig:convergence_analysis_all}
\end{figure*}

\begin{table*}[t]
	\caption{SRE, RMSE, Ps, MPSNR, and MSSIM of the Ablation Experiments Using Synthetic Datasets.}
	\vspace{-1mm}
	\label{tab:results_ablation_synth}
	\centering
	\scalebox{0.775}{
		\begin{tabular}{cccccccccccccc} \toprule
			\multirow{3}{*}{ Image }  & \multirow{3}{*}{Metrics} & \multicolumn{4}{c}{\textit{Synth 1}} & \multicolumn{4}{c}{\textit{Synth 2}} & \multicolumn{4}{c}{\textit{Synth 3}} \\ 
			\cmidrule(lr){3-6}\cmidrule(lr){7-10}\cmidrule(lr){11-14}
			& & \Ours & \textbf{\Ours} & \textbf{\Ours} & \textbf{\Ours} & \Ours & \textbf{\Ours} & \textbf{\Ours} & \textbf{\Ours} & \Ours & \textbf{\Ours} & \textbf{\Ours} & \textbf{\Ours} \\ 
			& & -- & \textbf{(HTV)} & \textbf{(SSTV)} & \textbf{(HSSTV)} & -- & \textbf{(HTV)} & \textbf{(SSTV)} & \textbf{(HSSTV)} & -- & \textbf{(HTV)} & \textbf{(SSTV)} & \textbf{(HSSTV)} \\ 
			
			\midrule
			
			\multirow{6}{*}{Case 1} 
			& Setup 
			& \begin{tabular}{c} \hspace{-4mm} $\lambda_{1} = 10^{0}$, \hspace{-4mm} \\\hspace{-4mm} $\varepsilon = 0.98$ \hspace{-4mm}\end{tabular} & \begin{tabular}{c} \hspace{-4mm} $\lambda_{1} = 10^{0}$, \hspace{-4mm} \\\hspace{-4mm} $\lambda_{2} = 10^{1}$, \hspace{-4mm} \\\hspace{-4mm} $\varepsilon = 0.95$ \hspace{-4mm}\end{tabular} & \begin{tabular}{c} \hspace{-4mm} $\lambda_{1} = 10^{1}$, \hspace{-4mm} \\\hspace{-4mm} $\lambda_{2} = 10^{-2}$, \hspace{-4mm} \\\hspace{-4mm} $\varepsilon = 0.95$ \hspace{-4mm}\end{tabular} & \begin{tabular}{c} \hspace{-4mm} $\lambda_{1} = 10^{0}$, \hspace{-4mm} \\\hspace{-4mm} $\lambda_{2} = 10^{-2}$, \hspace{-4mm} \\\hspace{-4mm} $\varepsilon = 0.98$ \hspace{-4mm}\end{tabular} 
			& \begin{tabular}{c} \hspace{-4mm} $\lambda_{1} = 10^{0}$, \hspace{-4mm} \\\hspace{-4mm} $\varepsilon = 0.98$ \hspace{-4mm}\end{tabular} & \begin{tabular}{c} \hspace{-4mm} $\lambda_{1} = 10^{0}$, \hspace{-4mm} \\\hspace{-4mm} $\lambda_{2} = 10^{0}$, \hspace{-4mm} \\\hspace{-4mm} $\varepsilon = 0.98$ \hspace{-4mm}\end{tabular} & \begin{tabular}{c} \hspace{-4mm} $\lambda_{1} = 10^{0}$, \hspace{-4mm} \\\hspace{-4mm} $\lambda_{2} = 10^{-1}$, \hspace{-4mm} \\\hspace{-4mm} $\varepsilon = 0.98$ \hspace{-4mm}\end{tabular} & \begin{tabular}{c} \hspace{-4mm} $\lambda_{1} = 10^{0}$, \hspace{-4mm} \\\hspace{-4mm} $\lambda_{2} = 10^{-1}$, \hspace{-4mm} \\\hspace{-4mm} $\varepsilon = 0.98$ \hspace{-4mm}\end{tabular} 
			& \begin{tabular}{c} \hspace{-4mm} $\lambda_{1} = 10^{-1}$, \hspace{-4mm} \\\hspace{-4mm} $\varepsilon = 0.98$ \hspace{-4mm}\end{tabular} & \begin{tabular}{c} \hspace{-4mm} $\lambda_{1} = 10^{-1}$, \hspace{-4mm} \\\hspace{-4mm} $\lambda_{2} = 10^{0}$, \hspace{-4mm} \\\hspace{-4mm} $\varepsilon = 0.95$ \hspace{-4mm}\end{tabular} & \begin{tabular}{c} \hspace{-4mm} $\lambda_{1} = 10^{0}$, \hspace{-4mm} \\\hspace{-4mm} $\lambda_{2} = 10^{-1}$, \hspace{-4mm} \\\hspace{-4mm} $\varepsilon = 0.95$ \hspace{-4mm}\end{tabular} & \begin{tabular}{c} \hspace{-4mm} $\lambda_{1} = 10^{-1}$, \hspace{-4mm} \\\hspace{-4mm} $\lambda_{2} = 10^{-1}$, \hspace{-4mm} \\\hspace{-4mm} $\varepsilon = 0.95$ \hspace{-4mm}\end{tabular} \\ 
			& SRE  
			& \ValSecnd{  23.54} & \Valbest{  24.31} &   21.84 &   23.52 
			&   20.85 & \Valbest{  21.11} &   20.78 & \ValSecnd{  20.99} 
			&   27.08 & \Valbest{  30.26} &   26.38 & \ValSecnd{  27.58} \\ 
			& RMSE 
			& \ValSecnd{  0.0166} & \Valbest{  0.0153} &   0.0200 &   0.0167 
			&   0.0204 & \Valbest{  0.0198} &   0.0206 & \ValSecnd{  0.0201} 
			&   0.0024 & \Valbest{  0.0017} &   0.0026 & \ValSecnd{  0.0023} \\ 
			& Ps   
			& \Valbest{  1.00} & \Valbest{  1.00} & \Valbest{  1.00} & \Valbest{  1.00} 
			& \Valbest{  1.00} & \Valbest{  1.00} & \Valbest{  1.00} & \Valbest{  1.00} 
			& \Valbest{  1.00} & \Valbest{  1.00} & \Valbest{  1.00} & \Valbest{  1.00} \\ 
			& MPSNR 
			&   47.69 & \Valbest{  54.23} & \ValSecnd{  49.34} &   47.74 
			&   42.38 & \Valbest{  42.84} &   42.39 & \ValSecnd{  42.55} 
			&   46.04 & \Valbest{  56.00} &   46.05 & \ValSecnd{  46.49} \\ 
			& MSSIM 
			&   0.9885 & \Valbest{  0.9988} & \ValSecnd{  0.9937} &   0.9886 
			&   0.9891 & \Valbest{  0.9911} &   0.9891 & \ValSecnd{  0.9897} 
			&   0.9810 & \Valbest{  0.9997} &   0.9822 & \ValSecnd{  0.9834} \\ 
			\cmidrule(lr){1-2}\cmidrule(lr){3-6}\cmidrule(lr){7-10}\cmidrule(lr){11-14}
			
			\multirow{6}{*}{Case 2} 
			& Setup 
			& \begin{tabular}{c} \hspace{-4mm} $\lambda_{1} = 10^{0}$, \hspace{-4mm} \\\hspace{-4mm} $\varepsilon = 0.98$ \hspace{-4mm}\end{tabular} & \begin{tabular}{c} \hspace{-4mm} $\lambda_{1} = 10^{0}$, \hspace{-4mm} \\\hspace{-4mm} $\lambda_{2} = 10^{1}$, \hspace{-4mm} \\\hspace{-4mm} $\varepsilon = 0.95$ \hspace{-4mm}\end{tabular} & \begin{tabular}{c} \hspace{-4mm} $\lambda_{1} = 10^{1}$, \hspace{-4mm} \\\hspace{-4mm} $\lambda_{2} = 10^{-2}$, \hspace{-4mm} \\\hspace{-4mm} $\varepsilon = 0.95$ \hspace{-4mm}\end{tabular} & \begin{tabular}{c} \hspace{-4mm} $\lambda_{1} = 10^{0}$, \hspace{-4mm} \\\hspace{-4mm} $\lambda_{2} = 10^{-1}$, \hspace{-4mm} \\\hspace{-4mm} $\varepsilon = 0.98$ \hspace{-4mm}\end{tabular} 
			& \begin{tabular}{c} \hspace{-4mm} $\lambda_{1} = 10^{0}$, \hspace{-4mm} \\\hspace{-4mm} $\varepsilon = 0.98$ \hspace{-4mm}\end{tabular} & \begin{tabular}{c} \hspace{-4mm} $\lambda_{1} = 10^{0}$, \hspace{-4mm} \\\hspace{-4mm} $\lambda_{2} = 10^{0}$, \hspace{-4mm} \\\hspace{-4mm} $\varepsilon = 0.98$ \hspace{-4mm}\end{tabular} & \begin{tabular}{c} \hspace{-4mm} $\lambda_{1} = 10^{0}$, \hspace{-4mm} \\\hspace{-4mm} $\lambda_{2} = 10^{-1}$, \hspace{-4mm} \\\hspace{-4mm} $\varepsilon = 0.98$ \hspace{-4mm}\end{tabular} & \begin{tabular}{c} \hspace{-4mm} $\lambda_{1} = 10^{0}$, \hspace{-4mm} \\\hspace{-4mm} $\lambda_{2} = 10^{-1}$, \hspace{-4mm} \\\hspace{-4mm} $\varepsilon = 0.98$ \hspace{-4mm}\end{tabular} 
			& \begin{tabular}{c} \hspace{-4mm} $\lambda_{1} = 10^{0}$, \hspace{-4mm} \\\hspace{-4mm} $\varepsilon = 0.95$ \hspace{-4mm}\end{tabular} & \begin{tabular}{c} \hspace{-4mm} $\lambda_{1} = 10^{-1}$, \hspace{-4mm} \\\hspace{-4mm} $\lambda_{2} = 10^{0}$, \hspace{-4mm} \\\hspace{-4mm} $\varepsilon = 0.95$ \hspace{-4mm}\end{tabular} & \begin{tabular}{c} \hspace{-4mm} $\lambda_{1} = 10^{0}$, \hspace{-4mm} \\\hspace{-4mm} $\lambda_{2} = 10^{-1}$, \hspace{-4mm} \\\hspace{-4mm} $\varepsilon = 0.95$ \hspace{-4mm}\end{tabular} & \begin{tabular}{c} \hspace{-4mm} $\lambda_{1} = 10^{-1}$, \hspace{-4mm} \\\hspace{-4mm} $\lambda_{2} = 10^{-1}$, \hspace{-4mm} \\\hspace{-4mm} $\varepsilon = 0.98$ \hspace{-4mm}\end{tabular} \\ 
			& SRE  
			&   20.55 & \Valbest{  21.64} &   20.19 & \ValSecnd{  20.84} 
			&   17.15 & \Valbest{  17.50} &   17.23 & \ValSecnd{  17.38} 
			&   21.95 & \Valbest{  28.67} &   22.34 & \ValSecnd{  22.72} \\ 
			& RMSE 
			&   0.0234 & \Valbest{  0.0206} &   0.0240 & \ValSecnd{  0.0226} 
			&   0.0310 & \Valbest{  0.0297} &   0.0308 & \ValSecnd{  0.0302} 
			&   0.0043 & \Valbest{  0.0020} &   0.0041 & \ValSecnd{  0.0040} \\ 
			& Ps   
			& \Valbest{  1.00} & \Valbest{  1.00} & \Valbest{  1.00} & \Valbest{  1.00} 
			& \Valbest{  1.00} & \Valbest{  1.00} & \Valbest{  1.00} & \Valbest{  1.00} 
			& \Valbest{  1.00} & \Valbest{  1.00} & \Valbest{  1.00} & \Valbest{  1.00} \\ 
			& MPSNR 
			&   41.32 & \Valbest{  49.20} & \ValSecnd{  43.39} &   41.70 
			&   38.29 & \Valbest{  39.00} &   38.35 & \ValSecnd{  38.51} 
			&   40.18 & \Valbest{  52.34} &   40.45 & \ValSecnd{  42.56} \\ 
			& MSSIM 
			&   0.9530 & \Valbest{  0.9959} & \ValSecnd{  0.9735} &   0.9572 
			&   0.9726 & \Valbest{  0.9787} &   0.9731 & \ValSecnd{  0.9743} 
			&   0.9321 & \Valbest{  0.9991} &   0.9361 & \ValSecnd{  0.9624} \\ 
			\cmidrule(lr){1-2}\cmidrule(lr){3-6}\cmidrule(lr){7-10}\cmidrule(lr){11-14}
			
			\multirow{6}{*}{Case 3} 
			& Setup 
			& \begin{tabular}{c} \hspace{-4mm} $\lambda_{1} = 10^{0}$, \hspace{-4mm} \\\hspace{-4mm} $\varepsilon = 0.98$ \hspace{-4mm}\end{tabular} & \begin{tabular}{c} \hspace{-4mm} $\lambda_{1} = 10^{0}$, \hspace{-4mm} \\\hspace{-4mm} $\lambda_{2} = 10^{0}$, \hspace{-4mm} \\\hspace{-4mm} $\varepsilon = 0.98$ \hspace{-4mm}\end{tabular} & \begin{tabular}{c} \hspace{-4mm} $\lambda_{1} = 10^{1}$, \hspace{-4mm} \\\hspace{-4mm} $\lambda_{2} = 10^{-2}$, \hspace{-4mm} \\\hspace{-4mm} $\varepsilon = 0.95$ \hspace{-4mm}\end{tabular} & \begin{tabular}{c} \hspace{-4mm} $\lambda_{1} = 10^{0}$, \hspace{-4mm} \\\hspace{-4mm} $\lambda_{2} = 10^{-2}$, \hspace{-4mm} \\\hspace{-4mm} $\varepsilon = 0.98$ \hspace{-4mm}\end{tabular} 
			& \begin{tabular}{c} \hspace{-4mm} $\lambda_{1} = 10^{0}$, \hspace{-4mm} \\\hspace{-4mm} $\varepsilon = 0.98$ \hspace{-4mm}\end{tabular} & \begin{tabular}{c} \hspace{-4mm} $\lambda_{1} = 10^{0}$, \hspace{-4mm} \\\hspace{-4mm} $\lambda_{2} = 10^{0}$, \hspace{-4mm} \\\hspace{-4mm} $\varepsilon = 0.98$ \hspace{-4mm}\end{tabular} & \begin{tabular}{c} \hspace{-4mm} $\lambda_{1} = 10^{0}$, \hspace{-4mm} \\\hspace{-4mm} $\lambda_{2} = 10^{-1}$, \hspace{-4mm} \\\hspace{-4mm} $\varepsilon = 0.98$ \hspace{-4mm}\end{tabular} & \begin{tabular}{c} \hspace{-4mm} $\lambda_{1} = 10^{0}$, \hspace{-4mm} \\\hspace{-4mm} $\lambda_{2} = 10^{-1}$, \hspace{-4mm} \\\hspace{-4mm} $\varepsilon = 0.98$ \hspace{-4mm}\end{tabular} 
			& \begin{tabular}{c} \hspace{-4mm} $\lambda_{1} = 10^{-1}$, \hspace{-4mm} \\\hspace{-4mm} $\varepsilon = 0.98$ \hspace{-4mm}\end{tabular} & \begin{tabular}{c} \hspace{-4mm} $\lambda_{1} = 10^{-1}$, \hspace{-4mm} \\\hspace{-4mm} $\lambda_{2} = 10^{0}$, \hspace{-4mm} \\\hspace{-4mm} $\varepsilon = 0.95$ \hspace{-4mm}\end{tabular} & \begin{tabular}{c} \hspace{-4mm} $\lambda_{1} = 10^{0}$, \hspace{-4mm} \\\hspace{-4mm} $\lambda_{2} = 10^{-1}$, \hspace{-4mm} \\\hspace{-4mm} $\varepsilon = 0.95$ \hspace{-4mm}\end{tabular} & \begin{tabular}{c} \hspace{-4mm} $\lambda_{1} = 10^{-1}$, \hspace{-4mm} \\\hspace{-4mm} $\lambda_{2} = 10^{-1}$, \hspace{-4mm} \\\hspace{-4mm} $\varepsilon = 0.98$ \hspace{-4mm}\end{tabular} \\ 
			& SRE  
			& \ValSecnd{  23.75} & \Valbest{  23.92} &   21.06 &   23.73 
			&   20.53 & \Valbest{  20.80} &   20.46 & \ValSecnd{  20.71} 
			&   26.33 & \Valbest{  29.66} &   25.91 & \ValSecnd{  26.93} \\ 
			& RMSE 
			& \ValSecnd{  0.0162} & \Valbest{  0.0159} &   0.0218 & \ValSecnd{  0.0162} 
			&   0.0211 & \Valbest{  0.0204} &   0.0213 & \ValSecnd{  0.0207} 
			&   0.0026 & \Valbest{  0.0018} &   0.0028 & \ValSecnd{  0.0025} \\ 
			& Ps   
			& \Valbest{  1.00} & \Valbest{  1.00} & \Valbest{  1.00} & \Valbest{  1.00} 
			& \Valbest{  1.00} & \Valbest{  1.00} & \Valbest{  1.00} & \Valbest{  1.00} 
			& \Valbest{  1.00} & \Valbest{  1.00} & \Valbest{  1.00} & \Valbest{  1.00} \\ 
			& MPSNR 
			&   46.76 & \Valbest{  50.54} & \ValSecnd{  48.43} &   46.82 
			&   41.90 & \Valbest{  42.24} &   41.90 & \ValSecnd{  42.06} 
			&   45.07 & \Valbest{  55.27} &   45.23 & \ValSecnd{  47.14} \\ 
			& MSSIM 
			&   0.9858 & \Valbest{  0.9949} & \ValSecnd{  0.9922} &   0.9860 
			&   0.9883 & \Valbest{  0.9902} &   0.9883 & \ValSecnd{  0.9889} 
			&   0.9761 & \Valbest{  0.9996} &   0.9783 & \ValSecnd{  0.9870} \\ 
			\cmidrule(lr){1-2}\cmidrule(lr){3-6}\cmidrule(lr){7-10}\cmidrule(lr){11-14}
			
			\multirow{6}{*}{Case 4} 
			& Setup 
			& \begin{tabular}{c} \hspace{-4mm} $\lambda_{1} = 10^{0}$, \hspace{-4mm} \\\hspace{-4mm} $\varepsilon = 0.98$ \hspace{-4mm}\end{tabular} & \begin{tabular}{c} \hspace{-4mm} $\lambda_{1} = 10^{0}$, \hspace{-4mm} \\\hspace{-4mm} $\lambda_{2} = 10^{0}$, \hspace{-4mm} \\\hspace{-4mm} $\varepsilon = 0.98$ \hspace{-4mm}\end{tabular} & \begin{tabular}{c} \hspace{-4mm} $\lambda_{1} = 10^{1}$, \hspace{-4mm} \\\hspace{-4mm} $\lambda_{2} = 10^{-2}$, \hspace{-4mm} \\\hspace{-4mm} $\varepsilon = 0.95$ \hspace{-4mm}\end{tabular} & \begin{tabular}{c} \hspace{-4mm} $\lambda_{1} = 10^{0}$, \hspace{-4mm} \\\hspace{-4mm} $\lambda_{2} = 10^{-1}$, \hspace{-4mm} \\\hspace{-4mm} $\varepsilon = 0.98$ \hspace{-4mm}\end{tabular} 
			& \begin{tabular}{c} \hspace{-4mm} $\lambda_{1} = 10^{0}$, \hspace{-4mm} \\\hspace{-4mm} $\varepsilon = 0.98$ \hspace{-4mm}\end{tabular} & \begin{tabular}{c} \hspace{-4mm} $\lambda_{1} = 10^{0}$, \hspace{-4mm} \\\hspace{-4mm} $\lambda_{2} = 10^{0}$, \hspace{-4mm} \\\hspace{-4mm} $\varepsilon = 0.98$ \hspace{-4mm}\end{tabular} & \begin{tabular}{c} \hspace{-4mm} $\lambda_{1} = 10^{0}$, \hspace{-4mm} \\\hspace{-4mm} $\lambda_{2} = 10^{-1}$, \hspace{-4mm} \\\hspace{-4mm} $\varepsilon = 0.98$ \hspace{-4mm}\end{tabular} & \begin{tabular}{c} \hspace{-4mm} $\lambda_{1} = 10^{0}$, \hspace{-4mm} \\\hspace{-4mm} $\lambda_{2} = 10^{-1}$, \hspace{-4mm} \\\hspace{-4mm} $\varepsilon = 0.98$ \hspace{-4mm}\end{tabular} 
			& \begin{tabular}{c} \hspace{-4mm} $\lambda_{1} = 10^{-1}$, \hspace{-4mm} \\\hspace{-4mm} $\varepsilon = 0.98$ \hspace{-4mm}\end{tabular} & \begin{tabular}{c} \hspace{-4mm} $\lambda_{1} = 10^{-1}$, \hspace{-4mm} \\\hspace{-4mm} $\lambda_{2} = 10^{0}$, \hspace{-4mm} \\\hspace{-4mm} $\varepsilon = 0.95$ \hspace{-4mm}\end{tabular} & \begin{tabular}{c} \hspace{-4mm} $\lambda_{1} = 10^{0}$, \hspace{-4mm} \\\hspace{-4mm} $\lambda_{2} = 10^{-1}$, \hspace{-4mm} \\\hspace{-4mm} $\varepsilon = 0.95$ \hspace{-4mm}\end{tabular} & \begin{tabular}{c} \hspace{-4mm} $\lambda_{1} = 10^{-1}$, \hspace{-4mm} \\\hspace{-4mm} $\lambda_{2} = 10^{-1}$, \hspace{-4mm} \\\hspace{-4mm} $\varepsilon = 0.98$ \hspace{-4mm}\end{tabular} \\ 
			& SRE  
			&   23.27 & \Valbest{  23.59} &   21.89 & \ValSecnd{  23.31} 
			&   18.62 & \Valbest{  19.02} &   18.65 & \ValSecnd{  18.92} 
			&   25.52 & \Valbest{  30.92} &   25.52 & \ValSecnd{  26.68} \\ 
			& RMSE 
			& \ValSecnd{  0.0172} & \Valbest{  0.0166} &   0.0199 & \ValSecnd{  0.0172} 
			&   0.0263 & \Valbest{  0.0251} &   0.0262 & \ValSecnd{  0.0254} 
			&   0.0029 & \Valbest{  0.0016} &   0.0029 & \ValSecnd{  0.0025} \\ 
			& Ps   
			& \Valbest{  1.00} & \Valbest{  1.00} & \Valbest{  1.00} & \Valbest{  1.00} 
			& \Valbest{  1.00} & \Valbest{  1.00} & \Valbest{  1.00} & \Valbest{  1.00} 
			& \Valbest{  1.00} & \Valbest{  1.00} & \Valbest{  1.00} & \Valbest{  1.00} \\ 
			& MPSNR 
			&   45.24 & \Valbest{  47.92} & \ValSecnd{  47.58} &   45.63 
			&   40.78 & \Valbest{  41.30} &   40.80 & \ValSecnd{  40.97} 
			&   43.65 & \Valbest{  55.21} &   44.34 & \ValSecnd{  45.62} \\ 
			& MSSIM 
			&   0.9790 & \ValSecnd{  0.9893} & \Valbest{  0.9900} &   0.9810 
			&   0.9843 & \Valbest{  0.9871} &   0.9844 & \ValSecnd{  0.9852} 
			&   0.9665 & \Valbest{  0.9996} &   0.9730 & \ValSecnd{  0.9803} \\ 
			\cmidrule(lr){1-2}\cmidrule(lr){3-6}\cmidrule(lr){7-10}\cmidrule(lr){11-14}
			
			\multirow{6}{*}{Case 5} 
			& Setup 
			& \begin{tabular}{c} \hspace{-4mm} $\lambda_{1} = 10^{1}$, \hspace{-4mm} \\\hspace{-4mm} $\varepsilon = 0.95$ \hspace{-4mm}\end{tabular} & \begin{tabular}{c} \hspace{-4mm} $\lambda_{1} = 10^{0}$, \hspace{-4mm} \\\hspace{-4mm} $\lambda_{2} = 10^{1}$, \hspace{-4mm} \\\hspace{-4mm} $\varepsilon = 0.95$ \hspace{-4mm}\end{tabular} & \begin{tabular}{c} \hspace{-4mm} $\lambda_{1} = 10^{1}$, \hspace{-4mm} \\\hspace{-4mm} $\lambda_{2} = 10^{-2}$, \hspace{-4mm} \\\hspace{-4mm} $\varepsilon = 0.95$ \hspace{-4mm}\end{tabular} & \begin{tabular}{c} \hspace{-4mm} $\lambda_{1} = 10^{1}$, \hspace{-4mm} \\\hspace{-4mm} $\lambda_{2} = 10^{-2}$, \hspace{-4mm} \\\hspace{-4mm} $\varepsilon = 0.95$ \hspace{-4mm}\end{tabular} 
			& \begin{tabular}{c} \hspace{-4mm} $\lambda_{1} = 10^{0}$, \hspace{-4mm} \\\hspace{-4mm} $\varepsilon = 0.98$ \hspace{-4mm}\end{tabular} & \begin{tabular}{c} \hspace{-4mm} $\lambda_{1} = 10^{0}$, \hspace{-4mm} \\\hspace{-4mm} $\lambda_{2} = 10^{0}$, \hspace{-4mm} \\\hspace{-4mm} $\varepsilon = 0.98$ \hspace{-4mm}\end{tabular} & \begin{tabular}{c} \hspace{-4mm} $\lambda_{1} = 10^{0}$, \hspace{-4mm} \\\hspace{-4mm} $\lambda_{2} = 10^{-1}$, \hspace{-4mm} \\\hspace{-4mm} $\varepsilon = 0.98$ \hspace{-4mm}\end{tabular} & \begin{tabular}{c} \hspace{-4mm} $\lambda_{1} = 10^{0}$, \hspace{-4mm} \\\hspace{-4mm} $\lambda_{2} = 10^{-1}$, \hspace{-4mm} \\\hspace{-4mm} $\varepsilon = 0.98$ \hspace{-4mm}\end{tabular} 
			& \begin{tabular}{c} \hspace{-4mm} $\lambda_{1} = 10^{0}$, \hspace{-4mm} \\\hspace{-4mm} $\varepsilon = 0.98$ \hspace{-4mm}\end{tabular} & \begin{tabular}{c} \hspace{-4mm} $\lambda_{1} = 10^{0}$, \hspace{-4mm} \\\hspace{-4mm} $\lambda_{2} = 10^{1}$, \hspace{-4mm} \\\hspace{-4mm} $\varepsilon = 0.95$ \hspace{-4mm}\end{tabular} & \begin{tabular}{c} \hspace{-4mm} $\lambda_{1} = 10^{0}$, \hspace{-4mm} \\\hspace{-4mm} $\lambda_{2} = 10^{-1}$, \hspace{-4mm} \\\hspace{-4mm} $\varepsilon = 0.98$ \hspace{-4mm}\end{tabular} & \begin{tabular}{c} \hspace{-4mm} $\lambda_{1} = 10^{0}$, \hspace{-4mm} \\\hspace{-4mm} $\lambda_{2} = 10^{-1}$, \hspace{-4mm} \\\hspace{-4mm} $\varepsilon = 0.98$ \hspace{-4mm}\end{tabular} \\ 
			& SRE  
			&   19.96 & \Valbest{  21.38} & \ValSecnd{  21.02} &   19.96 
			&   21.30 & \Valbest{  21.52} &   21.27 & \ValSecnd{  21.42} 
			&   21.22 & \Valbest{  24.96} &   21.46 & \ValSecnd{  21.54} \\ 
			& RMSE 
			&   0.0244 & \Valbest{  0.0211} & \ValSecnd{  0.0218} &   0.0244 
			&   0.0194 & \Valbest{  0.0189} &   0.0195 & \ValSecnd{  0.0191} 
			&   0.0047 & \Valbest{  0.0031} &   0.0046 & \ValSecnd{  0.0045} \\ 
			& Ps   
			& \Valbest{  1.00} & \Valbest{  1.00} & \Valbest{  1.00} & \Valbest{  1.00} 
			& \Valbest{  1.00} & \Valbest{  1.00} & \Valbest{  1.00} & \Valbest{  1.00} 
			& \Valbest{  1.00} & \Valbest{  1.00} & \Valbest{  1.00} & \Valbest{  1.00} \\ 
			& MPSNR 
			&   45.97 & \Valbest{  50.17} &   45.92 & \ValSecnd{  45.99} 
			&   42.48 & \Valbest{  42.83} &   42.51 & \ValSecnd{  42.65} 
			&   40.99 & \Valbest{  47.52} &   41.18 & \ValSecnd{  41.51} \\ 
			& MSSIM 
			& \ValSecnd{  0.9893} & \Valbest{  0.9983} &   0.9886 & \ValSecnd{  0.9893} 
			&   0.9890 & \Valbest{  0.9909} &   0.9890 & \ValSecnd{  0.9895} 
			&   0.9484 & \Valbest{  0.9980} &   0.9507 & \ValSecnd{  0.9550} \\ 
			\cmidrule(lr){1-2}\cmidrule(lr){3-6}\cmidrule(lr){7-10}\cmidrule(lr){11-14}
			
			\multirow{6}{*}{Case 6} 
			& Setup 
			& \begin{tabular}{c} \hspace{-4mm} $\lambda_{1} = 10^{0}$, \hspace{-4mm} \\\hspace{-4mm} $\varepsilon = 0.95$ \hspace{-4mm}\end{tabular} & \begin{tabular}{c} \hspace{-4mm} $\lambda_{1} = 10^{0}$, \hspace{-4mm} \\\hspace{-4mm} $\lambda_{2} = 10^{0}$, \hspace{-4mm} \\\hspace{-4mm} $\varepsilon = 0.95$ \hspace{-4mm}\end{tabular} & \begin{tabular}{c} \hspace{-4mm} $\lambda_{1} = 10^{1}$, \hspace{-4mm} \\\hspace{-4mm} $\lambda_{2} = 10^{-2}$, \hspace{-4mm} \\\hspace{-4mm} $\varepsilon = 0.95$ \hspace{-4mm}\end{tabular} & \begin{tabular}{c} \hspace{-4mm} $\lambda_{1} = 10^{0}$, \hspace{-4mm} \\\hspace{-4mm} $\lambda_{2} = 10^{0}$, \hspace{-4mm} \\\hspace{-4mm} $\varepsilon = 0.95$ \hspace{-4mm}\end{tabular} 
			& \begin{tabular}{c} \hspace{-4mm} $\lambda_{1} = 10^{0}$, \hspace{-4mm} \\\hspace{-4mm} $\varepsilon = 0.98$ \hspace{-4mm}\end{tabular} & \begin{tabular}{c} \hspace{-4mm} $\lambda_{1} = 10^{0}$, \hspace{-4mm} \\\hspace{-4mm} $\lambda_{2} = 10^{0}$, \hspace{-4mm} \\\hspace{-4mm} $\varepsilon = 0.98$ \hspace{-4mm}\end{tabular} & \begin{tabular}{c} \hspace{-4mm} $\lambda_{1} = 10^{0}$, \hspace{-4mm} \\\hspace{-4mm} $\lambda_{2} = 10^{-2}$, \hspace{-4mm} \\\hspace{-4mm} $\varepsilon = 0.98$ \hspace{-4mm}\end{tabular} & \begin{tabular}{c} \hspace{-4mm} $\lambda_{1} = 10^{0}$, \hspace{-4mm} \\\hspace{-4mm} $\lambda_{2} = 10^{-2}$, \hspace{-4mm} \\\hspace{-4mm} $\varepsilon = 0.98$ \hspace{-4mm}\end{tabular} 
			& \begin{tabular}{c} \hspace{-4mm} $\lambda_{1} = 10^{0}$, \hspace{-4mm} \\\hspace{-4mm} $\varepsilon = 0.98$ \hspace{-4mm}\end{tabular} & \begin{tabular}{c} \hspace{-4mm} $\lambda_{1} = 10^{0}$, \hspace{-4mm} \\\hspace{-4mm} $\lambda_{2} = 10^{0}$, \hspace{-4mm} \\\hspace{-4mm} $\varepsilon = 0.98$ \hspace{-4mm}\end{tabular} & \begin{tabular}{c} \hspace{-4mm} $\lambda_{1} = 10^{0}$, \hspace{-4mm} \\\hspace{-4mm} $\lambda_{2} = 10^{-1}$, \hspace{-4mm} \\\hspace{-4mm} $\varepsilon = 0.98$ \hspace{-4mm}\end{tabular} & \begin{tabular}{c} \hspace{-4mm} $\lambda_{1} = 10^{0}$, \hspace{-4mm} \\\hspace{-4mm} $\lambda_{2} = 10^{-1}$, \hspace{-4mm} \\\hspace{-4mm} $\varepsilon = 0.98$ \hspace{-4mm}\end{tabular} \\ 
			& SRE  
			&   16.28 & \Valbest{  17.14} &   14.86 & \ValSecnd{  16.80} 
			& \ValSecnd{  17.82} & \Valbest{  17.96} &   17.55 &   17.81 
			&   19.98 & \Valbest{  24.39} & \ValSecnd{  20.38} &   20.36 \\ 
			& RMSE 
			&   0.0374 & \Valbest{  0.0339} &   0.0425 & \ValSecnd{  0.0350} 
			& \ValSecnd{  0.0288} & \Valbest{  0.0282} &   0.0296 & \ValSecnd{  0.0288} 
			&   0.0054 & \Valbest{  0.0033} & \ValSecnd{  0.0051} & \ValSecnd{  0.0051} \\ 
			& Ps   
			& \Valbest{  1.00} & \Valbest{  1.00} & \Valbest{  1.00} & \Valbest{  1.00} 
			& \Valbest{  1.00} & \Valbest{  1.00} & \Valbest{  1.00} & \Valbest{  1.00} 
			& \Valbest{  1.00} & \Valbest{  1.00} & \Valbest{  1.00} & \Valbest{  1.00} \\ 
			& MPSNR 
			&   38.88 &   41.22 & \Valbest{  42.52} & \ValSecnd{  42.03} 
			&   38.21 & \Valbest{  38.55} &   38.14 & \ValSecnd{  38.24} 
			&   37.98 & \Valbest{  47.02} &   38.23 & \ValSecnd{  38.86} \\ 
			& MSSIM 
			&   0.9229 &   0.9560 & \Valbest{  0.9741} & \ValSecnd{  0.9660} 
			&   0.9729 & \Valbest{  0.9798} &   0.9724 & \ValSecnd{  0.9732} 
			&   0.8974 & \Valbest{  0.9956} &   0.9029 & \ValSecnd{  0.9176} \\ 
			\cmidrule(lr){1-2}\cmidrule(lr){3-6}\cmidrule(lr){7-10}\cmidrule(lr){11-14}
			
			\multirow{6}{*}{Case 7} 
			& Setup 
			& \begin{tabular}{c} \hspace{-4mm} $\lambda_{1} = 10^{1}$, \hspace{-4mm} \\\hspace{-4mm} $\varepsilon = 0.95$ \hspace{-4mm}\end{tabular} & \begin{tabular}{c} \hspace{-4mm} $\lambda_{1} = 10^{0}$, \hspace{-4mm} \\\hspace{-4mm} $\lambda_{2} = 10^{1}$, \hspace{-4mm} \\\hspace{-4mm} $\varepsilon = 0.98$ \hspace{-4mm}\end{tabular} & \begin{tabular}{c} \hspace{-4mm} $\lambda_{1} = 10^{1}$, \hspace{-4mm} \\\hspace{-4mm} $\lambda_{2} = 10^{-2}$, \hspace{-4mm} \\\hspace{-4mm} $\varepsilon = 0.95$ \hspace{-4mm}\end{tabular} & \begin{tabular}{c} \hspace{-4mm} $\lambda_{1} = 10^{1}$, \hspace{-4mm} \\\hspace{-4mm} $\lambda_{2} = 10^{-1}$, \hspace{-4mm} \\\hspace{-4mm} $\varepsilon = 0.95$ \hspace{-4mm}\end{tabular} 
			& \begin{tabular}{c} \hspace{-4mm} $\lambda_{1} = 10^{0}$, \hspace{-4mm} \\\hspace{-4mm} $\varepsilon = 0.98$ \hspace{-4mm}\end{tabular} & \begin{tabular}{c} \hspace{-4mm} $\lambda_{1} = 10^{0}$, \hspace{-4mm} \\\hspace{-4mm} $\lambda_{2} = 10^{-1}$, \hspace{-4mm} \\\hspace{-4mm} $\varepsilon = 0.98$ \hspace{-4mm}\end{tabular} & \begin{tabular}{c} \hspace{-4mm} $\lambda_{1} = 10^{0}$, \hspace{-4mm} \\\hspace{-4mm} $\lambda_{2} = 10^{-1}$, \hspace{-4mm} \\\hspace{-4mm} $\varepsilon = 0.98$ \hspace{-4mm}\end{tabular} & \begin{tabular}{c} \hspace{-4mm} $\lambda_{1} = 10^{0}$, \hspace{-4mm} \\\hspace{-4mm} $\lambda_{2} = 10^{-1}$, \hspace{-4mm} \\\hspace{-4mm} $\varepsilon = 0.98$ \hspace{-4mm}\end{tabular} 
			& \begin{tabular}{c} \hspace{-4mm} $\lambda_{1} = 10^{0}$, \hspace{-4mm} \\\hspace{-4mm} $\varepsilon = 0.95$ \hspace{-4mm}\end{tabular} & \begin{tabular}{c} \hspace{-4mm} $\lambda_{1} = 10^{-1}$, \hspace{-4mm} \\\hspace{-4mm} $\lambda_{2} = 10^{0}$, \hspace{-4mm} \\\hspace{-4mm} $\varepsilon = 0.95$ \hspace{-4mm}\end{tabular} & \begin{tabular}{c} \hspace{-4mm} $\lambda_{1} = 10^{0}$, \hspace{-4mm} \\\hspace{-4mm} $\lambda_{2} = 10^{-1}$, \hspace{-4mm} \\\hspace{-4mm} $\varepsilon = 0.95$ \hspace{-4mm}\end{tabular} & \begin{tabular}{c} \hspace{-4mm} $\lambda_{1} = 10^{0}$, \hspace{-4mm} \\\hspace{-4mm} $\lambda_{2} = 10^{-1}$, \hspace{-4mm} \\\hspace{-4mm} $\varepsilon = 0.95$ \hspace{-4mm}\end{tabular} \\ 
			& SRE  
			&   17.47 & \Valbest{  17.90} &   16.66 & \ValSecnd{  17.48} 
			&   15.89 & \ValSecnd{  15.95} &   15.68 & \Valbest{  16.05} 
			&   19.53 & \Valbest{  25.57} & \ValSecnd{  20.01} &   20.00 \\ 
			& RMSE 
			& \ValSecnd{  0.0323} & \Valbest{  0.0307} &   0.0351 & \ValSecnd{  0.0323} 
			&   0.0356 & \ValSecnd{  0.0353} &   0.0362 & \Valbest{  0.0348} 
			&   0.0056 & \Valbest{  0.0029} & \ValSecnd{  0.0054} & \ValSecnd{  0.0054} \\ 
			& Ps   
			& \Valbest{  1.00} & \Valbest{  1.00} & \Valbest{  1.00} & \Valbest{  1.00} 
			& \Valbest{  1.00} & \Valbest{  1.00} & \Valbest{  1.00} & \Valbest{  1.00} 
			& \Valbest{  1.00} & \Valbest{  1.00} & \Valbest{  1.00} & \Valbest{  1.00} \\ 
			& MPSNR 
			&   39.68 & \Valbest{  47.34} &   39.58 & \ValSecnd{  39.78} 
			&   35.80 & \ValSecnd{  35.97} &   35.77 & \Valbest{  36.11} 
			&   36.85 & \Valbest{  48.64} &   37.09 & \ValSecnd{  38.06} \\ 
			& MSSIM 
			&   0.9392 & \Valbest{  0.9957} &   0.9378 & \ValSecnd{  0.9406} 
			&   0.9533 & \ValSecnd{  0.9553} &   0.9531 & \Valbest{  0.9570} 
			&   0.8653 & \Valbest{  0.9969} &   0.8716 & \ValSecnd{  0.8980} \\ 
			\cmidrule(lr){1-2}\cmidrule(lr){3-6}\cmidrule(lr){7-10}\cmidrule(lr){11-14}
			
			\multirow{6}{*}{Case 8} 
			& Setup 
			& \begin{tabular}{c} \hspace{-4mm} $\lambda_{1} = 10^{0}$, \hspace{-4mm} \\\hspace{-4mm} $\varepsilon = 0.95$ \hspace{-4mm}\end{tabular} & \begin{tabular}{c} \hspace{-4mm} $\lambda_{1} = 10^{0}$, \hspace{-4mm} \\\hspace{-4mm} $\lambda_{2} = 10^{0}$, \hspace{-4mm} \\\hspace{-4mm} $\varepsilon = 0.95$ \hspace{-4mm}\end{tabular} & \begin{tabular}{c} \hspace{-4mm} $\lambda_{1} = 10^{0}$, \hspace{-4mm} \\\hspace{-4mm} $\lambda_{2} = 10^{0}$, \hspace{-4mm} \\\hspace{-4mm} $\varepsilon = 0.95$ \hspace{-4mm}\end{tabular} & \begin{tabular}{c} \hspace{-4mm} $\lambda_{1} = 10^{0}$, \hspace{-4mm} \\\hspace{-4mm} $\lambda_{2} = 10^{0}$, \hspace{-4mm} \\\hspace{-4mm} $\varepsilon = 0.95$ \hspace{-4mm}\end{tabular} 
			& \begin{tabular}{c} \hspace{-4mm} $\lambda_{1} = 10^{0}$, \hspace{-4mm} \\\hspace{-4mm} $\varepsilon = 0.98$ \hspace{-4mm}\end{tabular} & \begin{tabular}{c} \hspace{-4mm} $\lambda_{1} = 10^{0}$, \hspace{-4mm} \\\hspace{-4mm} $\lambda_{2} = 10^{-1}$, \hspace{-4mm} \\\hspace{-4mm} $\varepsilon = 0.98$ \hspace{-4mm}\end{tabular} & \begin{tabular}{c} \hspace{-4mm} $\lambda_{1} = 10^{0}$, \hspace{-4mm} \\\hspace{-4mm} $\lambda_{2} = 10^{-2}$, \hspace{-4mm} \\\hspace{-4mm} $\varepsilon = 0.98$ \hspace{-4mm}\end{tabular} & \begin{tabular}{c} \hspace{-4mm} $\lambda_{1} = 10^{0}$, \hspace{-4mm} \\\hspace{-4mm} $\lambda_{2} = 10^{-2}$, \hspace{-4mm} \\\hspace{-4mm} $\varepsilon = 0.98$ \hspace{-4mm}\end{tabular} 
			& \begin{tabular}{c} \hspace{-4mm} $\lambda_{1} = 10^{0}$, \hspace{-4mm} \\\hspace{-4mm} $\varepsilon = 0.98$ \hspace{-4mm}\end{tabular} & \begin{tabular}{c} \hspace{-4mm} $\lambda_{1} = 10^{0}$, \hspace{-4mm} \\\hspace{-4mm} $\lambda_{2} = 10^{0}$, \hspace{-4mm} \\\hspace{-4mm} $\varepsilon = 0.98$ \hspace{-4mm}\end{tabular} & \begin{tabular}{c} \hspace{-4mm} $\lambda_{1} = 10^{0}$, \hspace{-4mm} \\\hspace{-4mm} $\lambda_{2} = 10^{-1}$, \hspace{-4mm} \\\hspace{-4mm} $\varepsilon = 0.98$ \hspace{-4mm}\end{tabular} & \begin{tabular}{c} \hspace{-4mm} $\lambda_{1} = 10^{0}$, \hspace{-4mm} \\\hspace{-4mm} $\lambda_{2} = 10^{-1}$, \hspace{-4mm} \\\hspace{-4mm} $\varepsilon = 0.98$ \hspace{-4mm}\end{tabular} \\ 
			& SRE  
			&   13.84 & \Valbest{  14.96} &   11.89 & \ValSecnd{  14.08} 
			& \ValSecnd{  14.63} & \Valbest{  14.67} &   14.26 &   14.60 
			&   18.29 & \Valbest{  21.12} &   18.67 & \ValSecnd{  18.69} \\ 
			& RMSE 
			&   0.0485 & \Valbest{  0.0427} &   0.0592 & \ValSecnd{  0.0468} 
			& \ValSecnd{  0.0403} & \Valbest{  0.0401} &   0.0415 &   0.0405 
			&   0.0064 & \Valbest{  0.0047} & \ValSecnd{  0.0062} & \ValSecnd{  0.0062} \\ 
			& Ps   
			& \Valbest{  1.00} & \Valbest{  1.00} & \Valbest{  1.00} & \Valbest{  1.00} 
			& \Valbest{  0.99} & \Valbest{  0.99} & \Valbest{  0.99} & \Valbest{  0.99} 
			& \Valbest{  1.00} & \Valbest{  1.00} & \Valbest{  1.00} & \Valbest{  1.00} \\ 
			& MPSNR 
			&   36.73 & \ValSecnd{  40.36} &   37.76 & \Valbest{  41.30} 
			&   35.17 & \Valbest{  35.40} &   35.12 & \ValSecnd{  35.21} 
			&   35.34 & \Valbest{  43.66} &   35.56 & \ValSecnd{  36.44} \\ 
			& MSSIM 
			&   0.8846 & \ValSecnd{  0.9518} &   0.9118 & \Valbest{  0.9653} 
			&   0.9519 & \Valbest{  0.9556} &   0.9517 & \ValSecnd{  0.9525} 
			&   0.8303 & \Valbest{  0.9925} &   0.8375 & \ValSecnd{  0.8696} \\ 
			
			\bottomrule
		\end{tabular}
	}
\end{table*}

\begin{table*}[t]
	\caption{SRE, RMSE, Ps, MPSNR, and MSSIM of the Ablation Experiments Using Real Datasets.}
	\vspace{-1mm}
	\label{tab:results_ablation_real}
	\centering
	\scalebox{0.775}{
		\begin{tabular}{cccccccccccccc} \toprule
			\multirow{3}{*}{ Image }  & \multirow{3}{*}{Metrics} & \multicolumn{4}{c}{\textit{Jasper Ridge}} & \multicolumn{4}{c}{Samson} & \multicolumn{4}{c}{Urban} \\ 
			\cmidrule(lr){3-6}\cmidrule(lr){7-10}\cmidrule(lr){11-14}
			& & \Ours & \textbf{\Ours} & \textbf{\Ours} & \textbf{\Ours} & \Ours & \textbf{\Ours} & \textbf{\Ours} & \textbf{\Ours} & \Ours & \textbf{\Ours} & \textbf{\Ours} & \textbf{\Ours} \\ 
			& & -- & \textbf{(HTV)} & \textbf{(SSTV)} & \textbf{(HSSTV)} & -- & \textbf{(HTV)} & \textbf{(SSTV)} & \textbf{(HSSTV)} & -- & \textbf{(HTV)} & \textbf{(SSTV)} & \textbf{(HSSTV)} \\ 
			
			\midrule
			
			\multirow{6}{*}{Case 1} 
			& Setup 
			& \begin{tabular}{c} \hspace{-4mm} $\lambda_{1} = 10^{0}$, \hspace{-4mm} \\\hspace{-4mm} $\varepsilon = 0.95$ \hspace{-4mm}\end{tabular} & \begin{tabular}{c} \hspace{-4mm} $\lambda_{1} = 10^{0}$, \hspace{-4mm} \\\hspace{-4mm} $\lambda_{2} = 10^{-2}$, \hspace{-4mm} \\\hspace{-4mm} $\varepsilon = 0.95$ \hspace{-4mm}\end{tabular} & \begin{tabular}{c} \hspace{-4mm} $\lambda_{1} = 10^{0}$, \hspace{-4mm} \\\hspace{-4mm} $\lambda_{2} = 10^{-2}$, \hspace{-4mm} \\\hspace{-4mm} $\varepsilon = 0.95$ \hspace{-4mm}\end{tabular} & \begin{tabular}{c} \hspace{-4mm} $\lambda_{1} = 10^{0}$, \hspace{-4mm} \\\hspace{-4mm} $\lambda_{2} = 10^{-2}$, \hspace{-4mm} \\\hspace{-4mm} $\varepsilon = 0.95$ \hspace{-4mm}\end{tabular} 
			& \begin{tabular}{c} \hspace{-4mm} $\lambda_{1} = 10^{0}$, \hspace{-4mm} \\\hspace{-4mm} $\varepsilon = 0.95$ \hspace{-4mm}\end{tabular} & \begin{tabular}{c} \hspace{-4mm} $\lambda_{1} = 10^{-1}$, \hspace{-4mm} \\\hspace{-4mm} $\lambda_{2} = 10^{1}$, \hspace{-4mm} \\\hspace{-4mm} $\varepsilon = 0.98$ \hspace{-4mm}\end{tabular} & \begin{tabular}{c} \hspace{-4mm} $\lambda_{1} = 10^{0}$, \hspace{-4mm} \\\hspace{-4mm} $\lambda_{2} = 10^{0}$, \hspace{-4mm} \\\hspace{-4mm} $\varepsilon = 0.95$ \hspace{-4mm}\end{tabular} & \begin{tabular}{c} \hspace{-4mm} $\lambda_{1} = 10^{-1}$, \hspace{-4mm} \\\hspace{-4mm} $\lambda_{2} = 10^{0}$, \hspace{-4mm} \\\hspace{-4mm} $\varepsilon = 0.98$ \hspace{-4mm}\end{tabular} 
			& \begin{tabular}{c} \hspace{-4mm} $\lambda_{1} = 10^{-1}$, \hspace{-4mm} \\\hspace{-4mm} $\varepsilon = 0.95$ \hspace{-4mm}\end{tabular} & \begin{tabular}{c} \hspace{-4mm} $\lambda_{1} = 10^{-1}$, \hspace{-4mm} \\\hspace{-4mm} $\lambda_{2} = 10^{0}$, \hspace{-4mm} \\\hspace{-4mm} $\varepsilon = 0.95$ \hspace{-4mm}\end{tabular} & \begin{tabular}{c} \hspace{-4mm} $\lambda_{1} = 10^{-1}$, \hspace{-4mm} \\\hspace{-4mm} $\lambda_{2} = 10^{-2}$, \hspace{-4mm} \\\hspace{-4mm} $\varepsilon = 0.95$ \hspace{-4mm}\end{tabular} & \begin{tabular}{c} \hspace{-4mm} $\lambda_{1} = 10^{-1}$, \hspace{-4mm} \\\hspace{-4mm} $\lambda_{2} = 10^{-2}$, \hspace{-4mm} \\\hspace{-4mm} $\varepsilon = 0.95$ \hspace{-4mm}\end{tabular} \\ 
			& SRE  
			& \ValSecnd{  19.08} &   19.07 & \Valbest{  19.33} &   19.06 
			&   15.90 & \Valbest{  19.51} &   16.48 & \ValSecnd{  18.03} 
			& \ValSecnd{  15.00} & \Valbest{  15.92} &   14.49 &   14.74 \\ 
			& RMSE 
			& \ValSecnd{  0.0294} & \ValSecnd{  0.0294} & \Valbest{  0.0286} &   0.0295 
			&   0.0368 & \Valbest{  0.0254} &   0.0345 & \ValSecnd{  0.0296} 
			& \ValSecnd{  0.0086} & \Valbest{  0.0078} &   0.0090 &   0.0088 \\ 
			& Ps   
			& \Valbest{  1.00} & \Valbest{  1.00} & \Valbest{  1.00} & \Valbest{  1.00} 
			& \Valbest{  1.00} & \Valbest{  1.00} & \Valbest{  1.00} & \Valbest{  1.00} 
			& \Valbest{  0.99} & \Valbest{  0.99} & \Valbest{  0.99} & \Valbest{  0.99} \\ 
			& MPSNR 
			&   45.44 & \ValSecnd{  45.45} & \Valbest{  45.47} & \ValSecnd{  45.45} 
			&   45.27 &   43.75 & \ValSecnd{  45.52} & \Valbest{  46.74} 
			& \ValSecnd{  41.84} &   37.93 &   41.74 & \Valbest{  41.91} \\ 
			& MSSIM 
			& \Valbest{  0.9882} & \Valbest{  0.9882} & \ValSecnd{  0.9881} & \Valbest{  0.9882} 
			&   0.9842 &   0.9836 & \ValSecnd{  0.9855} & \Valbest{  0.9907} 
			& \ValSecnd{  0.9857} &   0.9741 &   0.9854 & \Valbest{  0.9863} \\ 
			\cmidrule(lr){1-2}\cmidrule(lr){3-6}\cmidrule(lr){7-10}\cmidrule(lr){11-14}
			
			\multirow{6}{*}{Case 2} 
			& Setup 
			& \begin{tabular}{c} \hspace{-4mm} $\lambda_{1} = 10^{0}$, \hspace{-4mm} \\\hspace{-4mm} $\varepsilon = 0.98$ \hspace{-4mm}\end{tabular} & \begin{tabular}{c} \hspace{-4mm} $\lambda_{1} = 10^{0}$, \hspace{-4mm} \\\hspace{-4mm} $\lambda_{2} = 10^{0}$, \hspace{-4mm} \\\hspace{-4mm} $\varepsilon = 0.98$ \hspace{-4mm}\end{tabular} & \begin{tabular}{c} \hspace{-4mm} $\lambda_{1} = 10^{0}$, \hspace{-4mm} \\\hspace{-4mm} $\lambda_{2} = 10^{-2}$, \hspace{-4mm} \\\hspace{-4mm} $\varepsilon = 0.98$ \hspace{-4mm}\end{tabular} & \begin{tabular}{c} \hspace{-4mm} $\lambda_{1} = 10^{0}$, \hspace{-4mm} \\\hspace{-4mm} $\lambda_{2} = 10^{-2}$, \hspace{-4mm} \\\hspace{-4mm} $\varepsilon = 0.98$ \hspace{-4mm}\end{tabular} 
			& \begin{tabular}{c} \hspace{-4mm} $\lambda_{1} = 10^{0}$, \hspace{-4mm} \\\hspace{-4mm} $\varepsilon = 0.98$ \hspace{-4mm}\end{tabular} & \begin{tabular}{c} \hspace{-4mm} $\lambda_{1} = 10^{-1}$, \hspace{-4mm} \\\hspace{-4mm} $\lambda_{2} = 10^{1}$, \hspace{-4mm} \\\hspace{-4mm} $\varepsilon = 0.98$ \hspace{-4mm}\end{tabular} & \begin{tabular}{c} \hspace{-4mm} $\lambda_{1} = 10^{0}$, \hspace{-4mm} \\\hspace{-4mm} $\lambda_{2} = 10^{0}$, \hspace{-4mm} \\\hspace{-4mm} $\varepsilon = 0.98$ \hspace{-4mm}\end{tabular} & \begin{tabular}{c} \hspace{-4mm} $\lambda_{1} = 10^{0}$, \hspace{-4mm} \\\hspace{-4mm} $\lambda_{2} = 10^{1}$, \hspace{-4mm} \\\hspace{-4mm} $\varepsilon = 0.98$ \hspace{-4mm}\end{tabular} 
			& \begin{tabular}{c} \hspace{-4mm} $\lambda_{1} = 10^{-1}$, \hspace{-4mm} \\\hspace{-4mm} $\varepsilon = 0.95$ \hspace{-4mm}\end{tabular} & \begin{tabular}{c} \hspace{-4mm} $\lambda_{1} = 10^{-1}$, \hspace{-4mm} \\\hspace{-4mm} $\lambda_{2} = 10^{0}$, \hspace{-4mm} \\\hspace{-4mm} $\varepsilon = 0.95$ \hspace{-4mm}\end{tabular} & \begin{tabular}{c} \hspace{-4mm} $\lambda_{1} = 10^{-1}$, \hspace{-4mm} \\\hspace{-4mm} $\lambda_{2} = 10^{-2}$, \hspace{-4mm} \\\hspace{-4mm} $\varepsilon = 0.95$ \hspace{-4mm}\end{tabular} & \begin{tabular}{c} \hspace{-4mm} $\lambda_{1} = 10^{-1}$, \hspace{-4mm} \\\hspace{-4mm} $\lambda_{2} = 10^{-2}$, \hspace{-4mm} \\\hspace{-4mm} $\varepsilon = 0.95$ \hspace{-4mm}\end{tabular} \\ 
			& SRE  
			&   15.38 & \ValSecnd{  15.49} & \Valbest{  15.53} &   15.37 
			&   11.02 & \Valbest{  14.51} &   11.09 & \ValSecnd{  12.86} 
			&   11.36 & \Valbest{  12.62} &   10.90 & \ValSecnd{  11.37} \\ 
			& RMSE 
			&   0.0441 & \ValSecnd{  0.0436} & \Valbest{  0.0432} &   0.0441 
			&   0.0597 & \Valbest{  0.0435} &   0.0590 & \ValSecnd{  0.0504} 
			& \ValSecnd{  0.0126} & \Valbest{  0.0110} &   0.0131 & \ValSecnd{  0.0126} \\ 
			& Ps   
			& \Valbest{  0.99} & \Valbest{  0.99} & \Valbest{  0.99} & \Valbest{  0.99} 
			& \Valbest{  0.99} & \Valbest{  0.99} & \Valbest{  0.99} & \Valbest{  0.99} 
			& \Valbest{  0.98} & \Valbest{  0.98} & \Valbest{  0.98} & \Valbest{  0.98} \\ 
			& MPSNR 
			&   40.34 & \Valbest{  40.97} &   40.31 & \ValSecnd{  40.36} 
			&   40.10 & \Valbest{  40.84} & \ValSecnd{  40.45} &   40.44 
			& \ValSecnd{  35.89} &   35.25 &   35.77 & \Valbest{  36.05} \\ 
			& MSSIM 
			&   0.9639 & \Valbest{  0.9732} &   0.9634 & \ValSecnd{  0.9641} 
			&   0.9513 & \ValSecnd{  0.9719} &   0.9561 & \Valbest{  0.9725} 
			&   0.9459 & \Valbest{  0.9571} &   0.9448 & \ValSecnd{  0.9483} \\ 
			\cmidrule(lr){1-2}\cmidrule(lr){3-6}\cmidrule(lr){7-10}\cmidrule(lr){11-14}
			
			\multirow{6}{*}{Case 3} 
			& Setup 
			& \begin{tabular}{c} \hspace{-4mm} $\lambda_{1} = 10^{0}$, \hspace{-4mm} \\\hspace{-4mm} $\varepsilon = 0.95$ \hspace{-4mm}\end{tabular} & \begin{tabular}{c} \hspace{-4mm} $\lambda_{1} = 10^{0}$, \hspace{-4mm} \\\hspace{-4mm} $\lambda_{2} = 10^{-2}$, \hspace{-4mm} \\\hspace{-4mm} $\varepsilon = 0.95$ \hspace{-4mm}\end{tabular} & \begin{tabular}{c} \hspace{-4mm} $\lambda_{1} = 10^{0}$, \hspace{-4mm} \\\hspace{-4mm} $\lambda_{2} = 10^{-2}$, \hspace{-4mm} \\\hspace{-4mm} $\varepsilon = 0.95$ \hspace{-4mm}\end{tabular} & \begin{tabular}{c} \hspace{-4mm} $\lambda_{1} = 10^{0}$, \hspace{-4mm} \\\hspace{-4mm} $\lambda_{2} = 10^{-2}$, \hspace{-4mm} \\\hspace{-4mm} $\varepsilon = 0.95$ \hspace{-4mm}\end{tabular} 
			& \begin{tabular}{c} \hspace{-4mm} $\lambda_{1} = 10^{0}$, \hspace{-4mm} \\\hspace{-4mm} $\varepsilon = 0.95$ \hspace{-4mm}\end{tabular} & \begin{tabular}{c} \hspace{-4mm} $\lambda_{1} = 10^{-1}$, \hspace{-4mm} \\\hspace{-4mm} $\lambda_{2} = 10^{1}$, \hspace{-4mm} \\\hspace{-4mm} $\varepsilon = 0.98$ \hspace{-4mm}\end{tabular} & \begin{tabular}{c} \hspace{-4mm} $\lambda_{1} = 10^{0}$, \hspace{-4mm} \\\hspace{-4mm} $\lambda_{2} = 10^{0}$, \hspace{-4mm} \\\hspace{-4mm} $\varepsilon = 0.95$ \hspace{-4mm}\end{tabular} & \begin{tabular}{c} \hspace{-4mm} $\lambda_{1} = 10^{-1}$, \hspace{-4mm} \\\hspace{-4mm} $\lambda_{2} = 10^{0}$, \hspace{-4mm} \\\hspace{-4mm} $\varepsilon = 0.98$ \hspace{-4mm}\end{tabular} 
			& \begin{tabular}{c} \hspace{-4mm} $\lambda_{1} = 10^{-1}$, \hspace{-4mm} \\\hspace{-4mm} $\varepsilon = 0.95$ \hspace{-4mm}\end{tabular} & \begin{tabular}{c} \hspace{-4mm} $\lambda_{1} = 10^{-1}$, \hspace{-4mm} \\\hspace{-4mm} $\lambda_{2} = 10^{0}$, \hspace{-4mm} \\\hspace{-4mm} $\varepsilon = 0.95$ \hspace{-4mm}\end{tabular} & \begin{tabular}{c} \hspace{-4mm} $\lambda_{1} = 10^{-1}$, \hspace{-4mm} \\\hspace{-4mm} $\lambda_{2} = 10^{-2}$, \hspace{-4mm} \\\hspace{-4mm} $\varepsilon = 0.95$ \hspace{-4mm}\end{tabular} & \begin{tabular}{c} \hspace{-4mm} $\lambda_{1} = 10^{-1}$, \hspace{-4mm} \\\hspace{-4mm} $\lambda_{2} = 10^{-2}$, \hspace{-4mm} \\\hspace{-4mm} $\varepsilon = 0.95$ \hspace{-4mm}\end{tabular} \\ 
			& SRE  
			& \ValSecnd{  18.76} & \ValSecnd{  18.76} & \Valbest{  18.98} &   18.75 
			&   15.24 & \Valbest{  18.16} &   15.61 & \ValSecnd{  16.74} 
			& \ValSecnd{  14.33} & \Valbest{  15.29} &   13.79 &   14.11 \\ 
			& RMSE 
			& \ValSecnd{  0.0303} &   0.0304 & \Valbest{  0.0296} &   0.0304 
			&   0.0393 & \Valbest{  0.0294} &   0.0378 & \ValSecnd{  0.0341} 
			& \ValSecnd{  0.0092} & \Valbest{  0.0083} &   0.0097 &   0.0094 \\ 
			& Ps   
			& \Valbest{  1.00} & \Valbest{  1.00} & \Valbest{  1.00} & \Valbest{  1.00} 
			& \Valbest{  1.00} & \Valbest{  1.00} & \Valbest{  1.00} & \Valbest{  1.00} 
			& \Valbest{  0.99} & \Valbest{  0.99} & \Valbest{  0.99} & \Valbest{  0.99} \\ 
			& MPSNR 
			& \ValSecnd{  44.46} & \Valbest{  44.47} & \Valbest{  44.47} & \Valbest{  44.47} 
			&   44.12 &   43.09 & \ValSecnd{  44.34} & \Valbest{  45.45} 
			& \ValSecnd{  40.89} &   37.38 &   40.78 & \Valbest{  40.98} \\ 
			& MSSIM 
			& \ValSecnd{  0.9849} & \ValSecnd{  0.9849} &   0.9848 & \Valbest{  0.9850} 
			&   0.9791 & \ValSecnd{  0.9812} &   0.9807 & \Valbest{  0.9868} 
			& \ValSecnd{  0.9823} &   0.9712 &   0.9819 & \Valbest{  0.9831} \\ 
			\cmidrule(lr){1-2}\cmidrule(lr){3-6}\cmidrule(lr){7-10}\cmidrule(lr){11-14}
			
			\multirow{6}{*}{Case 4} 
			& Setup 
			& \begin{tabular}{c} \hspace{-4mm} $\lambda_{1} = 10^{0}$, \hspace{-4mm} \\\hspace{-4mm} $\varepsilon = 0.95$ \hspace{-4mm}\end{tabular} & \begin{tabular}{c} \hspace{-4mm} $\lambda_{1} = 10^{0}$, \hspace{-4mm} \\\hspace{-4mm} $\lambda_{2} = 10^{0}$, \hspace{-4mm} \\\hspace{-4mm} $\varepsilon = 0.95$ \hspace{-4mm}\end{tabular} & \begin{tabular}{c} \hspace{-4mm} $\lambda_{1} = 10^{0}$, \hspace{-4mm} \\\hspace{-4mm} $\lambda_{2} = 10^{-2}$, \hspace{-4mm} \\\hspace{-4mm} $\varepsilon = 0.95$ \hspace{-4mm}\end{tabular} & \begin{tabular}{c} \hspace{-4mm} $\lambda_{1} = 10^{0}$, \hspace{-4mm} \\\hspace{-4mm} $\lambda_{2} = 10^{-2}$, \hspace{-4mm} \\\hspace{-4mm} $\varepsilon = 0.95$ \hspace{-4mm}\end{tabular} 
			& \begin{tabular}{c} \hspace{-4mm} $\lambda_{1} = 10^{0}$, \hspace{-4mm} \\\hspace{-4mm} $\varepsilon = 0.95$ \hspace{-4mm}\end{tabular} & \begin{tabular}{c} \hspace{-4mm} $\lambda_{1} = 10^{-1}$, \hspace{-4mm} \\\hspace{-4mm} $\lambda_{2} = 10^{1}$, \hspace{-4mm} \\\hspace{-4mm} $\varepsilon = 0.98$ \hspace{-4mm}\end{tabular} & \begin{tabular}{c} \hspace{-4mm} $\lambda_{1} = 10^{0}$, \hspace{-4mm} \\\hspace{-4mm} $\lambda_{2} = 10^{0}$, \hspace{-4mm} \\\hspace{-4mm} $\varepsilon = 0.95$ \hspace{-4mm}\end{tabular} & \begin{tabular}{c} \hspace{-4mm} $\lambda_{1} = 10^{0}$, \hspace{-4mm} \\\hspace{-4mm} $\lambda_{2} = 10^{0}$, \hspace{-4mm} \\\hspace{-4mm} $\varepsilon = 0.95$ \hspace{-4mm}\end{tabular} 
			& \begin{tabular}{c} \hspace{-4mm} $\lambda_{1} = 10^{-1}$, \hspace{-4mm} \\\hspace{-4mm} $\varepsilon = 0.95$ \hspace{-4mm}\end{tabular} & \begin{tabular}{c} \hspace{-4mm} $\lambda_{1} = 10^{-1}$, \hspace{-4mm} \\\hspace{-4mm} $\lambda_{2} = 10^{0}$, \hspace{-4mm} \\\hspace{-4mm} $\varepsilon = 0.95$ \hspace{-4mm}\end{tabular} & \begin{tabular}{c} \hspace{-4mm} $\lambda_{1} = 10^{-1}$, \hspace{-4mm} \\\hspace{-4mm} $\lambda_{2} = 10^{-2}$, \hspace{-4mm} \\\hspace{-4mm} $\varepsilon = 0.95$ \hspace{-4mm}\end{tabular} & \begin{tabular}{c} \hspace{-4mm} $\lambda_{1} = 10^{-1}$, \hspace{-4mm} \\\hspace{-4mm} $\lambda_{2} = 10^{-2}$, \hspace{-4mm} \\\hspace{-4mm} $\varepsilon = 0.95$ \hspace{-4mm}\end{tabular} \\ 
			& SRE  
			&   18.26 & \Valbest{  18.31} & \ValSecnd{  18.29} &   18.26 
			&   14.09 & \Valbest{  16.09} &   14.57 & \ValSecnd{  15.24} 
			& \ValSecnd{  14.22} & \Valbest{  14.95} &   13.76 &   14.05 \\ 
			& RMSE 
			& \ValSecnd{  0.0321} & \Valbest{  0.0319} & \ValSecnd{  0.0321} & \ValSecnd{  0.0321} 
			&   0.0439 & \Valbest{  0.0364} &   0.0417 & \ValSecnd{  0.0390} 
			& \ValSecnd{  0.0093} & \Valbest{  0.0086} &   0.0097 &   0.0095 \\ 
			& Ps   
			& \Valbest{  1.00} & \Valbest{  1.00} & \Valbest{  1.00} & \Valbest{  1.00} 
			& \Valbest{  1.00} & \Valbest{  1.00} & \Valbest{  1.00} & \Valbest{  1.00} 
			& \Valbest{  0.99} & \Valbest{  0.99} & \Valbest{  0.99} & \Valbest{  0.99} \\ 
			& MPSNR 
			&   43.37 & \Valbest{  43.89} & \ValSecnd{  43.39} & \ValSecnd{  43.39} 
			&   43.18 &   42.80 & \ValSecnd{  43.41} & \Valbest{  44.25} 
			& \ValSecnd{  40.00} &   37.34 &   39.87 & \Valbest{  40.12} \\ 
			& MSSIM 
			&   0.9801 & \Valbest{  0.9836} & \ValSecnd{  0.9803} &   0.9802 
			&   0.9742 & \ValSecnd{  0.9805} &   0.9761 & \Valbest{  0.9815} 
			& \ValSecnd{  0.9782} &   0.9713 &   0.9776 & \Valbest{  0.9792} \\ 
			\cmidrule(lr){1-2}\cmidrule(lr){3-6}\cmidrule(lr){7-10}\cmidrule(lr){11-14}
			
			\multirow{6}{*}{Case 5} 
			& Setup 
			& \begin{tabular}{c} \hspace{-4mm} $\lambda_{1} = 10^{0}$, \hspace{-4mm} \\\hspace{-4mm} $\varepsilon = 0.95$ \hspace{-4mm}\end{tabular} & \begin{tabular}{c} \hspace{-4mm} $\lambda_{1} = 10^{0}$, \hspace{-4mm} \\\hspace{-4mm} $\lambda_{2} = 10^{0}$, \hspace{-4mm} \\\hspace{-4mm} $\varepsilon = 0.95$ \hspace{-4mm}\end{tabular} & \begin{tabular}{c} \hspace{-4mm} $\lambda_{1} = 10^{0}$, \hspace{-4mm} \\\hspace{-4mm} $\lambda_{2} = 10^{-2}$, \hspace{-4mm} \\\hspace{-4mm} $\varepsilon = 0.98$ \hspace{-4mm}\end{tabular} & \begin{tabular}{c} \hspace{-4mm} $\lambda_{1} = 10^{0}$, \hspace{-4mm} \\\hspace{-4mm} $\lambda_{2} = 10^{-2}$, \hspace{-4mm} \\\hspace{-4mm} $\varepsilon = 0.95$ \hspace{-4mm}\end{tabular} 
			& \begin{tabular}{c} \hspace{-4mm} $\lambda_{1} = 10^{0}$, \hspace{-4mm} \\\hspace{-4mm} $\varepsilon = 0.98$ \hspace{-4mm}\end{tabular} & \begin{tabular}{c} \hspace{-4mm} $\lambda_{1} = 10^{0}$, \hspace{-4mm} \\\hspace{-4mm} $\lambda_{2} = 10^{1}$, \hspace{-4mm} \\\hspace{-4mm} $\varepsilon = 0.98$ \hspace{-4mm}\end{tabular} & \begin{tabular}{c} \hspace{-4mm} $\lambda_{1} = 10^{0}$, \hspace{-4mm} \\\hspace{-4mm} $\lambda_{2} = 10^{-1}$, \hspace{-4mm} \\\hspace{-4mm} $\varepsilon = 0.98$ \hspace{-4mm}\end{tabular} & \begin{tabular}{c} \hspace{-4mm} $\lambda_{1} = 10^{0}$, \hspace{-4mm} \\\hspace{-4mm} $\lambda_{2} = 10^{0}$, \hspace{-4mm} \\\hspace{-4mm} $\varepsilon = 0.98$ \hspace{-4mm}\end{tabular} 
			& \begin{tabular}{c} \hspace{-4mm} $\lambda_{1} = 10^{0}$, \hspace{-4mm} \\\hspace{-4mm} $\varepsilon = 0.95$ \hspace{-4mm}\end{tabular} & \begin{tabular}{c} \hspace{-4mm} $\lambda_{1} = 10^{-1}$, \hspace{-4mm} \\\hspace{-4mm} $\lambda_{2} = 10^{1}$, \hspace{-4mm} \\\hspace{-4mm} $\varepsilon = 0.95$ \hspace{-4mm}\end{tabular} & \begin{tabular}{c} \hspace{-4mm} $\lambda_{1} = 10^{0}$, \hspace{-4mm} \\\hspace{-4mm} $\lambda_{2} = 10^{-2}$, \hspace{-4mm} \\\hspace{-4mm} $\varepsilon = 0.95$ \hspace{-4mm}\end{tabular} & \begin{tabular}{c} \hspace{-4mm} $\lambda_{1} = 10^{0}$, \hspace{-4mm} \\\hspace{-4mm} $\lambda_{2} = 10^{-2}$, \hspace{-4mm} \\\hspace{-4mm} $\varepsilon = 0.95$ \hspace{-4mm}\end{tabular} \\ 
			& SRE  
			& \ValSecnd{  16.19} & \Valbest{  16.36} &   16.16 & \ValSecnd{  16.19} 
			&   10.75 & \Valbest{  12.91} &   10.04 & \ValSecnd{  12.12} 
			& \ValSecnd{   9.61} & \Valbest{  10.38} &    9.59 &    9.59 \\ 
			& RMSE 
			&   0.0401 & \Valbest{  0.0394} &   0.0401 & \ValSecnd{  0.0400} 
			&   0.0607 & \Valbest{  0.0489} &   0.0648 & \ValSecnd{  0.0532} 
			& \ValSecnd{  0.0148} & \Valbest{  0.0138} &   0.0149 & \ValSecnd{  0.0148} \\ 
			& Ps   
			& \Valbest{  1.00} & \ValSecnd{  0.99} & \ValSecnd{  0.99} & \Valbest{  1.00} 
			& \ValSecnd{  0.99} & \Valbest{  1.00} & \ValSecnd{  0.99} & \Valbest{  1.00} 
			& \ValSecnd{  0.94} & \Valbest{  0.96} & \ValSecnd{  0.94} & \ValSecnd{  0.94} \\ 
			& MPSNR 
			&   41.47 & \Valbest{  42.26} & \ValSecnd{  41.63} &   41.49 
			&   41.05 & \Valbest{  42.74} &   40.95 & \ValSecnd{  42.48} 
			& \ValSecnd{  37.42} &   30.36 & \ValSecnd{  37.42} & \Valbest{  37.43} \\ 
			& MSSIM 
			&   0.9734 & \Valbest{  0.9791} & \ValSecnd{  0.9744} &   0.9735 
			&   0.9631 & \Valbest{  0.9811} &   0.9625 & \ValSecnd{  0.9759} 
			& \ValSecnd{  0.9677} &   0.8906 & \ValSecnd{  0.9677} & \Valbest{  0.9679} \\ 
			\cmidrule(lr){1-2}\cmidrule(lr){3-6}\cmidrule(lr){7-10}\cmidrule(lr){11-14}
			
			\multirow{6}{*}{Case 6} 
			& Setup 
			& \begin{tabular}{c} \hspace{-4mm} $\lambda_{1} = 10^{0}$, \hspace{-4mm} \\\hspace{-4mm} $\varepsilon = 0.95$ \hspace{-4mm}\end{tabular} & \begin{tabular}{c} \hspace{-4mm} $\lambda_{1} = 10^{0}$, \hspace{-4mm} \\\hspace{-4mm} $\lambda_{2} = 10^{0}$, \hspace{-4mm} \\\hspace{-4mm} $\varepsilon = 0.95$ \hspace{-4mm}\end{tabular} & \begin{tabular}{c} \hspace{-4mm} $\lambda_{1} = 10^{0}$, \hspace{-4mm} \\\hspace{-4mm} $\lambda_{2} = 10^{-2}$, \hspace{-4mm} \\\hspace{-4mm} $\varepsilon = 0.95$ \hspace{-4mm}\end{tabular} & \begin{tabular}{c} \hspace{-4mm} $\lambda_{1} = 10^{0}$, \hspace{-4mm} \\\hspace{-4mm} $\lambda_{2} = 10^{-1}$, \hspace{-4mm} \\\hspace{-4mm} $\varepsilon = 0.95$ \hspace{-4mm}\end{tabular} 
			& \begin{tabular}{c} \hspace{-4mm} $\lambda_{1} = 10^{0}$, \hspace{-4mm} \\\hspace{-4mm} $\varepsilon = 0.98$ \hspace{-4mm}\end{tabular} & \begin{tabular}{c} \hspace{-4mm} $\lambda_{1} = 10^{-1}$, \hspace{-4mm} \\\hspace{-4mm} $\lambda_{2} = 10^{1}$, \hspace{-4mm} \\\hspace{-4mm} $\varepsilon = 0.98$ \hspace{-4mm}\end{tabular} & \begin{tabular}{c} \hspace{-4mm} $\lambda_{1} = 10^{0}$, \hspace{-4mm} \\\hspace{-4mm} $\lambda_{2} = 10^{-1}$, \hspace{-4mm} \\\hspace{-4mm} $\varepsilon = 0.98$ \hspace{-4mm}\end{tabular} & \begin{tabular}{c} \hspace{-4mm} $\lambda_{1} = 10^{0}$, \hspace{-4mm} \\\hspace{-4mm} $\lambda_{2} = 10^{0}$, \hspace{-4mm} \\\hspace{-4mm} $\varepsilon = 0.98$ \hspace{-4mm}\end{tabular} 
			& \begin{tabular}{c} \hspace{-4mm} $\lambda_{1} = 10^{-1}$, \hspace{-4mm} \\\hspace{-4mm} $\varepsilon = 0.98$ \hspace{-4mm}\end{tabular} & \begin{tabular}{c} \hspace{-4mm} $\lambda_{1} = 10^{-1}$, \hspace{-4mm} \\\hspace{-4mm} $\lambda_{2} = 10^{0}$, \hspace{-4mm} \\\hspace{-4mm} $\varepsilon = 0.98$ \hspace{-4mm}\end{tabular} & \begin{tabular}{c} \hspace{-4mm} $\lambda_{1} = 10^{-1}$, \hspace{-4mm} \\\hspace{-4mm} $\lambda_{2} = 10^{-1}$, \hspace{-4mm} \\\hspace{-4mm} $\varepsilon = 0.98$ \hspace{-4mm}\end{tabular} & \begin{tabular}{c} \hspace{-4mm} $\lambda_{1} = 10^{-1}$, \hspace{-4mm} \\\hspace{-4mm} $\lambda_{2} = 10^{-1}$, \hspace{-4mm} \\\hspace{-4mm} $\varepsilon = 0.98$ \hspace{-4mm}\end{tabular} \\ 
			& SRE  
			&   14.17 & \Valbest{  14.39} &   14.03 & \ValSecnd{  14.19} 
			&    9.92 & \Valbest{  12.49} &    9.33 & \ValSecnd{  11.97} 
			&    8.13 & \Valbest{  10.02} &    8.59 & \ValSecnd{   8.69} \\ 
			& RMSE 
			& \ValSecnd{  0.0500} & \Valbest{  0.0489} &   0.0507 & \ValSecnd{  0.0500} 
			&   0.0655 & \Valbest{  0.0525} &   0.0689 & \ValSecnd{  0.0543} 
			&   0.0175 & \Valbest{  0.0144} &   0.0166 & \ValSecnd{  0.0165} \\ 
			& Ps   
			& \Valbest{  0.99} & \Valbest{  0.99} & \Valbest{  0.99} & \Valbest{  0.99} 
			& \ValSecnd{  0.99} & \ValSecnd{  0.99} &   0.98 & \Valbest{  1.00} 
			&   0.90 & \Valbest{  0.95} & \ValSecnd{  0.93} & \ValSecnd{  0.93} \\ 
			& MPSNR 
			&   37.86 & \Valbest{  38.82} &   37.78 & \ValSecnd{  38.04} 
			&   37.76 & \ValSecnd{  38.41} &   37.76 & \Valbest{  39.70} 
			&   33.25 & \Valbest{  34.34} &   33.87 & \ValSecnd{  34.18} \\ 
			& MSSIM 
			&   0.9407 & \Valbest{  0.9563} &   0.9396 & \ValSecnd{  0.9435} 
			&   0.9219 & \ValSecnd{  0.9559} &   0.9222 & \Valbest{  0.9573} 
			&   0.9109 & \Valbest{  0.9496} &   0.9227 & \ValSecnd{  0.9294} \\ 
			\cmidrule(lr){1-2}\cmidrule(lr){3-6}\cmidrule(lr){7-10}\cmidrule(lr){11-14}
			
			\multirow{6}{*}{Case 7} 
			& Setup 
			& \begin{tabular}{c} \hspace{-4mm} $\lambda_{1} = 10^{0}$, \hspace{-4mm} \\\hspace{-4mm} $\varepsilon = 0.98$ \hspace{-4mm}\end{tabular} & \begin{tabular}{c} \hspace{-4mm} $\lambda_{1} = 10^{0}$, \hspace{-4mm} \\\hspace{-4mm} $\lambda_{2} = 10^{0}$, \hspace{-4mm} \\\hspace{-4mm} $\varepsilon = 0.98$ \hspace{-4mm}\end{tabular} & \begin{tabular}{c} \hspace{-4mm} $\lambda_{1} = 10^{0}$, \hspace{-4mm} \\\hspace{-4mm} $\lambda_{2} = 10^{-2}$, \hspace{-4mm} \\\hspace{-4mm} $\varepsilon = 0.98$ \hspace{-4mm}\end{tabular} & \begin{tabular}{c} \hspace{-4mm} $\lambda_{1} = 10^{0}$, \hspace{-4mm} \\\hspace{-4mm} $\lambda_{2} = 10^{-1}$, \hspace{-4mm} \\\hspace{-4mm} $\varepsilon = 0.98$ \hspace{-4mm}\end{tabular} 
			& \begin{tabular}{c} \hspace{-4mm} $\lambda_{1} = 10^{0}$, \hspace{-4mm} \\\hspace{-4mm} $\varepsilon = 0.98$ \hspace{-4mm}\end{tabular} & \begin{tabular}{c} \hspace{-4mm} $\lambda_{1} = 10^{0}$, \hspace{-4mm} \\\hspace{-4mm} $\lambda_{2} = 10^{1}$, \hspace{-4mm} \\\hspace{-4mm} $\varepsilon = 0.98$ \hspace{-4mm}\end{tabular} & \begin{tabular}{c} \hspace{-4mm} $\lambda_{1} = 10^{0}$, \hspace{-4mm} \\\hspace{-4mm} $\lambda_{2} = 10^{1}$, \hspace{-4mm} \\\hspace{-4mm} $\varepsilon = 0.98$ \hspace{-4mm}\end{tabular} & \begin{tabular}{c} \hspace{-4mm} $\lambda_{1} = 10^{0}$, \hspace{-4mm} \\\hspace{-4mm} $\lambda_{2} = 10^{1}$, \hspace{-4mm} \\\hspace{-4mm} $\varepsilon = 0.98$ \hspace{-4mm}\end{tabular} 
			& \begin{tabular}{c} \hspace{-4mm} $\lambda_{1} = 10^{-1}$, \hspace{-4mm} \\\hspace{-4mm} $\varepsilon = 0.98$ \hspace{-4mm}\end{tabular} & \begin{tabular}{c} \hspace{-4mm} $\lambda_{1} = 10^{-1}$, \hspace{-4mm} \\\hspace{-4mm} $\lambda_{2} = 10^{0}$, \hspace{-4mm} \\\hspace{-4mm} $\varepsilon = 0.98$ \hspace{-4mm}\end{tabular} & \begin{tabular}{c} \hspace{-4mm} $\lambda_{1} = 10^{-1}$, \hspace{-4mm} \\\hspace{-4mm} $\lambda_{2} = 10^{-1}$, \hspace{-4mm} \\\hspace{-4mm} $\varepsilon = 0.95$ \hspace{-4mm}\end{tabular} & \begin{tabular}{c} \hspace{-4mm} $\lambda_{1} = 10^{-1}$, \hspace{-4mm} \\\hspace{-4mm} $\lambda_{2} = 10^{-1}$, \hspace{-4mm} \\\hspace{-4mm} $\varepsilon = 0.95$ \hspace{-4mm}\end{tabular} \\ 
			& SRE  
			&   13.34 & \Valbest{  13.67} &   13.27 & \ValSecnd{  13.35} 
			&    8.26 & \Valbest{  13.61} &    8.38 & \ValSecnd{  11.28} 
			&    8.12 & \Valbest{  10.55} &    7.97 & \ValSecnd{   8.32} \\ 
			& RMSE 
			&   0.0548 & \Valbest{  0.0531} &   0.0551 & \ValSecnd{  0.0547} 
			&   0.0767 & \Valbest{  0.0466} &   0.0759 & \ValSecnd{  0.0588} 
			&   0.0173 & \Valbest{  0.0136} &   0.0177 & \ValSecnd{  0.0171} \\ 
			& Ps   
			& \Valbest{  0.98} & \Valbest{  0.98} & \Valbest{  0.98} & \Valbest{  0.98} 
			&   0.95 & \Valbest{  1.00} &   0.98 & \ValSecnd{  0.99} 
			&   0.91 & \Valbest{  0.96} &   0.91 & \ValSecnd{  0.92} \\ 
			& MPSNR 
			&   37.02 & \Valbest{  38.22} &   36.94 & \ValSecnd{  37.25} 
			&   36.88 & \Valbest{  39.74} &   37.62 & \ValSecnd{  38.83} 
			& \Valbest{  33.51} &   32.38 &   32.70 & \ValSecnd{  33.19} \\ 
			& MSSIM 
			&   0.9265 & \Valbest{  0.9502} &   0.9250 & \ValSecnd{  0.9310} 
			&   0.9036 & \Valbest{  0.9651} &   0.9317 & \ValSecnd{  0.9612} 
			& \ValSecnd{  0.9141} & \Valbest{  0.9250} &   0.8999 &   0.9117 \\ 
			\cmidrule(lr){1-2}\cmidrule(lr){3-6}\cmidrule(lr){7-10}\cmidrule(lr){11-14}
			
			\multirow{6}{*}{Case 8} 
			& Setup 
			& \begin{tabular}{c} \hspace{-4mm} $\lambda_{1} = 10^{0}$, \hspace{-4mm} \\\hspace{-4mm} $\varepsilon = 0.95$ \hspace{-4mm}\end{tabular} & \begin{tabular}{c} \hspace{-4mm} $\lambda_{1} = 10^{0}$, \hspace{-4mm} \\\hspace{-4mm} $\lambda_{2} = 10^{0}$, \hspace{-4mm} \\\hspace{-4mm} $\varepsilon = 0.95$ \hspace{-4mm}\end{tabular} & \begin{tabular}{c} \hspace{-4mm} $\lambda_{1} = 10^{0}$, \hspace{-4mm} \\\hspace{-4mm} $\lambda_{2} = 10^{-2}$, \hspace{-4mm} \\\hspace{-4mm} $\varepsilon = 0.95$ \hspace{-4mm}\end{tabular} & \begin{tabular}{c} \hspace{-4mm} $\lambda_{1} = 10^{0}$, \hspace{-4mm} \\\hspace{-4mm} $\lambda_{2} = 10^{-1}$, \hspace{-4mm} \\\hspace{-4mm} $\varepsilon = 0.95$ \hspace{-4mm}\end{tabular} 
			& \begin{tabular}{c} \hspace{-4mm} $\lambda_{1} = 10^{0}$, \hspace{-4mm} \\\hspace{-4mm} $\varepsilon = 0.98$ \hspace{-4mm}\end{tabular} & \begin{tabular}{c} \hspace{-4mm} $\lambda_{1} = 10^{0}$, \hspace{-4mm} \\\hspace{-4mm} $\lambda_{2} = 10^{1}$, \hspace{-4mm} \\\hspace{-4mm} $\varepsilon = 0.95$ \hspace{-4mm}\end{tabular} & \begin{tabular}{c} \hspace{-4mm} $\lambda_{1} = 10^{0}$, \hspace{-4mm} \\\hspace{-4mm} $\lambda_{2} = 10^{-1}$, \hspace{-4mm} \\\hspace{-4mm} $\varepsilon = 0.95$ \hspace{-4mm}\end{tabular} & \begin{tabular}{c} \hspace{-4mm} $\lambda_{1} = 10^{-1}$, \hspace{-4mm} \\\hspace{-4mm} $\lambda_{2} = 10^{0}$, \hspace{-4mm} \\\hspace{-4mm} $\varepsilon = 0.98$ \hspace{-4mm}\end{tabular} 
			& \begin{tabular}{c} \hspace{-4mm} $\lambda_{1} = 10^{-1}$, \hspace{-4mm} \\\hspace{-4mm} $\varepsilon = 0.95$ \hspace{-4mm}\end{tabular} & \begin{tabular}{c} \hspace{-4mm} $\lambda_{1} = 10^{-1}$, \hspace{-4mm} \\\hspace{-4mm} $\lambda_{2} = 10^{0}$, \hspace{-4mm} \\\hspace{-4mm} $\varepsilon = 0.95$ \hspace{-4mm}\end{tabular} & \begin{tabular}{c} \hspace{-4mm} $\lambda_{1} = 10^{-1}$, \hspace{-4mm} \\\hspace{-4mm} $\lambda_{2} = 10^{-2}$, \hspace{-4mm} \\\hspace{-4mm} $\varepsilon = 0.95$ \hspace{-4mm}\end{tabular} & \begin{tabular}{c} \hspace{-4mm} $\lambda_{1} = 10^{-1}$, \hspace{-4mm} \\\hspace{-4mm} $\lambda_{2} = 10^{-2}$, \hspace{-4mm} \\\hspace{-4mm} $\varepsilon = 0.95$ \hspace{-4mm}\end{tabular} \\ 
			& SRE  
			&   12.20 & \Valbest{  12.46} &   12.12 & \ValSecnd{  12.21} 
			&    7.43 & \Valbest{  10.39} &    6.26 & \ValSecnd{   9.40} 
			&    7.25 & \Valbest{   8.92} &    6.64 & \ValSecnd{   7.32} \\ 
			& RMSE 
			&   0.0621 & \Valbest{  0.0604} &   0.0623 & \ValSecnd{  0.0619} 
			&   0.0827 & \Valbest{  0.0630} &   0.0922 & \ValSecnd{  0.0714} 
			&   0.0188 & \Valbest{  0.0160} &   0.0199 & \ValSecnd{  0.0186} \\ 
			& Ps   
			& \Valbest{  0.97} & \Valbest{  0.97} & \Valbest{  0.97} & \Valbest{  0.97} 
			&   0.90 & \Valbest{  0.99} &   0.84 & \ValSecnd{  0.96} 
			& \ValSecnd{  0.88} & \Valbest{  0.92} &   0.85 & \ValSecnd{  0.88} \\ 
			& MPSNR 
			&   35.74 & \Valbest{  36.93} &   35.67 & \ValSecnd{  35.97} 
			&   35.82 & \ValSecnd{  37.22} &   35.18 & \Valbest{  37.76} 
			&   31.41 & \Valbest{  31.82} &   31.34 & \ValSecnd{  31.62} \\ 
			& MSSIM 
			&   0.9067 & \Valbest{  0.9371} &   0.9055 & \ValSecnd{  0.9124} 
			&   0.8875 & \Valbest{  0.9447} &   0.8707 & \ValSecnd{  0.9405} 
			&   0.8736 & \Valbest{  0.9192} &   0.8727 & \ValSecnd{  0.8796} \\  
			\bottomrule
		\end{tabular}
	}
\end{table*}

Figures~\ref{fig:synth_legendre_0.1_0_0},~\ref{synth_0.05_0.05_0.05}, and~\ref{fig:synth_legendre_200_noniid_Noniid_0.05_0.05} show the estimated abundance maps for \textit{Synth 1} in Case 4, for \textit{Synth 2} in Case 5, and for \textit{Synth 3} in Case 8. 
All the abundance maps of CLSUnSAL, RSSUn-TV, LGSU, and RDSWSU include residual noise in Cases 2, 5, and 8 (Figs.~\ref{fig:synth_legendre_0.1_0_0},~\ref{synth_0.05_0.05_0.05}, and~\ref{fig:synth_legendre_200_noniid_Noniid_0.05_0.05} (b), (d), (e), and (h)).
JSTV remained the noise (Fig~\ref{fig:synth_legendre_0.1_0_0} (c)) or obtained the over-smooth abundance maps (the second abundance map for \textit{Synth 2} in Fig.~\ref{synth_0.05_0.05_0.05} (c)) due to the difficulty of adjusting the parameters balancing the sparsity and piecewise-smoothness of abundance maps. For \textit{Synth 3} (Fig.~\ref{fig:synth_legendre_200_noniid_Noniid_0.05_0.05} (c)), JSTV obtained the significantly lower abundance maps. This may be due to an algorithm issue.
The abundance maps of MdLRR are relatively exact in Case 2 (Fig.~\ref{fig:synth_legendre_0.1_0_0} (i)), but are affected by sparse and stripe noise in Cases 5 and 8 (Figs.~\ref{synth_0.05_0.05_0.05} and~\ref{fig:synth_legendre_200_noniid_Noniid_0.05_0.05} (i)).
UnDIP and EGU-Net erroneously estimated that the abundances were high for the endmembers that are not present in the HS images due to the insufficient ability to capture the sparsity of abundance maps, resulting in the generation of inappropriate abundance maps (Figs.~\ref{fig:synth_legendre_0.1_0_0},~\ref{synth_0.05_0.05_0.05}, and~\ref{fig:synth_legendre_200_noniid_Noniid_0.05_0.05} (f) and (g)).
In particular, since all the existing methods do not account for stripe noise, they produced the abundance maps that are strongly affected by stripe noise (see Figs.~\ref{synth_0.05_0.05_0.05} and~\ref{fig:synth_legendre_200_noniid_Noniid_0.05_0.05} (b), (d)-(g)) or the smoother results than true abundance (Figs.~\ref{synth_0.05_0.05_0.05} and~\ref{fig:synth_legendre_200_noniid_Noniid_0.05_0.05} (c)).
In contrast, \Ourss accurately estimated abundance maps regardless of what type of noise contaminates HS images (Figs.~\ref{fig:synth_legendre_0.1_0_0},~\ref{synth_0.05_0.05_0.05}, and~\ref{fig:synth_legendre_200_noniid_Noniid_0.05_0.05} (j)-(l)).

Figures~\ref{fig:synth_legendre_HSI_0.1_0_0},~\ref{synth_HSI_0.05_0.05_0.05}, and~\ref{fig:synth_legendre_200_noniid200_noniid_HSI_0.1_0.1} display the reconstructed HS images for \textit{Synth 1} in Case 2, for \textit{Synth 2} in Case 5, and for \textit{Synth 3} in Case 8, respectively.
All the existing methods resulted in Gaussian noise remaining in the reconstructed HS images in Case 2 (Fig.~\ref{fig:synth_legendre_HSI_0.1_0_0} (c)-(j)).
Moreover, in the reconstructed HS images in Cases 5 and 8 (Figs.~\ref{synth_HSI_0.05_0.05_0.05} and~\ref{fig:synth_legendre_200_noniid200_noniid_HSI_0.1_0.1} (c)-(j)), we can see that residual stripe noise remains.
On the other hand, \Ourss produced clean reconstructed HS images due to the image-domain regularizations.

\subsection{Experiments With Real HS Images}
Tables~\ref{tab:result_jasper},~\ref{tab:result_samson}, and~\ref{tab:result_real_Urban} show the SRE, RMSE, Ps, MPSNR, and MSSIM results for \textit{Jasper Ridge}, \textit{Samson}, for \textit{Urban}, respectively. 
The best and second-best results are highlighted in bold and underlined, respectively. 
In Case 1 of the \textit{Jasper Ridge} experiments, the unmixing performances of CLSUnSAL, RSSUn-TV, LGSU, RDSWSU, MdLRR, and \Ourss were almost equal.
For \textit{Urban}, the unmixing performance of JSTV was degraded similar to the results of \textit{Synth 3}. This may also have been an issue with the algorithm.
However, CLSUnSAL, JSTV, RSSUn-TV, LGSU, RDSWSU, and MdLRR performed worse in the other cases than \Ours.
Since UnDIP and EGU-Net cannot capture the sparsity of abundance maps, their unmixing performance was low regardless of whether HS images are synthetic or real, the size of HS images, and the size of endmember libraries.
CLSUnSAL achieved the best SRE and RMSE values for \textit{Samson} in almost all the cases.
RDSWSU achieved the best results for \textit{Urban} in almost all the cases. This is because the appropriate weight values for abundance maps were computed due to the proper segmentation.
In contrast, \Ourss achieved the best and second best SRE, RMSE, Ps, MPSNR, and MSSIM values in almost all the cases where the HS image is contaminated by noise that cannot be handled by the existing methods (Cases 3, 4, 5, 6, and 8 for CLSUnSAL, RSSUn-TV, LGSU, RDSWSU, and MdLRR, and Cases 5, 6, and 8 for JSTV). 
In the comparison of the image-domain regularizations, HTV performed better in almost all cases, and HSSTV performed better in the \textit{Samson} experiments when HS images were contaminated with non-i.i.d. Gaussian noise (Cases 7 and 8).

Figures~\ref{real_0.1_0_0},~\ref{real_samson_0.1_0.05_0.05}, and~\ref{real_urban_noniid_urban_Noniid_0.05_0.05} show the estimated abundance maps for \textit{Jasper Ridge} in Case 2, for \textit{Samson} in Case 6, and for \textit{Urban} in Case 8, respectively. 
Although CLSUnSAL, LGSU, RDSWSU, and MdLRR achieved good SRE, RMSE, and MPSNR in Case 2 of the real data experiments, they yielded the abundance maps with residual noise (see Fig.~\ref{real_0.1_0_0} (b), (d), (e), (h), and (i)).
JSTV obtained the over-smooth abundance maps (Figs.~\ref{real_0.1_0_0} and~\ref{real_samson_0.1_0.05_0.05} (c)).
In Cases 6 and 8, the abundance maps estimated by all the existing methods except JSTV include noise, especially stripe noise (see Figs.~\ref{real_samson_0.1_0.05_0.05} and~\ref{real_urban_noniid_urban_Noniid_0.05_0.05}, (b) (d)-(g)). 
This is because they do not handle the stripe noise.
In contrast, \Ourss exactly estimated the abundance maps even under the conditions assumed by the existing methods, e.g., when the observed HS images are only contaminated by Gaussian noise (Fig.~\ref{real_0.1_0_0}). 
Furthermore, \Ourss estimated the abundance maps by removing not only Gaussian and sparse noise but also stripe noise cleanly.

Figures~\ref{real_HSI_0.1_0_0},~\ref{real_samson_HSI_0.1_0.05_0.05}, and~\ref{real_urban_noniid_HSI_Noniid_0.05_0.05} display the reconstructed HS images for \textit{Jasper Ridge} in Case 2, for \textit{Samson} in Case 6, and for \textit{Urban} in Case 8.
All the existing methods resulted in noise remaining in the reconstructed HS images (Figs.~\ref{real_HSI_0.1_0_0},~\ref{real_samson_HSI_0.1_0.05_0.05}, and~\ref{real_urban_noniid_HSI_Noniid_0.05_0.05} (c)-(j)).
In particular, they cannot handle stripe noise, and thus did not completely remove it in Case 6 (Figs.~\ref{real_samson_HSI_0.1_0.05_0.05} and~\ref{real_urban_noniid_HSI_Noniid_0.05_0.05} (c)-(j)).
On the other hand, \Ourss reconstructed the HS image cleanly (Figs.~\ref{real_HSI_0.1_0_0} and~\ref{real_samson_HSI_0.1_0.05_0.05} (k)-(m)). 
This verifies the effectiveness of the image-domain regularization.

\subsection{Comparison of Computational Cost}
We measured the actual running times on a Windows 11 computer with an Intel Core i9-13900 1.0GHz processor, 32GB of RAM, and NVIDIA GeForce RTX 4090. 
In addition, we used MATLAB (R2023b), python 3.8, and python 3.7 for CLSUnSAL, JSTV, RSSUn-TV, RDSWSU, MdLRR, and our method, for UnDIP, and for EGU-Net, respectively.
The stopping criteria of the comparison methods were set to the values recommended in the papers.

Table~\ref{tab:results_times} shows the averages of the running times in all the noise cases for \textit{Jasper Ridge}, \textit{Samson}, and \textit{Urban}. The running time of \Ourss varies depending on which regularization was employed. When using HTV, SSTV, and HSSTV, \Ourss took 30 seconds, 50 seconds, and 13 minutes for \textit{Jasper Ridge}, \textit{Samson}, and \textit{Urban}, respectively. Compared with the existing methods, \Ourss is faster than UnDIP, is the same as LGSU and JSTV, and is slower than CLSUnSAL, RSSUn-TV, EGU-Net, RDSWSU, and MdLRR.

\subsection{Convergence Analysis}
In addition, we experimentally analyzed the convergence of our method.
Figure~\ref{fig:convergence_analysis_all} plots the relative error of abundance maps: $\|\MatAbun^{(\NumIter+1)} - \MatAbun^{(\NumIter)}\|_{F}/\|\MatAbun^{(\NumIter)}\|_F$, the objective function values, the Frobenius distance between $\MatHSIObsev$ and $\MatEndmember\MatAbun^{(\NumIter)}+\MatSpar^{(\NumIter)}+\MatStripe^{(\NumIter)}$, the $\ell_{1}$ norm of $\MatSpar^{(\NumIter)}$, and the mean absolute values (MAV) of $\mathbf{D}_{v}(\MatStripe^{(\NumIter)})$ for \textit{Jasper Ridge} and \textit{Samson}. 
The relative error of abundance maps decreased (Fig.~\ref{fig:convergence_analysis_all} (a)). 
While getting larger as the number of iterations increases, the objective function value asymptotically approaches a certain value (Fig.~\ref{fig:convergence_analysis_all} (b)).
This is often found when solving optimization problems involving hard constraints, such as a data fidelity constraint $\MatEndmember\MatAbun + \MatSpar + \MatStripe \in \SetConsFidel$, a sparsity constraint $\MatSpar \in \SetConsSpar$, and a flatness constraint $\mathbf{D}_v(\MatStripe)=\MatZero$.
The $\ell_{2}$ distance and the MAV become smaller, where we can see that the variables updated by P-PDS approach the solution of our constrained convex optimization.

\subsection{Ablation Experiments}
To demonstrate the effectiveness of the image-domain regularization (the third term of Eq.~\eqref{eq_problem}), we compared \Ourss performance with the performance when the image-domain regularization was removed (referred to as \Ourss (--)). The hyperparameters $\lambda_{1}$, $\lambda_{3}$, $\ParamFidel$, and $\ParamConsSpar$ were set to the same as in \Ours.

Tables~\ref{tab:results_ablation_synth} and~\ref{tab:results_ablation_synth} show the SRE, RMSE, Ps, MPSNR, and MSSIM results of the ablation experiments for the synthetic and real datasets, respectively.
The best and second-best results are highlighted in bold and underlined, respectively.
\Ourss with the image-domain regularization was superior to \Ourss without the image-domain regularization.
In particular, the image-domain regularization contributed to an improvement in SRE, RMSE, and Ps and a significant improvement in MPSNR and MSSIM.
This implies that the reconstructed HS image has the desirable spatio-spectral property, resulting in the estimation of more appropriate abundance maps.

\subsection{Summary}
We summarize the experimental discussion as follows.
\begin{itemize}
	\item From the results of experiments in Case 1, Case 2, Case 3, Case 4, and Case 7 and the ablation experiments, we see that image-domain regularizations improve the unmixing performance.
	\item The results of experiments in Case 5, Case 6, and Case 8 verify that \Ourss accurately estimates abundance maps if HS images are degraded by various types of noise.
	\item \Ourss achieves good unmixing performance in experiments using both synthetic and real HS images.
\end{itemize}


\section{Conclusion}
\label{sec:conclusion}
In this paper, we have proposed a new method for noise-robust unmixing.
\Ourss adopts the image-domain regularization and explicitly models three types of noises.
We have formulated the unmixing problem as a constrained convex optimization problem that includes the regularization, and have developed the optimization algorithm based on P-PDS.
Experiments on synthetic and real HS images have demonstrated the superiority of \Ourss over existing methods.
\Ourss will have strong impacts on the field of remote sensing, including the estimation of abundance maps from HS images taken in measurement environments with severe degradation.
For future work, we will combine \Ourss with a learning-based and a coarse abundance map-based weighted sparse unmixing approach to realize more noise-robust blind unmixing.

\section*{Acknowledgement}
We would like to thank Ms. Yuki Nagamatsu, who was a student of Tokyo Institute of Technology, for her cooperation in the experiments.



%





\ifCLASSOPTIONcaptionsoff
  \newpage
\fi



%
\bibliographystyle{IEEEtran}



%

%

\begin{IEEEbiography}[{\includegraphics[width=1in,height=1.25in,clip,keepaspectratio]{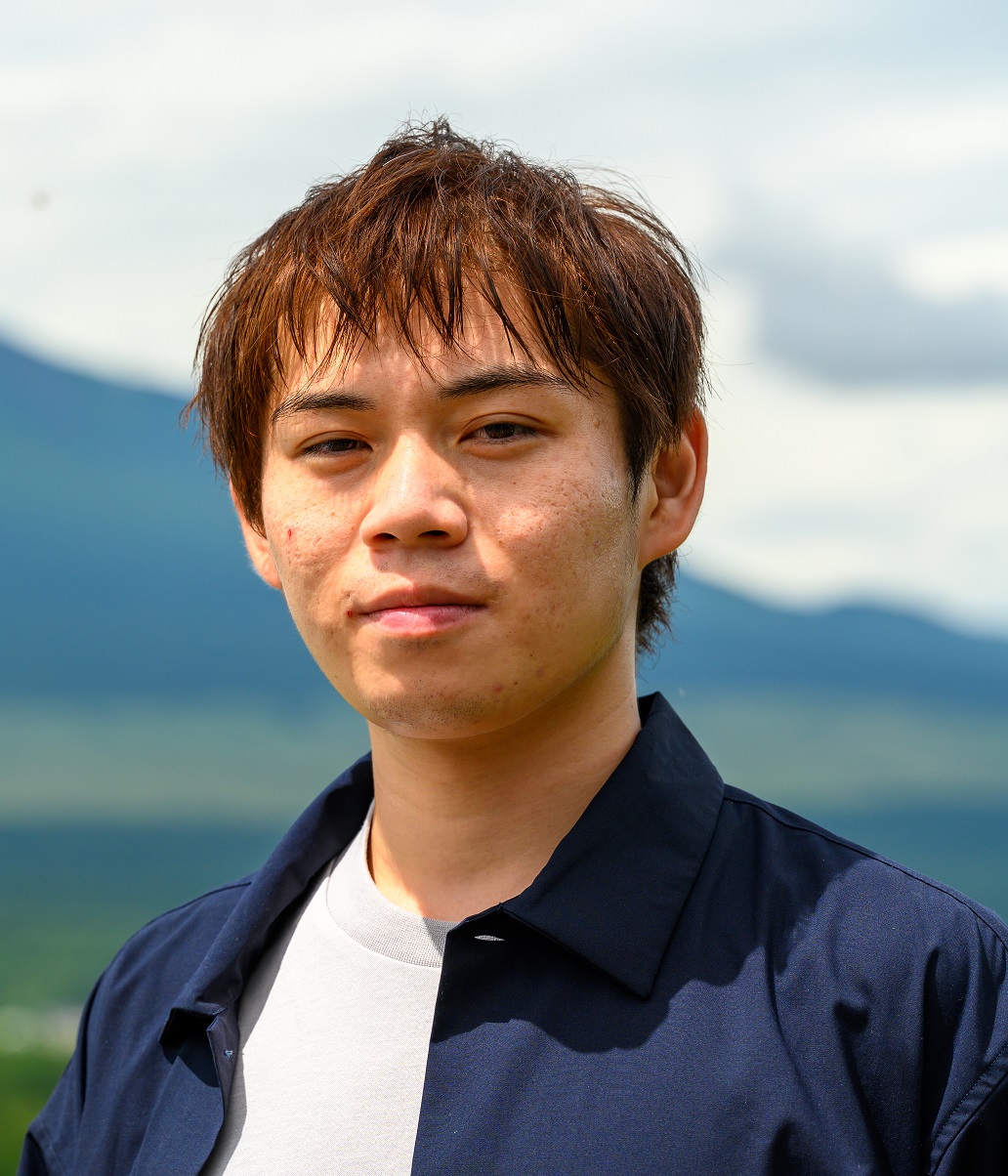}}]{Kazuki Naganuma}
	(S’21) received a B.E. degree and M.E. degree in Information and Computer Sciences in 2020 from the Kanagawa Institute of Technology and from the Tokyo Institute of Technology, respectively.
	
	He is currently pursuing an Ph.D. degree at the Department of Computer Science in the Tokyo Institute of Technology. 
	From April 2023 and October 2023 to present, he is a Research Fellow (DC2) of the Japan Society for the Promotion of Science (JSPS) and a Researcher of ACT-X of the Japan Science and Technology Corporation (JST), Tokyo, Japan. 
	His current research interests are in signal and image processing, remote sensing, and optimization theory.
	
	Mr. Naganuma received the Student Conference Paper Award from IEEE SPS Japan Chapter in 2023.
\end{IEEEbiography}

\begin{IEEEbiography}[{\includegraphics[width=1in,height=1.25in,clip,keepaspectratio]{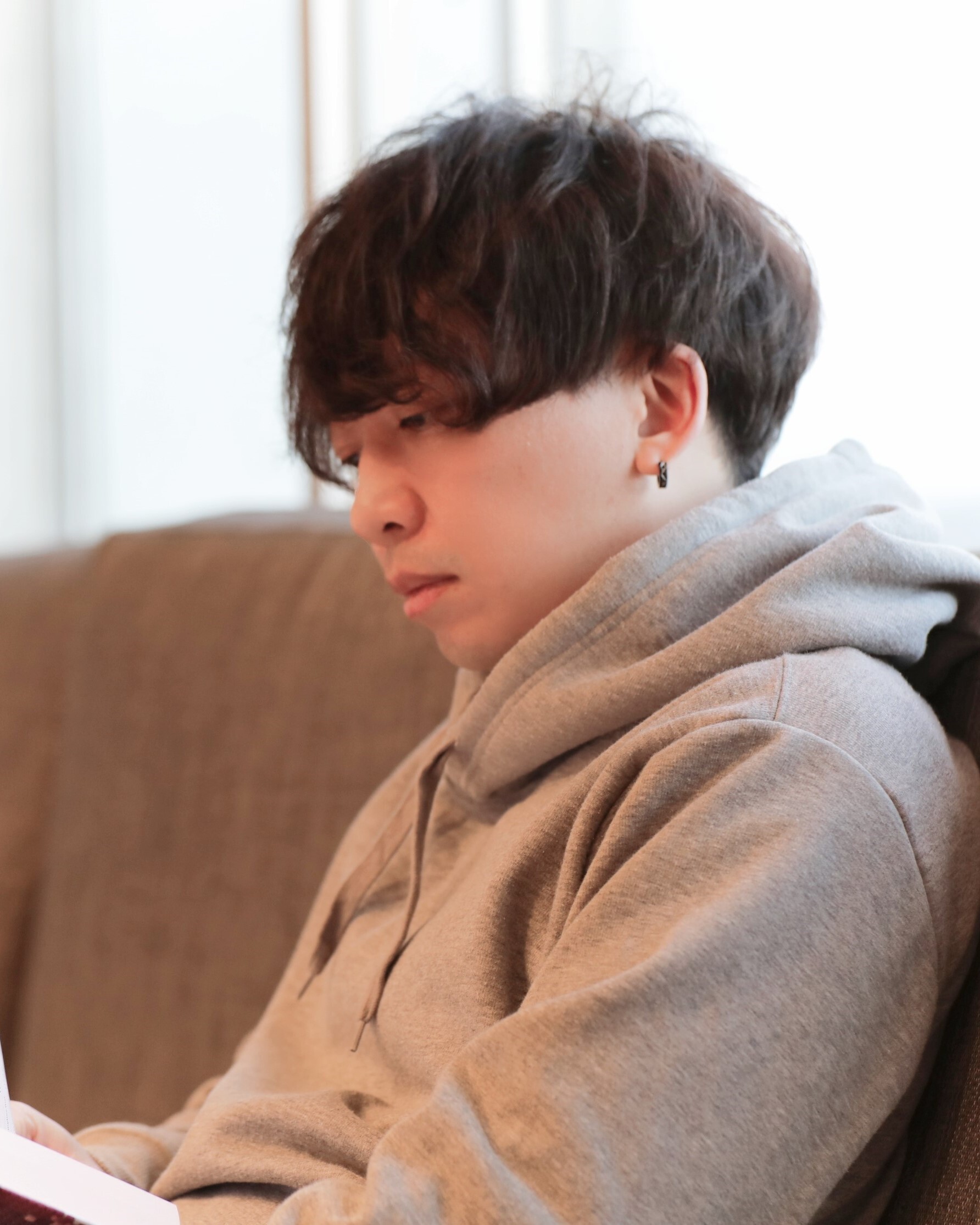}}]{Shunsuke Ono}
	(S’11–M’15–SM'23) received a B.E. degree in Computer Science in 2010 and M.E. and Ph.D. degrees in Communications and Computer Engineering in 2012 and 2014 from the Tokyo Institute of Technology, respectively.
	
	From April 2012 to September 2014, he was a Research Fellow (DC1) of the Japan Society for the Promotion of Science (JSPS). He is currently an Associate Professor in the Department of Computer Science, School of Computing, Tokyo Institute of Technology. From October 2016 to March 2020 and from October 2021 to present, he was/is a Researcher of Precursory Research for Embryonic Science and Technology (PRESTO), Japan Science and Technology Agency (JST), Tokyo, Japan. His research interests include signal processing, image analysis, remote sensing, mathematical optimization, and data science.
	
	Dr. Ono received the Young Researchers’ Award and the Excellent Paper Award from the IEICE in 2013 and 2014, respectively, the Outstanding Student Journal Paper Award and the Young Author Best Paper Award from the IEEE SPS Japan Chapter in 2014 and 2020, respectively, the Funai Research Award from the Funai Foundation in 2017, the Ando Incentive Prize from the Foundation of Ando Laboratory in 2021, the Young Scientists’ Award from MEXT in 2022, and the Outstanding Editorial Board Member Award from IEEE SPS in 2023. He has been an Associate Editor of IEEE TRANSACTIONS ON SIGNAL AND INFORMATION PROCESSING OVER NETWORKS since 2019.
\end{IEEEbiography}





\end{document}